\documentclass[11pt,a4paper]{report}

\usepackage[english]{babel}
\usepackage[mathscr]{euscript}
 \usepackage[cal=boondoxo]{mathalfa}
\usepackage{amssymb}
\usepackage{amsthm}
\usepackage{amsxtra}
\usepackage{bbm}
\usepackage{mathtools}
\usepackage{stmaryrd}
\usepackage[varg]{txfonts} \let \mathbb =\varmathbb 

\usepackage{color}
\usepackage{xcolor}
\usepackage[low-sup]{subdepth}
\usepackage{pgf}
\usepackage{tikz}
\usepackage{tkz-tab}
\usepackage{etoolbox}
\usepackage{caption}
 \usepackage{subcaption}
\usetikzlibrary{arrows,automata}
\usetikzlibrary{calc}
\usetikzlibrary{positioning}
\usepackage{extarrows}
 \usepackage{epic}
 \usepackage{layout}
 \usepackage{alltt}
\usepackage{xypic}
 \input xy
 \xyoption{all}

\usepackage{float}
 
 \tikzset{insert node/.style args={#1 at #2}{
    postaction=decorate,
    decoration={
      markings,
      mark= at position #2
        with
        {
         #1
        }
    }
  }
}

\DeclareMathAlphabet{\mathbfit}{OML}{cmm}{b}{it}

\usepackage[left=1.5in, right=1in, top=1in, bottom=1in, includefoot, headheight=13.6pt]{geometry}

\usepackage{xpatch}

\usepackage{abstract}
\addto\captionsenglish{\renewcommand{\abstractname}{}}    

\makeatletter
\xpatchcmd{\part}{\null\vfil}{\vspace*{.1\textheight}}{}{}

\providecommand{\abstractname}{}
\newenvironment{partwithabstract}
  {\begingroup\let\@endpart\relax\part@withabstract}
  {\endquotation\endgroup\@endpart}
\newcommand{\part@withabstract}{\@dblarg\part@@withabstract}
\def\part@@withabstract[#1]#2{%
  \part[#1]{#2}%
  \vfil
  \begin{center}\bfseries\abstractname\vspace{-.5em}\vspace{\z@}\end{center}
  \quotation
}
\makeatother

\usepackage{makeidx}
\makeindex

\usepackage{fancyhdr}                                     
\newcommand{\CHN}{}
\newcommand{\SHN}{}
\fancypagestyle{plain}{\fancyhf{}\fancyfoot[C]{\bfseries \thepage}}

\newcommand{\clearemptydoublepage}{\newpage{\pagestyle{empty}\cleardoublepage}}

\usepackage[nobottomtitles*]{titlesec}
\newcommand\chapformat[1]{\parbox[b]{0.9\textwidth}{ \filright\MakeUppercase #1}}
\titleformat{\chapter}[display]
     {\sc  \huge}
     {} 
     {0pt} {\titlerule[1pt] \vspace{1ex}%
  }[ \vspace{1ex}{\titlerule[1pt]}]   
\titleformat{\section}[hang]{\sc \filcenter}{\filcenter{\thesection}}{2ex}{\vspace{0ex}}[\vspace{0ex}]         
\titleformat{\subsection}[runin]{}{\thesubsection}{2ex}{\bf}[.]  
\titleformat{\subsubsection}[runin]{\sc \filcenter}{\filcenter{\thesubsubsection}}{2ex}{\vspace{0ex}}[.]  

\theoremstyle{plain}
 \newtheorem{theorem}{Theorem}
 \newtheorem{corollary}[theorem]{Corollary}
 \newtheorem{lemma}[theorem]{Lemma}
 \newtheorem{proposition}[theorem]{Proposition}
 \theoremstyle{definition}
 \newtheorem{definition}[theorem]{Definition}

 \newtheorem{remark}[theorem]{Remark}
 \newtheorem{example}[theorem]{Example}
\theoremstyle{remark}
 \newtheorem*{notation}{Notation}

\newcommand{\CC}{\ensuremath{\mathbbmss{C}}}
\newcommand{\MM}{\ensuremath{\mathbbmss{M}}}

\newcommand{\RR}{\ensuremath{\mathbbmss{R}}} 
\newcommand{\ZZ}{\ensuremath{\mathbbmss{Z}}} 
\newcommand{\HH}{\ensuremath{\mathbbmss{H}}}

\newcommand{\id}{\mathrm{id}}

\newcommand\GL{\operatorname{GL}}
\newcommand\SL{\operatorname{SL}}

\renewcommand{\O}{\operatorname{O}}
\newcommand{\sech}{\operatorname{sech}}
\newcommand{\SO}{\operatorname{SO}}
\newcommand{\End}{\operatorname{End}}

\newcommand{\tr}{\mathsf{tr}}
\newcommand{\Su}{\operatorname{S}}

\newcommand{\Ric}{\mathrm{Ric}}
\newcommand{\Rs}{\mathrm{R}}
\newcommand\grad{\operatorname{grad}}
\newcommand\rot{\operatorname{rot}}
\renewcommand\div{\operatorname{div}}

\renewcommand{\to}{\rightarrow}

\usepackage{amsmath}
\usepackage{accents}

\begin{document}
\pagenumbering{roman}
\setcounter{page}{1}

\begin{titlepage}

\newcommand{\HRule}{\rule{\linewidth}{0.5mm}} 

\center 
 



\HRule \\[0.7cm]
\textsc{\Huge \textbf{Relativity}} \\[0.4cm] 
\HRule \\[1.5cm]
 

Camilo \textsc{Arias Abad} \\[5pt]
Alexander \textsc{Quintero V\'elez}\\[5pt]
Juan Diego \textsc{V\'elez Caicedo}\\[50pt]

\begin{figure}[h]
	\centering
		\includegraphics[scale=0.09]{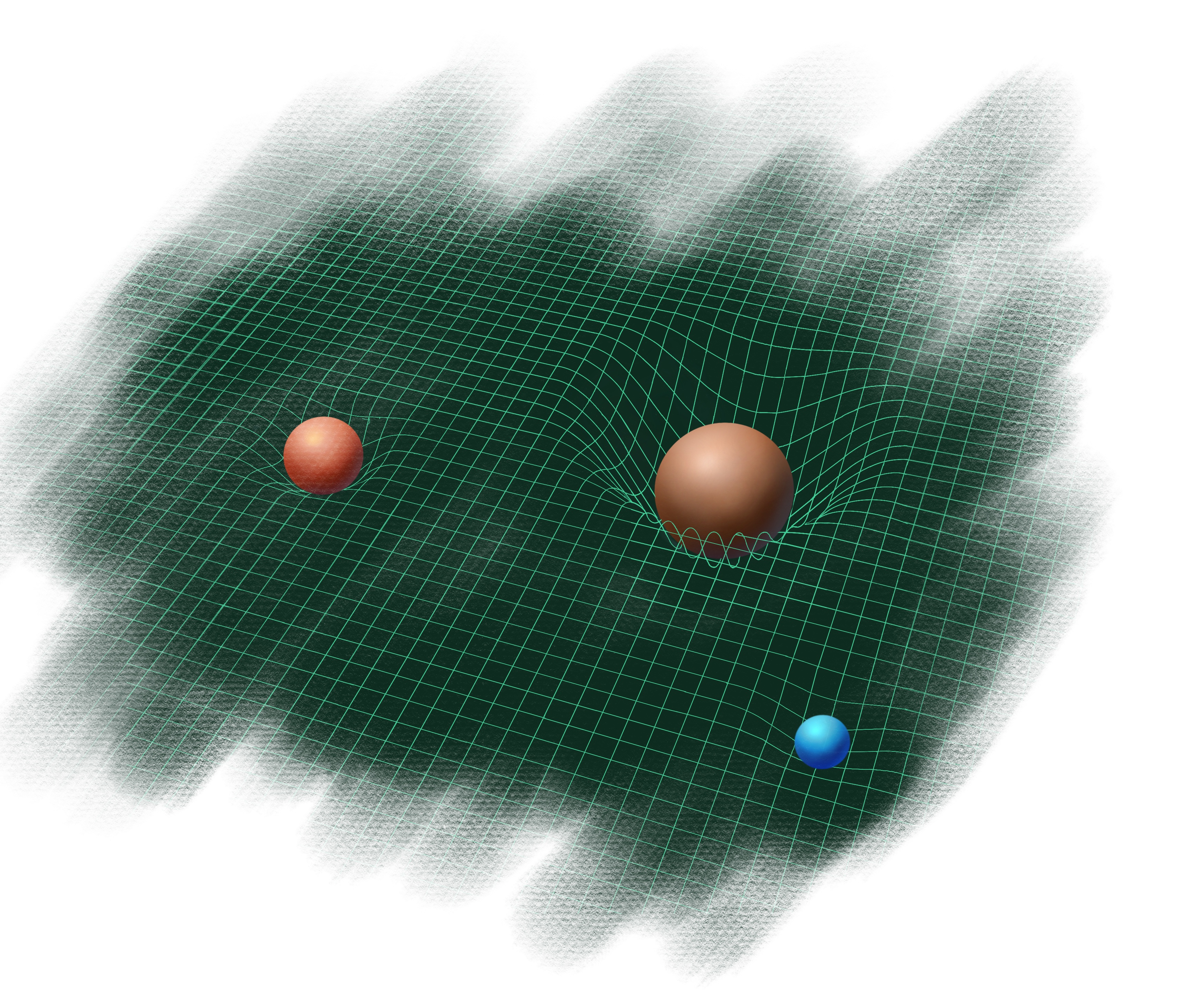}
\end{figure}

\vspace{50pt}





 

\vfill 

\end{titlepage}
 \clearemptydoublepage

\chapter*{Preface}

Starting in 2016, we ran a seminar at the Universidad Nacional de Colombia whose goal was to study Einstein's theory of relativity, and other related parts of physics. These notes are the report of what we learned. They are not written by experts, which we certainly are not, but by enthusiastic students. Our motivation was simply to fulfill our longstanding ambition of understanding Einstein's ideas on gravitation. We are mathematicians by training, interested in geometry. It feels to us as if, after years of cultivating bees, we just discovered honey.

People have been fascinated by relativity for over a century, and many have written about it.
Expositions of Einstein's theory can be found in all possible shapes, colors and levels of detail.
In studying special relativity, we found the books by Rindler \cite{rindler}, 
Schutz \cite{schutz}, as well and the illustrated book by Bais \cite{Bais}, to be specially clear.  For general relativity, we learned a lot from the books by Baez-Munian \cite{baez}, Carroll \cite{carroll}, Hartle \cite{hartle},  Hawking \cite{hawking}, Wald \cite{wald} and Weinberg 
\cite{weinberg}. The classical, and probably most complete reference for the
subject, is the book by Misner, Thorne and Wheeler \cite{missner}. The book by Choquet-Bruhat \cite{choquet}
is excellent for more mathematical aspects. Einstein' original papers \cite{SR} and \cite{GR} are amazing, and easier to read than we expected.

 As always, the internet was our best source of information.
The uncountable number of talks, lectures, discussions, blog posts, images and animations provided endless entertaintment, for which we are grateful. Lecture notes from courses in places far away often contained the explanation we were looking for. We found those by Tong \cite{Tong}, Blau \cite{Blau}, and Baez  \cite{baezbueno} to be particularly beautiful.

Differential geometry, the mathematics of relativity, is older than Einstein's gravity.
Even though it has not fascinated people quite to the same degree, excellent texts have been written about it.
We recommend those by Boothby \cite{boothby}, Tu \cite{Tu}, Do Carmo \cite{Docarmo}, Guillemin-Pollack \cite{GP}, Hirsch \cite{Hirsch}, Jost \cite{Jost}, Madsen-Thornehave \cite{madsen}, Morita \cite{morita}, Nakahara \cite{Nakahara}, O'neill \cite{Oneil}, Taubes \cite{taubs} and Warner \cite{warner}.

Clearly, there is no hole in the literature for these notes to fill. There are, however, differences in emphasis and notation between mathematicians and physicists, which sometimes make the road to relativity slower than it could be. Our hope is to provide, for a reader that shares our enthusiasm, as well as the weaknesses and strengths of a mathematical education, a path that is more familiar at some places. This is meant to be an introductory text, which explains in detail the fundamental ideas of the theory, and works out the most important examples and consequences. If you enjoy reading it half as much as we enjoyed writing it, we will have succeeded. If not, we recommend all of the sources above, because, even if you have to get there through a  long road, general relativity is one of the great stories that our species has to offer.
   \clearemptydoublepage
 
\tableofcontents
   \clearemptydoublepage
   
  
\chapter*{Introduction\markboth{Introduction}{}}\addcontentsline{toc}{chapter}{Introduction}

\section*{Special Relativity}

Special relativity is a theory about the relationship between time and space. Newton, as well as most people who have not studied physics, imagined this relationship to be captured by the following image

\begin{figure}[H]
\centering
\includegraphics[scale=0.33]{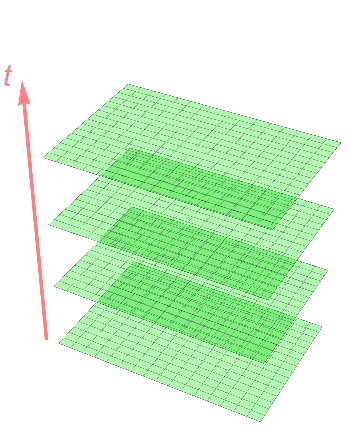}
\caption{Planes of simultaneous events.}
\end{figure}

\begin{figure}[H]
\centering
\includegraphics[scale=0.48]{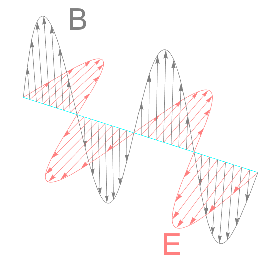}
\caption{Light is an electromagnetic wave.}\label{EMwave}
\end{figure}

In this description, there is a universal time that flows uniformly for all of space. Events are naturally ordered in time. Whether or not two events are simultaneous has a well defined answer. Either Beth was born before Alice, or Alice was born before Beth, or they were born at the same time. Most of us live our lives under these assumptions. However, according to special relativity, the relationship between time and space is more symmetric than it appears to be. This symmetry was discovered by  studying the behaviour of light.
In Maxwell's description, light is a wave of electric and magnetic fields. Electricity was discovered a long time ago.
The ancient greeks observed that, when amber is rubbed with a piece of cloth, a force is generated.
This observation lead them to conjecture the existence of what we now call charged particles, which were divided in two classes, positive and negative. Since the greek word for amber was elektron, these forces became known as electric forces. Opposite charges attract each other, and similar charges repel each other, according to Coulomb's law
\begin{equation}
F_E=\frac{1}{4\pi\varepsilon_{0}}\frac{qQ(y-x)}{\left\vert y-x\right\vert ^{3}},
\end{equation}
where, $q$ and $Q$ are the charges of the particles, measured in Coulombs, and $\varepsilon_0$ is a constant of nature
known as the permitivity of free space. Moving charges are subject to other forces, magnetic forces.
The electric and magnetic interactions experienced by a charged particle are determined by the electric and magnetic fields $E$ and $B$. 
In the nineteenth century, electromagnetism was studied experimentally by many physicists, including Ampere, Biot-Savart, Coulomb, Gauss, Faraday and Oersted. The properties of electric and magnetic fields are ultimately summarized by Maxwell's equations
\begin{eqnarray}
\div E&=& \frac{\rho}{\varepsilon_0}\label{M1},\\
\div B&=&0,\label{M2}\\
\rot E+\frac{\partial B}{\partial t}&=&0,\label{M3}\\
\rot B-\varepsilon_{0}\mu_{0}\frac{\partial E}{\partial t}\label{M4}&=&\mu_0 J.
\end{eqnarray}
The constant of nature $\mu_0$ is known as the permeability of the vacuum. Light is an electromagnetic wave,
such as the one depicted in Figure \ref{EMwave}.

\begin{figure}[H]
\centering
\begin{tabular}{cc}
  \vspace{0pt} \includegraphics[scale=0.4]{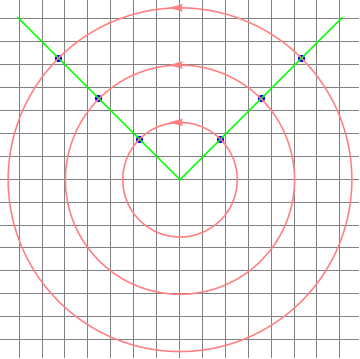} &
  \vspace{0pt} \includegraphics[scale=0.4]{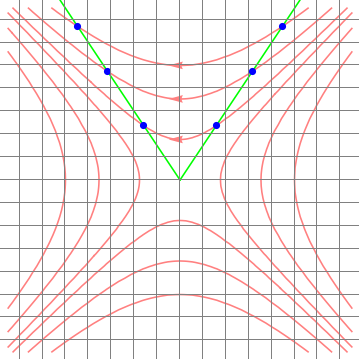}
\end{tabular}
\caption{Euclidean and hyperbolic rotations.}\label{ehrot}
\end{figure}

A remarkable feature of Maxwell's equations is that they imply that electromagnetic waves propagate with velocity 
\begin{equation}
c=\frac{1}{\sqrt{\mu_0 \varepsilon_0}}.
\end{equation}
This should be surprising. Intuitively, one expects that the speed of light emitted by a train traveling towards Alice is greater than that of light emitted by a train going away from her. Newtonian physics, and common sense, suggest that velocities should be added. Since Maxwell's equations predict that the speed of light is a constant $c$, it was assumed that the equations should only hold in a preferred reference frame, that of the ether, the hypothetical substance through which light was supposed to propagate. In 1887, Michelson and Morley attempted to measure the
relative speed of the Earth with respect to the ether at various points of the
Earth's orbit around the Sun. However, the experiments failed to measure such velocity. The results left no option but to conclude that the speed of light is independent of the state of motion of the observer. This posed a problem.
Suppose that Alice and Beth move with constant velocity $v$ with respect to each other.
Classically, it was assumed that Alice and Beth share a universal time $t$, and that the Galilean transformation
\begin{equation} \overline{x}=x- tv,\end{equation}
described the relationship between the positions they assign to an event. This formula implies that
\begin{equation}
 \dot{\overline{x}}=\dot{x}- v.\end{equation}
Therefore, if Alice measures the speed of light to be $c$, Beth will measure the speed of light to be $c-v$. In order for the speed of light to be constant, it was necessary to replace Galilean transformations.
Einstein postulated two simple rules from which the new transformations can be derived:
\begin{itemize}
\item {\textbf Postulate 1.} The speed of light is the same for all inertial observers.
\item {\textbf Postulate 2.} The equations of physics take the same form for all inertial observers.
\end{itemize}
It is easy to see that the only way to satisfy Einstein's postulates is to set
 
 \begin{equation}\label{Lorentz1} \overline{x}= \frac{ x -v t}{\sqrt{1-(v/c)^2}}, \quad \overline{t}=\frac{t-vx/c^2}{\sqrt{1-(v/c)^2}}.
\end{equation}
This rule is known as a Lorentz transformation. 
An important new feature is that, in contrast with the Newtonian description, Alice and Beth now have different time coordinates.
In classical mechanics one insists that there are no preferred directions in space. This means that all equations should remain invariant under Euclidean rotations. In special relativity there is an additional symmetry. Lorentz transformations are hyperbolic rotations that exchange space and time.
Figure \ref{ehrot} illustrates Euclidean and hyperbolic rotations.

\begin{figure}[H]
\centering
\begin{tabular}{cc}
\includegraphics[scale=0.4]{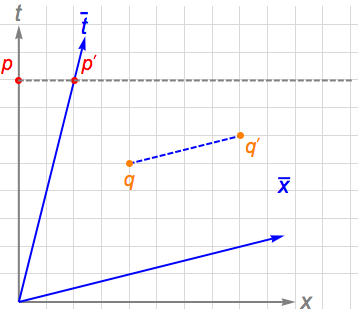}&
\includegraphics[scale=0.4]{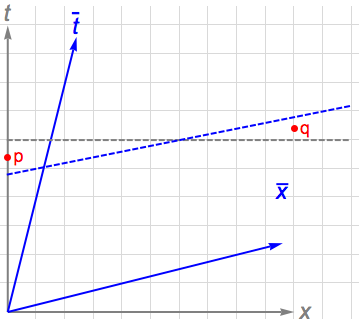}
\end{tabular}
\caption{The gray axes represent Alice's reference frame and the blue axes those of Beth. The figure on the left shows that Alice and Beth have different notions of simultaneity. The figure on the right shows that they disagree on the time ordering of events $p$ and $q$.}\label{order}
\end{figure}

This additional symmetry between time and space forces one to conclude that whether or not two events are simultaneous depends on the observer. Even more dramatically, it is possible for Alice to judge that event $p$ occurred before event $q$, and for Beth to believe the opposite.  Figure \ref{order} illustrates the situation.

\begin{figure}[h]
\centering
\includegraphics[scale=0.4]{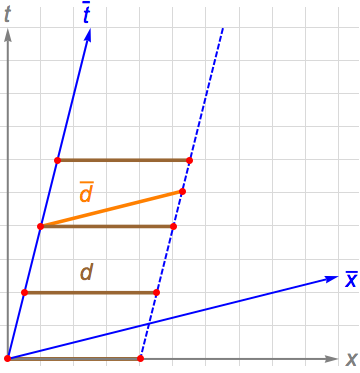}\caption{Length
contraction.}\label{flcont}
\end{figure}

Not only will Alice and Beth differ in the way they measure time. They will also disagree about the length of physical objects.
Suppose that Beth carries a ruler with her, so that Alice sees a ruler of length $d$ meters moving with velocity $v$. In this case, according to Alice, after $t$ seconds, the front end of the ruler will be in position $A(t)=d+vt $ and the back end of the ruler will be in position $B(t)=vt$.
Consider the position of the front end of the ruler after $t_0=\frac{dv}{c^2-v^2}$ seconds. Alice  will assign to this event the coordinates
\[ t_0=\frac{dv}{c^2-v^2},\quad x_0=\frac{d}{1-(v/c)^2}.\]
The coordinates that Beth will assign to this event are
\[ \overline{t}_0=0,\quad \overline{x}_0=\frac{d}{\sqrt{1-(v/c)^2}}=d \lambda_v.\]
Therefore, Beth will measure the length of the ruler to be equal to $d \lambda_v>d$. The fact that Alice sees the ruler moving causes her to perceive the length of the ruler to be contracted. Figure \ref{flcont} illustrates the situation.

\begin{figure}[h]
\centering
\includegraphics[scale=0.5]{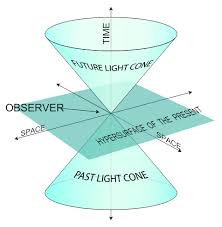}\caption{Different kinds of directions in Minkowski spacetime.}\label{lightcone}
\end{figure}

Since Lorentz transformations intertwine time and space, it becomes impossible to think of them separately. Instead, one is lead to consider a four dimensional spacetime. Just like Euclidean rotations are rigid motions that preserve distance, Lorentz transformations preserve a different notion of distance, that determined by the Minkowski metric
\[g=\begin{pmatrix}
-c^2&0&0&0\\
0&1&0&0\\
0&0&1&0\\
0&0&0&1
\end{pmatrix}.
\]
Spacetime has a definite geometry, given by the Minkowski metric, where the inner product between two vectors $v=(v^0,v^1,v^2,v^3)$ and $w=(w^0,w^1,w^2,w^3)$ is:
\begin{equation}
\langle v,w\rangle=-c^2v^0w^0+v^1w^1+v^2w^2+v^3w^3.
\end{equation}
In contrast with Euclidean geometry, in Minkowski geometry, the inner product of a vector with itself can be positive, zero and negative.
Therefore, directions in spacetime are classified in different types. A vector $v$ is called
\begin{itemize}
\item Timelike if $\langle v,v\rangle <0$.
\item Lightlike if $\langle v,v\rangle =0$.
\item Spacelike if $\langle v,v\rangle>0.$
\end{itemize}
Suppose that an object moves in space following the curve
$\alpha(t)=(x(t),y(t),z(t))$. Then, it traces a path in spacetime $\beta(t)=(t,x(t),y(t),z(t))$ which has four velocity:
\begin{equation}
\beta'(t)=(1,x'(t),y'(t),z'(t)).
\end{equation}
Then
\begin{align}
\langle \beta'(t),\beta'(t)\rangle =0&\Leftrightarrow |\alpha'(t)|=c,\\
\langle \beta'(t),\beta'(t)\rangle <0&\Leftrightarrow |\alpha'(t)|<c,\\
\langle \beta'(t),\beta'(t)\rangle >0&\Leftrightarrow |\alpha'(t)|>c.
\end{align}
Therefore, lightlike vectors are the four velocities of objects moving at the speed of light, timelike vectors are the four velocities of objects moving slower than light, and spacelike vectors are the four velocities of objects moving faster than light. As we will explain in a moment, in order to preserve causality, it is necessary to assume that physical objects travel slower than light. This means that the trajectories that they trace in spacetime point in timelike directions. The length of this trajectory is the amount of time that the observer will judge to have passed, the proper time. Lightike vectors form a cone, the light cone, illustrated in Figure \ref{lightcone}.\\

\subsubsection{Time travel and causality}

\begin{figure}[H]
\centering
\includegraphics[scale=0.4]{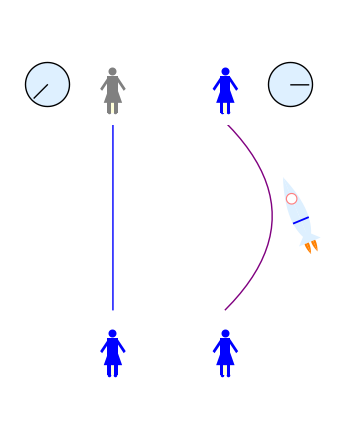}
 \caption{One twin ages faster than the other. }\label{figuretwin}
\end{figure}

We all travel in time at a rate of 1 second per second towards the future. This is true even in classical mechanics.
In special relativity, other kinds of time travel are possible, but not everything is allowed. The proper time that Alice and Beth will experience in going from $p$ to $q$ depends on the path they take. Therefore, by choosing different paths, they will experience different times. In Minkowski spacetime, a straight line is the trajectory that maximizes time from $p$ to $q$. If Alice stays on Earth while Beth travels at very high speed to a nearby star and comes back, then, Alice will have followed a straight line, while Beth will have not. Therefore, more time will have passed for Alice than for Beth. This is the twin paradox, illustrated in Figure \ref{figuretwin}.

\begin{figure}[h]
\centering
\includegraphics[scale=0.48]{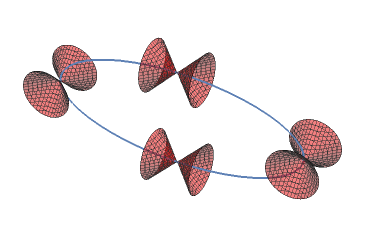}\caption{A closed timelike curve would violate causality.}\label{ncc}
\end{figure}

The twin paradox is an example of a kind of time travel that happens in special relativity, and contradicts our intuition. In this sense, time travel is possible. However, the real problem is going back to the past, which leads to all sorts of logical problems. If Alice travelled to the past and prevented her parents from meeting, then
she would not have been born, so she could not have travelled to the past, so her parents would have met, and she would have been born, and would have travelled...  It seems better to avoid this situation. If Alice travelled to the past, her world line would be a closed timelike curve. Luckily, it is a simple geometric property of Minkowski spacetime that there are no closed timelike curves. In this case, the geometry prevents logical problems. This is consistent with the observed fact that people tend not to travel to the past. Figure \ref{ncc} illustrates a closed timelike curve.

Einstein's special relativity provides a theory of time and space that is consistent with Maxwell's equations and the constancy and the speed of light. However, there is a new problem, Newton's theory of gravity is not compatible with special relativity. In Newton's theory, the gravitational forces depend on the distances between objects, but according to Einstein, these distances depend on the observers. It took Einstein ten more years to
develop his general theory of relativity, a geometric theory of gravity. In general relativity, gravity is not a force, but a consequence of the curvature of spacetime. The language in which general relativity is written is Riemannian geometry, the mathematics of curved spaces.

\section*{Geometry}
\begin{figure}[H]
\centering
\begin{tabular}{ccc}
\includegraphics[scale=0.33]{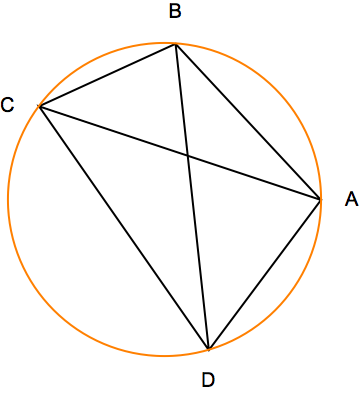}&
\includegraphics[scale=0.33]{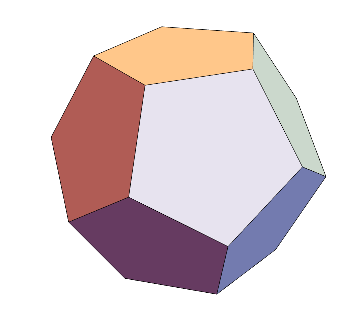}
\end{tabular}
\caption{Geometry in antiquity. The figure on the left represents Ptolemy's theorem $|AC||BD|=|AB||CD|+|BC||AD|$. The figure on the right is a dodecahedron, one of the five Platonic solids. }\label{fgreeks}
\end{figure}

Geometry, the study of shapes, has kept people busy for a long time. Pithagoras, Plato, Euclid, Archimedes and Ptolemy were interested in straight lines, circles, triangles and regular polyhedra. They made wonderful discoveries that are still studied today. Even though the ancient greeks considered the curved geometry of the sphere, the general methods for studying arbitrary curved surfaces are much more recent.
  Figure \ref{fgreeks} illustrates some highlights of old geometry.

In the nineteenth century, Gauss studied the geometry of curves and surfaces in three dimensional space. Riemann developed the formalism for describing arbitrary curved spaces.

\begin{figure}[h]
\centering
\includegraphics[scale=0.43]{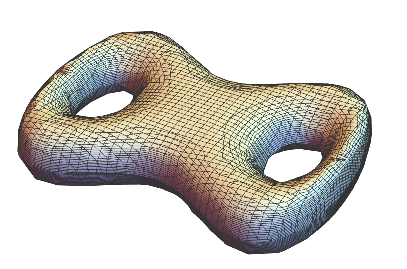}
\caption{A two dimensional surface in three dimensional space.}
\end{figure}

\begin{figure}[h]
\begin{tabular}{p{0.3\textwidth} p{0.3\textwidth} p{0.3\textwidth}}
  \vspace{0pt} \includegraphics[scale=0.32]{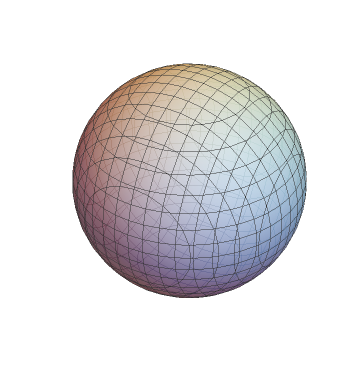} &
  \vspace{0pt} \includegraphics[scale=0.32]{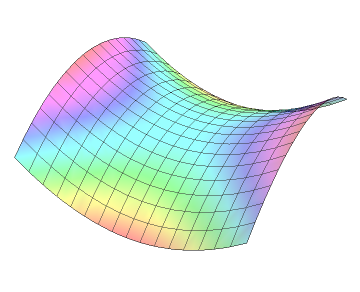}&
  \vspace{0pt} \includegraphics[scale=0.32]{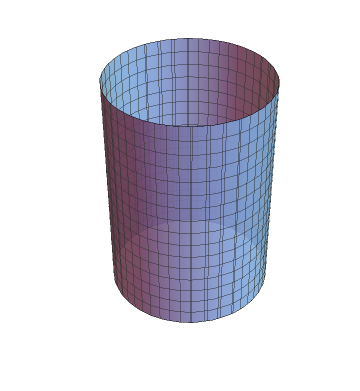}
\end{tabular}
\caption{Positive, negative and zero curvature.}\label{curvatures}
\end{figure}

In Riemannian geometry, the shape of a space is determined by a Riemannian metric, which is a rule for measuring lengths and angles. A Riemannian metric takes the form
\begin{equation}
g= \sum_{ij} g_{ij} dx^i \otimes dx^j,
\end{equation}
where the functions $g_{ij}$ give the inner product between the $i$-th and the $j$-th direction at each point. For instance, for ordinary Euclidean space, the metric is
\begin{equation}
g=dx \otimes dx + dy \otimes dy +dz \otimes dz.
\end{equation}
For the surface of the sphere, it is
\begin{equation}
g=d\theta \otimes d\theta+ \sin^2 \theta d\phi \otimes d\phi.
\end{equation}
In two dimensions, the curvature of a space $M$ is determined by a function, the Gaussian curvature
$K: M \to \RR$. The sphere has positive curvature, the plane has zero curvature and the saddle has negative curvature. These surfaces are illustrated in Figure \ref{curvatures}. In higher dimensions, since there are more degrees of freedom, measuring curvature is more complicated. The Riemannian metric determines the Levi-Civita connection, denoted $\nabla$, which is a rule for taking derivatives of vector fields. The expression
$\nabla_XY$ represents the covariant derivative of the vector field $Y$ in the direction of $X$. The covariant derivative is a version of the directional derivative that depends on the geometry of $M$. The curvature of $M$ is described by the Riemann curvature tensor, which is the quantity
\begin{equation}
R(X,Y)(Z)=\nabla_X \nabla_Y Z- \nabla_Y \nabla_X Z-\nabla_{[X,Y]}Z.
\end{equation}
Other measures of curvature are the Ricci tensor $\Ric$, which is the trace of the Riemann tensor, and the scalar curvature $R$, which is the trace of the Ricci tensor. 

\begin{figure}[h]
\begin{tabular}{p{0.3\textwidth} p{0.3\textwidth} p{0.3\textwidth}}
  \vspace{0pt} \includegraphics[scale=0.31]{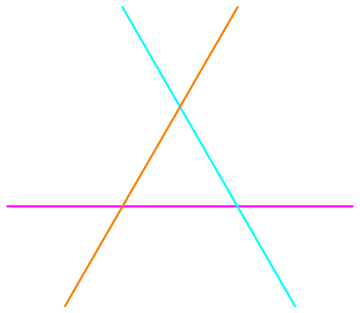} &
  \vspace{0pt} \includegraphics[scale=0.31]{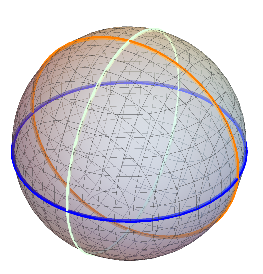}&
  \vspace{0pt} \includegraphics[scale=0.31]{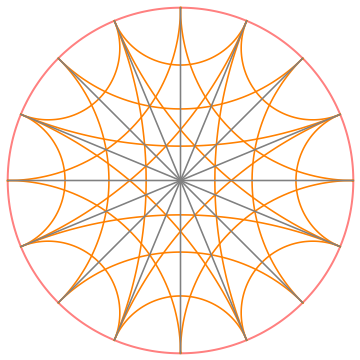}
\end{tabular}
\caption{Geodesics on the plane are straight lines, on the sphere are maximal circles, and on hyperbolic space  are circles orthogonal to the boundary.}\label{fgeo}
\end{figure}

In flat space there are special curves, straight lines, which give the shortest path between two points. A straight line is characterized by the property that its velocity is constant, so that its acceleration vanishes
\begin{equation}\gamma''(t)=0.
\end{equation}
This condition has an analogue on arbitrarily curved spaces, where it is expressed in terms of the Levi-Civita connection as
\begin{equation}\label{ingeo}
\nabla_{\gamma'(t)}\gamma'(t)=0.
\end{equation}
A curve $\gamma(t)$ that satisfies (\ref{ingeo}) is  known as a geodesic. They are the analogues of straight lines for curved spaces. Intuitively, they are those paths that have zero acceleration. For instance, on the sphere, geodesics are maximal circles. An ant walking on a sphere will move along a maximal circle unless it has a reason to deviate. Figure \ref{fgeo} illustrates some examples.

The spaces studied in Riemannian geometry are known as Riemannian manifolds. At the tangent space of each point in a Riemannian manifold there is an inner product that is equivalent to the usual inner product on Euclidean space. The way in which this inner product varies with the coordinates is what determines the geometry. Minkowski spacetime, which we encountered in Special Relativity, is not an example of a Riemannian manifold. This is because in Minkowski spacetime there are some vectors whose inner product with themselves is negative. A space that has an inner product of Minkowski type at each point is known as a Lorentzian manifold. Four dimensional Lorentzian manifolds model spacetime in general relativity. 

\section*{General Relativity}
Once Special Relativity was in place as a theory of spacetime, Einstein was left with the problem 
of finding a description of gravity that was compatible with relativity. The answer he found is geometric, and was motivated by a thought experiment. Imagine that Alice is in an elevator in empty space. Since there is no gravity, she will not be pushed to the floor. If she drops a ball, the ball will float with her. Einstein observed that, if the elevator was falling freely towards the Earth, pulled by the gravitational force, Alice would feel the same. The balls she dropped would still not fall to the ground. For Alice, the two situations would be equivalent.

\begin{figure}[h!]
\centering
\includegraphics[scale=0.26]{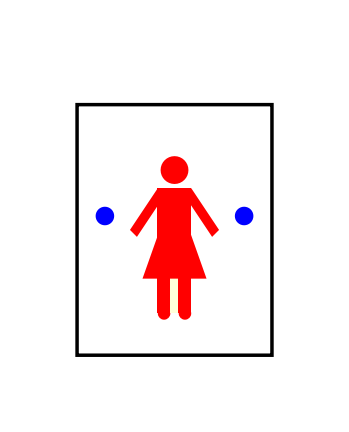}\caption{Alice feels weightless. She doesn't know whether she is at rest in empty space, or falling freely towards the Earth.}%
\end{figure}

Einstein also imagined that Beth was inside an elevator on the surface of the Earth. In this case, gravity makes Beth feel pushed against the floor. If she drops balls, the balls will fall.  He considered also the situation where there is no gravity, but the lift is being pulled up with constant acceleration. Again, Beth will feel heavy, balls will fall to the ground.
\begin{figure}[h]
\centering
\begin{tabular}{cc}
\vspace{0pt} \includegraphics[scale=0.26]{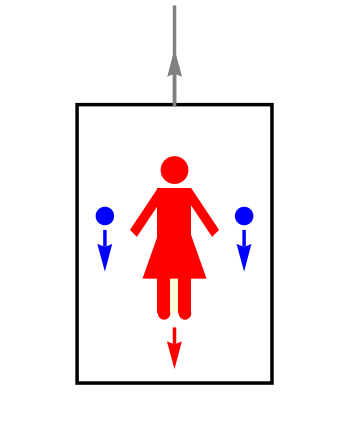}&
  \vspace{0pt} \includegraphics[scale=0.24]{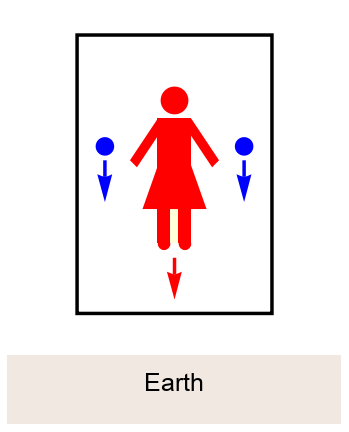} 
\end{tabular}
\caption{Beth feels heavy. She can't tell whether she is on the surface of the Earth or in empty space where her lift is being pulled up.}
\end{figure}
Einstein concluded that what Beth and Alice can detect inside their lifts is not whether or not there is a gravitational field, but whether or not they are moving in the way that is natural given the situation. Alice feels she is floating because in both cases she is following the natural kind trajectory. Beth feels heavy, because in both cases, she is deviating from the natural motion. This motivated Einstein to imagine that gravity is the curvature of spacetime. In the absence of gravity, spacetime is flat, and the natural motion that objects fall are straight lines, geodesics. In the presence of gravity, spacetime curves, and objects tend to move in the geodesics of curved spacetime.
What Beth experiences as gravity is her deviation from geodesic motion. Mathematically, this means that
spacetime should be modeled by a Lorentzian manifold, which may be curved. Minkowski spacetime is just the special case where there is no curvature. Gravity can be incorporated into special relativity by replacing Minkowski spacetime by a curved Lorentzian manifold. The following table describes this correspondence.

\begin{center}
\begin{tabular}{l c l}
\hline&&\\
 {\textbf Special Relativity}&$\to$& {\textbf General Relativity} \\  
\hline && \\
 Minkowski spacetime &$\to$ &Lorentzian manifold \\
 \hline &&\\
 Timelike straight lines &$\to$ &Timelike geodesics  \\ 
 \hline&&\\
Minkowski spacetime is flat &$\to$ & Curvature (Gravity)\\ 
\hline&&\\
Lorentz invariance &$\to $& Geometric character\\
\hline 
\end{tabular}
\end{center}

Newton's theory describes the gravitational force between two masses, which is proportional to the product of the masses, and inversely proportional to the square of the distance. From this law, the Poisson equation
\begin{equation}
\Delta \Phi=4 \pi G_{\mathrm{N}} \rho,
\end{equation}
can be deduced. It expresses the relationship between the gravitational potential $\Phi$, and the mass density function $\rho$. The analogue of the Poisson equation in General Relativity is Einstein's field equation, which describes the relationship between the mass and energy distribution and the curvature of spacetime. The field equation is
\begin{equation}\Ric-\frac{1}
{2}\Rs g=\frac{8\pi G_{\mathrm{N}}}{c^{4}} T .\end{equation}
In this equation, the left hand side is a geometric quantity, $\Ric$ is the Ricci curvature tensor, $R$ is the scalar curvature, and $g$ is the metric. The right hand side is proportional to the energy momentum tensor $T$, which describes the mass and energy distribution in spacetime. The relationship goes both ways. Mass and energy cause spacetime to curve. In turn, the curvature of spacetime determines the geodesics, the natural kind of motion that matter follows given a specific geometry.

\begin{figure}
\centering
\includegraphics[scale=0.4]{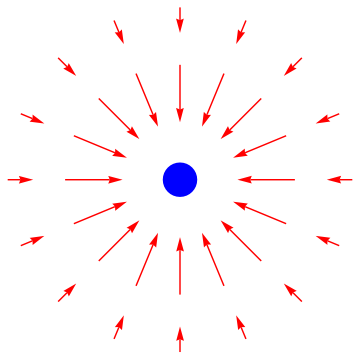}
\caption{Gravitational field generated by a point mass, according to Newtonian mechanics.}\label{fgfield}
\end{figure}

\begin{figure}[h]
\centering
\includegraphics[scale=0.4]{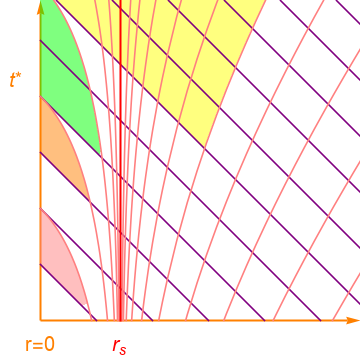}
\caption{The red and purple lines represent lightlike geodesics in a Schwarzschild spacetime in Eddington-Finkelstein coordinates. The future goes up. The region to the left of the red vertical line is the interior of the black hole. The green, orange, and pink regions represent the future cones of an event inside the black hole. They never reach the outside. The yellow region represents the future cone of an event outside the black hole.
}\label{fEFI}
\end{figure}

The simplest kind of gravitational field in Newtonian gravity is that generated by a point mass, depicted in Figure \ref{fgfield}. The relativistic description of this situation is provided by the geometry of Schwarzschild spacetime. The Schwarzschild metric is
\begin{equation} g=-\left(1-\frac{r_s}{r}\right)c^2 dt \otimes dt+\left(1-\frac{r_s}{r}\right)^{-1} dr \otimes dr+r^2\left(d\theta \otimes d\theta +\sin^2\theta d\phi \otimes d\phi\right) .\end{equation}
By modeling the gravity caused by a point mass with the Schwarzschild metric, General Relativity makes predictions that differ from those of Newton's theory. The precession of the perihelion of Mercury, the bending of light, gravitational time dilation and black holes are some of the fundamental predictions that have been experimentally tested and confirmed. Figure \ref{fEFI} illustrates lightlike geodesics in a Schwarzschild black hole.

 The Friedmann-Lemaitre-Robertson-Walker metrics are models for describing the universe as a whole. The fundamental assumption, known as the cosmological principle, is that, at the largest scale, space looks the same at all places and in all directions. It is homogeneous and isotropic. This symmetry condition leads to the FLRW metrics
\begin{equation}
g=-dt \otimes dt+ A^2(t)\left( \frac{1}{1-kr^{2}}dr\otimes dr+r^{2}
\big(d\theta\otimes d\theta+\sin^{2}\theta\text{ }d\phi\otimes d\phi\big)\right),
\end{equation}
where the scaling factor $A(t)$ gives the expansion of the universe, and the constant $k$ determines whether space has positive, negative or zero curvature. It is remarkable that such simple formula gives information about the entire universe. For a sense of scale, the speed of light is
$c\sim 300.000\text{ km/s}$,
the age of the universe is currently estimated at around $13.8$ billion years, and the size of the observable universe is $93$ billion light years. That is, the distance that light travels in 93 billion years. In kilometers this is
\[879854400000000000000000 \text{ km}.\]

\begin{figure}[H]
\centering
\includegraphics[scale=0.4]{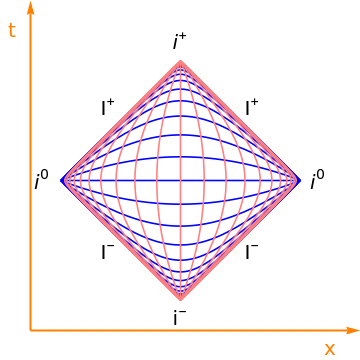}
\caption{The Penrose diagram for the Friedmann-Lemaitre-Robertson-Walker metric with scaling factor $A(t)=\epsilon t$ and $k=0$. In this diagram light travels at $\pi/4$ angles. Timelike curves come from $i^-$ and go to $i^+$. Lightlike curve come from $I^-$ and go to $I^+$. Spatial infinity is denoted by $i^0$.}
\end{figure}
   \clearemptydoublepage
 
\pagenumbering{arabic}
     \setcounter{page}{1}
\titleformat{\chapter}[display]
     {\sc  \huge}
     {\bf \Huge\thechapter \ \titlerule[1pt]} 
     {0pt} { \vspace{1ex}%
  \hspace{1ex} \chapformat }[ \vspace{1ex}{\titlerule[1pt]}]  
\numberwithin{theorem}{chapter}  
 \numberwithin{equation}{chapter}
 \begin{partwithabstract}{Geometry}
Einstein's theory of relativity is a geometric theory of gravity. Gravitation is the effect that
mass and energy have on the geometry of spacetime. The description of these geometric phenomena requires a mathematical
language to study curved spaces. Plane geometry, as studied by Euclid 23 centuries ago, provides an excellent description of flat 
space. However, Euclidean geometry lacks the tools to study curved surfaces. The development of analytic geometry by Descartes, and that of
calculus by Newton and Leibniz, allowed Gauss to study the geometry of curves and surfaces in three dimensional space. Later, Riemann introduced the
formalism for describing curved spaces in higher dimensions. This  formalism is known  as Riemannian geometry or, simply, differential geometry. As the name suggests, it uses differentiable calculus to study curved higher dimensional spaces. Riemannian geometry is the language in which Einstein's theory of gravitation is written. This book begins with an introduction to Riemannian geometry.

\begin{figure}[H]
\centering
\begin{tabular}{cc}
\includegraphics[scale=0.36]{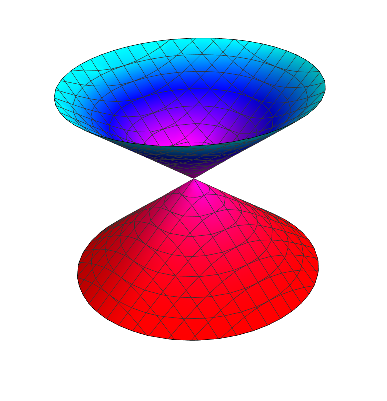}&
\includegraphics[scale=0.36]{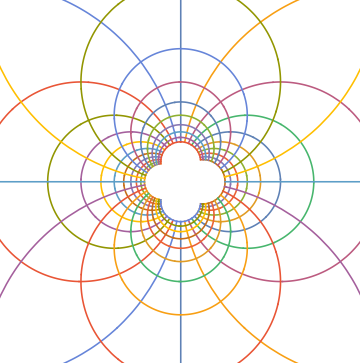}
\end{tabular}
\end{figure}
\end{partwithabstract}

 \chapter{Differentiable manifolds}

\vspace{3ex}

\section{Manifolds}

\begin{definition}
\label{1mtopo} A topological manifold of dimension $n$ is a Hausdorff, second
countable topological space which is locally isomorphic to $\RR^{n}.$
That is, given any point $p\in M$ there exists an open neighborhood $U_{p}$
that contains $p$ and a homeomorphism $\varphi:U_{p}\rightarrow V,$ for some
open subset $V\subseteq\RR^{n}$.
\end{definition}

If one wants to use the tools of calculus over the manifold $M$, it is
necessary to endow $M$ with an additional structure that allows for a notion of differentiability.
Let $M$ be a topological manifold. A \emph{chart }for $M$ is a pair $\left(
U,\varphi\right)  $, where $U$ is an open set in $M$ and $\varphi:U\rightarrow
V$ is a homeomorphism onto an open set $V$ of $\RR^{n}$. The chart
$(U,\varphi)$ assigns local coordinates to each point $p\in U$,
$x^{i}:U\rightarrow\RR,$ defined by $x^{i}\left(  p\right)
=u^{i}\left(  \varphi\left(  p\right)  \right)  ,$ where the functions $u^{i}$
denote the standard coordinates in $\RR^{n}$.

\begin{definition}
A smooth atlas for $M$ is a family $\left\{  \left(  U_{\alpha},\varphi
_{\alpha}\right)  \right\}  _{\alpha\in A}$of charts that satisfies the
following properties:

\begin{itemize}
\item The open sets $\left\{  U_{\alpha}\right\}  _{\alpha\in A}$ cover $M$.

\item For any $\alpha,\beta\in A$, the change of coordinates function
\[
h_{\beta\alpha}=\varphi_{\beta}\circ\varphi_{\alpha}^{-1}:\varphi_{\alpha
}\left(  U_{\alpha}\cap U_{\beta}\right)  \rightarrow\varphi_{\beta}\left(
U_{\alpha}\cap U_{\beta}\right)
\]
is a smooth function.
\end{itemize}
\end{definition}

\begin{figure}[h!]
	\centering
		\includegraphics[scale=0.56]{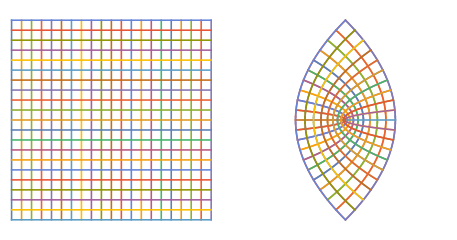}
	\caption{A smooth function from the plane to the plane.}
	
\end{figure}

A smooth atlas $(U_{\alpha}, \varphi_{\alpha})_{\alpha\in A}$ for $M$ is maximal if
it is not properly contained in another smooth atlas for $M$.

\begin{definition}
A manifold is a topological manifold together with a maximal atlas.
\end{definition}

It is an easy exercise to show that any atlas is contained in a unique maximal atlas. Therefore, any atlas, not necessarily maximal, gives a
topological manifold $M$ the structure of a manifold. We will say that a chart
$(U, \varphi)$ is smooth if it is contained in the maximal atlas defining $M$.

\begin{definition}
\label{1suave} A function $f:M\rightarrow\RR$ is smooth if for any
smooth chart $(U,\varphi)$, the map $f\circ\varphi^{-1}:V\rightarrow
\RR$ is smooth. The space of all smooth functions $f:M\rightarrow
\RR$ is denoted by $C^{\infty}(M)$ and has the structure of a
commutative ring with respect to pointwise multiplication.
\end{definition}

\begin{definition}
\label{1prop3}Let $M,$ $N$ be smooth manifolds. A function $f:M\rightarrow N$
is called smooth if for each $p\in M$\ there exist charts $\left(
U,\varphi\right)  $ and $\left(  W,\psi\right)  $ around $p$ and $f(p)$,
respectively, such that $f\left(  U\right)  \subseteq W$, and $\psi\circ
f\circ\varphi^{-1}:\varphi\left(  U\right)  \rightarrow\psi\left(  W\right)  $
is smooth .
\end{definition}

The function $\psi\circ f\circ\varphi^{-1}$ is called a\emph{ local
representation} of $f$ with respect to the charts $\left(  U,\varphi\right)  $
and $\left(  V,\psi\right)  $. It is a simple exercise to show that the
composition of smooth functions is smooth and that the identity function is
smooth. 

\begin{definition}
A smooth function $f:M\rightarrow N$ is called a diffeomorphism if it is
invertible and its inverse is smooth. The function $f$ is called a local
diffeomorphism if for each point $p\in M$ there is an open neighborhood $U$ of
$p$ such that the restriction of $f$ to $U$ is a diffeomorphism onto its image.
\end{definition}


\begin{example}
The topological space $\RR^{n}$ is a manifold of dimension $n$ with
respect to the atlas given by the identity map $(\RR^{n}, \mathrm{id}_{\RR^n})$.
\end{example}

\begin{example}
\label{1ejem1} The sphere of dimension $n$, denoted $S^{n}$, is the topological
subspace of $\RR^{n+1}$ defined as%
\[
S^{n}=\{(x^{1},\ldots,x^{n+1})\in\RR^{n+1}\mid(x^{1})^{2}+\cdots
+(x^{n+1})^{2}=1\}.
\]
$S^{n}$ inherits the topology from $\RR^{n+1}$ and becomes a Hausdorff
second countable space. We may endow $S^{n}$ with a smooth structure by means
of the stereographic projections. Let  $
N=\left(  0,\ldots,0,1\right)$ and $S=\left(  0,\ldots,0,-1\right)$ be the north and south poles of the sphere, and set $U_{S}=S^{n}\setminus\left\{  N\right\}$ and  $U_{N}=S^{n}\setminus\left\{  S\right\}$. Define $\varphi_{S}:U_{S}\rightarrow\RR^{n}$ and $\varphi_{N}%
:U_{N}\rightarrow\RR^{n}$ by
\[
\varphi_{S}\left(  x^{1},\ldots,x^{n+1}\right)  =\left(  \frac{x^{1}%
}{1-x^{n+1}},\ldots,\frac{x^{n}}{1-x^{n+1}}\right)  ,
\]
and
\[
\varphi_{N}\left(  x^{1},\ldots,x^{n+1}\right)  =\left(  \frac{x^{1}%
}{1+x^{n+1}},\ldots,\frac{x^{n}}{1+x^{n+1}}\right)  .
\]
Geometrically, $\varphi_{N}\left(  x^{1},\ldots,x^{n+1}\right)  $ is the point
of intersection of the straight line that passes through $N$ and $\left(
x^{1},\ldots,x^{n+1}\right)  $ with the plane $x^{n+1}=0.$ It is clear that
$\varphi_{S}$ and $\varphi_{N}$ are continuous functions, and it can be easily
proved that they are bijective with continuous inverses. In fact, the inverse
maps are given by
\[
\varphi_{S}^{-1}\left(  y^{1},\ldots,y^{n}\right)  =\left(  1+\left\vert
y\right\vert ^{2}\right)  ^{-1}\left(  2y^{1},\ldots,2y^{n},\left\vert
y\right\vert ^{2}-1\right),
\]
and
\[
\varphi_{N}^{-1}\left(  y^{1},\ldots,y^{n}\right)  =\left(  1+\left\vert
y\right\vert ^{2}\right)  ^{-1}\left(  2y^{1},\ldots,2y^{n},1-\left\vert
y\right\vert ^{2}\right),
\]
for each $y=\left(  y^{1},\ldots,y^{n}\right)  \in\RR^{n}$. Let us show
that $\left\{  (U_{S},\varphi_{S}),(U_{N},\varphi_{N})\right\}  $ is an atlas.
Obviously, $U_{S}\cup U_{N}=S^{n}$. The transition map
$\varphi_{S}\circ\varphi_{N}^{-1}:\RR^{n}\setminus\{0\}\rightarrow\RR%
^{n}\setminus\{0\}$ is given by
\[
\varphi_{S}\circ\varphi_{N}^{-1}\left(  y\right)  =\frac{(y^{1},\ldots,y^{n}%
)}{(y^{1})^{2}+\cdots+(y^{n})^{2}}.
\]
By symmetry, the map $\varphi_{N}\circ\varphi_{S}^{-1}$ is also smooth and we
conclude that $S^{n}$ is a manifold.
\begin{figure}[h]	
	\centering
	\includegraphics[scale=0.56]{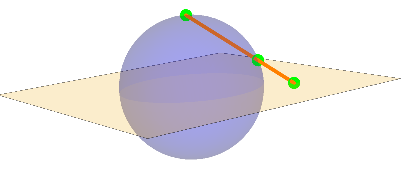}
	\caption{Stereographic projection.}	
\end{figure}

\end{example}

\begin{example}
\label{1ejem3}Let $M$ and $N$ be smooth manifolds of dimensions $m$ and $n$.
The cartesian product $M\times N$ can be endowed with the product topology,
and with a natural atlas $C=\left\{  \left(  U_{\alpha}\times V_{\beta
},\varphi_{\alpha}\times\psi_{\beta}\right)  \right\}  $ induced by two fixed
atlases $A=\left\{  \left(  U_{\alpha},\varphi_{\alpha}\right)  \right\}  $
and $B=\left\{  \left(  V_{\beta},\psi_{\beta}\right)  \right\}  ,$ for $M$
and $N$ respectively. Thus, the product of manifolds is a manifold in a natural way.
\end{example}

The diffeomorphism group $\mathrm{Diff}(M)$ of a manifold $M$ is the group:
\[ \mathrm{Diff}(M) := \{ \varphi: M \rightarrow M\mid \varphi \text{ is a diffeomorphism}\}.\]
The product operation in this group is given by composition of maps.  An action of a group $G$ on a manifold $M$ is a group homomorphism $\varphi: G \rightarrow \mathrm{Diff}(M)$. Equivalently, it is a function $\Phi:G \times M \rightarrow M$ such that
\begin{align*}
\Phi(e,p)&=p, \\
\Phi(g, \Phi(h,p))&=\Phi(gh,p).
\end{align*}
These two definitions are related by the condition $\Phi(g,p)=\varphi(g)(p)$. Two elements $p, p' \in M$ are in the same orbit if there exists $g \in G$ such that $gp=p'$. The relationship of being in the same orbit is an equivalence relation. Therefore,
an action of $G$ induces a partition of $M$ into orbits. The set of orbits of $M$ with respect to the action of $G$ is denoted by $M/G$. The space of orbits has a natural quotient topology and there is a continuous projection $\pi: M \to M/G$. In general, it is not the case that there is a manifold structure on $M/G$ such that $\pi$ is a smooth map. However, this happens for sufficiently well behaved actions. Many examples of manifolds arise as a quotient by a group action.

\begin{example}
The additive group $\mathbb{Z}^{2}$ acts on $M=\RR^{2}$ by
translations
 \[\left(  m,n\right)  \cdot\left(  x,y\right)  =\left(x+m,y+n\right)  ,\] 
for $m,n\in\mathbb{Z}$, and $x,y\in\RR$. The map
$f:\RR^{2}\rightarrow S^{1}\times S^{1}$ defined by $f(x,y)=(e^{2\pi
ix},e^{2\pi iy})$ is a local diffeomorphism. Moreover $f(x,y)=f(x^{\prime
},y^{\prime})$ if and only if $(x,y)-(x^{\prime},y^{\prime})\in\mathbb{Z}%
^{2}.$ We conclude that $f$ induces a bijection from $\RR%
^{2}/\mathbb{Z}^{2}$ to $S^{1}\times S^{1}$. 
\begin{figure}[H]
	\centering
	\includegraphics[scale=0.5]{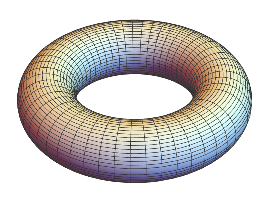}
	\caption{Torus.}
\end{figure}
\begin{example}
If one identifies pairs of edges of an octagon, the quotient space is a double torus. 
\end{example}
\begin{figure}[h!]
	\centering
	\includegraphics[scale=0.46]{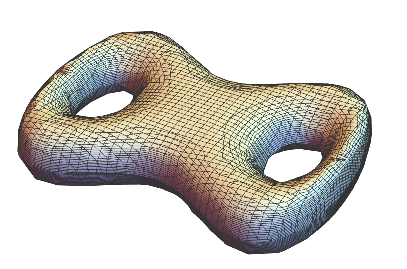}
	\caption{Double Torus.}
	
\end{figure} 
\end{example}

\begin{example}
The Mobius strip $M$ is the topological space defined as the quotient
$\RR^{2}/G$, where $G$ denotes the subgroup of diffeomorphism generated
by the map $\alpha\left(  x,y\right)  =\left(  x+1,-y\right)  .$ Let
$\pi:\RR^{2}\rightarrow M$ be the canonical map to the quotient. It can be shown
that $M$ admits a unique smooth structure such that $\pi$ is a local
diffeomorphism.%
\begin{figure}
	\centering
	\includegraphics[scale=0.55]{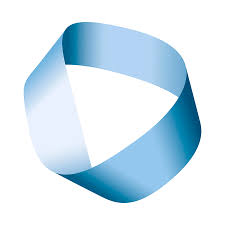}
	\caption{Moebius strip.}
	
\end{figure}

\end{example}

\begin{example}
\label{1Klein} The group $G=\mathbb{Z}/2\mathbb{Z}$ acts on the sphere $S^{2}$
by the antipodal map $(x,y)\mapsto(-x,-y).$ The quotient space $\RR P^2=S^{2}/G$ is
called the real projective plane.
\end{example}

\begin{example}
Let $G$ be the group of diffeomorphisms of the plane generated by $\phi$ and
$\rho$, where $\phi(x,y)=(x+1,y)$ and $\rho(x,y)=(-x,y+1)$. The quotient space
$K=\RR^{2}/G$ is called the Klein bottle. It is a good exercise to show that $K$ admits a unique
smooth structure such that the quotient map $\pi:\RR^{2}\rightarrow K$
is a local diffeomorphism.%
\begin{figure}[h!]
	\centering
	\includegraphics[scale=0.48]{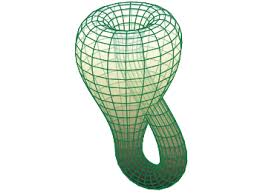}
	\caption{Klein bottle.}
	
\end{figure}

\end{example}

A partition of unity is a technical concept which is very useful in proving existence results in differential geometry.
Let $f: M \rightarrow \RR$ be a smooth function on a manifold. The support of $f$ is the set
\[ \mathrm{supp}(f) = \overline{\{ p \in M: f(p) \neq 0\}},\]
the closure of the set of points where $f$ is non-zero.

\begin{definition}
Let $\{U_{\alpha}\}$ be an open cover of a manifold $M$. A partition of unity
subordinate to $\{U_{\alpha}\}$ is a family of smooth functions $\rho_{\alpha
}:M\rightarrow\RR$ having the following properties:
\begin{itemize}
\item The support of $\rho_{\alpha}$ is contained in $U_\alpha$, $\mathrm{supp}(\rho_\alpha)\subseteq U_\alpha$.

\item Each function $\rho_{\alpha}$ takes only non-negative values:
$\rho_{\alpha}(p)\geq0$ for all $ p \in M$.

\item For each $p\in M$ there exists an open subset $V$ that contains $p$ such
that $\rho_{\alpha}\vert_{V}\neq0$ only for finitely many indices $\alpha$. 
Moreover: \[\sum_{\alpha}\rho_{\alpha}(q)=1\] for all $q\in V.$
\end{itemize}
\end{definition}

It can be proved that given any open cover, there exist a partition of unity subordinate to it. A proof of this technical fact can be found for example in chapter 13 of \cite{Tu}).

\section{The tangent space and the derivative}

A smooth structure on a topological manifold $M$ can be used to define the
tangent space at each point $p\in M$. This is a fundamental construction that
allows the use of the methods of calculus in the study of the topological properties of $M$. Before discussing the
general construction, let us consider an example.
 The tangent space to the sphere at a point $p\in S^{2}$ is the set of all vectors that
are perpendicular to $p:$
\[
T_{p}S^{2}=\{v\in\RR^{3}\;|\;\langle v,p\rangle=0\}.
\]
Note that $T_{p}S^{2}$ is a vector space of dimension two.%
\begin{figure}[h]
	\centering
	\includegraphics[scale=0.42]{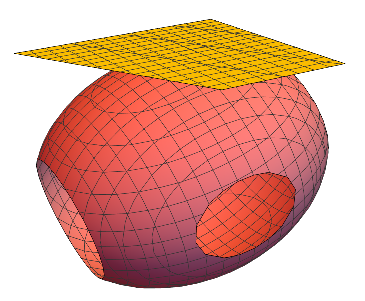}
	\caption{The tangent space of surface in space.}
	
\end{figure}

Intuitively, the tangent space at a point $p\in M$ is the vector space that
parametrizes all the possible velocities of an object moving in $M$ that
passes through the point $p$.

\begin{definition}
Let $p\in M$ be a point in $M$. A curve through $p$ is a smooth function
$\gamma:I\rightarrow M$ such that $\gamma(0)=p$, where $I$ is an
interval $(a,b)$ that contains $0$.
\end{definition}

There exists a natural equivalence relation on the set of all curves that
pass through $p\in M$. We say that two curves $\gamma$ and $\mu$ are
equivalent if and only if \[(\varphi\circ\gamma)^{\prime}(0)=(\varphi\circ
\mu)^{\prime}(0),\] for any choice of coordinates $\varphi:U\rightarrow V$.

\begin{definition}
The tangent space of $M$ at the point $p\in M$, denoted by $T_{p}M$, is the set of
equivalence classes of curves through $p$.%
\begin{figure}[H]
	\centering
	\includegraphics[scale=0.5]{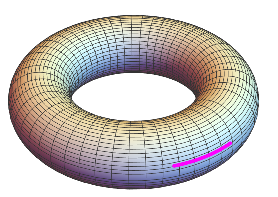}
	\caption{A vector tangent to the Torus.}
	
\end{figure}

\end{definition}

\begin{proposition}
The set $T_{p}M$ has a natural structure of a vector space of dimension
$m=\text{dim}(M)$.
\end{proposition}

\begin{proof}
Let us fix coordinates $\varphi : U\rightarrow V$. This choice determines a function
\[ F_\varphi:  T_pM \rightarrow  \RR^m, \qquad
\gamma  \mapsto  (\varphi \circ \gamma)'(0).
\]
The function $F_\varphi$ is bijective, with inverse given by
\[ (F_\varphi)^{-1}  :\RR^m  \rightarrow T_pM, \qquad
v  \mapsto (F_\varphi)^{-1} (v)=[\gamma]
\]
where 
\[\gamma(t)=\varphi^{-1}(tv+\varphi(p)).\]
This bijection gives $T_pM $ the structure of a vector space. It remains to show that this structure is independent of the choice of coordinates. It suffices to show that if  $\psi: U \rightarrow W$ is another choice of coordinates, then $F_\varphi \circ F_\psi^{-1}$ is a linear isomorphism. For this we compute:
\[F_\varphi \circ F_\psi^{-1}(v)=( \varphi \circ F_\psi^{-1}(v))'\vert_{t=0} = (\varphi \circ \psi^{-1}(tv+ \psi(p)))'\vert_{t=0}= D(\varphi \circ \psi^{-1})(v).\]
We conclude that the vector space structure on $T_pM$ is independent of the choice of coordinates.
\end{proof}

\begin{definition}
Let $f: M\rightarrow N$ be a smooth function. Given $p\in M$, the derivative
of $f$ at $p,$ denoted $Df(p)$, is the linear map
\[
Df(p): T_{p}M \rightarrow T_{f(p)}N,\qquad [\gamma] \mapsto[ f \circ\gamma] .
\]

\end{definition}

We leave it as an exercise for the reader to show the derivative of $f$ is a well defined linear map.

\begin{definition}
A function $f:M\rightarrow N$ is a submersion if for all $p\in M$, $Df(p)$ is
surjective. It is an immersion if for all $p\in M$,
$Df(p)$ is injective. It is an embedding if 
it is an immersion and a homeomorphism onto its image.
\end{definition}

Inclusions and projections are the canonical examples of immersions and submersions:

\begin{itemize}
\item The function $i:\RR^{k}\rightarrow\RR^{k+m}$ defined by
$(x_{1},\ldots,x_{k})\mapsto(x_{1},\ldots,x_{k},0\ldots,0)$ is an immersion.

\item The function $\pi:\RR^{k+m}\rightarrow\RR^{k}$ defined by
$(x_{1},\ldots,\dots,x_{k+m})\mapsto(x_{1},\ldots,x_{k})$ is a submersion.
\end{itemize}

\noindent The inverse function theorem can be used to show that, locally, these are all examples. That is, any immersion is locally 
isomorphic to an inclusion and any submersion is locally isomorphic to a projection. 

The inclusion $\iota: S^n \hookrightarrow \RR^{n+1}$ is an example of an embedding. One says that the sphere is an embedded submanifold of $\RR^{n+1}$. The curve $\gamma: \RR \to \RR^2$ given by $\gamma(t)= (t^3,t^2)$ is not an immersion because its derivative vanishes at $t=0$. This is reflected geometrically
as a singularity on the graph. The map $\gamma$ does induce a homeomorphism onto its image.
\begin{figure}[H]
	\centering
	\includegraphics[scale=0.58]{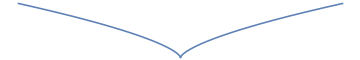}
	\caption{The map  $\gamma(t)= (t^3,t^2)$ is not an immersion.}
	
\end{figure}

The curve $\gamma: (0,\pi) \to \RR^2$ given by $\gamma(t)= (4 \sin(t),\cos(3t))$ is an immersion but it is not an embedding because it is not injective.
\begin{figure}[H]
	\centering
	\includegraphics[scale=0.48]{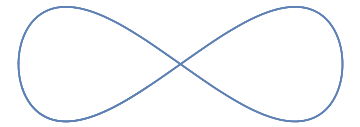}
	\caption{This is an immersion but not an embedding.}
	
\end{figure}


\section{Vector bundles}

We have seen that if $M$ is a manifold then for each point $p \in M$
there is a tangent space $T_{p}M$. Therefore, the tangent space
construction provides a family of vector spaces parametrized by the manifold
$M$. This is the fundamental example of a vector bundle.

\begin{definition}
A rank $k$ vector bundle over $M$ is a manifold $E$ together with a smooth map
$\pi:E\rightarrow M$ such that:

\begin{itemize}
\item For all $p\in M$, the set $E_{p}=\pi^{-1}(\left\{  p\right\}  )$ is a
vector space of dimension $k$.

\item There exists an open cover $\left\{  U_{\alpha}\right\}  _{\alpha
\in\mathcal{A}}$ of $M$ and diffeomorphisms
\[
\phi_{\alpha}:\pi^{-1}(U_{\alpha})\rightarrow U_{\alpha}\times\RR^{k}%
\]
such that the following diagram commutes:%
\[
\xymatrix{
\pi^{-1}(U_\alpha) \ar[d]_{\pi} \ar[r]^{\phi_{\alpha}} & U_\alpha \times \RR^k  \ar[d]^p\\
U_\alpha \ar[r]^{\mathrm{id}_{U_{\alpha}}}&  U_\alpha.
}
\]
Here $p$ denotes the natural projection.

\item The restriction of $\phi_{\alpha}$ to each fiber is a linear
isomorphism, that is, the function
\[
\phi_{\alpha}|_{\pi^{-1}(p)}:\pi^{-1}(p)\rightarrow\{p\}\times
\RR^{k},
\]
is a linear isomorphism.
\end{itemize}
\end{definition}

Given a vector bundle $\pi: E \rightarrow M$, the vector space $\pi^{-1}(p)$ is denoted by $E_{p}$, and called the fiber over $p$.

Intuitively, a vector bundle over $M$ is a family of vector spaces parametrized by $M$. It is a choice of vector space for each point in the space $M$.

\begin{example}
The manifold $M\times\RR^{k}$ together with the natural projection is a
vector bundle over $M$, called the trivial vector bundle.
\end{example}

\begin{example}
Let $E$ be the M\"{o}bius strip, regarded as the quotient space \[E=[0,1]\times
\RR/\sim,\] where $\sim$ denotes the equivalence relation that identifies $(0,v)$ with $(1,-v).$ Consider the map
$\pi:E\rightarrow S^{1}\subseteq\CC$ given by $(t,r)\mapsto e^{2\pi
it}$. Then, $\pi:E\rightarrow S^{1}$ is a vector bundle over the circle.
\end{example}

Let $\pi:E\rightarrow M$ be a vector bundle. A section of $E$ is a smooth map
$s:M\rightarrow E$ such that $\pi\circ s=\mathrm{id}_{M}$. We will
denote by $\Gamma(E)$ the set of all sections of $E$.
The set of sections $\Gamma(E)$ has the structure of a module over the ring $C^{\infty
}(M)$ with respect to the natural pointwise operations:
\begin{itemize}
    \item $(s+ s')(p)=s(p)+s'(p),$
    \item $(f*s)(p)=f(p)s(p),$
\end{itemize}
for $f \in C^{\infty
}(M)$ and $s,s' \in \Gamma(E)$.

Let $\pi:E\rightarrow M$ and $\pi:E^{\prime}\rightarrow M$ be vector bundles over $M$. An isomorphism from $E$ to $E^{\prime}$ is a diffeomorphism
$f:E\rightarrow E^{\prime}$ such that:
\begin{itemize}
\item The following diagram commutes:
\[
\xymatrix{
E \ar[r]^f \ar[d]_\pi& E' \ar[d]^\pi\\
M \ar[r]^{\mathrm{id}_M} & M
}
\]

\item The map $f|_{E_{p}}: E_{p} \rightarrow E^{\prime}_{p}$ is a linear isomorphism.
\end{itemize}

Clearly, if $f$ is an isomorphism from $E$ to $E^{\prime}$ then
$f^{-1}$ is an isomorphism from $E^{\prime}$ to $E$. We will say that
$E$ and $E^{\prime}$ are isomorphic if there is an isomorphism between them.
Let $\pi:E\rightarrow M$ be a vector bundle and $\left\{  U_{\alpha}\right\}
_{\alpha\in\mathcal{A}}$ an open cover of $M$, such that for each $\alpha$
there are local trivializations
\[
\phi_{\alpha}:\pi^{-1}(U_{\alpha})\rightarrow U_{\alpha}\times\RR^{k}.
\]
For each pair of indices $\alpha,\beta$, there are isomorphisms
\[
\phi_{\beta}\circ\phi_{\alpha}^{-1}:U_{\beta}\cap U_{\alpha}\times
\RR^{k}\rightarrow U_{\beta}\cap U_{\alpha}\times\RR^{k}.
\]
That is, for each $p \in U_{\beta}\cap U_{\alpha}$ we obtain a linear
automorphism of $\RR^{k}$. This defines smooth functions $f_{\beta,\alpha}: U_{\beta}\cap U_{\alpha}\rightarrow \mathrm{GL}(k,\RR)$ which satisfy the conditions
\begin{align*}
f_{\alpha\alpha}  &  =\mathrm{id}_{U_{\alpha}},\\
{f}_{\gamma\beta}\circ{f}_{\beta\alpha}  &  ={f}_{\gamma\alpha}.
\end{align*}
The vector bundle $E$ can be reconstructed from the data of these functions.

\begin{definition}
A family of cocycles is an open cover $\left\{  U_{\alpha}\right\}
_{\alpha\in\mathcal{A}}$ of $M$ together with smooth functions $f_{\beta\alpha
}:U_{\alpha}\cap U_{\beta}\rightarrow \mathrm{GL}(k,\RR)$ such that
\begin{enumerate}
\item[(1)] $f_{\alpha\alpha}=\mathrm{id}_{U_{\alpha}}$,

\item[(2)] ${f}_{\gamma\beta}\circ{f}_{\beta\alpha}={f}_{\gamma\alpha}$.
\end{enumerate}
\end{definition}

A family of cocycles $f_{\beta\alpha}$ determines a vector bundle $E$ as follows. As a set, one defines \ the total space as the disjoint union of the
sets $\coprod_{\alpha}U_{\alpha}\times\RR^{k}$ modulo the equivalence relation $\sim$ generated by $(p,v)\sim(p,f_{\beta\alpha}(v))$. That is
\[E=\coprod_{\alpha}U_{\alpha}\times\RR^{k}\ /\sim.\] The map
$\pi:E\rightarrow M$ is the projection onto the first factor. The topology and
the smooth structure on $E$ are characterized by the property that for each
$\alpha$, the natural function $U_{\alpha}\times\RR^{k}\rightarrow
\pi^{-1}(U_{\alpha})$ is a diffeomorphism.

\begin{remark}
Let $E$ be a vector bundle and $\phi_{\alpha}: \pi^{-1}(U_{\alpha})
\rightarrow U_{\alpha}\times\RR^{k}$ a family of local trivializations
for $E$ with corresponding cocycles $f_{\beta\alpha}$. The vector
bundle associated to the family of cocycles $f_{\beta\alpha}$ is naturally
isomorphic to $E$.
\end{remark}

The natural functors of linear algebra such as taking duals, tensor products  and exterior
powers can be used to construct new vector bundles out of given ones, as follows. Let $E,F$ be vector bundles over $M$ with local trivializations $\phi_{\alpha}:\pi^{-1}(U_{\alpha})\rightarrow U_{\alpha
}\times\RR^{k}$ and $\lambda_{\alpha}:\pi^{-1}(U_{\alpha})\rightarrow
U_{\alpha}\times\RR^{m},$ respectively. Let us denote by $f_{\beta
\alpha}$ and $g_{\beta\alpha}$ the corresponding families of cocycles. Then

\begin{itemize}
\item The family of cocycles $h_{\beta\alpha}$ given by \[h_{\beta\alpha
}(p)=f_{\alpha\beta}(p)^{\ast}\] defines a vector bundle $E^{\ast}$ whose fiber over $p$ is $E_{p}^{\ast}$.

\item The family of cocycles $h_{\beta\alpha}$ given by \[h_{\beta\alpha
}(p)=f_{\beta\alpha}(p)\oplus g_{\beta\alpha}(p)\] defines a vector bundle $E\oplus
F$ whose fiber over $p$ is $E_{p}\oplus F_{p}$.

\item The family of cocycles $h_{\beta\alpha}$ given by \[h_{\beta\alpha
}(p)=f_{\beta\alpha}(p)\otimes g_{\beta\alpha}(p)\] defines a vector bundle $E\otimes
F$ whose fiber over $p$ is $E_{p}\otimes F_{p}$.

\item For each $k\in\mathbb{N}$, the family of cocycles $h_{\beta\alpha}$
given by \[h_{\beta\alpha}(p)=\Lambda^{k}f_{\beta\alpha}(p)\] defines a vector
bundle $\Lambda^{k}E$ whose fiber over $p$ is $\Lambda^{k}E_{p}$.

\item For each $k\in\mathbb{N}$, the family of cocycles $h_{\beta\alpha}$
given by \[h_{\beta\alpha}(p)=f_{\beta\alpha}^{\otimes k}(p)\] defines a vector
bundle $E^{\otimes k}$ whose fiber over $p$ is $E_p^{\otimes k}$.
\end{itemize}

The constructions above are independent of
the choice of local trivializations for the original vector bundles $E$ and
$F$. That is, the natural functors of linear algebra can be applied in families to produce new vector bundles $E^{*}$, $E \oplus F$, $E
\otimes F$, $\Lambda^{k}E$, $E^{\otimes k}$ out of given ones. 

We have already mentioned that the tangent space provides the fundamental example of a vector bundle. Together with
the constructions above, one obtains many vector bundles naturally associated to any manifold. These vector bundles are fundamental tools
in the study of the topological and geometric properties of manifolds, as we will see below.

\section{The tangent bundle and vector fields}

Let us  describe the fundamental example of a vector bundle, the tangent bundle. As a set \[TM=\coprod_{p\in M}T_{p}M,\] is the disjoint union of
all tangent spaces.  The projection $\pi:TM\rightarrow M$ is given by
$\pi\lbrack\gamma]=p$ if $[\gamma]\in T_{p}M$. Let $\varphi:U\rightarrow
V\subseteq\RR^{m}$ be a coordinate chart, then $\varphi$ induces a
bijection
\[
F_\varphi:\pi^{-1}(U)\rightarrow U\times\RR^{m},\qquad F_\varphi([\gamma
])=(\pi([\gamma]),(\varphi\circ\gamma)^{\prime}(0)).
\]
Let us show that there exists a unique topology on $TM$ such that $\pi$ is
continuous and for any choice of coordinates $\varphi$, the bijection $F_\varphi$
is a homeomorphism. Since $\pi$ should be continuous we know that the sets
$\pi^{-1}(U)$ should be open.

Since $M$ can be covered with open sets that are the domain of coordinate
charts, it suffices to show that if $\varphi:U\rightarrow V$ and
$\varphi^{\prime}:U^{\prime}\rightarrow V^{\prime}$ are two charts then the
topologies induced on $\pi^{-1}(U\cap U^{\prime})$ are the same. It is enough
to prove that the function%
\[
F_\varphi \circ (F_{\varphi'})^{-1}:(U\cap U^{\prime})\times\RR^{m}\rightarrow(U\cap
U^{\prime})\times\RR^{m}%
\]
is a homeomorphism. This function is given by
\[
(p,v)\mapsto (p,D(\varphi \circ \varphi^{\prime-1})(\varphi'(p))(v)).
\]
We conclude that it is a homeomorphism and indeed a diffeomorphism which is
linear in the fibers. We define an atlas on $TM$ by declaring that the
functions $F_\varphi:\pi^{-1}(U)\rightarrow U\times\RR^{m}$ are smooth. It
only remains to show that $TM$ is a Hausdorff second countable space. We will
first show that it is Hausdorff. Let us take $[\gamma],[\mu]\in TM$. If
$p=\pi([\gamma])\neq\pi([\mu])=q$ then, since $M$ is Hausdorff, there exists
disjoint open sets $q\in U_{q},p\in U_{p}$ and since $\pi$ is continuous, the
open sets $\pi^{-1}(U_{p})$ and $\pi^{-1}(U_{q})$ separate $[\gamma]$ and
$[\mu]$. In case $p=q$ we consider the homeomorphism $\pi^{-1}(U)\cong
U\times\RR^{m}$ induced by the choice of local coordinates. Since $U$
is Hausdorff, this shows that $[\gamma]$ and $[\mu]$ can be separated in
$TM$. Finally, let us show that $TM$ is second countable. Consider a countable
basis $\left\{  U_{\alpha}\right\}  _{\alpha\in\mathcal{A}}$ for $M$ such that
each element of the basis is the domain of a coordinate chart and therefore
$\pi^{-1}(U_{\alpha})\cong U_{\alpha}\times\RR^{m}.$ For each $\alpha$
we take a countable basis $\left\{  W_{\beta}^{\alpha}\right\}  $ of $\pi
^{-1}(U_{\alpha})$ so that $\left\{  W_{\beta}^{\alpha}\right\}  $ is a
countable basis for $TM$.

\begin{definition}
The tangent bundle of a manifold $M$ is the vector bundle $TM$. A vector field
over $M$ is a section of the tangent bundle. The set of all vector fields over
$M$ is denoted by $\mathfrak{X}(M)=\Gamma(TM).$
\end{definition}

\begin{notation}
We have seen that given a chart $\varphi:U\rightarrow V\subset\RR^{n}$
there exists an identification
\[
D\varphi:TM|_{U}\cong TU\rightarrow U\times\RR^{m},
\]
which induces an isomorphism at the level of sections: \[\mathfrak{X}%
(U)\cong\Gamma(U\times\RR^{m})\simeq C^{\infty}(M,\RR^{m}).\]

\begin{figure}[H]
 \centering
 \includegraphics[scale=0.58]{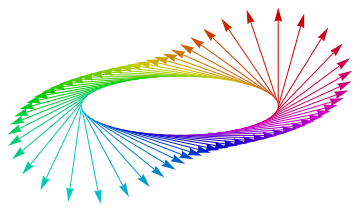}
 \caption{Vector field on an ellipse.}
\end{figure}

\begin{figure}[H]
 \centering
 \includegraphics[scale=0.5]{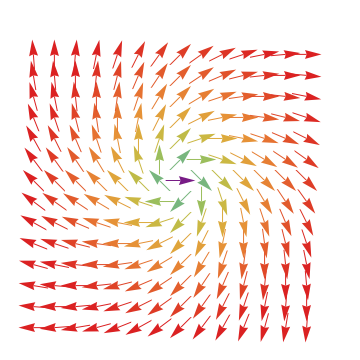}
 \caption{Vector field on the plane.}
\end{figure}

It is usual to denote by $\frac{\partial}{\partial x^{i}}$ the vector field that
corresponds to the constant function with value $e_{i}$ under this
isomorphism. Thus, we see that a vector field over $U$ can be written uniquely
in the form
\[X=\sum_{i}X^{i}\frac{\partial}{\partial x^{i}}.\]
We will also use the following shorthand notations for vector fields in local coordinates:
\[X=\sum_i X^{i} \frac{\partial}{\partial x^{i}}=\sum_i X^{i} \partial_{x^{i}}=\sum_i X^{i}\partial_{i}.\]
\end{notation}

Geometrically, a vector field is a smooth choice of a direction of movement for each
point in $M$. We have seen that, in general, the set $\Gamma(E)$ is a module
over the ring $C^{\infty}(M)$. In case $E=TM$, the space of sections has
an additional algebraic structure, $\mathfrak{X}(M)$ is a Lie algebra.

\begin{definition}
A Lie algebra is a vector space $\mathfrak{g}$ together with a bilinear map $[\,,\,]:\mathfrak{g}\otimes
\mathfrak{g}\rightarrow \mathfrak{g}$ such that:
\begin{itemize}
\item $[\,,\,]$ is skew symmetric, that is \[[x,y]+[y,x]=0.\]
\item $[\,,\,]$ satisfies the Jacobi identity, that is
\[[x,[y,z]]+[z,[x,y]]+[y,[z,x]]=0.\]
\end{itemize}
A subalgebra of a Lie algebra is a vector subspace that is closed with respect
to the bracket.
\end{definition}

\begin{example}
If $A$ is an associative algebra then there exists a Lie algebra,
$\mathrm{Lie}(A$), defined as follows. As a vector space $\mathrm{Lie}(A)=A$.
The bracket is given by the commutator, $[a,b]=ab-ba.$
\end{example}

\begin{example}
Let $V$ be a vector space. Then the space of endomorphisms of $V$,
$\mathrm{End}(V)$ is an associative algebra and therefore $\mathrm{Lie}%
(\mathrm{End}(V))$ is a Lie algebra. If $V=\RR^n$ the Lie algebra $\mathrm{End}(\RR^n)$ is denoted
$\mathfrak{gl}(n,\RR)$.
\end{example}

\begin{definition}
A derivation $D$ of an associative algebra $A$ is a linear function
$D:A\rightarrow A$ such that $D(ab)=D(a)b+aD(b).$ We denote by $\mathrm{Der}%
(A)$ the space of all derivations of $A$.
\end{definition}

\begin{proposition}
$\mathrm{Der}(A)\subseteq\mathrm{End}(A)$ is a Lie subalgebra.
\end{proposition}

\begin{proof}
Let us take  $D,D'\in \mathrm{Der}(A)$ and show that  $[D,D']=DD'-D'D$ is a derivation of $A$.
Indeed
\begin{align*}
(DD'-D'D)(ab)&=D(D'(ab))-D'(D(ab))\\
&=D(D'(a)b+aD'(b))-D'(D(a)b+aD(b))\\
&=D(D'(a))b+D'(a)D(b)+D(a)D'(b)\\
&\quad \,+aDD'(b)-D'(D(a))b-D(a)D'(b)\\
&\quad \, -D'(a)D(b)-aD'(D(b))\\
&=[D,D'](a)\;b+a[D,D'](b),
\end{align*}
as required.
\end{proof}

We will now show that the space $\mathfrak{X}(M)$ of vector fields on $M$
admits an algebraic description as the space of derivations of the algebra
$C^{\infty}(M)$.

\begin{lemma}
\label{extendf} Let $U\subseteq M$ be an open subset. There exists a
unique linear map $\rho:\mathrm{Der}(C^{\infty}(M))\rightarrow\mathrm{Der}(C^{\infty}(U))$ with the property that for any $\delta\in\mathrm{Der}(C^{\infty}(M))$ and
$g\in C^{\infty}(U)$:\[\rho(\delta)(g)(p)=\delta(\tilde{g})(p),\] for any
function $\tilde{g}\in C^{\infty}(M)$ which coincides with $g$ in a
neighborhood of $p$. Moreover, the linear map $\rho$ is a morphism of Lie algebras.
\end{lemma}

\begin{proof}
First we will show that given  $g \in C^{\infty}(U)$ and $p \in U$ there exists an open $p \in W \subseteq U$ and a function $\tilde{g} \in C^{\infty}(M)$ such that
\[\tilde{g}\vert_W=g\vert_W.\]
Choose an open $p \in U' \subseteq U$ and a chart $ \varphi: U' \rightarrow \RR^m$. Fix a function $\lambda \in C^{\infty}(\RR^m)$ such that
\[ \lambda(x)= \begin{cases} 1, & \text{ if }x\in [-1,1]^m,\\
0, & \text{ if } x \notin [-2,2]^m, \end{cases}\]
and we set
\[ \tilde{g}(x)= \begin{cases} g(x) \lambda(\varphi(x)), & \text{ if }x\in U',\\
0,& \text{ if } x \notin U'. \end{cases}\]
The function $\tilde{g}$ is smooth and coincides with $g$ on $ W:= \varphi^{-1} \big((-1,1)^m\big)$.
Now we define
\[ \rho(\delta)(g)(p):=\delta (\tilde{g}) (p).\]
Let us see that the definition is independent of $\tilde{g}$. It suffices to show that if $h\in C^{\infty}(M)$ is such that $h\vert_W=0$ then $\delta(h)\vert_W =0$. Fix a point $ p \in W$. As before,  there exists a smooth function $\chi \in C^{\infty}(M)$ such that $\chi(x)=1$ if $x \notin W$ y $\chi(x)=0$ in a neighbourhood of $p$. Then $ h =\chi h$ and:
\[ \delta(h)(p)= \delta(\chi h)(p)= (\delta(\chi)h+\delta(h)\chi)(p)=0.\]
We conclude that $\rho$ is well defined. Let us prove that it is a morphism of Lie algebras. We compute 
\begin{eqnarray*}
\rho([\delta, \delta'])(g)(p)&=& [\delta, \delta'](\tilde{g})(p)\\&=& \delta ( \delta'(\tilde{g}))(p)-\delta' ( \delta(\tilde{g}))(p)\\
&=&\rho(\delta) ( \rho(\delta')(g))(p)-\rho(\delta')(\rho(\delta)(g))(p)\\
&=&[\rho(\delta), \rho(\delta')](g)(p),
\end{eqnarray*}
as wanted.
\end{proof}

\begin{lemma}
There exists a linear map $L:\mathfrak{X}(M)\longrightarrow\mathrm{Der}%
(C^{\infty}(M))$ given by $X\mapsto L_{X}$, where:
\[
(L_{X}f)(p)=\dfrac{d}{dt}\bigg|_{t=0}f\circ\gamma(t),
\]
for a curve $\gamma$ such that $X(p)=[\gamma]\in T_{p}M.$ Moreover, $L$ is a
morphism of $C^{\infty}(M)$-modules.
\end{lemma}

\begin{proof}
We need to prove that the map $L$ is well defined. It suffices to observe that \[(L_Xf)(p)=Df(X(p))\] with $Df(p): T_pM \longrightarrow T_{f(p)} \RR \cong \RR$. 
Let us now see that $L_X$ is a derivation. We compute \begin{align*}
(L_X (fg))(p)&=\dfrac{d}{dt} \bigg|_{t=0}(fg)\circ \gamma (t)\\
&= \dfrac{d}{dt} \bigg|_{t=0}f(\gamma (t))g(\gamma (t))\\
&=\dfrac{d}{dt} \bigg|_{t=0}f (\gamma (t) ) g(p) + f(p)g(\gamma (t)),
\end{align*}
that is, 
\[
(L_X (fg))(p)=((L_Xf)g + f(L_Xg))(p).
\]
In order to show that $L$ is linear on functions we compute
\begin{eqnarray*}
(L(fX))(g)(p)&=& Dg(p)(fX(p))\\
&=&D(g)(p)(f(p)X(p))\\
&=& f(p) Dg(p)(X(p))\\
&=&(f L_Xg)(p),
\end{eqnarray*}
as wished.
\end{proof}

\begin{lemma}
The homomorphism $L:\mathfrak{X}(M)\rightarrow\mathrm{Der}(C^{\infty}(M))$
commutes with restrictions, i.e. for any open $U\subseteq M$ the following
identity holds $\rho\circ L=L\circ\rho$.
\end{lemma}

\begin{proof}
On the one hand we have
\[ (L_{\rho(X)} g)(p)= Dg(p)(X(p)).\]
On the other hand
\[ (\rho (L_X)g)(p)= L_X (\tilde{g}) (p)=D(\tilde{g})(p) (X(p))=Dg(p)(X(p)).\]
This shows the desired result.
\end{proof}

\begin{lemma}
\label{cover} If there exists an open cover $\{U_{\alpha}\}_{\alpha
\in\mathcal{A}}$ of $M$ such that $L_{\alpha}:\mathfrak{X}(U_{\alpha
})\rightarrow\mathrm{Der}(C^{\infty}(U_{\alpha}))$ is an isomorphism for all
$\alpha\in\mathcal{A}$ then $L:\mathfrak{X}(M)\rightarrow\mathrm{Der}%
(C^{\infty}(M))$ is an isomorphism.
\end{lemma}

\begin{proof}
Let us prove that $L$ is inyective. If $L_X=0$, then $L_X \vert_{U_\alpha}=0$ for all $\alpha$. Therefore $L_{X\mid_{U_\alpha}}=0$. Since $L_\alpha$ is inyective we conclude that $X \vert_{U_\alpha}=0$ for all $\alpha\in \mathcal{A}$. This implies that $X=0$. Let us now prove surjectivity. Consider a derivation $\delta \in \mathrm{Der}(C^{\infty}(M))$  and set  $\delta_\alpha = \delta \vert_{U_\alpha}$. By assumption there exist vector fields  $X_\alpha$ such that $\delta_\alpha = L_{X_\alpha}$. We define $X$ by \[X(p):= X_\alpha(p),\]
for any $\alpha$ such that $p \in U_\alpha$. It is easy to check that $X$ is well defined and $L_X=\delta$.
\end{proof}

\begin{theorem}
The linear map $L:\mathfrak{X}(M)\longrightarrow\mathrm{Der}(C^{\infty}(M))$
is an isomorphism of $C^{\infty}(M)$-modules.
\end{theorem}

\begin{proof}
In view of Lemma \ref{cover} it is enough to consider the case $M=\RR^m$. We have seen that in this case any vector field can be written uniquely in the form \[X=\sum_i X^i \dfrac{\partial}{\partial x^i},\]
with $X^{i} \in C^{\infty}(\RR^{m})$. Moreover
\[L_Xf= \sum_i X^i \dfrac{\partial f}{\partial x^i}.\] Let us show that $L$ is injective. If $L_X=0$, then \[\sum_i X^i \dfrac{\partial f}{\partial x^i} =0\] for any function $f \in C^{\infty}(M)$. Setting $f=x^i$, this implies that each function $X^i=0$, and therefore $X=0$. Let us now show that $L$ is surjective.  For a derivation $\delta \in \mathrm{Der}(C^{\infty}(\RR^m))$ we want to show that $\delta = L_Y$. Note that \[L_X x^j=\sum_i X^i \dfrac{\partial x^j}{\partial x^i}=X^j.\] Let us set $Y^i=\delta (x^i)$ and \[Y= \sum_i Y^i \dfrac{\partial}{\partial x^i}.\] We claim that $L_Y= \delta$.
Let us fix a function $f$,  a point $ p \in M$ and a path $\gamma(t)=(1-t)p+tx$. Using the fundamental theorem of calculus we compute.
\[ \int_0^1 (f\circ  \gamma)'(t)dt=  f(x)-f(p).\] Therefore
\[f(x)= f(p)+ \int_{0}^{1} \frac{d}{dt} f(tx + (1-t)p)dt.\] Expanding the derivative we obtain
\begin{eqnarray*}
f(x)&=& f(p) + \sum_i \int_{0}^{1} \dfrac{\partial f}{\partial x^i}(\gamma(t))(x^i - p^i)dt\\
&=& f(p)+ \sum_i (x^i - p^i) \int_{0}^{1}\dfrac{\partial f}{\partial x^i} (\gamma(t))dt.
\end{eqnarray*}
Applying $\delta$ on both sides, we obtain:
\begin{eqnarray*}
\delta (f)(x)&=& \sum_i \delta \Big((x^i -p^i) \int_{0}^{1} \dfrac{\partial f}{\partial x^i}(xt+(1-t)p)dt\Big) \\   &=&\sum_i Y^i \int_{0}^{1} \dfrac{\partial f}{\partial x^i} (\gamma(t))dt + \sum_i (x^i - p^i)\delta \Big( \int_{0}^{1} \dfrac{\partial f}{\partial x^i}(\gamma(t))dt \Big).
\end{eqnarray*}
Finally, we evaluate at $x=p$ to obtain:
\[\delta (f)(p)=\sum_i Y^i (p)\int_{0}^{1} \dfrac{\partial f}{\partial x^i} (p) dt= \sum_i Y^i (p) \dfrac{\partial f}{\partial x^i} (p)=(L_Y f)(p). \]
This completes the proof.
\end{proof}

\begin{corollary}
The isomorphism $L: \mathfrak{X} (M)\rightarrow\mathrm{Der} (C^{\infty}(M))$
gives the vector space $\mathfrak{X}(M)$ the structure of a Lie algebra.
\end{corollary}

The natural question arises of describing the bracket of vector fields more explicitly. This can be done as follows.

\begin{lemma}
The bracket of vector fields on $\RR^{m}$ is characterized by the
following properties:
\[\bigg[\dfrac{\partial}{\partial x^{i}},\dfrac{\partial}{\partial x^{j}%
}\bigg]=0, \qquad [X,fY]=f[X,Y]+(L_{X}f)Y.\]

\end{lemma}

\begin{proof}
Given two vector fields  \[X=\sum_i X^i\dfrac{\partial}{\partial x^i}, \qquad Y=\sum_j Y^j\dfrac{\partial}{\partial x^j},\] the conditions above imply:
\begin{eqnarray*}
[X,Y]&=& \sum_{i,j} \bigg[X^i\dfrac{\partial}{\partial x^i} , Y^j\dfrac{\partial}{\partial x^j}\bigg]\\
&=& \sum_{i,j} Y^j \bigg[X^i\dfrac{\partial}{\partial x^i} , \dfrac{\partial}{\partial x^j}\bigg] +X^i\dfrac{\partial Y^j}{\partial x^i} \dfrac{\partial}{\partial x^j}\\
&=& \sum_{i,j} -Y^j \dfrac{\partial X^i}{\partial x^j}\dfrac{\partial}{\partial x^i }+X^i\dfrac{\partial Y^j}{\partial x^i} \dfrac{\partial}{\partial x^j}\\
&=& \sum_i \left( \sum_j  \dfrac{\partial Y_i}{\partial x^j}X^j -\sum_j Y^j \dfrac{\partial X^i}{\partial x^j} \right) \dfrac{\partial}{\partial x^i}.
\end{eqnarray*}
This shows uniqueness. For existence, it suffices to show that the bracket induced by the isomorphism $L$ satisfies the conditions above.  The first condition is verified because partial derivatives commute.
For the second equation we compute
\begin{eqnarray*}
[L_X, L_{fY}]g&=& (L_X \circ L_{fY})g-(L_{fY} \circ L_X)g\\
&=&(L_X(f (L_{Y}g)))-f L_{Y}( L_X g)\\
&=&(L_Xf) (L_Yg)+f L_X (L_{Y}g)-f L_{Y}( L_X g)\\
&=&(L_Xf) (L_Yg)+f [L_X,L_Y]g,
\end{eqnarray*}
as required.
\end{proof}

\begin{remark}
For $X=\sum_{i}X^{i}\dfrac{\partial}{\partial x^{i}}$ and $Y=\sum_{j}%
Y^{j}\dfrac{\partial}{\partial x^{j}}$, the bracket between $X$ and $Y$ is
\[
\lbrack X,Y]=\sum_{i}\left(\sum_{j}\dfrac{\partial Y^{i}}{\partial x^{j} %
}X^{j}-\sum_{j} Y^{j}\dfrac{\partial X^{i}}{\partial x^{j}}\right)\dfrac{\partial}{\partial
x^{i}}.
\]

\end{remark}

We have seen that, in $U\subseteq\RR^{m}$, vector fields can be written
in the form \[X=\sum_{i}X^{i}\dfrac{\partial}{\partial x^{i}},\] with $X^{i}\in
C^{\infty}(U)$. Suppose that $M$ is a manifold and let $\varphi=(x^{1},\dots,x^{m})$ and $\bar{\varphi}=(\bar{x}^1,\dots,\bar{x}^{m})$ be two coordinates systems on $M$. A vector field $X\in\mathfrak{X}(M)$ can be written in two different ways: 
\[X=\sum_{i}X^{i}\dfrac{\partial}{\partial x^{i}}=\sum_{j}\bar{X}^{j}\dfrac{\partial}{\partial \bar{x}^{j}}.\] It is natural to ask what the
relationship is between the functions $X^{i}$ and $\bar{X}^{j}$. The chain rule
implies \[\dfrac{\partial}{\partial x^{i}}=\sum_{j}\dfrac{\partial \bar{x}^{j}%
}{\partial x^{i}}\dfrac{\partial}{\partial \bar{x}^{j}}.\]
Substituting in the equality above we obtain
\[
X=\sum_{i}X^{i}\dfrac{\partial}{\partial x^{i}}=\sum_{i}X^{i}\left(\sum_{j}\dfrac{\partial \bar{x}^{j}%
}{\partial x^{i}}\dfrac{\partial}{\partial \bar{x}^{j}}\right)=\sum_{j}\left(\sum_{i}\dfrac{\partial \bar{x}^{j}}{\partial x^{i}%
}X^{i}\right)\dfrac{\partial}{\partial \bar{x}^{j}}.
\]
One concludes that 
\[\bar{X}^{j}=\sum_{i}\dfrac{\partial \bar{x}^{j}}{\partial x^{i}} X^{i}.\]

\section{Vector fields and flows}

\begin{definition} A flow on $M$ is an action of the group $\RR$ on $M$, that is, a smooth function \[ \begin{array}{ccccl}
H : \RR \times M  &\rightarrow&  M ,\\ 
 (t,p) &\mapsto&  H(t,p)
\end{array}\] such that 
 $H(0,p)=p$
 and $H(s+t,p)=H(s,H(t,p)).$
\end{definition}

\begin{definition}
Given a flow $H$ on $M$, the vector field $X$ induced by $H$ is
 \[X(p)= \dfrac{d}{dt}\Bigr|_{t=0}H(t,p).\]
The vector field $X$ is called the infinitesimal generator of $H$. 
\end{definition}

Not every vector field is the infinitesimal generator of a flow. For instance, take  $M=\RR\setminus\{0\}$. 
Seen as a vector field on $\RR$, $X=\frac{\partial}{\partial x}$ generates the flow:
\[ H(t,x)=x+t.\]
This implies that $H(1,-1)=0 \notin M$. Therefore $X$ does not generate a flow on $M$. In this situation one says that the solution goes to infinity in finite time. It turns out that this is the only way in which a vector field can fail to generate a flow. In general, a vector field does generate a local flow.

\begin{definition}
A local flow on $M$ is an open subset $\Omega \subseteq \RR \times M$ that contains  $\lbrace 0\rbrace \times M$ and intersects each $\RR \times \lbrace p\rbrace $ in an interval, together with a smooth map 
\begin{eqnarray*}
H :\Omega  &\rightarrow&  M \\ 
  (t,p)  &\mapsto&  H(t,p)
\end{eqnarray*}
such that $H(0,p)=p$ and
 $H(s+t,p)=H(s,H(t,p))$, when both sides are defined.
\end{definition}

The infinitesimal generator of a local flow is defined in the same way as that of a flow.
The Picard Lindel\"of theorem discussed in \ref{PL} implies the following:
\begin{proposition}\label{unicidad} If $H$ and $\Gamma$ are two local flows which have the same infinitesimal generator then
they coincide in the intersection of their domains.
\end{proposition}

\begin{definition}
A local flow $H$ generated by $X$ is called maximal if any other local flow generated by $X$  has domain contained in that of $H$.
\end{definition}

By Proposition \ref{unicidad}, any local flow is contained in a unique maximal local flow. 

\begin{theorem}
The function that assigns to a maximal local flow its infinitesimal generator is a bijection between maximal local flows and vector fields.
\end{theorem}

\begin{proof}
By Proposition \ref{unicidad} the correspondence is injective. It remains to show that any vector field generates a local flow.
Since this is a local statement, it suffices to prove it for an open subset $U$ of $\RR^m$.
Let us consider a vector field $X$ on $U$. By the Picard-Lindel\"of theorem, there exists an open covering  $\{U_p\}_{p \in U}$ and numbers $\epsilon^p > 0$ such that for all $a\in U_p$ there exists a unique solution  $\gamma_a\colon (-\epsilon^p, \epsilon^p)\rightarrow U$, to the equations $\gamma_a(0) = a$ and $\dot{\gamma_a}=X(\gamma_a)$. 
Define $H$ as follows: Put $\Omega:= \bigcup_{p\in U} (-\epsilon^p, \epsilon^p)\times U_p$ and \[ H(t,a)= \gamma_a(t).\] 
The function $H$ is well defined by the uniqueness part of the Picard-Lindel\"of theorem. It remains to show that $H$ is a local flow. Clearly \[H(0,a)=\gamma_a(0)=a.\] It remains to show that  \[H(s+t,a)=H(s,H(t,a)).\] Fix $t,a$ and consider the following functions of $s$:  \[\eta(s) = H(s+t,a)); \, \omega (s)= H(s,H(t,a)) .\] We want to show that $\eta , \omega$ are integral curves of $X$ with the same initial conditions. On the one hand,  \[\eta (0) = H(t,a)= \omega(0).\] Next, we compute the derivatives  \[\dfrac{d}{ds} \omega(s)= \dfrac{d}{ds} H(s,H(t,a))=\dfrac{d}{ds}\gamma_{H(t,a)}(s)=X(\gamma_{H(t,a)}(s))= X(\omega(s)),\] 
and 
\[\dfrac{d}{ds} \eta(s)= \dfrac{d}{ds} H(s+t,a)= \dfrac{d}{ds} \gamma_a (s+t) = X(\gamma_a(t+a))=X(H(s+t,a))=X(\eta(s)).\]
This finishes the proof.
\end{proof}

Let us consider $M=\RR^2$. Consider the vector fields $E,X \in \mathfrak{X} (M)$ given by \[E=x\dfrac{\partial}{\partial x} + y \dfrac{\partial}{\partial y} \quad \text{and} \quad X=\dfrac{\partial}{\partial \theta}.\]
Recall that  $x=r\cos \theta$ and $y=r\sin \theta$ and therefore \[\dfrac{\partial}{\partial \theta}= \dfrac{\partial x}{\partial \theta} \dfrac{\partial}{\partial x} + \dfrac{\partial y}{\partial \theta} \dfrac{\partial}{\partial y} = -y \dfrac{\partial}{\partial x} + x \dfrac{\partial}{\partial y}.\]
We now compute $[E,X]$:  
\begin{align*}
[E,X]&= \left[x\dfrac{\partial}{\partial x} + y \dfrac{\partial}{\partial y} , -y \dfrac{\partial}{\partial x} + x \dfrac{\partial}{\partial y} \right] \\
&=-\left[x\dfrac{\partial}{\partial x} , y\dfrac{\partial}{\partial x}\right]+\left[x\dfrac{\partial}{\partial x}, x \dfrac{\partial}{\partial y}\right]-\left[y\dfrac{\partial}{\partial y}, y\dfrac{\partial}{\partial x}\right] + \left[y\dfrac{\partial}{\partial y}, x\dfrac{\partial}{\partial y}\right] \\
&=y\dfrac{\partial}{\partial x} + x \dfrac{\partial}{\partial y} - y\dfrac{\partial}{\partial x} - x\dfrac{\partial}{\partial y} \\
&=0.
\end{align*}
This can also be computed as follows. One observes that $E=r\dfrac{\partial}{\partial r}$, so that
\[
[X,E]=\left[\dfrac{\partial}{\partial \theta},r\dfrac{\partial}{\partial r}\right]=r\left[\dfrac{\partial}{\partial \theta} , \dfrac{\partial}{\partial r}\right] + \dfrac{\partial r}{\partial \theta} \dfrac{\partial}{\partial r}=0.
\]
We now consider the flows associated to $X$ and $E$. Using the identification $\RR^2\cong \CC$
we set $H^{E}: \RR \times \CC \rightarrow \CC$ by $H(t,z)=e^{t}z$ for each $(t,z)\in \RR \times \CC$. Observe that
\begin{align*}
 H(0,t)&=e^{0}z=z,\\
 H(t,H(s,z))&=e^{t}e^{s}z=e^{t+s}z,
\end{align*}
and therefore $H$ is a flow on $\CC\cong \RR^2$. Let us compute the generator of $H$. We have
\begin{align*}
 \dfrac{d}{dt}\bigg\vert_{t=0}e^{t}z = e^{0}z =z.
\end{align*}
We conclude that $H$ is generated by $E$.
Let us also define $F: \RR \times \CC \rightarrow \CC$ by setting $F(t,z)= e^{it}z$ for each $(t,z) \in \RR \times \CC$.
The flow $F$ is generated by
\begin{equation*}
\frac{d}{dt}\bigg\vert_{t=0} F(t,z) = \frac{d}{dt}\bigg\vert_{t=0}  e^{it}z = iz.
\end{equation*}
We conclude that the vector field $X$ generates $F$.
Notice that $H$ and $F$ commute, that is, \[F_{s}H_{t}(z)=e^{t+is}z=H_{t}F_{s}(z),\] 
for all $s,t \in \RR$. This is not a coincidence, we will see that given two vector fields $X$ and $Y$, the corresponding local flows commute precisely when $[X,Y]=0$.

\begin{figure}[H]
 \centering
 \includegraphics[scale=0.55]{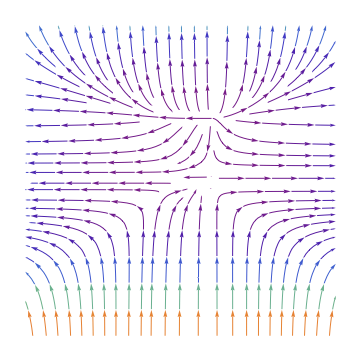}
 \caption{Flow lines of a vector field on the plane.}
\end{figure}

 Let $\phi: M \rightarrow N$ be a diffeomorphism and  $X\in \mathfrak{X}(M)$. The push forward of $X$ with respect to $\phi$, denoted $\phi_{*}X$, is the vector field on $N$ defined by \[(\phi_{*}X)(q)=D\phi \left(\phi^{-1}(q)\right)X\left(\phi^{-1}(q)\right).\]
Given a vector field $Y \in \mathfrak{X}(N)$ we define the pull-back, denoted $\phi^* Y$, as follows
\begin{align*}
& \phi^{*}Y=(\phi^{-1})_{*}Y.
\end{align*}
Given a smooth function $f\in C^{\infty}(M)$, the push forward of $f$ with respect to to $\phi$ is the function $\phi_{*}f=f\circ \varphi^{-1} \in C^{\infty}(N)$
For a smooth function $g\in C^{\infty}(N)$ we define the pull-back by $\phi^{*}g=g\circ \phi \in C^{\infty}(M)$.
Note that the pull-back of a function is defined for an arbitrary smooth function $\phi$ which is not necessarily a diffeomorphism.

\begin{lemma} Let $\phi: M \rightarrow N$ be a diffeomorphism, $X\in \mathfrak{X}(M)$ and $ f \in C^{\infty}(N)$.
Then \[(\phi_{*}X)g=\phi_{*}(X(\phi^{*}g))\in C^{\infty}(N).\]
\end{lemma}

\begin{proof}
Evaluating the left hand side at $q \in N$ one obtains
\begin{align*}
(\phi_{*}X)f(q)&=Df(q)\left((\phi_{*}X)(q)\right)\\
&= Df(p)\circ D\phi \left(\phi^{-1}(q)\right)(X\left(\phi^{-1}(q))\right).
\end{align*}
On the other hand, the right hand side at $q \in N$ is
\begin{align*}
\left(\phi_{*}(X(\phi^{*}f))\right)(q)&= X\left(\phi^{*}f\right)\left(\phi^{-1}(q)\right)\\
& = D\left(\phi^{*}f\right)\left(\phi^{-1}(q)\right)\left(X(\phi^{-1}(q))\right)\\
&=D(f\circ \phi) \left(\phi^{-1}(p)\right)\left(X(\phi^{-1}(q))\right)\\
&= Df(q)\circ D\varphi \left(\varphi^{-1}(q)\right)\left(X(\phi^{-1}(q))\right).
\end{align*}
Comparing these two equalities, we get the desired result. 
\end{proof}

Let $\phi: M \rightarrow N$ be a diffeomorphism and  $\delta \in \mathrm{Der}( C^{\infty}(M))$ a derivation of the algebra of functions on $M$. The push-forward of $\delta$, written $ \phi_*\delta$, is the element of $\mathrm{Der}( C^{\infty}(N))$ given by
\[  (\phi_*\delta) g= \phi_*( \delta(\phi^*g)),\]
for all $g \in C^{\infty}(N)$. Let us see that the identification between vector fields and derivations is compatible with the push-forward operation.
\begin{lemma}
Let $\phi: M \rightarrow N$ be a diffeomorphism and $X\in \mathfrak{X}(M)$. Then
\[ \phi_* L_X=L_{\phi_* X}.\]
\end{lemma}
\begin{proof}
Let us evaluate both sides of the equation on $g \in C^{\infty}(N)$. On the one hand,
\begin{eqnarray*}
(\phi_*L_X)(g)(q)&=& \phi_*( L_X(\phi^*g))(q)\\
&=& L_X(\phi^*g)(\phi^{-1}(q))\\
&=& D(\phi^*g) (\phi^{-1}(q)) (X(\phi^{-1}(q)))\\
&=&D(g \circ \phi) (\phi^{-1}(q)) (X(\phi^{-1}(q)))\\
&=& Dg (q) \circ D\phi( \phi^{-1}(q)) (X( \phi^{-1}(q))).
\end{eqnarray*}
On the other hand,
\begin{eqnarray*}
L_{\phi_* X}(g)(q)&=& Dg(q) \phi_*(X)(q)\\
&=& Dg(q) \circ D\phi ( \phi^{-1}(q)) (X( \phi^{-1}(q))).
\end{eqnarray*}
This proves the result.
\end{proof}

\begin{lemma}
Let $X$ and $Y$ be vector fields on $M$, let $H$ be the local flow that $X$ generates, and let  $f\in C^{\infty}(M)$ be a smooth function. Then
\begin{equation} \frac{d}{dt}\bigg\vert_{t=0} (H_{t}^{*}f)(p)=(Xf)(p),
\end{equation}
and
\begin{equation}
\frac{d}{dt}\bigg\vert_{t=0} (H_{t}^{*}Y)(p)=[X,Y](p).
\end{equation}
\end{lemma}

\begin{proof}
For the first statement, we compute:
\begin{align*}
\frac{d}{dt}\bigg\vert_{t=0} (H_{t}^{*}f)(p)&= \frac{d}{dt}\bigg\vert_{t=0}f(H(t,p))\\
&=Df(p)\left(\frac{d}{dt}\bigg\vert_{t=0}H(t,p)\right)\\
&= Df(p) (X(p))\\
&= (Xf)(p).
\end{align*}
For the second statement it suffices to show that the two vector fields induce the same derivation. Take a function $g\in C^{\infty}(M)$ and compute:
\begin{align*}
\frac{d}{dt}\bigg\vert_{t=0} (H_{t}^{*}Y)g&= \frac{d}{dt}\bigg\vert_{t=0}((H_{t}^{-1})_{*}Y)g\\
&= \frac{d}{dt}\bigg\vert_{t=0}((H_{-t})_{*}Y)g\\
&= \frac{d}{dt}\bigg\vert_{t=0} (H_{-t})_{*}\left(Y(H_{-t}^{*}g)\right)\\
&= \frac{d}{dt}\bigg\vert_{t=0}\,Y(H_{-t}^{*}g)+ \frac{d}{dt}\bigg\vert_{t=0} (H_{-t})_{*}(Yg)\\
&=  \frac{d}{dt}\bigg\vert_{t=0} Y (H_{-t}^{*}g) + \frac{d}{dt}\bigg\vert_{t=0} H_{t}^{*}(Yg)\\
&= -Y(Xg) + X(Yg)\\
&= [X,Y]g.
\end{align*}

\end{proof}

\begin{lemma}
	If $H$ is the local flow generated by $X$ then \[\frac{d}{dt}\bigg\vert_{t=t_{0}}H(t,p)= X(H(t_{0},p)).\]
\end{lemma}
\begin{proof}
By computing, using the properties of the flow $H$, we find
\begin{align*}
\frac{d}{dt}\bigg\vert_{t=t_0}\,H(t,p) &= \frac{d}{dt}\bigg\vert_{t=t_{0}}\, H(t-t_{0},H(t_{0},p))\\
&= \frac{d}{ds}\bigg\vert_{s=0}\,H(s,H(t_0,p))\\
&=X(H(t_0,p)),
\end{align*}
as wished.
\end{proof}
\begin{lemma}
If $H$ is the local flow generated by $X$  and $[X,Y]=0$, then
\[H_{t}^{*}Y=Y.\]
\end{lemma}

\begin{proof}
 It suffices to show that $\frac{d}{dt}\big\vert_{t=t_0}(H_{t}^{*}Y)(p)=0$. In fact,
 \begin{align*}
 \frac{d}{dt}\bigg\vert_{t=t_{0}}(H_{t}^{*}Y)(p)&= \frac{d}{dt}\bigg\vert_{t=t_{0}}((H_{t-t_{0}}\circ H_{t_{0}})^{*}Y)(p)\\
 &= \frac{d}{dt}\bigg\vert_{t=t_0}\left(H_{t_{0}}^{*}(H_{t-t_{0}}^{*}Y)\right)(p)\\
 &=\frac{d}{dt}\bigg\vert_{s=0}\left(H_{t_{0}}^{*}(H_{s}^{*}Y)\right)(p)\\
 &=D(H_{-t_0})\left(\frac{d}{ds}\bigg\vert_{s=0}(H_{s}^{*}Y)(p)\right)\\
 &= D(H_{-t_{0}})([X,Y](p)) \\
 &=0.
 \end{align*}
 
\end{proof}

\begin{theorem} Let $H^{X}$ y $H^{Y}$ be the local flows generated by $X$ y $Y$ respectively. Then $[X,Y]=0$ if and only if
\[H^{X}_{t}\circ H^{Y}_{s}=H^{Y}_{s}\circ H^{X}_{t},\]
for all $s,t \in \RR$ where both sides are defined.
\end{theorem}
\begin{proof}
Let us first assume that the flows commute. Then we have
\begin{align*}
[X,Y](p)&= \frac{d}{dt}\bigg\vert_{t=0}(H_{t}^{X*}Y)(p)\\
&= \frac{d}{dt}\bigg\vert_{t=0}((H_{-t}^{X})_{*}Y)(p)\\
&= \frac{d}{dt}\bigg\vert_{t=0}(DH^{X}_{-t})(H^{X}_{t}(p))(Y(H^{X}_{t}(p)))\\
&=\frac{d}{dt}\bigg\vert_{t=0}(DH^{X}_{-t})(H^{X}_{t}(p))\left(\frac{d}{ds}\bigg\vert_{s=0}H^{Y}_{s}(H^{X}_{t}(p))\right)\\
&=\frac{d}{dt}\bigg\vert_{t=0} \frac{d}{ds}\bigg\vert_{s=0} H^{X}_{-t}(H^{Y}_{s}(H^{X}_{t}(p))) \\
&=\frac{d}{dt}\bigg\vert_{t=0}\, \frac{d}{ds}\bigg\vert_{s=0}H^{Y}_{s}(p)=0.
\end{align*}
Let us now consider the other direction.
In the computation above, we showed that \[[X,Y](p)=\dfrac{d}{dt}\bigg\vert_{t=0}\dfrac{d}{ds}\bigg\vert_{s=0}H^{X}_{-t}(H^{Y}_{s}(H^{X}_{t}(p))).\] 
Let us assume that the vector fields commute. For this, fix $p$. We need to show that
\begin{equation}\label{flowcom}
H^{X}_{-t}(H^{Y}_{s}(H^{X}_{t}(H^{Y}_{-s}(p))))=p.
\end{equation}
It is clearly enough to show that
\begin{equation*}
\frac{d}{ds}\bigg\vert_{s=s_0}H^{X}_{-t}(H^{Y}_{s}(H^{X}_{t}(H^{Y}_{-s}(p))))=0.
\end{equation*}
We first compute
\begin{align*}
&\frac{d}{ds}\bigg\vert_{s=s_{0}} H^{X}_{-t}(H^{Y}_{s}(H^{X}_{t}(H^{Y}_{-s}(p)))) \\
&\qquad = \frac{d}{ds}\bigg\vert_{s=s_{0}} H^{X}_{-t}(H^{Y}_{s_0}(H^{X}_{t}(H^{Y}_{-s}(p))))+ \frac{d}{ds}\bigg\vert_{s=s_{0}}H^{X}_{-t}(H^{Y}_{s}(H^{X}_{t}(H^{Y}_{-s_0}(p)))).
\end{align*}
Lets now examine the first summand. We have 
\begin{align*}
\frac{d}{ds}\bigg\vert_{s=s_{0}} H^{X}_{-t}(H^{Y}_{s_0}(H^{X}_{t}(H^{Y}_{-s}(p))))&= DH^{X}_{-t}\left( DH^{Y}_{s_0}\left( DH^{X}_{t}\left(\frac{d}{ds}\bigg\vert_{s=s_0}H^{Y}_{-s}(p)\right) \right)\right)\\
&=-DH^{X}_{-t} (DH^{Y}_{s_0} (DH^{X}_{t}(Y(H^{Y}_{-s_0}(p)))))\\
&= -(H^{X}_{-t})_{*}( (H^{Y}_{s_0})_{*}( (H^{X}_{t})_{*}(Y(H^{X}_{-t}(H^{Y}_{s_0}(H^{X}_{t}(H^{Y}_{-s_0}(p)))))))\\
&=- Y(H^{X}_{-t}(H^{Y}_{s_0}(H^{X}_{t}(H^{Y}_{-s_0}(p))))).
\end{align*}
Next, we examine the second summand. Observe that, since $[X,Y]=0$, then $(H_{-t})_{*}Y=Y$. Then
\begin{align*}
\frac{d}{ds}\bigg\vert_{s=s_{0}}H^{X}_{-t}(H^{Y}_{s}(H^{X}_{t}(H^{Y}_{-s_0}(p))))&=DH^{X}_{-t}\left(\frac{d}{ds}\bigg\vert_{s=s_0}H^{Y}_{s}(H^{X}_{t}(H^{Y}_{-s_0}(p)))\right) \\
&= DH^{X}_{-t}\left(Y(H^{Y}_{s_{0}}(H^{X}_{t}(H^{Y}_{-s_0}(p))))\right)\\
&= Y(H^{X}_{-t}(H^{Y}_{s_0}(H^{X}_{t}(H^{Y}_{-s_0}(p))))).
\end{align*}
Putting these two equalities together, we get the desired result.
\end{proof}

\section{The cotangent bundle and tensor fields}\label{sec:cotangent}

The cotangent bundle of $M$, denoted $T^{*}M$, is the vector bundle dual to the
tangent bundle. A section of the cotangent bundle is called a $1$-form. We
will denote by $\Omega^{1}(M)=\Gamma(T^*M)$ the space of all differential $1$-forms on $M$.

If $f \in C^{\infty}(M)$ is a function, its derivative $Df$ at each point is a linear map from $T_pM$ to the real numbers. This defines a differential $1$-form
which is also denoted $df$.

Let $\varphi=(x^{1},\dots,x^{m}) \colon U \rightarrow V$ be a coordinate system on $M$. The choice of coordinates  induces an isomorphism of vector bundles $T^*U \to U \times \RR^m$ where $(p,\alpha)$ is mapped to $(p,\alpha_1,\dots,\alpha_m)$, where $\alpha = \sum_{j} \alpha_j dx^{j}$, $dx^{j}$ being the basis dual to $\frac{\partial}{\partial x^{j}}$, that is
$$
dx^{j} \left(\frac{\partial}{\partial x^{j}} \right) = \delta^{j}_{\phantom{j}i}.
$$
This means that any $1$-form $\alpha\in\Omega
^{1}(U)$ can be written uniquely in the form \[\alpha=\sum_{i}\alpha_{i}dx^{i}.\]
Again, one would like to know how the functions $\alpha_{i}$ change for different
choices of coordinates. Let $\varphi=(x^{1},\dots,x^{m})$ and
$\bar{\varphi}=(\bar{x}^{1},\dots,\bar{x}^{m})$ be two coordinate systems on $M$. Then, on the intersection of their domains,
\begin{equation*}
dx^{i}=\sum_{j}\frac{\partial x^{i}}{\partial\bar{x}^{j}}d\bar{x}^{j}, \label{oneform0}%
\end{equation*}
and therefore
\begin{equation*}
\alpha=\sum_{i}\alpha_{i}\left(\sum_{j}\frac{\partial x^{i}}{\partial \bar{x}^{j}}\partial\bar{x}^{j}%
\right)=\sum_{j}\left(\sum_{i}\alpha_{i}\frac{\partial x^{i}}{\partial\bar{x}^{j}}\right)d\bar{x}^{j}.
\end{equation*}
Thus
\begin{equation}\label{oneforms}
\bar{\alpha}_{j} = \sum_{i}\alpha_{i}\frac{\partial x^{i}}{\partial\bar{x}^{j}}.
\end{equation}

A tensor field $T$ of type $(p,q)$ is a section of the vector bundle
$TM^{\otimes p}\otimes T^{\ast}M^{\otimes q}$. The space of all such tensor fields will be  denoted by 
\[
\mathcal{T}^{(p,q)}(M)=\Gamma(TM^{\otimes p}\otimes T^{\ast}M^{\otimes q}).
\] 
As we have seen  before, the local coordinates $\varphi=(x^{1},\dots,x^{m})$ around a point in $M$ induce local trivializations on the vector  bundles $TM$ and $T^*M$. These trivializations give rise to a basis in $\mathcal{T}^{(p,q)}(M)$. An arbitrary tensor field $T \in \mathcal{T}^{(p,q)}(U)$ can be expressed in terms of this basis as
\[
T=\sum_{i_1,\dots,i_p}\sum_{j_1,\dots,j_q}T^{i_{1}\cdots i_{p}}_{\phantom{i_{1}\cdots i_{p}}j_{1}\cdots j_{q}}\frac{\partial}{\partial x^{i_1}}\otimes
\cdots\otimes\frac{\partial}{\partial x^{i_p}}\otimes dx^{j_{1}}\otimes\cdots\otimes
dx^{j_{q}}.
\]
The functions $T^{i_{1}\cdots i_{p}}_{\phantom{i_{1}\cdots i_{p}}j_{1}\cdots j_{q}}$
are called the local components of the tensor $T$ in the coordinates
$\varphi=(x^{1},\dots,x^{m})$. We can regard functions as tensor fields of type $(0,0)$, vector fields as tensor fields of type $(1,0)$ and $1$-forms as tensor fields of type $(0,1)$.

For each tensor field $T$ of type $(p,q)$ on $M$ we have a $C^{\infty}(M)$-multilinear map
$$
\tilde{T} \colon  \underbrace{\Omega^1(M) \times \cdots \times \Omega^1(M)}_{p} \times \underbrace{\mathfrak{X}(M)\times \cdots \times \mathfrak{X}(M)}_{q} \to C^{\infty}(M)
$$
defined, in terms of coordinates $\varphi=(x^{1},\dots,x^{m})$, by
$$
\tilde{T}(\alpha^1,\dots,\alpha^p,X_1,\dots,X_q) = \sum_{i_1,\dots,i_p} \sum_{j_1,\dots,j_q} T^{i_{1}\cdots i_{p}}_{\phantom{i_{1}\cdots i_{p}}j_{1}\cdots j_{q}} \alpha^{1}_{i_1} \cdots \alpha^{p}_{i_{p}} X^{j_1}_1 \cdots X^{j_q}_q,
$$
where $\alpha^{k} \in \Omega^1(M)$, $\alpha^{k} = \sum_i \alpha^{k}_{i} dx^{i}$, $X_{l} \in \mathfrak{X}(M)$, and $X_{l} = \sum_i X^{i}_{l}\partial /\partial x^{i}$. Conversely, each such $C^{\infty}(M)$-multilinear map arises from a unique tensor field in this way; Appendix \ref{0algebra basicos} contains a thorough discussion of the linear algebra involved in these manipulations. Hence we will not distinguish between the tensor field $T$ and the map $\tilde{T}$, and a tensor field of type $(p,q)$ on $M$ can be thought of as an operation on $p$ $1$-forms and $q$ vector fields yielding a smooth function on $M$.

It will be useful to have a formula that describes how the local components of a tensor transform for different choices of coordinates. Let $\varphi=(x^{1},\dots,x^{m})$ and $\bar{\varphi}=(\bar{x}^{1},\dots,\bar{x}^{m})$ be two coordinate systems on $M$ and $T$ a tensor field of type $(p,q)$. Then the components of $T$ in these two systems are  related by 
\begin{equation}\label{changetensor}
\bar{T}^{i_{1}\cdots i_{p}}_{\phantom{i_{1}\cdots i_{p}}j_{1}\cdots j_{q}} = \frac{\partial \bar{x}^{i_1}}{\partial x^{k_1}} \cdots \frac{\partial \bar{x}^{i_p}}{\partial x^{k_p}} T^{k_{1}\cdots k_{p}}_{\phantom{k_{1}\cdots k_{p}}l_{1}\cdots l_{q}}\frac{\partial x^{l_1}}{\partial \bar{x}^{j_1}}\cdots \frac{\partial x^{l_q}}{\partial \bar{x}^{j_q}}.
\end{equation}
This follows from the definition of components and the formulas
$$
\frac{\partial}{\partial \bar{x}^{i}} =\sum_{j} \frac{\partial x^{j}}{\partial \bar{x}^{i}} \frac{\partial}{\partial x^{j}}, \qquad d\bar{x}^{i} = \sum_{j} \frac{\partial \bar{x}^{i}}{\partial x^{j}} dx^{j}.
$$

Let $T$ be a tensor field  of type $(p,q)$ and $S$ a tensor field of type $(r,s)$. Then the tensor product $T \otimes S$ is the tensor field of type $(p+r,q+s)$ defined by
\begin{align*}
&T \otimes S \colon  \underbrace{\Omega^1(M) \times \cdots \times \Omega^1(M)}_{p} \times \underbrace{\mathfrak{X}(M)\times \cdots \times \mathfrak{X}(M)}_{q} \\
&\qquad \times  \underbrace{\Omega^1(M) \times \cdots \times \Omega^1(M)}_{r} \times \underbrace{\mathfrak{X}(M)\times \cdots \times \mathfrak{X}(M)}_{s} \rightarrow C^{\infty}(M)
\end{align*}
\begin{align*}
&(T\otimes S)(\alpha^{1},\dots,\alpha^{p},X_{1},\dots,X_{q},\beta^{1},\dots,\beta^{r},Y_1,\dots,Y_{s}) \\
&\qquad  =T(\alpha^{1},\dots,\alpha^{p},X_{1},\dots,X_{q})S(\beta^{1},\dots,\beta^{r},Y_1,\dots,Y_{s}).
\end{align*}
Thus, in terms of components,
$$
(T \otimes S)^{i_{1}\cdots i_{p}}_{\phantom{i_{1}\cdots i_{p}}j_{1}\cdots j_{q}}{}^{k_{1}\cdots k_{r}}_{\phantom{k_{1}\cdots k_{p}}l_{1}\cdots l_{s}}
= T^{i_{1}\cdots i_{p}}_{\phantom{i_{1}\cdots i_{p}}j_{1}\cdots j_{q}} S^{k_{1}\cdots k_{r}}_{\phantom{k_{1}\cdots k_{p}}l_{1}\cdots l_{s}}.$$

We mentioned before that diffeomorphisms act on functions and vector fields. More generally, they act on arbitrary tensor fields. In fact, let $f \colon M  \rightarrow N$ be a diffeomorphism. If $T$ is a tensor field of type $(p,q)$ on $M$, its push-forward $f_* T$ is a tensor field of type $(p,q)$ on $N$ defined by  
$$
(f_* T) (\alpha^1,\dots,\alpha^p,X_1,\dots,X_q) = T(f^* \alpha^1,\dots, f^* \alpha^{p}, f^{*} X_1,\dots, f^{*} X_q),
$$
where $\alpha^{k} \in \Omega^1(N)$ and $X_{l} \in \mathfrak{X}(M)$. The pull-back of a tensor field $S$ defined on $N$ is given by $f^* S = (f^{-1})_* S$. In coordinates we have the following relations which result from the definitions and the corresponding formulas for $1$-forms and vector fields: Letting $\varphi=(x^1,\dots,x^{m})$ and $\psi = (y^1,\dots,y^{m})$ coordinate systems on $M$ and $N$, we have
$$
(f_* T)^{a_{1}\cdots a_{p}}_{\phantom{a_{1}\cdots a_{p}}b_{1}\cdots b_{q}} =\sum_{i_1,\dots,i_p} \sum_{j_1,\dots,j_q} \frac{\partial f^{a_1}}{\partial x^{i_1}} \cdots \frac{\partial f^{a_p}}{\partial x^{i_p}} T^{i_{1}\cdots i_{p}}_{\phantom{i_{1}\cdots i_{p}}j_{1}\cdots j_{q}} \frac{\partial (f^{-1})^{j_1}}{\partial y^{b_1}} \cdots \frac{\partial (f^{-1})^{j_q}}{\partial y^{b_q}}
$$
and 
$$
(f^* S)^{i_{1}\cdots i_{p}}_{\phantom{i_{1}\cdots i_{p}}j_{1}\cdots j_{q}} =\sum_{a_1,\dots,a_p} \sum_{b_1,\dots,b_q} \frac{\partial (f^{-1})^{i_1}}{\partial y^{a_1}} \cdots \frac{\partial (f^{-1})^{i_p}}{\partial y^{a_p}} S^{a_{1}\cdots a_{p}}_{\phantom{a_{1}\cdots a_{p}}b_{1}\cdots b_{q}} \frac{\partial f^{b_1}}{\partial x^{j_1}} \cdots \frac{\partial f^{b_1}}{\partial x^{j_1}}.
$$

\begin{definition}
Let $X$ be a vector field on $M$ and let $H_t$ denote its flow. If $T$ is a tensor field of type $(p,q)$ on $M$, then the Lie derivative of $T$ with respect to $X$ is defined by
 \[L_X T=\frac{d}{dt}\bigg|_{t=0}H_{t}^{*}T.\]
\end{definition}

We compute the Lie derivative in coordinates for a few simple cases.

\begin{example}
Consider the Lie derivative of a function $f$ on $M$. In this case, $H_t^* f = f \circ H_t$ and therefore
$$
L_X f = \frac{d}{dt}\bigg|_{t=0}H_{t}^{*}f =  \frac{d}{dt}\bigg|_{t=0}f \circ H_t = df(X). 
$$
\end{example}

\begin{example}
Let $Y$ be a vector field on $M$. Then
$$
(H_t^* Y)(p) = \sum_{i,j} Y^{i}(H_t(p))\frac{\partial (H^{-1}_t)^{j}}{\partial x^{i}}(H_t(p)) \frac{\partial}{\partial x^{j}}\bigg\vert_p. 
$$
Using the formula 
$$
\frac{d}{dt}\bigg|_{t=0} \frac{\partial (H^{-1}_t)^{j}}{\partial x^{i}}(H_t(p)) = - \frac{\partial X^{j}}{\partial x^{i}}
$$
we find that
$$
L_X Y = \sum_{i}\left(\sum_{j}\frac{\partial Y^{i}}{\partial x^{j}} X^{j} - \sum_{j} Y^{j} \frac{\partial X^{i}}{\partial x^{j}} \right) \frac{\partial}{\partial x^{i}}
$$
or, in coordinate-free notation,
$$
L_X Y = [X,Y].
$$
\end{example}

\begin{example}
Finally, consider a $1$-form $\alpha$ on  $M$. Then 
$$
(H_t^* \alpha)(p) = \sum_{i,j} \alpha_i(H_t(p)) \frac{\partial H_t^{i}}{\partial x^{j}} dx^{j}(p).
$$
Using the formula
$$
\frac{d}{dt}\bigg|_{t=0} \frac{\partial H_t^{i}}{\partial x^{j}}(H_t(p)) =  \frac{\partial X^{i}}{\partial x^{j}}
$$
we obtain that
$$
L_X \alpha =\sum_{i,j} \left(\frac{\partial \alpha_i}{\partial x^{j}} X^{j} + \alpha_{j} \frac{\partial X^{j}}{\partial x^{i}} \right) dx^{i}.
$$
\end{example}

A general expression can be given for the Lie derivative of a tensor field of arbitrary type, namely,
\begin{align*}
    (L_X T)^{i_1 \cdots i_p}_{\phantom{i_1 \cdots i_p}j_1 \cdots j_q} &=\sum_k \bigg[ \frac{\partial}{\partial x^{k}}T^{i_1 \cdots i_p}_{\phantom{k i_2 \cdots i_p}j_1 \cdots j_q} X^{k} \\
    &\qquad\quad\: - T^{k i_2 \cdots i_p}_{\phantom{k i_2 \cdots i_p}j_1 \cdots j_q} \frac{\partial X^{i_1}}{\partial x^{k}} - \cdots - T^{i_1 \cdots i_{p-1} k}_{\phantom{i_1 \cdots i_{p-1} k}j_1 \cdots j_q} \frac{\partial X^{i_p}}{\partial x^{k}} \\
    &\qquad\quad\: + T^{i_1 \cdots i_p}_{\phantom{k i_2 \cdots i_p}k j_2 \cdots j_q}\frac{\partial X^{k}}{\partial x^{j_1}} + \cdots + T^{i_1 \cdots i_p}_{\phantom{k i_2 \cdots i_p}j_1 \cdots j_{q-1} k}\frac{\partial X^{k}}{\partial x^{j_{q}}}\bigg].
\end{align*}
This follows by using the computations of the preceding examples applied to each index. 
Some general properties of the Lie derivative are given next.
\begin{itemize}
\item If $f$ is a smooth function, then
$$
L_X df  = d(L_X f).
$$

    \item If $T$ and $S$ are tensor fields, then 
    $$
    L_X (T \otimes S) = L_X T \otimes S + T \otimes L_X S.
    $$
    
    \item If $T$ is a general tensor field and $\varphi$ is a diffeomorphism, then  
    $$
    \varphi^* (L_X T)= L_{\varphi^* X} (\varphi^* T). 
    $$
\end{itemize}

   \clearemptydoublepage
\chapter{Differential forms and integration}

\begin{center}
\parbox[b]{0.9\textwidth}{\small \sl In the absence of additional structure, there is no natural way to measure volumes or distances on 
a manifold. A differential $k$ form is a rule for measuring $k$ dimensional volumes at each point of $M$.
Differential forms can be integrated and come equipped with natural differential equations which are fundamental in the study of the topological properties of $M$. }
\end{center}

\vspace{3ex}

\section{Differential forms}

\begin{definition}
A $k$-form $\omega$ on $M$ is a section of the $k$-th exterior power of the
cotangent bundle $\Lambda^{k}T^{\ast}M$. The space of all $k$-forms on $M$
is denoted by $\Omega^{k}(M)=\Gamma\left(
\Lambda^{k}T^{\ast}M\right)$. In particular, $\Omega^{0}(M)$ is the space of smooth functions on $M$. 
\end{definition}

 We  have  already seen that, given  a local coordinate system $(x^{i})$ on $U$, for each point $p\in U$ there is a basis
$\{dx^{1}(p),\dots,dx^{m}(p)\}$ for $T^{\ast}_p M$. Therefore, the set
\[
\{dx^{i_{1}}(p)\wedge\cdots\wedge dx^{i_{k}}(p)\mid 1\leq i_{1}<\cdots
<i_{k}\leq m\}
\]
is a basis for $\Lambda^{k}T^{\ast}_p M$. Hence, on the neighborhood $U$, any element $\omega \in \Omega^k(M)$ can be uniquely represented as 
$$
\omega = \sum_{i_{1}<\cdots
<i_{k}} \omega_{i_1 \cdots i_k} dx^{i_1} \wedge \cdots \wedge dx^{i_k},
$$
with smooth functions $\omega_{i_1 \cdots i_k}$ on $U$. One can also sum over all $k$-tuples of indices  by introducing skew-symmetric coefficients:
$$
\omega = \frac{1}{k!} \sum_{i_1,\dots,i_k}\bar{\omega}_{i_1\cdots i_k} dx^{i_1} \wedge \cdots \wedge dx^{i_k},
$$
where the $\bar{\omega}_{i_1\cdots i_k}$ are the components of a skew-symmetric tensor and $\bar{\omega}_{i_1\cdots i_k} = \omega_{i_1\cdots i_k}$ for $i_1 < \cdots < i_k$. This skew-symmetric representation is often quite useful. 

We now observe that the graded vector space $\Omega^{\bullet}(M) = \bigoplus_{k \geq 0} \Omega^k(M)$ has a built-in graded algebra structure given by the wedge or exterior product. We multiply $\omega \in \Omega^k(M)$ by $\eta \in \Omega^l(M)$ to obtain $\omega \wedge \eta \in \Omega^{k+l}(M)$ defined as
$$
(\omega \wedge \eta)(p) = \omega(p) \wedge \eta(p),
$$
for each $p \in M$. This graded algebra is commutative, that is, $\omega \wedge \eta = (-1)^{kl} \eta \wedge \omega$.

We will now establish the existence and uniqueness of an operator 
$$
d \colon \Omega^{k}(M) \to \Omega^{k+1}(M)
$$
called the exterior derivative, which generalizes the gradient, divergence and rotational operations of vector calculus in the language of differential forms. 

\begin{proposition}\label{derext}
Given a manifold $M$, there is a unique degree $1$ linear operator $d$ on $\Omega^{\bullet}(M)$ such that:
\begin{enumerate}
\item[(i)]$d(d \omega)= 0$;  
    
\item[(ii)]$d(\omega \wedge \eta)= d \omega \wedge \eta + (-1)^{k} \omega \wedge d\eta$ for $\omega \in \Omega^{k}(M)$;

\item[(iii)]for functions $f \in \Omega^{0}(M)$, $df$ coincides with the differential of $f$ as defined in \S\ref{sec:cotangent}.
\end{enumerate}
\end{proposition}

The proof of Proposition \ref{derext} is a formal consequence of the following
two lemmas.

\begin{lemma}
\label{drlocal} Proposition \ref{derext} holds for $M=U\subseteq\RR%
^{m}$.
\end{lemma}

\begin{proof}
Since the algebra $\Omega^{\bullet}(U)$ is generated as an algebra by smooth functions on $U$ and the differential forms $dx^1, \dots ,dx^m$, there is at most one derivation satisfying the conditions of the proposition. Indeed, let
$$
\omega = \sum_{i_1 < \cdots < i_k} \omega_{i_1\cdots i_k} dx^{i_1} \wedge \cdots \wedge dx^{i_k}.
$$
Then since $ddx^{i} = 0$ we get
\begin{align*}
d \omega &=  \sum_{i_1 < \cdots < i_k} d \omega_{i_1\cdots i_k} \wedge dx^{i_1} \wedge \cdots \wedge dx^{i_k} \\
&= \sum_{i_1 < \cdots < i_k} \sum_j \frac{\partial \omega_{i_1\cdots i_k}}{\partial x^{j}} dx^{j} \wedge dx^{i_1} \wedge \cdots \wedge dx^{i_k}.
\end{align*}
Thus, to show that $d$ exists, define it by this formula. A simple computation, which we will leave to the reader, shows that the operator $d$ defined as above satisfies the required conditions.
\end{proof}

\begin{lemma}\label{sheaf}
Let $M$ be a manifold and $\{U_{\alpha}\}_{\alpha\in\mathcal{A}%
}$ an open cover of $M$. 
\begin{enumerate}
\item[(i)] If $D$ is a derivation of $\Omega^{\bullet}(M)$, then for each $\alpha\in\mathcal{A}$
there exists a unique derivation $D\vert_{U_{\alpha}}$ of $\Omega^{\bullet}(U_{\alpha})$ such
that
\[
(D\vert_{U_{\alpha}})(\omega\vert_{U_{\alpha}})=(D\omega)\vert_{U_{\alpha}},
\]
for any $\omega\in\Omega^{\bullet}(M)$. The derivation $D\vert_{U_{\alpha}}$ is called the restriction of $D$ to $U_{\alpha}$.

\item[(ii)] $D=0$ if and only if $D\vert_{U_{\alpha}}=0$ for all $\alpha\in\mathcal{A}$.

\item[(iii)] Given a family of derivations $D_{\alpha}$ on $\Omega^{\bullet}(U_{\alpha})$
such that
\[
D_{\alpha}\vert_{U_{\alpha}\cap U_{\beta}}= D_{\beta}%
\vert_{U_{\alpha}\cap U_{\beta}},
\]
there exists a unique derivation $D$ on $\Omega^{\bullet}(M)$ such that
$D_{\alpha}=D\vert_{U_{\alpha}}$.
\end{enumerate}
\end{lemma}

\begin{proof}
By the argument in the proof of Lemma \ref{extendf} one can show that given $\omega \in \Omega^{\bullet}(U_\alpha)$ and $p \in U_\alpha$ there exists an open neighborhood of $W$ of $p$ and a form $\tilde{\omega} \in \Omega^{\bullet}(M)$ such that $\omega\vert_W=\tilde{\omega}\vert_W$.
We then set
\[ ((D\vert_{U_{\alpha}})\omega)(p)= (D\tilde{\omega})(p),\]
It is easy to verify that $D\vert_{U_{\alpha}}$ is well defined and satisfies the required conditions.
Let us show the second statement. Obviously,  $D=0$ implies $D\vert_{U_{\alpha}}=0$. On the other hand, suppose that $D\vert_{U_{\alpha}}=0$ for all $\alpha \in \mathcal{A}$. Since $p\in U_\alpha$ for some  $\alpha\in \mathcal{A}$, we know that
\[(D\omega)(p)=(D\vert_{U_{\alpha}})(\omega\vert_{U_{\alpha}})(p)=0.\]
We conclude that $D=0$.
It remains to prove the last statement. For this one defines
\[ (D\omega)(p) = D_{\alpha}(\omega\vert_{U_\alpha})(p),\]
for any $\alpha$ such that $p \in U_\alpha$. This defines a derivation with the required properties.
\end{proof}

Endowed with the exterior derivative, the graded algebra $\Omega^{\bullet}(M)$ becomes a differential graded algebra. It is usually referred to as the de Rham complex of $M$. \\

We next consider the following situation: $M$ and $N$ are smooth manifolds and $f \colon M \to N$ is a smooth map. If $g$ is any smooth function on $N$, then we may combine this with $f$ to obtain a smooth function on $M$ which  we write 
$$
f^* g = g \circ f.
$$
Thus from $f$ we have constructed a new  induced map 
$$
f^* \colon C^{\infty}(N) \to C^{\infty}(M).
$$
We are now going to define the pull-back map $f^*$ taking $k$-forms on $N$ to $k$-forms on $M$:
$$
f^* \colon \Omega^k(N) \to \Omega^k(M).
$$
We first do this in local coordinates. So denote by $(x^{i})$ a local coordinate system on a  neighborhood $U $ of $M$ and by $(y^{a})$ a local coordinate system on a neighborhood $V$ of $N$. The basic idea is the substitution of coordinate  functions, replacing $dy^{a}$ by
$$
\sum_{i} \frac{\partial y^{a}}{\partial x^{i}} dx^{i}.
$$
Thus if 
$$
\omega = \sum_{a_1 < \cdots < a_k} \omega_{a_1\cdots a_k} dy^{a_1} \wedge \cdots \wedge dy^{a_k}$$
is a $k$-form on $V$, we set
\begin{align*}
f^*\omega &= \sum_{a_1 < \cdots < a_k}  f^*\omega_{a_1\cdots a_k} \left(\sum_{i_1}\frac{\partial y^{a_1}}{\partial x^{i_1}} dx^{i_1}\right) \wedge \cdots \wedge \left(\sum_{i_k}\frac{\partial y^{a_k}}{\partial x^{i_k}} dx^{i_k}\right)\\
&=\sum_{a_1 < \cdots < a_k} \sum_{i_1 < \cdots < i_k}(\omega_{a_1\cdots a_k} \circ f) \frac{\partial y^{a_1}}{\partial x^{i_1}} \cdots \frac{\partial y^{a_k}}{\partial x^{i_k}} dx^{i_1} \wedge \cdots \wedge dx^{i_k}.
\end{align*}
We now have $f^* \colon \Omega^{k}(V) \to \Omega^{k}(U)$. As a consequence of our study of coordinate changes in \S\ref{sec:cotangent}, the map $f^* \colon \Omega^k(N) \to \Omega^k(M)$ is defined by working out in each pair of coordinate systems on $M$ and $N$. For smooth maps $f \colon M \to N$ and $g \colon N \to P$, and differential forms $\omega, \eta \in \Omega^{k}(N)$, one can easily verify that $f^*(\omega \wedge \eta)= f^*\omega \wedge f^*\eta$ and $(g \circ f)^*  \omega = f^* (g^* \omega)$. 
Another basic property of this construction is the following. 

\begin{proposition}
If $f \colon M \to N$ is a smooth map and $\omega \in \Omega^k(N)$, then 
$$
f^* d\omega = d (f^* \omega).
$$
\end{proposition}
 
\begin{proof}
First we verify this for  functions $g$. But $f^* g= g \circ f$ and so, by the chain rule,
$$
d(f^* g) = d(g \circ f) = dg \circ Df  = f^* dg.
$$
In general, let
$$
\omega = \sum_{a_1 < \cdots < a_k} \omega_{a_1 \cdots a_k} dy^{a_1} \wedge \cdots \wedge dy^{a_k}.
$$
Since $f^{*} (\omega \wedge \eta)= f^* \omega \wedge f^* \eta$ and $d(f^* g) = f^* dg$,
$$
f^*\omega = \sum_{a_1 < \cdots < a_k} (\omega_{a_1\cdots a_k} \circ f)  df^{a_1} \wedge \cdots \wedge df^{a_k}.
$$
Using the properties of $d$,
\begin{align*}
d (f^*\omega) &= \sum_{a_1 < \cdots < a_k}\sum_{b}  \frac{\partial \omega_{a_1\cdots a_k}}{\partial y^{b}} \left(\sum_i \frac{\partial f^{b}}{\partial x^{i}} dx^{i}\right) \wedge df^{a_1} \wedge \cdots \wedge df^{a_k} \\
&= \sum_{a_1 < \cdots < a_k}\sum_{b} \frac{\partial \omega_{a_1\cdots a_k}}{\partial y^{b}} f^*(dy^b) \wedge f^* (dy^{a_1}) \wedge \cdots \wedge f^*(dy^{a_k}) \\
&= f^* \left( \sum_{a_1 < \cdots < a_k}\sum_{b} \frac{\partial \omega_{a_1\cdots a_k}}{\partial y^{b}} dy^b \wedge dy^{a_1} \wedge \cdots \wedge dy^{a_k} \right) \\
&= f^* d \omega,
\end{align*}
as required.
\end{proof}

This proposition may be rephrased by saying that the map $f^* \colon \Omega^{\bullet}(N) \to \Omega^{\bullet}(M)$ defines a homomorphism of differential graded algebras. We call it the pull-back homomorphism. 

Another important operation is the interior product $i_{X} \colon \Omega^{k}(M) \to \Omega^{k-1}(M)$ where $X \in \mathfrak{X}(M)$. In terms of the natural basis relative to a local coordinate system $(x^{i})$ on $M$, write $X = \sum_{j} X^{j} \partial / \partial x^{j}$ and $\omega = \sum_{i_1 < \cdots < i_k} \omega_{i_1 \cdots i_k} dx^{i_1}\wedge \cdots \wedge dx^{i_k}$, $\omega$ being a $k$-form on $M$. We define
$$
i_X \omega = \sum_{i_1 < \cdots < i_k}\sum_j\sum_{l=1}^k (-1)^{l-1} X^j\omega_{i_1 \cdots i_k} dx^{i_1}\wedge \cdots \wedge \frac{dx^{i_l}}{dx^j} \wedge \cdots \wedge dx^{i_k}.
$$

It is straightforward to see that this definition is independent of the local coordinate system. Even though the formula may seem complicated, the following properties characterize the contraction operation:
\begin{itemize}
    \item For all $X \in \mathfrak{X}(M)$ the operation $i_X$ is a derivation. This means that for $\omega, \eta \in \Omega^{k}(M)$ $$
    i_X (\omega \wedge \eta) = i_X \omega \wedge \eta + (-1)^{k} \omega \wedge i_X \eta.
    $$
    \item If $X=\frac{\partial}{\partial x^i}$ then:
    $$ i_X(dx^j)=\frac{\partial x^j}{\partial x^i}.$$
    \item The contraction operation is linear over functions:
  
    $$i_{fX}\eta=f i_X\eta,$$
    for any smooth function $f$.
\end{itemize}

We turn now to one more aspect of the calculus of differential forms. Consider the de Rham complex $\Omega^{\bullet}(M)$ of a smooth manifold $M$. The property $d \circ d = 0$ means that
$$
\mathrm{im}(d \colon \Omega^{k-1}(M) \to \Omega^{k}(M)) \subseteq \mathrm{\ker}(d \colon \Omega^{k}(M) \to \Omega^{k+1}(M))
$$
for every $k$, so we can take the quotient of these two vector spaces. The quotient space 
$$
H^{k}(M) = \frac{\mathrm{\ker}(d \colon \Omega^{k}(M) \to \Omega^{k+1}(M))}{\mathrm{im}(d \colon \Omega^{k-1}(M) \to \Omega^{k}(M))}
$$
is called the $k$th de Rham cohomology group of $M$. If $\omega \in \Omega^{k}(M)$ is such that $d\omega=0$, then its equivalence class 
$$
[\omega] = \omega + d\Omega^{k-1}(M) \in H^{k}(M)
$$
is called the cohomology class of $\omega$. The wedge product $\wedge \colon H^k(M) \otimes H^{l}(M) \to H^{k+l}(M)$ defined by
$$
[\omega] \wedge [\eta] = [\omega \wedge \eta],
$$
is  associative and commutative in the graded sense.
Therefore, the space 
$$
H^{\bullet}(M) = \bigoplus_{k \geq 0} H^{k}(M)
$$
has the structure of a graded commutative algebra. This algebra $H^{\bullet}(M)$ is called the de-Rham cohomology of $M$. 

\section{Classical vector calculus}\label{sec:vectorcalculus}
For the manifold $M=\RR^{3}$, the spaces of differential forms can be
identified with vector fields and smooth functions. Under these
identifications the exterior differential corresponds to the gradient, divergence
and curl. Let us recall the definitions of these operations.

\begin{definition}
\label{calc verctorial basico} We define the gradient, rotational and divergence
\begin{align*}
    \grad &\colon C^{\infty}(\RR^3) \to \mathfrak{X}(\RR^3), \\
    \rot &\colon \mathfrak{X}(\RR^3) \to \mathfrak{X}(\RR^3), \\
    \div &\colon \mathfrak{X}(\RR^3) \to C^{\infty}(\RR^3),
\end{align*}
by
\begin{align*}
\grad f &=\left( \frac{\partial f}{\partial x^{1}}, \frac{\partial f}{\partial x^{2}}, \frac{\partial f}{\partial x^{3}}\right), \\
\rot X &=\left( \frac{\partial X^{3}}{\partial x^{2}}- \frac{\partial X^{2}}{\partial x^{3}}, \frac{\partial X^{1}}{\partial x^{3}}- \frac{\partial X^{3}}{\partial x^{1}}, \frac{\partial X^{2}}{\partial x^{1}}- \frac{\partial X^{1}}{\partial x^{2}} \right), \\
\div Y &= \frac{\partial Y^{1}}{\partial x^{1}} + \frac{\partial Y^{2}}{\partial x^{2}} + \frac{\partial Y^{3}}{\partial x^{3}}.
\end{align*}
\end{definition}

\begin{center}

\begin{tabular}{p{0.4\textwidth} p{0.4\textwidth}}
  \vspace{0pt} \includegraphics[scale=0.46]{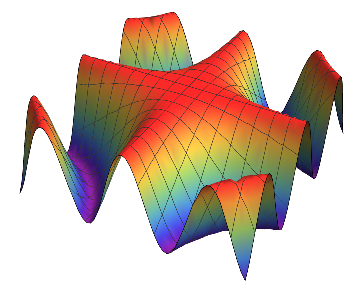}&
  \vspace{0pt} \includegraphics[scale=0.38]{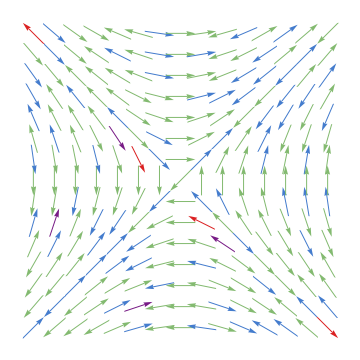}
\end{tabular}
 \captionof{table}{A function and its gradient.}
\end{center}

In order to give transparent formulas for the translation isomorphisms between the formalism of differential forms and that of vector calculus, we introduce the following notation. The vector-valued $1$-form and $2$-form
$$
dl = \left(\begin{array}{c} dx^{1} \\ dx^{2} \\ dx^{3} \end{array}  \right)
$$
and
$$
dS = \left(\begin{array}{c} dx^{2} \wedge dx^{3} \\ dx^{3} \wedge dx^{1} \\ dx^{1} \wedge dx^{2} \end{array} \right)
$$
are called the vectorial line element and the vectorial area element, respectively. The $3$-form
$$
dV = dx^{1} \wedge dx^{2} \wedge dx^{3}
$$
is called the volume element  of $\RR^3$. The usual translation isomorphisms are given by
\begin{alignat*}{3}
    \iota_1 &\colon \mathfrak{X}(\RR^3) \to \Omega^1(\RR^3), &&\qquad &&\iota_1(X) = X \cdot dl, \\
     \iota_2 &\colon \mathfrak{X}(\RR^3) \to \Omega^2(\RR^3), &&\qquad &&\iota_1(Y) = Y \cdot dS, \\
      \iota_3 &\colon C^{\infty}(\RR^3) \to \Omega^3(\RR^3), &&\qquad &&\iota_3(f) = f dV.
\end{alignat*}
Here the dot denotes the standard scalar product on $\RR^3$. 

Let us now use the above dictionary to translate the exterior derivative into the language of vector calculus. 

\begin{proposition}\label{prop:transVC}
For $f \in C^{\infty}(\RR^3)$ and $X,Y \in \mathfrak{X}(\RR^3)$, 
\begin{align*}
    df & = \grad f \cdot dl, \\
    d(X \cdot dl) &= \rot X \cdot dS, \\
    d(Y \cdot dS) &= (\div Y)dV.
\end{align*}
Hence the diagram
$$
\xymatrix{0 \ar[r] & \Omega^0(\RR^3) \ar[r]^-{d} & \Omega^1(\RR^3)\ar[r]^-{d} & \Omega^2(\RR^3)\ar[r]^-{d} & \Omega^{3}(\RR^3) \ar[r] & 0 \\
0 \ar[r]& C^{\infty}(\RR^3)\ar[r]^-{\grad} \ar[u]_-{\mathrm{id}} & \mathfrak{X}(\RR^3) \ar[u]_-{\iota_1}\ar[r]^-{\rot} & \mathfrak{X}(\RR^3)\ar[u]_-{\iota_2} \ar[r]^-{\div} & C^{\infty}(\RR^3) \ar[u]_-{\iota_3} \ar[r] & 0}
$$
is commutative. 
\end{proposition}

\begin{proof}
For $f \in C^{\infty}(\RR^3)$, we have
\begin{align*}
    df &= \frac{\partial f}{\partial x^{1}}dx^{1} + \frac{\partial f}{\partial x^{2}}dx^{2}+ \frac{\partial f}{\partial x^{3}}dx^{3} \\
    &= \left(\frac{\partial f}{\partial x^{1}}, \frac{\partial f}{\partial x^{2}}, \frac{\partial f}{\partial x^{3}} \right) \cdot dl \\
    &= \grad f \cdot dl,
\end{align*}
and for vector fields $X,Y \in \mathfrak{X}(\RR^3)$,
\begin{align*}
    d(X\cdot dl) &= d \left( X^{1} dx^{1}+ X^{2} dx^{2} + X^{3} dx^{3}\right) \\
    &= \left( \frac{\partial X^{3}}{\partial x^{2}}-\frac{\partial X^{2}}{\partial x^{3} } \right)dx^{2} \wedge dx^{3} + \text{cyclic permutations} \\
    &= \left(\frac{\partial X^{3}}{\partial x^{2}}-\frac{\partial X^{2}}{\partial x^{3} }, \frac{\partial X^{1}}{\partial x^{3}}-\frac{\partial X^{3}}{\partial x^{1} }, \frac{\partial X^{2}}{\partial x^{1}}-\frac{\partial X^{1}}{\partial x^{2} }\right) \cdot dS \\
    &= \rot X \cdot dS
\end{align*}
and 
\begin{align*}
    d(Y \cdot dS) &= \frac{\partial Y^{1}}{\partial x^{1}} dx^{1}  \wedge dx^{2} \wedge dx^{3} + \text{cyclic permutations} \\
    &= \left(\frac{\partial Y^{1}}{\partial x^{1}} + \frac{\partial Y^{2}}{\partial x^{2}} + \frac{\partial Y^{3}}{\partial x^{3}}  \right) dV,
\end{align*}
as was to be shown. 
\end{proof}

We close with the following corollary, which is a formal consequence of the
fact that $d^2=0$.

\begin{corollary}
$\rot \grad f = 0$ and $\div \rot X = 0$ for all smooth functions $f$ and all vector fields $X$.
\end{corollary}

The conclusion here is that the formalism of differential forms is an
extension of the three dimensional vector calculus which works in manifolds
of arbitrary dimension.

\section{Manifolds with boundary}

\begin{definition}
The upper half-space of dimension $m$ is defined as \[\HH^{m}%
=\{(x^{1},\dots,x^{m})\in\RR^{m}\mid x^{m}\geq0\}.\] The boundary of
$\HH^{m}$, denoted by $\partial\HH^{m}$, is the subspace
\[
\partial\HH^{m}=\{(x^{1},\dots,x^{m})\in\RR^{m}\mid x^{m}=0\}.
\]
\end{definition}

\begin{definition}
For an arbitrary subset $X\subseteq\RR^{m}$ we say that $f:X\rightarrow
M$ is smooth if for each $x\in X$ there exists an open subset $U_{x}%
\subseteq\RR^{m}$ and a smooth function $\tilde{f}_{x}:U_{x}\rightarrow
M$ such that \[\tilde{f}\vert_{U_{x}\cap X}=f\vert_{U_{x}\cap X}.\] If $X,Y$ are subsets of $\RR^{m}$, a function $f:X\rightarrow Y$ is
called a diffeomorphism if it is smooth, invertible and its inverse is smooth.
\end{definition}

\begin{lemma}
Let $U,V\subseteq\HH^{m}$ be open subsets and $\varphi:U\rightarrow V$
a diffeomorphism. Then \[\varphi(U\cap\partial\HH^{m})\subseteq
\partial\HH^{m}.\]
\end{lemma}

\begin{proof}
Suppose that there exists $p=(x^1,\dots,x^{m-1},0)\in U$ such that  $\varphi(p)=(y^1,\dots,y^m)$ with  $y^m>0.$ Consider the inverse function $\varphi^{-1}\vert_W:W\to U$ where $W\subseteq V$ is an open in $\RR^m$ with $\varphi(p) \in W$. Since $\varphi^{-1}$ is a diffeomorphism, its image is open in $\RR^m$. On the other hand $p \in \varphi^{-1} (W)\subseteq U$. This is imposible because any open in $\RR^m$ that contains $p$ also contains points whose last coordinate is negative.\end{proof}

\begin{definition}
A manifold with boundary $M$ of dimension $m$ is a Hausdorff second countable
topological space together with an atlas $(U_{\alpha}, \varphi_{\alpha
})_{\alpha\in\mathcal{A}}$, where $\{ U_{\alpha}\}_{\alpha\in A}$ is an open
cover of $M$ and $\varphi_{\alpha}:U_{\alpha}\to V_{\alpha}\subseteq
\HH^{m}$ are homeomorphisms such that the transition functions $
\varphi_{\beta}\circ\varphi^{-1}_{\alpha}:\varphi_{\alpha}(U_{\alpha}\cap
U_{\beta})\to\varphi_{\beta}(U_{\alpha}\cap U_{\beta})$ are diffeomorphisms.
\end{definition}

\begin{figure}[H]
\centering
\scalebox{.2}
	\centering
	\includegraphics[scale=0.42]{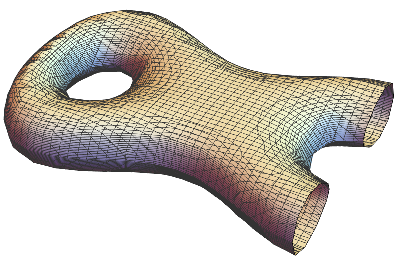}
	\caption{A surface with boundary.}
\end{figure}

The interior of a manifold with boundary $M$ is the subspace
\[
M^{\circ}=\{p\in M\mid \varphi_{\alpha}(p)\notin\partial\HH^{m}\text{ for
some chart }\varphi_{\alpha}\}.
\]
The boundary of $M$ is
\[
\partial M=\{p\in M\mid \varphi_{\alpha}(p)\in\partial \HH^{m}\text{ for
some chart }\varphi_{\alpha}\}.
\]
If $M$ is a manifold with boundary of dimension $m$, one can show that
\begin{itemize}
\item $M^{\circ} \cap\partial M= \emptyset$.

\item $M^{\circ}$ is a manifold of dimension $m$.

\item $\partial M$ is a manifold of dimension $m-1$.
\end{itemize}

\begin{example}
The closed disk
\[ D^m=\{ x\in \RR^m\mid |x| \leq 1\}\]
is a manifold with boundary. The cylinder
\[ C^m=\left\{ x\in \RR^m\mid \tfrac{1}{2} \leq |x| \leq 1\right\}\]
is also a manifold with boundary.
\end{example}

An embedding of a manifold with boundary $M$ into $N$ is an immersion which is a homeomorphism onto its image.

\section{Oriented manifolds}
\label{2sec_orientacion}

\begin{definition}
Let $V$ be a real vector space of dimension $k<\infty$. The vector space
$\Lambda^{k}V$ is one-dimensional and therefore, the topological space
$\Lambda^{k}V\setminus\{0\}$ has two connected components. An orientation of the vector space $V$ is a choice of one of these connected components.
\end{definition}

 An ordered basis $\{ v_{1},\dots,v_{k}\}$ for $V$ determines
an orientation which is the connected component of $v_{1}\wedge\dots\wedge v_{k}
\in\Lambda^{k}(V)$. It is a good exercise to show that two basis induce the same orientation if and
only if the change of base matrix has positive determinant.

An orientation on a one-dimensional vector space $L$ determines an orientation on $L^{\ast}$ by the condition
that if $v\in L$ and $\alpha\in L^{\ast}$ are oriented then $\alpha(v)>0.$

\begin{remark}
\label{exdual} If $V$ has dimension $k<\infty$ then there is a
natural isomorphism $
\Lambda^{k}(V^{\ast})\cong\big(\Lambda^{k}V\big)^{\ast}$, given by
\[
\big(\alpha_{1}\wedge\dots\wedge\alpha_{k}\big)(v_{1}\wedge\dots\wedge
v_{k})=\frac{1}{k!}\sum_{\sigma\in \mathfrak{S}_{k}}\alpha_{\sigma(1)}(v_{1}).\dots
\alpha_{\sigma(k)}(v_{k}),
\]
where $\mathfrak{S}_{k}$ denotes the symmetric group. From this, one concludes that an orientation on $V$ induces naturally an orientation on $V^{\ast}$.
\end{remark}

\begin{definition}
An orientation on a manifold $M$ is a choice of an orientation on each tangent
space $T_{p}M$ which is locally constant in the following sense. For each
point $p\in M$ there exist a local coordinate system $\varphi=(x^{1},\dots,x^{m}%
)\colon U\rightarrow V$ such that for all $q\in U$ the orientation on $T_{q}M$ is
given by \[\frac{\partial}{\partial x^{1}}\bigg\vert_q\wedge\dots\wedge\frac{\partial
}{\partial x^{m}}\bigg\vert_q.\]
A manifold is orientable if it admits an orientation.
An oriented manifold is a manifold together with a choice of orientation.
\end{definition}

\begin{definition}
An atlas $\{(U_{\alpha},\varphi_{\alpha}\})_{\alpha\in\mathcal{A}}$ on $M$ is said
to be oriented if for all $\alpha,\beta\in\mathcal{A}$ the transition
functions $
\varphi_{\beta}\circ\varphi_{\alpha}^{-1}:\varphi_{\alpha}(U_{\alpha}\cap
U_{\beta})\rightarrow\varphi_{\beta}(U_{\alpha}\cap U_{\beta})$ satisfy the condition 
$$
\det(D(\varphi_{\beta}\circ\varphi_{\alpha
}^{-1})(q))>0
$$
for all $q\in\varphi_{\alpha}(U_{\alpha}\cap U_{\beta})$.
\end{definition}

\begin{definition}
A volume form on a manifold $M$ of dimension $m$ is a differential form
$\omega\in\Omega^{m}(M)$ such that $\omega(p) \neq0$ for all $p \in M$.
\end{definition}

\begin{lemma}
Let $M$ be a manifold. Then
\begin{enumerate}
\item[(i)] an oriented atlas $\{(U_{\alpha}, \varphi_{\alpha})\}$ induces an
orientation on $M$;

\item[(ii)] all orientations are induced by an oriented atlas;

\item[(iii)] a volume form $\omega$ induces and orientation on $M$;

\item[(iv)] all orientations
are induced by a volume form.
\end{enumerate}
\end{lemma}

\begin{proof}
Let $(U_\alpha,\varphi_\alpha)$ be an oriented atlas. This defines an orientation on $M$ by declaring that
at each point $p \in M$
\[ \frac{\partial}{\partial x^{1}}\bigg\vert_p\wedge\dots\wedge\frac{\partial
}{\partial x^{m}}\bigg\vert_p\]
is oriented for any local coordinate system $\varphi=(x^1,\dots,x^m)$ in the atlas. Since the determinants of the derivatives of the transition functions are positive, this orientation is well defined. Conversely, given an orientation $\mathcal{O}$ on $M$, one can choose an oriented subatlas of the maximal atlas by requiring the condition that
\[ \frac{\partial}{\partial x^{1}}\bigg\vert_p\wedge\dots\wedge\frac{\partial
}{\partial x^{m}}\bigg\vert_p\]
is oriented. Let us now prove the third claim. Consider a volume form $\omega \in \Omega^m(M)$.
We define an orientation on each cotangent space $T^*M$ by declaring that
\[ \omega(p) \in \Lambda^mT_p^*M\]
lies in the positive connected component. Let us show that any orientation $\mathcal{O}$ can be defined in this manner. We consider an oriented atlas $\{(U_\alpha, \varphi_\alpha)\}$ inducing $\mathcal{O}$ and a partition of unity $\rho_\alpha$ subordinate to the cover $\{ U_\alpha\}$. Then we define a volume form $\omega \in \Omega^m(M)$ by
\[ \omega(p)= \sum_\alpha \rho_\alpha (p) dx_\alpha^1\wedge \dots \wedge dx_\alpha^m.\]
Here the sum is over all indices $\alpha$ such that $ p \in U_\alpha$. Since the partition of unity is locally finite, the sum is well defined. Since the atlas is oriented we know that, on the overlap $U_{\alpha} \cap U_{\beta}$,
\[dx_\alpha^1\wedge \dots \wedge dx_\alpha^m=\lambda dx_\beta^1\wedge \dots \wedge dx_\beta^m\]
for some $\lambda >0$ and therefore $\omega(p) \neq 0$.
\end{proof}

\begin{example}
The Klein bottle does not admit an orientation.
\end{example}
\begin{figure}[H]
\centering
\scalebox{.2}
	\centering
	\includegraphics[scale=0.44]{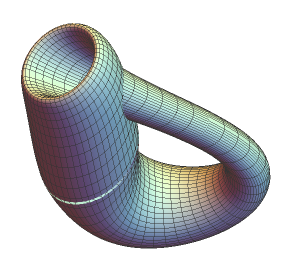}
	\caption{The Klein bottle is not orientable.}
\end{figure}

There is a `natural' way to orient the boundary of a manifold with a given orientation. To define it, we need the following lemma.

\begin{lemma}
\label{matrix} Let $U,V\subseteq\HH^{m}$ be open subsets, $p\in\partial
U$ and $\varphi:U\rightarrow V$ a diffeomorphism. The derivative matrix
$D\varphi(p)$ has the form
\[
D\varphi(p)=\left(
\begin{array}
[c]{c}%
\begin{array}
[c]{ccccc}%
\ast & \ast & \dots & \ast & \ast\\
\vdots & \vdots & & \vdots & \vdots\\
0 & 0 & \cdots & 0 & (\partial x^{m}/\partial y^{m})(p)%
\end{array}
\end{array}
\right),
\]
where, in addition, $(\partial x^{m}/\partial y^{m})(p)>0$.
\end{lemma}

\begin{proof}
We need to prove that $(\partial x^{k}/\partial y^{m})(p)=0$ for  $i<m$, and that $(\partial x^{m}/\partial y^{m})(p)>0$.
For $i<m$ the vector $\partial/\partial  y^{i}\vert_p$ is tangent to the boundary, and since $\varphi$ preserves the boundary, so is $D\varphi (p) (\partial/\partial  y^{i}\vert_p)$. We conclude that $(\partial x^{m}/\partial y^{m})(p)=0$. On the other hand, we know that $D\varphi (p)$ is not singular and therefore $(\partial x^{m}/\partial y^{m})(p)\neq 0.$ One also knows that \[\frac{\partial x^m}{\partial y^m}=\frac{d}{dt}\bigg|_{t=0}x^m (\varphi(p+ty^m))\] is a nonnegative number because $ x^m( \varphi(p+t y^m))>0$ for $t>0$.
\end{proof}

Let $M$ be a manifold with boundary, $p\in\partial M$ and $v\in T_{p}M$ such
that $v\notin T_{p}\partial M$. We say that $v$ points inside if for any
choice of coordinates $\varphi_{\alpha}:U_{\alpha}\rightarrow V_{\alpha}$, the
last component of $D\varphi_{\alpha}(p)(v)$ is positive. We say that $v$
points outside if it does not point inside. Note that, in view of the previous
lemma, if $\varphi_{\alpha}:U_{\alpha}\rightarrow V_{\alpha}$ and
$\varphi_{\beta}:U_{\beta}\rightarrow V_{\beta}$ are coordinates then the last
coordinate of $D\varphi_{\alpha}(p)(v)$ has the same sign as the last
coordinate of $D\varphi_{\beta}(p)(v)$.

\begin{definition}
Let $M$ be an oriented manifold with boundary. The manifold $\partial M$
acquires an orientation defined by the following rule. An ordered basis
$\{v_{1},\dots,v_{m-1}\}$ of $T_{p} \partial M$ is oriented if and only if the
ordered basis $\{e,v_{1},\dots,v_{m-1}\}$ for $T_{p} M$ is oriented, for any
vector $e \in T_{p}M$ that points outside.
\end{definition}

One can show that the orientation on the boundary does not depend on the vector $e \in T_{p}M$. We also want to point out the following.

\begin{remark}
If $dx^{1}\wedge\dots\wedge dx^{m}$ is an oriented volume form on
$\HH^{m}$ then $(-1)^{m}dx^{1}\wedge\dots\wedge dx^{m-1}$ is an
oriented volume form on $\partial\HH^{m}$.
\end{remark}

\section{Integration of forms}
\begin{definition}
Let $M$ be a smooth manifold of dimension $m$. The support of an $m$-form $\omega\in\Omega^{m}(M)$ is the set%
\[
\mathrm{supp}(\omega)=\overline{\{p\in M\mid \omega(p)\neq0\}}.
\]
The set of $m$-forms on $M$ with compact support will be denoted by $\Omega_{c}^{m}(M)$.
\end{definition}

\begin{definition}
Let $U\subseteq\HH^{m}$ be an open set and $\omega\in\Omega_{c}^{m}(U)$ a
form with compact support. Then $\omega$ can be written uniquely as
$\omega=f dx^{1}\wedge\cdots\wedge dx^{m}.$ The integral of $\omega$ over $U$ is
defined as
\[
\int_{U}\omega =\int_{U}f dx^{1}\cdots dx^{m},
\]
where the right hand side denotes the Riemann integral of the function $f$.
\end{definition}

\begin{lemma}
Let $\varphi\colon U\rightarrow V$ be an orientation preserving diffeomorphism
between open subsets of $\HH^{m}$ and $\omega\in\Omega_{c}^{m}(V)$. Then
\[\int_{U}\varphi^{\ast}\omega=\int_{V}\omega.\]
\end{lemma}

\begin{proof}
We write $ \omega = f \:dx^1 \wedge \dots \wedge dx^m$ and use the change of variable formula to compute
\begin{align*}
\int_U \varphi^*\omega &= \int_U (f \circ \varphi)  \det D\varphi \:dy^1  \dots dy^m \\
&= \int_U (f \circ \varphi)  \vert\det D\varphi\vert \:dy^1 \cdots dy^m\\
&= \int_V f \:dx^1 \cdots dx^m \\
&= \int_V \omega.
\end{align*}
This proves the result.
\end{proof}

In view of the previous lemma, the following definition makes sense.

\begin{definition}
Let $V$ be an oriented manifold which is diffeomorphic to an open subset of
$\HH^{m}$ and $\omega\in\Omega_{c}^{m}(V)$. We define \[\int_{V}\omega
=\int_{U}\varphi^{\ast}\omega,\] for any diffeomorphism $\varphi\colon U\subseteq
\HH^{m}\rightarrow V$ that preserves the orientation.
\end{definition}

Let now $M$ be an oriented $m$-dimensional manifold with boundary. 
We want to define the integral of any $n$-form $\omega$ with compact support over $M$. To this end, let $\{(U_{\alpha},\varphi_{\alpha})\}$ be a finite covering of an open subset of $M$ that contains the support of $\omega$ and $\{\rho_{\alpha}\}$ a partition of unity subordinate to the covering $\{U_{\alpha}\}$. Then $\omega$ can be written as a locally finite sum $\omega = \sum_{\alpha} \omega_{\alpha}$ where $\omega_{\alpha} = \rho_{\alpha} \omega$. We then define
$$
\int_M \omega = \sum_{\alpha} \int_{U_{\alpha}} \omega_{\alpha}.
$$
Let us show that the integral so defined is independent of the atlas and partition of unity employed. Consider another atlas $\{(V_{\beta},\psi_{\beta})\}$ which determines on $M$ the same orientation as $\{(U_{\alpha},\varphi_{\alpha})\}$ and let $\{\tau_{\beta}\}$ be a partition of unity subordinate to $\{V_{\beta}\}$. Then $U_{\alpha} \cap V_{\beta}$ will be a finite covering of an open set of $M$ that contains the support of $\omega$ and the family $\rho_{\alpha}\tau_{\beta}$ will be a partition of unity subordinate to $\{U_{\alpha} \cap V_{\beta}\}$. Thus
$$
\sum_{\alpha} \int_{U_{\alpha}} \rho_{\alpha}\omega = \sum_{\alpha} \int_{U_{\alpha}} \rho_{\alpha}\left( \sum_{\beta} \tau_{\beta}\right)\omega = \sum_{\alpha,\beta} \int_{U_{\alpha}\cap V_{\beta}} \rho_{\alpha} \tau_{\beta} \omega,
$$
where in the last equality it was used that, for each $\alpha$, the functions $\rho_{\alpha}\tau_{\beta}$ are defined in $U_{\alpha}$. Similarly,
$$
\sum_{\beta} \int_{V_{\beta}} \tau_{\beta}\omega = \sum_{\beta} \int_{V_{\beta}} \left( \sum_{\alpha} \rho_{\alpha}\right)\tau_{\beta}\omega = \sum_{\alpha,\beta} \int_{U_{\alpha}\cap V_{\beta}} \rho_{\alpha} \tau_{\beta} \omega,
$$
which proves the required independence. 


\section{Stokes' theorem}

Let us now discuss the higher dimensional generalization of the fundamental
theorem of calculus which expresses a relation between an integral over a manifold and one over its boundary. This generalization is called Stokes' theorem.

\begin{theorem}
Let $M$ be an $m$-dimensional oriented manifold with boundary and let
$\omega\in\Omega_{c}^{m-1}(M)$. Then \[\int_{\partial M}\iota^*\omega=\int_{M}%
d\omega,\]
where $\iota$ is the natural inclusion of $\partial M$ into $M$. 
\end{theorem}

\begin{proof}
We divide the proof in three steps of increasing generality:
\begin{enumerate}
\item The case $M=\HH^m$.
\item  The case where there is a coordinate chart $(U,\varphi)$ with $\mathrm{supp}(\omega) \subseteq U$.
\item The general case.
\end{enumerate}
In the first case, the differential form $\omega$ can be written in the form
\[\omega=\sum_i f_i\: dx^1\wedge\dots\wedge\widehat{dx^i}\wedge\dots\wedge dx^m,\]
where the notation $\widehat{dx^i}$ means that the factor is to be omitted. The two integrands $i^*\omega$ and $d\omega$ can be computed from the  definitions. Let us first compute $i^* \omega$. If we denote the standard coordinates of $ \RR^{n-1} \times \{0\} \subseteq \RR^{n}$ by $x^{1},\dots,x^{m-1}$, then
\begin{align*}
    \iota^*\omega &= \sum_{i} \iota^* f_{i}\: \iota^*dx^1\wedge\dots\wedge\widehat{\iota^*dx^i}\wedge\dots\wedge \iota^*dx^m \\
    &= \iota^*f_{m}\: dx^{1} \wedge \cdots \wedge dx^{m-1},
\end{align*}
since the inclusion $\iota \colon \RR^{m-1} \times \{0\} \to \RR^n$ satisfies $\iota^* dx^{i} = dx^{i}$ for $1 \leq i \leq m-1$ and $\iota^* dx^{m} = 0$. On the other hand,
\begin{align*}
    d\omega &= \sum_i df_{i} \wedge dx^1\wedge\dots\wedge\widehat{dx^i}\wedge\dots\wedge dx^n \\
    &=\sum_{i}\left(\sum_{j} \frac{\partial f_i}{\partial x^{j}} dx^{j} \right)\wedge dx^1\wedge\dots\wedge\widehat{dx^i}\wedge\dots\wedge dx^m \\
    &= \sum_i (-1)^{i-1} \frac{\partial f_i}{\partial x^{i}} dx^1\wedge\dots\wedge dx^{m}.
\end{align*}
We now turn to the integrals themselves. By definition, we have
$$
\int_{\partial \HH^{m}} \iota^* \omega =(-1)^{m} \int_{\RR^{m-1}} f_n(x^1,\dots,x^{m-1},0) dx^{1} \cdots dx^{m-1},
$$
and 
$$
\int_{\HH^m} d\omega = \sum_{i} \int_{\HH^m} (-1)^{i-1}\frac{\partial f_i}{\partial x^{i}} dx^1\dots dx^{m},
$$
as ordinary multiple integrals. Since the support of $\omega$ is compact, so is the support of $f_i$, and we obtain
$$
\int_{0}^{\infty} \frac{\partial f_n}{\partial x^{m}} dx^{m} = \big[ f_m\big]_{x^{m}=0}^{x^{m}=\infty} = - f_m(x^{1},\dots,x^{m-1},0)
$$
and, for $i \neq m$, 
$$
\int_{-\infty}^{\infty} \frac{\partial f_i}{\partial x^{i}} dx^{i} = \big[ f_i\big]_{x^{m}=-\infty}^{x^{m}=\infty} = 0.
$$
Hence
$$
\int_{\partial \HH^{m}} \iota^* \omega =(-1)^{m} \int_{\RR^{m-1}} f_m(x^1,\dots,x^{m-1},0) dx^{1} \cdots dx^{m-1} = \int_{\HH^{m}} d\omega,
$$
for our first case $M = \HH^n$.

For the second case, let $(U,\varphi)$ a coordinate chart on $M$ with $\mathrm{supp}(\omega) \subseteq U$. The definition of manifolds with boundary allows the two possibilities that $\varphi(U)$ is open in $\HH^n$ or in $\RR^n$. Without loss of generality we may assume the former here, since by the compactness of $\mathrm{supp}(\omega)$ we could always achieve it if necessary by translating and shrinking the chart domain. Extend $\varphi^{-1 \ast} \omega$ to a form $\omega'$ by setting it equal to zero outside $\varphi(U) \in \Omega^{m-1}(\HH^m)$, which is possible because $\mathrm{supp}(\varphi^{-1 \ast} \omega) = \varphi (\mathrm{supp}(\omega))$ is compact. Then, by the change-of-variables formula and the first case,
\begin{align*}
\int_{\partial M} \iota^* \omega &= \int_{\partial U} \iota^* \omega = \int_{\varphi(\partial U)} \iota^*\varphi^{-1 \ast} \omega  = \int_{\partial\HH^{m}} \iota^*\omega' = \int_{\HH^m} d\omega' \\
&= \int_{\varphi(U)} d(\varphi^{-1 \ast} \omega) =\int_{\varphi(U)} \varphi^{-1 \ast}  d\omega = \int_{U} d \omega = \int_{M} d \omega,
\end{align*}
and this completes the second step. 

Let us consider the last step. Since $\mathrm{supp}(\omega)$ is compact, we may choose a finite cover of it by coordinate charts $\{(U_{\alpha},\varphi_{\alpha})\}$. Let $\{\rho_{\alpha}\}$ be a partition of unity subordinate to $\{U_{\alpha}\}$. We may then write $\omega= \sum_{\alpha} \omega_{\alpha}$ with $\omega_{\alpha}= \rho_{\alpha} \omega$. Then, by the second case,
$$
\int_{\partial M} \iota^* \omega = \sum_{\alpha} \int_{\partial U_{\alpha}} \iota^* \omega_{\alpha} = \sum_{\alpha} \int_{U_{\alpha}} d \omega_{\alpha} = \int_{M} d \omega.
$$
This finishes the third case, and the proof of the theorem. 
\end{proof}

\begin{example}
Consider the interval $M=[a,b]\subseteq\RR$. A $0$-form is a smooth
function on $[a,b]$. Taking into account the orientation induced on the
boundary $\partial M=\{ a,b \}$, Stokes' theorem states that
\[
\int_{a}^{b}\frac{\partial f}{\partial t}dt=\int_{[a,b]}df=\int_{\partial
[a,b]}f=f(b)-f(a).
\]
This is of course the fundamental theorem of calculus.
\end{example}

\begin{example}
Given $a,b>0$, the area of the ellipse $M=\{(x,y)\in\RR^{2}%
\mid ax^{2}+by^{2}\leq1\}$ is 
$$
A(M)=\int_{M}dx\wedge dy.
$$
This integral can be computed using a change of variables
\[
\varphi:{D}^{2}\rightarrow M,\qquad\varphi(p,q)=\left(\frac{p}{\sqrt{a}},\frac
{q}{\sqrt{b}}\right).
\]
We find that
\begin{align*}
A(M)  &  =\int_{M}dx\wedge dy=\int_{{D}^{2}}\varphi^{\ast}dx\wedge\varphi
^{\ast}dy\\
&  =\int_{{D}^{2}}\frac{dp}{\sqrt{a}}\wedge\frac{dq}{\sqrt{b}}=\frac{1}%
{\sqrt{ab}}\int_{{D}^{2}}dp\wedge dq=\frac{\pi}{\sqrt{ab}}.
\end{align*}
On the other hand, we observe that \[\omega=\frac{1}{2}(xdy-ydx)\] satisfies
$d\omega=dx\wedge dy,$ so that Stokes' theorem gives \[A(M)=\int_{M}d\omega
=\int_{\partial M}\omega.\] Parametrising the boundary of the ellipse by the
function
$$
\gamma \colon [0,2\pi] \to \partial M, \qquad \gamma(t)= \left(\frac{\cos t}{\sqrt{a}},\frac{\sin t}{\sqrt{b}} \right)
$$
one obtains
\[
\gamma^{\ast}\omega=\frac{1}{2}(\gamma^{\ast}(xdy)-\gamma^{\ast}%
(ydx))=\frac{1}{2}\left(  \frac{\cos^{2}t}{\sqrt{ab}}+\frac{\sin^{2}t}%
{\sqrt{ab}}\right)  dt=\frac{dt}{2\sqrt{ab}}.
\]
Therefore
\[
A(M)=\int_{\partial M}\omega=\int_{0}^{2\pi}\gamma^{\ast}\omega=\int_{0}^{2\pi
}\frac{dt}{2\sqrt{ab}}=\frac{\pi}{\sqrt{ab}}.
\]

\end{example}

\section{The classical integral theorems}
We keep the notation introduced in \S\ref{sec:vectorcalculus}. Let us explain how Stokes' theorem looks as a theorem about vectors fields or functions on $\RR^3$. For this, we need the following preliminary lemma.

\begin{lemma}
\begin{enumerate}
\item[(i)]Let $C \subseteq \RR^3$ be a curve parametrized by $\gamma \colon [a,b] \to \RR^3$ and let  $T\colon C \to \RR^3$ denotes the positively oriented unit tangent field. If $\iota \colon C \to \RR^3$ denote the  inclusion, then
$$
\iota^* dl = T \,ds.
$$

\item[(ii)]Let $S \subseteq \RR^{3}$ be an oriented surface parametrized by $r \colon U \to \RR^3$ and let $n \colon S \to \RR^3$ denote the orienting unit normal field. If $\iota \colon S \to \RR^3$ denote the  inclusion, then
$$
\iota^* dS = n\, dA.
$$
\end{enumerate}
\end{lemma}

\begin{proof}
To prove (i), notice that $dl(T) =  T$ and $ds(T)=1$ at every point, so the first equation holds. To prove (ii), given an orthonormal basis $(v,w)$ of $T_p S$, then $n(p)$ extends this to a positively oriented orthonormal basis $(n,v,w)$ of $\RR^3$. Moreover, $dA(v,w)=1$, so $n\, dA (v,w)= n = v \times w = dS(v,w)$.
\end{proof}

We can now write the integral of a $1$-form $X \cdot dl$ associated to a vector field $X$ on $\RR^3$ over a parametrized curve $C \subseteq \RR^3$ as
$$
\int_{C} X \cdot dl = \int_{C} X \cdot T \, ds.
$$
Intuitively, this notation describes what happens to the vector field under integration, since $X(p) \cdot T(p)$ is the tangential component for the vector $X(p)$ at the point $p \in C$, and the contribution to the integral of a little piece of $C$ near $p$ is thus approximately the product $X(p) \cdot T(p) \, \Delta s$ of this tangential component and the arc length $\Delta s$ of the little piece. 
 
Similarly, we can write the integral of a $2$-form $Y \cdot dS$ associated to a vector field $Y$ on $\RR^3$ over a parametrized surface $S \subseteq \RR^3$ as
$$ 
\int_S Y \cdot dS = \int_S Y \cdot n \, dA,
$$
where $Y(p) \cdot n(p)$ is now the normal component of $Y$ at the point $p$ of the surface $S$. If $Y$ gives the strength and direction of a flux, then $Y \cdot n \, dA$ gives the infinitesimal rate of flow across $S$.  

The corollary result of Stoke's theorem that results for $\dim M = 3$ is called Gauss's integral theorem or the divergence theorem.

\begin{theorem}[Gauss's Integral Theorem]
If $U \subseteq \RR^3$ is open and $Y$ is a vector field on $U$, then
$$
\int_{V} \div Y \, dV = \int_{\partial V} Y \cdot n \, dA  
$$
for all compact $3$-dimensional submanifolds with boundary $V \subseteq U$. 
\end{theorem}

Here $V$ is thought of as canonically oriented by  $\mathbf{R}^3$, so by the orientation convention $n$ means the outward unit normal vector field on $\partial V$.

In the two-dimensional case we have the classical Stoke's theorem, for which the more general theorem is named.

\begin{theorem}[Stokes's Integral Theorem]
If $U \subseteq \RR^3$ is open and $X$ is a vector field on $U$, then
$$
\int_{S} \rot X \cdot n\, dA = \int_{\partial S} X \cdot T \, ds  
$$
for all oriented compact surfaces with boundary $S \subseteq U$. 
\end{theorem}

\begin{figure}[H]
\centering
\scalebox{.2}
	\centering
	\includegraphics[scale=0.64]{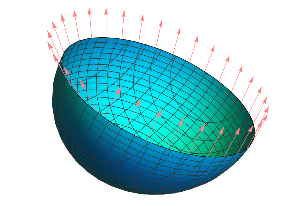}
	\caption{Stokes's integral theorem states that the integral of the vector field on the boundary is equal to the integral of the normal component of its rotational over the surface it bounds.}
\end{figure}

For completeness, we also mention the one-dimensional case, although it has no name of its own. 

\begin{theorem}
If $U \subseteq \RR^3$ is open and $f \colon U \to \RR$ is a smooth function, then
$$
\int_C \grad f \cdot T \, ds = f(q) - f(p)
$$
for all oriented curves $C \subseteq U$ from $p$ to $q$. 
\end{theorem}

\section{An application: conservation of mass}

Consider a domain  $D$ in $\RR^{3}$ which is contained in a region
that is filled with a fluid. For our immediate purposes,
by a fluid we mean a continuous distribution of matter that traverses a
well defined trajectory. Mathematically, the fluid is determined by two quantities.
\begin{itemize}
\item A density function $\rho(x,t),$ which specifies the density of the fluid at a point $x$ and time $t$.
\item A velocity vector field \[v=\sum_{k}
v^{k}(x,t)\partial_{x^k},\] 
which describes the movement of the fluid.
\end{itemize}
\begin{figure}[H]
	\centering
	\includegraphics[scale=0.54]{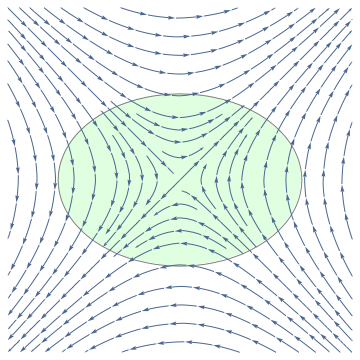}
	\caption{Motion of a fluid.}
	
\end{figure}
The density function has the property that, for each
$t$, the total mass contained in $D$ is equal to 
\[m(t,D)=\int_{D}\rho(x,t) dx^1 \wedge dx^2 \wedge dx^3.\] 
Hence, the rate of change of mass inside $D$ is
given by \[ \frac{dm}{dt}=\int_{D}\frac{d\rho(x,t)}{dt} dx^1 \wedge dx^2 \wedge dx^3.\]
On the other hand, the fluid flow rate across a small section of boundary
$\Delta A$ is given approximately by $\rho(p,t)
v(p,t)\cdot n \,\Delta A $.
\begin{figure}[H]
	\centering
	\includegraphics[scale=0.54]{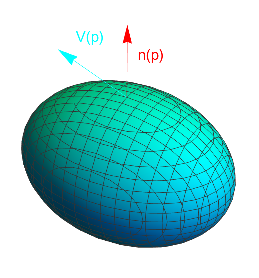}
	\caption{The normal vector and the velocity of a fluid.}
	
\end{figure}
Therefore, the total fluid crossing the boundary $\partial D$ at time $t$ is
\[\int\nolimits_{\partial D}\rho(x,t)v(x,t) \cdot n \,dA.\] 
By the principle of conservation of mass, the total fluid crossing the boundary
must be equal to the rate of change of mass, i.e.
\[\int\nolimits_{\partial D}\rho(x,t)v(x,t) \cdot n\, dA=-\int_{D}\frac{d\rho(x,t)}{dt} dx^1 \wedge dx^2 \wedge dx^3.\]
On the other hand, by Gauss' theorem,
\[\int\nolimits_{\partial D}\rho(x,t)v(x,t) \cdot n dA= \int_D \div (\rho v) dx^1\wedge dx^2 \wedge dx^3.\]
Since these equations are valid on an arbitrary domain $D$, one concludes:
\begin{equation}
\frac{d\rho}{dt}+\div(\rho v)=0. \label{CntE}%
\end{equation}
This equation is known as the continuity equation and expresses the conservation of mass for a fluid.
   \clearemptydoublepage
\chapter{The metric determines the geometry}

\vspace{3ex}

\section{The metric tensor}

So far we have considered only topological properties of smooth manifolds which,
by themselves, are flexible objects without any specific geometric structure.
In order to study geometric properties such as angles, distances and volumes,
additional structure is necessary. This structure is a Riemannian (or Lorentzian) metric.

Let $V$ be a finite dimensional vector space. A bilinear form $g:V\otimes
V\rightarrow\RR$ is symmetric if $g(v,w)=g(w,v)$. It is non degenerate
if $g(v,w)=0$ for all $w\in V$ implies $v=0$. As explained in Appendix
\ref{0algebra basicos}, given a non degenerate symmetric bilinear form $g$ there exists an orthonormal basis
$\{e_{1},\dots,e_{k}\}$ and a natural number $p\leq k$ such that
\[
g(e_{i},e_{j})=%
\begin{cases}
0 & \text{ if }i\neq j,\\
1 & \text{ if }i=j\leq p,\\
-1 & \text{ if }i=j>p.
\end{cases}
\]
Moreover, the number $p$ is well defined. The signature of $g$ is the pair of
numbers $(p,q)$ where $p+q=\dim V$.

A pseudo-Euclidean structure in a vector space $V$ of dimension $n$ is a
symmetric bilinear form $g:V\otimes V\rightarrow\RR$
which is symmetric and non degenerate. A Euclidean structure on $V$ is a
pseudo-Euclidean structure of signature $(n,0)$. A Lorentzian structure on $V$
is a pseudo-Euclidean structure of signature $(n-1,1).$

 If $V$ is a finite dimensional vector space with a Lorentzian structure then, in an orthonormal basis, the equation $g(v,v)=0$  takes the form
\[ (x^1)^2 + \dots + (x^{n-1})^2=(x^n)^2.\]
The solutions to this equation define a cone, which in relativity is known as the light cone. Vectors such that $g(v,v)<0$ are called time like vectors and vectors which satisfy $g(v,v)>0$ are space like vectors. 
\begin{figure}
 \centering
 \includegraphics[scale=0.56]{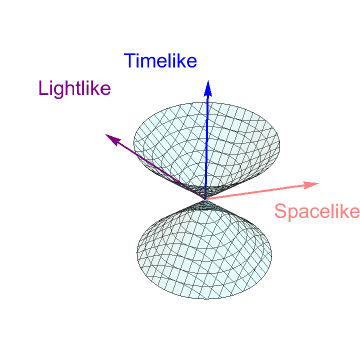}
 \caption{Lightcone.}
\end{figure}

A semi-Riemannian metric on $M$ is a section $g\in\Gamma(T^*M^{\otimes 2})$ such that for each $p\in M$ the bilinear form $ g_p:T_{p}M\otimes T_{p}M\rightarrow\RR$  
is a pseudo-Euclidean structure on $T_{p}M$.
A semi-Riemannian metric $g$ is Riemannian if for all $p \in M$ the bilinear
form $g_p$ is a Euclidean structure on $T_{p}M$. A semi-Riemannian metric
$g$ is Lorentzian if for all $p \in M$ the bilinear form $g_p$ is a
Lorentzian structure on $T_{p}M$. A Riemannian manifold is a manifold together
with a Riemannian metric. A Lorentzian manifold is a manifold together with a
Lorentzian metric.

Most of differential geometry is concerned with Riemannian manifolds. However,
we will focus mainly on the Lorentzian case because, in general relativity, spacetime
is modeled by a Lorentzian manifold. Observe that, in a Lorentzian manifold, each tangent space has 
a light cone that classifies vectors as light like, space like and time like. This asymmetry is the way in which
the difference between space and time is encoded in Einstein's theory. In what follows, a ``metric'' refers to either a Riemannian or Lorentzian metric.
We will only be specific when when the distinction is important.

In local coordinates, a metric $g$ can be written in the form \[g=\sum
_{ij}g_{ij}dx^{i}\otimes dx^{j},\] where the functions $g_{ij}$ are determined
by the property%

\[
g_{ij}(p)= g_p\left(\frac{\partial}{\partial x^{i}}\bigg\vert_p, \frac{\partial}{\partial x^{j}}\bigg\vert_p\right).
\]

\begin{example}
The standard Riemannian metric on the manifold $\RR^{m}$ which gives
each tangent space $T_{p}\RR^{m}\cong\RR^{m}$ the usual inner
product, is given by \[g=\sum_{i}dx^{i}\otimes dx^{i}.\]
\end{example}

\begin{example}
Minkowski spacetime is the Lorentzian manifold $\RR^{4}$ with metric
\[g=-dx^{0}\otimes dx^{0}+ \sum_{i=1}^3
dx^{i}\otimes dx^{i}.\]
\end{example}

\begin{example}
\label{4hiperbolico}
Consider the manifold $M=\HH_{+}^{n}$ defined by \[\HH_{+}%
^{n}=\{(x^{1},\ldots,x^{n})\in\RR^{n}\mid x^{n}>0\},\] with Riemannian metric
\[g=\frac{1}{(x^{n})^{2}}\sum_{i}dx^{i}\otimes dx^{i}.\] This manifold is called
the hyperbolic $n$-dimensional space.
\end{example}

Let $\iota:S\rightarrow M$ be an immersion and $g$ be a semi-Riemannian
metric on $M.$ The pull-back bilinear form $\iota^{\ast}g\in\Gamma
(T^*S^{\otimes 2})$ is defined by
\[
(\iota^{\ast}g)_p(v,w)=g_{\iota(p)}(D\iota(p)(v),D\iota(p)(w)),
\]
for $p\in S$ and $v,w\in T_{p}S$. The bilinear form $(\iota^{\ast}g)_p$ is
symmetric but in general it may fail to be non degenerate.
Let us consider local coordinates $\psi=(y^{a})$ around $p\in S$ and
$\varphi=(x^{i})$ around $\iota(p)\in M$, so that we can write \[g=\sum
_{i,j}g_{ij}dx^{i}\otimes dx^{j}.\]
The local expression of the pullback form in a neighborhood of $p$ is
\begin{equation}
\label{E10}\iota^{\ast}g=\sum_{a,b}\left(  \sum_{i,j}\iota^{*}g_{ij}\frac{\partial x^i}{\partial y^{a}}\frac{\partial x^{j}}{\partial y^{b}%
}\right)  dy^{a}\otimes dy^{b}.
\end{equation}

\begin{example}
Let $\iota:S^{2}\hookrightarrow\RR^{3}$ be the standard embedding of
the sphere in $\RR^{3}$ and \[g=\sum_{i}dx^{i}\otimes dx^{i}\] the
euclidean metric on $\RR^3$. Let $S$ be the southern hemisphere and consider coordinates
$\varphi:S\rightarrow\RR^{2}$ given by
\[
\varphi(x^{1},x^{2},x^{3})=\bigg(\frac{x^{1}}{1-x^{3}},\frac{x^{2}}{1-x^{3}%
}\bigg)
\]
which are defined by the stereographic projection from the north pole. The
inverse function $\varphi^{-1}:\RR^{2}\rightarrow S,$ with
$(y^{1},y^{2})\mapsto(x^{1},x^{3},x^{3})$ is given by
\begin{align*}
x^{1}(y^{1},y^{2})  &  =\frac{2y^{1}}{(y^{1})^{2}+(y^{2})^{2}+1},\ \\
x^{2}(y^{1},y^{2})  &  =\frac{2y^{2}}{(y^{1})^{2}+(y^{2})^{2}+1},\ \\
x^{3}(y^{1},y^{2})  &  =\frac{(y^{1})^{2}+(y^{2})^{2}-1}{(y^{1})^{2}%
+(y^{2})^{2}+1}.
\end{align*}
Thus, $\iota=\varphi^{-1}$ is an embedding of $\RR^{2}$ into
$\RR^{3}$. The Jacobian of $\iota$ is%
\[
D\iota=\frac{2}{((y^{1})^{2}+(y^{2})^{2}+1)^{2}}\left(
\begin{array}
[c]{cc}%
-(y^{1})^{2}+(y^{2})^{2}+1 & -2y^{1}y^{2}\\
-2y^{1}y^{2} & (y^{1})^{2}-(y^{2})^{2}+1\\
2y^{1} & 2y^{2}%
\end{array}
\right)
\]
We conclude that the pullback metric is
\[
\iota^{\ast}g=\frac{4}{((y^{1})^{2}+(y^{2})^{2}+1)^{2}}\left(dy^{1}\otimes
dy^{1}+dy^{2}\otimes dy^{2}\right).
\]
\end{example}

\begin{figure}[H]
 \centering
 \includegraphics[scale=0.5]{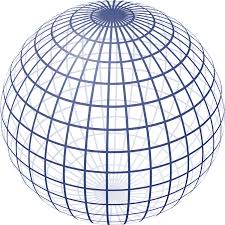}
 \caption{Sphere.}
\end{figure}

\begin{example}
In spherical coordinates $(\theta,\phi)$ for a sphere $S^{2}$ of
fixed radius $r>0$ the embedding $\iota:S^{2}\rightarrow\RR^{3}$ takes
the form
\begin{align*}
x^{1}  &  =r\sin\phi\cos\theta,\qquad 0<\phi<\pi,\\
x^{2}  &  =r\sin\phi\operatorname*{sen}\theta,\qquad
0<\theta<2\pi,\\
x^{3}  &  =r\cos\phi,
\end{align*}
and the metric is
\begin{equation}
\iota^{\ast}g=r^{2}\sin\nolimits^{2}\phi\text{ }d\theta\otimes
d\theta+r^{2}d\phi\otimes d\phi. \label{E13}%
\end{equation}

\end{example}

\begin{example}
Consider an embedding $\iota:S\hookrightarrow\RR^{3}$ of a surface in
$\RR^{3}$ and let $\varphi=(y^{1},y^{2})$ denote local coordinates for
$S$. In matrix notation, the  induced metric $\iota^{\ast}(g)$ is given by the
product $\iota^{\ast}g=D\iota^{\ast}D\iota$ where $D\iota$ is the Jacobian
matrix. It is common to use the notation
\[
\iota^{\ast}g=\left(
\begin{array}
[c]{cc}%
E & F\\
F & G
\end{array}
\right)  ,
\]
where
\begin{align*}
E   &=\sum_{i}\left(  \frac{\partial x^{i}}{\partial y^{1}}\right)
^{2},\\
G&=\sum_{i}\left(  \frac{\partial x^{i}}{\partial y^{2}}\right)
^{2},\\
F &=\frac{\partial x^{1}}{\partial y^{1}}\frac{\partial x^{1}}{\partial
y^{2}}+\frac{\partial x^{2}}{\partial y^{1}}\frac{\partial x^{2}}{\partial
y^{2}}+\frac{\partial x^{3}}{\partial y^{1}}\frac{\partial x^{3}}{\partial
y{2}}.
\end{align*}
It is also common to write
\[
\iota^{\ast}g=E \, dy^{1}\otimes dy^{1}+F\,dy^{1}\otimes dy^{2}+F\,dy^{2}\otimes
dy^{1}+G\,dy^{2}\otimes dy^{2}.
\]

\end{example}

\begin{example}
Let $N\subset\RR^{4}$ be the submanifold
\[
N=\{\left(  x^{a}\right)  \in\RR^{4}\mid (x^{0})^{2}-(x^{1})^{2}%
-(x^{2})^{2}-(x^{3})^{2}=0,\text{ }x^{0}>0\}.
\]
We fix local coordinates $\left(  t,\theta,\phi\right)  $ for $N$ such that
the inclusion $\iota:N\rightarrow\RR^{4}$ takes the form
\begin{align*}
x^{0}  &  =t,\\
x^{1}  &  =t\sin\phi\cos\theta,\\
x^{2}  &  =t\sin\phi\operatorname*{sen}\theta,\\
x^{3}  &  =t\cos\phi.
\end{align*}
For $t>0,$ $0<\theta<2\pi,$ and $0<\phi<\pi$. If $g$ is the Minkowski metric the $N$ is the part of the light cone
with positive first coordinate.
The induced bilinear form expressed in coordinates $\left(  t,\theta
,\phi\right)  =(y^{0},y^{1},y^{2})$ is
\[
\iota^{\ast}g=\sum_{a,b}\left(  g_{00}\frac{\partial x^{0}}{\partial y^{a}%
}\frac{\partial x^{0}}{\partial y^{b}}+g_{11}\frac{\partial x^{1}}{\partial
y^{a}}\frac{\partial x^{1}}{\partial y^{b}}+g_{22}\frac{\partial x^{2}%
}{\partial y^{a}}\frac{\partial x^{2}}{\partial y^{b}}+g_{33}\frac{\partial
x^{3}}{\partial y^{a}}\frac{\partial x^{3}}{\partial y^{b}}\right)
dy^{a}\otimes dy^{b}.
\]
which in this case simplifies to
\begin{align*}
\iota^{\ast}g  &  =dt\otimes dt-dt\otimes dt+t^{2}\operatorname*{sen}%
\nolimits^{2}\phi\text{ }d\theta\otimes d\theta+t^{2}d\phi\otimes d\phi\\
&  =t^{2}d\phi\otimes d\phi+t^{2}\operatorname*{sen}\nolimits^{2}\phi\text{
}d\theta\otimes d\theta.
\end{align*}
This form is degenerate and therefore does not define a metric on $N$.
\end{example}

\section{Length of a curve}

In a semi-Riemannian manifold not all directions are equal: some have
positive norm squared and others have negative norm squared. This asymmetry
allows for the distinction between different types of curves.
A curve $\gamma: I \rightarrow M$ on a semi-Riemannian manifold
is said to be spacelike if $g( \gamma^{\prime}(t),\gamma^{\prime}(t))>0$. It is said to be timelike if $g( \gamma^{\prime}(t),\gamma
^{\prime}(t))<0$. It is lightlike if $g( \gamma^{\prime}(t),\gamma
^{\prime}(t))=0$.

The length of a spacelike curve $\gamma:[a,b]\rightarrow M$ is \[L\left(
\gamma\right)  =\int_{a}^{b}\left\vert \gamma^{\prime}(s)\right\vert ds,\]
where the norm of $\gamma^{\prime}(s)$ is \[|\gamma^{\prime}(t)|=\sqrt
{g_{\gamma\left(  s\right)  }(\gamma^{\prime}(s),\gamma^{\prime}(s))}.\] The
length of a timelike curve $\gamma:[a,b]\rightarrow M$ is \[L\left(
\gamma\right)  =\int_{a}^{b}\left\vert \gamma^{\prime}(s)\right\vert ds,\]
where the norm of $\gamma^{\prime}(s)$ is \[|\gamma^{\prime}(t)|=\sqrt
{-g_{\gamma\left(  s\right)  }(\gamma^{\prime}(s),\gamma^{\prime}(s))}.\]
A lightlike curve has length zero.
Choosing local coordinates in $M$ we obtain the following formula for the
length of a spacelike curve.%
\[
L\left(  \gamma\right)  =\int_{a}^{b}\sqrt{
{\displaystyle\sum\limits_{i,j}}
g_{ij}(\gamma(s))\text{ }\frac{d\gamma^{i}}{ds}\frac{d\gamma^{j}}{ds}} ds.
\]
A reparametrization of a curve $\gamma$ is a curve $\gamma\circ\sigma$, where
$\sigma:[c,d]  \rightarrow[  a,b]$ is a
diffeomeorphism. One can show easily that the length of a curve is invariant under
reparame-trization, i.e., \[L(\gamma)=L(\gamma\circ\sigma).\]

\section{Isometries and Killing vector fields}

We have discussed above how vector fields generate flows, which are actions of the group $\RR$ by diffeomorphisms.
If the manifold $M$ is endowed with a metric, it is often interesting to consider diffeomorphisms that 
preserve the metric.

\begin{definition}
Let $(M,g)$ and $(N,h)$ be semi-Riemannian manifolds. A diffeomorphism $ \phi: M \rightarrow N$ is an isometry if
the derivative map $D\phi(p):T_pM \rightarrow T_{\phi(p)}N$ preserves the pseudo-Euclidean structure for all $p \in M$.
\end{definition}

\begin{example}
Let $M= \RR^m$ with the standard Riemannian metric. For any $v \in \RR^m$, the translation map $x \mapsto x+v$ is an isometry.
\end{example}

\begin{example}
Let $M= S^2$ with the standard Riemannian structure and consider a matrix $A \in \mathrm{GL}(3,\RR)$ such that
$A^{\mathrm{T}}A=I_3$ Then the map $\phi_A: S^2 \rightarrow S^2$ given by $x \mapsto Ax$ is an isometry. Indeed, take a point $p \in S^2$ and two tangent vectors 
$v,w \in T_p S^2 \subset \RR^3.$ Then
\[ \langle D\phi_A(p)(v), D\phi_A(p)(w) \rangle= \langle Av, A w \rangle =(Av)^{\mathrm{T}} \cdot Aw=v^{\mathrm{T}} A^{\mathrm{T}}A w=v^{\mathrm{T}}w= \langle v, w \rangle. \]
\end{example}

\begin{example}
The hyperbolic plane $\HH_2^+$ is the manifold \[ \HH^+_2=\{ z=x +i y \in \CC\mid y>0\}\]
with Riemannian metric
\[ g= \frac{1}{y^2} ( dx \otimes dx + dy \otimes dy).\]
Let $A= \left(\begin{array}{cc}a&b\\
c&d
\end{array}\right)$  be a matrix with $\det A=1$. The map
\[ \phi_A: \HH_2^+ \rightarrow\HH_2^+,\qquad z \mapsto \frac{az+b}{cz+d} \]
is an isometry of the hyperbolic plane known as a Moebius transformation. Let us first show that $\phi_A(z)$ belongs to the upper half plane. We have that
\begin{align*}
\phi_A(z)&= \frac{az+b}{cz+d}\\
&=\frac{(az+b)(c\overline{z}+d)}{|cz+d|^2}\\
&=\frac{ac|z|^2+adz+bc\overline{z}+bd}{|cz+d|^2}.
\end{align*}
From this, we conclude that
\[ \mathrm{Re}\,\phi_A(z)=\frac{ac|z|^2+adx+bcx+bd}{|cz+d|^2} \]
and
\[ \mathrm{Im}\,\phi_A(z)=\frac{ady-bcy}{|cz+d|^2}=\frac{y}{|cz+d|^2.}\]
In particular $\mathrm{Im}\,\phi_A(z)>0$.
Let us next show that
\[ \phi_A \circ \phi_{A'} = \phi_{AA'}.\]
Take $A'= \left(\begin{array}{cc}a'&b'\\
c'&d'
\end{array}\right)$
and compute
\[ (\phi_A \circ \phi_{A'}) (z)= \phi_A \left( \frac{az+b}{cz+d}\right)=\frac{(aa'+bc')z+ab'+bd')}{(ca'+dc')z+cb'+dd'}.\]
On the other hand,
\[ AA'= \left(\begin{array}{cc} aa' +bc'& ab'+bd'\\
ca'+dc'&cb'+dd'
\end{array}\right),\]
and therefore
\[ \phi_{AA'}(z)=\frac{(aa'+bc')z+ab'+bd'}{(ca'+dc')z+cb'+dd'}.\]
Thus the result holds true. This implies that $\phi_A$ is a diffeomorphism with inverse $\phi_{A^{-1}}$. It remains to show that the derivative of $\phi_A$ preserves the inner product.  Notice that
\[ D\phi_A(z)=\frac{1}{(cz+d)^2}.\]
Take vectors $v=\alpha+\beta i$ and $ w=\alpha'+ \beta' i$ in $T_z\HH^+_2\cong \CC$. In terms of the complex structure, the inner product can be computed as
\[  \langle v,w\rangle_z=\frac{\mathrm{Re}(v \overline{w})}{y^2}=\frac{\alpha \alpha'+\beta \beta'}{y^2}.\]
On the other hand:
\begin{align*} \langle D\phi_A(z)(v),D\phi_A(z)(w)\rangle_{\phi_A(z)}&=\left\langle \frac{v}{(cz+d)^2},\frac{w}{(cz+d)^2}\right\rangle_{\phi_A(z)}\\
&=\frac{1}{(\mathrm{Im} \phi_A(z))^2} \mathrm{Re}\left(\frac{v\overline{w}}{(cz+d)^2(c\overline{z}+d)^2}\right)\\
&=\frac{|cz+d|^4}{y^2}\frac{\mathrm{Re}(v\overline{w})}{|cz+d|^4}\\
&=\frac{\alpha \alpha'+\beta \beta'}{y^2}.
\end{align*}
One concludes that $\phi_A$ is indeed an isometry.
\end{example}

\begin{example}
Recall that Minkowski spacetime is the manifold $\RR^4$ with the metric:
\[ g= -dx^0 \otimes dx^0+ \sum_{i=1}^3 dx^i \otimes dx^i.\]
Given a number $0<u<1$ we set $\lambda_u=1/\sqrt{1-u^2}$, and define the matrix
\[ L= \left(\begin{array}{cccc}\lambda_u&-\lambda_u u&0&0\\
-\lambda_u u&\lambda_u&0&0\\
0&0&1&0\\
0&0&0&1
\end{array}\right).\]
The map $x \mapsto Lx$ is an isometry of Minkowski spacetime. 
Indeed, if we view the metric as a matrix:
\[ g= \left(\begin{array}{cccc}-1&0&0&0\\
0&1&0&0\\
0&0&1&0\\
0&0&0&1
\end{array}\right)\]
Then a simple computation shows that
\[ L^{\mathrm{T}} g L=g.\]
Take $v,w \in T_x\RR^4\cong \RR^4$ and compute
\[ \langle DL(x)(v), DL(x)(w)\rangle= \langle Lv ,Lw \rangle= v^{\mathrm{T}} L^{\mathrm{T}}gL w=v^{\mathrm{T}}gw=\langle v,w\rangle. \]
We conclude that $L$ is indeed an isometry of Minkowski spacetime. These transformations are known as Lorentz boosts.
\end{example}

Let $M$ be a manifold with metric $g$. A vector field $X \in \mathfrak{X}(M)$ is called a Killing vector field
if $L_Xg=0$. In local coordinates, 
the Lie derivative can be computed following the prescripticon of \S\ref{sec:cotangent}. If we write
$X= \sum_{i} X^{i} \partial/\partial x^i$
and
$g= \sum_{ij}g_{ij} dx^i \otimes dx^j$, we get
\[ L_X g=\sum_{ij} \sum_k \Big(X^{k}\frac{\partial g_{ij}}{\partial x^k}+ g_{kj} \frac{\partial X^{k}}{\partial x^i}+ g_{ik} \frac{\partial X^{k}}{\partial x^j}\Big) dx^i \otimes dx^j.\]
Therefore, the equations for a vector field to be a Killing vector field are
\begin{equation}
 \sum_k \Big(X^{k}\frac{\partial g_{ij}}{\partial x^k}+ g_{kj} \frac{\partial X^{k}}{\partial x^i}+ g_{ik} \frac{\partial X^{k}}{\partial x^j}\Big)=0.
\end{equation}

\begin{lemma}
Let $X$ and $Y$ be Killing vector fields in a Riemannian manifold $M$. Then $[X,Y]$ is also a Killing vector field.
\end{lemma}
\begin{proof}
The statement is a consequence of the fact that, for arbitrary vector fields $X$ and $Y$,
\[L_{[X,Y]} = L_{X}\circ L_{Y} - L_{Y} \circ L_{X}.\]
 We leave it as an exercise for the reader to verify this assertion. \end{proof}

\begin{proposition}
Let $X$ be a vector field on $M$ which has a metric $g$. If  $H_t$ is the local flow associated to $X$ then the vector field $X$ is Killing if and only
if $H_t$ is an isometry for all $t$.
\end{proposition}
\begin{proof}
Recall that the Lie derivative is defined by
\[  L_X g= \frac{d}{dt}\bigg\vert_{t=0} H_t^*g.\]
If $H_t$ is an isometry then $H_t^*g=g$ so that $L_Xg=0$. For the converse let us assume that
\[ \frac{d}{dt}\bigg\vert_{t=t_0} H_t^*g=0.\]
We compute
\begin{align*}
\frac{d}{dt}\bigg\vert_{t=t_0} H_t^*g&=  \frac{d}{ds}\bigg\vert_{s=0} H_{s+t_0}^*g\\
&=  \frac{d}{ds}\bigg\vert_{s=0} (H_{s}\circ H_{t_0})^*(g)\\
&=  \frac{d}{ds}\bigg\vert_{s=0} H_{t_0}^*( H_{s}^*g)\\
&= H_{t_0}^* \left(\frac{d}{ds}\bigg\vert_{s=0} H_{s}^*g\right)\\
\end{align*}
One concludes that $H_t^*g$ is independent of $t$. On the other hand,
$(H_0)^*(g)=g$ and therefore $H_t$ is an isometry for all $t$.
\end{proof}

\begin{example}
Let us consider the manifold $\RR^2$ with the standard Riemannian metric. The equations for a vector field
$ X = X^1 \partial/\partial x^1+ X^2 \partial/\partial x^2$
to be a Killing vector field are
\begin{equation}\label{Killing}
\frac{\partial X^1}{\partial x^1}= \frac{\partial X^2}{\partial x^2}= \frac{\partial X^1}{\partial x^2}+ \frac{\partial X^2}{\partial x^1}=0.
\end{equation}
For arbitrary constants $a,b,c \in \RR$, the vector field
\begin{equation}\label{Kplane}
X= a \frac{\partial}{\partial x^1}+ b \frac{\partial }{\partial x^2}+ c\left(x^2\frac{\partial}{\partial x^1}-x^1 \frac{\partial}{\partial x^2}\right)
\end{equation}
is a Killing vector field. Let us show that these are all Killing vector fields in the plane. Differentiating Equation \eqref{Killing} with respect to $x^1$ one obtains:
\[ 0= \frac{\partial^2 X^1}{\partial x^2 \partial x^1}+ \frac{\partial^2 X^2}{\partial x^1\partial x^1}=\frac{\partial^2 X^2}{\partial x^1\partial x^1}.\]
This implies that $X^2$ is a linear function of $x^1$ and, by symmetry, $X^1$ is a linear function of $x^2$. Using Equation \eqref{Killing} again one sees that the vector field has the form \eqref{Kplane}.

\end{example}

\begin{example}
Let us consider Minkowski spacetime $\RR^4$ with the metric
\[  g= -dx^0 \otimes dx^0+ \sum_{i=1}^3 dx^i \otimes dx^i.\]
The equations for a vector field
$X= \sum_{i=0}^3 X^{a}\partial/\partial x^a$
to be Killing are
\[ \frac{\partial X^0}{\partial x^0}=\frac{\partial X^i}{\partial x^i}=\frac{\partial X^i}{\partial x^j}+  \frac{\partial X^j}{\partial x^i}=\frac{\partial X^0}{\partial x^j}-  \frac{\partial X^j}{\partial x^0}=0\]
for all $i,j \geq 1$.The constant vector fields are Killing and generate translations. For $i,j\geq 1$ the vector field
\[ X= x^i \frac{\partial}{\partial x^j}- x^j \frac{\partial}{\partial x^i}\]
is Killing and generates rotations.
For $i \geq 1$ the vector field:
\[ X= x^0 \frac{\partial}{\partial x^i}+ x^i \frac{\partial}{\partial x^0}\]
is Killing and generates Lorentz boosts.
\end{example}

   \clearemptydoublepage
\chapter{Connections, parallel transport and geodesics}

\vspace{3ex}

\section{Connections}

Given a smooth function $f=(f^{1},\dots,f^{m}):M\rightarrow\RR^{m}$ and
a vector field $X\in\mathfrak{X}(M)$ it makes sense to consider the derivative
of the function $f$ in the direction of $X$, which at each point $p \in M$, is given by the element $T_{f(p)}\RR^m = \RR^{m}$ defined by the formula
\[
(Xf)(p)=((Xf^{1})(p),\dots,(Xf^{m})(p))=Df(p)(X(p)).
\]
On the other hand, if $\alpha\in\Gamma(E)$ is a section of a vector bundle
$E$, there is no natural way to differentiate $\alpha$ in the direction of a
vector field. A connection on a vector bundle $E$ is a choice that prescribes
such a differentiation rule.

\begin{definition}
Let $\pi:E\rightarrow M$ be a vector bundle. A connection $\nabla$ on $E$ is a
linear map $\nabla:\mathfrak{X}(M)\otimes\Gamma(E)\rightarrow\Gamma(E)$, written $(X,\alpha)\mapsto\nabla_{X}\alpha$, 
such that for any smooth function $f\in C^{\infty}(M)$, any vector field $X\in\mathfrak{X}(M)$
and any section $\alpha\in\Gamma(E)$, the following two conditions are satisfied:
\begin{enumerate}
\item[(1)] $\nabla_{fX}\alpha=f\nabla_{X}\alpha$.
\item[(2)] $\nabla_{X}\left(  f\alpha\right)  =\left(  X(f)\right)  \alpha
+f\nabla_{X}\alpha.$
\end{enumerate}
\end{definition}

In the case where $E=M \times\RR^{m}$ is the trivial bundle, a basic example of a connnection is provided by the directional derivative described above.

\begin{remark}
It is an easy exercise on partitions of unity to show that any vector bundle $\pi: E \rightarrow M$
admits a connection.
\end{remark}

If $E$ is a vector bundle with connection, we will say that a section
$\alpha\in\Gamma(E)$ is covariantly constant if $\nabla_{X} \alpha=0$ for any
vector field $X\in\mathfrak{X}(M)$. Of course, if the vector bundle is the trivial bundle and the connection is the directional derivative, covariantly constant sections are constant functions.

Let us consider the case where $E=TM$ and describe how a connection is expressed
in local coordinates $\varphi=(x^{1},\dots,x^{m})$. The Christoffel symbols $\Gamma_{ij}^{k}:M\rightarrow\RR$ are smooth functions characterized by
the condition \[\nabla_{\partial_{i}}{\partial_{j}}=\sum_{k}%
\Gamma_{ij}^{k}\partial_{k}.\] The connection $\nabla$ is
determined by the Christoffel symbols. In fact, given vector fields
$X=\sum_{i}X^{i}\partial_{i},$ $Y=\sum_{j}Y^{j}\partial_{j}$, one computes
\begin{align*}
\nabla_{Y}X   &=\sum_{i}Y^{i}\nabla_{\partial_{i}}\left(  \sum_{j}%
X^{j}\partial_{j}\right)  \\
&=\sum_{i,j}Y^{i}\nabla_{\partial_{i}}\left(
X^{j}\partial_{j}\right) \\
  &=\sum_{i,j}Y^{i}\left(  \frac{\partial X^{j}}{\partial x^{i}}\partial
_{j}+X^{j}\nabla_{\partial_{i}}\partial_{j}\right) \\
&=\sum_{i,j}Y^{i}\left(  \frac{\partial X^{j}}{\partial x^{i}}\partial
_{j}+X^{j}\sum_{k}\Gamma_{ij}^{k}\partial_{k}\right) \\
&  =\sum_{i,j}Y^{i}\frac{\partial X^{j}}{\partial x^{i}}\partial_{j}%
+\sum_{i,j}Y^{i}X^{j}\sum_{k}\Gamma_{ij}^{k}\partial_{k}\\
&  =\sum_{k}\left(  \sum_{i}Y^{i}\frac{\partial X^{k}}{\partial x^{i}} +\sum_{i,j}\Gamma_{ij}^{k}Y^{i}X^{j}\right)  \partial_{k}.
\end{align*}
 
As we have discussed before, given a vector bundle $E$, one can construct new bundles by
the usual operations of linear algebra such as taking duals and tensor
products. A connection $\nabla$ on $E$ induces connections
on all the bundles naturally associated to $E$. This is the content of the
following remark.

\begin{remark}
\label{associated} Let $\nabla, \nabla^{\prime}$ be connections on the vector
bundles $\pi:E\rightarrow M$ and $\pi:E^{\prime}\rightarrow M$, respectively. There are induced connections:
\begin{itemize}
 \item On $E^*$ given by 
    $$
    (\nabla_{X}\xi)(\alpha)=X(\xi(\alpha))-\xi(\nabla_{X}\alpha)
    $$

    \item On $E \oplus E'$ given by
    $$
    \nabla_{X}(\alpha+\beta)=\nabla_{X}\alpha+\nabla_{X}\beta.
    $$
    
    \item On $E \otimes E'$ given by 
    $$
    \nabla_{X}(\alpha\otimes\beta)=\nabla_{X}\alpha\otimes\beta+\alpha
\otimes\nabla'_{X}\beta.
    $$
    
    \item On $E^{\otimes k}$ given by
    $$
    \nabla_{X}(\alpha_{1}\otimes\dots\otimes\alpha_{k})=\sum_{i}\alpha_{1}%
\otimes\dots\otimes\nabla_{X}\alpha_{i}\otimes\dots\otimes\alpha_{k}
    $$
    
    \item On $\Lambda^k E$ given by
    $$
    \nabla_{X}(\alpha_{1}\wedge\dots\wedge\alpha_{k})=\sum_{i}\alpha_{1}%
\wedge\dots\wedge\nabla_{X}\alpha_{i}\wedge\dots\wedge\alpha_{k}
    $$
\end{itemize}
\end{remark}

\section{The Levi-Civita connection}

A  metric $g$ on a manifold $M$ induces a connection on the tangent bundle, called
the \emph{Levi-Civita Connection}. This means that, once the geometry of $M$ is fixed, there is a rule for 
covariantly differentiating vector fields on it.

\begin{definition}
Let $\nabla$ be a connection on $TM$. The torsion of $\nabla$ is the function $T: \mathfrak{X}(M) \times \mathfrak{X}(M) \rightarrow\mathfrak{X}(M)$ defined as
\begin{align*}
T(X,Y) = \nabla_{X} Y -\nabla_{Y} X - [X,Y].
\end{align*}
\end{definition}

It is easy to verify that, given vector fields $X,Y,Z\in\mathfrak{X}(M)$ and a function $f \in C^{\infty}(M)$, the
torsion satisfies the following properties:
\begin{itemize}
\item $T(fX,Y)=fT(X,Y)$ and 
$T(X,fY)=fT(X,Y)$.

\item $T\left(X,Y\right)  +T\left(  Y,X\right)  =0$.
\end{itemize}
In view of this, we can identify the torsion with a section $T \in \Gamma(\Lambda^2 T^*M \otimes TM)$, defined by
\[
T_p(v,w)=(\nabla_{X}Y)_p-(\nabla_{Y}X)_p-[X,Y]_p,
\]
for any choice of vector fields $X,Y$ such that $X_p=v$ and $Y_p=w$. 
With this in mind, a connection on $TM$ is called symmetric if its torsion is zero. It is easy to show that this a connection is symmetric if and only if for any choice of
coordinates, the Christoffel symbols satisfy $\Gamma_{ij}^{k}=\Gamma_{ji}
^{k}$.

Let now $M$ be a smooth manifold endowed with a metric. A connection on $TM$ is
compatible with $g$ if $g$ is covariantly constant, that is to say, $\nabla
_{X}g=0,$ for all $X\in\mathfrak{X}(M)$. Here $g$ is seen as a section of the vector bundle
$T^*M\otimes T^*M$ which has a connection induced by
$\nabla$. The condition that $g$ is covariantly constant is equivalent to
\[
{X}g(Y,Z)=g\left(  \nabla_{X}Y,Z\right)  +g\left(  Y,
\nabla_{X}Z\right) ,
\]
for all $X,Y,Z\in\mathfrak{X}(M)$.

The following is sometimes called the ``Fundamental Theorem of Riemannian Geometry'', and is based on the work of Levi-Civita.

\begin{theorem}
\label{4Teo3}Let $g$ be a metric on $M$. There exists a unique torsion free connection $\nabla$ which is
compatible with the metric. Moreover, this connection satisfies
\begin{align}\label{LC}
\begin{split}
g\left(  Z,\nabla_{Y}X\right)     =\frac{1}{2}\Big( & Xg\left(  Y,Z\right)
+Yg\left(  Z,X\right)  -Zg\left(  X,Y\right)   \\
    &-g\left(  \left[  X,Z\right]  ,Y\right)  -g\left(  \left[
Y,Z\right]  ,X\right)  -g\left(  \left[  X,Y\right]  ,Z\right)  \Big)  .
\end{split}
\end{align}

\end{theorem}

\begin{proof}
Any connection compatible with the metric satisfies
\begin{align*}
Xg\left(  Y,Z\right)  &=g\left(  \nabla_{X}Y,Z\right)  +g\left(  Y,\nabla
_{X}Z\right)  ,\\
Yg\left(  Z,X\right)  &=g\left(  \nabla_{Y}Z,X\right)  +g\left(  Z,\nabla
_{Y}X\right), \\
Zg\left(  X,Y\right)  &=g\left(  \nabla_{Z}X,Y\right)  +g\left(  X,\nabla
_{Z}Y\right)  .
\end{align*}
Adding the first two equations, subtracting the third and using the symmetry one obtains
\begin{align*}
&  Xg\left(  Y,Z\right)  +Yg\left(  Z,X\right)  -Zg\left(  X,Y\right) \\
& \qquad =g\left(  \left[  X,Z\right]  ,Y\right)  +g\left(  \left[  Y,Z\right]
,X\right)  +g\left(  \left[  X,Y\right]  ,Z\right)  +2g\left(  Z,\nabla
_{Y}X\right)  ,
\end{align*}
which implies equation \eqref{LC}. Since the metric is nondegenerate, this implies uniqueness.

In order to prove existence we define  $\nabla_{Y}%
X$
to be the unique vector field that satisfies equation \eqref{LC}. In order to prove that $\nabla$ defined in this way is a connection, the only nontrivial statement is
\[\nabla_X (fY)=f \nabla_X Y +(Xf) Y.\]
For this we compute
\begin{align*}
g\left(  Z,\nabla_{Y}\left(  fX\right)  \right)     =\frac{1}{2}\Big(
& fXg\left(  Y,Z\right)  +Yg\left(  Z,fX\right)  -Zg\left(  fX,Y\right)  
\\
&   -g\left(  \left[  fX,Z\right]  ,Y\right)  -g\left(  \left[
Y,Z\right]  ,fX\right)  -g\left(  \left[  fX,Y\right]  ,Z\right)  \Big)  .
\end{align*}
Using the equations
\begin{align*}
Yg\left(  Z,fX\right)  &=\left(  Yf\right)  g\left(  Z,X\right)  +fYg\left(
Z,X\right)  ,\\
Zg\left(  fX,Y\right)  &=\left(  Zf\right)  g\left(  X,Y\right)  +fZg\left(
X,Y\right)  ,\\
g\left(  \left[  fX,Z\right]  ,Y\right)  &=fg\left(  \left[  X,Z\right]
,Y\right)  -\left(  Zf\right)  g\left(  X,Y\right), \\
g\left(  \left[  fX,Y\right]  ,Z\right)  &=fg\left(  \left[  X,Y\right]
,Z\right)  -\left(  Yf\right)  g\left(  X,Z\right),
\end{align*}
one then obtains
\begin{align*}
g\left(  Z,\nabla_{Y}\left(  fX\right)  \right)   &  =fg\left(  Z,\nabla
_{Y}X\right)  +\frac{1}{2}\left(  2\left(  Yf\right)  g\left(  Z,X\right)
\right) \\
&  =g\left(  Z,f\nabla_{Y}X+\left(  Yf\right)  X\right)  ,
\end{align*}
as required. We leave it as an exercise to the reader to prove that $\nabla$ is symmetric and compatible with the metric.
\end{proof}

The connection defined by Theorem \ref{4Teo3} is called the Levi-Civita
connection on $(M,g)$. Given  a semi-Riemannian manifold we shall use this connection unless special exception is made. 

Considering Equation \eqref{LC} in local coordinates $\varphi=(x^{1}%
,\dots,x^{m})$ and taking $Z=\partial_{k}$, $Y=\partial_{j}$ and
$X=\partial_{i},$ we see that
\begin{equation}
\sum_{l}\Gamma_{ij}^{l}g_{lk}=\frac{1}{2}\left(  \frac{\partial g_{jk}%
}{\partial x^{i}}+\frac{\partial g_{ki}}{\partial x^{j}}-\frac{\partial
g_{ij}}{\partial x^{k}}\right)  . \label{chistofel 2}%
\end{equation}
Since the matrix $(g_{ij})$\ is invertible we can write $(g^{ij})$ for its
inverse. Then
\[
\sum_{k,l}\Gamma_{ij}^{l}g_{lk}g^{kn}=\frac{1}{2}\sum_{k}g^{kn}\left(
\frac{\partial g_{jk}}{\partial x^{i}}+\frac{\partial g_{ki}}{\partial x^{j}%
}-\frac{\partial g_{ij}}{\partial x^{k}}\right)  .
\]
Each of the terms in parenthesis is called the Christoffel symbol of the
first kind and is denoted by
\[
\Gamma_{k}^{ij}=\frac{\partial g_{jk}}{\partial x^{i}}+\frac{\partial g_{ki}%
}{\partial x^{j}}-\frac{\partial g_{ij}}{\partial x^{k}}.
\]
On the other hand
\[
\sum_{l}\left(  \Gamma_{ij}^{l}\sum_{k}g_{lk}g^{kn}\right)  =\Gamma_{ij}^{n}.
\]
Therefore
\begin{equation}
\label{Chris}\Gamma_{ij}^{n}=\frac{1}{2}\sum_{k}g^{kn}\Gamma_{k}^{ij}=\frac
{1}{2} \sum_{k} g^{kn}\Big(\frac{\partial g_{jk}}{\partial x^{i}}%
+\frac{\partial g_{ki}}{\partial x^{j}}-\frac{\partial g_{ij}}{\partial x^{k}%
}\Big).
\end{equation}
This useful formula expresses the Christoffel symbols, and therefore the
connection, in terms of the metric.

By Remark \ref{associated} we know that the Levi-Civita connection induces
connections on the vector bundles $TM^{\otimes p}\otimes T^{\ast}M^{\otimes
q}$. In other words, one can define the covariant derivative $\nabla_X T$ of a tensor field $T$ of type $(p,q)$ along a vector field $X$. The coordinate expression for $\nabla_X T$ may be worked out exactly as for vector fields.  Writing $X = \sum_j X^{j} \partial_j$ and $T =\sum_{i_1,\dots,i_p}\sum_{j_1,\dots,j_q} T^{i_1 \cdots i_p}_{\phantom{i_1 \cdots i_p} j_1 \cdots j_p}\partial_{i_1} \otimes \cdots \otimes \partial_{i_1} \otimes dx^{j_1} \otimes \cdots \otimes dx^{j_q}$, we have
\begin{align*}
 &\nabla_X T =\sum_{i_1,\dots,i_p}\sum_{j_1,\dots,j_q}  \sum_{k} X^k \nabla_{k} T^{i_1 \cdots i_p}_{\phantom{i_1 \cdots i_p} j_1 \cdots j_q} \partial_{i_1} \otimes \cdots \otimes \partial_{i_1} \otimes dx^{j_1} \otimes \cdots \otimes dx^{j_q},
\end{align*}
where here we have put
\begin{align*}
   \nabla_{k} T^{i_1 \cdots i_p}_{\phantom{i_1 \cdots i_p} j_1 \cdots j_q} =  \frac{\partial T^{i_1 \cdots i_p}_{\phantom{i_1 \cdots i_p} j_1 \cdots j_q}}{\partial x^k} + \sum _{l} T^{l i_2 \cdots i_p}_{\phantom{l i_2 \cdots i_p} j_1 \cdots j_q} \Gamma_{l k}^{i_1} + \cdots + \sum_{l} T^{i_1 \cdots i_{p-1} l}_{\phantom{i_1 \cdots i_{p-1} l} j_1 \cdots j_q} \Gamma_{l k}^{i_p}.
\end{align*}
We refer to $\nabla_{k} T^{i_1 \cdots i_p}_{\phantom{i_1 \cdots i_p} j_1 \cdots j_q}$ as the covariant derivative of $T$ with respect to $x^{i}$. 

Let $T$ be a tensor field of type $(p,q)$ and $S$ a tensor field of type $(r,s)$. Then for any vector field $X$, one has
$$
\nabla_X (T \otimes S) = \nabla_X T \otimes S + T \otimes \nabla_X S. 
$$
Thus, in terms of components, this becomes
$$
\nabla_k (T^{i_1 \cdots i_p}_{\phantom{i_1 \cdots i_p} j_1 \cdots j_q} S^{k_1 \cdots k_r}_{\phantom{k_1 \cdots k_r} l_1 \cdots l_s}) = \nabla_k T^{i_1 \cdots i_p}_{\phantom{i_1 \cdots i_p} j_1 \cdots j_q} S^{k_1 \cdots k_r}_{\phantom{k_1 \cdots k_r} l_1 \cdots l_s} + T^{i_1 \cdots i_p}_{\phantom{i_1 \cdots i_p} j_1 \cdots j_q} \nabla_k S^{k_1 \cdots k_r}_{\phantom{k_1 \cdots k_r} l_1 \cdots l_s}. 
$$
This justifies the extensive use of local expressions in the physics literature.

Now we want to define the notion of divergence of a tensor field. For this we need some terminology. 
Fix $p,q >0$ and let $1 \leq i \leq p$ and $1 \leq j \leq q$. Then there is a contraction map $C^{i}_{j} \colon \mathcal{T}^{(p,q)}(M) \to \mathcal{T}^{(p-1,q-1)}(M)$, which is defined by
\begin{align*}
&C^{i}_{j}(X_{1} \otimes \cdots \otimes X_p \otimes \alpha^{1} \otimes \cdots \otimes \alpha^{q} )\\
&\quad = \alpha^{i}(X_{j})X_{1} \otimes \cdots \otimes X_{j-1} \otimes X_{j+1} \otimes \cdots \otimes X_{p} \otimes \alpha^{1} \otimes \cdots \otimes \alpha^{i-1} \otimes \alpha^{i+1} \otimes \cdots \otimes \alpha^{q}  \end{align*}
It is straightforward to show that $C^{i}_{j}$ commutes with the connection, that is, $\nabla_X \circ C^{i}_{j} = C^{i}_{j} \circ \nabla_X$ for every vector field $X$.

Next let $T$ be a tensor field of type $(p,q)$ on $M$. If we let $X$ be any vector field, the above local expression for $\nabla_X T$ shows that $(\nabla_X T)(\alpha^1,\dots,\alpha^{p},X_1,\dots,X_q)$ depends only on the point values of $X$. Consequently, one gets a tensor field $\nabla T$ of type $(p,q+1)$ on $M$ with components $(\nabla T)^{i_1 \cdots i_p}_{\phantom{i_1 \cdots i_p}j_1 \cdots j_q k} = \nabla_k T^{i_1 \cdots i_p}_{\phantom{i_1 \cdots i_p}j_1 \cdots j_q}$.

With this background in mind, the divergence of a tensor field $T$ of type $(p,q)$ is a tensor field of type $(p-1,q)$ obtained by contracting the last contravariant and covariant indices of $\nabla T$:
$$
\div T = C^{p}_{q+1}( \nabla T).
$$
Written out explicitly in components, this is
$$
(\div T)^{i_1 \cdots i_{p-1}}_{\phantom{i_1 \cdots i_{p-1}}j_1 \cdots j_q} = \sum_{k} \nabla_{k} T^{i_1 \cdots i_{p-1} k}_{\phantom{i_1 \cdots i_{p-1}k}j_1 \cdots j_q}.
$$
For example, for a vector field, the following formula is easy to check:
$$
\div X = \sum_{i} \nabla_i X^{i} =\sum_{i} \frac{1}{\sqrt{\det g}} \frac{\partial}{\partial x^{i}} (\sqrt{\det g} X^{i}).
$$

The metric $g$ also induces an isomorphism $g^{\sharp}:TM\rightarrow T^{\ast}M$ which is given by
\[
g^{\sharp}(X)(Y)=g(X,Y).
\]
Since the metric $g$ is covariantly constant
we know that
\[
g^{\sharp}(\nabla_{Z}X)(Y)=g(\nabla_{Z}X,Y)=Zg(X,Y)-g(X,\nabla_{Z}%
Y)=\nabla_{Z}(g^{\sharp}(X))(Y).
\]
Thus we conclude that
\[
g^{\sharp}(\nabla_{Z}X)=\nabla_{Z}(g^{\sharp}(X)).
\]
This means that the isomorphism $g^{\sharp}$ preserves the connection. We also
conclude that the inverse map $g^{\flat} = (g^{\sharp})^{-1} \colon T^*M \rightarrow TM$ preserves the connection.

\begin{example}
Let $\varphi=(r,\theta)$ be the  polar coordinates in $\RR%
^{2},$ and $x^{1}=r\cos\theta,$ $x^{2}=r\sin\theta$, the
euclidean coordinates. The Jacobian matrix for the change of variables is
\[
J=\left(
\begin{array}
[c]{cc}%
\cos\theta & -r\sin\theta\\
\sin\theta & r\cos\theta
\end{array}
\right)
\]
Hence, in polar coordinates, the Euclidean metric is given by
\[
g=\left(
\begin{array}
[c]{cc}%
1 & 0\\
0 & r^{2}%
\end{array}
\right)
\]
Which can also be written as $g=dr\otimes dr+r^{2}d\theta\otimes d\theta.$
Using formula (\ref{Chris}) we see that the Christoffel symbols for the
standard connection $\nabla$ on $\RR^{2}$ are:
\[
\Gamma_{\theta\theta}^{r}=-r, \quad \Gamma_{r\theta}^{\theta}=\Gamma_{\theta
r}^{\theta}=1/r,
\]
and all other coefficients are zero. 

\end{example}

\section{The pullback of bundles and connections}

Let $\nabla,\nabla^{\prime}$ be connections on $E$. Then there exists a
differential form
\[
\theta\in\Omega^{1}(M,\mathrm{End}(E))=\Gamma(T^{\ast}M\otimes\mathrm{End}%
(E)),
\]
defined by \[\theta(X,\alpha)=\nabla_{X}\alpha-\nabla_{X}^{\prime}\alpha.\]
On the other hand, for any $\theta\in\Omega^{1}(M,\mathrm{End}(E))$ the
expression \[\nabla_{X}^{\prime}\alpha=\nabla_{X}\alpha+\theta(X,\alpha)\]
defines a connection on $E$. We conclude that the space $\mathrm{Conn}(E)$ has
the structure of an affine space modeled over the vector space $\Omega
^{1}(M,\mathrm{End}(E))$.

In local coordinates $\varphi=(x^{1},\dots,x^{m})$ where the bundle is
trivialized with a frame of sections $\{\alpha_{1},\dots,\alpha_{k}\}$ there
is a natural connection determined by the condition \[\nabla_{\partial_{i}%
}^{\prime}\alpha_{j}=0.\] Therefore any other connection $\nabla$ on $E|_{U}$
is determined by a differential form $\theta\in\Omega^{1}(U,\mathrm{End}(E))$
such that $\nabla_{X}\alpha=\nabla_{X}^{\prime}\alpha+\theta(X)(\alpha).$

Let $f:N\rightarrow M$ a smooth function and $\pi:E\rightarrow M$ a vector
bundle. Then the set $f^{\ast}E=\coprod_{p\in M}E_{f(p)},$ admits a
unique structure of a vector bundle over $M$ such that:
\begin{enumerate}
\item The projection $\pi:f^{\ast}E\rightarrow M$ is given by $v\in
E_{f(p)}\mapsto p$. 
\item The map $\tilde{f}:f^{\ast}E\rightarrow E$ given by $v\in E_{p}\mapsto
v\in E_{f(p)}$ is smooth.

\item The diagram
\[
\xymatrix{
f^*E \ar[r]^{\tilde{f}} \ar[d]_{\pi}& E\ar[d]^{\pi}\\
M \ar[r]^{f}& N}
\]
commutes and is a linear isomorphism on each fiber.

\item If $h:S\rightarrow M$ is another smooth map then there is a
natural isomorphism $h^{\ast}(f^{\ast}E)\cong(f\circ h)^{\ast}E.$
\end{enumerate}
The
vector bundle $f^{\ast}E$ is called the pullback of $E$ along $f$. We will
now see that a connection on $E$ induces one on $f^{\ast}E$.

\begin{proposition}
Let $\nabla$ be a connection on $\pi:E\rightarrow M$ and $f:N\rightarrow M$ a
smooth function. Then there exists a unique connection $f^{\ast}\nabla$ on
$f^{\ast}E$ such that for any $\alpha\in\Gamma(E)$, $X\in\mathfrak{X}(N)$
and $Y\in\mathfrak{X}(M)$ satisfying $Df(p)(X(p))=Y(f(p))$ the following holds:
\begin{equation}
\big((f^{\ast}\nabla)_{X}(f^{\ast}\alpha)\big)(p)=(\nabla_{Y}\alpha)(f(p)).
\label{pull0}%
\end{equation}

\end{proposition}

\begin{proof}
Since connections are local operators, it suffices to consider the case of a trivializable bundle. Let $\{ \alpha_1, \dots ,\alpha_k\}$ be a frame of local sections on $E$.
This defines a new connection $\nabla'$ in $E$ determined by
\[ \nabla'_Y \alpha_i=0.\]
We define the connection
\[ f^*\nabla=\nabla'' + f^*\theta,\]
where $\nabla''$ is the connection on $f^*E$ determined by the condition
\[ \nabla''_X(f^*(\alpha_i))=0,\]
and $\theta$ is the differential $1$-form
\[ \theta= \nabla-\nabla'.\]
Let us verify that the connection $f^*\nabla$ satisfies Equation (\ref{pull0}). Since $\{\alpha_1, \dots ,\alpha_k\}$ are a frame for $E$, it is enough to consider a section $\alpha$ of the form $\alpha= h \alpha_i$. Then we compute:
\begin{eqnarray*}
(f^*\nabla)_X(f^*(h\alpha_i))(p) &=& (f^*\nabla)_X(f^*(h) f^*(\alpha_i))(p)\\
&=&X( f^*(h)) f^*(\alpha_i)(p)+ f^*(h)f^*(\nabla_i)_X( f^*(\alpha_i))(p)\\
&=& X( f^*(h)) f^*(\alpha_i)(p)+ h(f(p))\theta(f(p))( Df(p)(X(p), \alpha_i(f(p))\\
&=& Y(h) (\alpha_i)(p)+ h(f(p))\theta(f(p))(Y(f(p)), \alpha_i( f(p)))\\
&=& \nabla_Y(h \alpha_i)( f(p)).
\end{eqnarray*}
It remains to prove the uniqueness of $f^*\nabla$. Let $\tilde{\nabla}$ be other connection on $f^*E$ with the required properties. Since $\{f^*\alpha_1, \dots ,f^*\alpha_k\}$ is a frame for $f^*E$ it suffices to show that:
\[ (f^*\nabla)_X (f^*\alpha_i)(p)= \tilde{\nabla}_X(f^*\alpha_i)(p).\]
This is the case because Equation (\ref{pull0}) guarantees that both sides are equal to $ (\nabla_Y\alpha_i)(p)$.
\end{proof}

\begin{remark}
\label{pullback} One can show that the pullback of connections is compatible with
composition of functions. That is, if $\nabla$ is a connection on
$\pi:E\rightarrow N$, and $f:M\rightarrow N$ and $h:S\rightarrow M$ are smooth
functions then $(f\circ h)^{\ast}\nabla=h^{\ast}(f^{\ast}\nabla)$.
\end{remark}

\section{Parallel transport}

Recall that we say that a section $\alpha\in\Gamma(E)$ of a vector bundle with
connection is covariantly constant if $\nabla_{X}(\alpha)=0,$ for any vector
field $X\in\mathfrak{X}(M)$. By imposing this conditions on vector bundles
over an interval one obtains the notion of parallel transport along a path.

\begin{proposition}
Let $\nabla$ be a connection on a vector bundle $\pi:E\rightarrow I$, where
$I=[a,b]$ is an interval. Given a vector $v\in E_{a}$ there exists a unique
covariantly constant section $\alpha\in\Gamma(E)$ such that $\alpha(a)=v.$
Moreover, the function $P_{a}^{b}:E_{a}\rightarrow E_{b}$ given by $P_{a}%
^{b}(v)=\alpha(b)$ is a linear isomorphism. The function $P_{a}^{b}$ is called
the parallel transport of the connection $\nabla$.
\end{proposition}

\begin{proof}
Since all vector bundles over an interval are trivializable, we may choose a frame $\{\alpha_1,\dots, \alpha_k\}$ for $E$.
There exists a one form $\theta \in \Omega^1(I, \End(E))$ such that:
\[ \nabla_X (\alpha_i)= \theta(X, \alpha_i).\]
Let us fix $v = \sum_i \lambda_i \alpha_i(a) \in E_a$. A section $ \alpha= \sum_i f_i \alpha_i$  is covariantly constant if it satisfies the differential equation:
\[ \sum_i \nabla_{\partial_t} (f_i \alpha_i)=0,\]
which is equivalent to
\[ \sum_i \frac{\partial f_i}{\partial t} \alpha_i+ f_i\theta( \partial_t, \alpha_i)=0.\]
The Picard-Lindel\"of theorem, see Appendix \ref{topan}, guarantees the existence and uniqueness of a solution of this equation. In order to show that $P_a^b$ is linear it is enough to observe that if $ \alpha$ and $\beta$ are covariantly constant, so are $\alpha + \beta $ and $\lambda \alpha$. It remains to show that  $P_a^b$ is an isomorphism. Suppose that  $v\in E_a$  is such that $P_a^b(v)=0$.
By symmetry we know that there exists a unique section  $\alpha \in \Gamma(E)$ such that $\alpha(b)=0$. This section is the zero section and we conclude that $v=0$.
\end{proof}

\begin{definition}
Let $\nabla$ be a connection on $\pi:E\rightarrow M$ and $\gamma
:[a,b]\rightarrow M$ a smooth curve. The parallel transport along $\gamma$
with respect to $\nabla$ is the linear isomorphism:
\[
P_{\nabla}(\gamma):E_{\gamma(a)}\rightarrow E_{\gamma(b)};\quad P_{\nabla
}(\gamma)(v)=P_{a}^{b}(v),
\]
where $P_{a}^{b}$ denotes the parallel transport associated with the vector
bundle $\gamma^{\ast}(E)$ over the interval $I=[a,b]$ with respect to the
connection $\gamma^{\ast}(\nabla)$.
\end{definition}

\begin{lemma}
Let $\gamma:[a,c]\rightarrow M$ be a curve and $b\in(a,c)$. Set $\mu
=\gamma|_{[a,b]};\quad\sigma=\gamma|_{[b,c]}.$ Then $P_{\nabla}(\gamma
)=P_{\nabla}(\sigma)\circ P_{\nabla}(\mu).$
\end{lemma}

\begin{proof}
It is enough to observe that if $\alpha \in \Gamma(\gamma^*(E))$ is covariantly constant then $ \alpha\vert_{[a,b]}$ and $\alpha\vert_{[b,c]}$ are also covariantly constant.
\end{proof}

\begin{lemma}
Parallel transport is parametrization invariant. That is, if $\nabla$ is a
connection on $\pi:E\rightarrow M$, $\gamma:[a,b]\rightarrow M$ is a curve and
$\varphi:[c,d]\rightarrow\lbrack a,b]$ is an orientation preserving
diffeomorphism then $P_{\nabla}(\gamma)=P_{\nabla}(\gamma\circ\varphi).$
\end{lemma}

\begin{proof}
In view of Exercise \ref{pullback} we know that:
\[ (\gamma \circ \varphi)^*(\nabla)= \varphi^*(\gamma^*(\nabla)).\]
Note that if $\alpha \in \Gamma(\gamma^*(E))$ is covariantly constant then $\varphi^*(\alpha) \in \Gamma(\varphi^*( \gamma^*(E)))=\Gamma( (\gamma \circ \varphi)^*(E))$ is also covariantly constant.
\end{proof}
\begin{figure}[H]
	\centering
	\includegraphics[scale=0.5]{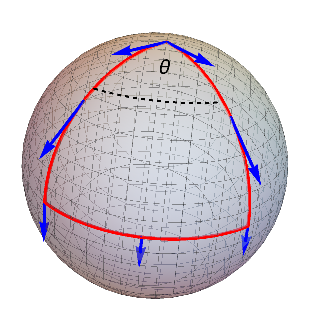}
	\caption{Parallel transport on the sphere.}
	
\end{figure}

\section{Geodesics}

In flat space, the distance between two points is minimized by a straight line. Objects moving in the absence of forces move along straight lines. On curved spaces, the notion of a straight
has to be replaced by that of a geodesic. These are preferred trajectories that minimize distances and prescribe the motion in the absence of forces, just like straight lines do in flat space.

\begin{definition}
Let $g$ be a metric on $M$ with Levi-Civita connection $\nabla$. A curve $\gamma:[a,b]\rightarrow M$ is a geodesic if its velocity
 $\gamma^{\prime} \in \Gamma( \gamma^*(TM))$ is covariantly constant with respect to the
connection $\gamma^{\ast}(\nabla)$.
\end{definition}

In local coordinates $\varphi=(x^{1},\dots,x^{m})$ where $\gamma=(u_{1}%
,\dots,u_{m})$ and $\nabla$ has Christoffel symbols $\Gamma_{ij}^{k}$ one has
$\gamma^{\prime}(t)=\sum_{i}u_{i}^{\prime}(t)\partial_{i},$
and the geodesic equation takes the form:
\begin{align*}
\gamma^{\ast}(\nabla)_{\partial_{t}}(\gamma^{\prime}(t))  &  =\sum_{i}%
\gamma^{\ast}(\nabla)_{\partial_{t}}(u_{i}^{\prime}(t)\partial_{i})\\
&  =\sum_{i}\Big(u_{i}^{\prime\prime}(t)\partial_{i}+u_{i}^{\prime}%
(t)\gamma^{\ast}(\nabla)_{\partial_{t}}\partial_{i}\Big)\\
&  =\sum_{i}\Big(u_{i}^{\prime\prime}(t)\partial_{i}+u_{i}^{\prime}(t)\sum
_{j}u_{j}^{\prime}(t)\nabla_{\partial_{j}}\partial_{i}\Big)\\
&  =\sum_{i}\Big(u_{i}^{\prime\prime}(t)\partial_{i}+u_{i}^{\prime}%
(t)\sum_{j,k}u_{j}^{\prime}(t)\Gamma_{ij}^{k}\partial_{k}\Big).
\end{align*}
We conclude that $\gamma$ is a geodesic precisely when it satisfies the system
of differential equations:
\begin{equation}
u_{i}^{\prime\prime}(t)+\sum_{j,k}u_{j}^{\prime}(t)u_{k}^{\prime}%
(t)\Gamma_{jk}^{i}=0,
\end{equation}
for $i=1,\dots,m$.

\begin{example}
On Euclidean space $\RR^{m}$ the Christoffel symbols are $\Gamma
_{ij}^{k}=0,$ and therefore the differential equation for a geodesic is just
$u_{i}^{\prime\prime}(t)=0.$ We conclude that geodesics in euclidean space are
straight lines. The same is true on Minkowski spacetime.
\end{example}

\begin{theorem}
\label{existencias de geodesicas}
Let $\nabla$ be the Levi-Civita connection on $TM$. Given $v\in T_{p}M$, there
exists an interval $\left(  -\epsilon,\epsilon\right)  $ for which there is a
unique geodesic $\gamma:\left(  -\epsilon,\epsilon\right)  \rightarrow M$ such
that $\gamma\left(  0\right)  =p$ and $\gamma^{\prime}(0)=v$.
\end{theorem}

\begin{proof}
Let $\varphi =(x^1,\dots, x^m)$  be local coordinates such that $\varphi(p)=0$. We  write $\gamma(t)=(u_1(t),\dots,u_m(t))$ and want to solve the system of equations:
\[ u''_i(t) + \sum_{j,k} u'_j(t) u'_k(t) \Gamma_{jk}^i  =0.\]
This is a second order ordinary differential equation. The existence and uniqueness of solutions is guaranteed by the Pickard-Lindel\"of theorem discussed in Apendix \ref{topan}.
\end{proof}

\begin{definition}
Let $M$ be a Lorentzian manifold and $\gamma:[a,b]\rightarrow M$ a
curve that is either timelike or spacelike. We say that $\gamma$ is
parametrized by arclength if \[\int_{a}^{s}|\gamma^{\prime}(t)|dt=s-a.\] Here,
as before, $|\gamma^{\prime}(t)|=\sqrt{g(\gamma^{\prime}(t),\gamma^{\prime
}(t))}$, if the curve is spacelike, and $|\gamma^{\prime}(t)|=\sqrt
{-g(\gamma^{\prime}(t),\gamma^{\prime}(t))}$ if the curve is timelike.
\end{definition}

It is easy to verify that if $\gamma:I \rightarrow M$ is a geodesic then $g(\gamma^{\prime
}(t),\gamma^{\prime}(t))$ is a constant function. One concludes that if $\gamma$ is
either spacelike or timelike then it can be parametrized by arclength.

\begin{figure}[H]
	\centering
	\includegraphics[scale=0.5]{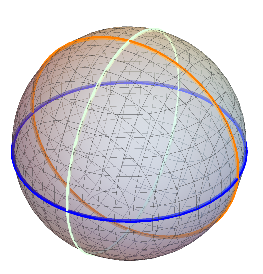}
	\caption{Geodesics on the sphere are maximal circles.}
\end{figure}

\begin{example}
The hyperbolic plane is the Riemannian manifold%
\[
\HH_{+}^{2}=\{(x,y)\in\RR^{2}\mid y>0\},\]
with metric
\[g=\frac{dx\otimes dx+dy\otimes dy}{y^{2}}.
\]
The components of the metric are \[g_{11}=g_{22}=\frac{1}{y^{2}},\quad
g_{12}=g_{21}=0.\] The components of the inverse matrix are:
\[
g^{11}=g_{22}=y^{2};\quad g^{12}=g^{21}=0.
\]

Using Equation (\ref{Chris}) we obtain:
\[
\Gamma_{11}^{1}=\Gamma_{22}^{1}= \Gamma_{12}^{2}=0; \quad\Gamma_{12}%
^{1}=\Gamma_{22}^{2}=\frac{-1}{y}; \quad\Gamma_{11}^{2}=\frac{1}{y}.
\]
The equations for a geodesic take the form \[\ddot{x}y=2\dot{x}\dot{y}%
;\quad\ddot{y}y=\dot{y}^{2}-\dot{x}^{2}.\]The first of these equations is
equivalent to \[\frac{d}{dt}\left(\frac{\dot{x}}{y^{2}}\right)=0,\] and we conclude
that
\begin{equation}
\dot{x}=cy^{2}. \label{c}%
\end{equation}
If $c=0,$ then $x$ is constant and one obtains geodesic that are vertical
lines. In case $c\neq0$, if we assume that the curve is parametrized by
arclength, we obtain $(\dot{x}^{2}+\dot{y}^{2})/y^{2}=1.$ Using Equation
\ref{c} we get:
\[
\frac{dy}{dx}=\frac{\dot{y}}{\dot{x}}=\sqrt{\frac{y^{2}-c^{2}y^{4}}{c^{2}%
y^{4}}}.
\]
This implies \[dx=\frac{cydy}{\sqrt{1-c^{2}y^{2}}},\] which has as solution
\[c(x-a)=-\sqrt{1-c^{2}y^{2}}.\] We conclude that geodesics in the hyperbolic
plane are vertical lines, as well as half circles centered at the $x$ axis.
\begin{figure}[H]
	\centering
	\includegraphics[scale=0.6]{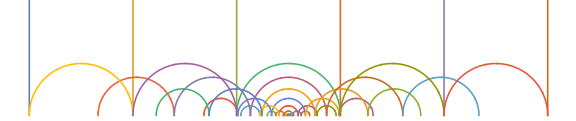}
	\caption{Geodesics on the hyperbolic plane.}
	
\end{figure}

\end{example}

\begin{remark}
\label{geodesica dual} Let $\gamma:I\rightarrow M$ be a curve. We define
$\theta\in\gamma^{\ast}(T^{\ast}M))$ by the formula $\theta(s)(v)=g(\gamma
^{\prime}(s),v),$ for any $v\in T_{\gamma(s)}M$. One can show that:

\begin{enumerate}
\item The curve $\gamma$ is a geodesic if and only if $\theta$ is covariantly
constant, i.e. $\nabla_{X}(\theta)=0.$

\item In local coordinates, the condition for $\theta$ to be covariantly
constant is:
\begin{equation}
\frac{d\theta_{k}}{ds}=\frac{1}{2}\sum_{i,j}\frac{\partial g_{ij}}{\partial
x^{k}}v^{i}v^{j}. \label{E69}%
\end{equation}
Here the functions $v^{l}$ are the coefficients of $\gamma^{\prime}(s)$
\[\gamma^{\prime}(s)=\sum_{l}v^{l}\partial_{l}.\]
\end{enumerate}
\end{remark}
   \clearemptydoublepage
\chapter{Curvature}

\vspace{3ex}

\section{The Riemann curvature tensor\label{4CURVATURA}}

A metric $g$ on a manifold $M$ determines geometric quantities such as angles and lengths. It
also determines the curvature of the space, which is a local quantity that measures how $M$ differs from flat space.

\begin{definition}
Let $\nabla$ be a connection on a vector bundle $\pi: E \rightarrow M$. The
curvature of $\nabla$ is the function:
\[
R: \mathfrak{X}(M)\otimes\mathfrak{X}(M)\otimes\Gamma(E) \rightarrow\Gamma(E)
\]
Defined by:
\[
R\left(  X,Y,\alpha\right)  =\nabla_{X}\nabla_{Y}\alpha-\nabla_{Y}\nabla
_{X}\alpha-\nabla_{\left[  X,Y\right]  }\alpha.
\]
It is more common to write $R\left(  X,Y\right)  (\alpha)$ instead of
$R\left(  X,Y,\alpha\right)  $.
\end{definition}
\noindent
One can check that:
\begin{itemize}
\item The curvature is skew symmetric on $X$ and $Y$.
\item The curvature is linear with respect to functions in each of the
variables .
\end{itemize}
\noindent
One concludes that the curvature $R$ is a tensor:
\[
R\in\Omega^{2}(M,\End(E))=\Gamma(\Lambda^{2}(T^{\ast}M)\otimes
\End(E)).
\]

\begin{proposition}
\label{4Prop2} Let $\nabla$ be a connection on $TM$ and $X,Y,Z\in
\mathfrak{X}(M)$. We denote by $\sum_{\mathrm{cyc}}$ the sum over cyclic permutations. The following identities hold.

\begin{align}
\sum_{\mathrm{cyc}}R\left(  X,Y\right)  Z-\sum_{\mathrm{cyc}} \left(  \nabla_{X}T\right)
\left(  Y,Z\right)  +T\left(  T\left(  X,Y\right)  ,Z\right)&=0 ,\\
\sum_{\mathrm{cyc}} \left(  \nabla_{X}R\right)  \left(  Y,Z\right)  +R\left(  T\left(
X,Y\right)  ,Z\right)  &=0.
\end{align}
Here, $T$ denotes the torsion of the connection $\nabla$, which is defined by:
\[ T(X,Y):= \nabla_XY-\nabla_Y X-[X,Y].\]
\end{proposition}

\begin{proof}
In order to prove the first identity we observe that:
\[
\left(  \nabla_{X}T\right)  \left(  Y,Z\right)  =\nabla_{X}\left(  T\left(
Y,Z\right)  \right)  -T\left(  \nabla_{X}Y,Z\right)  -T\left(  Y,\nabla
_{X}Z\right)  .
\]
From the definition of $T$ obtain:
\begin{align*}
T\left(  T\left(  X,Y\right)  ,Z\right)   &  =T\left(  \nabla_{X}Y-\nabla
_{Y}X-\left[  X,Y\right]  ,Z\right) \\
&  =T\left(  \nabla_{X}Y,Z\right)  +T\left(  Z,\nabla_{Y}X\right)  -T\left(
\left[  X,Y\right]  ,Z\right).
\end{align*}
Which implies:
\[
\sum_{\mathrm{cyc}}T\left(  T\left(  X,Y\right)  ,Z\right)
=\sum_{\mathrm{cyc}}\left(  \nabla_{X}\left(  T\left(  Y,Z\right)
\right)  -\left(  \nabla_{X}T\right)  \left(  Y,Z\right)  -T\left(  \left[
X,Y\right]  ,Z\right)  \right).
\]
Therefore:
\begin{align*}
&  \text{ \ \ \ }\sum_{\mathrm{cyc}}\left(  \left(  \nabla
_{X}T\right)  \left(  Y,Z\right)  +T\left(  T\left(  X,Y\right)  ,Z\right)
\right)= \sum_{\mathrm{cyc}}\left(  \nabla_{X}\left(  T\left(  Y,Z\right)
\right)  -T\left(  \left[  X,Y\right]  ,Z\right)  \right) \\
&  =\sum_{\mathrm{cyc}}\left(  \nabla_{X}\nabla_{Y}Z-\nabla_{X}%
\nabla_{Z}Y-\nabla_{X}\left[  Y,Z\right]  -\nabla_{\left[  X,Y\right]
}Z+\nabla_{Z}\left[  X,Y\right]  +\left[  \left[  X,Y\right]  ,Z\right]
\right) \\
&  =\sum_{\mathrm{cyc}}\left(  \nabla_{X}\nabla_{Y}Z-\nabla_{Y}%
\nabla_{X}Z-\nabla_{\left[  X,Y\right]  }Z\right)  =\sum_{\mathrm{cyc}}R\left(  X,Y\right)  Z.%
\end{align*}
For the second identity we compute:
\begin{align*}
\sum_{\mathrm{cyc}}R\left(  T\left(  X,Y\right)  ,Z\right)   &
=\sum_{\mathrm{cyc}}R\left(  \nabla_{X}Y-\nabla_{Y}X-\left[
X,Y\right]  ,Z\right)  \text{
\ \ \ \ \ \ \ \ \ \ \ \ \ \ \ \ \ \ \ \ \ \ \ \ }\\
&  =\sum_{\mathrm{cyc}}\left(  R\left(  \nabla_{X}Y,Z\right)
+R\left(  Z,\nabla_{Y}X\right)  -R\left(  \left[  X,Y\right]  ,Z\right)
\right).
\end{align*}
Also:
\[\sum_{\mathrm{cyc}}\left(  \nabla_{X}R\right)  \left(  Y,Z\right)=
\sum_{\mathrm{cyc}}\left(  \nabla_{X}\left(  R\left(  Y,Z\right)
\right)  -R\left(  \nabla_{X}Y,Z\right)  -R\left(  Y,\nabla_{X}Z\right)
-R\left(  Y,Z\right)  \nabla_{X}\right)  .
\]
Therefore:
\begin{align*}
&\sum_{\mathrm{cyc}}\left(  \left(  \nabla_{X}R\right)  \left(
Y,Z\right)  +R\left(  T\left(  X,Y\right)  ,Z\right)  \right)
  =\sum_{\mathrm{cyc}}\left(  \nabla_{X}\left(  R\left(  Y,Z\right)
\right)  -R\left(  Y,Z\right)  \nabla_{X}-R\left(  \left[  X,Y\right]
,Z\right)  \right) \\
&  =\sum_{\mathrm{cyc}}\left(  \nabla_{X}\nabla_{Y}\nabla_{Z}%
-\nabla_{X}\nabla_{Z}\nabla_{Y}-\nabla_{X}\nabla_{\left[  Y,Z\right]  }%
-\nabla_{Y}\nabla_{Z}\nabla_{X}+\nabla_{Z}\nabla_{Y}\nabla_{X}\right. \\
&  \text{ \ \ \ \ }+\nabla_{\left[  Y,Z\right]  }\nabla_{X}-\nabla_{\left[
X,Y\right]  }\nabla_{Z}+\nabla_{Z}\nabla_{\left[  X,Y\right]  }+\nabla
_{\left[  \left[  X,Y\right]  ,Z\right]  })=0.
\end{align*}
\end{proof}

Let $\left(  M,g\right)  $ be a semi-Riemannian manifold. The curvature $R$
of the Levi-Civita connection is called the Riemann curvature tensor. From
Proposition \ref{4Prop2} and the fact that the Levi-Civita connection is
torsion free we obtain%

\begin{equation}
\label{Bianchi1}R\left(  X,Y\right)  Z+R\left(  Z,X\right)  Y+R\left(
Y,Z\right)  X=0
\end{equation}

and
\begin{equation}
\label{Bianchi2}\left(  \nabla_{X}R\right)  \left(  Y,Z\right)  +\left(
\nabla_{Z}R\right)  \left(  X,Y\right)  +\left(  \nabla_{Y}R\right)  \left(
Z,X\right)  =0.
\end{equation}
These relations are known as the first and second Bianchi identities, respectively.

\begin{proposition}\label{simcurvature}
Let $\left(  M,g\right)  $ be a semi-Riemannian manifold and $X,Y,Z,V\in
\mathfrak{X}(M)$. The following identities hold.
\begin{align}
g\left(  R\left(  X,Y\right)  Z,V\right)  +g\left(  R\left(
Z,X\right)  Y,V\right)  +g\left(  R\left(  Y,Z\right)  X,V\right)  &=0 \label{B1}\\
g\left(  R\left(  X,Y\right)  Z,V\right)  +g\left(  R\left(  Y,X\right)
Z,V\right)  &=0 \label{B2}\\
g\left(  R\left(  X,Y\right)  Z,V\right)  +g\left(  R\left(  X,Y\right)
V,Z\right)  &=0 \label{B3}\\
g\left(  R\left(  Z,X\right)  Y,V\right)  -g\left(  R\left(  Y,V\right)
Z,X\right)  &=0 \label{B4}.
\end{align}

\end{proposition}

\begin{proof}
Property (\ref{B1}) follows from the first Bianchi identity. Equation  (\ref{B2}) holds because $R$ is skewsymmetric in the first to variables. Property  (\ref{B3}) is equivalent to \[g\left(  R\left(  X,Y\right)  Z,Z\right)  =0,\] which can be proved as follows:
\begin{align*}
g\left(  R\left(  X,Y\right)  Z,Z\right)   &  =g\left(  \nabla_{X}%
\nabla_{Y}Z-\nabla_{Y}\nabla_{X}Z-\nabla_{\left[  X,Y\right]
}Z,Z\right) \\
&  =g\left(  \nabla_{X}\nabla_{Y}Z,Z\right)  -g\left(  \nabla_{Y}%
\nabla_{X}Z,Z\right)  -g\left(  \nabla_{\left[  X,Y\right]  }%
Z,Z\right) \\
&  =Xg\left(  \nabla_{Y}Z,Z\right)  -g\left(  \nabla_{Y}Z,\nabla
_{X}Z\right)  -Yg\left(  \nabla_{X}Z,Z\right) \\
&  +g\left(  \nabla_{X}Z,\nabla_{Y}Z\right)  -\frac{1}{2}\left[
X,Y\right]  g\left(  Z,Z\right) \\
&  =\frac{1}{2}YXg\left(  Z,Z\right)  -\frac{1}{2}XYg\left(  Z,Z\right)
-\frac{1}{2}\left[  X,Y\right]  g\left(  Z,Z\right) \\
&  =\frac{1}{2}\left(  YX-YX-\left[  X,Y\right]  \right)  g\left(  Z,Z\right)
=0.
\end{align*}
To prove (\ref{B4}), we observe that (\ref{B1}) implies:
\begin{align*}
g\left(  R\left(  X,Y\right)  Z,V\right)  +g\left(  R\left(  Z,X\right)
Y,V\right)  +g\left(  R\left(  Y,Z\right)  X,V\right)   &  =0,\\
g\left(  R\left(  Y,Z\right)  V,X\right)  +g\left(  R\left(  V,Y\right)
Z,X\right)  +g\left(  R\left(  Z,V\right)  Y,X\right)   &  =0,\\
g\left(  R\left(  Z,V\right)  X,Y\right)  +g\left(  R\left(  X,Z\right)
V,Y\right)  +g\left(  R\left(  V,X\right)  Z,Y\right)   &  =0,\\
g\left(  R\left(  V,X\right)  Y,Z\right)  +g\left(  R\left(  Y,V\right)
X,Z\right)  +g\left(  R\left(  X,Y\right)  V,Z\right)   &  =0.
\end{align*}
Adding the identities above and using (\ref{B3}), we find:
\[
2g\left(  R\left(  Z,X\right)  Y,V\right)  +2g\left(  R\left(  Y,V\right)
X,Z\right)  =0.
\]
Using  (\ref{B3}) again one obtains
\[
g\left(  R\left(  Z,X\right)  Y,V\right)  =g\left(  R\left(  Y,V\right)
Z,X\right).
\]
\end{proof}

Given local coordinates $\varphi=(x^{1},\dots,x^{m})$ we define the functions
$R_{ijk}^{l}$ by the property \[R(\partial_{j},\partial_{k})(\partial_{i}%
)=\sum_{l}R_{ijk}^{l}\partial_{l}.\]  We also define $R_{lijk}=g\left(
R\left(  \partial_{j},\partial_{k}\right)  \partial_{i},\partial_{l}\right)
.$ One can directly compute:
\begin{align*}
R\left(  \partial_{i},\partial_{j}\right)  \partial_{k}  &  =\nabla
_{\partial_{i}}\nabla_{\partial_{j}}\partial_{k}-\nabla_{\partial_{j}}%
\nabla_{\partial_{i}}\partial_{k}-\nabla_{\left[  \partial_{i},\partial
_{j}\right]  }\partial_{k}\text{
\ \ \ \ \ \ \ \ \ \ \ \ \ \ \ \ \ \ \ \ \ \ \ \ \ \ \ \ \ \ \ \ \ \ \ \ \ \ \ \ \ \ \ \ \ \ \ \ \ \ \ \ \ }%
\\
&  =\nabla_{\partial_{i}}\left(  \sum_{l}\Gamma_{jk}^{l}\partial_{l}\right)
-\nabla_{\partial_{j}}\left(  \sum_{l}\Gamma_{ik}^{l}\partial_{l}\right)
\end{align*}%
\[
=\sum_{l}\left(  \frac{\partial\Gamma_{jk}^{l}}{\partial x^{i}}\partial
_{l}+\Gamma_{jk}^{l}\sum_{n}\Gamma_{il}^{n}\partial_{n}\right)  -\sum
_{l}\left(  \frac{\partial\Gamma_{ik}^{l}}{\partial x^{j}}\partial_{l}%
+\Gamma_{ik}^{l}\sum_{n}\Gamma_{jl}^{n}\partial_{n}\right)
\]

\[
=\sum_{n}\left(  \frac{\partial\Gamma_{jk}^{n}}{\partial x^{i}}-\frac
{\partial\Gamma_{ik}^{n}}{\partial x^{j}}+\sum_{l}\Gamma_{jk}^{l}\Gamma
_{il}^{n}-\sum_{l}\Gamma_{ik}^{l}\Gamma_{jl}^{n}\right)  \partial_{n}.
\]
We conclude that:
\begin{align*}
R_{ijk}^{l}  &  =\frac{\partial\Gamma_{ik}^{l}}{\partial x^{j}}-\frac
{\partial\Gamma_{ji}^{l}}{\partial x^{k}}+\sum_{n}\Gamma_{jn}^{l}\Gamma
_{ik}^{n}-\sum_{n}\Gamma_{kn}^{l}\Gamma_{ij}^{n},\\
R_{nijk}  &  =\sum_{l}R_{ijk}^{l}g_{ln}.
\end{align*}

The Bianchi identities are equivalent to:
\begin{align}
R_{ijk}^{l}+R_{kij}^{l}+R_{jki}^{l}  &  =0\\
\left(  \nabla_{\partial_{i}}R\right)  _{jkl}^{n}+\left(  \nabla_{\partial
_{l}}R\right)  _{jik}^{n}+\left(  \nabla_{\partial_{k}}R\right)  _{jli}^{n}
&  =0
\end{align}
The Christoffel symbols for  Euclidean space $\RR^{m}$  vanish and therefore $R=0$. The same is true for Minkowski space. It is a good exercise to show that in dimension $d=2$ the only nonzero components of the curvature tensor are
\[ R_{1212}=R_{2121}=-R_{1221}=-R_{2112},\]
and to compute the Riemann tensor for the hyperbolic plane and for the two
dimensional sphere in the coordinates provided by the stereographic projection.

\section{The Ricci tensor and scalar curvature }

The Ricci tensor, denoted by
$\mathrm{Ric}\in\Gamma((TM\otimes TM)^{\ast})$ is the tensor defined by:
\[
\mathrm{Ric}(X,Y)(p)=\tr\big(R(p)(X(p),-)(Y(p))\big).
\]
Here $X,Y\in\mathfrak{X}(M)$ are vector fields on $M$ and $R(p)(X(p),-)(Y(p))$
is the function from $T_{p}M$ to $T_{p}M$ defined by
\[
Z\mapsto R(p)(X(p),Z)(Y(p)).
\]
\noindent
One can verify that $\mathrm{Ric}(X,Y)=\mathrm{Ric}(Y,X)$ and that  the functions
\[\mathrm{Ric}_{ij}=\sum_{l}R_{jli}^{l},\]
are the components of the Ricci tensor.
\noindent
By raising the indices one obtains the $(0,2)$ tensor ${\mathrm{Ric}^{\sharp}%
}$ with components \[({\mathrm{Ric}}^{\sharp})^{ij}=\sum_{kl}g^{ik}%
g^{jl}\mathrm{Ric}_{kl}.\]

A straightforward computation shows that the following identities hold:

\begin{align} 
R_{ijkl}&=-R_{jikl}=-R_{ijlk}.\\
R_{ijkl}&=R_{klij}.\\
\mathrm{Ric}_{ij}&=\mathrm{Ric}_{ji}.
\end{align}

The scalar curvature of a semi-Riemannian manifold $(M,g)$, denoted by $S\in
C^{\infty}(M)$ is the function:
\[
S(p)=\mathrm{tr}\big(\mathrm{Ric}^{g}(p)\big),
\]
where $\mathrm{Ric}^{g}(p):T_{p}M\rightarrow T_{p}M$ is the linear function
characterized by:
\[
g(p)\Big(\mathrm{Ric}^{g}(p)(X),Y\Big)=\mathrm{Ric}(p)(X,Y).
\]
In local coordinates the scalar curvature is given by: \[\mathrm{S}=\sum
_{ij}g^{ij}\mathrm{Ric}_{ij}.\]

\begin{definition}
Einstein's tensor $G$ is the $(2,0)$ tensor defined by \[\mathrm{G}%
=\mathrm{Ric}-\frac{g\mathrm{S}}{2}.\] By raising the indices one obtains a tensor
${\mathrm{G}}^{\sharp}$ of type $(0,2)$ with components \[({\mathrm{G}}^{\sharp})
^{ij}=\sum_{kl}g^{ik}g^{jl}G_{kl}.\]
\end{definition}

\begin{proposition}
\label{necesaria} The following identities hold:%

\begin{equation}
\sum_{s}(\nabla_{\partial_{s}}R)_{kjl}^{s}+(\nabla_{\partial_{l}}%
\mathrm{Ric})_{kj}-(\nabla_{\partial_{j}}\mathrm{Ric})_{kl}=0, \label{G1}%
\end{equation}%
\begin{equation}
2\sum_{s}\left(  \nabla_{\partial s}\mathrm{Ric}\right)  _{j}^{s}%
-\nabla_{\partial j}\mathrm{S}=0, \label{G2}%
\end{equation}%
\begin{equation}
\sum_{s}(\nabla_{\partial_{s}}G^\sharp)^{si}=0. \label{G3}%
\end{equation}

\end{proposition}

\begin{proof}
We know that contracting indices commutes with covariant differentiation and therefore \[(\nabla_{\partial_{i}}\mathrm{Ric})_{jk}=
{\textstyle\sum_{s}}
(\nabla_{\partial_{i}}R)^{s}_{jsk}.\] On the other hand, the second Bianchi identity gives:
\[
(\nabla_{\partial_{i}}R)^{s}_{kjl}+(\nabla_{\partial_{l}}R)^{s}_{kij}+(\nabla_{\partial_{j}}R)^{s}_{kli}=0.
\]
Using the skew-symmetry of the Riemann tensor and summing over $i=s$ one obtains:
\[\sum_{s}
(\nabla_{\partial_{s}}R)^{s}_{kjl}+
\sum_{s}(\nabla_{\partial_{l}}R)^{s}_{ksj}-\sum_{s}
(\nabla_{\partial_{j}}R)^{s}_{ksl}=0.
\]
which is precisely:
\[
\sum_{s}
(\nabla_{\partial_{s}}R)^{s}_{kjl}+(\nabla_{\partial_{l}}%
\mathrm{Ric})_{kj}-(\nabla_{\partial_{j}}\mathrm{Ric})_{kl}=0,
\]
as required.
Let  us now prove the second identity. Multiplying Equation $(\ref{G1})$ by $g^{kr}$ and summing over $k$ one obtains:
\[
\sum_{s,k}
(g^{kr}\nabla_{\partial_{s}}R)^{s}_{kjl}+\sum_k g^{kr}(\nabla_{\partial_{l}}%
\mathrm{Ric})_{kj}-\sum_kg^{kr}(\nabla_{\partial_{j}}\mathrm{Ric})_{kl}=0.
\]
This can also be written:
\[
\sum_{s}
(\nabla_{\partial_{s}}R)^{sk}_{jl}+ (\nabla_{\partial_{l}}%
\mathrm{Ric})^k_{j}-(\nabla_{\partial_{j}}\mathrm{Ric})^k_{l}=0.
\]
We now contract the indices $j$ and $k$ to obtain:
\[
\nabla_{\partial_l}(S)-2 \sum_s \left( \nabla_{\partial_s} \mathrm{Ric}\right)^s_l,
\]
which is equivalent to $(\ref{G2})$. Finally, in order to prove $(\ref{G3})$ we multiply $(\ref{G2})$ by $g^{kj}$ and sum over $j$ to obtain:
\[2 \sum_{s,j} g^{jk}\left(\nabla_{\partial s}\mathrm{Ric}\right)^s_j - \sum_j g^{jk}\nabla_{\partial j}S=0.
\]
This is the same as:
\[2 \sum_{s} \left(\nabla_{\partial s}\mathrm{Ric}\right)^{sk}- \sum_j\nabla_{\partial j}(S)g^{jk}=0,
\]
which can also be written:
\[2 \sum_{s} \left(\nabla_{\partial s}\mathrm{Ric}\right)^{sk}- \sum_s\nabla_{\partial s}(S{g^{\sharp}})^{sj}= 2\sum_s( \nabla_{\partial_s}{G^{\sharp}})^{sj}=0.
\]
\end{proof}

\begin{figure}[H]
\begin{tabular}{p{0.3\textwidth} p{0.3\textwidth} p{0.3\textwidth}}
  \vspace{0pt} \includegraphics[scale=0.34]{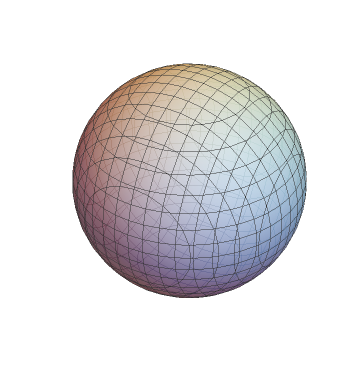} &
  \vspace{0pt} \includegraphics[scale=0.34]{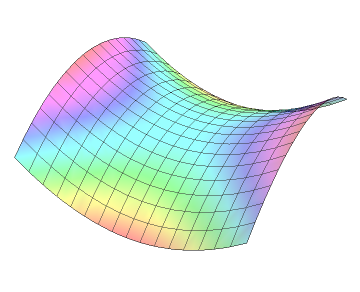}&
  \vspace{0pt} \includegraphics[scale=0.34]{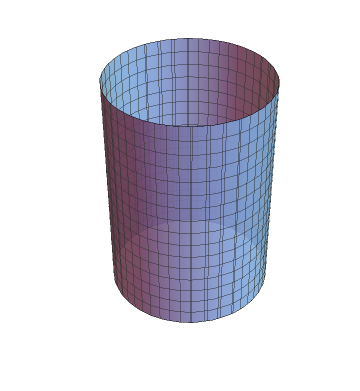}
\end{tabular}
\caption{Positive, negative and zero curvature.}
\end{figure}

\section{Sectional curvature }

We will now describe another local invariant of a semi-Riemannian manifold: the sectional curvature.
Let $(M,g)$ be a semi-Riemannian manifold and $\Pi \subset T_pM$ a two dimensional vector subspace of the tangent space at $p$ such that
the metric $g$ restricted to $\Pi$ is non-degenerated. The sectional curvature $K$ of $(M,g)$ evaluated at $\Pi$ is the number:
\[ K(p)(\Pi)=\frac{\langle R(X,Y)(Y),X\rangle}{\langle X,X\rangle \langle Y,Y \rangle - \langle X,Y \rangle^2 },\]
where the vectors $X$ and $Y$ generate $\Pi$. Note that the hypothesis that the metric is nondegenerated on $\Pi$ implies that the denominator is nonzero.
Let us show that the right hand side depends only on the vector subspace $\Pi$. The Bianchi identities imply that the numerator is symmetric on $X$ and $Y$. One concludes that the whole expression also is. It is also clear that the number does not change if $X$ or $Y$ are multiplied by a nonzero scalar. Finally, the antisymmetry of the Riemann tensor implies that the right hand side does not change if $X$ is replaced by $X'=X+ \lambda Y$. The quantity $K$ is known as the sectional curvature of $(M,g)$. A semi-Riemannian manifold is said to have constant sectional curvature if $K(p)(\Pi)$ is a constant quantity.

\begin{example}
Let $(\Sigma,g)$ be a Riemannian manifold of dimension $d=2$. At each point $p \in \Sigma$ there is a unique two dimensional subspace of $T_p\Sigma$, namely the whole tangent space. Therefore, in this case, the sectional curvature is a smooth function:
\[ K: \Sigma \rightarrow \RR.\]
Let us see that in this case $K$ is one half of the scalar curvature, $K= S/2$. This quantity is also known as the Gaussian curvature of the surface. If $X$ and $Y$ are an orthonormal basis for $T_p\Sigma$ then:
\begin{align*}
S(p)=\mathrm{tr}(\mathrm{Ric})=\mathrm{Ric}(X,X)+ \mathrm{Ric}(Y,Y)=2\langle R(X,Y)(Y),X\rangle
=2 K(p).
\end{align*}
\end{example}

\begin{example}
The $m$-dimensional sphere of raduis $R$:
\[S^m= \{ v \in \RR^{m+1}\mid |v|=R\}\]
has constant sectional curvature $K=1/R^2$.
\end{example}

\begin{example}
Recall that the $m$ dimensional hyperbolic space $\HH^m$ is the manifold:
\[\HH^m= \{ (x^1, \cdots,x^m) \in \RR^{m+1}: x^m>0\}\]
with metric:
\[ g=\frac{dx^1 \otimes dx^1+ \cdots + dx^m \otimes dx^m}{(x^m)^2}.\]
It is a good exercise to show that hyperbolic space has constant sectional curvature $K=-1$.\\
\end{example}

\begin{example}
Let $\RR^{1,4}$ be the $5$ dimensional Minkowski space. That is, the smooth manifold $\RR^{5}$ with metric:
\[ g= -dx^0 \otimes dx^0 + \sum_{i=1}^4 dx^i \otimes dx^i.\] 
De Sitter space is the submanifold:
\[ dS_4=\Big \{ (x^0, \cdots, x^4)\in \RR^{1,4}: -(x^0)^2+ \sum_{i=1}^4 (x^i)^2=\alpha^2 \Big \}.\]
One can prove that de Sitter space is diffeomorphic to $ \RR\times S^{3}$ and that the Minkowski metric induces a metric of Lorentz signature on $dS_4$. 
Moreover, the Riemann tensor satisfies:
\[ R_{abcd}=\frac{1}{\alpha^2} \big( g_{ac}g_{bd} - g_{ad}g_{bc}\big),\]
the Ricci tensor is proportional to the metric:
\[ \mathrm{Ric}=\frac{3 g}{\alpha^2}\]
and the sectional curvature of de Sitter space is $K=1/\alpha^2$.
\end{example}

\begin{figure}[H]
\centering
	\includegraphics[scale=0.4]{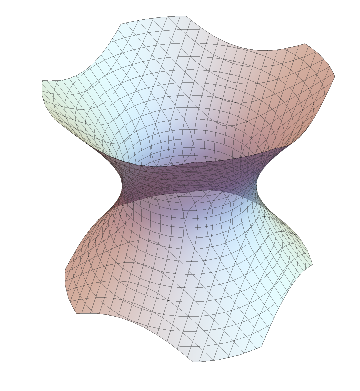}
	\caption{Two dimensional de Sitter space.}
\end{figure}

\begin{example}
Let $\RR^{2,3}$ be the smooth manifold $\RR^{5}$ with metric:
\[ g= -dx^0 \otimes dx^0 - dx^1 \otimes dx^1+ \sum_{i=2}^4 dx^i \otimes dx^i.\] 
Anti-de Sitter space is the submanifold:
\[ \mathrm{AdS}_4=\Big \{ (x^0, \cdots, x^4)\in \RR^{2,3} \mid -(x^0)^2-(x^1)^2+ \sum_{i=2}^4 (x^i)^2=-\alpha^2 \Big \}.\]
Anti-de Sitter space is diffeomorphic to $ \RR^{3}\times S^{1}$ and the Minkowski metric induces a metric of Lorentz signature on $AdS_4$.  The Riemann tensor satisfies:
\[ R_{abcd}=-\frac{1}{\alpha^2} \big( g_{ac}g_{bd} - g_{ad}g_{bc}\big).\]
The Ricci curvature is proportional to the metric:
\[ \mathrm{Ric}=-\frac{3g}{\alpha^2},\] and the sectional curvature of Anti-de Sitter space is $K=-1/\alpha^2.$

\end{example}

\begin{definition}
One says that a Riemannian manifold $(M,g)$ is locally isotropic at $p \in M$ if for every pair of unitary tangent vectors $u,v \in T_pM$ there exist open subsets $U,V \subseteq M$ and an isometry $\varphi: U \rightarrow V$ such that $\varphi(p) =p$ and $D\varphi(p)(v)=w$. 
\end{definition}

\begin{proposition}
\label{nice} Let $(M,g)$ be a $3$-dimensional Riemannian manifold which is locally isotropic at $p \in M$. Then $M$ has constant sectional
curvature at $p$. This means that  $K(p)(\Pi)=K(p)(\Pi')$ for any two planes $\Pi, \Pi' \subset T_pM$.
\end{proposition}
\begin{proof}
Given $\Pi$ and $\Pi'$ consider unitary vectors $v$, $v'$ which are orthogonal to $\Pi$ and $\Pi'$ respectively.
Fix a local isometry such that $D\varphi(p)(v)=v'$. This implies that $D\varphi(p)(\Pi)=\Pi'$.
Therefore:
\[ K(p)(\Pi)=K(\varphi(p))(D\varphi (H))=K(p)(\Pi').\]
\end{proof}

The proof of the following result can be found in Appendix \ref{SchurLemma}.
\begin{theorem}[Schur's Lemma]
\label{schur}Let $(M,g)$ be a connected Riemannian manifold of dimension $\geq3$. If there exists a function
$f: M \rightarrow \RR$ such that $f(p)=K(p)(\Pi)$, for all $\Pi \in T_pM$, then $f$ is constant.
\end{theorem}

Note that the condition that $d\geq 3$ is necessary. In dimension $d=2$ the statement is false since the Gaussian curvature is typically not constant.

A Riemannian manifold $(M,g)$ is called geodesically complete if the domain
of every geodesic can be extended to the whole real line. The following remarkable theorem, known as the Killing-Hopf theorem, provides a classification of manifolds of constant curvature. The proof can be found in Appendix \ref{KH}.

\begin{theorem}(Killing-Hopf)\label{remarkable}
Let $(M,g)$ be a geodesically complete simply connected Riemannian manifold of constant curvature $K$.
\begin{itemize}
\item If $K=0$ then $(M,g)$ is isometric to the Euclidean space $\RR^n$.
\item If $K=1$ then $(M,g)$ is isometric to the sphere $S^n$.
\item  If $K=-1$ then $(M,g)$ is isometric to the hyperbolic space $\HH^n$.
\end{itemize}
\end{theorem}

\begin{remark}
Let $(M,g)$ be a Riemannian manifold with sectional curvature $K$ and $C>0$ a positive constant. The sectional curvature $K_{Cg}$ of the manifold $(M,Cg)$ is given by:
\[ K_{Cg}=K/C.\]
This implies that any geodesically complete, connected, simply connected Riemannian
manifold of constant curvature $K=C$ is obtained by rescaling the metric of one of the model spaces above.
\end{remark}

\section{Curvature and parallel transport}

Given a vector bundle $ \pi: E \rightarrow M$ with a connection $\nabla$ and a path:
$\gamma: I \rightarrow M$, there is an associated linear isomorphism: 
\[ P_{\nabla}(\gamma):E_{\gamma(0)} \rightarrow E_{\gamma(1)},\]
called the parallel transport along the curve $\gamma$ with respect to the connection $\nabla$. In case the connection is flat, the parallel transport depends only on the homotopy class of the path relative to the endpoints. That is, if $\gamma$ and $\beta$ are two paths in $M$ which are homotopic relative to their endpoints, then $ P_{\nabla}(\gamma)= P_{\nabla}(\beta).$ Let us first prove some preparatory lemmas.

\begin{lemma}\label{constant}
Let $M$ be a connected manifold and $\pi: E \rightarrow M$ be a vector bundle with a connection $\nabla$.
If $\alpha, \beta \in \Gamma(E)$ are covariantly constant sections of $E$ such that $\alpha(p)=\beta(p)$ for some $p \in M$, then
$\alpha=\beta$.
\end{lemma}
\begin{proof}
Let $q \in M$ be some other point. Since $M$ is connected, we may choose a path $\gamma: I \rightarrow M$
such that $\gamma(0)=p$ and $\gamma(1)=q$. Since $\alpha$ and $\beta$ are covariantly constant sections, so are 
$\gamma^*\alpha$ and $\gamma^*\beta$. Thus $\gamma^*\alpha$ and $\gamma^*\beta$ satisfy the same ordinary differential equation with the same initial condition. One concludes that $\gamma^*\alpha=\gamma^*\beta$ and, in particular, $\alpha(q)= \beta(q)$.
\end{proof}

\begin{lemma}\label{local}
Let $\pi: E \rightarrow M$ be a vector bundle with a flat connection $\nabla$ over $M=[0,1]\times[0,1]$. For any $p=(t,s)\in M$ we set
\[ A_r(p):=P_{\nabla}(a_r(p)),\quad  B_r(p):=P_{\nabla}(b_r(p)),\]
where $a_r(p):[0,r]\rightarrow M $ denotes the path $l \mapsto (t+l,s)$ and 
$b_r(p):[0,r]\rightarrow M $ denotes the path $l \mapsto (t,s+l)$. Then
\[ A_t(0,s)B_s(0,0)=B_s(t,0)A_t(0,0).\]
\end{lemma}

\begin{proof}
By subdividing and reparametrizing the square if necessary, we may assume that $E=M \times V$ is a trivial vector bundle.
Consider the functions $G,F:M \rightarrow \End(V)$ defined by
\[ F(t,s)=A_t(0,s)B_s(0,0)\]
and
\[ G(t,s)=B_s(t,0)A_t(0,0).\]
We need to prove that $F=G$.
Let us fix an arbitrary vector $v \in V$ and define the functions $f,g: M \rightarrow V$ by:
\[ f(t,s):=F(t,s)(v),\]
and
 \[ g(t,s):= G(t,s)(v).\]
 It suffices to show that $f=g$. By construction, $f(0,0)=v=g(0,0)$. Therefore, in view of Lemma \ref{constant}, in order to prove that $f=g$ it is enough
 to show that $f$ and $g$ are covariantly constant. Since the situation is symmetric, it is enough to show that $f$ is covariantly constant. Denote by $X$ the vector field $\partial_t$ and by $Y$ the vector field $\partial_s$. We need to prove that $ \nabla_X f=\nabla_Y f=0$. Since the vector bundle $M \times V$ is trivial, there exists a one form $\theta \in \Omega^1(M, \End(V))$ such that:
 \[ \nabla_Z(W)=Z(W)+ \theta(Z)(W).\]
 Since $A_t(p)$ is given by parallel transport, it satisfies the differential equation
 \[ \frac{\partial}{\partial t} A_t= -\theta(X)A_t.\]
 With this we compute:
 \begin{eqnarray*}
 \nabla_X f(t,s)&=&\frac{\partial f}{\partial t} (t,s)+ \theta(X)f(t,s)\\
 &=& \frac{\partial}{\partial t}(A_t(0,s)) B_s(0,0)(v)+ \theta(X)(f)(t,s)\\
 &=&-\theta(X)A_t(0,s)B_s(0,0)(v)+\theta(X)(f)(t,s)\\
 &=&-\theta(X)(f)(t,s)+\theta(X)(f)(t,s)=0.
 \end{eqnarray*}
 Since the connection is flat we know that
 \begin{equation}\label{commute}
  \nabla_X \nabla_Y f=\nabla_Y \nabla_X f=0.
  \end{equation}
It only remains to show that $\nabla_Yf=0$. Let us begin by computing $\nabla_Y f(0,s)$.
\begin{eqnarray*}
\nabla_Y f(0,s)&=&\frac{\partial f}{\partial s}(0,s)+ \theta(Y) (f)(0,s)\\
&=& \frac{\partial}{\partial s}B_s(0,0)(v)+\theta(Y)(f)(0,s)\\
&=&-\theta(Y) B_s(0,0)(v)+ \theta(Y)(f)(0,s)\\
&=&-\theta(Y)(f)(0,s)+\theta(Y)(f)(0,s)\\
&=&0.
\end{eqnarray*}
We conclude that the function $\nabla_Y f$ vanishes on $(0,s)$. Fix $s \in [0,1]$ and consider the path $\gamma: [0,t] \rightarrow M$ defined by $\gamma(l)=(l,s)$. Equation (\ref{commute}) implies that $\gamma^*(\nabla_Y f)$ is covariantly constant. Moreover:
\[ \gamma^*( \nabla_Y f)(0)=\nabla_Yf(0,s)=0.\]
Therefore $\gamma^*(\nabla_Y f)=0$ and in particular:
\[ 0=\gamma^*(\nabla_Y f)(t)=\nabla_Y f(t,s).\]
\end{proof}

\begin{theorem}\label{flathol}
Let $\nabla$ be a flat connection on the vector bundle $\pi: E \rightarrow M$ and $\gamma, \beta:I \rightarrow M$ be 
paths which are homotopic with respect to their endponts. Then
\[ P_\nabla(\gamma)=P_\nabla(\beta).\]
\end{theorem}
\begin{proof}
Let $H: [0,1] \times [0,1] \rightarrow M$ be a homotopy between $\gamma$ and $\beta$. That is, assume that
\[ H(t,0)=\gamma(t),\quad H(t,1)=\beta(t),\quad  H(0,s)=H(0,0),\quad H(1,s)=H(1,0).\]
Consider the vector bundle $H^*(E)$ with the pullback connection $H^*(\nabla)$. Using the notation and the conclusion from Lemma \ref{local}, we know that
\[ A_1(0,1)B_1(0,0)=B_1(1,0)A_1(0,0).\]
Since the homotopy $H$ fixes the endpoints we know that \[B_1(0,0)=B_1(1,0)=\id,\]
so we are left with $A_1(0,1)=A_1(0,0)$. Finally, using the naturality of parallel transport with respect to pullback, one computes:
\begin{eqnarray*}
 P_{\nabla} (\gamma)&=&P_{\nabla}(H^*(a_1(0,0)))=P_{H^*(\nabla)}(a_1(0,0))=A_1(0,0)\\&=&A_1(0,1)
 =P_{H^*(\nabla)}(a_1(0,1))=P_{\nabla}(H^*(a_1(0,1)))\\&=&P_{\nabla}(\beta).
 \end{eqnarray*}
\end{proof}

\begin{corollary}\label{flats}
Let $\nabla$ be a flat connection on a vector bundle $\pi: E \rightarrow M$ on a simply connected manifold $M$.
Given a point $p \in M$ and a vector $v \in E_p$ there exists a unique covariantly constant section $\alpha \in \Gamma(E)$ such that $\alpha(p)=v$.
\end{corollary}
\begin{proof}
The uniqueness is guaranteed by Lemma \ref{constant}. Let us prove the existence. We define $\alpha$ at a point $q \in M$ by:
\[ \alpha(q):= P_\nabla(\gamma)(v),\]
where $\gamma$ is any path from $p$ to $q$. Since $M$ is simply connected, Theorem \ref{flathol} guarantees that $\alpha$ is well defined. It remains to show that it is covariantly constant. By construction, given any path $\gamma: I \rightarrow M$, the section $\gamma^*(\alpha)$ is covariantly constant and therefore:
\[ 0=\nabla_{\partial_t} \gamma^*(\alpha)= \nabla_{D\gamma( \partial_t)} \alpha.\]
Since $\gamma$ is arbitrary, one concludes that $\alpha$ is covariantly constant.
\end{proof}

\section{Geodesic deviation and Jacobi fields}\label{geodesic deviation}

On Euclidean space, straight lines which are parallel stay parallel. This does not happen on curved spaces. The curvature tensor can be interpreted as a measure of the deviation between
geodesics.  Let us fix a semi-Riemannian manifold $(M,g)$. A family of
geodesics in $M$ is an embedding $\sigma:I\times (-\epsilon, \epsilon)\rightarrow M,$ where  for each fixed $s$, the curve $\sigma_{s}(t)=\sigma(t,s)$ is a geodesic. One can define
vector fields $X,Y$ on the surface $S=\mathrm{im}(\sigma)$ by \[Y=\sigma_{\ast
}\partial_{s};\quad X=\sigma_{\ast}\partial_{t}.\] Since each of the
curves $\sigma_{s}(t)$ is a geodesic, we know that \[\nabla_{X}(X)=0.\]
Moreover, since the vector fields $\partial_{s}$ and $\partial_{t}$
commute, we have \[[Y,X]=0.\] Therefore, the fact that $\nabla$ is torsion free
 implies that the curvature satisfies:
\begin{equation}\label{jdev}
R(X,Y)(X)=\nabla_{X}\nabla_{Y}(X)-\nabla_{Y}\nabla_{X}(X)=\nabla_{X}\nabla_{X}(Y).
\end{equation}
Thus, the curvature is the second derivative of the vector $Y$ in the
direction of the geodesic. If we choose local coordinates and we write:
\[
\nabla_{X}\nabla_{X}Y=\sum_{i}A^{i}\partial_{i};\quad X=\sum_{j}X^{j}%
\partial_{j};\quad Y=\sum_{k}Y^{k}\partial_{k}.
\]
Then:
\begin{equation}
A^{i}=\sum_{j,k,l}X^{j}Y^{k}X^{l}R_{ljk}^{i}. \label{EE31}%
\end{equation}

Consider the two dimensional sphere. There is a one parameter family of geodesics which start at the equator and travel north. These geodesics start parallel and converge to the north pole.

\begin{figure}[H]
\centering
	\includegraphics[scale=0.4]{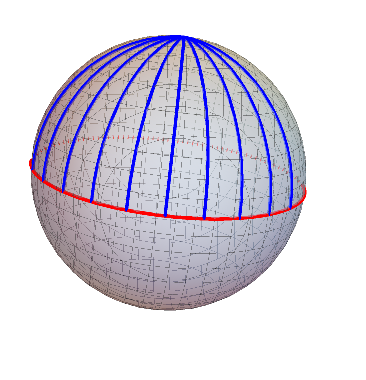}
	\caption{Geodesic deviation on the sphere.}
\end{figure}

\begin{definition}
Let $\gamma: I \to M$ be a geodesic. a vector field $V$ along $\gamma$ is called a Jacobi field if it satisfies the equation:
\begin{equation} \nabla_{\gamma'(t)} \nabla_{\gamma'(t)}(V)=R(\gamma'(t),V)(\gamma'(t)).\end{equation}
\end{definition}

The discussion above shows that if $\sigma: I \times (-\epsilon,\epsilon) \to M$ is a family of geodesics then
the vector field \[V=\frac{\partial}{\partial s} \Big \vert_{s=0} \sigma(t,s)\] is a Jacobi field.
The Picard-Lindel\"of theorem guarantees that given a geodesic $\gamma: I \to M$ and tangent vectors
$v,w \in T_{\gamma(0)}M$ there is  a unique Jacobi field $V$ such that:
\[ V(0)=v ; \,\,\, \nabla_{\gamma'(t)} V(0)=w.\]

\begin{definition}
Given a point $p$ in $M$ and a tangent vector $v \in T_pM$ the Picard-Lindel\"of theorem  guarantees that there is a geodesic $\gamma$ such that
$\gamma(0)=p $ and $\gamma'(0)=v$. Moreover, any two geodesics with these properties coincide on the intersection of their domains. The exponential map, defined on a sufficiently small open neighborhood $U$ of zero in $T_pM$ is the map:
\[ \exp_p: U \rightarrow M\]
that sends a vector $v \in U$ to $\gamma(1)$ where $\gamma$ is a geodesic as above. Let us compute the derivative of the exponential map at zero:

\[ D(\exp_p)(0)(v)=\frac{d}{dt}\Big \vert_{t=0} \exp(tv)=\frac{d}{dt}\Big \vert_{t=0} \gamma(t)=v.\]
One concludes that the derivative of the exponential map at zero is the identity. By the inverse function theorem, the exponential map is a local diffeomorphism near zero.
Fixing a basis on $T_pM$ one obtains coordinates around $p$, which are known as normal coordinates.
\end{definition}

It is an interesting exercise to prove that, in normal coordinates around $p$,  the Christoffel symbols $\Gamma^k_{ij}$ vanish at $p$.\\

\noindent
Jacobi fields can be used to describe the derivative of the exponential map, as the following result shows.

\begin{lemma}\label{dexp}
Let $M$ be a semi-Riemannian manfilold and $U$ a neighborhood of zero in $T_pM$ sufficiently small so that the exponential map restricted to $U$ is a diffeomorphism onto its image. Given $v \in U$ and $ w \in T_pM$ we set:
\[ \gamma(t)=\exp_p(tv)\]
and denote by $V$ the Jacobi field along $\gamma(t)$ that satisfies
\[ V(0)=0, \quad \nabla_{\gamma'(t)}V(0)=w.\]
Then:
\[ D(\exp_p)(v)(w)=V(1).\]
\end{lemma}
\begin{proof}
Consider the map $\sigma(t,s)= \exp_p(t(v+sw))$. Since $\sigma$ is a family of geodesics, 
we know that the vector field 
\[ W= \frac{d}{ds}\Big \vert_{s=0} \sigma(t,s)\]
is a Jacobi field. Moreover, since $\sigma(0,s)=p$ we know that $ W(0)=0$. Using the fact that the Levi-Civita connection is torsion free we compute:
\[ \nabla_{\gamma'(t)}W(0)=\nabla_{\gamma'(t)}\frac{d}{ds} \sigma(t,s)\Big\vert_{s=0} =\nabla_{W}\frac{d}{dt} \sigma(t,s)\Big\vert_{t=0}=\frac{d}{ds}\Big\vert_{s=0} (v+s w)=w.\]
Since $W$ and $V$ are Jacobi fields over $\gamma(t)$ with the same initial conditions, we conclude that
$V=W$. Finally we compute:
\[V(1)=W(1)= \frac{d}{ds}\Big \vert_{s=0} \sigma(1,s)=\frac{d}{ds}\Big \vert_{s=0} \exp_p(v +sw)=
D(\exp_p)(v)(w). \]
\end{proof}

\section{Gauss' lemma and curvature }

The tangent space $T_pM$ on a semi-Riemannian manifold $M$ is itself a semi-Riemannian manifold
with the constant metric induced by the value of $g$ at $p$. Since this metric is constant, the
 manifold $T_pM$ is flat. The exponential map $\exp_p: U \to M$ is in general not an isometry onto its image. Gauss' lemma is the statement that it is however a radial isometry. 

\begin{lemma}[Gauss]
Let $M$ be a Riemannian manfilold and $U$ a neighborhood of zero in $T_pM$ sufficiently small so that the exponential map restricted to $U$ is a diffeomorphism onto its image. Given $v \in U$ and $ w \in T_pM$ the following holds:
\begin{equation}\label{EGauss}
\langle v,w\rangle = \langle  D(\exp_p)(v)(v), D(\exp_p)(v)(w) \rangle .
\end{equation}
\end{lemma}
\begin{proof}
Consider the geodesic $\gamma(t)=\exp_p(tv)$ and let $W$ be the unique Jacobi field along $\gamma(t)$ such that
\[ W(0)=0,\quad \nabla_{\gamma'(t)}W(0)=0.\]
By lemma \ref{dexp} we know that $W(1)=D(\exp_p)(v)(w)$.
We  set $V=t \gamma'(t)$ and compute:
\[ \nabla_{\gamma'(t)} \nabla_{\gamma'(t)} V=\nabla_{\gamma'(t)} \nabla_{\gamma'(t)} t\gamma'(t)=\nabla_{\gamma'(t)} \gamma'(t)=0 =R(\gamma'(t), V)(\gamma'(t)).\]
We conclude that $\gamma'(t)\big\vert_{t=1}=V(1)=D(\exp_p)(v)(v)$. 
Since both sides of (\ref{EGauss}) are continuous functions of $v$ and $w$, it is enough to prove the statement for $v$ such that $ \langle v,v\rangle \neq 0$. By linearity, it suffices to prove the statement in the case where $w=v$ and in the case where $\langle v, w \rangle=0$. In the first case,  since the norm of the derivative of a geodesic is constant, we have
\[ \langle  D(\exp_p)(v)(v), D(\exp_p)(v)(w) \rangle=\langle V(1), W(1)\rangle =\langle \gamma'(t) \Big \vert_{t=1},  \gamma'(t) \Big \vert_{t=1}\rangle=\langle v,v\rangle.\]
Let us now consider the case where $w$ is orthogonal to $v$. Since $W(0)=0$, it suffices to show that
\begin{equation}\label{wgamma} \frac{d}{dt} \langle W, \gamma'(t)\rangle =0.\end{equation}
For this we compute:
\[ \frac{d}{dt} \langle W, \gamma'(t)\rangle= \langle \nabla_{\gamma'(t)}W, \gamma'(t)\rangle+ \langle W, \nabla_{\gamma'(t)} \gamma'(t)\rangle= \langle \nabla_{\gamma'(t)}W, \gamma'(t)\rangle.\]
By construction $\nabla_{\gamma'(t)}W(0)=w$ is orthogonal to $v=\gamma'(t) \big \vert_{t=0}$. Therefore,
it suffices to show that:
\[ \frac{d}{dt}\langle \nabla_{\gamma'(t)}W, \gamma'(t)\rangle=0.\]
Using that $W$ is a Jacobi field and $\gamma(t)$ is a geodesic, one computes:
\[  \frac{d}{dt}\langle \nabla_{\gamma'(t)}W, \gamma'(t)\rangle=\langle \nabla_{\gamma'(t)} \nabla_{\gamma'(t)}W, \gamma'(t)\rangle= \langle R(\gamma'(t), W)(\gamma'(t)), \gamma'(t)\rangle=0.\]
In the last step we have used proposition \ref{simcurvature}.
\end{proof}

The geometric meaning of Gauss' lemma is the following. Let $S_\epsilon$ be a small sphere of radius $\epsilon$ centered at $0 \in T_pM$. Then $\exp_p(S_\epsilon)$ is orthogonal to the radial geodesics of the form $\gamma(t)=\exp_p(tv)$.
\begin{figure}[H]
\centering
	\includegraphics[scale=0.22]{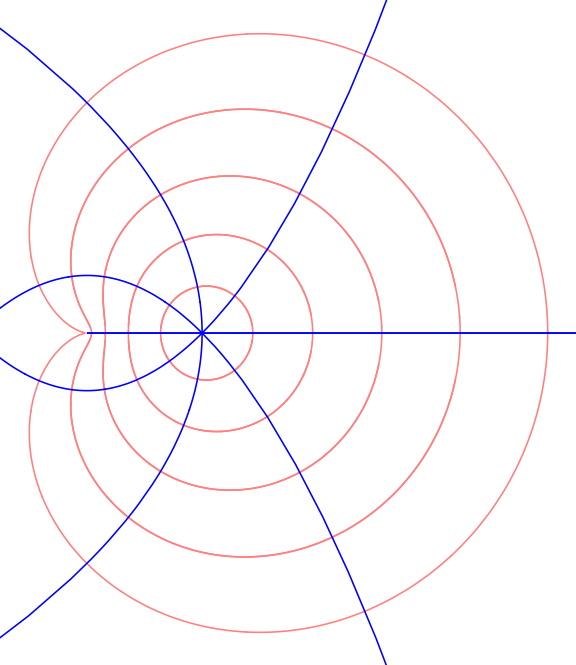}
	\caption{Gauss' lemma.}
\end{figure}

The following result shows that the curvature is precisely the obstruction to the exponential map being an isometry.

\begin{theorem}
Let $M$ be a semi-Riemannian manfilold and $U$ a convex neighborhood of zero in $T_pM$ sufficiently small so that the exponential map restricted to $U$ is a diffeomorphism onto its image. The exponential map:
\[ \exp_p: U \to \exp_p(U)\]
is an isometry if and only if the curvature of the Levi-Civita connection vanishes on $\exp_p(U)$.
 \end{theorem}
\begin{proof}
Since the metric on $U$ is constant, it is flat. Therefore, if the exponential map is an isometry, then the curvature vanishes on $\exp_p(U)$. Let us prove the converse. We assume that the curvature is zero and consider a point $q =\exp_p(v) \in \exp_p(U)$.
We fix vectors $w,z \in T_pM$ and consider parallel vectors $W,Z$ along $\gamma(t)=\exp_p(tv)$
such that
\[ W(0)=w, \quad Z(0)=z.\]
Then:
\[ \nabla_{\gamma'(t)} \nabla_{\gamma'(t)} (tW)=\nabla_{\gamma'(t)} (W)=0 =R(\gamma'(t),tW)(\gamma'(t)).\]
We conclude that $tW$ is a Jacobi field such that $tW(0)=0$ and $\nabla_{\gamma'(t)}W(0)=w$. By lemma
\ref{dexp} we conclude that $D(\exp_p)(v)(w)=(tW)(1)=W(1).$ By the same argument one also has that $D(\exp_p)(v)(z)=(tZ)(1)=Z(1).$ On the other hand, since $V, W$ are parallel, the quantity
$\langle W(t), V(t)\rangle$ is independent of $t$. We conclude that
\[ \langle D(\exp_p)(v)(w), D(\exp_p)(v)(z) \rangle= \langle W(1), Z(1)\rangle=\langle W(0),Z(0)\rangle =\langle w,z\rangle.\] 
\end{proof}
   \clearemptydoublepage

 \begin{partwithabstract}{Electromagnetism and Special Relativity}
The ancient Greeks observed that when amber is rubbed with a piece of cloth a force is generated.
This observation lead them to conjecture the existence of what we now call charged particles, which were divided in 
two classes: positive and negative. Since the Greek word for amber was elektron, these forces are known as electric forces.
It was also observed that moving charges are subject to other forces, which were given the name magnetic forces. A moving charged particle is subject to forces that are described by electric and magnetic fields. The fundamental equations satisfied by these fields are Maxwell's equations for electromagnetism. The incompatibility between Maxwell's equations and 
classical Newtonian mechanics made special relativity necessary.

\begin{center}
\begin{tabular}{cc}
  \vspace{0pt} \includegraphics[scale=0.5]{Figures/EB} &
  \vspace{0pt} \includegraphics[scale=0.5]{Figures/lightcone}
\end{tabular}
\end{center}
\end{partwithabstract}

\chapter{Electricity and Magnetism}\label{ch:EM}

\vspace{3ex}

\section{Coulomb's and Lorentz force laws}
The electric force between two charged particles is described by Coulombs's
law. This states that a particle of charge $Q$, measured in Coulombs, located at a place
$x \in\RR^{3}$ exerts over a charge $q$ located at $y \in\RR^{3}$
a force, measured in Newtons, given by
\begin{equation}
f_{\mathrm{e}}=\frac{1}{4\pi\varepsilon_{0}}\frac{qQ(y-x)}{\left\vert y-x\right\vert ^{3}},
\label{C}%
\end{equation}
where $\varepsilon_{0}=8.854\times10^{-12}\:\mathrm{N}^{-1}\mathrm{m}%
^{-2}\mathrm{C}^{2}$ is a constant of nature known as the \emph{permittivity
of free space}. The formula above takes the signs of the charges into account: if $q$ and $Q$
have opposite signs then the force is attractive while it is repulsive if the
signs are equal. One can describe the situation by postulating that the charge $Q$ determines an electric field
\[ E=\frac{1}{4\pi\varepsilon_{0}}\frac{Q(y-x)}{\left\vert y-x\right\vert ^{3}}\]
which determines the electric force caused by $Q$ on a charged particle. More generally,
an electric field $E$ defined on a region $U\subseteq\RR^{3}$ is a
vector field such that a particle of charge $q$ Coulombs
located at $y \in U$ is subject to an electric force 
\begin{equation}f_{\mathrm{e}}=q E.\end{equation}
The electric field
$E$ is measured in units of Newton per Coulomb $(\mathrm{N/C})$.

Magnetic fields exert a force on a charged particle only if
the particle is moving. A particle of charge $q$ Coulombs at the
place $y \in\RR^{3}$ moving with velocity $v$ in presence of a magnetic field $B$ is subject to a magnetic
force: \begin{equation}f_{\mathrm{m}}=qv \times B.\end{equation}
The magnetic field $B$ is therefore measured
in units \[\mathrm{T}=\frac{\mathrm{N\cdot s}}{ \mathrm{C\cdot m}}.\] The unit $\mathrm{T}$ is called a Tesla. A magnetic
field of magnitude 1 $\mathrm{T}$ exerts on a particle of charge 1 Coulomb
moving with speed of 1 $\mathrm{m/s}$ a magnetic force of magnitude 1 Newton which is perpendicular to the velocity and to the magnetic field.
In summary, a moving charged particle in the presence of
electric and magnetic fields is subject to a total force given by the Lorentz Law:
\begin{equation}
\label{Lorenz}f=q\left(E+v\times B\right).
\end{equation}
According to Newton's second law, the equation of motion for a charged particle of mass $m$ is
\begin{equation}
\label{f=ma}
m \ddot{y}=q \left(E(y)+v\times B(y) \right).
\end{equation}

\begin{figure}[h!]
	\centering
		\includegraphics[scale=0.52]{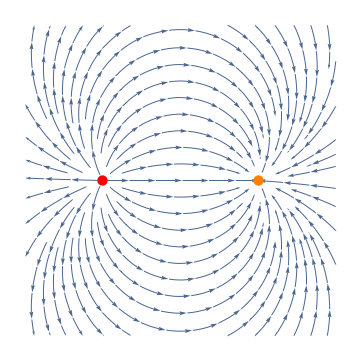}
	\caption{Electric Field generated by two charged particles.}
	
\end{figure}

\section{Electrostatics: charges at rest}

Suppose there are particles with charges $q_{1},\ldots,q_{n}$ located at
positions $x_{1},\ldots,x_{n}$. Coulomb's law implies that the electric field
generated by these particles is given by
\[
E(y)=\frac{1}{4\pi\varepsilon_{0}}{\textstyle\sum\limits_{i=1}^{n}}
\frac{q_{i} (y-x_{i})}{\vert y-x_{i}\vert ^{3}}.
\]

In the continuous limit, the charge is distributed
according to a time independent charge density function $\rho(x)$. The total amount of charge in a region $U$ is given by
\[Q=\int_{U}\rho(x)\, dV.\] In this situation the electric field is given by
\begin{equation}
E(y)=\frac{1}{4\pi\varepsilon_{0}}\int_{U}\frac{\rho(x)(y-x)}{\left\vert
y-x\right\vert ^{3}}dV. \label{complicada}%
\end{equation}

Let us consider a charge $Q$ located at a point $y $ in a region $U$ with boundary $S=\partial U$. We want to compute the total flux of the electric field across the surface $S$. 
\begin{figure}[H]
	\center
	\includegraphics[scale=0.42]{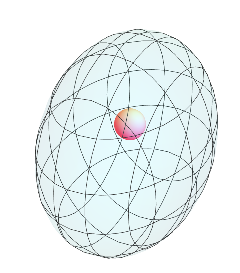}
	\caption{A charge in the interior of a region.}
	\label{Ampere}
\end{figure}Let us consider a little ball $B$ centered at $y$ with radius $r$.
A simple computation shows that outside the ball $B$ we have
\[ \div E=0.\]
If we set $V=U-B$, Stokes' theorem gives
\[ 0=\int_V \div E \,dV=\int_{\partial U} E\cdot n \,dA- \int_{\partial B} E\cdot n\, dA.\]
On the other hand, one can compute
 \[ \int_{\partial B} E\cdot n \, dA=\frac{Q}{4\pi \varepsilon_0}\int_{\partial B} \frac{dA}{r^2} =\frac{Q}{\varepsilon_0}.\]
We conclude that the total flux across the boundary of $U$ is proportional to the total charge inside of $U$. By linearity, this also holds for an arbitrary number of charges inside $U$. In the continuous limit one obtains
\begin{equation}
\int_{\partial U}E\cdot n\,dA=\frac{1}{\varepsilon_{0}}\int_{U}\rho(x)\,dV=\frac{Q}{\varepsilon_0}.
\label{Gaussint}%
\end{equation}
One concludes that for any ball $B$
\begin{equation}\label{ec:GaussFlux} \int_B \div E \,dV=\int_{\partial B}E\cdot n\,dA=\frac{1}{\varepsilon_{0}}\int_{B}\rho(x)\,dV.\end{equation}
We call the surface integral $\int_{\partial B}E\cdot n\,dA$ the \emph{electric flux} through $\partial B$. Then, equation \eqref{ec:GaussFlux} implies that the electric flux is equal to $1/\varepsilon_0$ times the total charge enclosed by the surface $\partial D$. This assertion is known as \emph{Gauss' flux theorem}. Since the ball $B$ is arbitrary, one concludes
\begin{equation}\label{Gaussdiff}
\div E= \frac{\rho}{\varepsilon_0}.
\end{equation}

The electric field $E$ generated by a charge density function $\rho(y)$ is the
gradient of a function $\phi(y)$, called the \emph{electric potential}:
\begin{equation}
\phi(y)=\frac{1}{4\pi\varepsilon_{0}}\int_{U}\frac{\rho(x)}{\left\vert
y-x\right\vert }dV. \label{potencial}%
\end{equation}
A short computation shows that $E = -\grad \phi$. In particular, we conclude that 
an electric field generated by static charges satisfies 
\begin{equation}
\rot E=0.
\end{equation}
Also, for a static field, the governing equation \eqref{Gaussdiff} reduces to the Poisson equation
\begin{equation}
\Delta \phi =-\frac{\rho}{\varepsilon_0}.%
\end{equation}

It should be noted that the formulas \eqref{complicada} and \eqref{potencial} are valid at all points $y$ in space, including the points inside the domain $U$. Indeed, if $\rho$ is a bounded field, say if $\rho$ is continuous on the closure of $U$, then $\rho(x)/\lvert y-x \rvert$ is an integrable function even though it has a singularity at the point $y=x \in U$, where $\lvert y-x \rvert = 0$.

\section{Electrodynamics: moving charges}

A moving charge is known as a current, and can be described mathematically by
the current density vector field 
\[j(t,x)=\rho(t,x)v(t,x),\] 
where $v(t,x)$ is a
time dependent vector field that describes the current's velocity. Consider an
oriented surface $S$. The current passing through $S$ at time $t$ is
defined by
\[
I(t)=\int_{S}j(t,x)\cdot n\,dA.
\]
Current is measured in units of Coulomb per second, known as an Ampere:
\[1 \:\mathrm{A}=1 \:\mathrm{C/s}.\] Consider a region $D$ and define the
function $Q(t)$ as the total amount of charge inside $D$ at time $t$:
\[Q(t)=\int_{D}\rho(t,x)\, dV.\] It is a fundamental fact that the charge is
conserved. Thus, the charge leaving the region $D$ must be equal to the change
in $Q(t)$ and we conclude
\[
\int_{\partial D}j(t,x)\cdot n\,dA=-\frac{\partial Q}{\partial
t}=-\int_{D}\frac{\partial\rho}{\partial t}\, dV.
\]
Since this equation holds for an arbitrary region $D$, one concludes that the charge density satisfies the continuity equation
\begin{equation}
\frac{\partial\rho}{\partial t} + \div j=0,\label{CC}%
\end{equation}
which is the field equation for the \emph{law of conservation of charge}.

\section{The Law of Biot-Savart}

Ampere, Biot and Savart were among the first to
measure the intensity of a magnetic field. By 1820, Oersted had discovered
that these fields could be generated by making an electric current
circulate through a conductor. The law of Biot-Savart describes the magnetic field induced by a stationary current. The term stationary means that the
current is a constant function of time, i.e., $j(x,t)=\rho(x)v(x)$.\\

\begin{figure}[H]
	\center
	\includegraphics[scale=0.4]{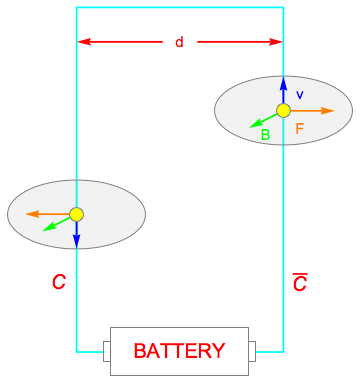}
	\caption{Ampere's experiment}
	\label{Amperee}
\end{figure}

We will  consider a classical experiment. A wire of constant cross
section $A$ is connected to a battery, so that a stationary current
circulates through it. Two segments $C$ and $\overline{C}$ are separated a
distance $d$, as shown in Figure \ref{Amperee}. 
It can be observed, though the effect is barely noticeable, that $C$ and
$\overline{C}$ repel each other. This indicates the presence  magnetic
fields $B_1,B_2$ which, according to Lorentz law, would exert a force
 on the opposite segment. Once this force is measured,
it is discovered that its intensity follows a square inverse law. These
experiments culminated in the law of Biot and Savart, which can be formulated
as follows. If $J(x)$ is a stationary current that circulates inside certain
region $U,$ then the total magnetic field this current induces at a point $y$ is given by the sum of the contributions
$\Delta B$ of all the small regions $\Delta R$ inside $U$. Each $\Delta B$
points in the direction of the vector $v\times(y-x),$ with magnitude equal
to \[\frac{K_{\mathrm{m}}\Delta q\left\vert v\right\vert}{ \left\vert y-x\right\vert ^{2}}.\]
Here $K_{\mathrm{m}}$ is a constant that for historical reasons is written as
$K_{\mathrm{m}}=\mu_{0}/4\pi$. The constant \[\mu_{0}=4\pi \times 10^{-7}\frac{\mathrm{Kg} \cdot \mathrm{m}}{\mathrm{C}^2} \] is the so called
\emph{permeability of the vacuum}.  The quantity $\Delta q=\rho(x)\mathrm{Vol}(\Delta
R)$ is the total amount of charge contained in $\Delta R$. That is
\[
\Delta B=\frac{\mu_{0}}{4\pi}%
j(x)\times(y-x)\frac{\mathrm{Vol}(\Delta R)}{\left\vert y-x\right\vert ^{3}}.%
\]%
Thus, the sum of all the $\Delta B$ in a region $U$ is equal to
\begin{align}
B(y)  &  =\frac{\mu_{0}}{4\pi}\int_{U}\frac{j(x)\times
(y-x)}{\left\vert y-x\right\vert ^{3}}dV.\label{bs}\nonumber
\end{align}
\begin{figure}[H]
	\centering
	\includegraphics[scale=0.49]{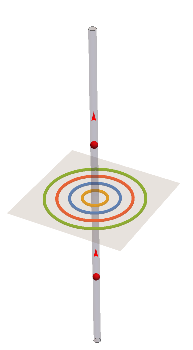}
	\caption{Flow lines of the magnetic field induced by a current on a straight wire.}	
\end{figure}

\section{There are no magnetic monopoles}

We will now see that a magnetic field $B$ induced by a stationary current, as described by the law of Biot-Savart, has zero divergence.
This fact follows from the existence of a \emph{vector potential} for
$B$. Let us set  \[A(y)=\frac{\mu_0}{4\pi}\int\limits_{U}\frac{j(x)}{\left\vert y-x\right\vert }dV.\]
A simple calculation shows that
\[
\rot\frac{j(x)}{\left\vert y-x\right\vert }=j(x)\times\frac
{y-x}{\left\vert y-x\right\vert ^{3}}.
\]
Thus,
\[
\rot A(y)=\frac{\mu_0}{4\pi} \int\limits_{U}\rot\frac{j(x)}{\left\vert
y-x\right\vert }dV=\frac{\mu_0}{4\pi}\int\limits_{U}j(x)\times\frac{y-x}{\left\vert
y-x\right\vert ^{3}}dV=B(y).
\]
This implies
\begin{equation}
\div B=\div(\rot A)=0.
\label{antesegunda}%
\end{equation}

The universality of this law for any magnetic field, not only one given by the law of Biot-Savart, is a fundamental law of nature. No one has ever observed a
\emph{monopole}, the magnetic equivalent of an electric particle. It is therefore assumed that an arbitrary magnetic field satisfies
\begin{equation}
\div B=0. \label{segunda}%
\end{equation}


\section{Magnetostatics}

We will now discuss the case where the electric and magnetic fields, the charge density and the current density functions are independent of time
\[ \frac{\partial E}{\partial t}=\frac{\partial B}{\partial t}=\frac{\partial \rho}{\partial t}=\frac{\partial j}{\partial t}=0.\]
Let us consider a closed circuit $C$  determined by a wire of
constant cross section $A$ through which an stationary current $I$ circulates. Suppose $C^{\prime}$ is a closed curve that is linked to $C$, as
shown below:

\begin{figure}[H]
	\centering
	\includegraphics[scale=0.41]{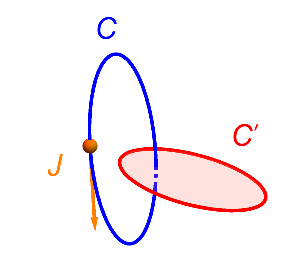}
	\caption{Linked wires.}	
\end{figure}
Fix parametrisations $\alpha(t)$, $\beta(s)$  for $C$ and
$C^{\prime}$, respectively. If the current density of $I$ is given by \[j=\rho(x)v(x),\]
then in a small segment of $C$ of length $\Delta l$, the density of the charge
would be \[\Delta q=\rho(x)A\Delta l,\] where $x=\alpha(t)$. 
The total amount of current passing across a section of $C$ at any time is
given by \[I=\frac{A j\cdot v(x)}{\left\vert v(x)\right\vert} =A\rho(x)\left\vert
v(x)\right\vert .\] Thus, \[\Delta q=\frac{I\Delta l}{\left\vert v(x)\right\vert} .\]
Using the law of Biot-Savart one can compute the magnetic field at a
point $y=\beta(s)$  as follows
\begin{align*}
\Delta B  &  =\frac{\mu_{0}}{4\pi|y-x|^3}\Delta q(v(x)\times (y-x))=\frac
{\mu_{0}I}{4\pi}\left(  \frac{v(x)}{\left\vert v(x)\right\vert }\times
\frac{(y-x)}{\left\vert y-x\right\vert ^{3}}\right)  \Delta l\\
&  =\frac{\mu_{0}I}{4\pi}\left(  \frac{\alpha^{\prime}(t)}{\left\vert
\alpha^{\prime}(t)\right\vert }\times\frac{(\beta(s)-\alpha(t))}{\left\vert
\beta(s)-\alpha(t)\right\vert ^{3}}\right)  \left\vert \alpha^{\prime
}(t)\right\vert \Delta t=\frac{\mu_{0}I}{4\pi}\alpha^{\prime}(t)\times
\frac{(\beta(s)-\alpha(t))}{\left\vert \beta(s)-\alpha(t)\right\vert ^{3}%
}\Delta t.
\end{align*}
From this one obtains
\[
B(y)=\frac{\mu_{0}I}{4\pi}\int\limits_{a}^{b}\alpha^{\prime}%
(t)\times\frac{(y-\alpha(t))}{\left\vert y-\alpha(t)\right\vert
^{3}}dt.
\]
The circulation of $B$ along $C^{\prime}$ is defined as
\[L=\int_c^dB(\beta(s)) \cdot \beta'(s) ds.\]
Using the formula for the magnetic field one obtains
\[L=\frac{\mu_{0}I}{4\pi}\int\limits_{c}^d \int\limits_{a}^{b}\Big(\alpha^{\prime}%
(t)\times\frac{(\beta(s)-\alpha(t))}{\left\vert \beta(s)-\alpha(t)\right\vert
^{3}}\Big)\cdot \beta'(s)dt ds.\]
Recall that if $a,b,c$ are arbitrary
vectors, then \[(a\times b)\cdot c=\det(a,b,c).\] From this we
see that the term inside the integral is equal to \[\frac{\det\Big(\alpha'(t),\beta
(s)-\alpha(t),\beta'(s)\Big)}{\left\vert
\beta(s)-\alpha(t)\right\vert ^{3}}.\] Consequently,
\[
L=\mu_0I \int\limits_{c}^{d}\int\limits_{a}^{b}\frac
{\det(\alpha^{\prime}(t),\text{ }\beta(s)-\alpha(t),\text{ }\beta^{\prime
}(s))}{4\pi\left\vert \beta(s)-\alpha(t)\right\vert ^{3}}dtds.
\]
We conclude that $L=-\mu_0 I L(C,C')$, where $L(C,C')$ is the linking number of $C$ and $C'$.
 The linking number is an integer which is a topological invariant of a configuration of two circles in space. More information regarding the linking number can be found in Appendix \S \ref{link}.
 
Let $S$ be a surface whose boundary is $C^{\prime}$. The flux trough $S$
is \[I=\int\limits_{S}j\cdot n\,dA,\] where $\mathbf{n}$
denotes the exterior normal vector to $S$. Let us now assume that $C'$ is a small circle which is simply linked to $C$ so that $L(C,C')=1$ then
\begin{equation}
L=\int\limits_{C^{\prime}}B(\beta(t)))\cdot \beta'(t) dt=\mu_{0}\int_{S}j\cdot n \text{
}dA. \label{ampere integral}%
\end{equation}
By Green's theorem: \[\int\limits_{C^{\prime}}B(\beta(t))\cdot \beta'(t) dt=\int\limits_{S}%
\rot B \cdot n \,dA.\] Therefore
\[
\int\limits_{S}\rot B\cdot n\,dA=\mu_{0}\int_{S}%
j\cdot n \,dA.
\]
Since $S$ is an arbitrary surface, one obtains Ampere's law for a static current:
\begin{equation}
\label{ley de ampere}
\rot B=\mu_{0}j. 
\end{equation}

\section{Varying electric fields}

Ampere's law for static currents  cannot possibly hold  for arbitrary currents. Taking the divergence on both sides, and using the equation for conservation of charge, one obtains
\[0 = \div(\rot B )=\mu_0\div j=-\mu_0\frac{\partial \rho}{\partial t}.\]
This shows that equation \eqref{ley de ampere} implies that the distribution of charge is constant. In the general case, a new term has to be added for the equation
to be consistent with the conservation of charge. This is Ampere's law
\begin{equation}
\rot B=\mu_{0}j+ \mu_0 \varepsilon_0 \frac{\partial E}{\partial t}, \label{ley de ampere general}%
\end{equation}
where the term $\mu_0 \varepsilon_0 \partial E / \partial t$ is called the \emph{displacement current}. In this case, by taking the divergence on both sides one imposes no additional restriction on the fields:
\begin{align*}
0 &=\div(\rot B)=\mu_{0}\div j+ \mu_0 \varepsilon_0 \div\left(\frac{\partial E}{\partial t}\right) \\
&=
-\mu_{0}\frac{\partial \rho}{\partial t}+ \mu_0 \varepsilon_0 \frac{\partial}{\partial t}\div E \\
&=-\mu_{0}\frac{\partial \rho}{\partial t}+\mu_{0}\frac{\partial \rho}{\partial t} \\
&=0.
\end{align*}
In the absence of currents, Ampere's law states that a time dependent electric field $E(t,x)$
induces a magnetic field $B(t,x)$ such that if $S$ is a surface with boundary $C$ then
\begin{equation}\label{amperenc}
\int\limits_{C}B(t,x) \cdot dl =\mu_{0}\varepsilon_{0}\frac{d}{dt}%
\int\limits_{S}E(t,x)\cdot n\, dA.
\end{equation}

\begin{figure}[H]
	\centering
	\includegraphics[scale=0.36]{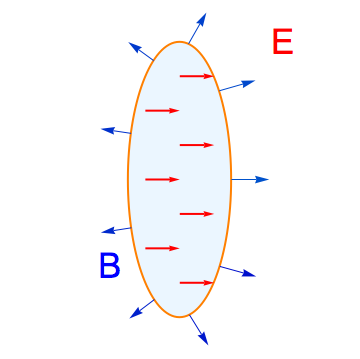}
	\caption{The flux of the electric field is the integral of the magnetic field on the boundary of the surface.}
	
\end{figure}

\section{Faraday's law of induction\label{donfara}}

Consider a closed wire moving with constant velocity $v$ with respect to some reference frame $O$, so that its
position at time $t$ is given by a map:
\[ \sigma_t(s): S^1 \rightarrow \RR^3; \,\, s \mapsto \sigma_0(s) + tv.\]

\begin{figure}[H]
	\centering
	\includegraphics[scale=0.46]{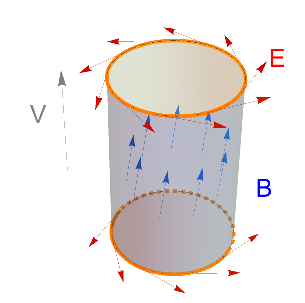}
	\caption{Faraday's law of induction.}
	
\end{figure}

Suppose there is a constant magnetic field $B$ which, in accordance with our previous discussions, satisfies
\[ \div B=0.\]
We denote by $S_t$ the surface whose boundary is $C_t$, the image of the  curve $\sigma_t$. Let $\Phi(t)$ be the flux of $B$ across $S_t$:
\[\Phi(t)=\int\limits_{S_t}B \cdot n\, dA.\]
We want to compute the rate of change of $\Phi(t)$ with respect to $t.$ Let us first estimate
 $\Phi(t+\Delta t)-\phi(t)$, for a small increment $\Delta
t$. Since the divergence of $B$ is zero, Stokes' theorem implies that:
\[
0=\int\limits_{D}\div B \, d V=\Phi(t+\Delta t)-\Phi(t)+\int
\limits_{S^{\prime}}B\cdot n\,dA,
\]
where $D$ is the region between $S_t$ and $S_{t+\Delta t}$ and $S'$ is the lateral part of the boundary. 
The expression in the last integral can be computed as follows:
\begin{eqnarray*}
B(\sigma_t(s))\cdot n\,dA  &=&B(\sigma_t(s))\cdot \left(\frac{\partial
\sigma}{\partial s} \times \frac {\partial\sigma}{\partial t}\right)ds dt\\
&=&\left( \frac{\partial
\sigma}{\partial t} \times B(\sigma_t(s)) \right) \cdot \frac{ \partial \sigma}{\partial s} ds dt\\
&  =& \Big(v \times B(\sigma_t(s)) \Big) \cdot \frac{ \partial \sigma}{\partial s} ds dt.
\end{eqnarray*}
Therefore
\[
\int\limits_{S^{\prime}}B\cdot n \,dA=\int\limits_{0}^{2\pi}%
\int\limits_{t}^{t+\Delta t} \Big(v \times B(\sigma_t(s)) \Big) \cdot \frac{ \partial \sigma}{\partial s} ds dt.\]
Since $\Delta t$ is small, this integral can be approximated by
\[\Delta t \int\limits_{0}^{2\pi}%
 \Big(v \times B(\sigma_t(s)) \Big) \cdot \frac{ \partial \sigma}{\partial s} ds.\]
Thus
\[
\frac{\Phi(t+\Delta t)-\Phi(t)}{\Delta t}\approx -\int\limits_{0}^{2\pi}
 \Big(v \times B(\sigma_t(s)) \Big) \cdot \frac{ \partial \sigma}{\partial s} ds.\]
One concludes that
\begin{equation}\label{work1}
  \frac{\partial \Phi}{\partial t}=-\int\limits_{C_t}(v \times
B) \cdot dl.
\end{equation}
Let us consider a test particle with charge $q=1C$ moving with velocity $w$ along the wire. Using the Lorentz force law, one computes the work done by the force in moving the charge once around the wire $C_t$ is
\begin{equation}\label{work2}
W_t=\int\limits_{C_t}F \cdot dl =\int\limits_{C_t}(v\times B) \cdot dl
+\int\limits_{C_t}(w\times B) \cdot dl. %
\end{equation}
The second integral is zero, since $w$ is tangent to the wire. Therefore
\begin{equation}W_t=\int\limits_{C_t}(v\times B)\cdot dl=-\frac{\partial \Phi}{\partial t}.
\end{equation}

Let us now analyse the situation from the point of view of an observer $\overline{O}$ that moves with the wire. 
For this observer, the magnetic field $B$ need no longer be stationary. Observer $\overline{O}$ also sees the test particle move
along the wire. She also uses the Lorentz force law to compute the work:

\[\overline{W}_t=\int\limits_{C_t}\overline{F} \cdot dl=\int\limits_{C_t}(w\times \overline{B})\cdot dl+\int\limits_{C_t}\overline{E}\cdot dl. \]
As before, the first integral is zero because $w$ is tangent to the curve. Observer $\overline{O}$ concludes that the
work is done by an electric field $\overline{E}$ which is induced by the varying magnetic field. The resulting equation is
known as the integral form of Faraday's law:
\begin{equation}
\int\limits_{C}E \cdot dl=-\frac{d }{d t}\int\limits_{S}%
B \cdot n\text{ }dA, 
\end{equation}
\begin{figure}[H]
	\centering
	\includegraphics[scale=0.36]{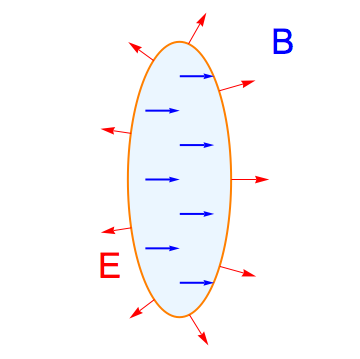}
	\caption{}
\end{figure}

By Stokes' theorem the integral form of Faraday's law is equivalent to
\begin{equation}
\rot E=-\frac{\partial B}{\partial t},
\label{tercera}%
\end{equation}
which shows that $E$ is generally not a conservative field.

\section{Conservation of energy}\label{EEM}
From Lorentz's formula \eqref{Lorenz} we see that the force acting on a moving charge due to a magnetic field is perpendicular to the velocity field . Hence the power density on the moving charge is produced entirely by the electric field. On any domain $U$ with current density $j$, the total power is given by the integral
\begin{equation}\label{ec:power}
\int_U j \cdot E \, dV.
\end{equation}
This power represents the rate of conversion of electromagnetic energy into other forms of energy such as thermal energy. Assuming that the energy is balanced in $U$, we equate the power with a rate of decrease of electromagnetic energy in $U$ together with the energy flux through the boundary $\partial U$. Such a balance principle was considered first by Poynting.

To obtain an expression for the power \eqref{ec:power} in terms of the electromagnetic field in $V$, we use Maxwell's equation \eqref{M4} to determine the current density $j$:
\begin{equation}\label{ec:7.35}
\int_U j \cdot E \, dV = \int_U \left[ \frac{1}{\mu_0} E \cdot \rot B - \varepsilon_0 E \cdot \frac{\partial E}{\partial t}  \right] dV.
\end{equation}
Now using the vector identity
\begin{equation}
\div (E \times B) = B \cdot \rot E - E \cdot \rot B,
\end{equation}
together with the field equation \eqref{M2}, we can rewrite \eqref{ec:7.35} as
\begin{equation}\label{ec:7.37}
\int_U j \cdot E \, dV = - \int_U \left[ \frac{1}{\mu_0} \div(E \times B) + \varepsilon_0 E \cdot \frac{\partial E}{\partial t} + \frac{1}{\mu_0} B \cdot \frac{\partial B}{\partial t}\right] dV.
\end{equation}
Applying the divergence theorem to the first term on the right-hand side, we obtain
\begin{equation}\label{ec:7.38}
\int_U j \cdot E \, dV = - \int_U \left[ \varepsilon_0 E \cdot \frac{\partial E}{\partial t} + \frac{1}{\mu_0} B \cdot \frac{\partial B}{\partial t}\right] dV - \frac{1}{\mu_0} \int_{\partial U} (E \times B) \cdot n \, dA,
\end{equation}
where $n$ denotes the outward unit normal on $\partial U$. 

Poynting observed that the volume integral on the right-hand side may be regarded as the rate of decrease of the energy of the electromagnetic field in $U$, while the surface integral may be regarded as the energy flux through $\partial U$. Indeed, we may define the electromagnetic field energy of the domain $U$ by the integral
\begin{equation}\label{ec:7.39}
\int_U \left( \frac{\varepsilon_0}{2}\lvert E\rvert^2 + \frac{1}{2\mu_0}\lvert B\rvert^2\right) dV.
\end{equation}
Then \eqref{ec:7.38} may be rewritten as
\begin{equation}\label{ec:7.40}
- \frac{d}{dt}\left[ \int_U \left( \frac{\varepsilon_0}{2}\lvert E\rvert^2 + \frac{1}{2\mu_0}\lvert B\rvert^2\right) dV \right] = \int_U j \cdot E \, dV +  \int_{\partial U} S \cdot n \, dA,
\end{equation}
where $S$ is defined by
\begin{equation}\label{eq:Poyntingvector}
S = \frac{1}{\mu_0} E \times B,
\end{equation}
and is called the \emph{Poynting vector}. 

Poynting regarded $S$ as the energy flux associated with the electromagnetic fields. Thus \eqref{ec:7.37} may be interpreted as a balance principle which asserts that the rate of decrease of the field energy in $U$ is equal to the rate of conversion of energy in $U$ plus the rate of energy flux through $\partial U$. We call this assertion \emph{Poynting's principle}. Since it is valid for all domains $U$, one obtains the equation
\begin{equation}\label{ec:7.42}
\varepsilon_0 E \cdot \frac{\partial E}{\partial t} + \frac{1}{\mu_0} B \cdot \frac{\partial B}{\partial t} +  \div S + j \cdot E  = 0,
\end{equation} 
which is known as \emph{Poynting's equation}. 
It should be mentioned that Poynting's principle is really an identity which is satisfied by all solutions of Maxwell's equations. In this sense, Poynting's principle is not a new axiom for electromagnetism but a theorem in the context of Maxwell's equations.

\section{Conservation of linear momentum }
We will derive a balance principle for the linear momentum similar to Poynting's principle. We regard the Lorentz force on the charge and the current in $U$ as a rate of conversion of the electromagnetic field momentum into mechanical momentum. Then this rate must be balanced by a rate of decrease of the electromagnetic field momentum in $U$ together with the linear momentum flux through $\partial U$. Following the same procedure as before, we write the momentum conversion rate in $U$ as the integral
\begin{equation}
\int_U \left( \rho E + j \times B\right) dV.
\end{equation} 
Now using Maxwell's equations \eqref{M1} and \eqref{M4} to determine the charge density $\rho$ and the current density $j$, we get
\begin{equation}
\int_U \left( \rho E + j \times B\right) dV = \int_U \left[ \varepsilon_0 (\div E) E + \frac{1}{\mu_0}\rot B \times B - \varepsilon_0 \frac{\partial E}{\partial t} \times B\right] dV.
\end{equation}
We can now rewrite the right-hand side as a sum of a rate of change of a volume integral and a surface integral.

Using the product rule and the system of field equations (\ref{M1}-\ref{M4}), we replace the left-hand side by
\begin{align}\label{ec:7.45}
\begin{split}
&\int_U \left( \rho E + j \times B\right) dV \\
& \quad = - \frac{d}{dt} \int_U \frac{1}{\mu_0} (E \times B) dV + \int_{U} \left[ \varepsilon_0(\div E) E + \frac{1}{\mu_0} \rot B \times B - \varepsilon_0 E \times \rot E \right] dV.
\end{split}
\end{align} 
The integrand of the second term on the right-hand side is the divergence of the \emph{Maxwell stress tensor} $\Theta$, which is defined by
\begin{equation}\label{eq:Maxwellstresstensor}
\Theta = \left(\frac{\varepsilon_0}{2} \lvert E\rvert^2 + \frac{1}{2\mu_0}  \lvert B\rvert^2 \right) I - \varepsilon_0 E \otimes E - \frac{1}{\mu_0} B \otimes B.
\end{equation}
We can verify the formula 
\begin{equation}\label{ec:3.47}
-\div \Theta = \varepsilon_0 (\div E) E + \frac{1}{\mu_0} (\div B) B -  \frac{1}{\mu_0} B \times \rot B -  \varepsilon_0 E \times \rot E
\end{equation}
by a direct calculation. Substituting \eqref{ec:3.47} into \eqref{ec:7.45} and using the divergence theorem, we obtain
\begin{equation}\label{ec:7.48}
- \frac{d}{dt} \int_U \varepsilon_0 (E \times B) dV = \int_U \left( \rho E + j \times B\right) dV + \int_{\partial U} \langle \Theta, n \rangle dA.
\end{equation} 
This identity has a form similar to \eqref{ec:7.40}.
As before, we regard the left-hand side of \eqref{ec:7.48} as the rate of decrease of the electromagnetic field momentum in $U$ and the second term on the right-hand side as the momentum flux through $\partial U$. Then \eqref{ec:7.48} becomes a balance principle, which asserts that the rate of decrease of the field momentum in $U$ is equal to the rate of momentum conversion in $U$ plus the momentum flux through $\partial U$. The field equation for this balance principle is
\begin{equation}\label{ec:7.49}
\varepsilon_0 \frac{\partial}{\partial t}(E \times B)+ \div \Theta + \rho E + j \times B  = 0.
\end{equation}
Like Poynting's equation \eqref{ec:7.42}, the balance equation \eqref{ec:7.49} is really an identity which is satisfied by all solutions of Maxwell's equations. Hence this identity does not place any additional restrictions on the electromagnetic field.
\section{Maxwell's equations and waves}

The relations between the electric and magnetic fields are sumarized in the following set of equations,  known as \emph{Maxwell's equations}:
\begin{eqnarray}
\div E&=& \frac{\rho}{\varepsilon_0}\label{M1},\\
\div B&=&0,\label{M2}\\
\rot E+\frac{\partial B}{\partial t}&=&0,\label{M3}\\
\rot B-\varepsilon_{0}\mu_{0}\frac{\partial E}{\partial t}\label{M4}&=&\mu_0 j.
\end{eqnarray}
\noindent
Equation (\ref{M1}) imposes the conservation of charge. Equation (\ref{M2}) is the nonexistence of magnetic monopoles. Equation (\ref{M3}) is the Maxwell-Faraday equation that states that a time dependent magnetic field is accompanied by an electric field . Equation (\ref{M4}) is Ampere's law. In the special case where there are no charges, so that $j=\rho=0$, the equations are known as the
vacuum Maxwell equations.

Let us consider solutions to the vacuum Maxwell equations of the form
$(0,E,0)$ and $(0,0,B)$. Maxwell's equations become
\begin{align}
\frac{\partial E}{\partial y}=
\frac{\partial B}{\partial z}=
 \frac{\partial E}{\partial z}= 
 \frac{\partial B}{\partial y} &=0,\\
\frac{\partial E}{\partial x}+ \frac{\partial B}{\partial t}&=0, \label{wave2}\\
\frac{\partial B}{\partial x}+ \varepsilon_0 \mu_0 \frac{\partial E}{\partial t}&=0 \label{wave3}. \,\, 
\end{align}
Therefore,  $E$ and $B$ depend only on
$x$ and $t$. Differentiating Equation (\ref{wave2}) with respect to $x$ one obtains
 \[\frac{\partial^{2}E}{\partial x^{2}}=-\frac{\partial^{2}B}
{\partial x\partial t}.\] Differentiating Equation (\ref{wave3}) with respect to $t$ we obtain: 
\[-\frac{\partial^{2}B}{\partial t\partial
x}=\varepsilon_{0}\mu_{0} \frac{\partial^2 E}{\partial t^2}.\] 
Hence, if we set $c=1/\sqrt
{\varepsilon_{0}\mu_{0}}$, then
\begin{equation}
\frac{\partial^{2}E}{\partial x^{2}}=\frac{1}{c^{2}}\frac
{\partial^{2}E}{\partial t^{2}}.\label{ss1}%
\end{equation}
Similarly, 
\begin{equation}\frac{\partial^{2}B}{\partial x^{2}}=\frac{1}{c^{2}}
\frac{\partial^{2}B}{\partial t^{2}} \label{ss2}.
\end{equation} 
\begin{figure}[H]
	\centering
	\includegraphics[scale=0.45]{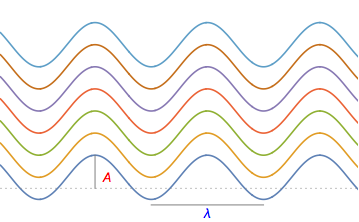}
	\caption{Wavelength and amplitude of a wave.}
\end{figure}
 
If $E_0$ and $B_0$ are constants such that $E_0=c B_0$  and $ck=\omega$ then \[E=E_{0}\sin(kx-\omega t);\,\,\,B=B_{0}
\sin(kx-\omega t),\] 
are solutions to Maxwell's equations. The wavelength is $\lambda=2\pi/k$ and the frequency is $f=\omega/2\pi$.
\begin{figure}[H]
	\centering
	\includegraphics[scale=0.5]{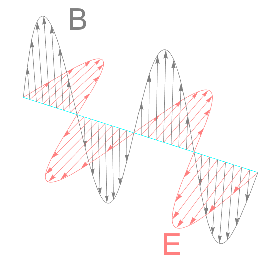}
	\caption{Electric and Magnetic fields.}
\end{figure}

 There are other solutions where $E$ can rotate in the $yz$-plane. For instance, the fields \[E=E_{0}\Big(0,\cos(kx-\omega t),
\sin(kx-\omega t)\Big),\,\,\,B=B_{0}\Big(0,-\sin(kx-\omega
t), \cos(kx-\omega t\Big)\] describe a circularly polarized
wave.
\begin{figure}[H]
	\centering
	\includegraphics[scale=0.5]{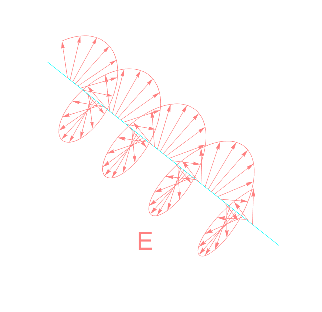}
	\caption{Circular polarization.}
\end{figure}

The  wave equation is derived from Maxwell's equations as follows. 
The general relations of vector calculus imply that
 \[\rot(\rot
E)=\grad(\div E)-\Delta E,\]
where $\Delta E$  is the vector laplacian applied to $E$. In cartesian coordinates this is the result of applying the
laplacian to the components of $E$. Since
\[ \div E=0 \quad \text{ and } \quad \rot E=-\frac{\partial
B}{\partial t},\] one concludes that \[\frac{\partial}{\partial t}(\rot
B)=\Delta E.\] 
On the other hand, \[\rot B=\varepsilon_{0}\mu
_{0}\frac{\partial E}{\partial t}.\] Therefore \[\Delta E=\varepsilon_{0}\mu
_{0}\frac{\partial^{2}E}{\partial t^{2}}.\] Equivalently,
\[
\Delta E^{i}=\frac{1}{c^{2}}\frac{\partial^{2}E^{i}}{\partial t^{2}}. \]
\noindent
Similarly, for the components of the magnetic field on obtains
\[
\Delta B^{i}=\frac{1}{c^{2}}\frac{\partial^{2}B^{i}}{\partial t^{2}}.
\]

One concludes that the electric and magnetic fields satisfy wave equations and propagate at velocity $c\approx 3\times 10^8 \: \mathrm{m/s}$. Depending on their frequency, electromagnetic waves are called by different names. 
The following table describes the type of wave that corresponds to a frequency measured in Herz $(1 \:\mathrm{Hz}=1\: \mathrm{s}^{-1})$.
\begin{center}
\begin{tabular}{ |c|c|c| } 
 \hline
 {\textbf Frequency (Hz)}& {\textbf Type of wave} \\ 
\hline $\sim 30000 $& radio waves \\ 
\hline $\sim 3 \times 10^8 $ & microwaves  \\ 
 \hline $\sim 3 \times 10^{12}$ & infrared \\
 \hline $\sim 3 \times 10^{14}$ & visible light\\
 \hline $\sim 3\times 10^{15}$& ultraviolette radiation\\
 \hline $\sim 10^{18}$& X rays\\
 \hline $\sim 10^{20} $& gamma rays\\
 \hline
\end{tabular}
\end{center}
The human eye can see a small part of the electromagnetic spectrum, roughly between $400\:\mathrm{THz}$ and $780\:  \mathrm{THz}$. We experience the different frequencies
as colors.

\begin{figure}[H]
	\centering
	\includegraphics[scale=0.45]{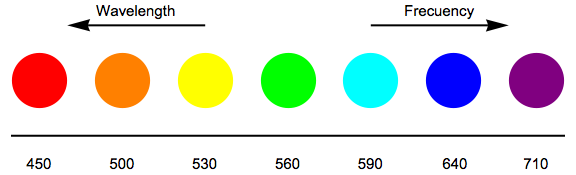}
	\caption{The frequency of different colors measured in THz.}
\end{figure}

\section{Galilean transformations and the speed of light}

Maxwell's equations imply that electromagnetic waves propagate with velocity \[c=\frac{1}{\sqrt{\varepsilon_0 \mu_0}},
\]
where $\varepsilon_0$ and $\mu_0$ are some universal constants of nature. This implies that the speed of light is independent of the reference frame, which
is in contradiction with the Galilean transformations of classical mechanics. Suppose that in a reference frame $O$, an electromagnetic wave propagates in the $x$ direction with electric and magnetic fields 
 \[E(t,x)=E_{0}\Big(0,\sin(kx-\omega t,0\Big), \quad B(t,x)=B_{0}
\Big(0,0,\sin(kx-\omega t)\Big).\] 
In a reference frame $\overline{O}$ which is moving in the $x$ direction with constant velocity $v$ with respect to $O$, the position is
\[ \overline{x}=x-vt.\]
Therefore, in the reference frame $\overline{O}$, the electric field $E$ is 
\[ E(t,\overline{x})=E_{0}\Big(0,\sin(k\overline{x}-(\omega-kv)t),0\Big).\]
From the point of view of $\overline{O}$, the electric field propagates with velocity
\[\frac{ \omega- kv}{k}=c+v,\]
which contradicts Maxwell's equations.  
The compatibility between Maxwell's equations and Galilean transformations could be restored by assuming the Maxwell's equations hold only with respect
to some preferred reference frame, that of the {\textit  ether}. The {\textit luminiferous aether}, whose existence was postulated by Robert Boyle in the 17th century, was supposed to be the medium in which light waves propagate. In 1887, A. Michelson and E. Morley performed an experiment that failed to detect the existence of the ether. A more radical change was necessary to make Maxwell's equations valid for different observers. The assumption that there is a universal time for all events in the universe had to be removed, and a more symmetric relation between space and time was discovered. In 1905, Einstein published the new theory for the electrodynamics of moving bodies. This theory is known as special relativity.
   \clearemptydoublepage 
\chapter{Special Relativity}

\begin{center}
\parbox[b]{0.9\textwidth}{\small \sl The equations for classical electrodynamics lead to very different interpretations depending on the reference frame used in the analysis.
A charged particle moving with constant velocity in presence of a magnetic field will experience a magnetic force, and deviate from its trajectory. In a reference frame in which it is at rest, 
it does not experience magnetic forces. In this case, the deviation must be be caused by an electric force.  
Eintein wrote:

\begin{quote}
Thus the existence of the electric field was a relative one, according
to the state of motion of the coordinate system used, and only the electric
and magnetic field together could be ascribed a kind of objective reality,
apart from the state of motion of the observer or the coordinate system. The
phenomenon of magneto-electric induction compelled me to postulate the
special principle of relativity.
 \end{quote}
 }
\end{center}

\vspace{3ex}

\section{The Michelson-Morley experiment}

Maxwell and other  other physicists of his time were bothered by
 an interesting feature of the mathematical description of electromagnetism.
According to Maxwell's equations, the speed of an electromagnetic wave in the vacuum is
$c=1/\sqrt{\varepsilon_{0}\mu_{0}}$, depending only on $\varepsilon_{0}$ and $\mu_{0}%
$, the permittivity and permeability of empty space. Classically, this can not hold for all reference frames.
 It was suggested,
initially, that this could be explained by the existence of a natural
medium, the\emph{ luminous ether}, an unidentified substance permeating space,
a sort of fluid that would 
vibrate in the presence of an electromagnetic field. Light would propagate at a constant speed $c$ with respect to
an observer at rest in the ether, as sound propagates at 340 \textrm{m/s}
with respect to the surrounding air. Even if the nature of this mysterious
medium would be difficult to establish, the relative motion of the Earth with
respect to the ether should be observable. It would be impossible for the
Earth to stay at rest with respect to the ether all year long.
   \begin{figure}[H]
\centering
\includegraphics[scale=0.4]{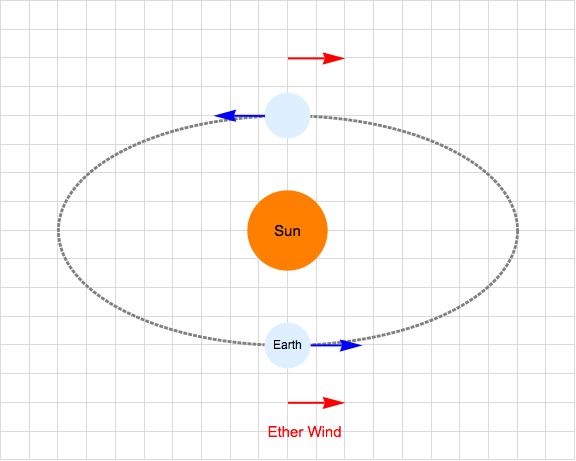}\caption{Movement with respect to ether.}%
\end{figure}
Two american
physicists, Albert A. Michelson and Edward Morley, attempted to measure the
relative speed of the Earth with respect to the ether at various points of the
Earth's orbit around the Sun. To accomplish this, Michelson and Morley
constructed an interferometer.

\begin{figure}[H]
\centering
\includegraphics[scale=0.5]{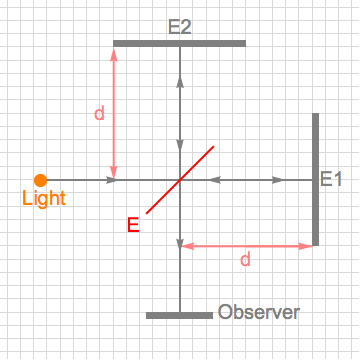}\caption{Michelson
and Morley's interferometer.}\label{MM}%
\end{figure}

The apparatus consisted of a source of light, three mirrors, and a
ocular lens. A beam of monochromatic light would split at a central
half-silvered mirror into two beams traveling at right angles to equally
distant mirrors $E_{1}$ and $E_{2}$. The light was then
reflected on each mirror, and recombined at $E,$ where it was directed
towards an observer $O$. The set up is described in Figure \ref{MM}.
If the laboratory moved with respect to the ether, an
interference pattern should appear, since the time it would take light to
travel both paths, of equal length $d$, would have to be different.

Let us consider the situation more precisely. Suppose that the ether fills empty space, and that the solar system moves through this
medium at some unknown speed $u.$ There must be
at least one point on the Earth's orbit where the velocity of our planet relative to
the ether is not zero. This is because, if $v$ is the Earth's speed in
its orbit around the Sun,  at diametrically opposed
points of the orbit the velocities relative to the ether are $u+v,$ and $u-v$. 

\begin{figure}[tbh]
\centering
\includegraphics[scale=0.5]{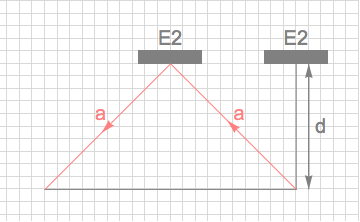}\caption{Trajectory of light in the interferometer.}\label{trainter}
\end{figure}

Consider a point  $P$ on the Earth's orbit where the Earth moves with respect to the ether with velocity $w\neq 0$.
An observer $O$ on Earth whose interferometer is oriented  opposite to the
direction of motion -the segment connecting the light source to the mirror
$E_{1}$ would point contrary to the Earth's direction of motion- could assume
that her laboratory is at rest while the ether would be moving in the opposite
direction. Then, the ether wind would
drag light coming from the source, so
that the velocity of a beam of light traveling towards $E_{1}$ would 
be $c+w$. One concludes that the time it takes a photon to move from $E$ to
$E_{1}$ would be $d/(c+w)$. Similarly, the time it takes the photon
to travel against the flow of ether 
from $E_{1}$ to $E$ would be $d/(c-w)$. Hence, the total time to go
from $E$ to $E_{1}$ and back must be:
\[
T=\frac{d}{c+w}+\frac{d}{c-w}=\frac{2cd}%
{c^{2}-w^{2}}=\frac{2d/c}{1-(w/c)^{2}}.
\]
On the other hand, we denote by $\overline{T}$ the time it takes light to go from
$E$ to $E_{2}$ and back. Let us compute $\overline{T}$ from the
stand point of an observer, Alice,  who is at rest relative to the ether.  Alice will see the light following the trajectory  described in Figure \ref{trainter}.

According to Alice, each photon should take $\overline{T}=2a/c$ to
travel back and forth. On the other hand,
\[
a^{2}=\Big(\frac{w \overline{T}}{2}\Big)^{2}+d^{2}=\frac{w^{2}\overline{T}^{2}}{4}+d^{2}.
\]
Substituting $\overline{T}=2a/c$ into this equation one obtains
$a^{2}=w^{2}a^{2}/c^{2}+d^{2}$. Solving for $a$ we see that
$a=cd/\sqrt{(c^{2}-w^{2})}.$ Thus,
\[
\overline{T}
=\frac{2d/c}{\sqrt{1-w^{2}/c^{2}}}.
\]
The time difference is
\[
T-\overline{T}=\frac{2d/c}{1-w^{2}/c^{2}}-\frac
{2d/c}{\sqrt{1-w^{2}/c^{2}}}=\frac{2d\Big(1-\sqrt{1-w^{2}/c^{2}}\Big)}{c(1-w^{2}/c^2)}.%
\]
Notice that if $w<c$ then $T>\overline{T}$, which would cause an interference pattern to appear. This
interference pattern was never observed by Michelson and Morley,
even though measurements were performed at different points diametrically opposed
along the Earth's orbit. Modern experiments like that performed by Brillet and Hall in 1978 \cite{Brillet} have
corroborated the result of Michelson-Morley with much higher precision. The results of the Michelson-Morley experiment left no option but to conclude that the speed of
light is independent of the state of motion of the observer. The incompatibility of this law with Newtonian mechanics led Einstein to postulate his Special Theory of Relativity.

\section{Lorentz tranformations}

In Newtonian mechanics it was assumed that there exists a universal time  that flows
regularly for all events in the universe. In particular, the question of whether or not
two events are simultaneous was supposed to have a well defined answer. The fact that the speed of light is the same for inertial observers,
implied by Maxwell's equations, together with the failure to detect the ether in which light should propagate, required a radical departure from this assumption.

  \begin{figure}[H]
\centering
\includegraphics[scale=0.35]{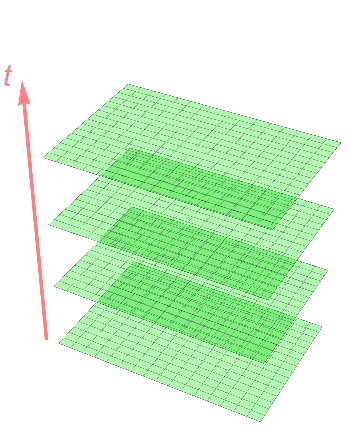}\caption{Planes of simultaneous events, as imagined in Newtonian mechanics.}%
\end{figure}

Einstein postulated the following two principles, from which he derived a new way of transforming the measurements obtained by different observers.
Suppose that  two observers  move at constant velocity with respect to each other. Without loss of generality one may assume they move along the $x$ axis of the coordinate system of each observer. Then:
\begin{itemize}
\item The speed of light in vacuum is the same for both observers.
\item The equations of physics take the same form in both systems of coordinates.
\end{itemize}
\begin{figure}[H]
\centering
\includegraphics[scale=0.5]{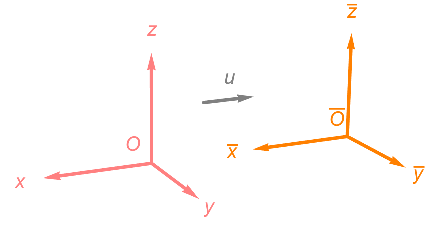}\caption{Observer moving at constant velocity}%
\end{figure}

Classically it was assumed that $O$ and $\overline{O}$ share a universal time $t$, and that Galilean transformations
\[ \overline{x}=x- tu\] 
describe the relationship between the positions that the two observers will assign to an event. This contradicts the postulate that the speed of light is the same for both observers. A different transformation rule can be derived from the Einstein's postulates. In the absence of forces, both observers
see that objects move in straight lines. Therefore, the transformation rule should send straight lines to straight lines. If we also assume that both observers
start their clocks at the same time and place, then the transformation must be linear. We write
\begin{align} \label{lambda}
\overline{x}=\lambda_u(x-ut),\\
x=\overline{\lambda}_{u}(\overline{x}+u\overline{t}) .\nonumber
\end{align}
\noindent
Since there are no preferred directions in space, reversing the direction of the $x$ axis should leave the transformation rule invariant. This implies that $\lambda_u=\overline{\lambda}_u$. Suppose that, at the moment when the clocks are started, a light pulse is emitted. Observer $O$ registers
the position of the light pulse after $t_0$ seconds to be $x_0 =ct_0$. By symmetry, also $\overline{x}_0=c\overline{t}_0$. One concludes
\[ x_0 \overline{x_0}=c^2 t_0 \overline{t}_0.\]
Using (\ref{lambda})  and the fact that $\lambda_u=\overline{\lambda}_u$ we obtain
 \[ \lambda_u^2(x_0 \overline{x}_0 -u^2 t_0 \overline{t}_0 +u x_0 \overline{t}_0 -u \overline{x}_0 t_0)=c^2 t_0 \overline{t}_0.\]
Using again the fact that both observers measure the speed of light to be $c$, we get
 \[ \lambda_u^2(t_0 \overline{t}_0c^2 -u^2 t_0 \overline{t}_0 +u ct_0 \overline{t}_0 -uc \overline{t}_0 t_0)=c^2 t_0 \overline{t}_0.\]
Dividing on both sides by $t_0 \overline{t}_0$, this is
 \[ \lambda_u^2(c^2 -u^2 )=c^2,\]
 which implies
\begin{equation} \lambda_u=\frac{1}{\sqrt{1-u^2/c^2}}.
 \end{equation}
 One can then solve for $\overline{t}$ in (\ref{lambda}) to obtain
 \[ \overline{t}= \lambda_u\left(t-\frac{ux}{c^2}\right).\]
 
 In conclusion, the two postulates imply the following relation between the coordinates of both observers
 \begin{align}\label{Lorentz1}
 \begin{split}
   \overline{t}&=\frac{t-ux/c^2}{\sqrt{1-u^2/c^2}}, \\
   \overline{x}&= \frac{ x -u t}{\sqrt{1-u^2/c^2}}, \\
   \overline{y}&=y, \\
   \overline{z}&=z.
 \end{split}
 \end{align} 
This rule is known as a Lorentz transformation or a Lorentz boost, and replaces the Galilean transformations of classical mechanics.  Lorentz transformations can also be written in the form
\[
\begin{pmatrix}
c\overline{t}\\
\overline{x}
\end{pmatrix}=\frac{1}{\sqrt{1-u^2/c^2}}
\begin{pmatrix}
1& -u/c\\
-u/c& 1
\end{pmatrix}
\begin{pmatrix}
ct\\
x
\end{pmatrix}.
\]
Since
\[ \lambda^2_u-(-u \lambda_u/c)^2=1,\]
one can use hyperbolic functions to write
\begin{align*}
  \cosh \phi &=\frac{1}{\sqrt{1-u^2/c^2}},  \\
  \sinh \phi &= \frac{-u/c}{\sqrt{1-u^2/c^2}}.
\end{align*}
So that the Lorentz transformation takes the form
\[
\begin{pmatrix}
c\overline{t}\\
\overline{x}
\end{pmatrix}=\begin{pmatrix}
\cosh \phi&-\sinh \phi\\
-\sinh \phi& \cosh \phi
\end{pmatrix}
\begin{pmatrix}
ct\\
x
\end{pmatrix}.
\]
\noindent
The formula above shows that Lorentz transformations are hyperbolic rotations. In Euclidean geometry, a rotation moves a point in the plane along a circle.
A hyperbolic rotation slides the points in the plane along a hyperbola.

\begin{figure}[H]
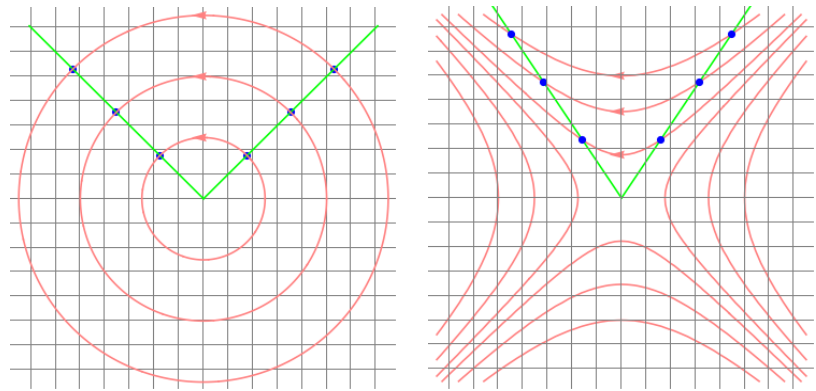

\centering
\begin{tabular}{cc}
  \vspace{0pt} \includegraphics[scale=0.4]{Figures/erot} &
  \vspace{0pt} \includegraphics[scale=0.4]{Figures/hrot}
\end{tabular}
\caption{Euclidean and hyperbolic rotations.}
\end{figure}
\remark{In order to simplify the formulas we will often measure time in new units so that the speed of light becomes $c=1$. For this one chooses as a new unit of time, the {\textit short second}, the time it takes light to travel 1 meter. This unit will be denoted by $\mathrm{ss}$.}

If $c=1$, the change of coordinates between $O$ and $\overline{O}$ can be visualized by drawing a standard Cartesian plane $(x,t)$ for $O$, and  skewed coordinates $(\overline{x}, \overline{t})$ for $\overline{O}$ as shown in Figure \ref{fig:8.7} below. 
\begin{figure}[H]
\centering
\includegraphics[scale=0.48]{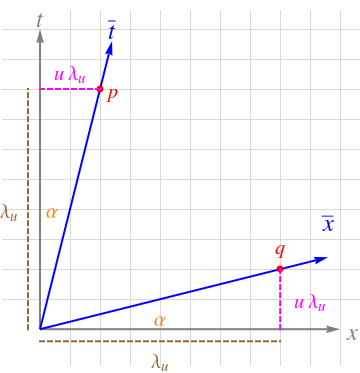}\caption{ Lorentz transformations. In the blue system of coordinates $p=(1,0)$ and $q=(0,1)$. In the gray system of coordinates $p=(\lambda_u, u\lambda_u)$ and $q=(u\lambda_u, \lambda_u)$.}
\label{fig:8.7}
\end{figure}

\begin{figure}[H]
\centering
\includegraphics[scale=0.45]{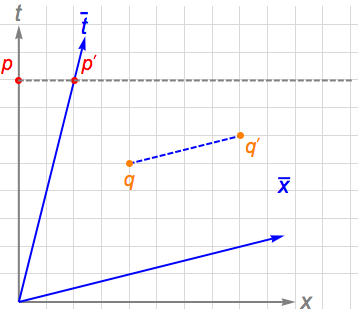}\caption{Different observers have different notions of simultaneity.}
\label{fig:8.8}
\end{figure}

In classical mechanics one insists that there are no preferred directions in space. This means that all equations should remain invariant under euclidean rotations. In special relativity there is an additional symmetry between time and space. Lorentz boosts are hyperbolic rotations that exchange space and time. This additional symmetry forces us to conclude that whether or not two events are simultaneous depends on the observer. In Figure \ref{fig:8.8}, the red points are simultaneous according to the gray observer, and the orange points are simultaneous with respect to the blue observer.

\begin{figure}[H]\label{causality}
\centering
\includegraphics[scale=0.42]{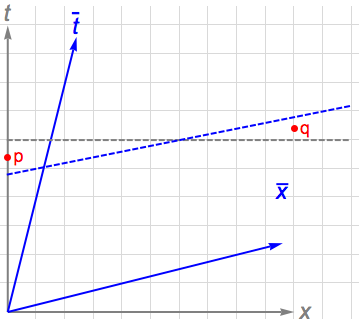}\caption{Causality}
\label{fig:8.9} 
\end{figure}

Even more dramatically, given two events, different observers may disagree on which event occurred first. In Figure \ref{fig:8.9}, the gray observer believes that  $p$ occurred before  $q$. The blue observer believes the opposite.

Most of us are used to the idea that causes should precede consequences. The fact that Alice exists is a consequence of her parents having met. In case Alice's parents met after she was born, it would be hard to imagine how their having met could have caused her existence. If the order in which events occur is 
not well defined, causality relations appear to be impossible.  In special relativity, there is a geometric condition that is necessary for two events to 
be causally related. The restriction that objects do not travel faster than the speed of light allows for a notion of causality that does not run into logical contradictions.
Imposing an absolute limit on the velocities at which objects can travel contradicts intuition and, again, Newtonian mechanics.

Suppose that a train travels
with velocity $v$ with respect to an observer on the tracks. Inside the train, a girl is running with velocity $u$ with respect to the train. In classical mechanics one assumes that the observer on the tracks will see the girl moving with velocity $u+v$. Since this process can be iterated, it is clear that there can be no limit
for the velocity that can be achieved. Let us now examine the situation relativistically. We denote by $(t,x),(\overline{t},\overline{x}),(\hat{t},\hat{x})$ the coordinates used by an observer on the tracks, the train and the girl, respectively. Since the train is moving with velocity $v$ with respect to the tracks, we know that
\begin{equation}\label{tt}
\begin{pmatrix}
c\overline{t}\\
\overline{x}
\end{pmatrix}=\begin{pmatrix}
\cosh \phi&-\sinh \phi\\
-\sinh\phi& \cosh\phi
\end{pmatrix}
\begin{pmatrix}
ct\\
x
\end{pmatrix},
\end{equation}
where
\begin{align*}
  \sinh\phi &=\frac{-v/c}{\sqrt{1-v^2/c^2}}, \\
  \cosh \phi &=\frac{1}{\sqrt{1-v^2/c^2}}.
\end{align*}
Similarly, since the girl is moving with velocity $u$ with respect to the train, we know that
\begin{equation}\label{gt}
\begin{pmatrix}
c\widehat{t}\\
\widehat{x}
\end{pmatrix}=\begin{pmatrix}
\cosh \psi&-\sinh \psi\\
-\sinh \psi& \cosh \psi
\end{pmatrix}
\begin{pmatrix}
c\overline{t}\\
\overline{x}
\end{pmatrix},
\end{equation}
where
\begin{align*}
 \sinh\psi &=\frac{-u/c}{\sqrt{1-u^2/c^2}}, \\
 \cosh\psi &=\frac{1}{\sqrt{1-u^2/c^2}}.
\end{align*}
 The hyperbolic functions satisfy the following identities for the sum of angles
\begin{align*}
\cosh(\phi + \psi)= \cosh\phi \cosh \psi+ \sinh\phi\sinh\psi,\\
\sinh(\phi + \psi)= \cosh\phi \sinh\psi+ \cosh\psi\sinh\phi.
\end{align*}
Equations (\ref{tt}) and (\ref{gt}) together with these imply
\begin{equation}\label{gtr}
\begin{pmatrix}
c\widehat{t}\\
\widehat{x}
\end{pmatrix}=\begin{pmatrix}
\cosh(\phi+\psi)&-\sinh(\phi+\psi)\\
-\sinh(\phi+\psi)& \cosh(\phi+\psi)
\end{pmatrix}
\begin{pmatrix}
ct\\
x
\end{pmatrix}.
\end{equation}
\noindent
If the girl travels with velocity $w$ with respect to the tracks, one should also have:
\begin{equation}\label{gtr}
\begin{pmatrix}
c\widehat{t}\\
\widehat{x}
\end{pmatrix}=\begin{pmatrix}
\cosh\theta&-\sinh\theta\\
-\sinh\theta& \cosh\theta
\end{pmatrix}
\begin{pmatrix}
ct\\
x
\end{pmatrix},
\end{equation}
where
\begin{align*}
 \sinh\theta &=\frac{-w/c}{\sqrt{1-w^2/c^2}}, \\
 \cosh\theta &=\frac{1}{\sqrt{1-w^2/c^2}}.
\end{align*}
Therefore, consistency requires that
\begin{align*}
 \cosh\phi \cosh\psi+ \sinh\phi\sinh\psi=\cosh\theta,\\
 \cosh\phi \sinh\psi+ \cosh\psi\sinh\phi=\sinh\theta.
\end{align*}
This is equivalent to
\begin{eqnarray}
\frac{1+uv/c^2}{\sqrt{(1-u^2/c^2)(1-v^2/c^2)}}=\frac{1}{\sqrt{1-w^2/c^2}},\\
\frac{-(u+v)/c}{\sqrt{(1-u^2/c^2)(1-v^2/c^2)}}=\frac{-w/c}{\sqrt{1-w^2/c^2}}.
\end{eqnarray}
\noindent
These relations are satisfied precisely when
\begin{equation}\label{lorentzvel}
w= \frac{v+u}{1+uv/c^2}.
\end{equation}
The observer on the tracks sees the girl moving with velocity $w$, which is not the sum of $u$ and $v$. One can verify that, as long as $v$ and $u$ don't exceed the speed of light, neither does $w$. Suppose for example that
$u=v=2c/3$. Classically, the observer on the tracks would see the girl moving at speed $4c/3>c$. Relativistically, the girl is seen traveling
with velocity
\[ w=\frac{12c}{13}<c.\]

 \begin{figure}[H]\label{causality}
\centering
\includegraphics[scale=0.45]{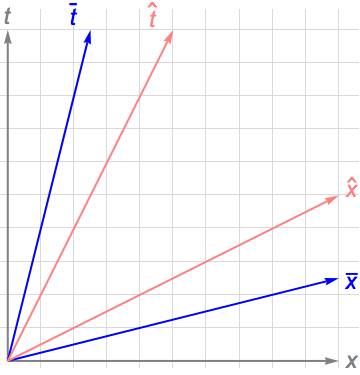}\caption{A red girl running in a blue train. The gray coordinates correspond to an observer on the tracks. The blue ones, to one sitting inside the train. The red ones are those of a girl running inside.}%
\end{figure}
Let us now consider a lantern that is turned on inside the train. A passanger in the train will see the light
traveling with velocity $c$. According to (\ref{lorentzvel}), the observer on the tracks will see the light with velocity
\begin{equation*}
w= \frac{v+c}{1+cv/c^2}= \frac{v+c}{1+v/c}=\frac{c(v+c)}{c+v}=c.
\end{equation*}
In accordance with Einstein's postulates, both observers see the light traveling with the same speed.

\section{Minkowski spacetime}
Minkowski spacetime, denoted by $\MM$, is the Lorentzian manifold $\RR^4$ with the metric which, in  coordinates $(t,x,y,z)$, where we assume $c=1$, takes the form
\[g(t,x,y,z)=\begin{pmatrix}
-1&0&0&0\\
0&1&0&0\\
0&0&1&0\\
0&0&0&1
\end{pmatrix}.\]
Special relativity can be naturally formulated in terms of the geometry of Minkowski spacetime. We know that two observers that move with relative velocity $v$ in the direction of $x$ have systems of coordinates related by a Lorentz boost:
\begin{equation}\label{lb4}
\begin{pmatrix}
\overline{t}\\
\overline{x}\\
\overline{y}\\
\overline{z}
\end{pmatrix}=\begin{pmatrix}
\cosh\phi&-\sinh\phi&0&0\\
-\sinh\phi& \cosh\phi&0&0\\
0&0&1&0\\
0&0&0&1
\end{pmatrix}
\begin{pmatrix}
t\\
x\\
y\\
z
\end{pmatrix},
\end{equation}
where
\begin{align*}
 \sinh\phi &=\frac{-v}{\sqrt{1-v^2}}, \\
 \cosh\phi &=\frac{1}{\sqrt{1-v^2}}.
\end{align*}
The first hint of the relationship between the geometry of Minkowski spacetime and special relativity is the fact that Lorentz boosts are 
isometries of $\MM$.  The condition for a linear isomorphism $w \mapsto Lw$ to be an isometry of Minkowski spacetime is that:
\[ \langle Lv, Lw \rangle =\langle v,w \rangle\]
for all $v,w \in \RR^4=T_0\MM$. This condition is equivalent to
\begin{equation}\label{isom}
L^{\mathrm{T}} g L =g.
\end{equation}
If $L$ is the Lorentz boost in (\ref{lb4}) we compute:
\begin{align*} \begin{pmatrix}
\cosh\phi&-\sinh\phi&0&0\\
-\sinh\phi& \cosh\phi&0&0\\
0&0&1&0\\
0&0&0&1
\end{pmatrix}\begin{pmatrix}
-1&0&0&0\\
0&1&0&0\\
0&0&1&0\\
0&0&0&1
\end{pmatrix}&\begin{pmatrix}
\cosh\phi&-\sinh\phi&0&0\\
-\sinh\phi& \cosh\phi&0&0\\
0&0&1&0\\
0&0&0&1
\end{pmatrix} \\
&\qquad\qquad\qquad= \begin{pmatrix}
-1&0&0&0\\
0&1&0&0\\
0&0&1&0\\
0&0&0&1
\end{pmatrix}.
\end{align*}
One concludes that $L$ is an isometry of $\MM$. More generally,
for a fixed vector $v=(v_1,v_2,v_3) \in \RR^3$ such that $|v|<1$, there is a Lorentz boost in the direction of $v$
\[ L_v=\begin{pmatrix}
\lambda_v&-\lambda_v v_1& -\lambda_v v_2&-\lambda_v v_3\\
-\lambda_v v_1& 1+\frac{(\lambda_v -1)v^2_1}{|v|^2}& \frac{(\lambda_v-1)v_1v_2}{|v|^2}& \frac{(\lambda_v-1)v_1v_3}{|v|^2}\\
-\lambda_v v_2& \frac{(\lambda_v -1)v_1v_2}{|v|^2}&1+ \frac{(\lambda_v-1)v^2_2}{|v|^2}& \frac{(\lambda_v-1)v_2v_3}{|v|^2}\\
-\lambda_v v_3& \frac{(\lambda_v-1)v_1v_3}{|v|^2}&\frac{(\lambda_v-1)v_2v_3}{|v|^2}&1+ \frac{(\lambda_v-1)v^2_3}{|v|^2}\\
\end{pmatrix}\]
The general Lorentz boost is also an isometry of $\MM$. It describes the relationship between coordinate systems of observers that
move with relative velocity $v$. Another type of isometry of $\MM$ is a translation by a constant vector $w \mapsto w+ a$,
where $a \in \RR^4$ is a constant vector. The derivative of this map at any point is the identity, which clearly satisfies condition (\ref{isom}).
Translations relate the coordinates of observers that are at rest with respect to each other.

Given an isometry of ordinary euclidian $3$ dimensional space $A \in O(3)$, there is an isometry of $\MM$ given by:
\[ L= \begin{pmatrix}
1&0\\
0& A
\end{pmatrix}.\]
These space rotations relate the coordinates of observers that put their coordinate axes in different directions.
The Lorentz group, denoted $O(3,1)$, is the group of linear isometries of $\MM$. The Lorentz group is a Lie group of dimension $d=6$. It turns out that all isometries of Minkowski spacetime are the composition of a linear isometry and a translation.

\begin{remark}
The Poincar\'e group is the group of isometries $P\colon \MM \rightarrow \MM$ of the form:
\[ Pw=Lw+ a,\]
where $L \in O(3,1)$ and $a$ is a constant vector. One can show that the Poincar\'e group is the group of isometries of Minkowski spacetime. From this, it follows that the Poincar\'e group is a Lie group of dimension $d=10$.
\end{remark}

At every point $p$ of Minkowski spacetime the tangent space $T_p\MM$ is naturally identified with $\RR^4$. The Minkowski metric breaks the symmetry in this vector space. Not all vectors have the same properties. A vector $v \in T_p \MM \simeq \RR^4$ is:
\begin{itemize}
\item \emph{timelike} if $\langle v, v \rangle <0$;
\item \emph{spacelike} if $ \langle v, v \rangle >0$;
\item \emph{lightlight} if $\langle v, v \rangle =0$.
\end{itemize}
We say that a timelike vector $v=v_t\partial_t+ v_x\partial_{x}+v_y \partial_y +v_z \partial_z$ \textit{points to the future} if $v_t>0$; it \emph{points to the past} if $v_t<0$. Lightlike vectors form two cones, pointing to the past and the future, respectively. 
The fact that we have a global system of coordinates for Minkoswki spacetime allows us to consistently define at all points a notion of future cone and past cone. In curved spacetime this is not necessarily the case, and it is sometimes required as a condition on spacetime.  
\begin{figure}[h]
\centering
\includegraphics[scale=0.5]{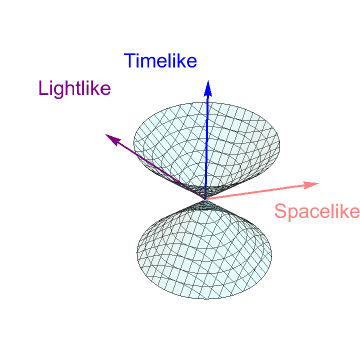}\caption{Causal
cones}%
\end{figure}

\section{Motion of particles and observers in Minkowski Spacetime}
\label{observa}
An even in special relativity is a point in Minkowski spacetime.
The series of events that encompasses the whole existence of a material particle $P$ can be described by the image of certain curve in Minkowski spacetime $\gamma:I\rightarrow\MM%
^{4}$, called its \textit{worldline}.
Material particles cannot travel at a speed higher than the speed of light. This is reflected in the fact that $\gamma^{\prime}(s)$ must be a
timelike vector. We must also demand that $P$ moves towards its future, so that $t(s)$ is an increasing function. Write $\gamma$ in the standard coordinates of
$\RR^{4}$, $\gamma(s)=(t(s),x^{i}(s))$, where we shall use $t(s)$ to
represent the time coordinate of $\gamma,$ and Latin superindices $x^{i}$, $i=1,2,3,$ to denote the spatial coordinates of $\gamma$.

Let us assume units where $c=1$. We can verify that the condition that the curve is timelike means that the observer moves at a speed less than $1$. In fact:
\[
\frac{dt}{ds}>\sqrt{%
{\sum_{i}}
\left(  \frac{dx^{i}}{ds}\right)  ^{2}}.
\]
This implies that the magnitude of its \emph{3-velocity}, $u(t)$, is less than $1$. That is:
\begin{align*}
\left\vert u(t)\right\vert  &  =\sqrt{%
{\sum_{i}}
\left(  \frac{dx^{i}}{dt}\right)  ^{2}}=\sqrt{%
{\sum_{i}}
\left(  \frac{dx^{i}}{ds}\frac{ds}{dt}\right)  ^{2}}\\
&  =\frac{ds}{dt}\sqrt{%
{\sum_{i}}
\left(  \frac{dx^{i}}{ds}\right)  ^{2}}< 1.
\end{align*}
A particle moving in a timelike curve in Minkowski spacetime can also be regarded as an \textit{observer}.

\begin{definition}
By an \emph{observer} in Minkowski spacetime we mean any \emph{timelike} curve that can be
written in standard coordinates as
$\gamma (s)=(t(s),x^{i}(s)),$ 
with $t^{\prime}(s)>0.$ The unitary tangent vector $\mathbf{u}=\gamma^{\prime}(s_{0})/\left\vert \gamma^{\prime}(s_{0})\right\vert $ is
called the observer's $4$-\emph{velocity } at $p=\gamma(s_{0}).$
\end{definition}
Whenever we want to emphasize that a particular curve $\gamma(s)$ represents the worldline of an observer, we will denote his worldline curve as $O(s)$, $\overline{O}(s)$, $O'(s)$, etc.

\begin{definition}
The time experienced by an observer $O=\gamma(s)$ as she goes from event $p=\gamma(a)$ to event $q=\gamma(b)$, called her \textit{proper time}, is given by
\[ \tau(a,b)= \int \limits_a^b\sqrt{-\langle \gamma'(s), \gamma'(s)\rangle} ds.\]
We will say that $\gamma(s)$ is parametrized by proper time if $ \langle \gamma'(s),\gamma'(s)\rangle=-1$. This is equivelent to the condition
$\tau(a,s)=s-a$.
\end{definition}
The proper times is the time a clock moving with $O$ will measure.
Notice that the proper time is independent of the parametrization of the worldline. Indeed, if $\varphi\colon [\overline{a},\overline{b}]\rightarrow [a,b]$ is an orientation preserving diffeomorphism and we set $\lambda = \gamma \circ \varphi$ then
 \begin{align*}
  \int \limits_{\overline{a}}^{\overline{b}}\sqrt{-\langle \lambda'(l), \lambda'(l)\rangle} dl &= \int \limits_{\overline{a}}^{\overline{b}}\varphi'(l)\sqrt{-\langle \gamma'(\varphi (l)), \gamma'(\varphi(l))\rangle} dl\\
   &=\int \limits_a^b\sqrt{-\langle \gamma'(s), \gamma'(s)\rangle} ds. 
  \end{align*}
  
 \begin{remark}
 Given a worldline $\gamma(s):I \to \MM$, there exists a unique orientation preserving diffeomorphism $\varphi: [0,T] \to I$ such that
 $\gamma(\varphi(\tau))$ is parametrized by proper time.
 The inverse diffeomorphism $\varphi^{-1}:I \to [0,T]$ is defined by setting:
 \[ \tau(s)= \int \limits_0^s\sqrt{-\langle \gamma'(s), \gamma'(s)\rangle} ds.\]
 \end{remark}
 
It is customary to denote by $\tau$ the \emph{proper time parameter}. Instead of a universal time that runs uniformly for all observers, special relativity postulates that
the time that a clock measures depends on the worldline of the clock. Different worldlines going from $p$ to $q$  will measure different times.

\begin{figure}[H]
\centering
\includegraphics[scale=0.4]{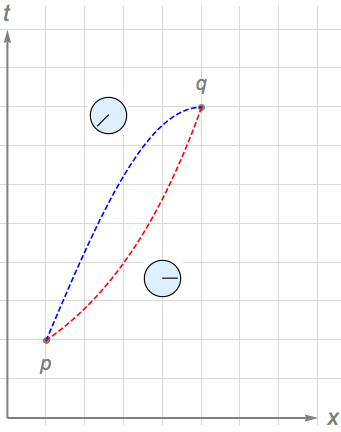}\caption{Proper time depends on the length of a curve.}%
\end{figure}

The chronological future of an event $p$ is the set of events that can be reached from $p$ following a curve $\gamma(s)$ whose derivative lies in the future cone.
The chronological future is composed of those events 
which can be affected by the event $p$. It is important that, if $q$ is in the chronological future of $p$, then $p$ is not in the chronological future of $q$. Alice's birth  is in the chronological future of the party where her parents met. However, this party is not in the chronological future of Alice's birth. There is nothing Alice can do to prevent her parents from meeting.

  \begin{figure}[H]
\centering
\includegraphics[scale=0.45]{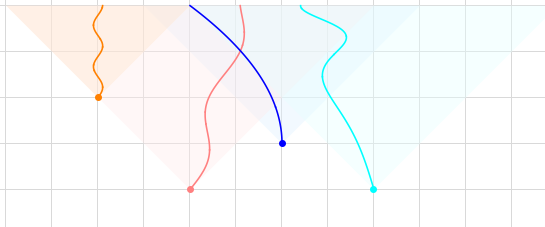}\caption{Worldlines stay in the causal future.}%
\end{figure}
We say that $O(s)$ is an \emph{inertial observer} if the curve $O(s)$ is a straight line, when we write it in the standard coordinates of $\MM$. At each fixed point of his worldline $p=\gamma(\tau_0)$, each inertial observer can choose a system of coordinates  determined by its 4-velocity $u_0(\tau_0)=\mathbf{u}(\tau_0)$ and by any collection of spatial vectors $u_i(\tau_0)$ so that the set $\{u_a(\tau_0)\mid a=0,1,2,3\}$ is an orthonormal frame, also known as a \emph{Lorentz frame} at $p$.

For instance, the standard coordinates of Minkowski space time $(t,x^i)$ correspond to the system of coordinates for the observer $O(s)=(s,0,0,0).$ The observer $\overline O$ that moves a constant velocity $v$ in the direction of the $x$-axis of $O$ is represented by the curve $ \overline O(s)=(s,vs,0,0).$ A frame at $p=O(s)$ is given by expressing each point of $\RR^4$ as a linear combination of its 4-velocity $\mathbf{u}(s)= \lambda_{u} \partial_t+u \lambda_{u}\partial_x$ and the orthogonal spatial vectors $u_{1}(s)=u \lambda_{u}\partial_t+\lambda_{u}\partial_x$, $u_{2}=\partial_y$, $u_{3}=\partial_z$. This amounts to a change of bases given by the matrix \[
L=\left(
\begin{array}
[c]{cccc}%
\lambda_{u} & \lambda_{u}u & 0 & 0\\
\lambda_{u}u & \lambda_{u} & 0 & 0\\
0 & 0 & 1 & 0\\
0 & 0 & 0 & 1
\end{array}
\right) ,
\] 
with $0<u<1.$ The change of coordinates
associated to $A$ will then be the Lorentz boost discussed before, that is:
\begin{align}\label{Lorentz 1.1}
\begin{split}
\overline{t}   &=\lambda_{u}(t-ux)\\
\overline{x}&=\lambda_{v}(x-ut),\\
\overline{y}  &  =y,\\
\overline{z}&=z.
\end{split}
\end{align}

\section{Twins}

As we have already mentioned, the time that a clock measures as it travels from an event $p$ to an event $q$ depends on its trajectory in spacetime. One learns in Euclidean geometry that the shortest path between two points is a straight line. In special relativity, straight lines maximize proper time. Suppose that an event $q$ is in the causal future of $p$, so that the straight path $\gamma(s)=p +s(q-p)$ is timelike. Among all the possible worldlines going from $p$ to $q$, the straight path has the longest proper time $T_\gamma$. Since $q$ is in the causal future of $p$, one can make a Lorentz transformation so that in the new coordinates \begin{align*}
  p&=(t_0,x_0,y_0,z_0),  \\
  q&=(t_0+ r,x_0,y_0,z_0).
\end{align*}
Any worldline from $p$ to $q$ can be parametrized in the form:
\[ \beta(s)=\gamma(s)+ \alpha(s),\]
where $\alpha(0)=\alpha(1)=0$, and $\alpha'(s)$ is spacelike and orthogonal to $\gamma'(s)$. The proper time of $\beta(s)$ is 
\begin{align*}
T_\beta =\int_0^1 \sqrt{-\langle \gamma'(s),\gamma'(s)\rangle- \langle \alpha'(s),\alpha'(s)\rangle}ds\leq \int_0^1 \sqrt{-\langle \gamma'(s),\gamma'(s)\rangle}ds=T_\gamma.
 \end{align*}
 The conclusion is that, of all possible ways of going to an event in one's causal future, straight lines take the longest time.  Straight lines in Minkowski spacetime
 can be characterized geometrically without any reference to the coordinates. They are precisely the geodesics in $\MM$. This observation becomes relevant in General Relativity, where spacetime is curved so there are no straight lines, but there still are geodesics.
 
\begin{figure}[H]
\centering
\includegraphics[scale=0.45]{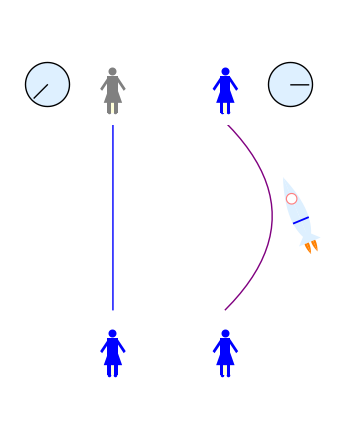}
 \caption{One twin ages faster than the other. }%
\end{figure}

 Suppose that two twins, Alice and Beth, are traveling together at constant speed. Alice boards a rocket, accelerates to a distant planet and comes back 
to meet Beth when Beth's clock has measured 20 years. Beth has stayed on a geodesics trajectory in Minkowski spacetime. Alice has not. Therefore, Alice's proper time is shorter than that of Beth.  Alice will appear younger than Beth, since less time has passed for her.

   \begin{figure}[H]
\centering
\includegraphics[scale=0.35]{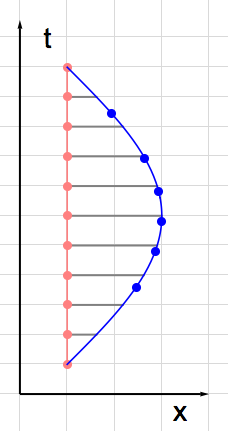}\caption{The dots mark time for Beth and Alice.}\label{color}%
\end{figure}

 In Figure \ref{color} the red dots mark equal time intervals according to Beth. The blue dots mark equal time intervals according to Alice.
 \section{Time travel and causality}

We all travel in time at a rate of 1 second per second towards the future. This is true even in classical mechanics.
In special relativity, other kinds of time travel are possible, but not everything is allowed. 
\subsection*{Going to the future fast}
\begin{figure}[H]
\centering
\includegraphics[scale=0.4]{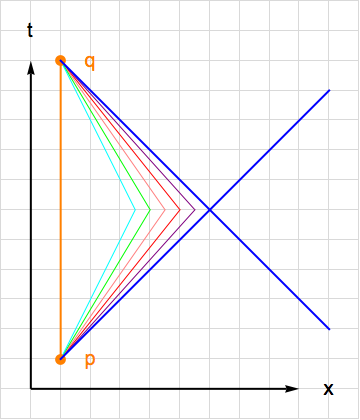}\caption{Proper time can be made arbitrarily short.}\label{future}%
\end{figure}

The condition that she cannot travel faster than the speed of light
restricts the events towards which Alice can travel. Alice can only hope to travel to events in her chronological future. Suppose that $p$ is the event that represents Alice's birth, and $q$ is an event in the chronological future of $p$. As we discussed before, the straight line from $p$ to $q$ is the worldline that takes the longest possible time. On the other hand,  by traveling at speeds close to that of light, Alice can make the time from $p$ to $q$ arbitrarily small. This means that, in principle,
if Alice is interested in what happens to the Earth one million years from now, by traveling very far away at high speed and coming back, she could find out.
This means that not only is it possible to travel to the future at 1 second per second, as in classical mechanics, it is possible to travel to the future arbitrarily fast.
In Figure \ref{future} the orange worldline maximizes the time from $p$ to $q$. The other trajectories make the time arbitrarily short.
There is an asymmetry regarding travel to the future. Given an event $q$ in Alice's chronological future, she can travel to it in an arbitrarily short amount of time.
However, there is an upper bound on the time that she can spend traveling to $q$. This means that, while it is possible to find out what will happen to the Earth in a Million years, it is not possible to factor a very large number and show the answer to someone on Earth \textit{tomorrow}.

\section*{No way to the past}

The possibility of traveling to the past leads to all sorts of logical contradictions. If Alice travelled to the past and prevented her parents from meeting, then
she would not have been born, so she could not have travelled to the past, so her parents would have met, and she would have been born, and would have travelled...  It seems better to avoid this situation. This is dealt with in special relativity by defining Alice's causal future to be those events that she can reach by traveling more slowly than light. If you define the future to be the events that Alice can go to, then obviously she cannot go anywhere but to the future.

\begin{figure}[H]
\centering
\includegraphics[scale=0.5]{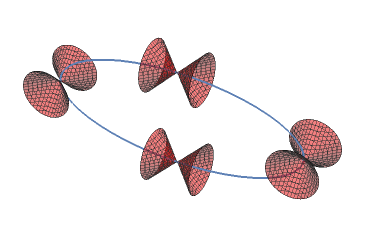}\caption{Causality violation}%
\end{figure}

The situation is more subtle than that. It is true that, tautologically, Alice cannot go anywhere but \emph{ her} future. However, Bob, who is moving with respect to Alice,
has a different way of ordering events in time. As the following diagram shows, there are events $p$ and $q$ which happen in different orders for Alice and Bob.

\begin{figure}[H]
\centering
\includegraphics[scale=0.42]{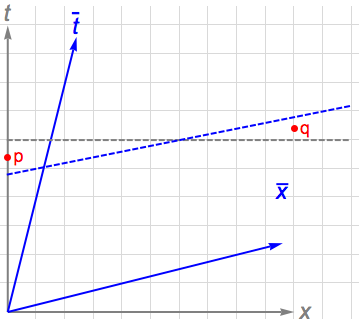}\caption{Time ordering depends on the observer.}%
\end{figure}

This raises a natural question. Suppose that $p$ represents Alice's birth, and $q$ is an event in the chronological future of $p$.  Is it possible that Bob judges $q$ to have occurred before $p$? If so, then, from Bob's point of view, Alice would be able to travel to the past.
 Since $q$ is in the chronological future of $p$, there is a Lorentz transformation such that, in the new coordinates
 \begin{align*}
  p&=(t_0,x_0,y_0,z_0), \\
  q&=(t_0+ r,x_0,y_0,z_0).
 \end{align*}
The following diagram shows lines of simultaneity for the different speeds at which Bob may be traveling. No matter what speed Bob is traveling at, he will also judge 
$p$ to have occurred before $q$.
\begin{figure}[H]
\centering
\includegraphics[scale=0.45]{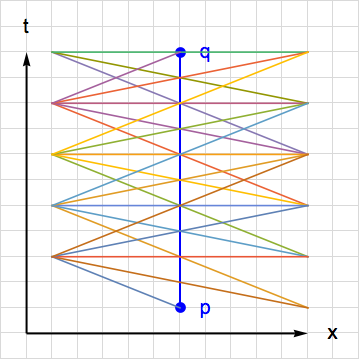}\caption{The event $p$ occurs before $q$ for all observers.}%
\end{figure}
Conveniently, Alice is not able to travel to the past. Not even from Bob's point of view.

\section{Length contraction}
According to Special Relativity, an observer at rest will perceive the length of a moving object as being shorter than the length measured in the object's reference frame.
This phenomenon is known as \textit{Lorentz contraction}. 
Let us now analyze the way in which the two observers $O$ and $\overline{O}$
measure distances. Suppose a bar moves along with $\overline{O}$ at constant velocity $u$. Two flashing lights are set at both ends of the bar, and they are
synchronized in such a way that they keep flashing \emph{simultaneously}, according to $O$. The world line of the bar is shown in Figure \ref{fig:8.20} below.

\begin{figure}[H]
\centering
\includegraphics[scale=0.42]{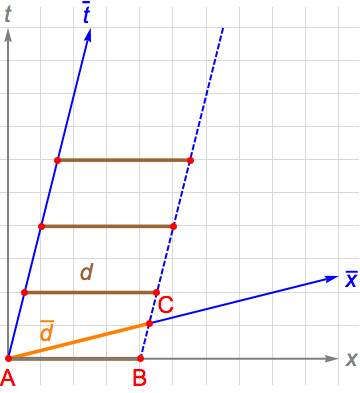}\caption{Length contraction. The length of the ruler is shorter for the observer that sees it moving. }%
\label{fig:8.20}
\end{figure}

Suppose that the length of the bar as measured by $O$ is $d=\Delta x$, the difference of the $x$ coordinates corresponding to events $A$ and
$B$, in $O$'s coordinates. See figure \ref{fig:8.20}. Since these two events are simultaneous from $O$'s perspective, the length of the bar is $d$.
Notice that these same events have  coordinates $(0,0)$ and $(-\lambda_u du,$ $\lambda_ud)$ in  $\overline{x}$-coordinates, and therefore are not simultaneous. 

For $\overline{O}$, on the other hand, the length of the bar would
be $\overline{d}=\Delta\overline{x}$, the
$\overline{x}$-coordinate difference  between events $A$ and $C$, which in $\overline{O}$'s system of coordinates are simultaneous:  
$A=(0,0)$ and $C=(0,\lambda_u d).$ Hence, $\lambda_ud=\overline{d},$ and consequently 
$$d=\overline{d} \sqrt{1-u^{2}}
 <\overline{d}.
$$ 
One concludes that $O$ measures a shorter length
for the bar as compared with the measurements performed by $\overline{O}$.

\section*{An example: A train in a tunnel}

\begin{figure}[H]
\centering
\includegraphics[scale=0.42]{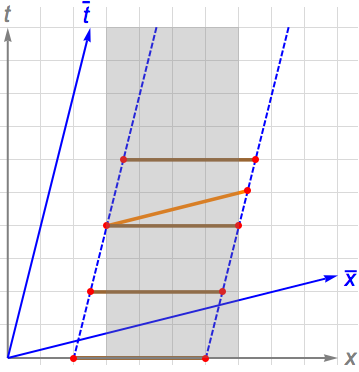}\caption{Does a moving train fit in a tunnel?}\label{train}%
\end{figure}

Imagine a train that passes through a tunnel. Oscar, an observer on the tracks, sees that the train fits precisely in the tunnel, so that there is one moment in which the whole train is inside the tunnel. Since the train is moving with respect to Oscar, its length will appear contracted. John, an observer traveling in the train will
judge the train to be longer. John  will believe that at no moment is the train completely contained in the tunnel. The apparent contradiction arises from the implicit assumption that whether or not two events are simultaneous is independent of the observer. Let us consider precisely what it means to say that the train
fits exactly in the tunnel. This sentence means that the event $p$, when the front leaves the tunnel, is simultaneous with the event $q$, when the back enters the tunnel. This precise formulation makes it clear that the statement Oscar makes is one about simultaneity. Since simultaneity is dependent on the observer, John and Oscar reach different conclusions. In Figure \ref{train}, the dotted lines represent the worldlines of the back and front ends of the train. The gray region represents the tunnel.

\section{Velocities under Lorentz Transformations}

In this section we want to generalize Formula (\ref{lorentzvel}). Suppose $O$ and $\overline{O}$ are two observers in Minkowski spacetime
where $\overline{O}$ moves in the direction of the $x$-axis of $O$ at constant
speed $u$. Their coordinates $(t,x^{i})$ and $(\overline{t}%
,\overline{x}^{i})$, respectively, are related by the matrix equation
\begin{equation}
\left(
\begin{array}
[c]{cccc}%
\lambda_u & \lambda_uu & 0 & 0\\
\lambda_uu & \lambda_u & 0 & 0\\
0 & 0 & 1 & 0\\
0 & 0 & 0 & 1
\end{array}
\right)  \left(
\begin{array}
[c]{c}%
\overline{t}\\
\overline{x}^{1}\\
\overline{x}^{2}\\
\overline{x}^{3}%
\end{array}
\right)  =\left(
\begin{array}
[c]{c}%
t\\
x^{1}\\
x^{2}\\
x^{3}%
\end{array}
\right)  \label{rr}%
\end{equation}
where $\lambda_u=1/\sqrt{1-u^{2}}$. Let $\alpha:I\rightarrow\RR^{4}$ be a timelike or a null curve that
describes the world line of a particle $P$. Write
\[
\alpha^{0}(s)=t(\alpha(s)),\text{ }\alpha^{i}(s)=x^{i}(\alpha(s))
\]
in $O$'s frame of reference, and
\[
\overline{\alpha}^{0}(s)=\overline{t}(a(s)),\text{ }\overline{\alpha}%
^{i}(s)=\overline{x}^{i}(\alpha(s))
\]
in $\overline{O}$'s frame. Hence, the 3-velocity of $P$ at $s=s_{0},$ as
measured by $O,$ is given by $v=%
{\textstyle\sum_{i}}
v^{i}\partial_{x^{i}}$, where
\[
v^{i}=\left.\frac{  d\alpha^{i}}{d\alpha^{0}}\right\vert _{s=s_{0}}=\left.\frac{
d\alpha^{i}/ds}{d\alpha^{0}/ds}\right\vert _{s=s_{0}}.
\]
Similarly, the 3-velocity of $P$ measured by $\overline{O}$ would be
$\overline{v}=%
{\textstyle\sum_{i}}
\overline{v}^{i}\partial_{\overline{x}^{i}}$, with 
$$
\overline{v}%
^{i}=\left.  \frac{d\overline{\alpha}^{i}}{d\overline{\alpha}^{0}}\right\vert
_{s=s_{0}}=\left.  \frac{d\overline{\alpha}^{i}/ds}{d\overline{\alpha}%
^{0}/ds}\right\vert _{s=s_{0}}.
$$
On the other hand, equation (\ref{rr}) says
that
\[
\left(
\begin{array}
[c]{cccc}%
\lambda_u & \lambda_uu & 0 & 0\\
\lambda_uu & \lambda_u & 0 & 0\\
0 & 0 & 1 & 0\\
0 & 0 & 0 & 1
\end{array}
\right)  \left(
\begin{array}
[c]{c}%
d\overline{\alpha}^{0}/ds\\
d\overline{\alpha}^{1}/ds\\
d\overline{\alpha}^{2}/ds\\
d\overline{\alpha}^{3}/ds
\end{array}
\right)  =\left(
\begin{array}
[c]{c}%
d\alpha^{0}/ds\\
d\alpha^{1}/ds\\
d\alpha^{2}/ds\\
d\alpha^{3}/ds
\end{array}
\right)
\]
Henceforth,
\[
\frac{d\alpha^{0}}{ds}=\lambda_u\frac{d\overline{\alpha}^{0}}{ds}+\lambda_uu\frac{d\overline{\alpha}^{1}}{ds}.
\]
Also,
\[
\frac{d\alpha^{1}}{ds}=\lambda_u u\frac{d\overline{\alpha}^{0}}{ds}+\lambda_u\frac{d\overline{\alpha}^{1}}{ds},
\]
and consequently
\[
v^{1}=\frac{d\alpha^{1}}{d\alpha^{0}}=\frac{d\alpha^{1}/ds}{d\alpha^{0}/ds}%
=\frac{\lambda_uu(d\overline{\alpha}^{0}/ds)+\lambda_u(d\overline{\alpha}^{1}/ds)}%
{\lambda_u(d\overline{\alpha}^{0}/ds)+\lambda_uu(d\overline{\alpha}^{1}/ds)}.
\]
Dividing each term by $\lambda_u(d\overline{\alpha}^{0}/ds)$ one obtains%
\begin{equation}
v^{1}=\frac{u+d\overline{\alpha}^{1}/d\overline{\alpha}^{0}}{1+u(d\overline
{\alpha}^{1}/d\overline{\alpha}^{0})}=\frac{u+\overline{v}^{1}}{1+u\overline
{v}^{1}}. \label{cv1}%
\end{equation}
Similarly one gets%
\begin{align}
v^{2}  &  =\frac{d\alpha^{2}}{d\alpha^{0}}=\frac{d\alpha^{2}/ds}{d\alpha^{0}/ds}%
=\frac{d\overline{\alpha}^{2}/ds}{\lambda_u(d\overline{\alpha}^{0}/ds)+\lambda_uu(d\overline
{\alpha}^{1}/ds)}\nonumber\\
&  =\frac{d\overline{\alpha}^{2}/d\overline{\alpha}^{0}}{\lambda_u(1+u\text{
}d\overline{\alpha}^{1}/d\overline{\alpha}^{0})}=\frac{\overline{v}^{2}%
}{\lambda_u(1+u\overline{v}^{1})}, \label{cv2}%
\end{align}%
and
\begin{equation}
v^{3}=\frac{d\alpha^{3}}{d\alpha^{0}}=\frac{d\overline{\alpha}^{3}/d\overline{\alpha
}^{0}}{\lambda_u(1+u\text{ }d\overline{\alpha}^{1}/d\overline{\alpha}^{0})}%
=\frac{\overline{v}^{3}}{\lambda_u(1+u\overline{v}^{1})}. \label{cvv3}%
\end{equation}
Formulas \ref{cv1}, \ref{cv2} and \ref{cvv3} give the relationship between
the components of the velocity of $P$ as measured by $O$ and $\overline{O}.$ In standard units, these formulas read%
\begin{align}\label{ccv1}
\begin{split}
v^{1}  &  =\frac{u+\overline{v}^{1}}{1+(u/c^{2})\overline{v}^{1}%
}\\
v^{2}  &  =\frac{\overline{v}^{2}}{\lambda_u(1+(u/c^{2})\overline{v}^{1}%
)},\text{ }\\
v^{3}  &  =\frac{\overline{v}^{3}}{\lambda_u(1+(u/c^{2})\overline{v}^{1}%
)},
\end{split}
\end{align}
with $\lambda_u   =1/\sqrt{1-u^2/c^{2}}$. We see that at non-relativistic velocities, that is, if $u<<\mathrm{c}$, one has
$u^2/c^{2}\approx0,$ and $\lambda_u\approx1.$ Formulas \eqref{ccv1} tend to the
classical Galilean addition of velocities.

\section{Bell's spaceship paradox}
  
\begin{figure}[H]
\centering
\includegraphics[scale=0.42]{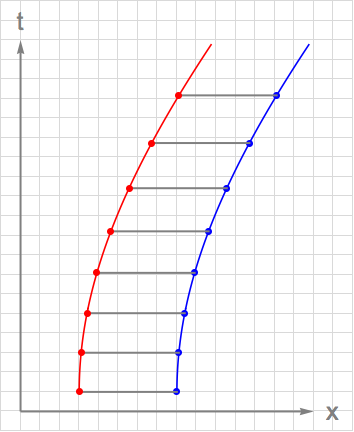}\caption{Edward believes the distance between the rockets stays constant.}\label{bellrest}%
\end{figure}

 The following thought experiment was proposed by Dewan and Beran \cite{Dewan}. It became known as Bell's spaceship paradox after 
 Bell \cite{Bell} introduced a variation that is now more popular. Suppose that an experimenter, Edward, which is at rest, programs  two rockets so that 
 they accelerate in such a way that, in his frame, they remain at constant distance. Before the rockets accelerate, Edward links them with a delicate string. Edward will see that the length of the string stays constant as the rockets accelerate away from the lab. Figure \ref{bellrest} illustrates the situation when the rockets move along hyperbolic trajectories in Minkowski spacetime.

\begin{figure}[H]
\centering
\includegraphics[scale=0.42]{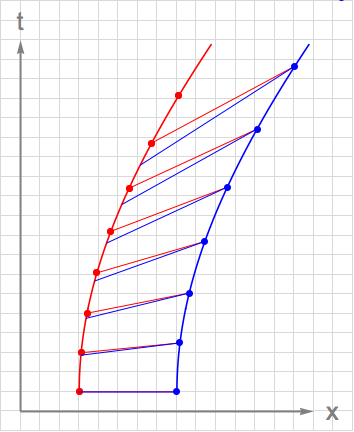}\caption{Both pilots believe the rockets are separating, but at different rates.}\label{Bellmoves}%
\end{figure}

  Let us now consider the situation from the point of view of the pilots. Since the rockets are moving on accelerated trajectories, they do not have a constant inertial frame. However, for each point $p$ in their worldlines, there is a reference frame in which the time direction is tangent to the curve. At the event $p$, they will use this moving reference frame to decide which events are simultaneous.  Figure \ref{Bellmoves} shows the lines of simultaneity for each of the pilots. Both of them
  believe that the rockets are separating. However, they differ in their perception of the situation. The red and blue lines are not parallel. 

\begin{figure}[H]
\centering
\includegraphics[scale=0.42]{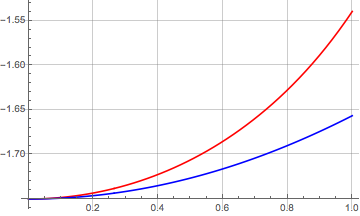}\caption{Distance between the rockets as a function of proper time. The blue line corresponds to the distance according the pilot in front, the red
line is the distance according to the trailing pilot.}\label{RB}%
\end{figure}

Figure \ref{RB} is a plot of the factors by which each of the pilots sees the distance change, as a function of proper time. The trailing pilot will judge the distance to be increasing more rapidly.

The question posed by Dewan and Beran is whether or not the string will break. The situation can be modelled at different levels of detail. A more realistic situation will take into account the forces that the string exerts on the rockets, which will depend on the Hooke constant of the string and other parameters of that type. The simplest analysis, where this force not taken into account, and the string is supposed to be inelastic, leads to the conclusion that it will break.

\section{The Doppler effect}
\label{sec: Doopler effect} 
 
\begin{figure}[H]
\centering
\includegraphics[scale=0.5]{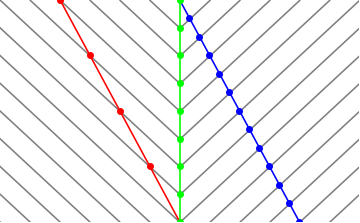}\caption{Doppler effect.}\label{dopler}%
\end{figure}

 The pitch of an ambulance's siren is higher when the ambulance is approaching and lower when it is going away. This is the classical Doppler effect for mechanical waves. In Figure \ref{dopler}, the green line represents an object Green that is emitting a wave. The blue line represents an object Blue approaching Green and the red line, an object Red going away from Green. The image shows that Blue will encounter the pulses more frequently and Red less frequently. For a sound wave, this means that the pitch of the sound perceived by Blue will be higher than that perceived by Red. For a light-wave, this means that Blue will perceive the color shifted to the blue and Red will perceive the color shifted to the red.

  Let us consider precisely the change in the period of the wave. Suppose that Green emits a pulse of light at intervals of $t_0$ seconds. If Blue is approaching with velocity $v$, in Green's reference frame, the time that passes between two events in which Blue receives consecutive pulses is
  \[ \frac{t_0}{1+v}.\]
  This is the change in the period of the wave that is predicted by the classical analysis. The relativistic version takes into account Blue's reference frame, which is related to that of Green by a Lorentz boost. In Blue's reference frame, the time that passes between two consecutive pulses is
\[\overline{t}_0=t_0\sqrt{\frac{1-v}{1+v} }. \]
This means that the frequencies of the waves perceived by Green and Blue are related by
\[ \frac{\overline{f}}{f}=\sqrt{\frac{1+v}{1-v} }.\]
Even though Figure \ref{dopler} represents both the classical and the relativistic situations, the Doppler factors by which the period changes are different.
  \begin{figure}[H]
\centering
\includegraphics[scale=0.55]{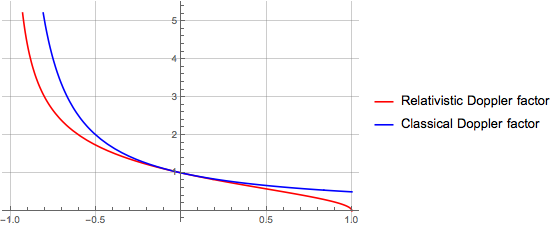}\caption{The blue line corresponds to the classical Doppler factor. The red one to the relativistic Doppler factor.}%
\end{figure}

The relativistic Lorentz factor transforms in a way that is consistent with the rule for adding velocities in special relativity. Suppose that there 
is another observer, Purple, which moves with velocity $u$ with respect to Blue. Then, the velocity $w$ between Green and Purple is
\[ w =\frac{v+u}{1+uv}.\] 
Therefore, Purple will perceive the frequency of the wave to be
\[ \hat{f} =f \sqrt{\frac{1+w}{1-w}}.\]
This can also be expressed as follows
\begin{align*}
\hat{f} =f \sqrt{\frac{1+\frac{v+u}{1+uv}}{1-\frac{v+u}{1+uv}}}=f \sqrt{\frac{1+uv+v+u}{1+uv-v-u}}=f \sqrt{\frac{1+u}{1-u}} \sqrt{\frac{1+v}{1-v}}.
\end{align*}
This is the consistency condition necessary for the Lorentz invariance of the Doppler factors.

\section {Aberration of Light}

Assume that $\overline{O}$ is an observer that moves with velocity $u$ in the positive direction of the $x$-axis of an inertial observer $O$. Suppose that there is also a rod that moves with constant velocity $u$ with respect to $O$ in the direction of $x$, along the line $y=1$. As we discussed before, the length $l$ that $O$ will measure for the rod will be smaller that the length $\overline{l}$ that $\overline{O}$ will measure for it.
However, \textit{$O$ actually sees the rod as if it were as long as the rod
$\overline{O}$ measures}. To understand this apparent paradox we have to
clarify what we mean by \textit{seeing} instead of measuring.
Imagine that we have a large piece of photographic paper that acts as a projection
screen. Parallel rays of light coming from an object $B$ and impinging
the paper perpendicularly at the same time (according to the observer carrying the paper with him) would print an image on
its surface. The size of this image is what we will call the
\textit{apparent} size of $B.$  In the following discussion we assume that $c=1$.
\begin{figure}[H]
\centering
\includegraphics[scale=0.6]{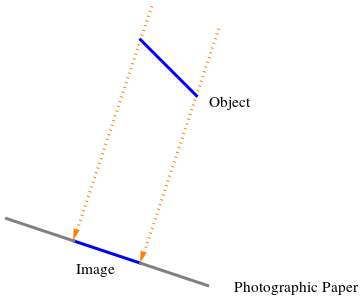}
\end{figure}

Assume the rod has unit length, as measured by
$\overline{O}.$ As we already calculated, $O$ would measure $\lambda_u^{-1}$ for its length.
Let $A$ and $B$ be the simultaneous events, according to $\overline{O}$, corresponding to the emission
of two parallel beams of light coming from the tail and the front of the rod,
and moving downwards as perceived by $\overline{O}.$ 

\begin{figure}[H]
\label{aberracion}
\centering
\includegraphics[scale=0.5]{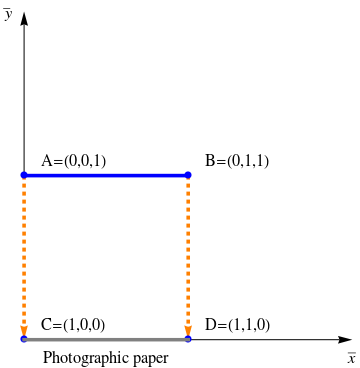}\caption{The events $A,B,C$ and $D$ correspond to the emission and reception of the rays of light. The coordinates are the $(\overline{t},\overline{x},\overline{y})$ coordinates used in the reference frame of $\overline{O}$.}%
\end{figure}

With respect to $\overline{O}$, the event $A$ has coordinates $\overline{t}=\overline{x}=0,$ and $\overline{y}=1$. The event $B$ has coordinates $\overline{t}=0 $ and $\overline{x}=\overline{y}=1$. In $O$'s reference frame, the event $A$ has coordinates
$t=x=0,$ $y=1$ and $B$ has coordinates $t=u\lambda_u$, $x=\lambda_u$, $y=1$. We notice $A$ and $B$ are not simultaneous according to $O$.
Let $C$ and $D$ label the events corresponding to the arrival of both rays of
light at $\overline{O}$'s photographic paper. These events have coordinates
$\overline{t}=1$, $\overline{x}=\overline{y}=0,$ and $\overline{t}=\overline
{x}=1$, $\overline{y}=0,$ respectively. 
According to $O$, the event $C$ has coordinates $t=\lambda_u,$ $x=u\lambda_u,$ $y=0$, and $D$
has coordinates $t=\lambda_u+\lambda_uu,$ $x=\lambda_u+u\lambda_u,$ $y=0.$ 
\begin{figure}[H]
\centering
\includegraphics[scale=0.6]{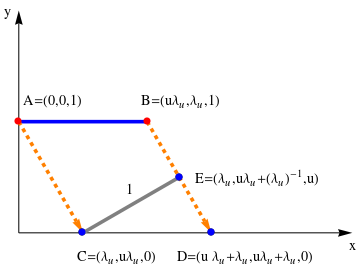}\caption{The events $A,B,C$ and $D$ correspond to the emission and reception of the rays of light. The coordinates are the $(t,x,y)$ coordinates used in the reference frame of $O$.}%
\label{aberracion}
\end{figure}

Suppose that, at the event $C$, the observer $O$ places a photographic paper \textit{of
length one} at an angle $\alpha=\arctan(u\lambda_u)$ with respect to the $x$-axis. In this case, the spatial coordinates of the front end of the
paper, denoted by $E$ in Figure \ref{aberracion}, are $x=u \lambda_u+ \lambda^{-1}_u$ and $y=u$. If we denote by $\beta$ the angle that the beams of light form with the $x$ axis, then
\[ \tan \beta=(u\lambda_u)^{-1}=(\tan \alpha)^{-1},\]
and one concludes that the rays of light meet the photographic paper orthogonally.  We claim that \textit{both rays reach the
opposite sides of the plate at the same time, as measured by }$O$. We know that the beam of light at the back of the rod meets the paper at the event 
$C$, for which the time coordinate is $t=\lambda_u$. On the other hand,
the beam that leaves the front of the rod has worldline
\[ \overline{t}(\tau)=\tau,\qquad \overline{x}(\tau)=1,\qquad \overline{y}(\tau)=1-\tau.\]
Therefore, in the reference frame of $O$, the worldline is
\[ t(\tau)=\lambda_u(\tau +u),\qquad x(\tau)=\lambda_u(1+ u\tau),\qquad y(\tau)=1-\tau.\]
At $\tau=1-u$, one obtains the event $E$, with coordinates
\[t=\lambda_u,\qquad x=u \lambda_u +\lambda^{-1}_u, \qquad y=u.\]
We conclude that, according to $O$, both beams of light meet the paper
at $t=\lambda_u$. Moreover, the length of the image in the photographic paper is $l=1$. The same as the length that observer $\overline{O}$ measures for the rod.


\section{Muon Decay: An experimental test for Special Relativity}

\begin{figure}[H]
\centering
\includegraphics[scale=0.48]{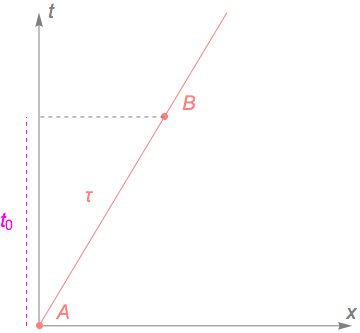}\caption{Muon decay.}
\label{fmuon}
\end{figure}

One of the most dramatic examples of time dilation predicted by Einstein's
Special Relativity takes place at the subatomic level.
Muons are particles that decay into neutrinos and electrons after a period
that is intrinsic to the particle, called its \textit{lifetime},
denoted by $\tau$. A particle's lifetime is its proper time between
its birth and decay. At rest, the lifetime of a Muon is $\tau\approx
\mathrm{2.2}\times10^{-6}\: \mathrm{s}$. In a series of famous experiments
performed at CERN in 1970 (\cite{hartle}, Page 65), Muons were accelerated to
velocities of the order of $v=0.9994c$. For these ultra-rapid Muons
scientists measured a lifetime equal to $t_0=\mathrm{64.419}\pm\mathrm{0.58}%
\times10^{-6} \: \mathrm{s}$.
Let $A$ be the event corresponding to the
crossing of the particle through the laboratory and let $B$ the event
corresponding to its decay, as illustrated in Figure \ref{fmuon}.

If $t_0$ is the time it takes for the particle to decay from the standpoint of an observer in the laboratory, and $\tau$ is its lifetime, then 

\[
t=\frac{1}{\sqrt{1-v^{2}/c^{2}}}\tau\text{ }=\mathrm{28.871}%
\times\mathrm{2.2}\times10^{-6}=\mathrm{63.51}\times10^{-6},
\]
a theoretical prediction in great agreement with the experiments!

\section{Energy, Momentum and Mass}

In this section we want to discuss the dynamics of a particle from a relativistic view point. We will see that when objects move at low speed, Einstein's dynamics reproduces Newton's picture of the world. 
\noindent
Let us start by analyzing the collision of two identical spheres $B$ and
$\overline{B}$ in Minkowski spacetime. We assume that associated to any particle
$P$ whose world-line is timelike there is a nonzero scalar called its
\emph{rest mass}, that we measure in \textrm{kg}, and that we will denote by
$m_{0}(P)$. In relativistic mechanics the \emph{total mass} of $P$ is a scalar
that depends on the observer, and that can be identified with the \emph{total
energy }of the particle, as measured by that particular observer. A precise
definition can be given after we introduce the notion of $4$-momentum. In this
section, however, we will refer to the mass of a particle as a scalar $m(P)$
determined by each inertial observer, and which must coincide with $m_{0}(P)$ when the particle is seen to be at rest.

The purpose of the thought experiment we will discuss next is to determine the
mass that an inertial observer $O$ would measure for a particle that moves
along the $x$-axis at constant speed $u$\textrm{.} We will consider elastic collisions, that is,  we assume
the conservation of classical momentum.

\begin{figure}[tbh]
\centering
\includegraphics[scale=0.6]{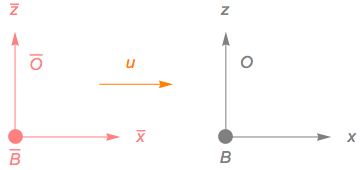}\caption{Two inertial observers before the collision.}\label{momfram}
\end{figure}

We consider two inertial frames of reference. The first, denoted by
$(t,x,y,z)$, corresponds to an observer $O$ for whom $B$ stays at rest
at the origin. The second, $(\overline{t}%
,\overline{x},\overline{y},\overline{z})$ corresponds to an observer
$\overline{O}$ that moves in the $x$ direction with
constant speed $u$ with respect to $O$, and sees the particle $\overline{B}$
 at rest. The situation is illustrated in Figure \ref{momfram}.

\begin{figure}[tbh]
\centering
\includegraphics[scale=0.6]{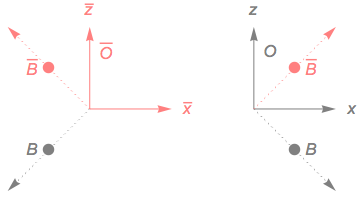}\caption{The trajectories of the particles after the collision.}
\label{col}
\end{figure}

After $B$ and $\overline{B}$ collide the two
particles move in the $x$-$z$ plane as indicated in Figure \ref{col}.

By symmetry, the magnitude of the $z$-component
of the velocities of $B$ and $\overline{B}$ should be the same as measured by
$O,$ and by $\overline{O}$, respectively. Denote these quantities by
$v^{3}(B)$ and $\overline{v}^{3}(\overline{B})$, respectively. Hence,
$|v^{3}(B)|=|\overline{v}^{3}(\overline{B})|.$ On the other hand, by formula
(\ref{cvv3}) ones has
\begin{equation}\label{2vel}
v^{3}(\overline{B})=\frac{\overline{v}^{3}(\overline{B})}{\lambda_u(1+(u%
/c^{2})\overline{v}^{1}(\overline{B}))},
\end{equation}
where $\overline{v}^{1}(\overline{B})$ denotes the $\overline{x}$-component of
the velocity, as measured by $\overline{O}$. Denote by $m_{0}(B)=m_{0}(\overline{B})$ the rest
masses of $B$ and $\overline{B}$, and by $m(B)$ and
$m(\overline{B})$ the post-collision masses of $B$ and $\overline{B}$ as
measured by $O$. By the classical conservation of momentum in the
frame of reference of $O$ one must have $m(B)v^{3}(B)+m(\overline{B}%
)v^{3}(\overline{B})=0$. Using (\ref{2vel}), the right hand side of this
equation can be written as
\[
m(\overline{B})v^{3}(\overline{B})=\frac{m(\overline{B})\overline{v}%
^{3}(\overline{B})}{\lambda_u(1+(u/c^{2})\overline{v}^{1}%
(\overline{B}))}.
\]
Thus, we may write
\begin{equation}
m(B)v^{3}(B)=\frac{-m(\overline{B})\overline{v}^{3}(\overline{B})}%
{\lambda_u(1+(u/c^{2})\overline{v}^{1}(\overline{B}))}.\text{ }
\label{limite}%
\end{equation}
Since $v^{3}(B)=-\overline{v}^{3}(\overline{B})$ one gets%
\begin{equation}
m(B)=\frac{m(\overline{B})}{\lambda_u(1+(u/c^{2})\overline{v}%
^{1}(\overline{B}))}.
\label{limite1}
\end{equation}
As we consider more and more glancing collisions, the quantity $\overline
{v}^{1}(\overline{B})$ approaches zero while $v^{1}(\overline{B})$ approaches
$u$. In the limit $\overline{B}$ and $B$ will just touch
tangentially, and henceforth the $z$-component of the velocities of both balls
will be equal to zero. In the limit, the velocity in the $x$-direction would
be $v^{1}(B)=\overline{v}^{1}(\overline{B})=0,$ $v^{1}(\overline
{B})=u$. Since  $B$ stays still from $O$'s view point, he
would deduce that $m(B)=m_{0}(B).$ Henceforth, Formula
\eqref{limite1} becomes
\begin{equation}
m_{0}(B)=\frac{m(\overline{B})}{\lambda_u}. \label{E40}%
\end{equation}
We conclude $$m(\overline
{B})=\lambda_u m_{0}(B)=\frac{m_{0}(\overline{B})}{\sqrt{1-(u/c)^{2}}}.$$

Hence, from $O$'s perspective, mass increases with velocity by a factor of
$1/\sqrt{1-(u/c)^{2}}$. In fact, when $u \rightarrow c$ the post-collision mass of $\overline{B}$ approaches infinity. This implies that no particle with nonzero rest
mass can ever reach the speed of light! From this last formula Einstein was
able to deduce in a way that is characteristic of his thinking what
is perhaps the most celebrated formula in all of physics.

The series $(1-x^{2})^{-1/2}$ is convergent for $\left\vert
x\right\vert <1$, and the first two terms in Taylor's expansion around zero
are $1+x^{2}/2$. Therefore, for $x=u/c$ this series
converges. One concludes that
\begin{equation}
m(\overline{B})= m_{0}(\overline{B})+\frac{m_{0}(\overline{B})}
{2} \frac{u^{2}}{c^{2}}+\dots = m_{0}(\overline{B})+\frac{E_K}{c^{2}}+\dots,
\label{E41}%
\end{equation}
where $E_{K}=\frac{1}{2}m_{0}(\overline{B})u^{2}$ is the kinetic
energy of $\overline{B},$ as measured by $O$. From his point of view, if
$\Delta m(\overline{B})=m(\overline{B})-m_{0}(\overline{B})$ denotes the mass
increment, formula (\ref{E41}) tells us that $E_{K}\approx \Delta m(\overline
{B})c^{2}$. Einstein observes that the mass increment $\Delta
m(\overline{B})$ is indistinguishable from an increase in kinetic energy
$E_{K}$. From this, he conjectures that \emph{mass and energy are just two
manifestations of one single entity}. Strictly speaking, this reasoning
leads one to postulate the equivalence of mass and energy not as a theorem,
but rather as a heuristic law. 

\begin{definition}
\label{total energy}The \emph{total energy} of a particle with rest mass
$m_{0}\neq0,$ as measured by an inertial observer $O,$ is defined to be\emph{ }$E=c%
^{2}m_{0}/\sqrt{1-u^2/c^{2}}$. Its rest energy is
defined as $E_{0}=m_{0}c^{2}.$
\end{definition}

When $u\ll c$, the total energy $E$ can be approximated
as $$E=E_{K}+m_{0}(\overline{B})c^{2}= \text{ kinetic energy }+ \text{ rest energy
of } \overline{B} .$$

 \subsection*{4-Momentum and 4-acceleration\label{proper acceleration}}

As we discussed in the previous section, associated to each particle $P$ there is a non negative
scalar $m_{0}>0$ called its \emph{rest mass}. Suppose $\beta(\tau)$ represents the worldline of $P$. At each point $q=\beta(\tau_0)$ its \emph{ 4-momentum} is defined to be
its rest mass times its 4-velocity at $q$, $\mathbf{u}=\beta^{\prime}(\tau)$
\begin{equation}
\label{four momentum}
\mathbf{p}=m_{0}\mathbf{u}=m_0 \beta'(\tau).
\end{equation}
 On the other hand, we define its \emph{proper acceleration} or \emph{4-acceleration} as $$\mathbf{a}(\tau)=\nabla_{\mathbf{u(\tau)}} \mathbf{u(\tau).}$$
 In any inertial system the 4-acceleration can be computed as
 $\mathbf{a}(\tau)=\beta^{\prime\prime}(\tau).$ The \emph{4-force} that accounts for such acceleration is
 defined as $\mathbf{f}(\tau)=m_{0}\mathbf{a}(\tau).$

We notice that $\mathbf{f}(\tau)=\mathbf{p}^{\prime}(\tau)$ is the change in momentum, as one would expect. Since $\beta$ is assumed to
be parametrized by proper time, one has $\left\langle \beta^{\prime}(\tau),\beta^{\prime\prime
}(\tau)\right\rangle =0.$ Thus, $\mathbf{a}(\tau)$ and $\mathbf{u}(\tau)$ \emph{are orthogonal
vectors}. 

Let $O$ be an observer that measures the momentum of $P$ at the point $q$. Choose $x=(x^{a})$ a Lorentz frame for $O$ at $q$ and let
$b^{a}(\tau)=x^{a}(\beta(\tau))$ be the coordinates of the world line of $P$ in this
frame of reference. Since $x^0$ is the time coordinate according to $O's$ clock, we also denote it by $t$. 
In these coordinates $O$'s 4-velocity $\mathbf{u}$ is equal to $\partial_{t}$:

\begin{figure}[h]
\centering
\includegraphics[scale=0.5]{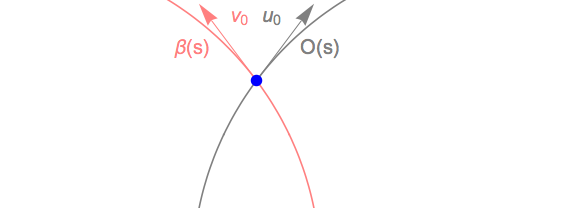}\caption{The $4$-momentum of intersecting observers.}%
\end{figure}

In the basis $\{\partial_{t}=\mathbf{u},$ $\partial
_{x^{i}}\}$ one can write the 4-velocity of $P$ at $q$ as
\begin{eqnarray*}
\mathbf{u} &=&\frac{dt}{d\tau }\partial _{t
}+\sum\limits_{i}\frac{dx^{i}}{d\tau }\partial _{x^{i}} \\
&=&\frac{dt}{d\tau }\partial _{t }+\sum\limits_{i}\frac{dx^{i}}{dt}%
\frac{dt}{d\tau }\partial _{x^{i}} \\
&=&\frac{dt}{d\tau}\partial _{t }+\frac{dt}{d\tau }\sum\limits_{i}%
\frac{dx^{i}}{dt}\partial _{x^{i}}.
\end{eqnarray*}%
Since 
$\left\langle \beta ^{\prime }(\tau ),\beta ^{\prime }(\tau )\right\rangle
=-1$ we see that
\[
-\left( \frac{dt}{d\tau }\right) ^{2}+\left( \frac{dt}{d\tau }\right)
^{2}\sum\limits_{i}\left( \frac{dx^{i}}{dt}\right) ^{2}=-1.
\]%
Solving for $dt/d\tau $ one obtains:%
\[
\frac{dt}{d\tau }(q)=\frac{1}{\sqrt{1-\left\vert v(t(q))\right\vert ^{2}}}
\]%
where $$v(t(q))=\sum\limits_{i} \frac{dx^{i}}{dt}(t(q))$$ is the \emph{3-velocity} of $P$ at time $t(q),$ 
measured using the Lorentz coordinates of $O$. As usual, we write $$\lambda_v=\frac{1}{\sqrt{1-\left\vert v(t(q))\right\vert ^{2}}}$$
Hence, the 4-velocity can be written as
\begin{equation}\label{eq:4-vel}
\mathbf{u}(\tau) = \lambda_v \left(\partial_t + v(t(\tau)) \right).
\end{equation}
Thus, for the $4$-momentum we get the expression
\begin{align}
\begin{split}
\label{estaba bien}
\mathbf{p}(\tau)  &  =m_0 \lambda_v \left(\partial_t + v(t(\tau)) \right) \\
&  =m(t(\tau))\partial_{t}+m(t(\tau))v(t(\tau)),
\end{split}
\end{align}
where the term $m(t(\tau))=m_{0}\lambda_{v(t(\tau))}$ corresponds to the \emph{relativistic mass} of the observer, which is the same as the \emph{total energy} $E$ of $P$ (\ref{total energy}). The spatial term of the 4-momentum, on the other hand, corresponds to the \emph{relativistic 3-momentum} of $P$, as measured by $O$
\begin{equation}
\label{3 momentum}
p=m(t(\tau))v(t(\tau)).
\end{equation}
When $v$ is small compared with the speed of light so that $\lambda_v \approx 1$, we see that $m(t(\tau))\approx m_{0}$, and $p=m_0v$ is approximately $P$'s  classical 3-momentum.

By taking derivatives again we see that the 4-force at $q$ is given
by
\begin{align*}
\mathbf{f}(\tau)&=\frac{d }{d\tau}m(t(\tau))\partial_{t}+\frac{d}{d\tau}(m(t(\tau))v(t(\tau))).
\end{align*}
Its spatial component acting on $P$, as measured by $O$, corresponds to the second term in the previous equation, and it is equal to
\begin{equation}
\frac{d}{d\tau}(m(t(\tau))v(t(\tau))=\frac{d(m(t(\tau))v(t(\tau))}{dt}\frac{dt}{d\tau}=\lambda_v\frac{d(m(t(\tau))v(t(\tau))}{dt}. \label{ef2}%
\end{equation}
The term $$f(t(\tau))=\frac{d(m(t(\tau))v(t(\tau))}{dt}$$ is called \emph{the relativistic 3-force}. We see from (\ref{3 momentum}) that this coincides with the rate of change of the 3-momentum $dp/dt=f$. On the other hand, the temporal component of $\mathbf{f}(\tau)$ can be written in another way. We have already remarked that the $4$-acceleration $\mathbf{a}$ is orthogonal to $\mathbf{u}$. Therefore, $\langle  \mathbf{f}(\tau),\mathbf{u}(\tau)\rangle =  0$ and this yields
$$
\frac{d}{d\tau}m(t(\tau)) = \lambda_{v}f(t(\tau)) \cdot v(t(\tau))
$$
Consequently, we get the following expression for the $4$-force:
\begin{equation}\label{ef1}
\mathbf{f}(\tau) = \lambda_{u} \left(f(t(\tau)) \cdot v(t(\tau)) \:\partial_t + f(t(\tau)) \right).    \end{equation}
As a conclusion, the change in mass of $P$ is equal to the classical work done on the particle, which classically is the energy imparted to $P$. This is in keeping with the equivalence of mass and energy. Again, when $v\ll c= 1$ one has that 
\[
f(t(\tau))\approx m_{0}\frac{dv(t(\tau))}{dt}=m_{0}a(t(\tau)),
\]
where $a(t(\tau))=dv(t(\tau))/dt$ is the \textit{3-acceleration of $P$}.  This is Newton's
second law. 

Notice that while the 4-velocity, 4-acceleration and the 4-force are geometric objects, their corresponding 3-counterparts are not intrinsically defined but \emph{depend on the coordinates one chooses to measure them}. However, the total energy $E$ will only depend on the 4-velocity of $O$. In fact,
$O$ can compute $E$ as: 
\begin{equation}
E=p^{0}=-\left\langle \mathbf{p},\mathbf{u}\right\rangle
=-m_{0}\left\langle \mathbf{v},\mathbf{u}\right\rangle .
\label{energy formula}
\end{equation}
On the other hand, $$\left\langle \mathbf{p},\mathbf{p}\right\rangle
=\left\langle m_{0}\mathbf{v}\mathbf{,}m_{0}\mathbf{v}\right\rangle
=m_{0}^{2}\left\langle \mathbf{v},\mathbf{v}\right\rangle =-m_{0}%
^{2}.$$ Thus,
\[
-m_{0}^{2}=\left\langle \mathbf{p},\mathbf{p}\right\rangle
=-(p^{0})^{2}+\sum_{i}\left\langle p^{i},p^{i}\right\rangle =-E^{2}+\left\vert
p\right\vert ^{2},
\]
where $\left\vert p\right\vert ^{2}$ denotes the norm of the
relativistic $3$-momentum of $p$. From this
one gets
\begin{equation}
E=\sqrt{m_{0}^{2}+\left\vert p\right\vert ^{2}}. \label{energy 2}%
\end{equation}
Notice that when the $3$-momentum is zero one finds the rest energy of the
particle, $E=m_{0}$, which in standard units is written as $E=m_{0}%
c^{2}$.

\subsection*{Conservation of Momentum\label{CMomentum}}

Suppose $\gamma_{1}:I\rightarrow \MM$ and $\gamma_{2}:I\rightarrow \MM$ are the worldlines
of two particles $B_1$ and $B_2$ which collide at a certain point $q$ on the worldline of an
observer $O$. If $\mathbf{p}_{1}$, $\mathbf{p}_{2}$ denote their corresponding
momenta at $q$ before collision and $\widetilde{\mathbf{p}}_{1}$,
$\widetilde{\mathbf{p}}_{2}$ are their momenta at $q$ afterwards, then a
fundamental law of physics says that $$\mathbf{p}_{1}+\mathbf{p}_{2}%
=\widetilde{\mathbf{p}}_{1}+\widetilde{\mathbf{p}}_{2}$$ in $T_{q}\MM$. This
law is known as the \emph{conservation of the 4-momentum}.

Choose $x=(x^{a})$ a Lorentz
frame for $O$ at $q$. Let 
$\{\mathbf{u},\partial_{x^{i}}\}$ be an orthonormal frame at $q$, where $\mathbf{u}$ is $O$'s $4$-velocity. The $4$-momentum $\mathbf{p}$
of a particle $B$ with rest mass $m_{0}$ decomposes as 
\[\mathbf{p}=p^{0}\mathbf{u}+\sum_ip^i \partial_{x^i} .\]
Moreover, as we have seen before (\ref{estaba bien}) the 4-momentum of $B$ can be separated as the total energy plus the relativistic 3-momentum, that is,
$p^{0}=E$, $p^{i}=Ev^{i}$, where $$E=\frac{m_{0}}{\sqrt{1-\left\vert v%
\right\vert ^{2}}},$$ is the total energy of $B$ and $v$, its $3$-velocity, as measured by $O$. Applying this decomposition to the particles above,  the conservation of the 4-momentum can be written as:
$$p_{1}^{0}+p_{2}^{0}  =\widetilde{p}_{1}^{0}+\widetilde{p}_{2}^{0},$$ and 
$$p^i_{1}+p^i_{2} =\widetilde{p}^i_{1}+\widetilde
{p}^i_{2}.$$
The first equation is equivalent to $$E_{1}+E_{2}=\widetilde{E}_{1}+\widetilde{E}_{2},$$ where $E_{i}$ and $\widetilde{E}_{i}$ denote the energy of the particle $B_{i}$, with $i=1,2$, before and after the collision, respectively. Since the second equation is just the conservation of the relativistic 3-momentum, \emph{the conservation of the $4$-momentum is equivalent to the conservation of energy plus the conservation of the relativistic $3$-momentum}.

\subsection*{Particles with Zero Rest Mass\label{zero mass}}

Guided by thermodynamic considerations, Max Planck postulated around 1900 that radiant energy is emitted in definite \emph{quanta} of energy $E=h\nu$, where $\nu$ is the frequency of the radiation, and $h$ a universal constant  whose accepted value is $6.626\times10^{-34}$ Joul-Hrz$^{-1}$. This motivated Einstein to postulate that light could be regarded as a beam of particles, \emph{photons}, with energy given by Planck's formula, that in terms of the angular frequency $\omega=2\pi \nu$ of the photon could be expressed as $E=\hslash\omega$, where $\hslash=h/(2\pi)$ (pronounced \textquotedblleft h bar\textquotedblright) is called the \emph{reduced Planck constant}.

Some particles, for example  photons, have no mass. For these particles the worldlines will be lightlike geodesics, not timelike curves, and therefore it makes no sense to parametrize them by proper time. 
For instance, a photon $P$ moving in spacetime has a worldline that is a lightlike geodesic 
$\beta:I\rightarrow \MM$. One can reparametrize the curve by any linear change of parameter $s=a\tau+b$. The constant $b$ can be fixed by choosing an arbitrary origin on the worldline of $P$, but the constant $a>0$ is arbitrary. Any such parameter is called an \emph{affine parameter}.

On the other hand, having no mass, it makes no sense to define the 4-momentum of a photon as in (\ref{four momentum}). 
In order to extend this notion for massless particles, we start by noticing that one could have defined the 4-momentum of a particle with rest mass $m_0$ equal to its four velocity, if we had chosen to parametrize its worldline as $\gamma (m_0\tau).$ In analogy, one could think that each photon's 4-momentum is equal to its 4-velocity, once we have fixed a particular parametrization for its worldline. This motivates the following definition.    
\begin{definition}
A photon is represented mathematically as a lightlike geodesic curve $\beta(s)$ with a \emph{given parametrization}. Its 4-momentum is defined as its 4-velocity $\beta'(s)$ so that its energy as measured by an inertial observer $O$ with 4-velocity $\mathbf{u}$ is given by $$E=-\left\langle \beta'(s),\mathbf{u}\right\rangle.$$ 
\end{definition}

The way to choose a particular parametrization for $\beta$ depends on having some information about the total energy $E_0$ of the photon as measured by some particular observer $O_0$. This is because once the total energy (equivalently, the frequency of $P$) is determined by $O_0$, an observer with 4-velocity $\mathbf{u}_0$, one can choose a unique affine parameter $s$ for $\beta$ such that
$$E_0=-\left\langle \beta'(s),\mathbf{u_0}\right\rangle.$$

  \section{Electromagnetism and Special Relativity}\label{sec:relMaxwell}

The reason for the invention of Special Relativity was the incompatibility between classical physics and electromagnetism.
The least one should ask of special relativity is that it fixes these inconsistencies. Fortunately, Maxwell's theory can be 
naturally formulated in a Lorentz invariant manner. Classically, the distribution of charge is described by a charge density function $\rho$, and a current density function $j$. They are required to satisfy the conservation of charge equation
\begin{equation}\label{cc}
\div j+\frac{\partial \rho}{\partial t}=0.
\end{equation}
The first question that arises is how to describe the charge distribution from the point of view of a moving observer. That is, to describe a transformation rule that relates the densities measured by observers in relative motion. This  is resolved by interpreting the charge and current densities as components of a vector field in Minkowski spacetime
\[{\mathbf j}=\rho \partial_t+ j_x \partial_x+ j_y \partial_y+ j_z \partial_z.\]
This interpretation as a vector field immediately provides a transformation rule for arbitrary diffeomorphisms of Minkowski spacetime. Suppose that an observer, Alice,
perceives the charge distribution as a static charge, so that $j=0$ and $\rho$ is independent of time. An second observer, Beth, is moving with relative velocity $v$ in the $x$ direction so that their coordinates are related by
\begin{align*}
  \overline{t}&= \lambda_v(t -\frac{vx}{c^2} ), \\
 \overline{ x}&=\lambda_v(x-vt). 
\end{align*} Then
\begin{align*}
{\mathbf j}=\rho \partial_t=\rho \lambda_v\partial_{\overline{t}} -v\rho\lambda_v \partial_{\overline{x}}.
\end{align*}
This means that, while Alice believes that the charge is static and there are no currents, Beth thinks that there is a current in the $\overline{x}$ direction. This is not surprising, since moving charges generate currents.
The conservation of charge \ref{cc} also takes an invariant form.  Let us compute the Lie derivative of the volume form 
in the direction on the vector field ${\mathbf j}$:
\begin{align*}
L_{{\mathbf j}}\left(cdt\wedge dx\wedge dy \wedge dz\right)&=\left(c L_{{\mathbf j}}dt\wedge dx\wedge dy \wedge dz\right)+\left(cdt\wedge L_{{\mathbf j}}dx\wedge dy \wedge dz\right)\\
&\quad \, +\left(cdt\wedge dx\wedge L_{\mathbf j}dy \wedge dz\right)+\left(cdt\wedge dx\wedge dy \wedge L_{\mathbf j}dz\right)\\
&=\left( \div j+\frac{\partial \rho}{\partial t}\right)cdt\wedge dx\wedge dy \wedge dz.
\end{align*}
One concludes that the conservation of charge is the condition that the vector field ${\mathbf j}$ preserves the volume form. 
Once the charge distribution is expressed in an invariant form, it is natural to do the same with the electric and magnetic fields.
Suppose that, in Alices' reference frame, there are electric and magnetic fields \[E=(E_x,E_y,E_z),\quad \,\,B=(B_x,B_y,B_z).\] 
The components of these fields can be put together to define a differential form $\mathbf F$ on Minkowski spacetime
\begin{equation}\label{eq:Faradayform}
{\mathbf F}=B_x dy \wedge dz+B_y dz \wedge dx + B_z dx \wedge dy+ E_x dx \wedge dt + E_y dy \wedge dt + E_z dz \wedge dt.\end{equation}
Maxwell's equations can be written in the following simple form
\begin{align}\label{YM}
\begin{split}
  d{\mathbf F}&=0, \\
d\!\star\! {\mathbf F}&=\mu_0\!\star\! {\mathbf j}^\flat.  
\end{split}
\end{align}
In the expressions above, $\star$ denotes the Hodge star operator (see Appendix \ref{Hodge}), and ${\mathbf j}^\flat$ is the differential form dual to ${\mathbf j}$ with respect to the Minkowski metric.
The equations (\ref{YM}) are written in an invariant form that is independent of any choice of coordinates. Let us expand them to recover
Maxwell's equations: 
\begin{align*}
d{\mathbf F}&=\div B\, dx\wedge dy \wedge dz+ \left(\frac{\partial E_y}{\partial x}-\frac{\partial E_x}{\partial y}+\frac{\partial B_z}{\partial t}\right)dx \wedge dy \wedge dt \\
&\quad \,+\left(\frac{\partial E_x}{\partial z}-\frac{\partial E_z}{\partial x}+\frac{\partial B_y}{\partial t}\right)dz \wedge dx \wedge dt+\left(\frac{\partial E_z}{\partial y}-\frac{\partial E_y}{\partial z}+\frac{\partial B_x}{\partial t}\right)dy \wedge dz \wedge dt.
\end{align*}
Therefore, the condition $d\,{\mathbf F}=0 $ is equivalent to the equations
\begin{align*}
\div B&=0\\
\frac{\partial B}{\partial t}+ \rot E&=0.
\end{align*}
In order to compute the second condition, we first notice that
\[\star {\mathbf F}=cB_x dt \wedge dx +cB_y dt \wedge dy + cBz dt \wedge dz+ \frac{1}{c}E_x dy \wedge dz + \frac{1}{c} E_y dz \wedge dx + \frac{1}{c} E_z dx\wedge dy.\]
Therefore
\begin{align*}
d\!\star\!{ \mathbf F}&=\frac{1}{c}\div E\, dx\wedge dy \wedge dz+ \left(c\frac{\partial B_x}{\partial y}-c\frac{\partial B_y}{\partial x}+\frac{1}{c}\frac{\partial E_z}{\partial t}\right)dx \wedge dy \wedge dt \\
&\quad\,+\left(c\frac{\partial B_z}{\partial x}-c\frac{\partial B_x}{\partial z}+\frac{1}{c}\frac{\partial E_y}{\partial t}\right)dz \wedge dx \wedge dt+\left(c\frac{\partial B_y}{\partial z}-c\frac{\partial B_z}{\partial y}+\frac{1}{c}\frac{\partial E_x}{\partial t}\right)dy \wedge dz \wedge dt.
\end{align*}
Also,
\[ \mu_0{\mathbf j}^{\flat} =-\frac{\rho dt}{\varepsilon_0}+ \mu_0j_x dx+ \mu_0j_y dy+ \mu_0j_z dz,\]
so that,
\[ \mu_0\! \star\!{\mathbf j}^{\flat} =c \mu_0\rho dx\wedge dy \wedge dz- c\mu_0j_x dy \wedge dz \wedge dt- c\mu_0j_y dz \wedge dx \wedge dt- c\mu_0j_z dx \wedge dy \wedge dt.\]
One concludes that $d\!\star\! {\mathbf F}=\mu_0\!\star\! {\mathbf j}^\flat $ is equivalent to the equations
\begin{align*}
\div E&= \frac{\rho}{\varepsilon_0},\\
\rot B-\mu_0 \varepsilon_0\frac{\partial E}{\partial t}&=\mu_0 j.
\end{align*}

The conclusion is that equations (\ref{YM}) are intrinsically defined on $\MM$, without any reference to particular coordinates. In any system of coordinates where the Minkowski metric takes the standard form, they are equivalent to Maxwell's equations. Thus, the equations for electromagnetism are naturally invariant under the symmetries of Minkowski spacetime. This suggests that spacetime has a definite geometry which plays a  role in the laws of physics. We saw before that Maxwell's equations cannot be made compatible with Galilean transformations. In contrast, they are manifestly invariant with respect to Lorentz transformations. Suppose that in Alice's reference frame the electromagnetic two form is
\[{\mathbf F}=B_x dy \wedge dz+B_y dz \wedge dx + B_z dx \wedge dy+ E_x dx \wedge dt + E_y dy \wedge dt + E_z dz \wedge dt. \]
Beth is moving with velocity $v$ with respect to Alice, so that their coordinates are related by a Lorentz boost
\begin{align*}
 t&= \lambda_v\left(\overline{t} +\frac{v\overline{x}}{c^2} \right),  \\
 x&=\lambda_v\left(\overline{x}+v\overline{t}\right).
\end{align*}
The form ${\mathbf F}$ con be expressed in Beth's reference frame as follows:
\begin{align*}
{\mathbf F}&=B_x dy \wedge dz+B_y dz \wedge dx + B_z dx \wedge dy+ E_x dx \wedge dt + E_y dy \wedge dt + E_z dz \wedge dt. \\
&=B_x d\overline{y} \wedge d\overline{z}+\lambda_vB_y d\overline{z} \wedge (d\overline{x} +vd\overline{t})+ \lambda_vB_z (d\overline{x}+vd\overline{t}) \wedge d\overline{y}\\
&\quad\,+ \lambda^2_vE_x (d\overline{x}+vd\overline{t}) \wedge \left(d\overline{t}+\frac{v}{c^2}d\overline{x}\right) + \lambda_vE_y d\overline{y} \wedge \left(d\overline{t}+\frac{v}{c^2}d\overline{x}\right) + \lambda_vE_z d\overline{z} \wedge \left(d\overline{t}+\frac{v}{c^2}d\overline{x}\right)\\
&= B_x d\overline{y} \wedge d\overline{z}+\lambda_v\left(B_y+\frac{vE_z}{c^2}\right) d\overline{z} \wedge d\overline{x} + \lambda_v\left(B_z-\frac{vE_y}{c^2}\right) d\overline{x} \wedge d\overline{y}+E_x d\overline{x} \wedge d\overline{t}\\
& \quad\,+\lambda_v\left(E_y-v B_z\right) d\overline{y} \wedge d\overline{t} +\lambda_v \left(E_z+vB_y\right) d\overline{z} \wedge d\overline{t}.
\end{align*}
One concludes that
\begin{align*}
\overline{E}_x&=E_x,\\
\overline{E}_y&=\lambda_v\Big(E_y-v B_z\Big),\\
\overline{E}_z&=\lambda_v \Big(E_z+vB_y\Big),\\
\overline{B}_x&= B_x ,\\
\overline{B}_y&=\lambda_v\Big(B_y+\frac{vE_z}{c^2}\Big) ,\\
\overline{B}_z&=\lambda_v\Big(B_z-\frac{vE_y}{c^2}\Big). \\
\end{align*}
The transformation rule mixes the electric and magnetic components of the form ${\mathbf F}$. Beth will believe that there are magnetic fields in a situation where Alice only sees an electric field. Consider the situation where, from Alice's point of view, there is a point charge $Q$ resting at the origin. In this case $B=0$, and
\begin{align*}
 E_x&=\frac{Qx}{4\pi \varepsilon_0r^3},\\   E_y&=\frac{Qy}{4\pi \varepsilon_0r^3}, \\
 E_z&=\frac{Qz}{4\pi \varepsilon_0r^3}.
\end{align*}
Therefore,
\[ \overline{E}=\frac{Q}{4\pi \varepsilon_0r^3}\begin{pmatrix}
x\\
\lambda_v y\\
\lambda_v z
\end{pmatrix} \quad \text{and} \quad  \overline{B}=\frac{Qv\lambda_v}{4\pi \varepsilon_0r^3c^2}\begin{pmatrix}
0\\
z\\
-y
\end{pmatrix}.\]
When $\overline{t}=0$, Beth will see the electric field
\[ \overline{E}=\frac{\lambda_vQ}{4\pi \varepsilon_0\big(\lambda^2_v \overline{x}^2+\overline{y}^2+\overline{z}^2\big)^{3/2}}\begin{pmatrix}
\overline{x}\\
\lambda_v\overline{y}\\
\lambda_v\overline{z}
\end{pmatrix}.\]
Note that, according to Beth, the electric field is not symmetric with respect to rotations. It will appear weaker in the direction of $\overline{x}$ as a consequence of Lorentz contraction.

\begin{figure}[H]
\centering
\begin{tabular}{cc}
  \vspace{0pt} \includegraphics[scale=0.4]{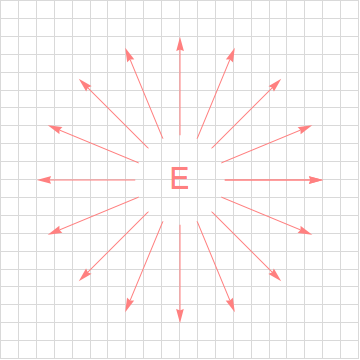} &
  \vspace{0pt} \includegraphics[scale=0.4]{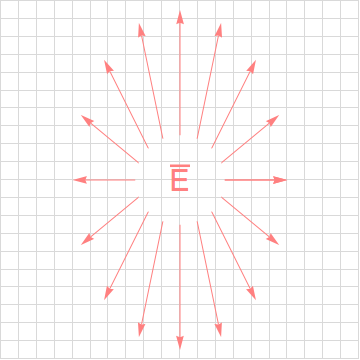}
\end{tabular}
\caption{The electric fields as seen by Alice and Beth.}
\end{figure}

So far, we have given a description of the electromagnetic field as a $2$-form on Minkowski spacetime. It remains to specify a force law that describes the effect  that the electromagnetic field has on a charged particle. Consider again Alice's reference frame. To the Lorentz force \eqref{Lorenz} one associates the $4$-force (see Equation \eqref{ef1})
\begin{align*}
\mathbf{f} &= \lambda_v \left( f \cdot v \: \partial_t + f \right) \\
&= \lambda_v q \left( E \cdot v \: \partial_t +  E + v \times B\right).
\end{align*}
Using the Minkowski metric, we have the associated $1$-form:
\begin{align*}
\mathbf{f}^{\flat} = \lambda_v q \big(&- E \cdot v \: dt + E_x dx + E_y dy + E_z dz \\
& + i_{v} (B_x dy \wedge dz + B_y dz \wedge dx + B_z dx \wedge dy)\big).\end{align*}
Here the symbol $i_v$ denotes contraction with the $3$-velocity $v$. On the other hand, introducing the $4$-velocity $\mathbf{u} = \lambda_v (\partial_t + v)$ (as in Equation \eqref{eq:4-vel}), one can check by a straightforward calculation that
\begin{align*}
i_{\mathbf{u}} \mathbf{F} = \lambda_v \big(&- E \cdot v \: dt + E_x dx + E_y dy + E_z dz \\
&+ i_{v} (B_x dy \wedge dz + B_y dz \wedge dx + B_z dx \wedge dy)\big).
\end{align*}
Thus, we see that
$$
\mathbf{f}^{\flat}  = q i_{\mathbf{u}} \mathbf{F},
$$
This means that the intrinsic $4$-velocity $\mathbf{u}$ along the world line of the charged particle and the intrinsic $4$-force $\mathbf{f}$ along the world line are linearly related by means of the $2$-form $\mathbf{F}$.  Therefore, the relativistic form of the Lorentz-force law is expressible as
\begin{equation}
\frac{d \mathbf{p}}{d \tau} = q (i_{\mathbf{u}} \mathbf{F})^{\sharp},
\end{equation}
where $\mathbf{p}$ is the $4$-momentum and $(i_{\mathbf{u}} \mathbf{F})^{\sharp}$ is the vector field dual to the $1$-form $i_{\mathbf{u}} \mathbf{F}$. In conclusion, this law is intrinsically defined, in spite of its initial coordinate expression. Notice also that the temporal part of this law is the statement that the change in energy is the work done by the electric field, while the spatial part reduces to
\begin{equation}\label{frel}
\frac{d p}{d t} = q \left(  E + v \times B\right),
\end{equation}
where $p$ is the $3$-momentum. 

As one would expect, for $v\ll c$ the Lorentz factor $\lambda_v \approx1$  and \eqref{frel} approximates the classical force law. However, the relativistic version is the correct force law that allows for a bound on the speed of the particle. Consider the situation of a constant electric field $E$ in the $x$ direction. According to the classical Lorentz force law, a particle will accelerate to reach arbitrarily high velocity. Relativistically, the equations of motion are
\begin{align*}
\frac{d^2(ct)}{d\tau^2}&=\frac{q\lambda_vEv }{mc},\\
\frac{d^2x}{d\tau^2}&=\frac{q\lambda_vE }{m}.\\
\end{align*}
Therefore
\[\left(\frac{d^2x}{d\tau^2}\right)^2-\left(\frac{d^2(ct)}{d\tau^2}\right)^2=\frac{q^2 E^2}{m^2}\left(\lambda^2_v-\frac{v^2 \lambda^2_v}{c^2}\right)=\frac{q^2 E^2}{m^2}.\]
This has solution
\begin{align*}
  ct&=\frac{qE}{m}\sinh(\tau),\\
  x&=\frac{qE}{m}\cosh(\tau).
\end{align*}
The velocity is then
\[ v=\frac{dx}{dt}=\frac{dx}{d\tau} \frac{d\tau}{dt}=c \tanh \tau.\]
Since $\tanh \tau=\sqrt{1-\sech^2\tau}<1$,
 one concludes that $v=c\tanh \tau<c$. The particle never goes faster than the speed of light. As the particle accelerates it becomes more massive, and therefore, it is more and more difficult to increase the velocity. 

\begin{figure}[H]
\center
 \includegraphics[scale=0.45]{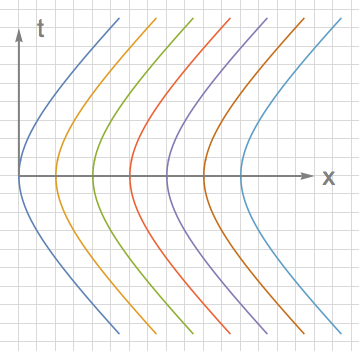} 
 \caption{Hyperbolic motion due to a constant electric field.}
 \end{figure}

   \clearemptydoublepage 
   
 \begin{partwithabstract}{Gravity and Curvature}
According to Special Relativity, the relationship between time and space is more symmetric than common sense and classical physics indicate. The geometry of Minkowski spacetime provides a precise description of these symmetries. 
Maxwell's equations for electromagnetism take a geometric form as tensor equations on Minkowski spacetime. Gravity arises in General Relativity as the curvature of spacetime. Energy and matter cause spacetime to bend according to Einstein's field equation
\[\Ric-\frac{1}
{2}\Rs g=\frac{8\pi G_{N}}{c^{4}} T ,\]
where the left hand side is a geometric quantity that depends on the metric, and the right hand side describes the distribution of energy and matter. In turn, the geometry of spacetime determines the trajectories of matter, which moves along geodesics.

\begin{figure}[H]
\centering
\begin{tabular}{ccc}
\vspace{0pt} \includegraphics[scale=0.32]{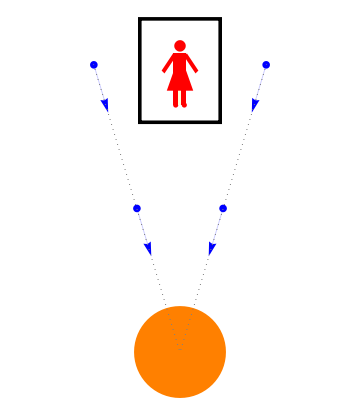}&
\vspace{0pt}\includegraphics[scale=0.32]{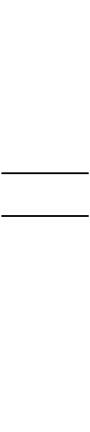}&
  \vspace{0pt} \includegraphics[scale=0.42]{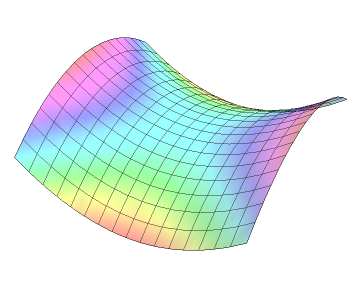} 
\end{tabular}
\end{figure}
\end{partwithabstract}

 \chapter{From Minkowski to curved spacetimes}

\begin{center}
\parbox[b]{0.9\textwidth}{\small \sl In this chapter we describe how many of the constructions that  appear in Special Relativity are still available when Minkowski spacetime
is replaced by a possibly curved Lorentzian manifold. }
\end{center}

\vspace{3ex}

\section{Light cones and causality}

We consider a spacetime manifold $M$ of dimension $d=4$,  which is a Lorentzian manifold with metric  $g$. At any point $p\in M$, the tangent space $T_pM$ is endowed with a Lorentzian inner product. Therefore,  the vector space $T_pM$ is decomposed into vectors of different types:
\begin{itemize}
\item A vector $v \in T_pM$ is \emph{timelike} if $\langle v,v\rangle<0$.  We will denote by $\tilde{C}_p$ the space of timelike vectors in $T_pM$.
\item A vector $v \in T_pM$ is \emph{spacelike} if $\langle v,v\rangle>0$. 
\item A vector $v \in T_pM$ is \emph{lightlike} if $\langle v,v\rangle =0$.
\item A vector $v \in T_pM$ is \emph{causal} if $\langle v,v\rangle \leq 0$.
\end{itemize}
In Minkowski spacetime, timelike vectors either point to the past or the future, depending on the sign of the time component in the standard coordinates.
For a general spacetime, timelike vectors fall in two different classes, but there is no natural way to distinguish between the past and the future. We say that two timelike vectors  $v,w \in T_pM$ point in the same direction, and write  $v \sim w$, if $\langle v,w\rangle<0$.

\begin{figure}[h!]
\centering
\includegraphics[scale=0.5]{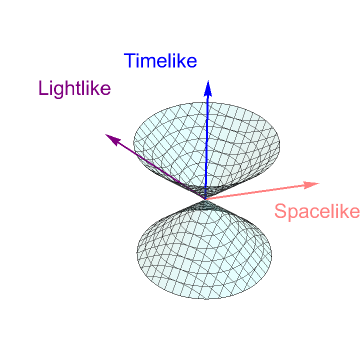}\caption{Light cones in the tangent space of spacetime.}%
\end{figure}
\begin{lemma}
The relation $\sim$ is an equivalence relation on $\tilde{C}_p$. Moreover, this equivalence relation has exactly two equivalence classes, which are the path connected components of the set of timelike vectors. Each of these equivalence classes is open and convex.
\end{lemma}
\begin{proof} The relationship is clearly symmetric and reflexive.
Let us prove that it is transitive. We first observe
that if $v,w$ are timelike vectors, then $\left\langle v,w\right\rangle
\neq0$. Suppose the contrary. We may assume that $v,w$ have norm one, and therefore, it would be possible to find an orthonormal basis $ \{ v,w,z,u\}$ for $T_pM$ such that $\langle v,v\rangle=\langle w,w\rangle =-1.$ This would contradict the fact that the metric has Lorentzian signature. In order to prove that the relation is transitive, it is enough to show that $v \sim w$ if and only if $v$ and $w$ are in the same path connected component of the space of timelike vectors. Suppose that $\langle v,w\rangle<0$ and consider the straight path $\theta(s)=sv+(1-s)w$. We claim that $\theta(s)$ is timelike for all $s \in [0,1]$. One computes
\[ \langle \theta(s),\theta(s)\rangle=s^2 \langle v, v\rangle + (1-s)^2 \langle w, w\rangle + 2 s(1-s) \langle v, w\rangle <0,\]
and concludes that $v $ and $w$ are in the same path connected component and that the path connected components are convex. On the other hand, suppose that $\beta(s)$ is a path of lightlike vectors from $v$ to $w$, and consider the continous function $f(s)=\langle v, \beta(s) \rangle$. Clearly, $f(0)=\langle v, v\rangle <0$. If $f(1)=\langle v,w \rangle >0 $, there 
would be some $s$ such that  $\langle v, \beta(s) \rangle =0$, which would contradict the statement above. One concludes that $v \sim w$. Let us show that there are exactly two equivalence classes. Since $v$ does not point in the same direction as $-v$, there are at least two classes. On the other hand, since $w$ must be related to either $v$ or $-v$, there are at most two equivalence classes. Since the function $\psi: T_pM \to \RR$  given by $v\mapsto \langle v,v \rangle$ is continuous, then $\tilde{C}_p= \psi^{-1}(-\infty,0)$ is open, and therefore its path connected components are open. 
\end{proof}

For a general spacetime manifold, the difference between the past and the future is an additional structure that needs to be specified. A \emph{time orientation} on $M$ is a locally constant choice of a future cone for each point $p \in M$. \emph{Locally constant} means that, for each $p \in M$, there is an open neighbourhood $U$ that contains $p$, and a vector field $X$ defined on $U$, such that $X(q)$ lies in the future cone for all $q \in U$. In case a connected Lorentzian manifold admits a time orientation, it admits exactly two of them. 

From now on  the word \emph{spacetime} will mean a Lorentzian manifold of dimension four with a fixed time orientation.
The existence of a time orientation on a Lorentzian manifold is a topological condition that is not always satisfied. Consider for example the cylinder $ T= S^1 \times \RR$ with Lorentzian metric:
\[ g(t,\theta)= \begin{pmatrix}
\sin^2(\frac{\theta}{2})-\cos^2(\frac{\theta}{2})&-2\cos(\frac{\theta}{2})\sin(\frac{\theta}{2})\\
-2\cos(\frac{\theta}{2})\sin(\frac{\theta}{2})&\cos^2(\frac{\theta}{2})-\sin^2(\frac{\theta}{2})
\end{pmatrix}.\]
It is a good exercise to check that this Lorentzian manifold is not time orientable.
\begin{figure}[H]
\centering
\includegraphics[scale=0.40]{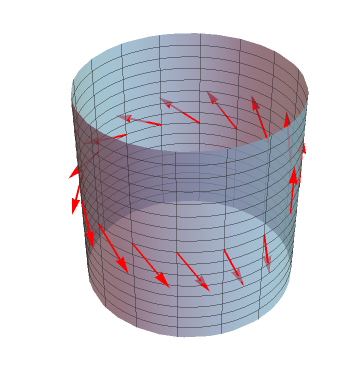}\caption{A cylinder with a Lorentzian metric that is not time orientable.}%
\end{figure}

 The \emph{worldline} of an object in $M$ is a curve $\gamma: I \rightarrow M$ such that $\gamma'(\tau)$ is timelike and belongs to the future cone in
$T_{\gamma(\tau)}M$. We will assume that the worldline is parametrized by proper time so that
\[ \langle \gamma'(\tau), \gamma'(\tau)\rangle =-c^2.\]
The\emph{ chronological future} of  an event $p\in M$, denoted $C^{+}_p(M),$ is the set of all points
that can be reached from $p$ along a piecewise smooth timelike curve that goes to the future. Similarly, the\emph{ chronological past} of $p$, $C^{-}_p(M),$ is set of all points $q$ that can be reached
from $p$ along a piecewise smooth timelike curve that goes to the past. The \emph{causal future} and \emph{causal past} of $p$, denoted $I^+_p(M)$ and $I^-_p(M)$, are the sets of events that can be reached from $p$ along piecewise smooth causal curves going to the future and past, respectively. Naturally, the time ordering is transitive:
\begin{itemize}
\item If $q \in C^{\pm}_p(M),$ then $C^{\pm}_q(M)\subseteq C^{\pm}_p(M).$
\item If $q \in I^{\pm}_p(M),$ then $I^{\pm}_q(M)\subseteq I^{\pm}_p(M).$
\end{itemize}

In Special Relativity, the geometry of Minkowski spacetime rules out the possibility of traveling to the past. There are no closed timelike curves in Minkowski spacetime. This avoids logical paradoxes that appear, for instance, once people are allowed to prevent their own birth. In order stay away from logical problems, it is natural to impose causality conditions on spacetime manifolds.
 There are several  causality conditions that are often imposed on a spacetime manifold. 
Some of the most common are the following:
\begin{itemize}
\item $M$ is \emph{chronological} if it does not admit closed timelike curves. 
\item $M$ is \emph{causal} if it does not admit closed causal curves. 
\item $M$ is \emph{strongly causal} if, for any $p\in M$ and any open neighbourhood $U$ that contains $p$, there is an open neoighboorhood of $p$,
$V \subseteq U$, such that any causal curve that starts and ends in $V$ is contained in $U$. 
\end{itemize}
The strong causality condition requires that causal curves are far from being closed. A causal curve that goes sufficiently far has to stay away from a neighborhood of the event where it started.
Clearly, a strongly causal spacetime is causal, and a causal spacetime is chronological.
We will always assume that spacetime manifolds are causal. 
\begin{figure}[H]
\centering
\includegraphics[scale=0.6]{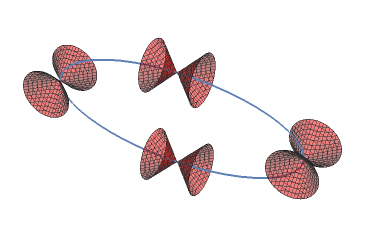}\caption{It seems better to avoid travel to the past.}%
\end{figure}
The condition of being chronological imposes strong restrictions on the spacetimes that arise in General Relativity.  In particular, we will see that compact spacetimes are not chronological.

\begin{lemma}
A Manifold $M$ admits a time orientable Lorentz metric if and only if it admits a non-vanishing vector field.
\end{lemma}
\begin{proof}
Suppose that $g$ is a time oriented Lorentz metric on $M$. There exists a covering $\{U_\alpha\}_{\alpha \in \mathcal{A}}$, and vector fields $X_\alpha$ defined on $U_\alpha$, such that
$X_\alpha(q)$ points to the future for all $q\in U_\alpha$. Choose a partition of unity $\rho_\alpha$, subordinate to the covering $\{U_\alpha\}_{\alpha \in \mathcal{A}}$, and define the vector field $X$ on $M$ by
\[ X(p)= \sum_{\{ \alpha: p\in U_\alpha \}}\rho_{\alpha}(p) X_\alpha(p).\]
Since the future cone at the tangent space of each point $p\in M$ is closed under addition, the vector field $X$ is non-vanishing.

Let us now prove the converse. Suppose that $X$ is a non-vanishing vector field on $M$. Fix a Riemannian metric $h$ on $M$ and define $H$ to be the distribution orthogonal to $X$ with respect to $h$. There is a unique Lorentz metric $g$  on $M$ for which $X$ is orthogonal to $H$,  $\langle X,X \rangle=-1$, and the restriction of $g$ and $h$ to $H$ coincide. This Lorentz metric is time orientable since one can declare that $X$ points to the future.
\end{proof}

It is a theorem of Heinz Hopf \cite{Hopf} that, for a compact manifold, the existence of a non-vanishing vector field is equivalent to the vanishing of the Euler characteristic. One concludes that compact manifolds  with vanishing Euler characteristic admit time orientable Lorentz structures. The proof of the following technical result can be found in Appendix \ref{openfuture}.
\begin{proposition}\label{open}
Let $M$ be a time oriented Lorentzian manifold. For any point $p \in M$, the sets $C^+_p(M)$ and $C^-_p(M)$ are open.
\end{proposition}

\begin{proposition}
A time oriented Lorentzian manifold that is compact is not chronological.
\end{proposition}
\begin{proof}
By proposition \ref{open}, the sets $C^+_p(M)$ form an open cover of $M$. Since $M$ is compact, there are points $p_1, \dots, p_n$ such that
$C^+_{p_1}(M), \dots ,C^+_{p_n}(M)$ cover $M$. We may assume that $n$ is minimal with that property. If $p_1 \in C^+_{p_j}(M) $ with $j>1$, then,
$C^+_{p_1}(M) \subseteq C^+_{p_j}(M)$, which would contradict the minimality of $n$. One concludes that $p_1 \notin  C^+_{p_2}(M)\cup \dots  \cup C^+_{p_n}(M)$. This implies that $p_1 \in C^+_{p_1}(M)$, so that there is a timelike closed curve in $M$.

\end{proof}

\section{Proper time, velocity and momentum }\label{proper time}

Suppose that Alice's worldline is the curve $\gamma(\tau): I \to M$, parametrized so that 
\[ \langle \gamma'(\tau),\gamma'(\tau)\rangle =-c^2.\]
 Alice's proper time is
\[ T=\frac{1}{c}\int_a^b  \sqrt{-\langle \gamma'(\tau),\gamma'(\tau)\rangle}d\tau=b-a. \]
It is the time that her clock will measure as she goes from $p=\gamma(a)$ to $q=\gamma(b)$.
Alice's $4$-velocity is the tangent vector to the worldline 
\begin{equation}\mathbf{u}(\tau)= \frac{d\gamma}{d\tau}=\gamma'(\tau).\end{equation}
Alice's $4$-acceleration is the covariant derivative of the velocity vector with respect to the Levi-Civita connection
\begin{equation} \mathbf{a}(\tau)= \nabla_{\gamma'(\tau) }\gamma'(\tau).\end{equation}
The  $4$-acceleration vanishes precisely when the worldline is a geodesic. This corresponds to the fact that, in the absence of forces, objects move
along geodesics in spacetime. Using the fact that the norm of the $4$-velocity is constant, we compute
\[ 0=\nabla_{\gamma(\tau)} \langle \gamma'(\tau),\gamma'(\tau)\rangle=2 \langle \nabla_{\gamma'(\tau)}\gamma'(\tau),\gamma'(\tau)\rangle=2\langle \mathbf{a}(\tau),\mathbf{u}(\tau)\rangle.\]
One concludes that the $4$-acceleration is orthogonal to the $4$-velocity. Since the $4$-velocity is timelike, this implies that the $4$-acceleration is not a timelike vector.
If Alice has rest mass $m$, then her $4$-momentum is
\begin{equation}
\mathbf{p}(\tau) =m \mathbf{u}(\tau)=m \gamma'(\tau).
\end{equation}
In the absence of forces, $\gamma(\tau)$ is a geodesic, and therefore
\begin{equation}
\nabla_{\gamma'(\tau)}\mathbf{p}(\tau)=0,
\end{equation}
the momentum is covariantly constant.

\label{coordenadas de fermi}
 \section{Geodesic motion and Fermi coordinates}
 
 In Special Relativity, the coordinate systems for different inertial observers are related by Lorentz transformations. In particular, any inertial observer has a system of coordinates
 where it is at rest, and the Minkowski metric takes the standard form. Let us discuss how such coordinates are described geometrically. Suppose that Alice moves along a timelike geodesic $\gamma(\tau)$ in Minkowski spacetime, and starts her clock at the event $p=\gamma(0)$. The vector $v_0 =\gamma'(\tau)$ is timelike and satisfies $\langle v_0, v_0 \rangle=-c^2$. Denote by $H$ the orthogonal complement to $v_0$ in  $T_p \MM$. Since $v_0$ is timelike, the restriction of the Minkowski metric to $H$ is euclidean. One can fix an orthonormal basis for $v_1, v_2 ,v_3$ for $H$. Using the vector space structure on $\MM$, the choice of bases $v_0,v_1,v_2, v_3$ provides coordinates $\varphi=(t,x)$ on $\MM$ by \[\varphi(p+tv_0+ x^1 v_1+ x^2 v_2+x^3 v_3) = (t,x^1,x^2,x^3).\]  In these coordinates, the Minkowski metric takes the standard form and Alice is at rest. Suppose that $(\overline{t},\overline{x})$ are other coordinates with the same properties. Since the event $p$, where the clock was started, corresponds to the origin in both systems of coordinates, they are related by a linear transformation $A$. Set $\overline{v}_0=A(v_0)$. Since Alice is at rest in both systems of coordinates, then
 \[ \gamma(\tau)=p+\tau v_0=p+\tau \overline{v}_0.\]
 One concludes that $v_0=\overline{v}_0$. Therefore the linear transformation $A$ takes the form:
 \[ A= \begin{pmatrix}1&0\\
 0& B
 \end{pmatrix},\]
 where $B$ is a linear isometry of $H$, the orthogonal complement to $v_0$. This means that the coordinate system is determined by the choice of an orthonormal basis on the
 orthogonal complement to the tangent space of the worldline. Once Alice fixes an event $p$ where she starts her clock, and an orthonormal basis for $H$, there is a unique set of coordinates where she is at rest and the Minkowski metric takes the standard form.\\
 
Let us now consider the situation in a general spacetime $M$ with metric $g$. Alice is moving along a timelike goedesic $\gamma(\tau)$, and she starts her clock at $p=\gamma(0)$. The velocity vector $\gamma'(0) \in T_pM$ is timelike, and therefore, the restriction of $g$ to $H= \gamma'(0)^\perp$ has euclidean signature. Given an orthonormal basis $v_1, v_2, v_3$ for $H$, there are unique vector fields $V_1(\tau), V_2(\tau),V_3(\tau)$ along $\gamma(\tau)$ such that
\[ \nabla_{\gamma'(\tau)}V_i(\tau)=0,\qquad V_i(0)=v_i.\]
 Moreover
 \[ \nabla_{\gamma'(\tau)}\langle  V_i(\tau),V_j(\tau)\rangle=\langle \nabla_{\gamma'(\tau)}V_i(\tau),V_j(\tau) \rangle +\langle V_i(\tau), \nabla_{\gamma'(\tau)}V_j(\tau) \rangle=0,\]
 which implies that \[\langle V_i(\tau), V_j(\tau)\rangle =\langle v_i, v_j \rangle =\delta_{ij}.\]
 Also, since $\gamma(\tau)$ is a geodesic, then
  \[ \nabla_{\gamma'(\tau)}\langle V_i(\tau),\gamma'(\tau)\rangle=\langle \nabla_{\gamma'(\tau)}V_i(\tau),\gamma'(\tau)\rangle +\langle V_i(\tau), \nabla_{\gamma'(\tau)}\gamma'(\tau) \rangle=0,\]
  so that $\langle V_i(\tau), \gamma'(\tau)\rangle=0$. One concludes that the metric takes the standard form on the frame $\gamma'(\tau), V_1(\tau),V_2(\tau), V_3(\tau)$. This frame can be used to construct coordinates, as follows. Let $W'$ be a sufficiently small neighborhood of the origin in $\RR^4$, and define the map
  $\phi: W' \rightarrow M$ by \begin{equation}
  \label{extension geodesica}
   \phi(t,x^1,x^2,x^3)= \exp(\gamma(t))(x^1 V_1(t)+x^2 V_2(t)+x^3 V_3(t)).
   \end{equation}
  The derivative of $\phi$ at the origin satisfies
  \[ D\phi(0)(\partial_t)=\gamma'(0), \qquad D\phi(0)(\partial_{x^i})=v_i.\]
  In particular, $D\phi(0)$ is nonsingular. By the implicit function theorem, there exists a neighborhood of zero $W \subseteq W'$ such that $\phi\vert_W$ is a diffeomorphism onto its image. We denote by $\varphi: U \rightarrow W$ the inverse function of $\phi$. These coordinates $\varphi=(x,t)$ are called \emph{Fermi coordinates} around $p$. 
  
\begin{figure}[H]
\centering
\includegraphics[scale=0.45]{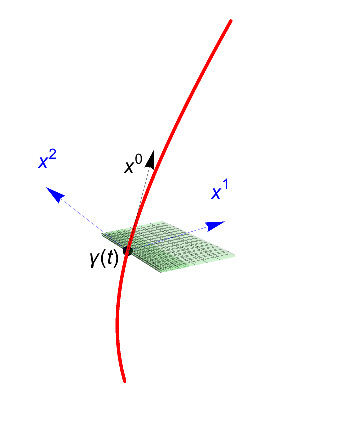}\caption{Fermi coordinates. The red line is a geodesic worldline. On each point of the worldline there is an orthonormal frame. The green surface represents the exponential of the vector space generated by the spatial components of the frame, which is the surface with constant $t=x^0$ coordinate.}
\label{fermicon}
\end{figure}

   The Fermi coordinates can be described in words as follows. One starts with a geodesic worldline $\gamma(\tau)$ which is parametrized by proper time. At a fixed point $p=\gamma(\tau_0)$ one fixes an orthonormal frame $\{\gamma'(\tau_0),v^1,v^2,v^3\}$, which is parallel transported to every other point of $\gamma(\tau)$. The point $q \in M$ with Fermi coordinates $(t,x^1,x^2,x^3)$ is determined as follows:
   First, one moves along this geodesic from $p$ to the point $z=\gamma(\tau_0+t)$. Then, by (\ref{existencias de geodesicas}), there is a unique geodesic $\alpha(s)$ with $\alpha(0)=z,$ and such that $$\alpha'(0)=x^1V_1(\tau_0+t)+x^2V_2(\tau_0+t)+x^3V_3(\tau_0+t).$$ The point $q$ is then determined by $\alpha(1)=q$. Figure \ref{fermicon} illustrates this construction. 
  
\noindent  
In Fermi coordinates, the worldline takes the form $\gamma(\tau)=(\tau ,0,0,0)$, so that Alice is at rest. Moreover, 
  \[ \partial_t(\gamma(\tau))=\gamma'(\tau), \qquad \partial_{x^i}(\gamma(\tau))=V_i(\tau),\]
so that, in Fermi coordinates,  the metric takes the standard form along the worldline.  
 In the flat case, the exponential map identifies 
 a neighborhood of $p$ with an open in Minkowski spacetime and the Fermi coordinates coincide with the inertial system of Special Relativity.
 Note however that, in the presence of curvature, there is no control on the form of the metric away from the worldline.\\

 The properties of Fermi coordinates can be summarized as follows
   
 \begin{lemma}
 Suppose that Alice moves along a timelike geodesic $\gamma(\tau)$. In a neighborhood of $p =\gamma(0)$, there are Fermi coordinates $\varphi=(t,x)$ such that:
 \begin{itemize}
  \item Alice is at rest in Fermi coordinates. This means that:
  \[ \varphi(\gamma(\tau))=(\tau,0,0,0).\]
  \item The metric takes the standard form on the worldline. That is:
   \[ g(\gamma(\tau))=\begin{pmatrix}
   -c^2&0&0&0\\
   0&1&0&0\\
   0&0&1&0\\
   0&0&0&1
   \end{pmatrix}.\]
   \item The Christoffel symbols vanish on the worldline:
   \[ \Gamma^a_{bc} (\gamma(\tau))=0.\]
   \end{itemize}
 \end{lemma}
 \begin{proof}
The construction of the coordinates makes clear that the first two properties are satisfied. Let us prove the last one. We will use the convention $x^0=t$. Fix $\tau$ and fix arbitrary real numbers $n^1,n^2,n^3$ and consider the path $\theta:I \rightarrow M$ given, in Fermi coordiantes, by
\[ \theta(s)=(\tau,n^1 s, n^2 s, n^3 s).\]
From the construction of the coordinates it follows that $\theta(s)$ is a geodesic. Therefore, the geodesic equations reduce to
\[ \sum_{i,j>0}\Gamma^k_{ij}(\theta(s)) n^i n^i=0.\]
These hold for all $s$ and all values of the $n^{i}$. Now set $s=0$. Then
$$
\sum_{i,j>0}\Gamma^k_{ij}(\gamma(s)) n^i n^i=0.
$$
Since the Christoffel symbols are independent of the $n^{i}$, we can conclude that
$$
\Gamma^k_{ij}(\gamma(\tau))=0
$$
for every $\tau$ and for $i,j > 0$. To deal with the remaining symbols, note that the vector fields $V_i(\tau)=\partial_{x^i}(\tau)$ are parallel along $\gamma(\tau)$,  and therefore
 \[ 0= \nabla_{\gamma'(\tau)} V_i(\tau)=\nabla_{\partial x^0} \partial_{x^i}(\gamma(\tau))=\sum_k \Gamma_{0i}^k(\tau(\gamma)) \partial_{x^k}.\]
 One concludes that
 \[ \Gamma_{0i}^k(\tau(\gamma))=0,\]
 as required.
\end{proof}
 
 \label{coordenadas generales}
 \section{Acceleration and Fermi-Walker coordinates}

 The construction of Fermi coordinates depends strongly on the fact that Alice was moving along a geodesic.
 The fact that $\gamma(\tau)$ is a geodesic guarantees that the frame $\gamma'(\tau), V_1(\tau),V_(\tau),V_3(\tau)$, obtained by parallel transport, remains 
 orthonormal for all $\tau$. In case $\gamma(\tau)$ is not a geodesic, the condition \[\langle V_i(\tau),V_j(\tau)\rangle=\delta_{ij}\] still holds, since parallel transport preserves angles.
 However, in general, there is no reason for the velocity vector $\gamma'(\tau)$ to remain orthogonal to $V_i(\tau)$. We see that, in the accelerated case, parallel transport does not provide an orthonormal frame along the worldline. The Fermi-Walker transport allows to construct such a frame for accelerated observers. Suppose that $\gamma: I \to M$ is a timelike curve parametrized by proper time. The Fermi-Walker connection on $\gamma^* TM$ is defined by
 \begin{equation}
 \label{FW transport}
 \nabla^{\mathrm{FW}}_{\gamma'(\tau)}V(\tau)=\nabla_{\gamma'(\tau)} V(\tau)-\frac{1}{c^2}\langle V(\tau), \mathbf{a}(\tau)\rangle \gamma'(\tau)+\frac{1}{c^2}\langle V(\tau) ,\gamma'(\tau)\rangle \mathbf{a}(\tau),
 \end{equation}
 for $V(\tau) \in \Gamma(\gamma^*TM)$. One says that $V(\tau)$ is Fermi-Walker parallel if it is covariantly constant with respect to the Fermi-Walker connection.
 In case $\gamma(\tau)$ is a geodesic, then $\nabla^{\mathrm{FW}}=\nabla$. The notion of Fermi-Walker transport is an alternative to the parallel transport that takes into account acceleration.
 \begin{lemma}
Let $\gamma$ be a timelike curve parametrized by proper time. Then:
\begin{enumerate}
\item The velocity vector $\gamma'(\tau)$ is Fermi-Walker parallel. 
\item Given a vector $v \in T_{\gamma(0)}M$, there is a unique Fermi-Walker parallel vector field $V(\tau)$, such that $V(0)=v$.
\item If $V(\tau)$ and $W(\tau)$ are Fermi-Walker parallel, then $\langle V(\tau),W(\tau)\rangle$ is constant.
\end{enumerate}
 \end{lemma}
 \begin{proof}
 For the first statement we compute
 \begin{align*}
 \nabla^{\mathrm{FW}}_{\gamma'(\tau)}\gamma'(\tau)&=\nabla_{\gamma'(\tau} \gamma'(\tau)-\frac{1}{c^2}\langle \gamma'(\tau), \mathbf{a}(\tau)\rangle \gamma'(\tau)+\frac{1}{c^2}\langle \gamma'(\tau) ,\gamma'(\tau)\rangle \mathbf{a}(\tau)\\
 &=\mathbf{a}(\tau)- \mathbf{a}(\tau)\\
 &=0.
 \end{align*}
The second statement is true for any connection. For the last statement we compute
\begin{align*}
\nabla_{\gamma'(\tau)}\langle V(\tau),W(\tau)\rangle  &=\langle \nabla_{\gamma'(\tau)}V(\tau),W(\tau)\rangle+\langle V(\tau),\nabla_{\gamma'(\tau)}W(\tau)\rangle\\
&=\langle \frac{1}{c^2}\langle V(\tau), \mathbf{a}(\tau)\rangle \gamma'(\tau)-\frac{1}{c^2}\langle V(\tau) ,\gamma'(\tau)\rangle \mathbf{a}(\tau) ,W(\tau)\rangle\\
&\quad\: +\langle V(\tau),\frac{1}{c^2}\langle W(\tau), \mathbf{a}(\tau)\rangle \gamma'(\tau)-\frac{1}{c^2}\langle W(\tau) ,\gamma'(\tau)\rangle \mathbf{a}(\tau)\rangle \\
&=0.\\
\end{align*}
 \end{proof}
 
 \begin{figure}[H]
\centering
\includegraphics[scale=0.4]{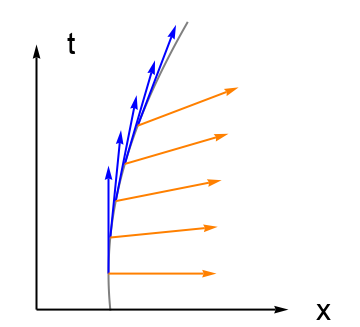}\caption{Fermi-Walker transport for hyperbolic motion.}%
\end{figure} 
 Using the Fermi-Walker connection one can imitate the construction of Fermi coordinates even in the case of accelerated motion.
 Suppose that Alice is moving along a timelike curve $\gamma(\tau)$, and she starts her clock at $p=\gamma(0)$. The velocity vector $\gamma'(0) \in T_pM$ is timelike, and therefore, the restriction of $g$ to $H= \gamma'(0)^\perp$ has euclidean signature. Given an orthonormal basis $v_1, \dots, v_3$ for $H$, there are unique Fermi-Walker parallel vector fields $V_1(\tau), V_2(\tau),V_3(\tau)$ such that \[ V_i(0)=v_i.\]
 Moreover, the metric takes the standard form on the frame $\gamma'(\tau), V_1(\tau),V_2(\tau), V_3(\tau)$. As in the geodesic case, the map
 \begin{equation}
 \label{FWC}
 \phi(t,x^1,x^2,x^3)= \exp(\gamma(t))(x^1 V_1(t)+x^2 V_2(t)+x^3 V_3(t))
 \end{equation}
 is a local diffeomorphism with inverse $\varphi: U \rightarrow W$. The coordinates $\varphi=(t,x)$ are called Fermi-Walker coordinates around $p$. 
 Unlike in the geodesic case, the Christoffel symbols are not all zero on the worldline. However, the following result holds.
 \begin{lemma}
 \label{FW lema}
 Suppose that Alice moves along a timelike curve $\gamma(\tau)$. In a neighborhood of $p =\gamma(0)$, the Fermi-Walker coordinates $\varphi=(t,x)$ satisfy the following properties:
 \begin{itemize}
  \item Alice is at rest in Fermi-Walker coordinates. This means that
  \[ \varphi(\gamma(\tau))=(\tau,0,0,0).\]
  \item The metric takes the standard form on the worldline. That is
   \[ g(\gamma(\tau))=\begin{pmatrix}
   -c^2&0&0&0\\
   0&1&0&0\\
   0&0&1&0\\
   0&0&0&1
   \end{pmatrix}.\]
   \item For $i,j>0$,
   \[ \Gamma^j_{0i} (\gamma(\tau))=\Gamma^0_{00} (\gamma(\tau))=\Gamma^k_{ij} (\gamma(\tau))=0.\]
   \end{itemize}
 \end{lemma}
 \begin{proof}
  The first two statements follow from the construction. Let us prove the last statement by the same computation as in the geodesic case. 
 Fix arbitrary real numbers $n^1,n^2,n^3$ and consider the path $\theta:I \rightarrow M$ given, in Fermi-Walker coordiantes, by
\[ \theta(s)=(\tau,n^1 s, n^2 s, n^3s).\]
From the construction of the coordinates it follows that $\theta(s)$ is a geodesic. From the construction of the coordinates it follows that $\theta(s)$ is a geodesic. Thus, the geodesic equations reduce to
\[ \sum_{i,j>0}\Gamma^k_{ij}(\theta(s)) n^i n^i=0.\]
Since these hold for all $s$ and all values of the $n^{i}$, setting $s=0$, we can conclude that
$$
\Gamma^k_{ij}(\gamma(\tau))=0
$$
for every $\tau$ and for $i,j > 0$.
Let us consider the remaining symbols. Using that the acceleration $\mathbf{a}(\tau)$ is orthogonal to the velocity we compute
\[ \mathbf{a}(\tau)=\nabla_{\gamma'(\tau)}\gamma'(\tau)=\sum_{k\geq 0}\Gamma_{00}^k (\gamma(\tau))V_i(\tau)=\sum_{k\geq 1}\Gamma_{00}^k (\gamma(\tau))V_i(\tau),\]
so that $\Gamma_{00}^0(\gamma(\tau))=0$. For $i>0$, the vector field $V_i$ is Fermi-Walker parallel and orthogonal to the velocity, therefore:
\[ \nabla_{\gamma'(\tau)}V_i(\tau)=\frac{1}{c^2}\langle V_i(\tau), \mathbf{a}(\tau)\rangle \gamma'(\tau).\]
We conclude that $\nabla_{\gamma'(\tau)}V_i$ is parallel to the velocity and therefore
\[ \Gamma^j_{0i} (\gamma(\tau))=0,\]
as required.
 \end{proof}
 \section{An observer moving with constant acceleration \label{acelerado}}

We want to analyze the dynamics of an observer $\overline{O}$ who moves in a
spaceship in Minkowski's spacetime in the direction of the $x$-coordinate of
the canonical observer $O$ with constant 4-acceleration. Let $\gamma (\tau
)=(t(\tau ),x(\tau ),0,0)$ be her worldline written in standard
coordinates. We fix a start point $p=\gamma (0)$ in such a way that the initial conditions are 
$t(0)=0$, $x(0)=a^{-1}$,  $t^{\prime }(0)=1$ and $x^{\prime }(0)=0$. The four acceleration of $\overline{O}$ is given by 
\begin{equation*}
\mathbf{a}(\tau )=t^{\prime \prime }(\tau )\partial _{t}+x^{\prime \prime
}(\tau )\partial _{x^{i}}.
\end{equation*}%
Since we are assuming that $\left\langle \mathbf{a}(\tau ),\mathbf{a}(\tau
)\right\rangle =a^{2}$ is constant, we deduce that 
\begin{equation*}
-t^{\prime \prime }(\tau )^{2}+x^{\prime \prime }(\tau )^{2}=a^{2}.
\end{equation*}%
On the other hand, since $\gamma (\tau )$ is parametrized by proper time we also have
that  $$t^{\prime }(\tau )^{2}-x^{\prime }(\tau )^{2}=1.$$  These conditions imply the
following system of ordinary differential equations%
\begin{align*}
t^{\prime }(\tau )^{2}-x^{\prime }(\tau )^{2}& =1, \\
-t^{\prime \prime }(\tau )^{2}+x^{\prime \prime }(\tau )^{2}& =a^{2}, \\
t(0)& =0, \\
x(0)& =a^{-1}, \\
t^{\prime }(0)& =1, \\
x^{\prime }(0)& =0.
\end{align*}
It easy to see that the solution of this system is given by 
\begin{equation}
\gamma (\tau )=\Big(\frac{\sinh (a\tau )}{a},\frac{\cosh (a\tau )}{a}\Big).
\end{equation}

\begin{figure}[tbh]
\centering
\includegraphics[scale=0.5]{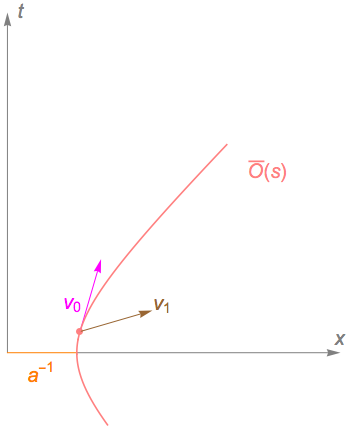}
\caption{Accelerated observer}
\end{figure}

Let us compute the Fermi-Walker coordinates for $\overline O$. The map $\phi $
is given by:
\begin{equation*}
\phi (t,x)=\exp(\gamma (t))\left(ax\gamma (t)\right)=\gamma (t)+ax\gamma (t),
\end{equation*}%
so that the standard coordinates $(t,x)$ are related to the Fermi-Walker
coordinates by 
\begin{align}
\begin{split}
t& =\frac{(1+\overline{x}a)\sinh (a\overline{t})}{a},  \label{fwacc} \\
x& =\frac{(1+a\overline{x})\cosh (a\overline{t})}{a}. 
\end{split}
\end{align}
Therefore: 
\begin{align*}
dt& =\sinh (a\overline{t})d\overline{x}+(1+a\overline{x})\cosh (a\overline{t}%
)d\overline{t}, \\
dx& =\cosh (a\overline{t})d\overline{x}+(1+\overline{x}a)\sinh (a\overline{t}%
)d\overline{t}.
\end{align*}%
In standard coordinates, the Minkowski metric is
\begin{equation*}
g=-dt\otimes dt+dx\otimes dx.
\end{equation*}%
So that, in Fermi-Walker coordinates
\begin{equation*}
g=-\left(1+\frac{\overline{x}}{a}\right)^{2}d\overline{t}\otimes d\overline{t}+d%
\overline{x}\otimes d\overline{x}.
\end{equation*}
As expected, the metric takes the standard form when $\overline{x}=0$. In
Fermi-Walker coordinates, the non-zero Christoffel symbols are
\begin{equation}
\Gamma _{01}^{0}=\Gamma _{10}^{0}=\frac{a+x}{a^{2}},\qquad \Gamma _{00}^{1}=%
\frac{(a+x)}{a^{2}}.
\end{equation}%
Let us consider the acceleration vector $\mathbf{a}(\tau )$ which, in
standard coordinates takes the form 
\begin{equation*}
\mathbf{a}(\tau )=a\sinh (a\tau )\partial _{t}+a\cosh (a\tau )\partial _{x}.
\end{equation*}%
Using equation (\ref{fwacc}), we conclude that 
\begin{equation*}
\frac{\partial t}{\partial \overline{x}}=\sinh (a\overline{t}),\qquad \frac{%
\partial x}{\partial \overline{x}}=\cosh (a\overline{t}).
\end{equation*}%
This implies that 
\begin{equation*}
\partial _{\overline{x}}=\frac{\partial t}{\partial \overline{x}}\partial
_{t}+\frac{\partial x}{\partial \overline{x}}\partial _{x}=\frac{\mathbf{a}%
(\tau )}{a},
\end{equation*}%
which is equivalent to 
\begin{equation*}
\mathbf{a}(\tau )=a\partial _{\overline{x}}.
\end{equation*}%
In a rocket that accelerates at a constant
rate $g=9.8 \,\mathrm{m /s^2}$ equal to the gravitation acceleration,  the
passengers will feel as if they were on the Earth's surface. This 3-force
would be locally indistinguishable from a fictitious gravitation
force. This remarkable observation is known as the \emph{%
equivalence} \emph{principle}, a fundamental principle that led Einstein to
the formulation of his General Theory of Relativity. We will come back to
this discussion in detail  in \S \ref{e.p.}.  
\begin{figure}[H]
\centering
\includegraphics[scale=0.4]{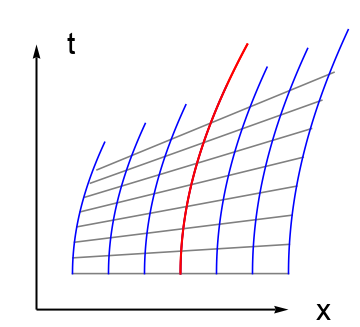}
\caption{Fermi-Walker coordinates for hyperbolic motion in Minkowski
spacetime.}
\end{figure}

\section{A Journey to Kepler 22-b\label{viaje}}

In a distant future humans may have developed the technology to
explore outer space beyond the limits of our own solar system. We may imagine, centuries from now, a scouting party
in search of a new home for humanity.  Kepler-22b, an
exoplanet discovered in 2011 by the Kepler space telescope, an Earth-like
celestial body located 600 light years away from our planet, is an
ideal place to settle down. Its sun, a yellow dwarf of the northern constellation of
Cygnus, provides the planet with light. Its size, 2.5 times that
of Earth, suggests that it holds an atmosphere. According to
density estimates, Kepler 22-b might also posses vast oceans of water. Its
temperature is estimated between 22 and 27 degrees Celsius. Years in that
remote place last 289 days.

The adventurous travel could go as follows. The
crew starts their trip at some space station located $1/a%
=c^{2}/g\simeq$ 0.97 light years away from
Earth. As usual, $g=9.8 \,\mathrm{m /s^2}$ denotes the
acceleration of gravity on the surface of the Earth. A few minutes after departure the spaceship would have reached a velocity of
several thousand kilometers per hour and continues accelerating steadily at
rate $g.$ In the first hour the
rocket will have gained a tremendous speed, around 120000 $\mathrm{km/h}$.
Inside the probe the crew experiences a comfortable atmosphere. They appear to
be motionless, everything seems to be at rest. The astronauts
experience no forces besides a fictitious gravity that feels identical to that
on Earth. According to plan, they will be reaching a maximum velocity of
99.99$\%$ the speed of light, 68.6 terrestrial years after
departure. But this amounts to only four years nine months and eighteen
days, as recorded in the spaceship logbook. By then, they will already be
67 light-years away from Earth. At this moment the powerful engines fed with
the little available hydrogen in interstellar space will stop, and will not be
ignited again until the final approach to the planet.

During the next two years measured in proper time, they will experience total weightlessness.  At that fantastic
speed normal light coming from the stars registers a frequency outside
the visible spectrum. But infrared radiation and other low frequency
electromagnetic waves coming from approaching celestial objects have now become
visible. This is also the case for ultraviolet and other high frequency
radiation coming from receding stars. This phenomenon is discussed in Section \S\ref{dopler}. One may ask if even at this
incredible speed the travelers would necessarily take more than six hundred years to
reach their destination. This would certainly be the span of time recorded on
terrestrial calendars. But not for them! Einstein's theory predicts that, when
arriving at their new home, each crew member will have aged only about 6.8 years.

The explorers will stay in Kepler-22b for a decade, building a space station and
the foundations of the new human colony. Once the mission is completed,
they will undertake their journey back home. When they get back to Earth, the
former young astronauts will be middle-aged adults after a long journey of 23.8 years, according to the spaceship's calendar. However, more than
12 hundred years will have elapsed here on Earth.

Let us provide some calculations to support the story. In the coordinates of the space center on
Earth, $(t,x,y,z),$ and in classical units, the equation of motion will be
\[
x^{2}-c^{2}t^{2}=\left(\frac{g}{c^{2}}\right)^{-2}=\left(\frac{c^{2}}{g}\right)^{2}.
\]
 From this we obtain $x(t)=\sqrt{c^{2}t^{2}+(c%
^{2}/g)^{2}}$. On the other hand, the spaceship's velocity measured
from Earth is \[v(t)=\frac{dx}{dt}=\frac{c^{2}t}{\sqrt{c^{2}%
t^{2}+(c^{2}/g)^{2}}}.\] Solving for $t$, one observes that
the spaceship will reach a speed of \textrm{0.9999}$c,$ when $t_{0}=$
\textrm{2.16}$\times10^{9}$ seconds, which is \textrm{68.63} years.
However, the proper time for the crew will just be
\[%
{\int_{0}^{t_{0}}}
\sqrt{1-\frac{v(s)^2}{c^2}}ds\approx\text{\textrm{1.52}}%
\times10^{8}\text{ seconds,}%
\]
equivalent to $\mathrm{4.8}$ years in the spaceship's calendar. By then they
will have traveled $x(t_{0})=\mathrm{6.4}\times10^{17}$ \textrm{m}, which is $\mathrm{67.6}$ light-years. The total time $t_{1}$ it takes to
reach Kepler-22b will approximately be $\mathrm{1.89}\times10^{10}$ seconds.
That amounts to $\mathrm{600.96}$ years. But the proper time will just be
\[%
{\int_{0}^{t_{1}}}
\sqrt{1-\frac{v(s)^2}{c^2}}ds\approx\mathrm{2.14}\times10^{8}\text{
seconds,}%
\]
or $\mathrm{6.8}$ years. Hence, when they return to Earth, each member of the
crew will be $\mathrm{23.8}$ years older.

\section{Redshift and blueshift}

In this section we will examine the discussion in \S\ref{sec: Doopler effect} from a different perspective.
We claimed that, during most of the journey to Kepler 22-b 
the sky would look strange to a passenger
on the spaceship. To see why this is the case, consider a photon $P$
whose world-line is given by $\beta:I\rightarrow\MM$, where
$\beta(s)=(-b+\hslash\omega_{0}s,\hslash\omega_{0}s,0,0),$ $b>0.$ The
scalar $\omega_{0}$ represents the angular frequency as measured by an
inertial observer on Earth $O(\tau)=(\tau,0,0,0)$ at $q_{0}$. Hence, the energy of the photon, as
measured by $O$, is equal to $E_{0}=-\left\langle \beta^{\prime}(0),\mathbf{u}%
\right\rangle =\hslash\omega_{0}$, where $\mathbf{u}=\partial_{t}$.

On the other hand, the energy measured by the accelerated observer
$\overline{O}$ at $q$ would be $E_{1}=-\left\langle \beta^{\prime}(s
_{0}),\mathbf{v}\right\rangle $, where $s_{0}$ is the value of the
parameter for which $\beta(s_{0})=q,$ and $\mathbf{v}$ is the
$4$-velocity of $\overline{O}$ at $q$. Thus,
\[
E_{1}=\hslash\omega_{0}(\cosh(as_{0})-\sinh(as_{0}))=\hslash\omega
_{0}e^{-as_{0}},
\]
where $\overline{O}(s_{0})=q.$ Therefore, the frequency measured at $q$ by
$\overline{O}$ would be $\omega_{1}=E_{1}/\hslash=\omega_{0}e^{-(\mathrm{a}%
s_{0})}$. For $s_{0}>0$ this frequency is less that $\omega_{0}$. Hence, light
from a star that is moving away from the spaceship will look red shifted. On
the contrary, when the spaceship approaches a star the light would be blue
shifted, as $\omega_{1}>\omega_{0}.$

The value of $s_{0}$ can be determined by solving the system $-b+\hslash
\omega_{0}s=a^{-1}\sinh(as)$, $\hslash\omega_{0}s=a^{-1}\cosh(as)$.
\ For this, we note that%
\[
(\hslash\omega_{0}s)^{2}-(-b+\hslash\omega_{0}s)^{2}=a^{-2}(\cosh
^{2}(as)-\sinh^{2}(as))=a^{-2}.
\]
One obtains
\begin{equation}
s_{0}=\frac{1+a^{2}b^{2}}{2a^{2}b\omega_{0}},\text{ \ }s_{0}=\frac{1}%
{a}\mathrm{arccosh}\left(  \frac{a^{2}b^{2}+1}{2ab}\right).  \label{E42}%
\end{equation}

The spectrum of frequencies of visible light varies in the range
of $\omega_{R}=\mathrm{3.8}\times10^{14},$ the frequency of red light, and
$\omega_{V}=\mathrm{7.16}\times10^{14}$, the frequency of violet light$.$ As
the astronauts travel farther away, the light of our sun will become dimmer,
and each time more shifted toward the red side of the spectrum. It will remain
visible while $\omega_{V}$, the maximum frequency visible light, does not drop
below the value $\omega_{R}$, that is, while $\omega_{V}e^{-(\mathrm{a}s_{0}%
)}\geq\omega_{R}$, or equivalently, while $e^{-(\mathrm{a}s_{0})}\geq1/2.$
From this, we need $s_{0}\leq\ln(2)/\mathrm{a}$. For light coming from
approaching stars we similarly see that $-s_{0}\geq-\ln(2)/\mathrm{a.}$ From Equation (\ref{E42}) we conclude that only while $ s_{0}
\leq\mathrm{6.3}\times10^{15}$\textrm{ss} will the astronauts be able to see
the light which as emitted in the usual visible spectrum. Measured in years, this corresponds to
$0.67$ years. Roughly after eight months of travel to Kepler 22-b, measured in proper time, the light they will \emph{see} coming from the Earth will have been emitted
as ultraviolet radiation.


\section{Fermi-Walker transport for circular motion}
We will now calculate the Fermi-Walker transport for an observer $O
$ who rotates in a circle of radius $r=1$ with constant angular velocity $0<\omega <1$, where we shall assume $c=1$. In standard coordinates $O$'s worldline is given by
 $\gamma (t )=(t,\cos (t\omega ),\sin (t\omega ),0).$

Our first step is to construct at each point $p=\gamma(\tau)$ an orthonormal frame $L(\tau )=\{u_{0}(\tau ),u_{1}(\tau ),u_{2}(\tau ),u_{3}(\tau )\}.$
In terms of his proper time $\tau $ we have that $t(\tau )=\lambda \tau ,$
where $\lambda =1/\sqrt{1-\omega ^{2}}$ so that the four velocity
\[
u_0(\tau )=(\lambda ,-\lambda \omega \sin (\lambda \omega \tau
),\lambda \omega \cos (\lambda \omega \tau ),0)
\]%
has norm $-\lambda ^{2}+\lambda ^{2}\omega ^{2}=-1$. On the other hand, the four acceleration
is: 
\[
\mathbf{a}(\tau )=(0,-\lambda ^{2}\omega ^{2}\cos (\lambda \omega \tau
),-\lambda ^{2}\omega ^{2}\sin (\lambda \omega \tau )).
\]%
Let's denote by $u_2(\tau)$ the normalized 4-acceleration. Since $u_{3}(\tau )=(0,0,0,1)$ can always be taken as part of $L(\tau )$, by
taking a cross product (taking into account the Lorentz signature) of this vector with  $u_{2}(\tau )$ we may construct a spacelike unitary vector $$u_{1}(\tau
)=(\lambda \omega ,-\lambda \sin (\lambda \omega \tau ),\lambda \cos
(\lambda \omega \tau ),0)$$ so that \begin{align}\label{vectores}
\begin{split}
u_{0}(\tau ) &=(\lambda ,-\lambda \omega \sin (\lambda \omega \tau
),\lambda \omega \cos (\lambda \omega \tau ),0) ,  \\
u_{1}(\tau ) &=(\lambda \omega ,-\lambda \sin (\lambda \omega \tau
),\lambda \cos (\lambda \omega \tau ),0), \\
u_{2}(\tau ) &=(0,-\cos (\lambda \omega \tau ),-\sin (\lambda \omega \tau ),0),  \\
u_{3}(\tau ) &=(0,0,0,1), 
\end{split}
\end{align}
is a Lorentz frame at each point of $\gamma (\tau ).$

Let $V(\tau )$ be any spatial vector that we want to transport along $\gamma
(\tau )$. If we express $V(\tau )=\sum\limits_{i}V^{i}(\tau )u_{i}(\tau )$
the equation (\ref{FW transport}) becomes:%
\begin{equation}
\frac{dV(\tau )}{d\tau }=\sum\limits_{i}\frac{d V^{i}(\tau )}{d\tau }%
u_{i}(\tau )+\sum\limits_{i}V^{i}(\tau )\frac{du_{i}(\tau )}{d\tau }=\left\langle V(\tau ),\mathbf{a}(\tau )\right\rangle u_{0}(\tau ),
\label{eeee1}
\end{equation}%
since the covariant derivative coincides with the ordinary derivative in
Minkowski flat spacetime. Now, a straightforward computation shows that 
\begin{eqnarray*}
\frac{du_{0}(\tau )}{d\tau } &=&\lambda ^{2}\omega ^{2}u_{2}(\tau ), \\
\frac{du_{1}(\tau )}{d\tau } &=&\lambda ^{2}\omega u_{2}(\tau ), \\
\frac{du_{2}(\tau )}{d\tau } &=&\lambda ^{2}\omega ^{2}u_{0}(\tau )-\omega
\lambda ^{2}u_{1}(\tau ).
\end{eqnarray*}%
Therefore (\ref{eeee1}) becomes%
\begin{eqnarray*}
\frac{dV(\tau )}{d\tau } &=&\omega ^{2}\lambda ^{2}V^{2}(\tau )u_{0}(\tau
)+\left( \frac{dV^{1}(\tau )}{d\tau }-\omega \lambda ^{2}V^{2}(\tau )\right)
u_{1}(\tau ) \\
&&+\left( \frac{dV^{2}(\tau )}{d\tau }+\omega \lambda ^{2}V^{1}(\tau )\right)
u_{1}(\tau )+\frac{dV^{3}(\tau )}{d\tau }u_{3}(\tau ) \\
&=&\left\langle V(\tau ),\mathbf{a}(\tau )\right\rangle u_{0}(\tau ) \\
&=& \lambda
^{2}\omega ^{2}V^{2}(\tau )u_{0}(\tau ).
\end{eqnarray*}
From this we get the following system of differential equations:%
\begin{eqnarray*}
\frac{dV^{1}(\tau )}{d\tau } &=&\omega \lambda ^{2}V^{2}(\tau ), \\
\frac{dV^{2}(\tau )}{d\tau } &=&-\omega \lambda ^{2}V^{1}(\tau ), \\
\frac{dV^{3}(\tau )}{d\tau } &=&0.
\end{eqnarray*}%
The first two equations can be solved by taking the derivative of the first
equation and substituting it into the second, and then solving the
corresponding second order linear equation. In this way one can obtain as general
solution%
\begin{eqnarray*}
V^{1}(\tau ) &=&c_{1}\sin (\omega \lambda ^{2}\tau )+c_{2}\cos (\omega
\lambda ^{2}\tau ), \\
V^{2}(\tau ) &=&c_{1}\cos (\omega \lambda ^{2}\tau )-c_{2}\sin (\omega
\lambda ^{2}\tau ), \\
V^{3}(\tau ) &=&c_{3}.
\end{eqnarray*}%
When $V(\tau )$ corresponds to each one of the spatial vectors $u_{i}(\tau)$, in each case we determine the constant $c_{1},c_{2},c_{3}$ using
the initial condition $V(0)=u_{i}(0)$. From this we calculate the the
Fermi-Walker transport $U_{i}(\tau )$ of each $u_{i}(\tau )$:
\begin{eqnarray*}
U_{1}(\tau ) &=&\cos (\omega \lambda ^{2}\tau )u_{1}(\tau )-\sin (\omega
\lambda ^{2}\tau )u_{2}(\tau ), \\
U_{2}(\tau ) &=&\sin (\omega \lambda ^{2}\tau )u_{1}(\tau )+\cos (\omega
\lambda ^{2}\tau )u_{2}(\tau ), \\
U_{3}(\tau ) &=&u_{3}(\tau ).
\end{eqnarray*}%
By substituting the values in (\ref{vectores}) one gets in the canonical
coordinates of Minkowski spacetime the expressions 
\begin{eqnarray*}
U_{1}(\tau ) &=&\big(\lambda \omega \cos (\omega \lambda ^{2}\tau ),
-\lambda \cos (\omega \lambda ^{2}\tau )\sin (\omega \lambda \tau )+\sin
(\omega \lambda ^{2}\tau )\cos (\omega \lambda \tau ),\\&&\:\: \lambda \cos
(\omega \lambda ^{2}\tau )\cos (\omega \lambda \tau ) +\sin (\omega \lambda
^{2}\tau )\sin (\omega \lambda \tau ),0\big), \\
U_{2}(\tau ) &=&\big(-\lambda \omega \sin (\omega \lambda ^{2}\tau ),%
\lambda \sin (\omega \lambda ^{2}\tau )\sin (\omega \lambda \tau )+\cos
(\omega \lambda ^{2}\tau )\cos (\omega \lambda \tau ),\\&&\:\: \sin (\omega
\lambda \tau )\cos (\omega \lambda ^{2}\tau )-\lambda \cos (\omega \lambda
\tau )\sin (\omega \lambda \tau ),0\big), \\
U_{3}(\tau ) &=&\big(0,0,0,1\big).
\end{eqnarray*}%
In standard units we may rewrite these vectors by replacing $\omega 
$ by $\omega /c$ and $\lambda $ by $\lambda =1/\sqrt{1-(\omega /c)^{2}}.$ 
\begin{figure}[H]
\centering
\includegraphics[scale=0.47]{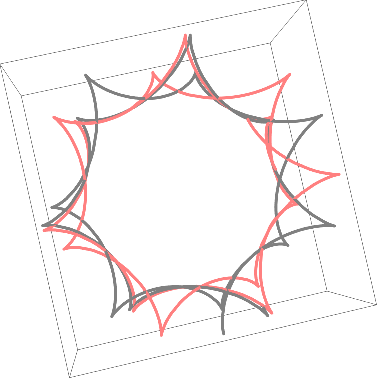}\caption{The pink and gray lines represent the Fermi-Walker transport of spacelike orthogonal vectors for circular motion.}%
\end{figure}

\section{The physical meaning of coordinates}

The physical meaning of coordinates in Relativity is a subtle issue. Not
every system of coordinates for a spacetime 4-manifold provides
\emph{ true  time and spatial coordinates}
in the sense that these numbers correspond to
measurements an observer would assign to events in a neighborhood of his lab. 

By a lab we mean a clock that the observer \emph{carries with him} to measure
his proper time, and three mutually perpendicular rods (three spatial
axes) that will determine the spatial coordinates of events in his neighborhood. By using light rays and his clock he can calibrate his spatial units. This
is done by setting a unit of distance as the length traveled by any photon
in a unit of time, according to his clock. The observer
verifies that every photon that crosses his lab in any spatial
direction must also register a speed equal to one. We will consider coordinates where the observer remains at the origin, so that his worldline is constant in space.

There is one more desirable condition his coordinates must have. To
understand this, let us imagine that our scientist is locked inside a building
that sits at the north pole of the Earth. Suppose he has already chosen three calibrated rods that are fixed to the walls of his lab, where he has set a Foucault pendulum that swings in his lab's $x$-$z$ plane. After a few minutes he would observe how the pendulum's oscillating
plane changes slowly. He attributes this motion to some \emph{unknown
forces}. One night he decides to step outside of his
lab. He looks at the sky and notices \ that the firmament is slowly rotating around the northern star, and that his pendulum is actually swinging in a fixed plane with respect to the distant stars. It is then that he
realizes that those mysterious forces are indeed fictitious, due to the rotation of Earth that is dragging his $x$ and $y$ axes. To avoid this nuisance, he decides to choose a new set of mutually perpendicular roads that are not attached to the walls but  articulated at the origin so that they can rotate freely. In order to keep their axes motionless with respect to the fixed stars he uses three gyroscopes that keep each of his spatial axis pointing in the same direction in space.

Attaching gyroscopes to each axis is a physical procedure to transport his frame of reference along his worldline in such a way that his spatial axes are only allowed to \emph{change in the direction of time}. This is because the angular momentum of each axis is preserved (and so it is the orthogonality of the spatial coordinates). If an unexpected earthquake were to  momentarily shake his lab, the gyroscopes ensure his spatial axes would stay still with respect to an inertial observer. 

By only changing in the time direction we mean the following. Once he fixes an initial frame at a point $p=\gamma(\tau_0)$, say, $$F=\{\gamma'(\tau_0), u_{i}(p)\},$$  he transports $F$ along his worldline in such a way that
\[
\nabla _{\gamma'(\tau)}u_{i}(\tau )=\eta (\tau )\gamma'(\tau).
\]%
We shall see below that this last condition is equivalent to using the Fermi-Walker transport (\ref{FW transport}) to transport $F$, if we guarantee that all vectors in each transported frame remain pairwise orthogonal. 

On the other hand, the orthogonality of each frame can be physically interpreted as choosing coordinates so that the metric looks at every point of $\gamma(\tau)$ like that of an inertial observer. This mathematical property, on the other hand, corresponds to choosing coordinates where the speed of any photon is equal to 1.  The following definition summarizes the properties of those coordinates that naturally originate as the time and space measurements an observer performs in a neighborhood of his lab.

\begin{definition}
\label{observer def}Let $O$ be any observer in space-time whose worldline we
denote by $\gamma (\tau )$. We will say that coordinates $x=(x^{a})$ defined
in a neighborhood $U_{p}$ of an event $p=\gamma(\tau _{0})$ have \emph{physical
meaning} if the following conditions are satisfied:

\begin{itemize}
\item The observer $O$ moves forward in time but he does not move spatially
with respect to his frame. This means that
\[
t(\gamma (\tau ))=\tau ,\qquad x^{i}(\gamma (\tau ))=0,
\]%
where $\tau $ is his proper time.

\item The spatial axes are orthogonal, that is, if $u_{i}(\gamma (\tau ))=\partial _{x^{i}}$, then $
\left\langle u_i,u_j\right\rangle =\delta_{ij}$ for $
i,j > 0$

\item Each spatial vector $v(\gamma(\tau))=\sum_{i}v^{i}u_i(\gamma(\tau))$ in $T_{\gamma
(\tau )}M$ is spacelike. 
\item The speed of any photon that crosses the laboratory is $c=1$. That is, if $\beta (s)$ represents the
wolrdline of this photon, and if $\beta (0)=p,$ $t(s)=x^{0}(\beta (s))$ and $%
\beta ^{i}(s)=x^{i}(\beta (s)),$ then its speed, as measured by $O,$ is
equal to $1$:

\[
c=1=\sqrt{\sum\limits_{i}\left( \frac{d\beta ^{i}}{dt}(0)\right) ^{2}}.
\]

\item Spatial directions only change in
the direction of time. 
\[
\nabla _{\gamma'(\tau)}u_{i}(\tau )=\eta_i (\tau )\gamma'(\tau ).
\]%

\end{itemize}
\end{definition}


\begin{lemma}
\label{lema largo}

Suppose $x=(x^{a})$ are coordinates for $O$ around a point $p=\gamma(\tau_0)$ satisfying the conditions of lemma \ref{observer def}. Then it is always possible for $O$
to send a light signal in any spatial direction $v=\sum_{i}v^{i}u_i(\gamma(\tau))$ of his choice. This
means that there is a null geodesic $\beta (s)=(t(s),b^{i}(s)),$ with $\beta
(0)=p,$ such that $v^{i}=\frac{d\beta ^{i}}{ds}(0).$ Moreover, we can choose $\beta $
such that $t^{\prime }(0)>0.$
\end{lemma}

\begin{proof}
We recall that given any null vector $w$ there is a unique null geodesic (up
to affine reparametrization) $\beta $ with $\beta (0)=p$ and $\beta ^{\prime
}(0)=w$. The first condition in (\ref{observer def}) implies that $\gamma ^{\prime
}(\tau)=\partial_t(\gamma(\tau))$ and therefore $\left\langle \partial_t(p),\partial_t(p)\right\rangle =-1.$ Hence, it suffices to find a
null vector of the form $w=\zeta \partial_t(p)+v$, with $\zeta >0.$

We are looking for a positive real $\zeta $ such that $\left\langle
\zeta \partial_t(p)+v,\zeta \partial_t(p)+v\right\rangle
=0$. This means that we must solve the following quadratic equation for $%
\zeta $: 
\begin{equation}
-\zeta ^{2}+2\zeta \left\langle \partial_t(p),v\right\rangle
+\left\vert v\right\vert ^{2}=0.  \label{ecuadratic}
\end{equation}%
But 
\[
4\left\langle \partial_t(p),v\right\rangle ^{2}+4\left\vert v\right\vert
^{2}>0,
\]%
thus (\ref{ecuadratic}) has two different real roots, $\zeta _{1},\zeta
_{2}.$ Since their product equals $-\left\vert v\right\vert ^{2},$ $\zeta
_{1}$ and $\zeta _{2}$ must have opposite signs and consequently we can
choose a positive root for (\ref{ecuadratic}).
\end{proof}

\begin{lemma}
\label{final}Let $O$ be any observer. Let $x=(x^{a})$ be a system of
coordinates defined in a neighborhood $U_{p}$ of an event $p=\gamma (\tau
_{0})$ satisfying the conditions in lemma \ref{observer def}. In these
coordinates, the matrix representing the metric at $p$ takes the standard
form: 
\[
g(p)=\left( 
\begin{array}{cccc}
-1 & 0 & 0 & 0 \\ 
0 & 1 & 0 & 0 \\ 
0 & 0 & 1 & 0 \\ 
0 & 0 & 0 & 1%
\end{array}%
\right) .
\]
\end{lemma}

\begin{proof}
By the previous lemma $O$ can send a light signal that crosses $O$ in the
direction of $\partial_{x^i}(p)$. That is, there is a null geodesic $\beta
(s)=(t(s),b^{i}(s))$ with $\beta (0)=p,$ and such that 
\[
\beta ^{\prime }(0)=\zeta \gamma'(\tau_0)+\partial_{x^i}(p),
\]
with  $\zeta > 0$. Since $\beta ^{\prime }(0)$ is a null vector 
\begin{equation}
0=\left\langle \beta ^{\prime }(0),\beta ^{\prime }(0)\right\rangle =\zeta
^{2}g_{00}+2\zeta g_{0i}+g_{ii}.  \label{d1}
\end{equation}%
On the other hand, the speed of this photon, as measured by $O$ is equal to $c=1$, and therefore 
\[
\left\vert \frac{d\beta ^{i}}{dt}(0)\right\vert =1.
\]%
By the chain rule 
\[
\frac{d\beta ^{i}}{dt}(0)=\frac{(d\beta ^{i}/ds)(0)}{(dt/ds)(0)}=\frac{1}{%
\zeta }.
\]%
Thus, $\zeta =1.$ 
Similarly, a photon that crosses $p$ in the spatial direction of $-\partial_{x^i}(p)$ moves along a geodesic $%
\alpha (s)$, with $\alpha (0)=0,$ and $\alpha ^{\prime }(0)=\zeta ^{\prime
}\gamma'(p)-\partial_{x^i}(p)$, with $\zeta ^{\prime }=\left\vert
-\partial_{x^i} (p)\right\vert =\zeta=1 .$ Hence, as in (\ref{d1}) 
\begin{equation}
(\zeta ^{^{\prime }})^{2}g_{00}-2\zeta ^{^{\prime }}g_{0i}+g_{ii}=0.
\label{d2}
\end{equation}%
Subtracting (\ref{d1}) from (\ref{d2}) one obtains $g_{0i}=g_{i0}=0.$ 
 The fact that $g_{ij}=0,$
  for $i\neq j$ is the orthogonality condition of the
spatial axes. 
\medskip

Finally, we show that $g_{00}=-1$. For this we notice that
\[ \langle \partial_t(\gamma(\tau)),\partial_t(\gamma(\tau))\rangle=\langle \gamma'(\tau),\gamma'(\tau)\rangle=-1,\]
from which the result follows.
\end{proof}

Now we are ready to provide a physical characterization of the Fermi-Walker
coordinates.

\begin{proposition}\label{charFW}
Let $O$ be an observer whose worldline is given by $\gamma (\tau )$. Let $%
x^{a}$ be a system of coordinates with physical meaning for $O$ defined in a
neighborhood of a point $p=\gamma (\tau _{0})$. Then, the first four conditions in lemma \ref%
{observer def} imply that the metric takes the Minkowski form at every point of the worldline. The last condition implies that in a neighborhood of $p$ the orthonormal
frame $F=\{\partial_{x^a}\}$ is Fermi-Walker transported along the worldline. 

\begin{proof}
We use that $\langle \partial_{x^i}(\gamma(\tau)),\gamma'(\tau)\rangle =0$, to compute:
\begin{eqnarray*}
0 &=& \frac{d}{d\tau} \langle \partial_{x^i}(\gamma(\tau)),\gamma'(\tau)\rangle \\
&=&\langle \nabla _{\gamma'(\tau )}\partial_{x^i}(\gamma(\tau) ),\gamma'(\tau
)\rangle +\langle \partial_{x^i}(\gamma(\tau) ),\nabla _{\gamma'(\tau )}\gamma'(\tau)\rangle  \\
&=&\eta_i (\tau )\left\langle \gamma'(\tau ),\gamma'(\tau )\right\rangle
+\langle \partial_{x^i}(\gamma(\tau) ),\nabla _{\gamma'(\tau )}\gamma'(\tau
)\rangle ,
\end{eqnarray*}%
and consequently 
\begin{equation}\label{isfw}
\eta_i (\tau )=\langle \partial_{x^i}(\gamma(\tau) ),\nabla _{\gamma'(\tau )}\gamma'(\tau
)\rangle.
\end{equation}%
Since, in this case, $\langle \partial_{x^i}(\gamma(\tau) ),\gamma'(\tau)\rangle =0$ the Fermi-Walker transport equation (\ref{FW transport}) takes the form 
\begin{equation}
\nabla _{\gamma'(\tau)}\partial_{x^i}(\gamma(\tau) )=\langle
\partial_{x^i}(\gamma(\tau) ),\mathbf{a}(\tau )\rangle \gamma'(\tau),
\label{FW-simple}
\end{equation}%
which, by (\ref{isfw}), is satisfied.
\end{proof}
\end{proposition}

It is clear that the Fermi-Walker coordinates have physical meaning. Moreover, by proposition \ref{charFW}, for every set of coordinates $x^a$ with physical meaning, the frame $\{\partial_{x^a}\}$ along the worldline coincides with that given by the Fermi-Walker construction. 

\section{The equivalence principle and tidal forces}\label{e.p.}

Special relativity was motivated by the observation that motion is relative. It makes sense to say that Alice moves with constant velocity with respect to Beth. But it is pointless to try to decide which one of them is at rest. There is no experiment that Alice can do to decide whether or not she is moving. If she drops balls, she will see them floating. The condition that there is no preferred inertial frame of reference, together with the constancy of the speed of light, forced the introduction of Lorentz transformations and special relativity.

\begin{figure}[H]
\centering
\includegraphics[scale=0.3]{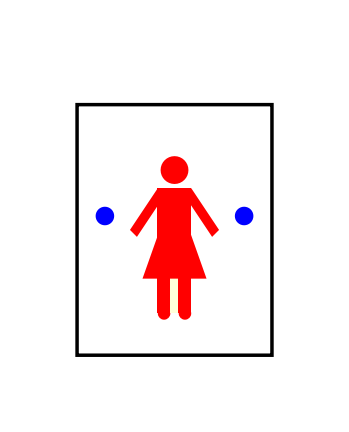}\caption{Alice moving with constant speed inside a lift in empty space.}%
\end{figure}

Acceleration, on the other hand, can be detected by Alice. Suppose she is inside a lift that is being pulled up with constant acceleration. Alice will feel the floor pushing her up. If she drops balls, they will fall to the ground. Mathematically, this corresponds to the fact that the world line of an accelerated object is not a geodesic in Minkowski spacetime. This deviation from geodesic motion is what Alice can detect inside her lift.

\begin{figure}[h]
\centering
\includegraphics[scale=0.3]{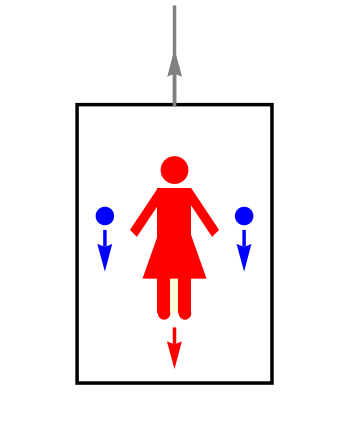}\caption{Alice in an accelerated lift in empty space.}%
\end{figure}

Consider also the situations where Beth is at rest inside her lift close to the surface of the Earth. She feels she is pushed against the floor. If she drops balls, they fall to the ground. Einstein thought that this situation is locally indistinguishable from the one where Alice's lift is being pulled up. According to general relativity, the Earth causes the geometry of spacetime to change, so that Beth's  world line is no longer geodesic. Again, what Beth detects inside her lift is the deviation from geodesic motion.

\begin{figure}[H]
\centering
\includegraphics[scale=0.3]{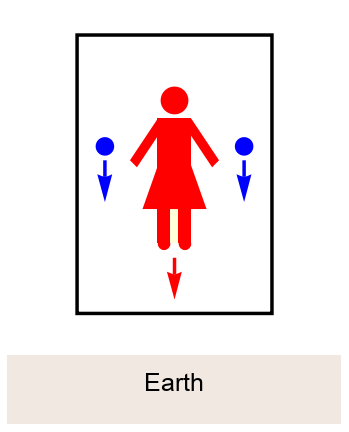}\caption{Alice in a lift  in presence of a gravitational field.}%
\end{figure}

Suppose  that the rope that sustains Beth's lift is cut, so that it starts falling freely towards the Earth. Beth no longer feels the floor pushing her up. If she drops balls, she will see them float. This situation is locally indistinguishable from
uniform motion. According to general relativity, Beth is now moving along a geodesic on a spacetime that is curved due to the presence of the Earth. Since there is no deviation from geodesic motion, there is nothing Alice can detect that would distinguish her situation from rest in empty space. Mathematically, this corresponds to the fact that her reference frame is given by Fermi coordinates, so that, very close to the world line, she is at rest in Minkowski spacetime.

\begin{figure}[H]
\centering
\includegraphics[scale=0.3]{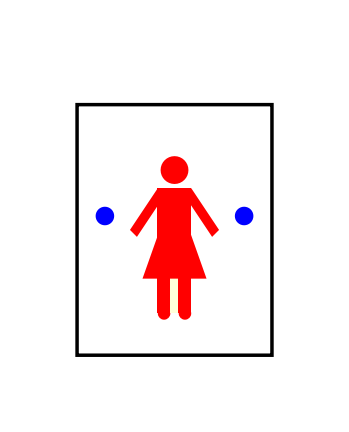}\caption{Beth falling freely towards the Earth.}%
\end{figure}

The equivalence principle in special relativity can be summarized as follows.
\begin{center}
\begin{tabular}{ccc}
\hline &&\\
\multicolumn{3}{c}{\textbf Equivalence Principle in Special Relativity} \\  
&&\\
\hline && \\
 {\textbf Physics}&$\Leftrightarrow$& {\textbf Mathematics} \\  
&&\\
\hline && \\
There is no experiment  that can &&Given a straight world line in\\ distinguish an observer at rest from one
&& Minkowski spacetime, there are \\  in uniform motion. As a consequence, && coordinates where the observer is  \\ the concept of being at rest is meaningless.&&at rest and the metric takes\\  &&the standard form.  \\
&&\\
\hline
\end{tabular}
\end{center}

General relativity goes further. Even in curved spacetime, only deviations from geodesic motion can be detected in a small laboratory.
In this sense, the existence of the gravitational field is relative. The gravitational effects manifest themselves as the curvature of spacetime. An observer in geodesic motion uses Fermi coordinates, so that the derivatives of the metric vanish on the world line. Therefore, for  local experiments, the metric is well approximated by the Minkowski metric. 

\begin{center}
\begin{tabular}{ccc}
\hline &&\\
\multicolumn{3}{c}{\textbf Equivalence Principle in General Relativity} \\  
&&\\
\hline && \\
 {\textbf Physics}&$\Leftrightarrow$& {\textbf Mathematics} \\  
&&\\
\hline && \\
There is no experiment that  can &&There are Fermi coordinates\\ distinguish a small laboratory falling
&& around timelike geodesics \\ freely under the effect of gravity from  && in curved spacetime. \\ one at rest in empty space. &&\\  
&&\\
\hline
\end{tabular}
\end{center}

Throughout our discussion on the equivalence principle, we insisted that Alice and Beth are confined to a small lift or laboratory. The reason we made this assumption is that, in a large laboratory, it is possible to measure the effect of \emph{tidal forces}. Consider again the situation where Beth is falling towards the Earth. The gravitational field of the earth is radial, so that  balls that start very far apart come together as they fall. This will allow Beth to know that she is not at rest in empty space. Note, however, that this effect will be hard to measure if the laboratory is very small. Mathematically, this corresponds to the fact that, in Fermi coordinates, the metric is well approximated by the Minkowski metric only very close to the world line! As you move away from the world line the metric changes and the effects of curvature (gravity) become measurable.

\begin{figure}[H]
\centering
\includegraphics[scale=0.45]{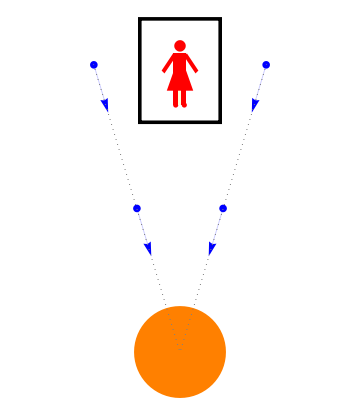}\caption{Tidal forces.}%
\end{figure}

\section{Tidal forces: newtonian analysis}
Tidal forces allow an observer that is falling freely towards a massive object
to distinguish her situation from that of an observer that is floating in empty space.

\begin{figure}
\centering
\includegraphics[scale=0.4]{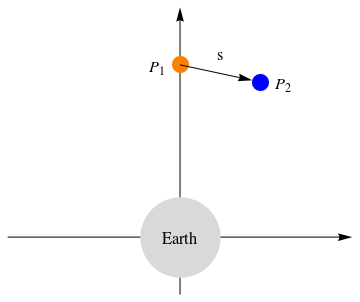}
\caption{Tidal forces}\label{tidalforces1}
\end{figure}

Suppose that one releases two test particles of mass $m=1$ that at
time $t=0$ are separated a short distance $s_{0},$ and that fall freely
towards the center of an object of mass $M$, say the Earth. We set a coordinate system with the $x^{3}$ axis pointing
upwards, as shown in Figure \ref{tidalforces1}. Let $s(t)$ be the separation vector at time $t$
between the test particles $P_{1}$ and $P_{2}$. Assume that particle $P_{1}$ is
originally at position $(0,0,d)$, where $d$ is the distance between $P_{1}$
and the origin of coordinates.

We suppose that, compared with the size of the massive body, the original
separation distance $s_{0}=\lvert s(0) \rvert$ is very small. Say $1$\textrm{m}
compared with the radius of the Earth $R\approx 6.37\times 10^{6}\textrm{m}$. If $\alpha (t)=(a^{i}(t)),$ and $\beta (t)=(b^{i}(t))$ are the
trajectories of each particle, then according to Newton's Second Law, $$%
\frac{d^{2}a^{k}}{dt^{2}}=-\frac{\partial \Phi} {\partial x^{k}}(\alpha (t)),$$ 
$$\frac{d^{2}b^{k}}{dt^{2}}=-\frac{\partial \Phi} {\partial x^{k}}(\beta (t)).$$ Here $\Phi$
denotes the gravitational potential  $\Phi (x)=\frac{-G_NM}{|x|}$, $G_N=6.674\times 10^{-11}$ is the \emph{gravitational constant} and $M$ is the mass of the body
responsible for the gravitational field. 

We write $s(t)=\beta (t)-\alpha (t)$
for the separation vector, and define $h_{k}(x)=\frac{\partial \Phi}
{x^{k}}(x).$ The linear approximation of $h_{k}$ gives
\[
h_{k}(x+s)\approx h_{k}(x)+\sum_{i}\frac{\partial
h_{k}}{\partial x^{i}}(x)s^{i}.
\]
Thus,%
\[
\frac{\partial \Phi (\beta (t))}{\partial x^{k}}\approx\frac{\partial \Phi }{%
\partial x^{k}}(\alpha (t)+s(t))\approx\frac{\partial \Phi (\alpha (t))}{\partial
x^{k}}+\sum_{i}\frac{\partial ^{2}\Phi (\alpha (t))}{\partial x^{i}\partial
x^{k}}s^{i}(t).
\]%
From this one concludes:
\begin{align}
\frac{d^{2}s^k}{dt^{2}} \approx \frac{d^{2}b^k(t)}{dt^{2}}-\frac{d^{2}a^k(t)}{dt^{2}}   
 \approx-\sum_{i}\frac{\partial ^{2}\Phi (\alpha (t))}{\partial x^{i}\partial
x^{k}}s^{i}(t).  \label{E49}
\end{align}%
A computation shows 
\[
\frac{\partial ^{2}\Phi }{\partial x^{i}\partial x^{k}}=\frac{G_{N}M}{\left\vert
r(x)\right\vert ^{3}}\left( \delta _{ik}-\frac{3x^{i}x^{k}}{%
\left\vert r(x)\right\vert ^{2}}\right) \text{, where }\delta
_{ik}\text{ denotes Kronecker's function.}
\]%
Thus, one gets from (\ref{E49}) the system of equations:%
\begin{equation}
\frac{G_NM}{r(t)^{3}}\left( 
\begin{array}{ccc}
-1 & 0 & 0 \\ 
0 & -1 & 0 \\ 
0 & 0 & 2%
\end{array}%
\right) \left(
\begin{array}{c}
s^{1}(t) \\ 
s^{2}(t) \\ 
s^{3}(t)%
\end{array}%
\right) =\left( 
\begin{array}{c}
d^{2}s^{1}/dt^{2} \\ 
d^{2}s^{2}/dt^{2} \\ 
d^{2}s^{3}/dt^{2}%
\end{array}%
\right) .  \label{E50}
\end{equation}%
We have used the fact that the test particle $P_{1}$ moves down the $x^{3}$
axis so that $$a^{1}(t)=a^{2}(t)=0,$$ $$a^{3}(t)=r(t)=d-1/2gt^{2}.$$ 

The separation vector compresses in the horizontal direction
due to attractive tidal forces, while it stretches in the vertical
direction due to tidal repulsion. The term\emph{\ tidal}\ comes from the
fact that these are the precisely the forces
responsible for the daily tides.

\begin{figure}[tbh]
\centering
\includegraphics[scale=0.45]{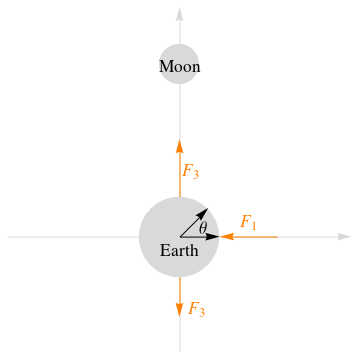}
\caption{Tides}\label{tides}
\end{figure}

Let us analyze the acceleration of the separation vector $s(t)=\beta
(t)-\alpha (t),$ at $t=0,$ of a region of water in the ocean of mass $m$ with respect to a
particle $t_{1}$ located at the center of the Earth when they
 fall towards the Moon. We set a
coordinate system at the center of the Moon, as shown in figure \ref{tides}. The coordinates
of both particles at $t=0$ are $\beta (0)=(R\sin \theta ,$ $0,$ $d-R\cos
\theta ),$ and $\alpha (0)=(0,0,d)$, respectively, where $d$ is the distance
between the centers of the Earth and the Moon and $R$ denotes the Earth's
radius. Thus, $s(0)=(R\sin \theta ,0,-R\cos \theta )$, and the tidal forces
at $t=0$ are given by:
\begin{align*}
F_{1}& =\frac{-G_{N}mM}{d^3}R\sin \theta,  \\
F_{2}& =0, \\
F_{3}& =\frac{-2G_{N}mM}{d^{3}}R\cos \theta ,
\end{align*}
where $M$ is the mass of the Moon.
The force $F_{1}$ is responsible for compressing the oceans. On the other
hand, when $\theta =0,$ the force $$F_{3}=-\frac{2G_{N}mM}{d^{3}}R$$ is
directed towards the Moon, while if $\theta =\pi,$ $$F_{3}=\frac{2G_{N}mM}{d^{3}}R$$  is repulsive. Thus, $F_{3}$ is responsible for
pulling the ocean away form the center of the Earth on both sides of the $x^{3}$-axis. 

\section{Time dilation due to acceleration}
\label{dilatacion del tiempo}
We have already  discussed the effect of time dilation between an inertial observer and a constantly accelerated observer in a hypothetical trip of a space ship to an exoplanet.  In this section we want to analyze a similar situation for the case of two observers that are being constantly accelerated inside the ship, but separated some distance $h.$

Suppose that  Alice and Beth, who are twins, travel in an accelerating rocket. Alice sits in the back of the ship, and Beth, in front. The length of the rocket is $h$ and its constant acceleration is $a$.

\begin{figure}[H]
\centering
\includegraphics[scale=0.35]{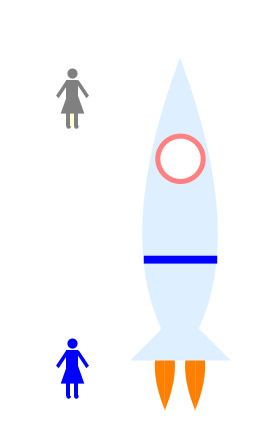}\caption{Beth is in the front, Alice in the back.  }%
\end{figure}

The twins will move along hyperbolas in Minkowski spacetime
\[ \gamma_A(\tau)=\Big(\frac{\sinh(a \tau)}{a},\frac{\cosh(a\tau)}{a}\Big),\qquad\gamma_B(\tau)=\Big(\frac{\sinh(a \tau)}{a},\frac{\cosh(a\tau)}{a}+h\Big).\]
Suppose that Alice sends light rays towards Beth at constant time intervals $\Delta (\tau_A)$. We want to compute the rate at which Beth receives the signals. 
These signals correspond to successive crests of an electromagnetic
wave so that $f_A=1/\Delta (\tau_A)$ is the frequency of the wave, measured by Alice.
The world line of a ray of light that is sent at $\gamma_A(\tau)$ is
\[ \theta_\tau(s)=\gamma_A(\tau)+s(1,1).\]
This ray intersects Beth's world line at a point $\gamma_B(\overline{\tau})$. 
\begin{figure}[H]
\centering
\includegraphics[scale=0.5]{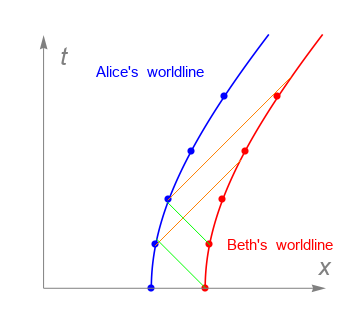}\caption{Alice sends blue light and Beth receives red light. Or no light at all. The orange lines are rays of light sent by Alice and the green lines are rays of light sent by Beth.}%
\end{figure}
Since the velocity of the ray of light is $1$, we know that
$\Delta t = \Delta x$
and therefore

\[ \frac{\sinh(a\overline{\tau})-\sinh(a \tau)}{a}=\frac{1}{a}\Big(\cosh(a\overline{\tau})+ah-\cosh(a\tau)\Big). \]
Equivalently, using that $\cosh(z)-\sinh(z)=e^{-z}$, one obtains
\begin{equation}\label{hyper-id}e^{-a\overline{\tau}}=e^{-a\tau}-ah,\end{equation}

\begin{equation}\overline{\tau}=\frac{-\ln \Big(e^{-a\tau}-ah\Big)}{a}.
\end{equation}
In standard units this equation can be written as:
\begin{equation}\overline{\tau}=-c/a\ln \Big(e^{-a\tau/c}-ah/c^2\Big).
\end{equation}
The expression inside the parenthesis tends to $-ah/c^2$ when $\tau \to \infty$, and therefore the logarithm is not defined for $\tau>>0$.
This means that after a long time the rays of light that Alice sends to Beth will never reach her. Even if Alice keeps sending signals forever, Beth will only receive finitely many of them.

Moreover, the same could happen even at the beginning of their trip ($\tau=0$) since the expression inside the parenthesis could also be negative. For instance, this could happen if  $a$ is very large. By the equivalence principle, we may think that Alice sits on the surface of a very massive body with an enormous gravitational force. Then light emanating from the surface will never reach any point located at a distance beyond $h=c^2/a.$

Consider a value of $\tau$ that is sufficiently small so that $\overline{\tau}$ is defined. We are interested in the rate at which Beth receives the signals. 
In standard units, Equation \ref{hyper-id} becomes
\[e^{-a\overline{\tau}/c}=e^{-a\tau/c}-\frac{ah}{c^2}.\]
Using implicit differentiation one obtains:
\[\frac{d\overline{\tau}}{d\tau}=1+\frac{ah}{c^2e^{-a\overline{\tau}/c}}\]
For values of $\overline{\tau}$ which are very small with respect to $c$ one
obtains the approximation:

\begin{equation}
\label{dilatacion del rojo}
\Delta(\overline{\tau}_B) \approx \Delta (\tau_A)(1+\frac{ah}{c^2})>\Delta (\tau_A).
\end{equation}

Beth will measure the time between consecutive receptions of the signal to be more than $\Delta(\tau_A)$. She perceives the light to be lower frequency, shifted to the red.

It is interesting to compare the distortions in the perception of time that occurs in the accelerating rocket with the Doppler effect. If Alice and Beth are moving apart form each other at constant speed, and they both emit blue light, then the other 
will see the light shifted to the red. The Doppler effect is symmetric for Alice and Beth. The rocket situation is not. Alice will see light emitted by Beth shifted to the blue. The asymmetry arises because, in order to meet Beth, Alice needs to go in the direction of the acceleration while, in order to meet Alice, Beth needs to go against the acceleration.
According to the equivalence principle, the situation in the rocket
should be locally equivalent to one where Alice and Beth are at rest in the presence of gravity. This effect, known as gravitational time dilation, is discussed in the following section.

\section{Gravitational time dilation and redshift}\label{GTD}

Time runs more slowly in the presence of a gravitational potential. If a clock $A$ is on the surface of the Earth, and an identical clock $B$ is 1 km above the surface, then, after a million years,
$B$ will be $3$ seconds faster than $A$.  Amazingly precise experiments have been made by Wineland et.al 
 \cite{Wineland}, where this effect was measured for a difference in height of less than a meter.
The situation can be analyzed using the two dimensional version of the Schwartzshild metric,
which describes the geometry of spacetime around a massive object. 
The Schwartzschild radius of an object of mass $M$ is:
\[ r_s= \frac{2G_{N}M}{c^2}.\]
The $c^2$ in the denominator makes this radius very small. For instance, the Schwartzschild radius of the Earth is $r_s\approx 0.88$mm. For $r>r_s$ the geometry caused by gravity due to the mass, is given by the Schwartzschild metric
\[ g=-Lc^2 dt \otimes dt + \frac{1}{L}dr \otimes dr,\]
where 
\[ L= 1-\frac{r_s}{r}.\]
For simplicity, let us consider units where $c=1$. The light cones for the Schwartzschild metric are described as follows.
If $v=(x,y)$ is a lightlike  vector then:
\[ 0=\langle v,v \rangle=\begin{pmatrix}x&y\end{pmatrix} \begin{pmatrix}-L&0\\
0& L^{-1}\end{pmatrix} \begin{pmatrix}x\\y\end{pmatrix}=-Lx^2+\frac{y^2}{L}=0.\]
Equivalently,
\[ \frac{y}{x}=\pm L.\]
This means that the vector fields
\[ V= L^{-1}\partial_t + \partial_r;\,\,\,  W =-L^{-1}\partial_t +\partial_r\]
are light like. The vector field $V$ is tangent to the world line of a ray of light going away from the mass, and $W$  is tangent to the trajectory of 
a ray of light going towards the mass.
Suppose that $\gamma(z)=(t(z),z)$ is an integral curve of $V$. Then:
\[ t'(z)=L^{-1}=\frac{z}{z-r_s},\]
and therefore:
\[ t(z)=z +r_s \ln(z-r_s)+K.\]
Similarly, if
 $\beta(z)=(t(z),z)$ is an integral curve of $W$. Then:
\[ t'(z)=-L^{-1}\]
so that
\[ t(z)=-z -r_s \ln(z-r_s)+K.\]
One concludes that the trajectories of rays of light in the Schwartzschild metric are:
\[ \gamma(z)=(z +r_s \ln(z-r_s)+K,z);  \,\,\,\beta(z)=(-z -r_s \ln(z-r_s)+K,z).\]

The figure bellow depicts the trajectories of light in the Schwarzschild metric. The gray curves correspond to light going away from the mass, and the red curves, to light going towards the mass.

\begin{figure}[H]
\centering
\includegraphics[scale=0.4]{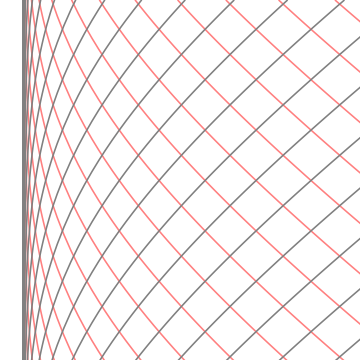}\caption{Light like geodesics in the Schwartzschild metric.}%
\end{figure}

\begin{figure}[H]
\centering
\includegraphics[scale=0.5]{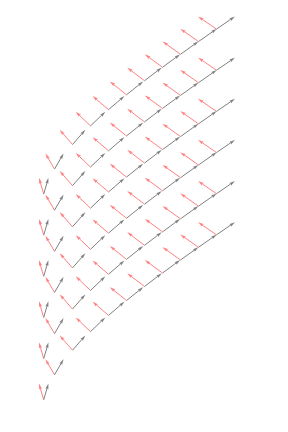}
\caption{Light cones in Schwartzschild metric.}\label{lcsm}
\end{figure}

For large values of $r$, the Schwartzschild metric tends to the Minkowski metric, so that the light cones have the usual slope.
However, when $r\to r_s$, $L$ tends to zero, so that the light cones become more and more vertical. This is depicted in figure \ref{lcsm}

\begin{figure}[H]
\centering
\includegraphics[scale=0.4]{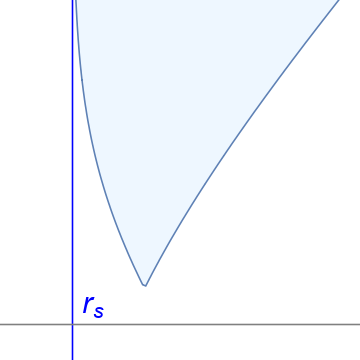}\caption{Causal future of an event in the Schwartzschild metric.}%
\end{figure}

Consider a couple of twins, Alice and Beth. Alice lives in the valley and Beth lives on top of a mountain.
Let $r_A$ and $r_B$ be the distances from the center of the Earth for Alice and Beth, respectively.

\begin{figure}[H]
\centering
\includegraphics[scale=0.35]{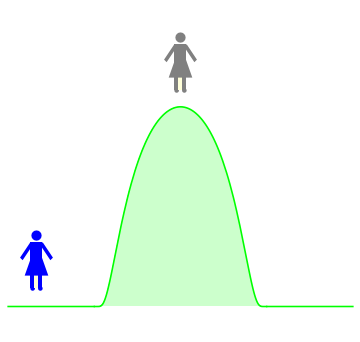}\caption{Alice and Beth are twins. Alice lives in the valley, and Beth on top of the mountain. Beth is older than Alice.}%
\end{figure}
\noindent
Alice sends Beth a ray of light every second. This means that the proper time that Alice measures between two consecutive emissions of light is:
\[ \Delta(\tau_A)= \Delta t \sqrt{L(r_A)}.\]
The Schwartzschild metric does not depend on $t$ and therefore, the trajectories of the two rays of light are parallel. This implies that the proper time that Beth will measure between two receptions of light is:
\[ \Delta(\tau_B)= \Delta t \sqrt{L(r_B)}.\]

Therefore:
\[\Delta(\tau_B)=\Delta(\tau_A) \sqrt{\frac{L(r_B)}{L(r_A)}}.\]

Since $L(r_B)>L(r_A)$, this means that, while Alice sends light every $\Delta(\tau_A)=1$ second, Beth receives light every $\Delta(\tau_B)>1$ seconds. Beth will judge the frequency of light to be less than the frequency Alice will assign to it. The light is shifted to the red.
\begin{figure}[H]
\centering
\includegraphics[scale=0.4]{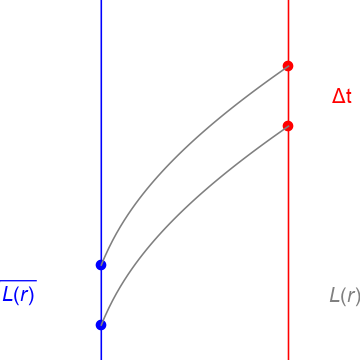}\caption{Alice sends blue light and Beth receives red light.}%
\end{figure}

Suppose that the height of the mountain is $1$km, and that Alice continues to send light for a million years. Then:

$$\tau_A=3154\times 10^{10} s.$$
 
The time that Beth will measure between the first and the last reception of light is:

\[\tau_B=\tau_A \sqrt{\frac{L(r_B)}{L(r_A)}}=\tau_A\sqrt{\frac{r_A(r_B-r_s)}{r_B(r_A-r_s)}}.\]

Since $r_s=8.8 \times 10^{-3}$ m,  and $r_A$ 
\[ r_A=6371 \times 10^3 m,\,\,\, r_B=6372 \times 10^3 m.\]
One concludes that 
\[ \tau_B-\tau_A \approx 3.4 \,s.\]
After a million years, the difference in the clocks is about $3.4$ seconds.\\

Let us see what happens in the limit $r_B \to \infty$, where Beth is very far away from the Earth. In this case:
\[\tau_B=\tau_A \sqrt{\frac{1}{L(r_A)}}=\tau_A\sqrt{\frac{r_A}{r_A-r_s}}\approx \tau_A (1.0000000006984775).\]

This means that, on the surface of the Earth, time runs more slowly than on empty space by about one second every 45 years.
\subsection{Comparing the two situations}

The Equivalence Principle allows one to compare the time dilation due to acceleration
with that due to gravitation.
Recall that Equation (\ref{dilatacion del rojo}) gives
\begin{equation}
\Delta(\overline{\tau}_B) \approx \Delta (\tau_A)\Big(1+\frac{ah}{c^2}\Big)
\end{equation}
The frequencies of a light signal sent by Alice are related by: 
\begin{equation}
\label{frequencias}
f_B\approx f_A\Big(1+\frac{ah}{c^2}\Big)^{-1} \approx f_A\big(1-\frac{ah}{c^2}\big)
\end{equation}
As we shall discuss in detail later, the difference of gravitational
potential between Beth and Alice would be then equal to $\Phi (x_{B})-\Phi
(x_{A})$, with $\Phi (x)=-G_NM/x$, where $M$ is the mass of the body responsible for the gravitational field, $x_{A}=R,$ the radius of $M$, and $%
x_{B}=R+h$. Now,%
\begin{align*}
\Phi (x_{B})-\Phi (x_{A})& =\frac{-G_N\text{ }M}{R+h}+\frac{G_N\text{ }M%
}{R}= \\
G_NM(\frac{1}{R}-\frac{1}{R+h})& =\frac{h\text{ }G_N\text{ }M}{R(R+h)}%
\simeq h\frac{G_N\text{ }M}{R^{2}}.
\end{align*}%
Taking $M=5.97\times 10^{24}$ \textrm{kg}, the mass of the Earth, and using $R=6.37\times 10^{6}$ for the radius of the Earth, we get $%
G_NM_{E}/R_{E}^{2}=\mathrm{9.8}$ \textrm{m/s}$^{2},$ which is the
acceleration of gravity on the Earth's surface.
\noindent

One then has $\Phi (x_{B})-\Phi (x_{A})\simeq hg$, and therefore formula
(\ref{frequencias}
) can be written as 
\[
f_{B}=f_{A}(1-\frac{hg}{c^{2}})=f_{A}\left( 1+\frac{\Phi (x_{A})-\Phi (x_{B})%
}{c^{2}}\right) .
\]%
This equation suggests that \emph{in the presence of a gravitational field
the frequency of a light signal will be red shifted} by a factor of $1+(\Phi
(x_{A})-\Phi (x_{B}))/c^{2}$  as the signal climbs the
gravitational potential. In general, one would expect%
\begin{equation}
f_{B}=f_{A}\left( 1+\frac{\Phi (x_{A})-\Phi (x_{B})}{c^{2}}\right) 
\label{E55}
\end{equation}


   \clearemptydoublepage 
\chapter{The Energy-Momentum tensor}

\begin{center}
\parbox[b]{0.9\textwidth}{\small \sl In this chapter we discuss how the distribution of energy, matter and momentum is described relativistically by the energy momentum tensor.
We explain the basic equations for classical fluids, and the issues that arise when they are treated relativistically. We also consider  the energy momentum tensor for the electromagnetic field. Based on those examples, we identify the properties that a general energy momentum tensor is expected to have. }
\end{center}

\vspace{3ex}

\section{The equation of continuity}\label{sec:cont}
Fluid flow is an intuitive physical notion which is represented mathematically by a continuous transformation of the three-dimensional Euclidean space $\RR^3$  onto itself. The parameter $t$ describing the transformation is identified with the time, and we may suppose its range to be $-\infty < t < \infty$. In order to describe the transformation analytically we introduce a fixed system of Cartesian coordinates $(x^1,x^2,x^3)$. With these coordinates we specify a particular position in the fluid. By simplicity, we denote $(x^1,x^2,x^3)$ by $x$.  

The mathematical description of the state of a moving fluid is consists of three quantities:
\begin{itemize}
\item A time-dependent vector field $v(t,x)$ which gives the distribution of the fluid velocity.

\item A function $\rho(t,x)$ which gives the density distribution of the fluid.

\item A function $p(t,x)$ which gives the pressure of the fluid.
\end{itemize}
The time-dependent vector field $j = \rho v$ is called \emph{mass flux density}.

Given five quantities, namely, the three components of the velocity $v$, the density $\rho$ and the pressure $p$, the state of the moving fluid is completely determined. 
\begin{figure}[H]
	\centering
	\includegraphics[scale=0.4]{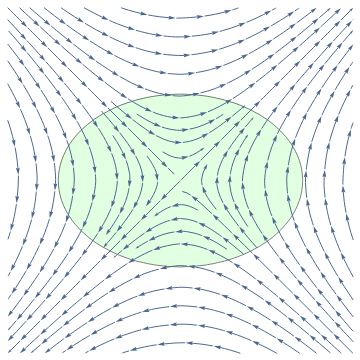}
	\caption{Motion of a fluid.}
\end{figure}

We will derive the fundamental equations of fluid dynamics. Let us begin with the equation which expresses the conservation of matter. We consider some region $D$ of space. The mass of the  fluid in this region is
$$
 \int_{D} \rho(t,x) \, dV. 
$$ 
Hence, the decrease per unit time in the mass of fluid in the region $D$ can be written as
$$
-\frac{d}{dt} \int_D \rho \, dV = -\int_{D} \frac{\partial \rho}{\partial t} \, dV. 
$$
On the other hand, the mass flowing in unit time through a sruface element $dA$ along the outward normal $n$ of the surface $\partial D$ bounding $D$ is $\rho(t,x) v(t,x) \cdot n \, dA$.  
\begin{figure}[H]
	\centering
	\includegraphics[scale=0.5]{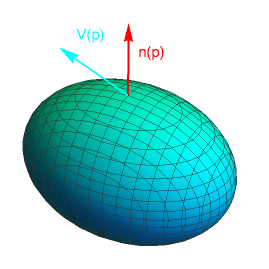}
	\caption{The normal vector and the velocity of a fluid.}	
\end{figure}
The total mass of fluid flowing out of the region $D$ in the unit time is therefore
$$
\int_{\partial D} \rho(t,x) v(t,x) \cdot n \, dA. 
$$
The principle of conservation of mass is expressed by equating the two expressions, that is
$$
\int_{D} \frac{\partial \rho}{\partial t} \, dV = - \int_{\partial D} \rho v \cdot n \, dA.
$$
The surface integral can be transformed by the divergence theorem to a volume integral
$$
 \int_{\partial D} \rho v \cdot n \, dA = \int_D \div (\rho v) \, dV.
$$
Thus
$$
\int_D \left[ \frac{\partial \rho}{\partial t} + \div (\rho v)\right] dV = 0. 
$$
Since this equation must hold for any region $D$, the integrand must vanish, i.e.
\begin{equation}\label{Cont}
\frac{\partial\rho}{\partial t}+\div(\rho\mathrm{v})=0. %
\end{equation}
This is the \emph{equation of continuity}. Expanding the expression $\div(\rho v)$, we can also write  \eqref{Cont} as
\begin{equation}\label{eq:cont}
\frac{\partial\rho}{\partial t} + \rho \div v + v \cdot \grad \rho = 0. 
\end{equation}

An important special case is that of an \emph{incompressible flow}. This means that the density may be supposed invariable, i.e. constant throughout the volume of the fluid and throughout its motion. Mathematically, this means that if $x(t)$ is the trajectory of a particle moving with the fluid then $\rho(t,x(t))$ is constant. Differentiating this condition with respect to $t$, we get
\begin{equation}\label{eq:compr}
\frac{\partial \rho}{\partial t} + v \cdot \grad \rho = 0. 
\end{equation}
Combining \eqref{eq:cont} and \eqref{eq:compr}, we see that the equation of continuity takes the simple form
\begin{equation}
\div v = 0. 
\end{equation}
An incompressible flow is thus one for which velocity vector field $v(t,x)$ is divergenceless. 

\section{Euler's equation}
We consider now the dynamics of fluid motion. The intention is to derive the equations which governs the action of forces, external and internal, upon the fluid. For our purposes, we will only deal with \emph{perfect fluids}. These are characterized by the fact that no shear forces are possible. What this means is that the force exerted by the surrounding fluid on a surface element $dA$ with unit outward normal $n$ is $-p(t,x) n \, dA$. 

We will use the following result, which is a formal consequence of the divergence theorem. 

\begin{lemma}\label{normal}
Let  $D$ be a compact region of space with bounding surface $\partial D$ and let $f$ be a function defined in an open set that contains $D$. Then
$$
\int_D \grad f \, dV = \int_{\partial D} f n \, dA,
$$
where $n$ is the unit outward normal. 
\end{lemma}

\begin{proof}
In the divergence theorem, let $X = f e$ where $e$ is a constant vector. Then
$$
\int_D \div (fe) \, dV = \int_{\partial D} f e \cdot n \, dA.
$$ 
Since $\div (fe) = \grad f \cdot e = e \cdot \grad f$ and $fe \cdot n = e \cdot (f n)$,
$$
\int_D e \cdot \grad f \, dV = \int_{\partial D} e \cdot (f n) \, dA.
$$
Taking $e$ outside the integrals,
$$
e \cdot \int_D \grad f \, dV = e \cdot \int_{\partial D} f n \, dA,
$$
and since $e$ is an arbitrary constant vector,
$$
\int_D \grad f \, dV = \int_{\partial D} f n \, dA,
$$
as was to be shown. 
\end{proof}

Let us consider some region $D$ in space. By imposing the above perfect fluid condition, the total force on the volume occupied by the fluid in $D$ is equal to the integral
$$
- \int_{\partial D} p n \cdot dA
$$
of the pressure, taken over the surface $\partial D$ bounding the region $D$. Transforming it to a volume integral, by means of Lemma \ref{normal}, we have
$$
- \int_{\partial D} p n \cdot dA = - \int_D \grad p \, dV.
$$
We see that the fluid surrounding any volume element $dV$ exerts on that element a force $- \grad p \, dV$. In other words, we can say that a force  $-\grad p$ acts on unit volume of the fluid. 
 
On the other hand, let $x(t)$ be the trajectory followed by a particle moving with the fluid. The \emph{acceleration} of this particle is given by 
$$
a= a(t,x(t)) = \frac{d v(t,x(t))}{dt}. 
$$
Using the chain rule, we can calculate it by the formula
\begin{equation}\label{eq:matder}
a = \frac{\partial v}{\partial t} + (v \cdot \grad) v.
\end{equation}
where we have denoted by $(v \cdot \grad)$ the operator $\sum v^i \partial/ \partial {x^i}$ applied to $v$:
$$(v \cdot \grad) v= \left (\sum v^i \frac{\partial v^1}{\partial {x^i}},\\ \sum v^i \frac{\partial v^2}{\partial {x^i}},\\ \sum v^i \frac{\partial v^3}{\partial {x^i}}\right ).$$
Thus, $a$ is the rate of change of the velocity of a given fluid particle as it moves about in space.\footnote{This time derivative should not be confused with the partial derivative with respect to $t$ at a fixed position $x$.}

We can now write down the equation of motion of a volume element in the fluid by equating the force $-\grad p$ to the product of the mass per unit volume $\rho$ and the acceleration $a$:
$$
\rho a  = -  \grad p.
$$  
By means of \eqref{eq:matder}, this may be written in the form
\begin{equation}\label{eu}
\rho \frac{\partial v}{\partial t} + \rho (v \cdot \grad) v = - \grad p.
\end{equation}
This is called \emph{Euler's equation} and is one of the fundamental equations of fluid dynamics. As we have seen above, it is just a reformulation of Newton's second law for perfect fluids.

\section{The momentum flux}\label{sec:momflux}
Let us choose some region $D$ in space, and find how the momentum of the fluid contained in $D$ varies with time. The momentum rate of change in $D$ is
$$
\frac{d}{dt} \int_D \rho v \, dV = \int_D \frac{\partial}{\partial t}(\rho v) \, dV = \int_D \left( \rho \frac{\partial v}{\partial t} + \frac{\partial \rho}{\partial t} v \right) dV.
$$
Using the equation of continuity \eqref{Cont} and Euler's equation \eqref{eu}, we obtain
$$
\frac{d}{dt} \int_D \rho v \, dV = -\int_D \left[ \grad p + \rho (v \cdot \grad) v + \div(\rho v) v \right] dV.
$$
We claim that the integrand on the right-hand side is the divergence of a symmetric rank-two tensor $\Pi$,  defined by
$$
\Pi = p I + \rho v \otimes v.
$$
Indeed, $\Pi$ has components
$$
\Pi^{ij} = p \delta^{ij} + \rho v^{i} v^{j}. 
$$
Thus
\begin{align*}
\sum_{j}\frac{\partial \Pi^{ij}}{\partial x^{j}} &=\sum_{j} \frac{\partial p}{\partial x^{j}} \delta^{ij} + \sum_{j} \frac{\partial}{\partial x^{j}}(\rho v^{i} v^{j})  \\
&= \frac{\partial p}{\partial x^{i}} + \sum_{j} \left[ \rho v^{j} \frac{\partial v^{i}}{\partial x^{j}} + v^{i} \frac{\partial (\rho v^{j})}{\partial x^{j}}\right]. 
\end{align*}
From this follows we get the formula
$$
\div \Pi =  \grad p + \rho (v \cdot \grad) v + \div(\rho v) v.
$$
In view of this last equation, the rate of change of the momentum contained in $D$ is expressible as
\begin{equation}\label{eq:divmomflux}
\frac{d}{dt} \int_D \rho v \, dV = -\int_D \div \Pi \, dV.
\end{equation}
Applying the divergence theorem to the integral on the right-hand side, we obtain
\begin{equation}\label{eq:momflux}
\frac{d}{dt} \int_D \rho v \, dV  = - \int_{\partial D} \langle \Pi,n \rangle dA,
\end{equation}
where $n$ denotes the outward unit normal on $\partial D$. The surface integral on the right is therefore the amount of momentum flowing out through the bounding surface $\partial D$ in unit time. The tensor $\Pi$ is called the \emph{momentum flux  density tensor}. The vector $\langle \Pi,n \rangle$ gives the momentum flux in the direction of $n$, i.e. through a surface perpendicular to $n$. 

We regard \eqref{eq:momflux} as a balance principle, which asserts that the rate of decrease of the momentum in $D$ is equal to the momentum efflux trough $\partial D$. We call this assertion the \emph{principle of conservation of linear momentum}. Since it is valid for all regions $D$, from \eqref{eq:divmomflux} we obtain the equation
\begin{equation}\label{eq:euV2}
\frac{\partial}{\partial t}(\rho v)  + \div \Pi = 0, 
\end{equation}
which sometimes is also called Euler's equation. 

\section{Perfect fluid energy-momentum tensor}
\begin{figure}[H]
	\centering
	\includegraphics[scale=0.45]{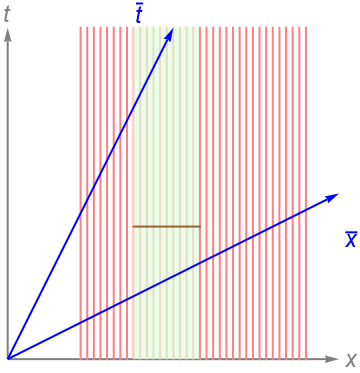}
	\caption{A swarm of identical particles according to Alice.}\label{swarmid}
\end{figure}

Consider a swarm of identical noninteracting particles that in Alice's reference frame are at rest. Assume that they are uniformly distributed over space, with $n$ particles per unit volume in this reference frame, and have individual rest mass $m$. The product $\rho = n m$ of the individual mass by the particle density gives the density of the swarm. The situation is represented in Figure \ref{swarmid}. The red lines are the world lines of the swarm of particles,
the green region is the world line of a box of volume one which is static according to Alice. The density $\rho$ is proportional to the number of red lines that intersect the brown box.
Let us calculate the density from the point of view of Beth, which is moving with constant velocity $v$ with respect fo Alice. In the image below, the green region represents the worldline of a box of volume one which is static according to Beth. Again, the density that Beth will observe is proportional to the number of times the red lines intersect the brown box.
\begin{figure}[H]
	\centering
	\includegraphics[scale=0.45]{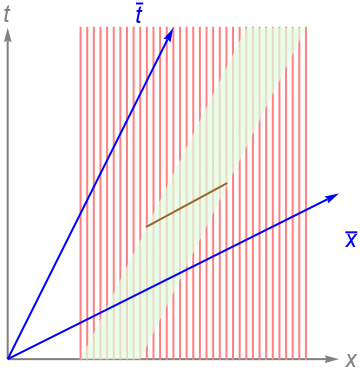}
	\caption{The swarm of identical particles according to Beth.}	
\end{figure}
The density of the swarm according to Beth will be again the product $\bar{\rho} = \bar{n} \bar{m}$ of the particle density $\bar{n}$ by the individual mass $\bar{m} = \lambda_v m$, where as usual $\lambda_v = 1 / \sqrt{1 - v^2/c^2}$. To determine $\bar{n}$ in Beth's frame, note that, by the Lorentz contraction, a region containing $n$ particles occupies the smaller volume which undergoes contraction by the factor $1/ \lambda_v$. Hence there are $\bar{n} = \lambda_v n$ particles per unit volume according to Beth, and 
$$
\bar{\rho} = \bar{n} \bar{m} = \lambda_v^2 nm = \lambda_v^2 \rho.
$$

In the classical description of the perfect fluid we reviewed in the previous sections, the density $\rho$ is a function of time and position. The analysis above shows that, relativistically, the density cannot be regarded as a function. The fact that $\lambda_v$ appears quadratically in the expression for $\bar{\rho}$ suggests that $\rho$ is the component of a rank-two tensor. Consider the symmetric rank-two tensor $T$ which in Alice's coordinates $(ct,x,y,z)$ is
$$
T = \rho c^4 \, dt \otimes dt.
$$
In other words, the matrix $(T_{ab})$ of $T$ in $(ct,x,y,z)$-coordinates is
$$
(T_{ab}) = \left( \begin{array}{cccc}\rho c^4 & 0 & 0 & 0 \\ 0 & 0 & 0 & 0 \\  0 & 0 & 0 & 0 \\ 0 & 0 & 0 & 0 \end{array}\right).
$$
In Beth's coordinates $(c\bar{t},\bar{x},\bar{y},\bar{z})$ the tensor $T$ is
\begin{align*}
T &= \lambda_{v}^2 \rho c^4 \, d \bar{t} \otimes d \bar{t} + \lambda_{v}^2 \rho c^2 v \, d\bar{x} \otimes d \bar{t} + \lambda_{v}^2 \rho c^2 v \, d\bar{t} \otimes d \bar{x} + \lambda_{v}^2 \rho v^2 \, d\bar{x} \otimes d \bar{x} \\
&=  \bar{\rho} c^4 \, d \bar{t} \otimes d \bar{t} + \bar{\rho} c^2 v \, d\bar{x} \otimes d \bar{t} + \bar{\rho} c^2 v \, d\bar{t} \otimes d \bar{x} + \bar{\rho} v^2 \, d\bar{x} \otimes d \bar{x},
\end{align*}
so that the matrix $(\bar{T}_{ab})$ of $T$ in these coordinates has the form
$$
(\bar{T}_{ab}) =  \left( \begin{array}{cccc}  \bar{\rho} c^4 & \bar{\rho} c^2 v & 0 & 0 \\ \bar{\rho} c^2 v & \bar{\rho} v^2 & 0 & 0 \\  0 & 0 & 0 & 0 \\ 0 & 0 & 0 & 0 \end{array}\right).
$$
Thus, we see that $T_{00}/c^2 = \rho c^2$ and $\bar{T}_{00}/c^2 = \bar{\rho} c^2$ are actually the densities of the relativistic energy of the swarm measured by Alice and Beth, respectively. 

The tensor $T$ we have just introduced can be formulated in a more intrinsic way by introducing the $4$-velocity $\mathbf{u}$ of the swarm of particles. In Alice's frame, we have $\mathbf{u} = \partial_t$ and the tensor $T$ is then
$$
T = \rho c^4 \, dt \otimes dt = \rho \, \partial_t^{\flat} \otimes \partial_t^{\flat} = \rho \, \mathbf{u}^{\flat} \otimes \mathbf{u}^{\flat},
$$
or, what is equivalent,
$$
T^{\sharp} = \rho \, \mathbf{u} \otimes \mathbf{u}. 
$$
This description now serves to determine the components of $T$ in a frame of reference that moves with $3$-velocity $-v$ with respect to Alice. In such frame, we have $\mathbf{u} = \lambda_{v} (\partial_t  - v)$ and therefore
\begin{align*}
T = \lambda_v^2 \rho c^4 \,  dt \otimes dt + \lambda_v^2 \rho c^2 \, dt \otimes v^{\flat} +  \lambda_v^2 \rho c^2 \, v^{\flat} \otimes dt + \lambda_v^2 \rho \, v^{\flat} \otimes v^{\flat}.
\end{align*}
Thus, $T$ has matrix $(T_{ab})$ of components
$$
(T_{ab}) = \lambda_v^2 \rho \left( \begin{array}{cccc}  c^4 &  c^2 v^{1} &  c^2 v^{2}  & c^2 v^{3}  \\ c^2 v^{1} & v^{1}v^{1} & v^{1}v^{2} & v^{1}v^{3} \\  c^2 v^{2} & v^{2}v^{1} & v^{2}v^{2} & v^{2}v^{3} \\ c^2 v^{3} & v^{3}v^{1} & v^{3}v^{2} & v^{3}v^{3} \end{array}\right).
$$

Let us now see what specific physical significance one ascribes to the components of $T$. As we have already noted, the component $T_{00}/c^2$ is the total density of energy of the swarm in the observer's Lorentz frame:
$$
T_{00}/c^2 = \lambda_v^2 \rho c^2 = \text{density of energy}. 
$$
The components $T_{i0}/c^2$ can be interpreted by observing that, in the observer's Lorentz frame, there are $\lambda_v n$ particles per unit volume and the $i$-component of $3$-momentum of the swarm is $p^{i}=\lambda_v m  v^{i}$. Thus
$$
T_{i0}/c^2 = \lambda_v^2 \rho v^i = \lambda_v n p^{i} = \text{density of $i$-component of momentum}.
$$
The components $T_{ij}$ can be interpreted by considering a $2$-surface of area $A$ at rest in the observer's frame with positive normal pointing in the $k$-direction. During a lapse of time $\Delta t$, the number of particles crossing $A$ is $\lambda_v n v^{j} A \Delta t$. Thus
$$
T_{ij} = \lambda v^2 \rho v^{i}v^{j} = (\lambda_v n v^{j}) p^{i} = \text{$j$-component of flux of $i$-component momentum}.
$$

Because of this interpretation we call $T$ the \emph{energy-momentum tensor} of the swarm. The whole information can be summarized as follows.
\begin{center}
\begin{tabular}{ccc}
\hline && \\
\multicolumn{3}{c}{\textbf{Energy-momentum tensor of the swarm}} \\  
&&\\
\hline && \\
 $T_{00}/c^2$ &$\to$ & density of energy \\ 
 &&\\
 $T_{i0}/c^2$ &$\to$ & density in $i$-component of momentum  \\ 
 &&\\
 $T_{ij}$ &$\to$ & $j$-component of flux of $i$-component momentum\\ 
 &&\\
\hline
\end{tabular}
\end{center}

We will now consider a slightly more general situation where, in the rest frame, the swarm of identical particles may form a fluid that exerts internal pressure $p$. We incorporate this pressure $p$ explicitly into the energy-momentum by writing, in the rest $(ct,x,y,z)$-coordinates,
$$
T = \rho c^4 \, dt \otimes dt + p \, dx \otimes dx + p \, dy \otimes dy + p \, dz \otimes dz,
$$
so that the matriz $(T_{ab})$ of $T$ takes the form
$$
(T_{ab}) = \left( \begin{array}{cccc}\rho c^4 & 0 & 0 & 0 \\ 0 & p & 0 & 0 \\  0 & 0 & p & 0 \\ 0 & 0 & 0 & p \end{array}\right).
$$
In a frame-independent geometric language this is
$$
T = p g + \left( \rho + \frac{p}{c^2} \right) \mathbf{u}^{\flat} \otimes \mathbf{u}^{\flat},
$$
with $g$ being the Minkowski metric. Indeed, in the rest $(ct,x,y,z)$-coordinates, $\mathbf{u} = \partial_t$ and thus
\begin{align*}
&p g + \left( \rho + \frac{p}{c^2} \right) \mathbf{u}^{\flat} \otimes \mathbf{u}^{\flat}  \\
&\quad = - pc^2 \, dt \otimes dt + p \, dx \otimes dx + p \, dy \otimes dy + p \, dz \otimes dz + \left( \rho + \frac{p}{c^2} \right) c^4 \, dt \otimes dt \\
&\quad = \rho c^4 \, dt \otimes dt + p \, dx \otimes dx + p \, dy \otimes dy + p \, dz \otimes dz.
\end{align*}
Note that this $T$ generalizes the momentum flux density tensor $\Pi$ introduced in \S\ref{sec:momflux}. For this reason, one refers to this situation as a \emph{relativistic perfect fluid}. We will see in the next section that this relativistic perfect fluid is entirely defined by the energy-momentum tensor $T$. 

The preceding discussion motivates the following formal generalization. The flow of a fluid could be described literally by a vast swarm of particles in a spacetime $M$. Instead of this discrete model it is easier to deal with a smooth model, where the $4$-velocity of the flow is given by a timelike unit vector field $\mathbf{u}$ on $M$. Intuitively, the integral curves of $\mathbf{u}$ are the average worldlines of the ``particles'' of the fluid. Moreover the fluid is characterized by two smooth functions $\rho$ and $p$ on $M$ which respectively represent the mass density and the pressure for observers whose $4$-velocity is $\mathbf{u}$. These determine a geometric, frame-independent expression for the fluid's energy-momentum tensor $T$.

This discussion can be summarized rigorously as follows. A \emph{relativistic perfect  fluid} on a spacetime $M$ is a triple $(\mathbf{u},\rho,p)$ where:
\begin{itemize}
\item $\mathbf{u}$ is a timelike future-pointing unit vector field on $M$ called the \emph{flow vector field}.

\item $\rho$ is a smooth  function on $M$ called the \emph{mass density function}.

\item $p$ is a smooth function on $M$ called the \emph{pressure function}.

\item The energy-momentum tensor is

\begin{equation}
\label{perfecto fluido}
T = p g + \left( \rho + \frac{p}{c^2} \right) \mathbf{u}^{\flat} \otimes \mathbf{u}^{\flat}.
\end{equation}

\end{itemize}

Evidently this formula for $T$ is equivalent to the following three equations for $X, Y \in \mathbf{u}^{\perp}$:
$$
T(\mathbf{u},\mathbf{u}) = \rho c^4, \qquad T(X,\mathbf{u}) = T(\mathbf{u},X) = 0, \qquad T(X,Y) =  p \langle X,Y\rangle.$$

\item For $O$ any observer with four velocity $\mathbf{v}$ at a point $p$ in her worldline one has: 

\begin{itemize}
\item The energy density measured by $O$ at $p$ is given by $T_p(\mathbf{v},\mathbf{v})/c^2$. 

In a Lorentz's frame of reference for $O$ at $p:$ \item $T_{p}(\mathbf{v},\partial_{x^i})/c^2$ represent the density of the $i$-component of momentum.
\item $T_{p}(\partial_{x^i},\partial_{x^j})$ represent the $j$-component of flux of $i$-component momentum.
\end{itemize}

\section{Conservation of energy-momentum}
 At this point the natural question arises of formulating a relativistic analogue of the conservation laws for a perfect fluid. Classically, as we saw in \S\ref{sec:cont} and \S\ref{sec:momflux}, these are given by the equation of continuity and the conservation of linear momentum, the latter being just a rewriting Euler's Equation. We shall see below that the relativistic analogue of these conservation laws can be combined into one elegant law expressing that the energy-momentum tensor of the relativistic perfect fluid is a conserved quantity. 
 
Let $(\mathbf{u},\rho,p)$ a relativistic perfect fluid on an spacetime $M$ and let $T$ be the corresponding energy-momentum tensor. We need to clarify what it means to assert that $T$ is a conserved quantity. By definition, we shall say that $T$ obeys the \emph{conservation law} if $T$ has divergence zero:
$$
\div T^{\sharp} = 0.
$$
This condition has the following consequence.

\begin{proposition}\label{prop:eqperffluid}
The conservation law for $T$ is equivalent to
\begin{align}
\mathbf{u} \rho  -\left( \rho + \frac{p}{c^2}\right) \div \mathbf{u} &= 0,  \label{eq:relcont}\\
\left( \rho + \frac{p}{c^2}\right) \nabla_{\mathbf{u}}\mathbf{u} + \grad_{\perp} p &= 0 \label{eq:relmomflux},
\end{align}
where $\grad_{\perp} p$ is the component of $\grad p$ orthogonal to $\mathbf{u}$. 
\end{proposition}

\begin{proof}
Writing $T^{\sharp}$ in terms of coordinates, 
$$
T^{ab} = p g^{ab} + \left( \rho + \frac{p}{c^2}\right) u^{a}u^{b}. 
$$
The divergence is then
\begin{align*}
\sum_{b} \nabla_{b} T^{ab} = \sum_{b} \left\{ \nabla_b p g^{ab} + \nabla_b\left( \rho + \frac{p}{c^2}\right) u^{a}u^{b} + \left( \rho + \frac{p}{c^2}\right) \nabla_b u^{a}u^{b} + \left( \rho + \frac{p}{c^2}\right) u^{a}\nabla_b u^{b}  \right\}.
\end{align*}
Expressed invariantly this is the vector field
$$
\div T^{\sharp} = \grad p + \mathbf{u} \left( \rho + \frac{p}{c^2}\right) \mathbf{u} + \left( \rho + \frac{p}{c^2}\right) \nabla_{\mathbf{u}} \mathbf{u} + \left( \rho + \frac{p}{c^2}\right) (\div \mathbf{u})\mathbf{u}.
$$
But $\div T^{\sharp} = 0$, and since $\mathbf{u}$ is a unit vector field, $\nabla_{\mathbf{u}} \mathbf{u}$ is perpendicular to $\mathbf{u}$. Hence the second equation  is obvious, and $\langle \div T^{\sharp},\mathbf{u} \rangle = 0$ gives the first  equation. 
\end{proof}

 The first equation is the formula for the rate of change of energy density as measured by an observer whose $4$-velocity is $\mathbf{u}$. The second equation is an analogue of Newton's second law, with force replaced by spatial pressure gradient, and mass replaced by $\rho + p/c^2$, while $\nabla_{\mathbf{u}} \mathbf{u}$ is indeed the spatial acceleration of the particles fo the flow as self-measured. 
 
Let us now consider the non relativistic limit of the conservation law for $T$. As anticipated above, under this limit approximation, it will be possible to recover the conservation laws for a perfect fluid, namely the equation of continuity and the conservation of linear momentum, from the equations derived in Proposition \ref{prop:eqperffluid}. So assume that the underlying spacetime of the relativistic fluid $(\mathbf{u},\rho,p)$ is the Minkowski spacetime $\mathbb{M}$. Consider a frame of reference that moves with $3$-velocity $v$ with respect to a rest frame. Then, in this frame we have $\mathbf{u}=\lambda_v (\partial_t + v)$. Thus, equation \eqref{eq:relcont} becomes
\begin{equation}\label{eq:relcontV2}
\frac{\partial}{\partial t}(\lambda_v \rho) -  \frac{1}{c^2} \frac{\partial \lambda_v}{\partial t} p + \lambda_v v \cdot \grad \rho - \left( \rho + \frac{p}{c^2}\right)\div(\lambda_v v) = 0.
\end{equation}
On the other hand, after some algebraic manipulation, the component form of equation \eqref{eq:relmomflux} simplifies to
\begin{equation}\label{eq:relmomfluxV2}
\frac{\partial}{\partial t} \left(\left( \rho + \frac{p}{c^2}\right) v^{i} \lambda_v^2 \right)  + \sum_{j}  \frac{\partial}{\partial x^{j}} \left( p \delta^{ij}+ \left( \rho + \frac{p}{c^2}\right) v^{i} v^{j} \lambda_v^2 \right) = 0.
\end{equation}
Now we are in a position to take the non-relativistic limit. This corresponds to taking the limit where $\rho/c^2 \approx 0$, and $\lambda_v \approx 1$ so that $\rho + p/c^2 \approx \rho$. Using this, we can rewrite \eqref{eq:relcontV2} as
$$
\frac{\partial \rho}{\partial t}  + v \cdot \grad \rho -\rho\div v = 0,
$$
which coincides with the equation of continuity in the form \eqref{eq:cont}. Similarly, equation \eqref{eq:relmomfluxV2} reduces to
$$
\frac{\partial}{\partial t} (\rho v^{i})  + \sum_{j}  \frac{\partial}{\partial x^{j}} ( p \delta^{ij}+ \rho v^{i} v^{j} ) = 0,
$$
or, recalling the definition of the momentum flux density tensor $\Pi$,
$$
\frac{\partial}{\partial t} (\rho v^{i})  + \sum_{j}  \frac{\partial \Pi^{ij}}{\partial x^{j}} = 0.
$$
This indeed coincides with Euler's equation in the form \eqref{eq:euV2}, expressing the conservation of linear momentum. 

The relation between the conservation law for the energy-momentum tensor $T$ and the classical conservation laws for a perfect fluid is summarized in the following table.
\begin{center}
\begin{tabular}{ccc}
\hline &&\\
{\textbf Relativistic law} &$ $&{\textbf Classical law}\\  
&&\\
\hline && \\
Time component of $\div(T^\sharp)=0$ &$\to $ & $\displaystyle\frac{\partial \rho}{\partial t}+ \div (\rho v)=0$ (continuity)   \\ 
 &&\\
 Space components of   $\div(T^\sharp)=0$& $\to$ &   $\displaystyle\frac{\partial}{\partial t}(\rho v) + \div \Pi = 0$ (linear momentum) \\  
&&\\
\hline
\end{tabular}
\end{center}

\section{Electromagnetic energy-momentum tensor}
In this section we shall show that the energy and linear momentum production of the electromagnetic field may be obtained from a certain tensor field $T$, known as the electromagnetic energy-momentum tensor. As in the relativistic perfect fluid case, this energy-momentum tensor is shown to have zero divergence. This condition allows us to recover the balance equations for energy and momentum of electromagnetic fields obtained in \S \ref{ch:EM}. 

We have pointed out that in \S\ref{sec:relMaxwell}, that we may identify the electric and magnetic $3$-vectors with a $2$-form on Minkowski spacetime:
$$
F  = (E^1 \,  dx^1  +  E^2 \, dx^2  + E^3 \,  dx^3) \wedge dt +  B^1 \, dx^2 \wedge dx^3 + B^2 \, dx^3 \wedge dx^1 + B^3 \, dx^1 \wedge dx^2.
$$
Using this $2$-form, we have shown that the source-free Maxwell's equations are just the coordinate representations in a Lorentz coordinate system for the following equations:
\begin{align*}
d F &= 0, \\
\delta F &= 0. 
\end{align*}
An important property of these equations is that they are entirely independent of the choice of the Lorentz coordinate system. As a result, the source-free Maxwell's equations are valid in all Lorentz coordinate systems. 

To generalize the theory of electromagnetism from Minkowski spacetime to and arbitrary spacetime, we proceed as follows. Formally, an \emph{electromagnetic field} on a spacetime $M$ is a $2$-form $F$ on $M$. In the end, only this formal definition is essential. If $F$ is an electromagnetic field on $M$ we shall say that $F$ obeys the \emph{source-free Maxwell's equations} if $F$ is closed and co-closed. Notice that the source-free Maxwell's equations become conditions that help determine $F$. 

We now introduce the \emph{energy-momentum tensor} $T$ of an electromagnetic field $F$ on $M$. This is by definition the symmetric rank-two tensor defined as
$$
T(X,Y) = \kappa\left( \langle i_X F,i_{Y}F \rangle - \frac{1}{2} \langle F,F \rangle \langle X,Y \rangle \right), 
$$
where $X$ and $Y$ are arbitrary vector fields on $M$, and where $\kappa$ is some constant. Given an arbitrary local frame $\{e_a\}$, the components of $T$ are 
\begin{align*}
T_{ab} &= \kappa\left( \langle i_{e_a} F,i_{e_b}F \rangle - \frac{1}{2} \langle F,F \rangle \langle e_a,e_b \rangle \right) \\
&=\kappa \sum_{c,d} \left(F_{ac} F_{bd} g^{cd}- \frac{1}{4} F_{cd}F^{cd} g_{ab} \right) .
\end{align*}
We would like to establish the basic link between the conservation law for the energy-momentum tensor $T$ and the source-free Maxwell's equations obeyed by $F$. For this purpose, we need some notation. Given a $q$-vector $X_1 \wedge \cdots \wedge X_q$ and a $k$-form $\omega$, with $q \leq k$, we write
$$
i_{X_1 \wedge \cdots \wedge X_q} \omega := \ i_{X_1} \cdots i_{X_q} \omega = \omega (X_1,\dots, X_q,\dots).
$$
Analogously, if $\eta$ is a $q$ form and $\omega$ is a $k$-form, with $q \leq k$, we write
$$
i_{\eta} \omega := i_{\eta^{\sharp}} \omega,
$$
where $\eta^{\sharp}$ is the $q$-vector metrically equivalent to $\eta$. With this understood, we can prove the following.

\begin{proposition}
The energy momentum tensor $T$ of an electromagnetic field $F$ on $M$ satisfies
$$
\div T^{\sharp} = \kappa \{(i_{\delta F} F)^{\sharp} - (i_{F} dF)^{\sharp}\}. 
$$
In particular, of $F$ obeys the source-free Maxwell's equations, then $\div T^{\sharp} = 0$. 
\end{proposition}

\begin{proof}
In terms of local coordinates $\{x^{a}\}$, it is not difficult to show that $i_{\delta F} F$ and $i_{F} dF$ have components
\begin{align*}
(i_{\delta F} F)_a &=\sum_{b,c} \nabla_b F^{b c} F_{a c}, \\
(i_{F} dF)_a &=\sum_{b,c} \left( \frac{1}{2} F^{bc} \nabla_a F_{bc} + F^{bc} \nabla_{b} F_{c a} \right).
\end{align*}
On the other hand, one finds the coordinate component of $\div T^{\sharp}$ to be
\begin{align*}
\sum_b  \nabla_b T^{a b} &=  \kappa \sum_{b,c,d}  \nabla_b \left( F^{ac} F^{bd} g_{cd}- \frac{1}{4} F_{cd}F^{cd} g^{ab}  \right) \\
&=   \kappa \sum_{b,c,d} \left(  F^{ac} \nabla_b F^{bd} g_{cd} + \nabla_b F^{ac} F^{bd} g_{cd}- \frac{1}{2} F^{cd} \nabla_{b} F_{cd} g^{ab} \right) \\
&=  \kappa  \left( \sum_{b,c}  \nabla_b F^{bc} F^{a}_{\phantom{a}c} - \sum_{b,c}\left( F^{bc} \nabla_b F_{c}^{\phantom{c}a} + \sum_{d} \frac{1}{2} F^{cd} \nabla_{b} F_{cd} g^{ab}\right) \right).
\end{align*}
On comparing with the above equalities we deduce that
$$
\sum_b  \nabla_b T^{a b} = \kappa \{ (i_{\delta F} F)^a -  (i_{F} dF)^a \}.  
$$
Expressed invariantly this is the vector field
$$
\div T^{\sharp} = \kappa \{(i_{\delta F} F)^{\sharp} - (i_{F} dF)^{\sharp}\},
$$
as was to be shown. 
\end{proof}

Now we go back to electromagnetic field $F$ on Minkowski spacetime $\mathbb{M}$. Our task is to show that the vanishing of the divergence of the energy-momentum tensor $T$ of $F$ implies the balance equations \eqref{ec:7.42} and \eqref{ec:7.49} in the classical theory. But first we need to pick the proportionality constant $\kappa$ appropriately. It turns out that $\kappa = c^2/\mu_0$ does the job. Thus, the expression for the components of $T$ relative to an arbitrary Lorentz frame are
$$
T^{ab} = \frac{c^2}{\mu_0} \sum_{c,d} \left(F^{ac} F^{bd} g_{cd} - \frac{1}{4} F_{cd}F^{cd} g^{ab} \right),
$$
or more explicitly
\begin{align*}
T^{00} &= \frac{\varepsilon_0}{2} \lvert E  \rvert^2 + \frac{1}{2 \mu_0} \lvert B  \rvert^2 , \\
T^{0i} &= T^{i0} = \frac{1}{\mu_0}(E \times B)^{i}, \\
T^{ij} &= c^2 \varepsilon_0  \left( \frac{1}{2} \lvert E  \rvert^2 \delta^{ij} - E^{i} E^{j} \right) +  \frac{c^2}{\mu_0}\left( \frac{1}{2} \lvert B  \rvert^2 \delta^{ij} - B^{i} B^{j} \right).
\end{align*}
From the definitions of the of the Poynting vector \eqref{eq:Poyntingvector} and the Maxwell stress tensor \eqref{eq:Maxwellstresstensor} it follows immediately that $T^{0i} =S^{i}$ and $T^{ij} = c^2 \Theta^{ij}$ and hence $T$ has component matrix
 $$
 (T^{ab}) = \left( \begin{array}{cccc} \frac{\varepsilon_0}{2} \lvert E  \rvert^2 + \frac{1}{2 \mu_0} \lvert B  \rvert^2 & S^1 & S^2 & S^3 \\ c^2 \varepsilon_0  (E \times B)^{1} &c^2 \Theta^{11} & c^2 \Theta^{12} & c^2 \Theta^{13} \\  c^2 \varepsilon_0  (E \times B)^{2} & c^2 \Theta^{21} & c^2 \Theta^{22} & c^2 \Theta^{23} \\ c^2 \varepsilon_0  (E \times B)^{3} & c^2 \Theta^{31} & c^2 \Theta^{32} & c^2 \Theta^{33}  \end{array} \right).
 $$
 We are now able to prove the following. 
 
 \begin{proposition}
 The conservation law for $T$ is equivalent to 
 \begin{align}
 \varepsilon_0 E  \cdot \frac{\partial E}{\partial t} + \frac{1}{\mu_0} B \cdot \frac{\partial B}{\partial t} + \div  S &= 0, \label{eq:consenergyV2}\\
 \varepsilon_0 \frac{\partial}{\partial t} (E \times B) + \div \Theta &= 0.   \label{eq:consmomentumV2}
 \end{align}
 \end{proposition}
 
 \begin{proof}
The conservation law for $T$ takes the form of a set of divergence conditions
$$
\sum_{b} \frac{\partial T^{ab}}{\partial x^{b}} = 0
$$
on the components $T^{ab}$ of $T$. Taking the free index $a$ to be $0$, we obtain
$$
\frac{\partial T^{00}}{\partial t} + \sum_{i} \frac{\partial T^{0i}}{\partial x^{i}} = 0,
$$
which by our above remarks gives
$$
\varepsilon_0 E  \cdot \frac{\partial E}{\partial t} + \frac{1}{\mu_0} B \cdot \frac{\partial B}{\partial t} + \sum_{i} \frac{\partial S^{i}}{\partial x^{i}} = 0,
$$
This, of course, is precisely the component form of the first equation. Similarly, restricting the free index $a$ to be a spatial index $i$, we obtain
$$
\frac{\partial T^{i0}}{\partial t} + \sum_{j} \frac{\partial T^{ij}}{\partial x^{j}} = 0,
$$
that is
$$
c^2 \varepsilon_0 \frac{\partial}{\partial t}(E \times B)^{i} + c^2 \sum_{j} \frac{\partial \Theta^{ij}}{\partial x^{j}} = 0,
$$
or, equivalently
$$
 \varepsilon_0 \frac{\partial}{\partial t}(E \times B)^{i} + \sum_{j} \frac{\partial \Theta^{ij}}{\partial x^{j}} = 0.
$$
This then gives us the component form of the second equation. Thus, the proposition is proved. 
 \end{proof}

\section{The symmetry of $T$}

As we saw in the examples of fluids and electromagnetism, the energy momentum tensor is symmetric. Let us consider what would happen if this were not the case. Suppose that $T_{12}(p)\neq T_{21}(p)$  and consider a very small cube $C$ of side length $L\ll1$ located at $p$. Recall that $T_{ij}$ is the $i$-th momentum flow density in the $j$-th direction.
The $z$ component $\tau_z$ of the torque on the cube is the sum of contributions from the $4$ faces of the cube that do not intersect the $z$ axis. The area of each face is $L^2$ and therefore
\[ \tau_z=\frac{L^3}{2}\Big( T_{12}-T_{21}+T_{12}-T_{21}\Big)=L^3(T_{12}-T_{21}).\]
On the other hand, since $L$ is very small, the mass density of the cube is approximately constant and equal to $\rho$. Therefore, the moment of inertia $I_z$ of the cube with respect to the $z$ axis is
\[ I_z\approx \int_C \rho (x^2+y^2)dV\approx \frac{1}{6}\rho L^5.\]
The angular acceleration with respect to the $z$ axis is then
\[ \alpha_z\approx \frac{\tau_z}{ I_z}\approx (T_{12}-T_{21}
)\frac{L^3}{6L^5}\approx (T_{12}-T_{21}
)\frac{1}{6L^2}. \]
One concludes that if $T_{12}\neq T_{21}$, then, when $L\to 0$, the cube will have arbitrarily large angular acceleration. This is not a reasonable physical behavior, so the tensor $T$ must be symmetric.
\begin{figure}[H]
	\center
	\includegraphics[scale=0.4]{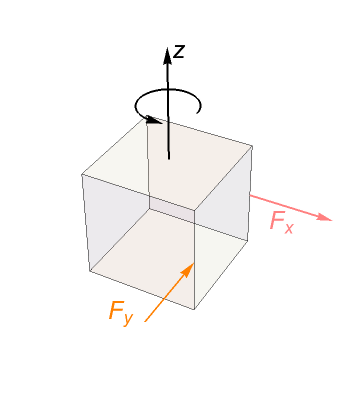}
	\caption{If $T$ were not symmetric, small cubes would rotate very fast.}
	\label{Cube}
\end{figure}

   \clearemptydoublepage 
\chapter{The Field equation\label{supa}}

\begin{center}
\parbox[b]{0.9\textwidth}{\small \sl Einstein's field equation
\[\Ric-\frac{1}
{2}\Rs g=\left(\frac{8\pi G_{N}}{c^{4}}\right) T ,\]
describes the relationship between the mass and energy distribution and the geometry of spacetime. The right hand side is proportional to the energy momentum tensor, while the left hand side is a tensor constructed from the metric. In this chapter we give a heuristic deduction of this equation. }
\end{center}

\vspace{3ex}

\section{Newton's law and the Poisson equation}\label{sec:11.1}

Newton's law of gravitation states that the gravitational force between two objects is
proportional to the product of their masses, and inversely proportional to the square of the distance between them. According to this law, if objects of masses $M$ and $m$ are located at $x$ and $y$, then, the second object will experience a force in Newtons
\[ f_{\mathrm{gr}}=G_N\frac{Mm(x-y)}{\left\vert x-y\right\vert ^{3}}.\]
The constant $G_N$ is the gravitational constant
$$G_N\approx 6.674
\times10^{-11}\mathrm{m^3kg^{-1}s^{-2}}.$$
The situation is described by postulating that the mass $M$ induces a gravitational field
\begin{equation}\label{gfield} g(y)=G_N\frac{M(x-y)}{\left\vert x-y\right\vert ^{3}},
\end{equation}
which determines the gravitational acceleration that other masses will experience.
More generally, a gravitational field $g$ is a
vector field such that a particle of mass $m$ 
located at $y $ is subject to a gravitational force 
\[f_{\mathrm{gr}}(y)=g(y)m.\] 
According to Newton's second law, the equation of motion for an object of mass $m$ in the presence of a gravitational field is
\begin{equation}
\label{f=ma}
m \ddot{y}(t)=f_{\mathrm{gr}}(y(t)).
\end{equation}
Equivalently,
\begin{equation}
\ddot{y}(t)=g(y(t)),
\end{equation}
so that the gravitational field gives the acceleration of any object moving under the effect of gravity.

Suppose that there are objects with masses $M_{1},\ldots,M_{n}$ located at
positions $x_{1},\ldots,x_{n}$. The gravitational field they generate is
\[
g(y)=G_N\sum_{i}\frac{M_i(x_i-y)}{\left\vert x_i-y\right\vert ^{3}}.
\]
In the continuous limit the mass is distributed
according to a density function $\rho(x)$ so that the total amount of mass in a region $U$ is
\[M=\int_{U}\rho(x)\, dV.\] In this situation the gravitational field is given by 
\begin{equation}
g(y)=G_N\int_{U}\frac{\rho(x)(x-y)}{\left\vert
x-y\right\vert ^{3}}\, dV. %
\end{equation}

 Consider an object of mass $M$ located at a point $x $ inside a region $U$ with boundary $S=\partial U$. Let us compute the flux of the gravitational field across the boundary.
\begin{figure}[H]
	\center
	\includegraphics[scale=0.4]{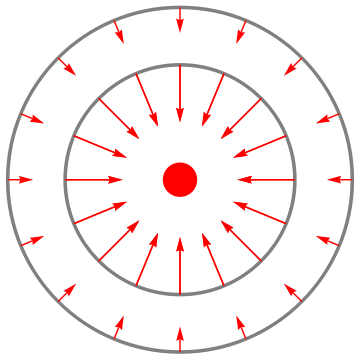}
	\caption{The flux of the gravitational field is equal for both spheres.}
	\label{Ampere}
\end{figure}
Let $B$ be a small ball centered at $x$ with radius $r$.
Outside of $B$, the gravitational field (\ref{gfield}) has no divergence
\[ \div g=0.\]
If  $V=U\setminus B$, Stokes' theorem gives
\[ 0=\int_V \div g \,dV=\int_{\partial U} g\cdot n \, dA- \int_{\partial B} g\cdot n \, dA.\]
On the other hand
 \[ \int_{\partial B} g\cdot n \, dA=\int_{\partial B}-\vert g \vert dA=-\frac{G_NM}{r^2}\int_{\partial B}dA=-4 \pi MG_N .\]
 One concludes that the flux across the boundary of $U$ is proportional to the mass inside of $U$. By linearity, this also holds for an arbitrary number of particles inside of $U$. In the continuous limit one obtains the gravitational version of Gauss' law
\begin{equation}
\int_{\partial U}g\cdot n \,dA=-4\pi G_N M.
\end{equation}
Since the region $U$ is arbitrary, the above implies
\begin{equation}\label{divg}
\div g=-4\pi G_N \rho.
\end{equation}

The gravitational field $g$ generated by a mass density function $\rho(x)$ is the negative gradient
 of the gravitational potential $\Phi(y)$, given by 
\begin{equation}
\Phi(y)=-G_N \int_{U}\frac{\rho(x)}{\left\vert
y-x\right\vert }dV. \label{potencial}%
\end{equation}
In fact, it is easy to check that \[g=-\grad \Phi.\]
In view of (\ref{divg}), and recalling that the Laplacian es defined as $\Delta = \div \circ \grad$, one obtains the Poisson equation for the potential
\begin{equation}\label{eq:Poissongrav}
\Delta \Phi=4 \pi G_N \rho.
\end{equation}

As an example, consider the interior of a spherical shell of uniform density.
It turns out that the gravitational field vanishes inside the shell. Let $S$ be a sphere of radius $r$ inside the shell, and $p$ a point in $S$.
If the tangential component of $g(p)$ were not zero, by spherical
symmetry, it would have to be nonzero at any other point in $S$. This
tangential component would be a non vanishing vector field on the sphere $S$. A theorem of topology, known as the hairy ball theorem, states that such a vector field does not exist. A proof of this theorem can be found, for instance, in the book by Guillemin and Pollack \cite{GP}. One concludes that $g(p)$ is normal to the sphere $S$. By symmetry, it has to be normal and of equal magnitude $C$ at any
other point in $S$. This implies that the flux of $g$ across $S$ is
\[\int_S g \cdot n \,dA=4\pi r^{2}C .\] 
On the other hand, if $D$ is the ball with boundary $S$, then
\[\int_S g \cdot n \,dA= \int_D \div g \, dV=-4 \pi G_N  \int_D \rho \, dV=0.\] 
One concludes that $C=0$, and that the gravitational field vanishes inside the shell.
\begin{figure}[H]
	\center
	\includegraphics[scale=0.4]{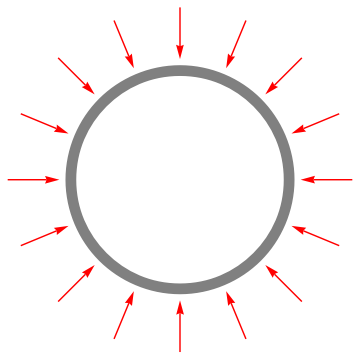}
	\caption{The gravitational field caused by a spherical shell.}
	\label{Ampere}
\end{figure}

\section{Units and dimensions}

 As discussed above, in Newton's theory, the relationship between the mass density and
 the gravitational potential is given by the Poisson equation
$\Delta\Phi=  4 \pi G_N \rho$. Relativistically, the mass and energy density is not described by a scalar function, but by the energy momentum tensor $T$. Also, the effect of gravity should be a change in the geometry, so that
the gravitational potential is replaced by the metric. Therefore, the analogue of the Poisson equation
should be an expression of the form:
$G\propto T,$
stating that the energy momentum tensor is proportional to some tensor $G$ constructed from the metric and its partial derivatives. The tensor $G$ must be a rank two symmetric tensor with zero divergence, since $T$ is. The dimensions of the quantities involved also give important clues as to the nature of $G$.

For this analysis we use coordinates of spacetime that have dimensions  of length, for instance, in Minkowski spacetime $(x^0,x^1,x^2,x^3)$ where $x^0=ct$. We write $[x^a]=[\mathrm{L}].$
The other fundamental dimensions are time $[\mathrm{T}]$ and mass $[\mathrm{M}]$.

\begin{center}
\begin{tabular}{ccc}
\hline &&\\
{\textbf Quantity} &&{\textbf Dimensions}\\  
&&\\
\hline && \\
Length& & $[\mathrm{L}]$ \\ 
 &&\\ \hline
 &&\\
Time&  &$[\mathrm{T}]$   \\  
&&\\ \hline
&&\\
Mass&& $[\mathrm{M}]$ \\  
&&\\ \hline
&&\\
Force&  &$[\mathrm{M}][\mathrm{L}][\mathrm{T}]^{-2} $ \\  
&&\\ \hline
&&\\
Energy &  &$  [\mathrm{M}][\mathrm{L}]^2[\mathrm{T}]^{-2}$\\  
&&\\ \hline
&&\\
Energy  density&  &$  [\mathrm{M}][\mathrm{L}]^{-1}[\mathrm{T}]^{-2}$\\  
&&\\
\hline
\end{tabular}
\end{center}

All tensorial quantities acquire dimensions as follows. Since the coordinates have dimension of length, so does the tensor $dx^a$. So that 
\[ [dx^a]=[L]\]
and 
\[[ \partial_{x^a}]=[L]^{-1}.\]
If a scalar function $f$ has dimensions $[f]=[L]^i[M]^j[T]^k$ then the tensor \[S=f \partial_{x^{a_1}}\otimes \cdots \otimes  \partial_{x^{a_r}}\otimes {dx^{b_1}}\otimes \dots \otimes {dx^{b_s}}\]
has dimension
\[ [S]=[\mathrm{L}]^{i+s-r}[\mathrm{M}]^j[\mathrm{T}]^k.\]
The dimensions of a quantity specify how the quantity changes when the units of measurement change. For instance, suppose that the metric tensor 
\[g=\sum_{ab}g_{ab} dx^a \otimes dx^b\]
is used to measure the length of a curve in meters $(\mathrm{m})$, and 
\[\overline{g}=\sum_{ab}\overline{g}_{ab} dx^a \otimes dx^b\]
is used to measure length in new meters $(\overline{\mathrm{m}})$, where $\epsilon \:\overline{\mathrm{m}} =1\: \mathrm{m}$.
Then, the length of a curve is
\[ L(\gamma(\tau))=\left( \int_0^1 \sqrt{g(\gamma'(\tau),\gamma'(\tau))}d\tau \right) \mathrm{m}=\left( \int_0^1 \sqrt{\overline{g}(\gamma'(\tau),\gamma'(\tau))}d\tau \right) \overline{\mathrm{m}}.\]
This implies that
\[\left( \int_0^1 \sqrt{g(\gamma'(\tau),\gamma'(\tau))}d\tau \right)\epsilon =\left( \int_0^1 \sqrt{\overline{g}(\gamma'(\tau),\gamma'(\tau))}d\tau \right).\]
Since the path $\gamma(\tau)$ is arbitrary, one concludes that
\[ \overline{g}= \epsilon^2 g ,\]
so that the metric tensor has dimensions of length squared $ [g]=[\mathrm{L}]^2$.
This implies that the components of the metric $g_{ab}$ are dimensionless. Since $[\partial_{x^a}]=[\mathrm{L}]^{-1}$ then the partial derivatives of the components of the metric have dimensions of negative length 
\[ [\partial_{x^c}g_{ab}]=[\mathrm{L}]^{-1},\quad  [\partial_{x^d}\partial_{x^c}g_{ab}]=[\mathrm{L}]^{-2}.\]
The components of the energy momentum tensor have dimensions of energy density and therefore
\[ [T]=[\mathrm{M}][\mathrm{L}][\mathrm{T}]^{-2}.\]
The universal constant of gravitation $G_N$ has dimensions
\[ [G_N]=[\mathrm{M}]^{-1}[\mathrm{L}]^3[\mathrm{T}]^{-2}.\]
Therefore, the tensor $(G_N/c^4)T$
is adimensional. One concludes that the relativistic analogue of the Poisson equation takes the form
\begin{equation}\label{rpoisson} G \propto  \frac{G_N}{c^4} T
\end{equation}
where $G$ is a dimensionless tensor.

\section{The Einstein tensor}
We are looking for the tensor $G$ that goes on the left hand side of $(\ref{rpoisson})$.
This tensor should have the following properties:
\begin{itemize}
\item $G$ has rank two and is symmetric $G_{ab}=G_{ba}$
\item $G$  has no divergence, $\div G^\sharp=0$.
\item $G$ is dimensionless.
\end{itemize}

It is a theorem due to Lovelock \cite{Lovelock}, that the conditions above, together with the requirement that the tensor is natural, completely determine $G$. We will not formalize the idea that the tensor is natural. Intuitively, it means that the way it is expressed in terms of the metric and its derivatives is the same in all coordinate systems.
Since $G$ should be dimensionless, its components $G_{ab}$ should have dimensions $[\mathrm{L}]^{-2}$, therefore, they should be linear in the second derivatives of the metric or quadratic in the first derivatives. There are two natural symmetric tensors associated to the metric: the Ricci tensor $\Ric$ and the metric itself.
The Ricci tensor is adimensional and symmetric, but in general it may have divergence. The metric is symmetric and has no divergence, but has units of $[\mathrm{L}]^2$. On the other hand, the scalar curvature \[\Rs=\sum_{ab}g^{ab}\Ric_{ab}\] is obtained by contracting the Ricci tensor with the inverse of the metric tensor, and therefore, it has units of $[L]^{-2}$. This means that
$\Rs g$ is a symmetric adimensional tensor. It is then natural to look for $G$ of the form:
\[ G=\lambda \Ric + \mu \Rs g,\]
for constants $\lambda,\mu$ that make $G$ divergenceless. We have proved in Proposition \ref{necesaria} that if $\mu=-\lambda/2$ then $\div(G^{\sharp})=0$. Let us consider a more general solution
\[ G=\lambda \Ric -\frac{\lambda}{2}\Rs g+\nu \Rs g.\]
In this case,
 \[ \gamma\div(\Rs g^{\sharp})=0.\]
Since the metric is covariantly constant, then
\begin{eqnarray*}
 \nu\sum_a g^{ab} \partial_{x^{a}} \Rs = 0.
\end{eqnarray*}
Choosing coordinates where the metric is diagonal at the point $p \in M$, one concludes that, if $\nu\neq 0$, then $\Rs$ is locally constant. Assuming that $M$ is connected, $\Rs$ is constant. Therefore, if $\nu \neq 0$, taking the trace of $G$, one obtains
\[ \sum_a T^a_a\propto \sum_a G^a_a=\lambda \Rs -2\lambda\Rs +4\nu \Rs=(4\nu-\lambda )\Rs=\text{constant}. \]
The condition that the trace of the energy momentum tensor is a constant is too restrictive. The conclusion is that $\nu=0$ and therefore, setting $\lambda=1$ one obtains
\begin{equation} G=\Ric -\frac{1}{2}\Rs g.\end{equation}
The tensor $G$ is known as the Einstein tensor. As required, it is a rank two, symmetric,  dimensionless tensor with vanishing divergence. Therefore, the equation we are looking for takes the form
\begin{equation}\label{alpha}\Ric-\frac{1}
{2}\Rs g=\alpha\left( \frac{G_{N}}{c^{4}} T\right) ,\end{equation}
for some dimensionless proportionality constant $\alpha$. This constant can be specified by considering the newtonian limit of (\ref{alpha}), which should recover Newton's equations.

\section{Newtonian limit and the value of $\alpha$}\label{svalpha}

In situations where velocities are small compared to that of light and objects are not too massive, Newton's theory  correctly describes gravity. Therefore, Einstein's theory should approximate classical gravity in that non relativistic regime. The two fundamental equations in Newton's gravity are the Poisson equation
\[
\Delta \Phi=4 \pi G_N \rho,
\]
that describes the gravitational field in terms of the mass distribution, and the equation of motion
\begin{equation}\label{fma} f=ma\end{equation}
that describes the motion of objects in the presence of gravity. The value of the proportionality constant $\alpha$ in Einstein's equation (\ref{alpha}) can be found by requiring that, in the non-relativistic limit, one recovers the Poisson equation, and that the relativistic  equation of motion
\[ \nabla_{\gamma'(\tau)}\gamma(\tau)=0\]
 becomes $a=-\grad\Phi$.

The Newtonian limit refers to situations where relativistic effects are negligible. This occurs when objects are moving slowly and are not very massive. 
The precise assumptions are the following:
\begin{itemize}
\item We consider units where the speed of light is $c=1$ and massive objects are moving at relatively small velocities $v\ll c=1$.
\item As in \S \ref{section 00}, the spacetime manifold is $\RR^4$ with a metric
\[ g= \eta + \varepsilon,\]
where $\eta$ is the Minkowski metric 
\begin{equation*}
\eta= \begin{pmatrix}
-1&0&0&0\\
0&1&0&0\\
0&0&1&0\\
0&0&0&1
\end{pmatrix}
\end{equation*}
and $\varepsilon$ is a small perturbation so that $\varepsilon_{ab}\ll1$ and $\frac{\partial \varepsilon_{ab}}{\partial x^c}\ll1$. This is the assumption that gravity is weak, so that spacetime is approximately flat.
\item The energy momentum tensor $T$ takes the form
\[T= \begin{pmatrix}
\rho&0&0&0\\
0&0&0&0\\
0&0&0&0\\
0&0&0&0
\end{pmatrix}.\]
This is the assumption that matter is moving slowly so that the momentum is small and the energy momentum tensor is dominated by the energy density $\rho$.
\item The metric is independent of time $\partial_t \varepsilon_{ab}=0$.
\end{itemize}

In the following computations we will disregard terms that are quadratic or higher in $\varepsilon$ and its derivatives. We write $X \sim Y$ to mean that $X=Y$ modulo higher order terms.
Since $c=1$, the Einstein equation is
\begin{equation} \Ric-\frac{1}{2}\Rs g=\alpha G_N T.\end{equation}
Taking traces on both sides one obtains
\begin{equation}\label{atrace} \Rs =\alpha G_N \rho.\end{equation}
Since the only nonzero term in $T$ is $T_{00}$, we focus on the equation
\begin{equation} G_{00}=\alpha G_N \rho.\end{equation}
In view of (\ref{atrace}), this is equivalent to
\begin{equation}\label{roo}2 R_{00}= \alpha G_N \rho.\end{equation}
The Ricci tensor is the contraction of the curvature tensor,  therefore
\begin{equation}\label{oo}
R_{00}=\sum_a\left[\frac{\partial \Gamma^{a}_{00}}{\partial x^a}-\frac{\partial \Gamma^{a}_{a0}}{\partial x^0}+\sum_b \big( \Gamma^{a}_{ab} \Gamma^{b}_{00}- \Gamma^{a}_{0b} \Gamma^{b}_{0a}\big)\right].
\end{equation}
The Christoffel symbols are
\begin{equation}
\Gamma_{ab}^c=\frac{1}{2}\sum_d g^{dc}\Big( \frac{\partial \varepsilon_{ad}}{\partial x^b}+ \frac{\partial \varepsilon_{bd}}{\partial x^a}- \frac{\partial \varepsilon_{ab}}{\partial x^d}\Big).
\end{equation}
Neglecting higher order terms, this becomes
\begin{equation}\label{Cr}
\Gamma_{ab}^c\sim \frac{1}{2} g^{cc}\Big( \frac{\partial \varepsilon_{ac}}{\partial x^b}+ \frac{\partial \varepsilon_{bc}}{\partial x^a}- \frac{\partial \varepsilon_{ab}}{\partial x^c}\Big).
\end{equation}
Therefore, the terms in the second sum in (\ref{oo}) are quadratic in $\varepsilon$. Also, since the time derivatives of the metric vanish, the second term can be disregarded. One obtains:
\begin{equation}
R_{00}\sim \sum_{i>0}\frac{\partial \Gamma^{i}_{00}}{\partial x^i}\sim \frac{1}{2}\sum_{i>0}\Big( \frac{\partial \varepsilon_{0i}}{\partial x^i\partial x^0}+ \frac{\partial \varepsilon_{0i}}{\partial x^i\partial x^0}- \frac{\partial \varepsilon_{00}}{\partial x^i\partial x^i}\Big).\end{equation}
Since the partial derivatives with respect to $x^0=t$ vanish,  this becomes
\begin{equation}
2R_{00}\sim -\Delta g_{00}.
\end{equation}
Therefore, equation (\ref{roo}) tends to the Poisson equation as long as the potential is
\begin{equation}\label{pot} \Phi= K-\frac{4 \pi g_{00}}{\alpha},\end{equation}
for some constant $K$.
One can now use the requirement that the geodesic equation tends to 
\begin{equation}\label{ac}
a=-\grad \Phi
\end{equation} 
to determine $\alpha$. Consider the world line $\gamma(\tau)$ of an object, which, as usual, is parametrized by proper time so that \[\langle \gamma'(\tau),\gamma'(\tau)\rangle=-c^2=-1.\]
The geodesic equation is
\begin{equation}
\frac{d^2 x^a}{d\tau ^2}+\sum_{bc} \frac{d x^b}{d\tau}\frac{d x^c}{d\tau} \Gamma_{bc}^a=0.
\end{equation}
Recall that \[\frac{dt}{d\tau}=\frac{dx^0}{d\tau}=\lambda_v=\frac{1}{1-v^2/c^2},\] which, under our assumptions, tends to one.
On the other hand, for $i>0$:
 \[\frac{dx^i}{d\tau}=\frac{dx^i}{dt}\frac{dt}{d\tau}\sim\frac{dx^i}{dt} \ll 1. \] 
Therefore, the geodesic equation tends to
\begin{equation}
\frac{d^2 x^a}{d\tau ^2}\sim- \Gamma_{00}^a.
\end{equation}
Using (\ref{Cr}) and the fact that the metric is time independent, one obtains that, for $i>0$
\begin{equation}
\Gamma_{00}^i=-\frac{1}{2}\frac{\partial g_{00}}{\partial x^i}.
\end{equation}
Therefore, the geodesic equation becomes
\begin{equation}
a\sim \frac{1}{2} \grad  g_{00}.
\end{equation}
In order for this to be equal to (\ref{ac}) one needs to set 
\begin{equation}\label{valpha}\Phi=-\left(\frac{c^2}{2}+\frac{g_{00}}{2}\right),\end{equation}
where the integration constant is $-c^2/2$ since in the absence of gravity on should recover the Minkowski metric.
Replacing in (\ref{pot}) one gets
\[ \alpha=8\pi.\]

The conclusion is that, in order to recover Newton's equations in the non-relativistic limit,  the field equation must be
\begin{equation}
\boxed{\Ric -\frac{1}{2}\Rs g= \frac{8 \pi G_N}{c^4} T.}
\end{equation}

The relationship between Newton's  gravity and Einstein's theory can be summarized as follows:

\begin{center}
\begin{tabular}{ccc}
\hline &&\\
{\textbf Classical Gravity} &&{\textbf General Relativity}\\  
&&\\
\hline && \\
Energy density $\rho$ &$\to $ & Energy momentum tensor $T_{ab}$ \\ 
 &&\\
 Gravitational potential $\Phi$& $\to$ &  Metric $g$ \\  
 &&\\
Equation of motion $F=m a$ & $\to$ & Geodesic equation $\nabla_{\gamma'(t)}\gamma'(\tau)=0$ \\  
&&\\
Poisson Equation $\Delta\Phi=4\pi G_N \rho$& $\to$ &  Einstein equation

$G_{ab}=8\pi G_N/c^4T_{ab}$ \\  
&&\\
Conservation of energy and momentum& $\to$ &  $\div T^\sharp=0$ \\  
&&\\
\hline
\end{tabular}
\end{center}

\section{Estimate of $g_{00}$} \label{section 00}

Suppose that in Minkoswki spacetime there is a body $\mathcal{B}$ of mass $M$ and radius $R$, located at the
origin of the coordinate system. This body $\mathcal{B}$ induces a perturbation of the Minkowski
metric $\eta$ creating a new metric that for simplicity we assume to be  \emph{static}, and of the form $g=\eta+\varepsilon$. In this context, the word static means that the entries of $\varepsilon$ are functions of the
spatial coordinates alone. Moreover, we also assume that $%
\varepsilon_{ij}(x^{i})\rightarrow0$ when
 $\lvert x \rvert \to \infty $. This is a
reasonable hypothesis since the strength of the field must approach zero when we move far
away from $\mathcal{B}$. 

In this situation we want to estimate the metric coefficient $
g_{00}$. We start by considering the case of a observer $O$ at
a fixed distance $x$ in the direction of $x^{1}$ from the surface of $\mathcal{B}$. It has worldline $%
O(s)=(s,x,0,0)$ in the standard coordinates of $\RR^{4}.$ Notice that 
$O$ is not moving along a geodesic. It stays in a fixed location outside of $\mathcal{B}$. In this case $\langle O^{\prime
}(s),O^{\prime }(s)\rangle =g_{00}(O(s))$ and therefore, since $g$ does not depend on $t$,
\begin{equation}
\left\langle O^{\prime }(s),O^{\prime }(s)\right\rangle =g_{00}(x),
\label{E60}
\end{equation}%
so that 
\[
\left\vert O^{\prime }(0)\right\vert =\sqrt{-g_{00}(x)}.
\]%
The $4$-velocity is then
$$
\mathbf{u}=\frac{1}{\sqrt{-g_{00}(x)}%
}\partial _{t}.
$$
As in \S \ref{dilatacion del tiempo}, we want to compare the frequencies $f_{B}$
and $f_{A}$ of a pulse of light, as measured by two observers $A
$ and $B$ at fixed distances $x_{A}<x_{B}$ from the center of $\mathcal{B}$. 

Let $\gamma (s )
$ be a null geodesic corresponding to the worldline of the light signal
emitted by $B$ at $s=s_0$, and suppose it is received by $A$ at $s
=s_{1}.$ The energy of this signal measured by $B$, at its emission, and by $A$ when it is received, can be  calculated as $$%
E_{A}=-\left\langle \gamma ^{\prime }(s_{1}),\mathbf{u}_{A}\right\rangle $$ and $$E_{B}=-\left\langle
\gamma ^{\prime }(s_0)),\mathbf{u}_{B}\right\rangle ,$$ 
respectively. Since the energy of a circular wave is $\hslash $ times its
frequency, we see that 
\begin{align*}
f _{A}& =-\hslash ^{-1}\left\langle \gamma ^{\prime }(s_{1}),%
\mathbf{u}_{A}\right\rangle =-\frac{\hslash ^{-1}}{\sqrt{-g_{00}(x_{A})}}%
\left\langle \gamma ^{\prime }(s_{1}),\partial _{t}\right\rangle,  \\
f _{B}& =-\hslash ^{-1}\left\langle \gamma ^{\prime }(s_0),%
\mathbf{u}_{B}\right\rangle =-\frac{\hslash ^{-1}}{\sqrt{-g_{00}(x_{B})}}%
\left\langle \gamma ^{\prime }(s_{0}),\partial _{t}\right\rangle .
\end{align*}%

Since $\partial_t$ is a Killing vector field and $\gamma(s)$ is a geodesic, we have that:
 $\left\langle \gamma
^{\prime }(s),\partial _{t}\right\rangle $ is constant. Therefore:
\begin{equation}
f _{B}=f _{A}\frac{\sqrt{-g_{00}(x_{A})}}{\sqrt{%
-g_{00}(x_{B})}}.  \label{E56}
\end{equation}%
Recall we are assuming $g_{00}(x^{i})\rightarrow -1,$ as $\lvert x \rvert\to \infty$. If $f
_{\infty }$ denotes the limit $\lim_{x_{B}\rightarrow \infty }f _{B}$ we see
that
\begin{equation}
f _{\infty }=f _{A}\sqrt{-g_{00}(x_{A})}.  \label{E56-1}
\end{equation}%
On the other hand, for a gravitational potential that is not very strong we had calculated in equation (\ref{E55}) that, in standard units $$%
f _{B}=f _{A}\left ( 1+\frac{\Phi (x_{A})-\Phi (x_{B})}{c^{2}}\right).$$
Again, by letting $x_{B}\rightarrow \infty $, and since the gravitational
potential $\Phi (x_{B})$ approaches zero, in the limit one obtains%
\begin{equation}
f _{\infty }=f _{A}\left( 1+\frac{\Phi (x_{A})}{c^{2}}\right) .
\label{E57}
\end{equation}%
Comparing (\ref{E56-1}) with (\ref{E57}) one gets the estimate 
\[
g_{00}(x_{A})=-\left( 1+\frac{\Phi (x_{A})}{c^{2}}\right) ^{2}.
\]%
Now, for a celestial body like the Earth, or even the Sun, the term 
$$\frac{\Phi (x_{A})}{c^{2}}=\frac{-G_{N} M}{c^{2}x_{A}}$$ 
is very small. In the case of the Earth it is of order $10^{-10}$. So one has the approximation $$\left( 1+\frac{\Phi x_{A}}{c^{2}}\right) ^{2} \approx 1+2\frac{\Phi (x_{A})}{c^{2}}.
$$ 
This gives the following estimate for $g_{00}$:%
\begin{equation}
g_{00}(x)=- 1+2\frac{G_{N}M}{c^{2}x} .
\label{estimate new}
\end{equation}

\section{The cosmological constant}

We arrived at the Einstein tensor
\[ G=\Ric -\frac{1}{2}\Rs g\]
by looking for dimensionless tensors derived from the metric. If one is willing to introduce a new constant $\Lambda$, that depends on the units and is of dimension $[\mathrm{L}]^{-2}$, then the tensor
\[ \Lambda g\]
is rank two, symmetric, dimensionless and has zero divergence. By adding this term to $G$ one obtains the Einstein equation with cosmological constant $\Lambda$ 
\begin{equation}\label{cc}
\Ric -\frac{1}{2}\Rs g+ \Lambda g=\frac{8 \pi G_N}{c^4}T.
\end{equation} If 
\[ T^{\Lambda}=-\frac{\Lambda c^4}{8 \pi G_N}g,\]
then equation (\ref{cc}) can be rewritten as:
\begin{equation}\label{cc1}
\Ric -\frac{1}{2}Rg=\frac{8 \pi G_N}{c^4}(T+T^{\Lambda}),
\end{equation}
which is the usual Einstein equation with energy momentum tensor $T + T^{\Lambda}$.
Therefore, introducing a cosmological constant has the effect of assigning a nonzero energy density and pressure to empty spacetime. Taking traces in (\ref{cc}) with $T=0$ one obtains
\begin{equation}
 4\Lambda=\Rs.
\end{equation}
Therefore, the vacuum Einstein equation with cosmological constant $\Lambda$ becomes
\begin{equation}
\Ric =\Lambda g.
\end{equation}
A Riemannian or Lorentzian manifold is called an Einstein manifold if its Ricci curvature is proportional to the metric. Einstein manifolds are the solutions to the Einstein equations with cosmological constant in the vacuum.

The following are some simple examples in Euclidean signature.
 Euclidean space is an Einstein manifold with $\Lambda=0$.
 The  four dimensional sphere of radius $\alpha$ is an Einstein manifold with $\Lambda =3/\alpha^2$.
The $4$-dimensional hyperboloid
 \begin{equation}\label{hyperboloid}H=\{ (x^0,\dots,x^4) \in \RR^{5}\mid -(x^0)^2+(x^1)^2 + \dots + (x^4)^2=-\alpha^2\},\end{equation}
with the metric induced by the Minkowski metric \[g=-dx^0\otimes dx^0+\sum_{i=1}^5 dx^i \otimes dx^i\] on $\RR^5$, is an Einstein manifold with $\Lambda=-3/\alpha^2$. These examples have counterparts in Lorentzian signature. Of course, the Lorentzian analogue of Euclidean spacetime is Minkowski spacetime, which is Einstein with $\Lambda=0$.
The Lorentzian version of the sphere is de Sitter space
\[ \mathrm{dS}_4=\{ (x^0,\dots,x^5) \in \RR^{5}\mid -(x^0)^2+(x^1)^2 + \dots + (x^4)^2=\alpha^2\},\]
where, again, the metric is induced by the Minkowski metric. Just like the group of rotations $\mathrm{SO}(5)$ acts
on the four dimensional sphere, the isometry group $\mathrm{SO}(1,4)$ acts transitively by isometries on $\mathrm{dS}_4$. de Sitter space is an Einstein manifold with $\Lambda=3/\alpha^2$. There is a diffeomorphism $\varphi:\RR \times S^3\to \mathrm{dS}_4$,
given by
\[ \varphi(\tau, v)= \big(\alpha\sinh(\tau/\alpha), \alpha\cosh(\tau/ \alpha)v\big).\]
With respect to this diffeomorphism, the metric takes the form
\[ g= -d\tau \otimes  d\tau +\alpha^2 \cosh^2(\tau/\alpha) \omega,\]
where $\omega$ is the round metric on the sphere. 
\begin{figure}[H]
	\center
	\includegraphics[scale=0.38]{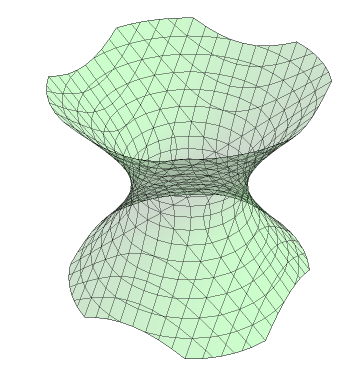}
	\caption{de Sitter spacetime is diffeomorphic to $\RR \times S^3$.}
	\label{Ampere}
\end{figure}

Anti de Sitter space corresponds to the hyperboloid (\ref{hyperboloid}) in Lorentzian signature. It is the space defined by
\[\mathrm{AdS}_4=\{ (x^0,\dots,x^5) \in \RR^{5}\mid -(x^0)^2-(x^1)^2 + (x^3)^2 +(x^3)^2 +(x^4)^2=-\alpha^2\},\]
where the metric is induced by the metric of signature $(2,3)$ on $\RR^5$:
\[g=-dx^0 \otimes dx^0-dx^1 \otimes dx^1+dx^2 \otimes dx^2+dx^3 \otimes dx^3+dx^4 \otimes dx^4.\]
Anti de Sitter space is an Einstein manifold with $\Lambda=-3/\alpha^2$. 
Rewriting the equation that defines $\mathrm{AdS}_4$ as
\[  (x^0)^2+(x^1)^2 = (x^3)^2 +(x^3)^2 +(x^4)^2+\alpha^2,\]
shows that there is a diffeomorphism $\phi:S^1 \times \RR^3\to \mathrm{AdS}_4$,
given by
\[ \phi(z, v)= \big( \sqrt{\alpha^2 +|v|^2}z,v\big).\]
The vector field  $\partial_\theta$ is timelike, and therefore, $\mathrm{AdS}_4$ is not chronological.

\begin{figure}[H]
	\center
	\includegraphics[scale=0.38]{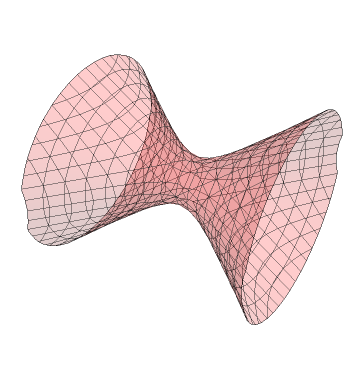}
	\caption{Anti de Sitter spacetime is diffeomorphic to $S^1 \times \RR^3$.}
	\label{Ampere}
\end{figure}

It will be no big surprise that the cosmological constant arises in cosmology. We will encounter it again in the discussion of the cosmological models provided by the Friedmann-Lemaitre-Robertson-Walker metric.

\section{The geometric meaning of Einstein's equation}
\label{Baez}
In this section we discuss the geometric meaning of Einstein's field equation. Our exposition is based on the beautiful paper  by Baez and Bunn \cite{baezbueno}, which we recommend.\\
Let us fix units so that the speed of light is $c=1$, and the gravitational constant is $G_N=1$. 
Einstein's equation is then
\begin{equation}\label{ein}
\Ric -\frac{1}{2}\Rs g= 8 \pi  T.
\end{equation}
 Taking traces on both sides one obtains
 \begin{equation}\label{trace}
\Rs=-8 \pi \mathrm{tr}\, T
\end{equation}
Substituting (\ref{trace}) back into (\ref{ein}) one gets
\begin{equation}\label{nein}
\Ric=8 \pi \left(T-\frac{1}{2} \mathrm{tr}\, T g\right),
\end{equation}
which is equivalent to Einstein's equation. We will use the following observation:
\begin{lemma}\label{Baezl}
The equation (\ref{nein})
holds if and only if, for each point $p \in M$, and all frames $v^0,\dots, v^3$ of $T_pM$ where the metric takes the standard Minkowski form, the condition
\begin{equation}\label{lein}\Ric_{00}(p)=4 \pi \big(T_{00}+T_{11}+T_{22}+T_{33})(p),
\end{equation}
holds.
\end{lemma}
\begin{proof}
First, suppose that (\ref{nein}) is satisfied. In this case, for any frame at $p$
\[\Ric_{00}(p)=8 \pi \big(T_{00}-\frac{1}{2} \mathrm{tr}\, T g_{00}\big)(p).\]
On the other hand, if the metric takes the Minkowski form, the right hand side can be computed
\[8 \pi \big(T_{00}+\frac{1}{2} \mathrm{tr}\, T \big)(p) =8\pi \big(T_{00}+\frac{1}{2} (-T_{00}+T_{11}+T_{22}+T_{33})\big)(p)=4\pi  \big(T_{00}+T_{11}+T_{22}+T_{33})(p). \]
Conversely, suppose that condition (\ref{lein}) holds for any Minkowski frame at $p$. Then
\[ \Ric(p)_{00} -8 \pi \big(T_{00}-\frac{1}{2} \mathrm{tr}\, T g_{00}\big)(p) \]
vanishes for all Minkowski frames at $p$. This implies that the corresponding quadratic form on $T_pM$ vanishes on timelike vectors, which form an open set (an open cone). A quadratic form that vanishes on an open set is necessarily zero, and one concludes that
\[ \Ric(p) -8 \pi \big(T-\frac{1}{2} \mathrm{tr}\, T g\big)(p)=0. \]
Since $p$ is arbitrary, equation (\ref{nein}) holds.
\end{proof}

In summary, so far we have seen that Einstein's equation is equivalent to the condition that (\ref{lein})
holds for any Minkowski frame at all points $p\in M $.\\

Consider an observer, Alice, that is falling freely with world line $\gamma(\tau)$, which is a timelike geodesic parametrized by proper time. Set $p =\gamma(0)$ and fix a Lorentz frame $v^0=\gamma'(0),v^1, v^2,v^3$ with corresponding Fermi coordinates $(x^0=t, x^1,x^2,x^3)$, so that the Christoffel symbols vanish along the worldline $\gamma(\tau)=(\tau,0,0,0)$. Suppose that Alice travels in a small ball which is also falling freely.  At time $t=0$ the radius of the ball is $\varepsilon>0$, and she paints dots at  $q_1,q_2,q_3$, the intersection of the ball with the coordinate axes. These points have coordinates:
\[ q_1=(0,\varepsilon,0,0)=\varepsilon e_1,\,\, q_2=(0,0,\varepsilon,0)=\varepsilon e_2,\,\, q_3=(0,0,0,\varepsilon)=\varepsilon e_3.\]
The following picture illustrates the situation.

\begin{figure}[H]
\centering
\includegraphics[scale=0.45]{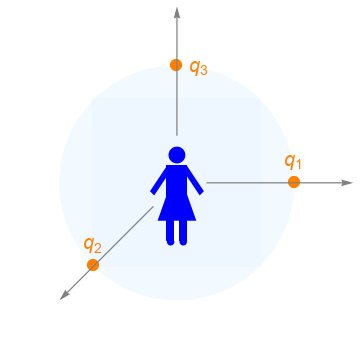}\caption{Alice's ball at time $t=0$.}%
\end{figure}

Tidal forces will cause the ball to deform, and Alice is interested in the way in which the volume of the sphere changes. She keeps track of the volume by following the trajectories of each of the points $q_i$.
The worldline of $q_i$ will be denoted $\gamma_i(\tau)$. We assume that, at time $t=0$, the points are at rest with respect to Alice, so that $\gamma'_i(0)=\partial_t(q_i)$. Since $q_i$ moves along a geodesic, this implies that its worldline is
\[ \gamma_i(\tau)=\exp(q_i)(\tau \partial_t).\]
Let $r_i(\tau)$ the $i$-th component of $\gamma_i(\tau)$. For each value of $\tau$, the volume of the deformed ellipsoid can be approximated by
\[ V(\tau) \sim \frac{4\pi}{3}r_1(\tau) r_2(\tau) r_3(\tau).\]
Since $\dot{r}_i(0)=0$, one obtains
\begin{equation}\label{volvol} \frac{\ddot{V}(0)}{V(0)}\sim \sum_i \frac{\ddot{r}_i(0)}{\varepsilon}.\end{equation}
\begin{figure}[H]
\centering
\includegraphics[scale=0.45]{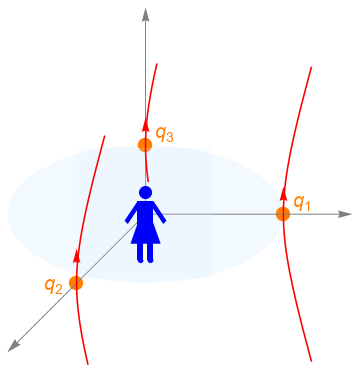}\caption{The ball deforms due to gravity.}%
\end{figure}
Let us compute 
\[ \lim_{\varepsilon \to 0} \frac{\ddot{r}_i(0)}{\varepsilon}.\]
The map $\sigma_i(\tau,s) :I \times (-\delta,\delta) \to M$, with $\varepsilon<\delta$, defined by
\[ \sigma_i (\tau, s)= \exp (s e_i)(\tau \partial_t)\]
is a one parameter family of geodesics. This implies that the vector field
\[ W_i(\tau) := \frac{d}{ds}\Big \vert_{s=0} \sigma_i(\tau, s),\]
is a Jacobi field. Therefore, it satisfies:
\[ \nabla_{\gamma'(\tau)}\nabla_{\gamma'(\tau)} W_i(\tau)= R(\gamma'(\tau),W_i(\tau))(\gamma'(\tau)).\]
Since the Christoffel symbols vanish on the worldline, this implies:
\[\frac{d^2}{d\tau^2} W_i(\tau)=R(\gamma'(\tau),W_i(\tau))(\gamma'(\tau)). \]
Evaluating at $\tau=0$, one obtains
\[\frac{d^2}{d\tau^2} \Big \vert_{\tau=0}W_i(\tau)=R(\partial_t,\partial_{x^i})(\partial_t)=\sum_lR^l_{00i}\partial_{x^l}=-\sum_lR^l_{0i0}\partial_{x^l}. \]
Recall that $r_i(\tau)$ is the $i$-th component of $\gamma_i(\tau)=\sigma_i(\tau,\varepsilon)$ and therefore
\begin{equation*}
\lim_{\varepsilon \rightarrow 0}\frac{\ddot{r}_{i}(0)}{\varepsilon }
\end{equation*}%
is the $i$-th component of 
\begin{eqnarray*}
\lim_{\varepsilon \rightarrow 0}\frac{\ddot{\gamma}_{i}(0)}{\varepsilon }
&=&\left. \frac{d}{d\varepsilon }\right\vert _{\varepsilon =0}\left. \frac{%
d^{2}}{d\tau ^{2}}\right\vert _{\tau =0}\sigma (\tau ,\varepsilon )=\left. 
\frac{d^{2}}{d\tau ^{2}}\right\vert _{\tau =0}\left. \frac{d}{d\varepsilon }%
\right\vert _{\varepsilon =0}\sigma (\tau ,\varepsilon ) \\
&=&\left. \frac{d^{2}}{d\tau ^{2}}\right\vert _{\tau =0}W_{i}(\tau
)=-\sum_{l}R_{0i0}^{l}\partial _{x^{l}}.
\end{eqnarray*}

We conclude that
\begin{equation}\label{rad} \lim_{\varepsilon \to 0} \frac{\ddot{r}_i(0)}{\varepsilon}=-R^i_{0i0}.
\end{equation}
Replacing (\ref{rad}) into (\ref{volvol}), and using that $R^0_{000}=0$, one obtains that
\begin{equation}\lim_{\varepsilon \to 0} \frac{\ddot{V}(0)}{V(0)}\sim \sum_{i>0} \frac{\ddot{r}_i(0)}{\varepsilon}=-\sum_{i>0}R^i_{0i0}=-\sum_{a\geq 0}R^a_{0a0}=-\Ric_{00}(p).\end{equation}
Therefore, Einstein's equation in the form (\ref{lein}) implies that, for very small $\varepsilon$
\begin{equation}\label{volume}
\ddot{V}\sim  -4 \pi V \big(T_{00}+T_{11}+T_{22}+T_{33}).
\end{equation}
This equation describes how the energy momentum tensor determines the change in volume of Alice's ball. The volume changes in such a way that its second derivative is proportional the negative volume times the sum of the energy density and the pressures at the three spatial directions, measured in Alice's frame.

As an example, consider the vacuum Einstein equation
with cosmological constant $\Lambda$. In units with $c=G_N=1$, the  tensor $T^{\Lambda}$ is
\[ T^{\Lambda}=-\frac{\Lambda }{8 \pi }g,\]
so that, in a Lorentz frame, it takes the form
\[ T^\Lambda=\frac{\Lambda }{8 \pi }\begin{pmatrix}
1&0&0&0\\
0&-1&0&0\\
0&0&-1&0\\
0&0&0&-1
\end{pmatrix}.\]
Equation (\ref{volume}) becomes in this case
\begin{equation}
\ddot{V}\sim  V \Lambda .
\end{equation}
One concludes that, if $\Lambda>0$, then the volume increases exponentially. On the other hand, if $\Lambda<0$, the volume decreases exponentially.

\section{The astonishing analogy: geodesic deviation and tidal forces}

In \S \ref{geodesic deviation} we considered a family of geodesics $%
\sigma :I\times (-\epsilon ,\epsilon )\rightarrow M$, i.e., for each fixed $\tau$, the curve $\sigma _{\tau}(t)=\sigma (t,\tau )$ is a geodesic. We
defined vector fields $X,Y$ on the surface $S =\mathrm{im}(\sigma) $ as 
\begin{align*}
X&=\sigma _{\ast }\partial _{t}, \\ Y&=\sigma _{\ast }\partial _{\tau}.
\end{align*}%
Since each of the curves $\sigma _{\tau }(t)$ is a geodesic, we know that $%
\nabla _{X}X=0.$ Moreover, since the vector fields $\partial _{\tau}$ and 
$\partial _{t}$ commute, we also have $[Y,X]=0.$ Therefore, the fact that $%
\nabla $ is torsion free implies that the curvature satisfies: 
\begin{equation}
R(X,Y)X=\nabla _{X}\nabla _{Y}X-\nabla _{Y}\nabla _{X}X=\nabla
_{X}\nabla _{X}Y.  \label{jdev}
\end{equation}%
Thus, the curvature is the second derivative of the vector $Y$ in the
direction of the geodesic. In local coordinates we write $X=\displaystyle{\sum_{j}}X^{j}\partial _{j}$, $Y=\displaystyle{\sum_{k}Y^{k}}\partial _{k}$ and $\nabla _{X}\nabla _{X}Y=\displaystyle{\sum_{i}}A^{i}\partial _{i}$, and equation  \eqref{jdev} becomes 
\begin{equation}
A^{i}=\sum_{j,k,l}X^{j}Y^{k}X^{l}R_{ljk}^{i}.  \label{EE31}
\end{equation}
Let us fix a timelike geodesic $\sigma _{0}(t)$ and denote by $\mathbf{u}(t)$
its 4-velocity at $\sigma _{0}(t).$ The vector field 
\[
A(t)=R(\mathbf{u}(t),Y(t),\mathbf{u}(t))=\nabla _{\mathbf{u}(t)}\nabla _{%
\mathbf{u}(t)}Y(t)
\]
represents the \emph{acceleration of the separation vector} $Y(t)$ between $%
\sigma _{0}(t)$ and an  infinitesimally
close timelike geodesic $\sigma _{1}(t).$  
\begin{figure}[tbh]
\centering
\includegraphics[scale=0.45]{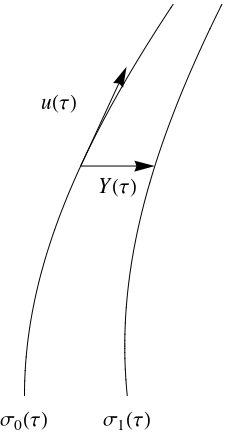}
\caption{Geodesic deviation}
\end{figure}
This equation strongly resembles equation (\ref{E49}) which appeared in
the newtonian analysis of tidal forces.  Let us assume that $\sigma
_{0}(\tau)=(t(\tau) ,a^{i}(\tau))$, and $\sigma _{1}(t(\tau) )=(t(\tau) ,b^{i}(\tau
))$ are geodesics that correspond to the worldlines of two falling
particles that move towards the center of a body $\mathcal{B}$ of mass $M$, say
the Earth. 
We assume that the particles are very close together, so that the separation vector 
\[
s(\tau )=\sigma _{1}(\tau )-\sigma _{0}(\tau )=(0,b^{i}(\tau )-a^{i}(\tau ))
\]%
can be approximated by the vector $Y(\tau ).$ 

We know that
\begin{equation}
A^i(\tau )=\sum_{abc}R^{i}_{cab}(\sigma _{0}(\tau ))\text{ }\mathbf{u}%
^{a}(\tau )Y^{b}(\tau )\mathbf{u}^{c}(\tau ).  \label{E52}
\end{equation}%
Assuming that the velocity is much less that the speed of light,
the only relevant terms correspond to $a=c=0$. Therefore:
\[
A^i(\tau )=\sum_{k}R^{i}_{00k}(\sigma _{0}(\tau ))\text{ }Y^{k}(\tau )%
\mathbf{=-}\sum_{k}R^{i}_{00k}(\sigma _{0}(\tau ))%
Y^{k}(\tau ).
\]%
Using the approximation $Y(\tau )\approx D(\tau )$,  one obtains
\begin{equation}
A^i(\tau )=-\sum_{k}R^{i}_{0k0}(\sigma _{0}(t))D^{k}(\tau ).
\label{EE51}
\end{equation}%
On the other hand, we had seen in (\ref{E49}) that%
\begin{equation}
\frac{d^{2}s^{i}}{d\tau ^{2}}=-\sum_{k}\frac{\partial ^{2}\Phi }{\partial
x^{k}\partial x^{i}}(\sigma _{0}(\tau ))s^{k}(\tau ).  \label{EEE51}
\end{equation}%
Since $A(\tau )\approx \frac{d^{2}s}{d\tau ^{2}}$ one sees that 
\[
R_{0k0}^{i}\approx \frac{\partial ^{2}\Phi }{\partial x^{k}\partial x^{i}}.
\]
If we think of this equation as a tensor equation, we can
contract  indices on both sides to
obtain
$$\Ric_{00}=\sum_{k}R_{0k0}^{k}=\sum_{i}\frac{\partial ^{2}\Phi }{(\partial x^{i})^2}=\Delta \Phi .
$$
Consequently, by virtue of \eqref{eq:Poissongrav}, one concludes
\begin{equation}
\Ric_{00}=4\pi G_{N}\rho .  \label{EE52}
\end{equation}%
In an empty spacetime the latter equation just becomes 
\begin{equation}\Ric_{00}=0.\label{riccif}\end{equation} 
By lemma  \ref{Baezl}, if equation (\ref{riccif}) holds at every point then  
$$\Ric=0.$$ This was the first field equation discovered by
Einstein. With that in hand he was able to explain the 
anomalous precession of the perihelion of Mercury, and was able to
predict the bending of a ray of light as it passes near a celestial
body. 

As we have discussed before, is reasonable to expect that the tensor of energy-momentum $T$ should be
the mathematical object replacing $\rho$. So one would expect that the
analogue of Newton's law would be given by an equation of the form $$
\Ric_{ab}=\kappa T_{ab}$$ for some suitable constant $\kappa$. We know that $%
\sum_{a}\nabla _{a}T^{ab}=0$. But is not
the case that in general $\sum_{a}\nabla _{a}\Ric^{ab}$ is
equal to zero. However,  the tensor $$G^{ab}=\Ric^{ab}-\frac{%
1}{2}\Rs g^{ab}$$ does satisfy $\sum_{a}\nabla _{a}G^{ab}=0$ (see Proposition \ref%
{necesaria}). So, as Einstein himself suggested, it would be reasonable to
seek for an equation of the form $$G^{ab}=\kappa T^{ab},$$
or equivalently, of the
form $$G_{ab}=\kappa T_{ab}.$$ That is 
\begin{equation}
\Ric_{ab}-\frac{1}{2}\Rs g_{ab}=\kappa T_{ab}.  \label{E53}
\end{equation}%
Let's know see how to determine the constant $\kappa$ by passing to the Newtonian limit.
First, we notice that form this equation one gets
\[
\sum_{a,b }g^{ab }\Ric_{ab }-\frac{1}{2}\Rs\sum_{a,b }%
g^{ab}g_{ab }=\kappa\sum_{a,b }g^{ab }T_{ab}.
\]%
If we set $\mathrm{tr}\, T =\sum_{a,b }g^{ab }T_{ab} $, and use the fact that  $\sum_{a,b} g^{ab}g_{ab} = 4$, this means that%
\[
\mathrm{R}=-\kappa \mathrm{tr}\, T.
\]%
Therefore \eqref{E53} can be rewritten as 
\begin{equation}
\Ric_{ab}=\kappa T_{ab}+\frac{1}{2}\Rs g_{ab}=\kappa \left(T_{ab}-\frac{1}{2}g_{ab}\mathrm{tr}\, T\right).
\label{E54}
\end{equation}%
For a perfect fluid at non relativistic velocities, like in the case of a weak gravitational field, the components of the tensor of energy
momentum tensor in the standard coordinates $(t,x^{i})$  reduce to $T^{00}=\rho ,$ the other components $T^{ab}=\rho
v^{a}v^{b}$ being very closed to zero. Also, under these hypotheses we
already know that $g_{00}\approx -1$, as we proved in \S \ref{section 00}.

Since we are assuming the entries of $\varepsilon $ to be very small,
neglecting terms of quadratic order one sees that $$(\eta +\varepsilon )(\eta
-\varepsilon )\approx \eta ^{2}= I_{4},$$ the $4\times 4$ identity matrix.
Hence, for our metric $g=\eta +\varepsilon$ we have $g^{ab}\approx \eta
_{ab}-\varepsilon _{ab}$. In particular, $$g_{00}g^{00}+%
\sum_{i}g_{0i}g^{i0}\approx 1.$$ Hence, $$g_{00}g^{00}-\sum_{i}\varepsilon
_{0i}\varepsilon _{i0}\approx 1,$$ and since $\sum_{i}\varepsilon _{0i}\varepsilon
_{i0}\approx 0,$ we get $g_{00}g^{00}\approx 1$. On the other hand, $$T_{00}=g_{00}g_{00}T^{00}=\rho. $$ Henceforth, $$g^{00}T_{00}\approx
-\rho .$$
Taking $a=b=0$ in Equation \eqref{E54} one obtains
\[
\Ric_{00}=\kappa\left(T_{00}-\frac{1}{2}g_{00}\mathrm{tr}\, T\right)\approx\kappa\left(\rho -\frac{1}{2}(-1)(-\rho
)\right)=\frac{1}{2}\kappa\rho ,
\]%
But Equation (\ref{EE52}) implies that $4\pi \rho =1/2\kappa\rho ,$ from which we
get $\kappa=8\pi G_{N}$. 

Summarizing, Einstein's field equations must be $$
G=8\pi 
G_{N}T.
$$
In nonstandard units of time ($\mathrm{ss}$), the
constant $G_{N}$ has units of $\mathrm{m}^{3}\cdot \mathrm{kg}^{-1}\cdot\mathrm{s}^{-2}$. Hence, in standard units $G_{N}$ must be divided by $c^{2}$. Similarly, $T_{ab}=\rho v^{a}v^{b}.$
Hence, in standard units one must also divide by $c^{2}$ and henceforth one
can write Einstein'
s equation in the usual form
\begin{equation}
G=\frac{8\pi G_{N}}{c^{4}}T.  
\end{equation}

\section{Einstein's  equation and variational principles}
In analytical mechanics it can be shown that the trajectories of a conservative system may be characterised either by a system of Lagrange's equations or by Hamilton's principle. The former is a system of differential equations, and the latter is a variational principle. Trajectories of the system are, on the one hand, solution curves of Lagrange's equations and, on the other hand, extremal curves satisfying certain boundary conditions.  

In much the same way, Einstein's theory of gravity may be characterised either by the field equation or by a variational principle. To keep things simple, we restrict ourselves to the vacuum Einstein's field equations.

\subsection*{Variational formulation of Newton's theory of gravity} 
 As a warm-up, we first consider Newton's theory of gravity. We have shown in \S\ref{sec:11.1}  that the newtonian potential  satisfies the Laplace equation
\begin{equation}\label{eq:Laplace}
 \Delta \Phi = 0
\end{equation}
 in a vacuum. We want to show that $\Phi$ may be caracterized by a variational principle.
Let $U$ be a bounded domain in an instantaneous space in Newtonian spacetime. We define the so-called \emph{Dirichlet action functional} $I_{\mathrm{D}}[\Phi]$ for smooth functions $\Phi$ on the closure $\bar{U}$ by
\begin{equation}
I_{\mathrm{D}}[\phi] = \int_U \grad \Phi \cdot \grad \Phi \, dV.
\end{equation}
Then we have the following variational principle: $\Phi$ is the newtonian potential in $U$ for vacuum if and only if it is an extremum for the Dirichlet action functional among all smooth functions having the same boundary values as $\Phi$.    

This variational principle may be proved easily in the following way. We consider a family of functions 
\begin{equation}
\phi_{\varepsilon} (x) = \Phi(x) + \varepsilon \eta(x), \qquad x \in U,
\end{equation}
where $\varepsilon$ is a parameter, and where $\eta(x)$ is a smooth function on $\bar{U}$ satisfying the boundary condition
\begin{equation}\label{eq:bdy}
\eta(x) = 0, \qquad x \in \partial U.
\end{equation}
When $\varepsilon = 0$, the function $\Phi_{0}$ reduces to the function $\Phi$. Thus $\Phi_{\varepsilon}$ corresponds to a $1$-parameter family of variations from the function $\Phi$ in the ``direction'' of the function $\eta$. Using the $1$-parameter family $\Phi_{\varepsilon}$, the variation of the Dirichlet action functional $I_{\mathrm{D}}$ is then
\begin{align*}
\frac{d}{d\varepsilon}\bigg\vert_{\varepsilon = 0} I_{\mathrm{D}}[\Phi_{\varepsilon}] &= \frac{d}{d\varepsilon}\bigg\vert_{\varepsilon = 0} \int_U \grad \Phi_{\varepsilon} \cdot \grad \Phi_{\varepsilon} \, dV  \\
&= \int_{U} \left[ 2 \grad \Phi \cdot \grad \Phi_{\varepsilon} \right]_{\varepsilon= 0} dV \\
&=  \int_{U} 2 \grad \Phi \cdot \grad \eta \, dV \\
&= \int_{U} 2 \div(\eta \grad \Phi)  \, dV - \int_U 2 \eta \Delta \Phi \, dV \\
&= \int_{\partial U} 2 \eta \grad \Phi \cdot n \, dA - \int_U 2 \eta \Delta \Phi \, dV \\
&= \int_{U} (-2 \Delta \Phi) \eta \, dV,
\end{align*}
where the surface integral vanishes by \eqref{eq:bdy}. Consequently, $\frac{d}{d\varepsilon}\big\vert_{\varepsilon = 0} I_{\mathrm{D}}[\Phi_{\varepsilon}] = 0$ for all $\eta$ if and only if $\Phi$ satisfies \eqref{eq:Laplace}. Thus the variational principle is proved. 

The preceding variational principle asserts that $I_{\mathrm{D}}[\Phi]$ is an extremum at $\Phi$ over the class of functions having the same boundary values as $\Phi$. In fact $I_{\mathrm{D}}[\Phi]$ is minimum among the values $I_{\mathrm{D}}[\Phi_{\varepsilon}]$, since 
\begin{align*}
I_{\mathrm{D}}[\Phi_{\varepsilon}] - I_{\mathrm{D}}[\Phi] &= \int_U \big( \grad (\Phi + \varepsilon \eta) \cdot \grad (\Phi + \varepsilon \eta) - \grad \Phi \cdot \grad \Phi \big) dV \\
&= \int_U 2 \varepsilon \grad \Phi \cdot \grad \eta \, dV + \int_U \varepsilon^2 \grad \eta \cdot \grad \eta \, dV \\
&= \int_U \varepsilon^2 \grad \eta \cdot \grad \eta \, dV,
\end{align*}
which is positive unless $\eta = \mathrm{const}=0$, where we have again used the boundary condition \eqref{eq:bdy} to determine the value of the constant. An extremum of an action functional in general need not be a minimum, of course. The fact that $I_{\mathrm{D}}[\Phi]$ is actually a minimum is asserted by the Dirichlet principle. 

\subsection*{Variational formulation of Einstein's theory of gravity} 
As discussed in \S\ref{svalpha}, in the general theory of relativity the lorentzian metric $g$ plays the role of the gravitational potential. In the vaccuum, the system of field equations is
\begin{equation}\label{eq:vacuum}
\mathrm{Ric}(g) - \frac{1}{2}g \mathrm{R}(g)  = 0.
\end{equation}
Notice that the left-hand side of \eqref{eq:vacuum} is formed by partial derivatives up to second order in the components $g_{ab}$ of the lorentzian metric $g$. Thus the system of field equations \eqref{eq:vacuum} is comparable to the field equation \eqref{eq:Laplace}. 

The field equations \eqref{eq:vacuum} may be derived from a variational principle. In order to  construct the action functional, it is necessary to introduce the appropriate functional space of field variables. For this, we let $\mathcal{M}$ be the set of all lorentzian metrics on the underlying spacetime manifold $M$. This is a Frech\'et manifold under the topology of $C^{\infty}$-uniform convergence on all compact domains in $M$. It also turns out to be an open cone in the space $\Gamma(\mathrm{S}^2 T^*M)$ of symmetric rank-$2$ tensor fields over $M$. The Frech\'et space $\Gamma(\mathrm{S}^2 T^*M)$ is hence the model for the manifold $\mathcal{M}$ itself, so that at each point $g \in \mathcal{M}$ the tangent space $T_{g}\mathcal{M}$ is isomorphic to $\Gamma(\mathrm{S}^2 T^*M)$ itself. This expresses the fact that all infinitesimal deformations of a lorentzian metric $g$ are symmetric tensors of the same rank. 

With this background in mind, we define the \emph{Einstein-Hilbert action functional} $I_{\mathrm{EH}}[g]$ for $g \in \mathcal{M}$ on a domain $U$ with compact closure in $M$, by
\begin{equation}\label{eq:EH}
I_{\mathrm{EH}}[g] = \int_{U} \mathrm{R}(g) \, \mathrm{vol}_{g},
\end{equation}
where $\mathrm{vol}_{g}$ denotes the volume element on $M$ determined by $g$ and the orientation of $M$. We require that the boundary values of $g$ and its first derivatives be held fixed. In other words, we consider the variation of the action functional $I_{\mathrm{EH}}[g]$ over the class of lorentzian metrics $g$ having the same boundary values and the same first derivatives on $\partial U$.\footnote{It can be verified easily from the transformation law of the components of the metric that if the boundary condition $g'_{ab}= g_{ab}$ and $\partial_{c} g'_{ab} = \partial_{c} g_{ab} $ are satisfied on $\partial U$ relative to any coordinate system $(x^{c})$ in $U$, then the same are satisfied relative to all  other coordinate systems in $U$. Thus the boundary conditions are actually conditions on the metrics $g$ and $g'$ independent of the choice of the coordinate system $(x^{c})$.} 

Before plunging into a detailed analysis of the field equations obtained by variation of $I_{\mathrm{EH}}[g]$, we shall make some general comments on the choice of the ``lagrangian density'' of $I_{\mathrm{EH}}[g]$. The general theory of relativity differs from other physical theories in the fact that from the lorentzian metric and its derivatives it is impossible to built a scalar whose square could play the role of the lagrangian density. Indeed, the components of the affine connection that are built from the first derivatives of the metric can be made to vanish (at a point) by a choice of a coordinate system, and so no scalar density of the schematic form ``square of the affine connection'' can be constructed. The simplest scalar that arises in lorentzian geometry is the scalar curvature, and this involves second derivatives of the metric. A lagrangian density linear in the scalar curvature is then possible, and can lead to second order field equations. On this account, it would be most natural to select $\mathrm{R}(g)$ as the lagrangian density of the action functional $I_{\mathrm{EH}}[g]$. 

Now we claim that $g \in \mathcal{M}$ satisfies the system of field equations \eqref{eq:vacuum} if and only if it is an extremum for the Einstein-Hilbert action functional over the class of lorentzian metrics having the same boundary values and the same first derivatives on the boundary as $g$. To obtain this result, we proceed as follows. Consider any $1$-parameter family of the form
\begin{equation}
g_{\varepsilon} = g + \varepsilon h,
\end{equation}
where $h \in \Gamma(\mathrm{S}^2 T^*M)$ is a symmetric rank-$2$ tensor field whose value and the value of its first derivatives on $\partial U$ are zero, so that the tensors $g_{\varepsilon}$ all satisfy the aforementioned boundary conditions. Also, by what we have said above, $g_{\varepsilon} \in \mathcal{M}$ for sufficiently small $\varepsilon$ and thus corresponds to a $1$-parameter family of variations from the lorentzian metric $g$ in the ``direction'' of $h$. Hence the condition that $I_{\mathrm{EH}}[g]$ be stationary at $g$ reads
\begin{equation}\label{eq:varprinciple}
\frac{d}{d\varepsilon}\bigg\vert_{\varepsilon = 0} I_{\mathrm{EH}}[g + \varepsilon h] = 0
\end{equation}
for all $h \in \Gamma(\mathrm{S}^2 T^*M)$ which satisfy the boundary conditions. 

Now let us calculate the left-hand side of \eqref{eq:varprinciple}. To do this, we need to have expressions for the variations of the Levi-Civita connection $\Gamma(g)$, the Ricci tensor $\mathrm{Ric}(g)$, the scalar curvature $\mathrm{R}(g)$ and the volume element $\mathrm{vol}_{g}$. To get a better feeling of what these expressions look like, we will work in local coordinates. We start with the following.

\begin{lemma}\label{lem:varmet}
The variation of the inverse metric is given by
\begin{equation}
\frac{d}{d\varepsilon}\bigg\vert_{\varepsilon = 0} (g + \varepsilon h)^{ab} = - g^{ac} g^{bd} h_{cd}. 
\end{equation}
\end{lemma}

\begin{proof}
From $(g + \varepsilon h)^{ac} (g + \varepsilon h)_{cd} = \delta^{a}_{\phantom{a}d}$, it follows that
$$
0 = \frac{d}{d\varepsilon}\bigg\vert_{\varepsilon = 0} (g + \varepsilon h)^{ac} (g + \varepsilon h)_{cd} =   \left(\frac{d}{d\varepsilon}\bigg\vert_{\varepsilon = 0} (g + \varepsilon h)^{ac}\right) g_{cd} + g^{ac} h_{cd}.
$$
Hence
$$
\frac{d}{d\varepsilon}\bigg\vert_{\varepsilon = 0} (g + \varepsilon h)^{ab} = \left(\frac{d}{d\varepsilon}\bigg\vert_{\varepsilon = 0}(g + \varepsilon h)^{ac}\right) g_{cd} g^{db} = - g^{ac} g^{bd} h_{cd},
$$
as asserted.
\end{proof}

Next we derive the variation formula for the Levi-Civita connection. 

\begin{lemma}\label{lem:varconn}
The variation of the Levi-Civita connection is given by
\begin{equation}\label{eq:varconn}
\frac{d}{d\varepsilon}\bigg\vert_{\varepsilon = 0} \Gamma(g + \varepsilon h)^{c}_{ab} = \frac{1}{2} g^{cd} (\nabla_{a} h_{bd} + \nabla_{b} h_{ad} - \nabla_{d} h_{ab}). 
\end{equation}
\end{lemma}

\begin{proof}
Recall that
$$
\Gamma(g + \varepsilon h)^{c}_{ab} = \frac{1}{2} (g + \varepsilon h)^{cd}(\partial_{a} (g + \varepsilon h)_{bd} + \partial_{b} (g + \varepsilon h)_{ad} - \partial_{d} (g + \varepsilon h)_{ab}).
$$
Hence, 
\begin{align*}
&\frac{d}{d\varepsilon}\bigg\vert_{\varepsilon = 0} \Gamma(g + \varepsilon h)^{c}_{ab}  \\
&= \frac{1}{2}\left(\frac{d}{d\varepsilon}\bigg\vert_{\varepsilon = 0} (g + \varepsilon h)^{cd}\right)(\partial_{a} g_{bd} + \partial_{b} g_{ad} - \partial_{d} g_{ab})  \\
&\quad \, + \frac{1}{2} g^{cd} \left\{\partial_{a} \left(\frac{d}{d\varepsilon}\bigg\vert_{\varepsilon = 0}  (g + \varepsilon h)_{bd}\right) + \partial_{b} \left(\frac{d}{d\varepsilon}\bigg\vert_{\varepsilon = 0}  (g + \varepsilon h)_{ad}\right) - \partial_{d} \left(\frac{d}{d\varepsilon}\bigg\vert_{\varepsilon = 0}  (g + \varepsilon h)_{ab}\right)\right\} \\
&= \frac{1}{2} \left(\frac{d}{d\varepsilon}\bigg\vert_{\varepsilon = 0} (g + \varepsilon h)^{cd}\right) (\partial_{a} g_{bd} + \partial_{b} g_{ad} - \partial_{d} g_{ab})  + \frac{1}{2} g^{cd} (\partial_a h_{bd} + \partial_b h_{ad} - \partial_d h_{ab}).
\end{align*}
In normal coordinates centered at $p \in M$, one has $\Gamma^{c}_{ab}(g) = 0$ at $p$. It follows that $\partial_{a} h_{bc} = \nabla_{a} h_{bc}$ at $p$ and, in particular, $\partial_{a} g_{bc} = 0$ at $p$ for all $a,b,c$. Thus, we obtain 
$$
\frac{d}{d\varepsilon}\bigg\vert_{\varepsilon = 0} \Gamma(g + \varepsilon h)^{c}_{ab} = \frac{1}{2} g^{cd} (\nabla_a h_{bd} + \nabla_b h_{ad} - \nabla_d h_{ab})
$$
at $p$. Since both sides of this equation are component of tensors, the result holds in any coordinate system and at any point. 
\end{proof}

Since the Riemann  curvature tensor is defined solely in terms of the Levi-Civita connection, we can readily compute its variation. 

\begin{lemma}\label{lem:varriem}
The variation of the Riemann curvature tensor is given by
\begin{align}\label{eq:varriem}
\begin{split}
\frac{d}{d\varepsilon}\bigg\vert_{\varepsilon = 0} \mathrm{Riem}(g + \varepsilon h)^{d}_{abc} = \frac{1}{2} g^{de} &\left( \nabla_{a} \nabla_{b} h_{ce} + \nabla_{a} \nabla_{c} h_{be} - \nabla_{a} \nabla_{e} h_{bc} \right.\\
&\,\,\left. - \nabla_{b} \nabla_{a} h_{ce} - \nabla_{b} \nabla_{c} h_{ae} +  \nabla_{b} \nabla_{e} h_{ac} \right).   
\end{split}
\end{align}
\end{lemma}

\begin{proof}
We have the standard formula
\begin{align*}
&\mathrm{Riem}(g + \varepsilon h)^{d}_{abc} \\
&\quad = \partial_{a} \Gamma(g + \varepsilon h)^{d}_{bc} - \partial_{b} \Gamma(g + \varepsilon h)^{d}_{ac} +  \Gamma(g + \varepsilon h)^{e}_{bc} \Gamma(g + \varepsilon h)^{d}_{ae} -  \Gamma(g + \varepsilon h)^{e}_{ac} \Gamma(g + \varepsilon h)^{d}_{be}.
\end{align*}
Thus, we compute
\begin{align*}
&\frac{d}{d\varepsilon}\bigg\vert_{\varepsilon = 0} \mathrm{Riem}(g + \varepsilon h)^{d}_{abc}\\
&\quad = \partial_{a} \left(\frac{d}{d\varepsilon}\bigg\vert_{\varepsilon = 0}\Gamma(g + \varepsilon h)^{d}_{bc} \right) - \partial_{b} \left(\frac{d}{d\varepsilon}\bigg\vert_{\varepsilon = 0}\Gamma(g + \varepsilon h)^{d}_{ac} \right) \\
&\quad\quad\, + \left( \frac{d}{d\varepsilon}\bigg\vert_{\varepsilon = 0}\Gamma(g + \varepsilon h)^{e}_{bc} \right)\Gamma(g)^{d}_{ae} + \Gamma(g)^{e}_{bc} \left( \frac{d}{d\varepsilon}\bigg\vert_{\varepsilon = 0}\Gamma(g + \varepsilon h)^{d}_{ae}\right) \\
&\quad\quad\, - \left(\frac{d}{d\varepsilon}\bigg\vert_{\varepsilon = 0}\Gamma(g + \varepsilon h)^{e}_{ac}\right) \Gamma(g)^{d}_{be} - \Gamma(g)^{e}_{ac} \left(\frac{d}{d\varepsilon}\bigg\vert_{\varepsilon = 0}\Gamma(g + \varepsilon h)^{d}_{be}\right).
\end{align*}
As in the proof of Lemma \ref{lem:varconn}, we use normal coordinates centered at $p \in M$ to calculate that 
$$
\frac{d}{d\varepsilon}\bigg\vert_{\varepsilon = 0} \mathrm{Riem}(g + \varepsilon h)^{d}_{abc} = \nabla_{a} \left(\frac{d}{d\varepsilon}\bigg\vert_{\varepsilon = 0}\Gamma(g + \varepsilon h)^{d}_{bc} \right) - \nabla_{b} \left(\frac{d}{d\varepsilon}\bigg\vert_{\varepsilon = 0}\Gamma(g + \varepsilon h)^{d}_{ac} \right)
$$
at $p$, and then observe that this formula holds everywhere. The present lemma follows directly  substituting \eqref{eq:varconn} into this equation.  
\end{proof}

As an immediate consequence, the variation formula for the Ricci tensor is found. 

\begin{lemma}\label{lem:varric}
The variation of the Ricci tensor is given by
\begin{equation}
\frac{d}{d\varepsilon}\bigg\vert_{\varepsilon = 0} \mathrm{Ric}(g + \varepsilon h)_{ab} = \frac{1}{2} g^{cd} \left( \nabla_{d} \nabla_{a} h_{bc} + \nabla_{d} \nabla_{b} h_{ac} - \nabla_{d} \nabla_{c} h_{ab}  - \nabla_{a} \nabla_{b} h_{cd} \right).   
\end{equation}
\end{lemma}

\begin{proof}
This follows by contracting $a = d$ in \eqref{eq:varriem} and relabeling the indices. 
\end{proof}

We also get the variation formula for the scalar curvature.  

\begin{lemma}
The variation of the scalar curvature is given by
\begin{equation}
\frac{d}{d\varepsilon}\bigg\vert_{\varepsilon = 0} \mathrm{R}(g + \varepsilon h) = -  g^{ac}  g^{bd} \left( \nabla_{a} \nabla_{c} h_{bd} - \nabla_{a} \nabla_{b} h_{cd} + h_{cd}  \mathrm{Ric}(g)_{ab}\right).
\end{equation}
\end{lemma}

\begin{proof}
From $\mathrm{R}(g + \varepsilon h) = (g + \varepsilon h)^{ab} \mathrm{Ric}(g + \varepsilon h)_{ab}$, and using Lemmas \ref{lem:varmet} and \ref{lem:varric}, we compute that
\begin{align*}
&\frac{d}{d\varepsilon}\bigg\vert_{\varepsilon = 0} \mathrm{R}(g + \varepsilon h)\\ &\quad= \left( \frac{d}{d\varepsilon}\bigg\vert_{\varepsilon = 0} (g + \varepsilon h)^{ab}\right) \mathrm{Ric}(g)_{ab} + g^{ab} \left( \frac{d}{d\varepsilon}\bigg\vert_{\varepsilon = 0} \mathrm{Ric}(g + \varepsilon h)_{ab}\right) \\
&\quad = -g^{ac} g^{bd} h_{cd} \mathrm{Ric}(g)_{ab} - \frac{1}{2} g^{ab} g^{cd} \left( \nabla_{d} \nabla_{a} h_{bc} + \nabla_{d} \nabla_{b} h_{ac} - \nabla_{d} \nabla_{c} h_{ab}  - \nabla_{a} \nabla_{b} h_{cd} \right) \\
&\quad = - g^{ac}  g^{bd} \left( \nabla_{a} \nabla_{c} h_{bd} - \nabla_{a} \nabla_{b} h_{cd} + h_{cd}  \mathrm{Ric}(g)_{ab}\right), 
\end{align*}
as required.
\end{proof}

Finally, we come to the variation formula of the volume element. 

\begin{lemma}
The variation of the volume element on $M$ is given by
\begin{equation}\label{eq:varvol}
\frac{d}{d\varepsilon}\bigg\vert_{\varepsilon = 0} \mathrm{vol}_{g + \varepsilon h} = \frac{\mathrm{tr} (g^{-1} h)}{2}  \, \mathrm{vol}_{g}.
\end{equation} 
\end{lemma}

\begin{proof}
The volume element $\mathrm{vol}_{g + \varepsilon h}$ corresponds to a $4$-form on $M$ given in local coordinates by
$$
\mathrm{vol}_{g + \varepsilon h} = \sqrt{\det (g + \varepsilon h)} \, dx^{0} \wedge dx^{1} \wedge dx^{2} \wedge dx^{3}.
$$ 
On taking the variation of the determinant we obtain
\begin{align*}
\frac{d}{d\varepsilon}\bigg\vert_{\varepsilon = 0} \det (g + \varepsilon h) &= \lim_{\varepsilon \rightarrow 0} \frac{\det (g + \varepsilon h) -  \det g}{\varepsilon} \\
&= \det g \lim_{\varepsilon \rightarrow 0} \frac{\det(g^{-1}(g + \varepsilon  h)) - 1}{\varepsilon} \\
&= \det g \lim_{\varepsilon \rightarrow 0} \frac{1 + \varepsilon \mathrm{tr} (g^{-1} h) + \mathrm{O}(\varepsilon^2) - 1}{\varepsilon} \\
&= \det g \, \mathrm{tr} (g^{-1} h)
\end{align*}
This implies that
$$
\frac{d}{d\varepsilon}\bigg\vert_{\varepsilon = 0} \sqrt{\det (g + \varepsilon h)} = \frac{1}{2} \frac{1}{\sqrt{\det g}} \frac{d}{d\varepsilon}\bigg\vert_{\varepsilon = 0} \det (g + \varepsilon h)  = \frac{\mathrm{tr} (g^{-1} h)}{2} \sqrt{\det g}.
$$
Therefore,
\begin{align*}
\frac{d}{d\varepsilon}\bigg\vert_{\varepsilon = 0} \mathrm{vol}_{g + \varepsilon h} &= \frac{d}{d\varepsilon}\bigg\vert_{\varepsilon = 0} \sqrt{\det (g + \varepsilon h)} \, dx^{0} \wedge dx^{1} \wedge dx^{2} \wedge dx^{3}  \\
&= \frac{\mathrm{tr} (g^{-1} h)}{2} \sqrt{\det g} \, dx^{0} \wedge dx^{1} \wedge dx^{2} \wedge dx^{3} \\
&= \frac{\mathrm{tr} (g^{-1} h)}{2} \mathrm{vol}_g,
\end{align*}
as we wished to show.
\end{proof}

Now we come back to the left-hand side of \eqref{eq:varprinciple}. To perform the calculation, we shall have to introduce  one more piece of notation. For $h, h' \in \Gamma(\mathrm{S}^2 T^* M)$ and $g$, the scalar $g^{ac} g^{bd} h_{bc} h'_{ad}$ induces a function on $M$ that we denote by $\langle h,h'\rangle_{g}$. In particular we note that $\mathrm{tr} (g^{-1} h) = \langle g,h\rangle_{g}$. With the further notation $\Delta = g^{ab} \nabla_{a} \nabla_{b}$, we may write the variation of the scalar curvature in the invariant form
\begin{equation}\label{eq:varscurv}
\frac{d}{d\varepsilon}\bigg\vert_{\varepsilon = 0} \mathrm{R}(g + \varepsilon h) = -\Delta h + \div (\div  h) - \langle \mathrm{Ric}(g),h \rangle_{g}.
\end{equation}
These observations taken together with the  preceding lemma yield the following.

\begin{proposition}
For every $h \in \Gamma(\mathrm{S}^2 T^* M)$ such that the values of $h$ and its first derivatives vanish on $\partial U$, we have
\begin{equation}
\frac{d}{d\varepsilon}\bigg\vert_{\varepsilon = 0} I_{\mathrm{EH}}[g + \varepsilon h] = - \int_U \left\langle \mathrm{Ric}(g) - \frac{1}{2} \mathrm{R}(g) g, h \right\rangle_{g} \mathrm{vol}_{g}.
\end{equation}
\end{proposition}

\begin{proof}
From \eqref{eq:varvol} and \eqref{eq:varscurv}, we find that
\begin{align*}
\frac{d}{d\varepsilon}\bigg\vert_{\varepsilon = 0} I_{\mathrm{EH}}[g + \varepsilon h] &= \frac{d}{d\varepsilon}\bigg\vert_{\varepsilon = 0} \int_{U} \mathrm{R}(g + \varepsilon h) \, \mathrm{vol}_{g + \varepsilon h} \\
&= \int_{U} \left\{\left( \frac{d}{d\varepsilon}\bigg\vert_{\varepsilon = 0} \mathrm{R}(g + \varepsilon h)\right) \mathrm{vol}_{g} + \mathrm{R}(g) \left( \frac{d}{d\varepsilon}\bigg\vert_{\varepsilon = 0} \mathrm{vol}_{g + \varepsilon h}\right) \right\} \\
&= \int_{U} \left\{\left(-\Delta h + \div (\div  h) - \langle \mathrm{Ric}(g),h \rangle_{g}\right) \mathrm{vol}_{g} + \mathrm{R}(g) \frac{\mathrm{tr} (g^{-1} h)}{2} \mathrm{vol}_{g}  \right\}\\
&= \int_U \left( -\Delta h + \div (\div  h) -\langle \mathrm{Ric}(g),h \rangle_{g} + \frac{1}{2} \mathrm{R}(g)\, \mathrm{tr} (g^{-1} h) \right) \mathrm{vol}_{g}. 
\end{align*}
Using the definitions, we see that the contribution of the first terms in the integrand may be expressed as a surface integral which vanishes by virtue of the boundary condition on the first derivatives of $h$. Thus, upon replacing $\mathrm{tr} (g^{-1} h)$ by $\langle g,h \rangle_{g}$, we get 
$$
\frac{d}{d\varepsilon}\bigg\vert_{\varepsilon = 0} I_{\mathrm{EH}}[g + \varepsilon h] = - \int_U \left\langle \mathrm{Ric}(g) - \frac{1}{2} \mathrm{R}(g) g, h \right\rangle_{g} \mathrm{vol}_{g},
$$
as was to  be shown. 
\end{proof}

From this proposition we see that the Einstein-Hilbert action functional $I_{\mathrm{EH}}[g]$ is stationary at $g$ if and only if $g$ satisfies the system of field equations \eqref{eq:vacuum}. Thus the variational principle is proved.  

\section{Predictions and tests}

We have now described the basic structure of General Relativity. At this point it is natural to ask whether this theory makes any predictions that would distinguish it from Newton's gravity. The answer is that it does, and its predictions have been verified to amazing accuracy. Throughout the text we mention several of those predictions, and the tests that have been made to confirm them. We list some of them here, as evidence of the fact that General Relativity correctly predicts phenomena that Newton's gravity does not account for.

\subsubsection{Perihelion of Mercury} 
The perihelion of a planet is the point in its orbit that is closer to the Sun. Due to the gravitational pull of other planets, the perihelion does not always occur at the same place, but shifts along the orbit. This shift is known as the precession of the perihelion.
An anomalous precession of the perihelion
of Mercury had been noticed since 1859. By analyzing observations of transits
of Mercury, french astronomer
Urbain Le Verrier found that the actual rate of precession of Mercury's
perihelion disagreed with that predicted by Newton's theory by 38" (arc
seconds) per century. Many ad-hoc
explanations were devised. The existence of another planet, Vulcan, was
postulated and rejected. 
Einstein \cite{EinsteinMer} used General relativity to correctly predict the precession of Mecury's perihelion. This is discussed in more detail in section \S \ref{Mercury}.

\subsubsection{Bending of light} 

According to General Relativity, a massive object causes spacetime to curve, and light around the object bends. The first observation of light deflection was performed by Arthur
Eddington and Frank Watson Dyson  during the total solar eclipse of May 29,
1919, when stars near the Sun could be observed. This experiment was the first experimental confirmation of Einsten's theory of gravity.
 We discuss light bending in more detail in section \S  \ref{bending}.

\subsubsection{Gravitational time dilation}
General Relativity predicts that time runs more slowly in the presence of a gravitational potential. It predicts that if a clock $A$ is on the surface of the Earth, and an identical clock $B$ is 1 km above the surface, then, after a million years,
$B$ will be $3$ seconds faster than $A$. 
If light with frequency $f$ is sent from $A$ to $B$, then, since time runs faster for $B$, an observer at $B$ will judge the light to have frequency $\overline{f}<f$. The light shifts to the red. This effect is known as gravitational redshift.
These predictions have been confirmed using atomic clocks traveling on airplanes. The effects are strong enough that the satellites for Global Positioning Systems take them into account. Amazingly precise experiments have been made by Wineland et.al 
 \cite{Wineland}, where this effect was measured for a difference in height of less than a meter. Gravitational time dilation and redshift are discussed in \S \ref{GTD}.
   \clearemptydoublepage 

 \begin{partwithabstract}{Solutions to Einstein's Equation}
The Schwarzschild metric is a solution to the vacuum Einstein equation that describes the curvature
of spacetime caused by a spherically symmetric mass. The geometry of Schwarzschild spacetime accounts for some of the basic predictions of general relativity, such as gravitational time dilation, the bending of light, the anomalous precession of the 
perihelion of Mercury and black holes. The Friedmann-Lemaitre-Robertson-Walker models describe the large scale properties of the universe, cosmology. They are determined by the cosmological principle, which states that space looks the same at all places and in all directions. The FLRW models account for the expansion of the universe, its age and diameter.

\begin{figure}[H]
\centering
\begin{tabular}{cc}
\vspace{0pt} \includegraphics[scale=0.4]{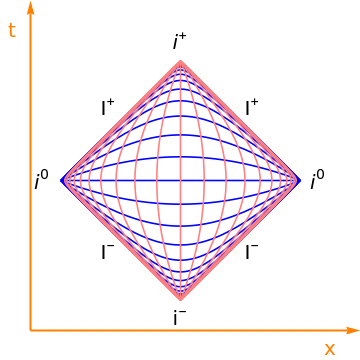}&
  \vspace{0pt} \includegraphics[scale=0.4]{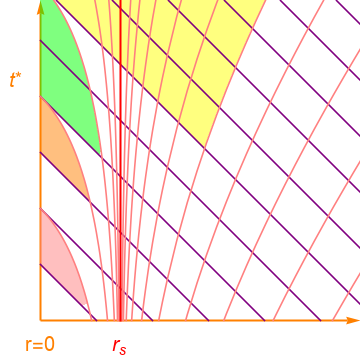} 
\end{tabular}
\end{figure}
\end{partwithabstract}

 \chapter{The Schwarzschild solution}
\label{ssss}

\begin{center}
\parbox[b]{0.9\textwidth}{\small \sl One of the first exact solutions of Einstein's field equations was discovered by Karl
Schwarzschild, in 1915, only a few months after Einstein introduced his
general theory of relativity. Schwarzschild discovered his
celebrated solution while serving in the German army during World War I. He
died the following year from a rare autoimmune disease, at the early age of
forty two. The Schwarzschild metric is a solution of the vacuum Einstein equation
$\Ric=0$. It is a metric on the manifold $\mathscr{S}=\RR \times \RR_{>r_s} \times S^2$, which, in coordinates
$(t,r,\theta,\varphi)$ takes the form
\begin{equation}
g=-\left(  1-\frac{r_s}{r}\right) c^2 dt\otimes dt+\left(  1-\frac
{r_s}{r}\right)  ^{-1}dr\otimes dr+r^{2}d\Omega, \label{metrica esferica}%
\end{equation}
where
\[ d\Omega=r^2\Big( d\theta \otimes d\theta+\sin^2\theta\, d\varphi \otimes d\varphi \Big)\]
is the round metric on the sphere. This metric describes the geometry of spacetime outside a spherically symmetric mass such as a non rotating star. Many important relativistic phenomena such as the bending of light, the anomalous perihelion of Mercury and gravitational time dilation arise in Schwarzschild spacetime.
The goal of this chapter is to discuss the basic geometric as well as physical properties of this solution. }
\end{center}

\vspace{3ex}

\section{Gravitational potential of a point mass}

Consider the Newtonian description of the gravitational field generated by a point mass $M$. According to Newton's law, the mass generates a gravitational field
\begin{equation}g=-G_NM \sum\limits_i \frac{x^i}{r^3}\partial_{x^i},\end{equation}
where $r=\sqrt{(x^1)^2+(x^2)^2+(x^3)^2}$.
\begin{figure}[H]
\centering
\includegraphics[scale=0.4]{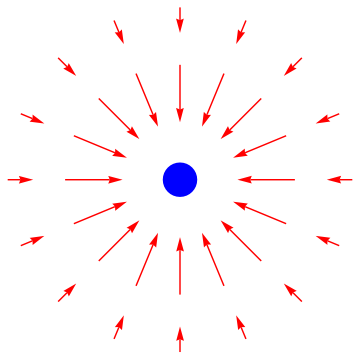}
\caption{Gravitational field generated by a point mass.}
\end{figure}

The gravitational field is the negative gradient of the  gravitational potential \[ \Phi(r)= -\frac{G_N M }{r}.\]

\begin{figure}[H]
\centering
\includegraphics[scale=0.5]{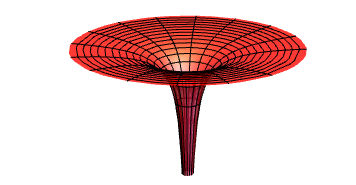}
\caption{Gravitational potential generated by a point mass.}
\end{figure}

It turns out that the time independent gravitational potential is determined by the Poisson equation, the spherical symmetry of the situation, and the condition that the potential goes to zero when $r \to \infty$. Since the mass is concentrated at the origin, the Poisson equation becomes the Laplace equation
\[ \Delta \Phi =0.\]
The situation is spherically symmetric, and therefore, the potential $\Phi$ is a function of the radius $\Phi=\Phi(r)$.
The Laplace equation becomes
\begin{align*}
0=\Delta \Phi &=\sum\limits_i \frac{\partial^2 \Phi}{(\partial x^i)^2}=\sum\limits_i \frac{\partial }{\partial x^i}\left(  \frac{\partial \Phi}{\partial r} \frac{\partial r}{\partial x^i}\right)\\
&=\sum\limits_{i}\left[  \frac{\partial^2 \Phi}{\partial x^i\partial r} \frac{\partial r}{\partial x^i}+  \frac{\partial \Phi}{\partial r} \frac{\partial^2 r}{(\partial x^i)^2}\right]\\
&=\sum\limits_{i}\left[  \frac{\partial^2 \Phi}{\partial r^2} \left(\frac{\partial r}{\partial x^i}\right)^2+  \frac{\partial \Phi}{\partial r} \frac{\partial^2 r}{(\partial x^i)^2}\right]\\
&=\sum\limits_{i}\left[  \frac{\partial^2 \Phi}{\partial r^2}\left( \frac{x^i}{r}\right)^2+  \frac{\partial \Phi}{\partial r} \left(\frac{1}{r}-\frac{(x^i)^2}{r^3}\right)\right]\\
&= \frac{\partial^2 \Phi}{\partial r^2}+ \frac{2}{r}\frac{\partial \Phi}{\partial r} .
\end{align*}
This has solutions
\begin{equation}
\Phi(r)=\frac{A}{r}+B,
\end{equation}
where $A,B$ are arbitrary constants. The assumption that $\Phi(r)\to 0$ when $r \to \infty$ implies that $B=0$. The free parameter $A$ depends on the mass $M$ and one concludes
\begin{equation}
\Phi(r)=-\frac{G_NM}{r}.
\end{equation}

This computation of the gravitational potential for a point mass has a relativistic analogue, known as Birkhoff's theorem. It characterizes the Schwarzschild metric as the unique spherically symmetric, asymptotically flat and static solution to the vacuum Einstein equation.

\section{Spherical symmetry and Birkhoff's theorem}
The goal of this section is to proof Birkhoff's theorem, which characterizes the Schwarzschild metric. Since we will be interested in spherical symmetry, it will be convenient to use spherical coordinates on $\RR^3$. Recall that spherical coordinates are related to Euclidean coordinates by
\begin{align}\label{scoor}
x&=r \sin \theta\cos \varphi,\nonumber\\
y&=r \sin\theta\sin \varphi,\\
z&=r \cos\theta,\nonumber
\end{align}
where $r>0$, $ \varphi \in(0,2\pi)$ and $\theta \in (0, \pi)$.
\begin{figure}[H]
\centering
\includegraphics[scale=0.5]{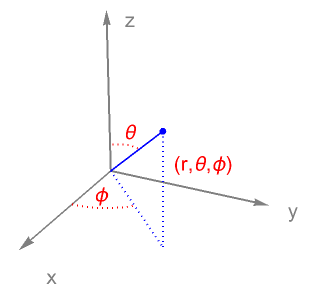}
\caption{Spherical coordinates.}
\end{figure}
In spherical coordinates, the  Euclidean metric $g= dx \otimes dx + dy \otimes dy + dz \otimes dz$ takes the form
\[ g=  dr \otimes dr +r^2\left( d\theta \otimes d\theta + \sin^2\theta\, d\varphi \otimes d\varphi \right).\]
In particular, the induced metric on a sphere $S_r^2$, of radius $r$, is
\[h=r^2\left( d\theta \otimes d\theta + \sin^2\theta\, d\varphi \otimes d\varphi \right).\]
We will sometimes write $d\Omega$ to denote the induced metric on the unit sphere.
Given a submanifold $U$ of $\RR^3$ which is invariant under the action of $\mathrm{SO}(3)$, and a metric $h $ on $U$, we say that $h$ is spherically symmetric if, for all $A$ in $\mathrm{SO}(3)$,
\[A^*h=h.\] 
The Euclidean metric $g$ is of course spherically symmetric.

\begin{figure}[H]
\centering
\includegraphics[scale=0.6]{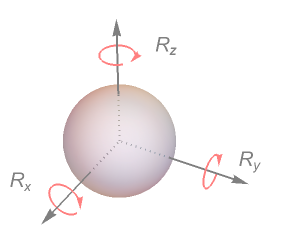}
\caption{Spherical Symmetry: rotations are isometries.}
\end{figure}

We will now describe all spherically symmetric metrics on $\RR^3\setminus\{0\}$.

\begin{lemma}
Let $S_r^2$ be the sphere of radius $r$ in $\RR^3$. If $h$ is a spherically symmetric metric on $S_r $, then, $h$ is a constant multiple of the metric induced by the Euclidean metric in $\RR^3$. Explicitly,
\begin{equation}\label{metrick} h=\beta r^2\left( d\theta \otimes d\theta + \sin^2\theta \, d\varphi \otimes d\varphi \right),\end{equation}
for some constant $\alpha>0$.
\end{lemma}

\begin{proof}
By rescaling, we may assume that $r=1$. Consider the point $p=(0,1,0)\in \RR^3$ which has spherical coordinates $\theta(p)=\varphi(p)=\pi/2$.  We claim that, with respect to the metric $h$, the vectors $\partial_\theta(p), \partial_\varphi(p)$ are orthogonal and have the same norm \[|\partial_\theta(p)|=| \partial_\varphi(p)|=\beta.\]
Using \eqref{scoor} one sees that
\[ \partial_\varphi (p)=-\partial_x(p),\qquad \partial_\theta(p)=-\partial_z(p).\]
Let $A$ be the rotation by $\pi/2$ around the $y$ axis. We  have
\[ DA(p)(\partial_\theta(p))=\partial_x(p)=-\partial_\varphi(p),\qquad DA(p)(\partial_\varphi(p))=-\partial_z(p)=\partial_\theta(p). \]
Since $A$ preserves the metric $h$, then
\[ \langle \partial_\theta(p),\partial_\varphi(p)\rangle =  \langle DA(p)(\partial_\theta(p)),DA(p)(\partial_\varphi(p))\rangle=- \langle \partial_\varphi(p),\partial_\theta(p)\rangle.\]
We conclude that $\partial_\theta(p), \partial_\varphi(p)$ are orthogonal and have the same norm, which we call $\beta$.  Both sides of \eqref{metrick} are metrics on the sphere that are invariant under the action of $\mathrm{SO}(3)$ and coincide at the point $p=(0,1,0)$. Since the group of rotations acts transitively on the sphere, two invariant metrics that coincide at a point are equal. This completes the proof.
\end{proof}

\begin{lemma}\label{sphere}
Let $g$ be a spherically symmetric Riemannian metric on  $N=\RR^{3}\setminus\{0\}$. Then, in spherical coordinates, the metric takes the form
\[
g=\alpha(r)dr\otimes dr+\beta(r)\left(d\theta\otimes d\theta+\sin^{2}\theta\, d\varphi\otimes d\varphi\right),
\]
where $\alpha(r)$ and $\beta(r)$ are positive valued functions.
\end{lemma}

\begin{proof}
Let us first prove that the vector field $\partial_r$ is orthogonal to $\partial_\theta$ and $\partial_\varphi$.
Fix a point $p \in N$ and consider the linear transformation $A \in \mathrm{SO}(3)$ which rotates by an angle of $\pi$ with respect to the axis spanned by $p$. The diffeomorphism $A$ preserves the vector field $\partial_r$. Moreover, since the metric $g$ is $\mathrm{SO}(3)$ invariant
\[ \langle \partial_r(p),\partial_\theta (p)\rangle =\langle DA(p)(\partial_r(p)),DA(p)(\partial_\theta (p))\rangle=\langle \partial_r(p),-\partial_\theta (p)\rangle,\]
and one concludes that
\[\langle \partial_r(p),\partial_\theta (p)\rangle=0.\]
The same argument shows that 
\[\langle \partial_r(p),\partial_\varphi (p)\rangle=0.\]
One concludes that the coefficients of $ dr \otimes d\theta$ and of $dr \otimes d\varphi$ vanish.
On the other hand, Lemma \ref{sphere} guarantees that the restriction of $g$ to each sphere $S_r^2$ is a multiple of 
the standard metric. Therefore
\[
g=\alpha(r,\theta,\varphi)dr\otimes dr+\beta(r)\left(d\theta\otimes d\theta+\sin^{2}\theta\, d\varphi\otimes d\varphi\right).
\]
It remains to show that $\alpha$ depends only on $r$. We know that
\[\beta(r)\left(d\theta\otimes d\theta+\sin^{2}\theta\, d\varphi\otimes d\varphi\right)\]
is invariant with respect to the action of $\mathrm{SO}(3)$. Since $g$ is spherically symmetric, we conclude that
\begin{equation}\label{alpha}\alpha(r,\theta,\varphi)dr\otimes dr\end{equation}
is also $\mathrm{SO}(3)$ invariant. The group of rotations fixes the coordinate $r$ and acts transitively on each sphere, therefore, in order for \eqref{alpha} to be $\mathrm{SO}(3)$ invariant it is necessary that $\alpha$ is independent of $\theta $ and $\varphi$.
\end{proof}

The Schwarzschild spacetime with radius $r_s\geq0$ is the manifold $M= \RR \times \RR_{>r_s} \times S^2$ with metric
\begin{equation}
g=-\left(  1-\frac{r_s}{r}\right)c^2  dt\otimes dt+\left(
1-\frac{r_s}{r}\right)  ^{-1}dr\otimes dr+r^{2}\left( d\theta \otimes d\theta + \sin^2\theta\,d\varphi \otimes d\varphi \right).
\label{forma de swch}%
\end{equation}
The Schwarzschild metric is a solution of the vacuum Einstein equation, and it is invariant with respect to the action of $\mathrm{SO}(3)$ on $S^2$. The parameter $r_s$ is called the Schwarzschild radius. Clearly, when $r_s=0$, one recovers Minkowski spacetime. Also, for $r\to \infty$, the metric tends to the Minkowski metric. For this reason, one says that the metric is asymptotically flat.

A stationary spacetime is a spacetime together with a timelike Killing vector field $T$ that generates a global flow by isometries. A stationary spacetime is called static if for any two vector fields $X, Y$ which are orthogonal to $T$, their Lie bracket $[X,Y]$ is also orthogonal to $T$. The Schwarzschild spacetime is a static spacetime with vector field $T=\partial_t$. This is clearly a timelike vector field and generates the flow
\[H_s(t,r,\theta,\varphi)=(t+s, r,\theta,\varphi).\]
For each fix $s \in \RR$, the map $H_s$ is an isometry, and therefore, $T$ is a Killing vector field. Let $X,Y$ be vector fields orthogonal to $T$. Then, they are of the form
\[ X= X^r \partial_r+ X^\theta \partial_\theta +X^\varphi \partial_\varphi,\qquad  Y= Y^r \partial_r+ Y^\theta \partial_\theta +Y^\varphi \partial_\varphi.\]
Since $\partial_t$ does not appear in either of these expressions, it does not appear in the expression
for $[X,Y]$. One concludes that $[X,Y]$ is orthogonal to $T$, and that the Schwarzschild spacetime is static. The condition that a spacetime is static implies that, around each point, there is a spacelike submanifold of dimension three that is orthogonal to $T$. The figure bellow depicts an example of a static spacetime.

\begin{figure}[H]
\centering
\includegraphics[scale=0.55]{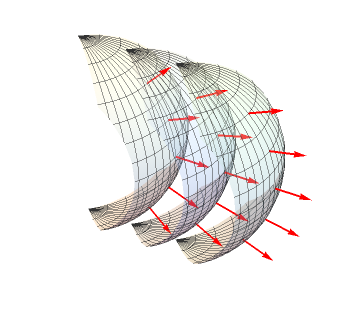}
\caption{Static spacetime.}
\end{figure}

\begin{center}
\begin{tabular}{ccc}
\hline &&\\
\multicolumn{3}{c}{\textbf  Properties of the Schwarzschild spacetime} \\  
&&\\
\hline && \\
Solution of the vacuum Einstein equation && $\Ric_{ab}-\frac{1}{2}\mathrm{R} g_{ab}=0.$ \\
&&\\
Spherically symmetric &&$  A^*g=g$ for all $A\in SO(3).$ \\
&&\\
Static &&$  \partial_t$ is a  Killing vector field.\\
&&\\
Asymptotically flat&& Tends to Minkowski spacetime when $r\to \infty$. \\
&&\\
\hline
\end{tabular}
\end{center}

The properties above characterize the Schwarzschild spacetime. In fact, the condition that the metric is static follows from the others. This result is referred to as Birkhoff's theorem. We will present a version of Birkhoff's theorem which is not the strongest possible, but will be enough for our purposes. For different, stronger formulations, the reader may consult \cite{Blau}, \cite{choquet} and \cite{plebansky}.

\begin{theorem}[Birkhoff] Let $g$ be a Lorentzian metric on  $M= \RR \times ( \RR^3\setminus\{0\})$ such that
\begin{enumerate}
\item[(1)] $g$ is invariant with respect to the action of $\mathrm{SO}(3)$ on $ \RR^3\setminus \{0\}$.
\item[(2)] $g$ is a solution of the vacuum Einstein equations.
\item[(3)] In polar coordinates, it takes the form\footnote{The most general spherically symmetric metric has the form 
\begin{equation}
g=-e^{2\alpha(r,t)}dt\otimes dt+\zeta(r,t) dr \otimes dt +e^{2\beta(r,t)}
dr\otimes dr+\eta(r,t)d\Omega.\end{equation}
We assume that $\zeta=0$ and $\eta=r^2$. These conditions can be obtained, at least locally, with appropriate changes of coordinates.}
\begin{equation}\label{fbir}
g=-e^{2\alpha(r,t)}dt\otimes dt+e^{2\beta(r,t)}
dr\otimes dr+r^2d\Omega.\end{equation}
\end{enumerate}
Then, there exists $r_s>0$ such that $(M,g)$ is isometric to the Schwarzschild spacetime with radius $r_s$. 
\end{theorem}

\begin{proof}
 Let us first prove that  $\alpha$ and $\beta$ can be chosen to be independent of $t$. We write $f'$ to denote the derivative of $f$ with respect to $r$. The Christoffel symbols for
 a metric of the form (\ref{fbir}) are
\begin{alignat*}{3}
 \Gamma_{00}^{0} &  =\frac{\partial \alpha}{\partial t},\quad &&\Gamma_{01}^{1}   =\frac{\partial \beta}{\partial t},\quad &&\Gamma_{11}^{0}   =\frac{e^{2\beta}}{e^{2\alpha}}\frac{\partial \beta}{\partial t},\\
\Gamma_{01}^{0} &  =\alpha'(r),\quad &&\Gamma_{00}^{1}   =\alpha'(r)e^{2\alpha(r)-2\beta(r)},\quad &&\Gamma_{11}^{1}   =\beta'(r),\\
\Gamma_{22}^{1} &  =-re^{-2\beta(r)},\quad &&\Gamma_{33}^{1}   =-r\sin^{2}\theta e^{-2\beta(r)}, \quad &&
\Gamma_{12}^{2}   =\frac{1}{r},\\
\Gamma_{33}^{2}&=-\sin\theta
\cos\theta,\quad &&
\Gamma_{13}^{3}   =\frac{1}{r}, \quad &&
\Gamma_{23}^{3}=\frac{\cos\theta
}{\sin\theta}.
\end{alignat*}
This implies that
\begin{align}
\Ric_{10}&=\frac{2}{r}\frac{\partial \beta}{\partial t},\label{r10} \\
\Ric_{22}&=-e^{-2\beta}\left( 1+r(\alpha'-\beta')\right)+1. \label{r22}
\end{align}
Since the metric satisfies the vacuum Einstein equation, the Ricci tensor is zero and therefore, equation (\ref{r10}) implies that $\beta$ is independent of $t$. Equation (\ref{r22}) then implies that 
\[ \alpha'=\frac{e^{2\beta}-1}{r}+\beta'\]
is also independent of $t$. Therefore,
\[ \alpha(r,t)=\mu(t)+ \nu(r).\]
We set 
\[ \tau=\int_0^t e^{\mu(s)}ds,\]
so that
\[ e^{2\nu(r)}d\tau \otimes d\tau=e^{2\nu(r)}\left(\frac{d\tau}{dt}\right)^2dt \otimes dt=e^{2\nu(r)}e^{2 \mu(t)}dt \otimes dt=e^{2\alpha(r,t)}dt \otimes dt.\]
Renaming the variable $\tau=t$, we are left with the case where the metric is independent of $t$, that is,
\begin{equation}
g=-e^{2\alpha(r)}dt\otimes dt+e^{2\beta(r)}
dr\otimes dr+r^2d\Omega.\label{segundom}%
\end{equation}
The nonzero Christoffel symbols are
\begin{alignat*}{3}
\Gamma_{01}^{0} &  =\alpha'(r),\quad &&
\Gamma_{00}^{1}   =\alpha'(r)e^{2\alpha(r)-2\beta(r)},\quad && 
\Gamma_{11}^{1}   =\beta'(r),\\
\Gamma_{22}^{1} &  =-re^{-2\beta(r)},\quad &&
\Gamma_{33}^{1}   =-r\sin^{2}\theta e^{-2\beta(r)},\quad &&
\Gamma_{12}^{2}   =\frac{1}{r},\\
\Gamma_{33}^{2}&=-\sin\theta
\cos\theta,\quad &&
\Gamma_{13}^{3}   =\frac{1}{r},\quad &&
\Gamma_{23}^{3}=\frac{\cos\theta
}{\sin\theta}.
\end{alignat*}
The components of the Ricci tensor that are not automatically zero are
\begin{align}\label{tabla 1}
\Ric_{00} &  =e^{2(\alpha(r)-\beta(r))}\left(\alpha^{\prime\prime
}(r)+\alpha^{\prime}(r)^{2}-\alpha^{\prime}(r)\beta^{\prime}(r)+\frac{2}%
{r}\alpha^{\prime}(r)\right)
\\
\Ric_{11} &  =-\alpha^{\prime\prime}(r)-\alpha^{\prime}(r)^{2}%
+\alpha^{\prime}(r)\beta^{\prime}(r)+\frac{2}{r}\beta^{\prime}(r)\\
\Ric_{22} &  =e^{-2\beta(r)}\left(r(\beta^{\prime}(r)-\alpha^{\prime
}(r))-1\right)+1\\
\Ric_{33} &  =\sin^{2}\theta \, \Ric_{22}.
\end{align}
The condition that the Ricci tensor vanishes implies
\begin{equation}
0=e^{-2(\alpha(r)-\beta(r))}\Ric_{00}+\Ric_{11}=\frac{2}%
{r}(\alpha^{\prime}(r)+\beta^{\prime}(r)),\label{first sw}%
\end{equation}
and one concludes that \begin{equation}\beta(r)=-\alpha(r)+C,\end{equation} for a constant $C$.
Rescaling the variable $t$ by $\overline{t}=e^{C}t$, we may assume that 
 \begin{equation}
 \alpha(t)=-\beta(t).
 \end{equation}
 With this assumption, the condition $\Ric_{22}=0$ implies that

\begin{equation}e^{2\alpha(r)}(2r\alpha^{\prime}(r)+1)=1,\end{equation} which is equivalent to
\begin{equation}(re^{2\alpha(r)})^{\prime}=1.\end{equation} One concludes that
 \begin{equation}re^{2\alpha(r)}=r-r_s,
\end{equation} and therefore \[e^{2\alpha(r)}=1-\frac{r_s}{r}.\] 
One can verify that, for this choice of $\alpha(r)$ and $\beta(r)$, all components of the Ricci tensor vanish.
In summary, the metric takes the form
\begin{equation}
g=-\left(  1-\frac{r_s}{r}\right)  dt\otimes dt+\left(
1-\frac{r_s}{r}\right)  ^{-1}dr\otimes dr+r^{2}d\Omega.\label{sw0}%
\end{equation}
Setting $t=c\overline{t}$, this becomes
\begin{equation}
g=-\left(  1-\frac{r_s}{r}\right)c^2  d\overline{t}\otimes d\overline{t}+\left(
1-\frac{r_s}{r}\right)  ^{-1}dr\otimes dr+r^{2}d\Omega,%
\end{equation}
as required.
\end{proof}

\begin{center}
\begin{tabular}{ccc}
\hline &&\\
\multicolumn{3}{c}{\textbf  Gravity caused by a point mass} \\  
&&\\
\hline && \\
 {\textbf Newtonian description}&& {\textbf Relativistic description} \\  
&&\\
\hline && \\
Gravitational potential  && Schwarzschild metric \\
&&\\
$\displaystyle\Phi=-\frac{G_N M}{r}$&&$ \displaystyle g=-\left( 1-\frac{r_{s}}{r}\right) c^2 dt^2+\left(
1-\frac{r_{s}}{r}\right)  ^{-1}dr^2+r^{2}d\Omega$ \\
&&\\
\hline
\end{tabular}
\end{center}
\section{The Schwarzschild metric}

 In this section we will describe some basic geometric properties of the Schwarzschild spacetime
\[ g=-\Big(1-\frac{r_s}{r}\Big)c^2 dt \otimes dt+\Big(1-\frac{r_s}{r}\Big)^{-1} dr \otimes dr+r^2\Big(d\theta \otimes d\theta +\sin^2\theta\,d\varphi \otimes d\varphi\Big) .\]
Consider the spacelike surface in Schwarzschild spacetime given by $\theta=\pi/2$ and 
$t=\text{constant}$. In units where $c=1$, the metric on this surface is
\[g =\Big(\frac{r}{r-r_s}\Big) dr \otimes dr +r^2 d\varphi \otimes d\varphi.\] 
This surface has the geometry of the Flamm paraboloid $F$, which is the graph of the function
\[ h(r,\varphi)=2 \sqrt{r_s(r-r_s)},\]
defined on set $U$ of points in the plane with $r>r_s$.  Consider the parametrization $\phi: U \rightarrow F$
of the Flamm paraboloid given by
\[ (r,\varphi)\mapsto (r\cos\varphi,r\sin\varphi, 2 \sqrt{r_s(r-r_s)}).\]
The derivative of $\phi$ is
\[ D\phi(r,\varphi)=\begin{pmatrix}
\cos\varphi&-r\sin\varphi\\
\sin\varphi&r \cos\varphi\\
\sqrt{\frac{r_s}{(r-r_s)}}&0
\end{pmatrix}\]
From this we compute 
 \begin{align*} \langle D\phi (\partial_r),D\phi (\partial_r)\rangle&=\frac{r}{r-r_s},\\
  \langle D\phi (\partial_\varphi),D\phi (\partial_\varphi)\rangle&=r^2,\\
   \langle D\phi (\partial_r,D\phi (\partial_\varphi)\rangle&=0.\\
\end{align*}
This shows that the Flamm paraboloid does have the geometry of a slice of Schwarzschild spacetime.
\begin{figure}[H]
\centering
\includegraphics[scale=0.65]{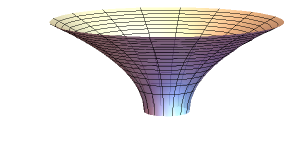}
\caption{The Flamm paraboloid has the geometry of the spacelike surface in Schwarzschild spacetime with $\theta=\pi/2$ and $t=\text{constant}$.}
\end{figure}

\begin{figure}
\centering
\includegraphics[scale=0.38]{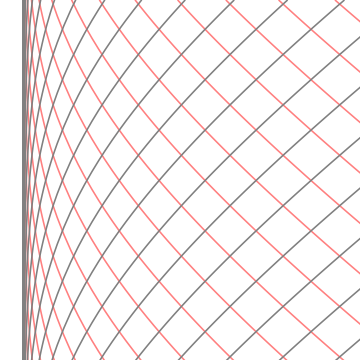}
\caption{Light like goedesics in Schwarzschild spacetime.}
\label{llgss}
\end{figure}

Let us also consider surfaces with $\theta=\text{constant}$ and $\varphi=\text{constant}$. In this case
the metric is
\[ g=-\Big(1-\frac{r_s}{r}\Big) dt \otimes dt+\Big(1-\frac{r_s}{r}\Big)^{-1} dr \otimes dr.\]
Then, according to the discussion in \S \ref{GTD}, the light like geodesics are depicted in Figure \ref{llgss}. Red lines correspond to rays of light going towards the mass, gray lines, to light going away from the mass.

Recall from \S\ref{section 00} that the relationship between the Newtonian potential $\Phi$ and the relativistic metric is
\begin{equation}\label{Phig} \Phi=-\left(\frac{c^2}{2}+\frac{g_{00}}{2}\right).\end{equation}
In the case of the Schwarzschild metric
\begin{equation}\label{g00} g_{00}=\left(\frac{r_s-r}{r}\right)c^2,\end{equation}
and the potential is
\begin{equation}\label{pot} \Phi(r)=-\frac{G_N M}{r},\end{equation}
where $M$ is the total mass. Replacing (\ref{g00}) and (\ref{pot}) into (\ref{Phig}),
one concludes that
\begin{equation} r_s=\frac{2G_N M}{c^2}.\end{equation}
The Schwarzschild radius is then proportional to the mass of the object. Naturally, when $M=0$, the Schwarzschild solution becomes Minkowski spacetime. The $c^2$ in the denominator makes the Schwarzschild radius of most human scale objects very small. It is only for very dense objects that the Schwarzschild radius is larger than the radius of the object itself. The Schwarzschild metric only accounts for the geometry of spacetime for $r>r_s$. The question of what happens for $r<r_s$ is an interesting one, which we postpone until the discussion on Eddington-Finkelstein coordinates and black holes.
\begin{table}[H]
\begin{center}
\begin{tabular}{ |c|c|c|c|} 
 \hline&&&\\
{\textbf Object} & {\textbf Mass (Kg)} & {\textbf Schwarzschild radius (m)} & {\textbf  Physical radius (m)}\\ 
 \hline &&&\\
Orange &$\sim  0.14 $& $2.08 \times 10^{-28}$& $\sim 3.5 \times 10^{-2}$ \\ 
\hline&&&\\
Human &$\sim 70$ &$ 1.04 \times 10^{-25}$& $\sim 1$ \\ 
 \hline&&&\\
Moon& $7.35 \times 10^{22}$ &$1.09 \times 10^{-4}$ &$1.73 \times 10^6$\\ 
 \hline&&&\\
Earth& $5.97 \times 10^{24}$& $8.87 \times 10^{-3}$ &$6.37 \times 10^{6}$\\ 
 \hline&&&\\
 Sun &$ 1.99 \times 10^{30}$& $2.95 \times 10^3$ &$6.9\times 10^{8}$\\ 
 \hline&&&\\
  Milky Way & $1.6 \times 10^{42}$& $2.4 \times 10^{15}$ &$5 \times 10^{20}$\\ 
 \hline
\end{tabular}
\end{center}
\caption{Schwartzschild radius of some familiar objects.}
\end{table}

\section{Planetary motion in Newtonian gravity}

Before discussing the corresponding computation in Schwarzschild spacetime, we
will  describe the basic equations of planetary motion in Newton's theory of gravity.
Suppose that a planet $P$ of mass $m$ moves under the action of a gravitational
force caused by a massive object of mass $M$. The trajectory of $P$ is the curve
 \[\alpha(t)=(r(t),\theta(t),\varphi(t)),\]
 and the force is \[f(t)=-\frac{G_NM
m}{| \alpha(t)| ^{3}}\alpha(t).\]
The angular momentum of $P$ is 
\[L(t)=m\alpha(t)\times\alpha^{\prime}(t).\]Since, by Newton's second law \[\alpha''
(t)=-\frac{MG_N}{| \alpha(t)|^{3}}
\alpha(t),\] one has \[\alpha(t)\times\alpha^{\prime\prime}(t)=0.\] This
implies
\[
L^{\prime}(t)=m\alpha'(t)\times \alpha'(t)+m
\alpha(t)\times\alpha''(t)=0,
\]
and therefore $L(t)=L$ must be constant. Since $\alpha(t)$ is at all
times perpendicular to $L$, it lies on a plane orthogonal to $L$.
By rotating coordinates if necessary, we may assume this is the 
plane $\theta=\pi/2$, so that the motion occurs on the $xy$ plane according to equations
\begin{equation}
x(t)=r(t)\cos \varphi(t),\qquad  y(t)= r(t) \sin\varphi (t),\qquad  z(t)=0.
\end{equation}
Then
\begin{align}\label{vel}
x'(t)&=r'(t)\cos\varphi(t)-r(t) \sin\varphi(t)\varphi'(t),\\ \nonumber
 y'(t)&= r'(t) \sin\varphi (t)+r(t)\cos \varphi(t) \varphi'(t),\\z'(t)&=0.\nonumber
\end{align}
One concludes that
\[ \alpha(t) \times \alpha'(t)=r^2(t) \varphi'(t) \partial_z\]
and therefore
\begin{equation}\label{call}
l=|L|=mr^2(t) \varphi'(t).
\end{equation}
On the other hand, the gravitational potential is
 \begin{equation}\Phi(t)=-\frac{G_N M}{r},\end{equation}
so that
the total work $W$ done by $F$ to move a particle from $p=\alpha(t_{1})$ to
$q=\alpha(t_{2})$ is 
\begin{equation}\label{W1}
W   =\int_{t_1}^{t_2}F(t)\cdot \alpha'(t)dt=-m\int_{t_{1}
}^{t_{2}}\nabla \Phi(\alpha(t))\cdot\alpha'(t)dt=-m\int_{t_{1}
}^{t_{2}} \frac{d}{dt}\Phi(\alpha(t))dt
=m\big(\Phi(p)-\Phi(q)\big).
\end{equation}
Since $F=m\alpha''(t)$, one can also compute the work as 
\begin{equation}\label{W2}
W   =\int_{t_1}^{t_2}F(t)\cdot \alpha'(t)dt=\int_{t_1}^{t_2}m \alpha''(t)\cdot \alpha'(t)dt
=\int_{t_1}^{t_2}\frac{d}{dt}K(t)dt=K(q)-K(p),\end{equation}
where $K(t)$ is the kinetic energy 
 \[K(t)=\frac{1}{2}m\vert \alpha^{\prime
}(t)\vert ^{2}.\]
The total energy of the particle is the sum of the potential and kinetic energy 
\begin{equation}
E(t)=m\Phi(\alpha
(t))+K(t).\end{equation}
Using \eqref{W1} and \eqref{W2}, one concludes that
\[ E(q)=E(p),\]
so that the total energy is constant.
Using (\ref{vel}) one computes that
\begin{equation}
K(\alpha(t))=\frac{1}{2}m\left(  r^{\prime}(t)^{2}+r^{2}%
(t)\varphi^{\prime}(t)^{2}\right)  . \label{kinetic}%
\end{equation}
Solving for $\varphi^{\prime}(t)$ in (\ref{call}), one obtains the equation of motion for $P$, given as
\begin{equation}
E=\frac{m}{2}\left(  \frac{dr}{dt}\right)  ^{2}+\left(
\frac{l^{2}}{2mr^{2}(t)}-\frac{G_{N}Mm}%
{r(t)}\right)  .\label{mov planetas}%
\end{equation}
Equation (\ref{mov planetas}) can be solved for $r$ in terms of $\varphi$, as
follows. One writes
\[
\frac{dr}{dt}=\frac{dr}{d\varphi}\frac{d\varphi}{dt}=\frac{dr}{d\varphi}\frac
{l}{mr^{2}(t)}.
\]
Define a new function $u(t)$ by \[u(t)=\frac{l^{2}}
{Mmr(t)}.\] In terms of $\varphi$ one has
\[
\frac{dr}{d\varphi}=\frac{dr}{du}\frac{du}{d\varphi}=-\frac{l^{2}
}{Mmu^2}\frac{du}{d\varphi},
\]
and therefore%
\[
\frac{dr}{dt}=-\frac{l}{mr^{2}(t)}\frac{l^{2}}{mMu^2%
}\frac{du}{d\varphi}=-\frac{l}{mr^{2}(t)}\frac{l^{2}}{mM
}\frac{M^2m^2r^2(t)}{l^4}\frac{du}{d\varphi}=-\frac{M}{l}\frac{du}%
{d\varphi}.
\]
We can write (\ref{mov planetas}) in terms of $u$ as follows:
\[
E=\frac{mM^2}{2 l^{2}}\left(  \frac
{du}{d\varphi}\right)^{2}+\left(  \frac{mM^{2}}{2 l^{2}}u^{2}%
-\frac{G_Nm^2M^2}{l^{2}}u\right)  .
\]
Multiplying both sides by $2 l^{2}/(mM^{2})$ one obtains
\[
\frac{2l^{2}E}{mM^{2}}=\left(  \frac{du}{d\varphi}\right)
^{2}+u^{2}-2mG_Nu.
\]
Taking derivatives with respect to $\varphi$ on both sides gives 
\begin{equation}
\frac{du}{d\varphi}\left(\frac{d^{2}u}{d\varphi^{2}}+u-mG_N\right)=0.
\end{equation}
Solutions to this equation are given by
 \[u(\varphi)=mG_N+\varepsilon\cos\varphi,\] 
 so that
 \[ r(\varphi)=\frac{p}{1+e\cos\varphi},\]
 whrere $e=\varepsilon/mG_N$ and $p=l^2/Mm^2G_N$. The value of $e$ determines the shape of the curve. Since
 $-e\cos(\varphi)=e\cos(\varphi+\pi),$
 by displacing the angle we may assume that $e\geq0$. The curves are the following:
 \[\frac{p}{1+e\cos(\varphi)}=\begin{cases}\ \text{Circle}& \text{ if } e=0,\\
\ \text{Ellipse}& \text{ if }  0<e<1,\\
\text{Parabola}& \text{ if } e=1,\\
 \text{Hyperbola}& \text{ if } e>1.\\
 \end{cases}
 \]
Figure \ref{polarconics} illustrates the possibilities for the shape of the orbit.
\begin{center}
\begin{figure}[H]
\centering
\includegraphics[scale=0.5]{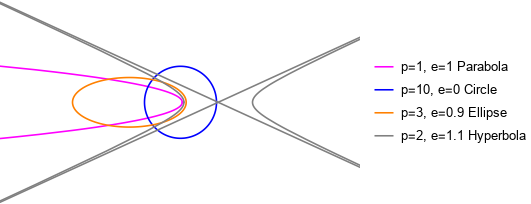}\caption{The polar plots of $r(\varphi)=\frac{p}{1+e\cos(\varphi)}$ are conic sections.}%
\end{figure}
\label{polarconics}
\end{center}

One concludes that all the trajectories in the Newtonian description of gravity due to a massive object are conic sections. The bounded orbits are ellipses and the unbounded orbits are
parabolae or hyperbolae. As the name indicates, these are the shapes formed by the intersection of a plane and a cone. The following figure illustrates the possibilities.

\begin{figure}[h]
\begin{center}
  \begin{tabular}{ccc}
  \includegraphics[scale=0.3]{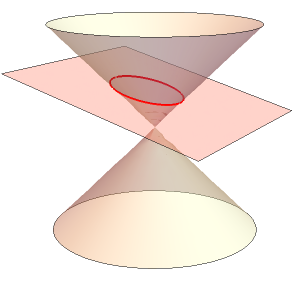}&
  \includegraphics[scale=0.3]{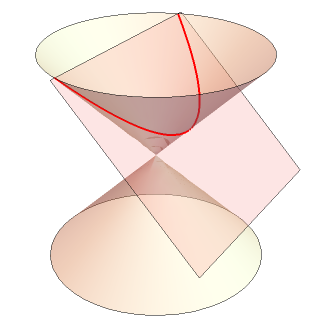}&
 \includegraphics[scale=0.3]{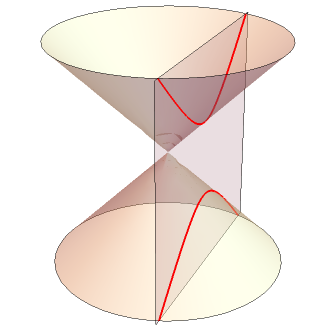}
\end{tabular}
\end{center}
\caption{Ellipse, parabola and hyperbola, the conic sections. }
\end{figure}

\begin{figure}[h]
\centering
\includegraphics[scale=0.75]{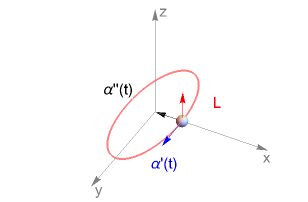}
\caption{Planetary orbit.}\label{tipor}
\end{figure}

The planets in the solar system move along bounded trajectories, and we conclude that their orbits are ellipses. Figure \ref{tipor} illustrates a typical orbit.

Using that 
\[l=m| \alpha(t) \times \alpha'(t)|=mr^{2}
(t)\varphi^{\prime}(t)\]
is constant, one can compute the area swept by a planet
during an interval of time $\Delta t=t_{2}-t_{1}$ 
\begin{equation}A=\frac{1}{2}\int
\limits_{t_{1}}^{t_{2}}r^{2}(t)\varphi^{\prime}(t)dt=\frac{l\Delta t}{2m}.\end{equation}
This shows that the area depends only on the time difference! The planet sweeps equal areas in equal time intervals. This fact is known as Kepler's law of planetary motion.
Table \ref{exentricity} describes the orbits of the planets of the solar system. The Perihelion is the shortest distance from the orbit to the Sun. The Aphelion is the largest distance from the orbit to the Sun. These are  measured in astronomical units (AU). An astronomical unit is the average distance from the Earth to the Sun,
$1 \text{AU}=1.5 \times 10^{8} \text{km}.$.
\begin{table}[h]
\begin{center}
\begin{tabular}{ |c|c|c|c| } 
 \hline
 {\textbf Planet} & {\textbf Eccentricity} & {\textbf Perihelion} (AU)&{\textbf Aphelion} (AU)\\ 
 \hline&&&\\
 Mercury & 0.206 & 0.31 & 0.47\\ 
 \hline&&&\\
Venus & 0.007 & 0.718 &0.728 \\ 
 \hline&&&\\
Earth & 0.017 & 0.98& 1.02 \\ 
\hline&&&\\
Mars & 0.093 & 1.38& 1.67 \\ 
 \hline&&&\\
Jupiter & 0.048 & 4.95& 5.45 \\ 
 \hline&&&\\
Saturn & 0.056& 9.02&10.0 \\ 
 \hline&&&\\
Uranus & 0.047 & 18.3&20.1 \\ 
 \hline&&&\\
Neptune & 0.009 & 30&30.3 \\ 
 \hline&&&\\
Pluto & 0.248& 29.7& 49.9 \\ 
 \hline
\end{tabular}
\caption{The orbits of the planets in the solar system.}
\label{exentricity}
\end{center}
\end{table}

\section{Timelike geodesics in Schwarzschild spacetime \label{nonzero mass in sw}}
In this section we will describe timelike geodesics in Schwarzschild spacetime, which are the trajectories of 
objects falling freely under the action of gravity. They are the relativistic counterparts of the Newtonian planetary orbits described in the previous section. We begin with a geometric lemma  which will be used in what follows.

\begin{lemma}\label{kill}
Let $\gamma(\tau)$ be a geodesic in a pseudo-Riemannian manifold $M$, and $X$ a Killing vector field
on $M$. Then the function 
\[ \langle \gamma'(\tau), X(\gamma(\tau))\rangle \]
is constant.
\end{lemma}
\begin{proof}
Using that $\gamma(\tau)$ is a geodesic we compute:
\begin{equation} \frac{d}{d\tau}  \langle \gamma'(\tau), X(\gamma(\tau))\rangle =  \langle \gamma'(\tau), \nabla_{\gamma'(\tau)}X(\gamma(\tau))\rangle. \end{equation}
On the other hand, since $X$ is Killing, it satisfies that for any two vector fields $Y,Z$, the equation
\[ \langle \nabla_Y X, Z\rangle+ \langle Y, \nabla_Z X\rangle=0.\]
Setting $Y=Z=\gamma'(\tau)$, this becomes
\[ 2 \langle \gamma'(\tau), \nabla_{\gamma'(\tau)} X(\gamma(\tau))\rangle=0.\]
\end{proof}

Consider a geodesic $\gamma(\tau):I\rightarrow  M$ in Schwarzschild spacetime that is parametrized by proper time. This is the worldline of an object $P$ that is falling freely. Since the Schwarzschild metric is independent of
the coordinates $t$ and $\varphi$, the vector fields $\partial_{t}$ and $\partial_{\varphi
}$ are Killing vector fields. Therefore, lemma \ref{kill} implies that $e=-\left\langle \gamma^{\prime}(\tau),\partial_{t}\right\rangle$ and $\ell=\left\langle
\gamma^{\prime}(\tau),\partial_{\varphi}\right\rangle$ are constant functions of $\tau$. If we write \[\gamma(\tau)=(t(\tau),r(\tau),\theta
(\tau),\varphi(\tau)),\] we see that
 \begin{align}e&=t'(\tau)\left(1-\frac{r_s}{r(\tau)}\right)c^2\label{e},\\
 \ell&=r^{2}(\tau)\sin^{2}\theta(\tau)\varphi'(\tau)\label{l}.\end{align}
As in the Newtonian case, the spatial trajectory
is contained in a plane in $\RR^{3}$. To show this,  note that it is possible to rotate the coordinates in such a way that $\varphi'(0)=0$, which implies $\ell=0$, and therefore, $\varphi'(\tau)=0$.
In this case, $P$ moves in a plane perpendicular to the equatorial plane
$\theta=\pi/2$. Rotating coordinates again if necessary, we may assume  that the plane of motion is the equatorial plane $\theta=\pi/2.$ In
these coordinates 
\begin{equation}\label{lvarphi}\ell=r^{2}(\tau)\varphi^{\prime}(\tau),\end{equation} 
which is the same expression as for the angular momentum of a planet of mass $m=1$ in the Newtonian formulation.
Since $\gamma(\tau)$ is parametrized by proper time, it satisfies
\begin{equation}
-\left(1-\frac{r_s}{r(\tau)}\right)t^{\prime}(\tau)^{2}c^2+\left(1-\frac{r_s}%
{r(\tau)}\right)^{-1}r^{\prime}(\tau)^{2}+r^{2}(\tau)\varphi^{\prime}(\tau)^{2}=-c^2. \label{eq material}%
\end{equation}
Solving for $t^{\prime}(\tau)$ and $\varphi^{\prime}(\tau)$ in \eqref{e} and \eqref{l}, and
substituting in \eqref{eq material} one obtains
\begin{equation}\label{amotion}
-\left(1-\frac{r_s}{r(\tau)}\right)^{-1}\frac{e^{2}}{c^2}+\left(1-\frac{r_s}%
{r(\tau)}\right)^{-1}r'(\tau)^{2}+\frac{\ell^{2}}{r(\tau)^{2}}=-c^2,
\end{equation}
which is equivalent to
\begin{equation}\label{rmotion}
\frac{e^2-c^4}{c^2}=r'(\tau)^2+\left(\frac{\ell^2}{r(\tau)^2}-\frac{r_s\ell^2}{r(\tau)^3}- \frac{r_sc^2}{r(\tau)}\right).
\end{equation}
If one sets
\[ E=\frac{m(e^2-c^4)}{2c^2},\]
and recalls that
\[ l=\ell m,\]
then \eqref{rmotion} can be written as
\begin{equation}\label{rmotion1}
E=\frac{m}{2}r'(\tau)^2+\left(\frac{l^2}{2mr(\tau)^2}- \frac{G_NmM}{r(\tau)}-\frac{G_N Ml^2}{mc^2r(\tau)^3}\right).
\end{equation}
This should be compared with the Newtonian counterpart \eqref{mov planetas} which is
\begin{equation*}
E=\frac{m}{2}r'(t)  ^{2}+\left(
\frac{l^{2}}{2mr^{2}(t)}-\frac{G_{N}Mm}
{r(t)}\right)  .
\end{equation*}
Thus, the relativistic equation differs from the Newtonian one by the addition of a term which is cubic in $1/r$.
Reassuringly, in the non relativistic limit  where $c \to \infty$, so that $dt/d\tau=\lambda_v \to 1$, the relativistic equation \eqref{rmotion1} tends to \eqref{mov planetas}. However, the qualitative behaviour of relativistic motion can be
quite different from the Newtonian orbits.  We will see that trajectories are not necessarily conics, or even
closed curves. Let us describe the behavior of the solutions.  The function
\begin{equation}V(r,l)=\frac{l^2}{2mr(\tau)^2}- \frac{G_NmM}{r(\tau)}-\frac{G_N Ml^2}{mc^2r(\tau)^3}\end{equation}
is called the effective potential.
In order to simplify the calculations, we choose units where $G_N=c=1,$ and assume that $m=1$. We also write the equations 
in terms of the new variables
\[ \rho=\frac{r}{M},\qquad h=\frac{l}{M}.\]
In these coordinates, the potential is
\begin{equation}V(\rho,h)=\frac{h^2}{2\rho^2}- \frac{1}{\rho}-\frac{ h^2}{\rho^3}.\end{equation}
For $h$ and $M$ fixed, the derivative of the potential is
\begin{equation}V'(\rho)=-\frac{h^2}{\rho^3}+ \frac{1}{\rho^{2}}+\frac{ 3h^2}{\rho^4}=\frac{1}{\rho^4}\left( \rho^{2}-\rho h^2+3h^2\right).\end{equation}
One concludes that the potential has critical points at
\begin{equation}
\rho=\frac{h^2 \pm\sqrt{h^4-12h^2}}{2}=\frac{h^2}{2}\Bigg(1 \pm \sqrt{1-\frac{12}{h^2}}\Bigg).
\end{equation}
By changing the orientation of the angle $\varphi$ if necessary, we may assume that $\ell=r^2(\tau)\varphi'(\tau)\geq0$ so that also $h\geq0$. Then
\[\# \text{ of critical points of } V(\rho)=\begin{cases}
0& \text{if } h< \sqrt{12},\\
1&  \text{if } h= \sqrt{12},\\
2& \text{if } h>\sqrt{12}.\\
\end{cases} \]
Figure \ref{Effpots} illustrates some examples.
\begin{center}
\begin{figure}[h]
\centering
\includegraphics[scale=0.6]{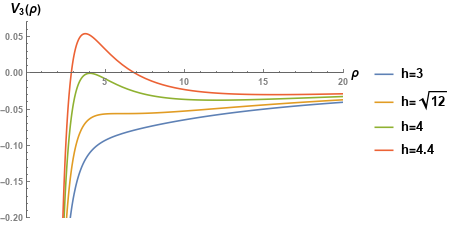}\caption{Effective
potentials for different values of $h$.}%
\end{figure}
\label{Effpots}
\end{center}

The equation of motion (\ref{rmotion1}) becomes
\begin{equation}\label{rmotion3}
\rho'(\tau)=\pm\frac{\sqrt{2}}{M} \sqrt{E-V(\rho(\tau))}.
\end{equation}
The shape of the orbit is determined by the potential and its relationship with $E$ as follows:
\begin{enumerate}
\item {$h\leq\sqrt{12}:$} In this case $V'(\rho)\geq0$. We claim that $\rho(\tau)$ can not have local extrema.
Suppose that $\tau_0$ is a local minimum so that $\rho''(\tau_0)>0$. For $\tau>\tau_0$ close to $\tau_0$, the function increases and therefore
\begin{equation*}\
\rho'(\tau)=\frac{\sqrt{2}}{M} \sqrt{E-V(\rho(\tau))},
\end{equation*}
differentiating both sides one gets
\begin{equation*}
\rho''(\tau)=-\frac{\sqrt{2}}{2M} \frac{\big(\frac{dV}{d\rho}\big)\rho'(\tau)}{\sqrt{E-V(\rho(\tau))}}=-\frac{1}{M^2} \Big(\frac{dV}{d\rho}\Big)\leq0.
\end{equation*}
Taking the limit $\tau \to \tau_0$ one concludes that $\rho''(\tau_0)\leq 0$, which is a contradiction. A similar argument shows that $\rho(\tau)$ does not have local maxima. The conclusion is that in this case, the orbits either plunge towards the mass or go off to infinity.
\item {$h>\sqrt{12}, E\gg V:$} In this case
\[ |\rho'(\tau)|=\frac{\sqrt{2}}{M} \sqrt{E-V(\rho(\tau))}>0,\]
so that $\rho'(\tau)$ cannot change signs. Again, the trajectory either goes to the mass or to infinity.
\item  {$h>\sqrt{12}, E\sim V:$} In this case $V(\rho)$ has a local minimum $A$ at $\rho_0$. If $ A<E<0$ and $\rho(0)=\rho_0$, then the orbit is bounded
\end{enumerate}

\subsubsection*{Circular orbits}
\begin{center}
\begin{figure}[H]
\centering
\includegraphics[scale=0.6]{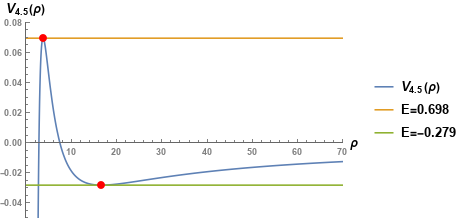}\caption{Circular orbit: the radius stays constant.}
\label{quieta}
\end{figure}
\end{center}

Suppose that $h>\sqrt{12}$.
If $\rho(\tau)$ is constant then $E=V(\rho(0))$. Differentiating the equation of motion one obtains
\begin{equation*}
0=\rho''(\tau)=-\frac{1}{M^2} \frac{dV}{d\rho}(\rho(0)),
\end{equation*}
so that $\rho(\tau)=\rho(0)$ is a critical point of $V(\rho)$.  There are two possibilities, either $V$ has a local maximum or a local minimum at $\rho_0$. 
The situation is depicted in Figure \ref{quieta}.
We see that for each value of $h>\sqrt{12}$ there are exactly two circular orbits.

\begin{center}
\begin{figure}[h]
\centering
\includegraphics[scale=0.35]{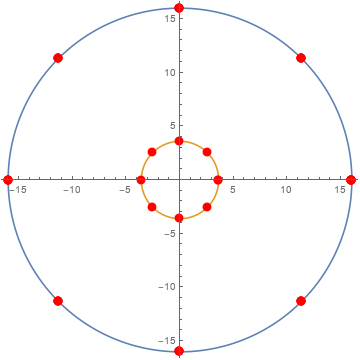}\caption{There are two circular orbits for each $h>\sqrt{12}.$}%
\end{figure}
\end{center}

\subsubsection*{Radial plunge orbits}

Another simple type of orbit is the radial free fall of a particle
$P$ originally at rest and coming from far away. This last condition is interpreted mathematically by imposing the condition that
\[\lim_{\tau \to -\infty}\rho(\tau)=\infty.\] 
In this situation, the  angle $\varphi(\tau)$ is constant, so that
\[h=\frac{l}{M}=\frac{lm}{M}=\frac{m r^2(\tau)\varphi'(\tau)}{M}=0.\]
Since the particle is originally at rest, then
\[\lim_{\tau \to -\infty}\rho'(\tau)=0.\] 
Making $\tau$ go to $-\infty $ in  \eqref{amotion} one concludes that $e=c^2$ so that $E=0$. 
The equation of motion becomes
\begin{equation}
\rho'(\tau)=-\frac{\sqrt{2}}{M} \sqrt{\frac{1}{\rho(\tau)}},
\end{equation}
which can be written in the form
\begin{equation}
\rho(\tau)^{1/2} \rho'(\tau)=-\frac{\sqrt{2}}{M}.
\end{equation}
This can be integrated to obtain
\begin{equation}\label{rhotau}
\rho(\tau)=\left(\frac{3}{\sqrt{2}M}\right)^{2/3}(C-\tau)^{2/3}.
\end{equation}
Also,
\[
\frac{dt}{d\rho}=\frac{dt/d\tau}{d\rho/d\tau}
= -\frac{M\rho^{3/2}}{\sqrt{2}(\rho-2)}.
\]
Integrating both sides one obtains
\begin{equation}
t(\rho)    =2M\left[ -\frac{2}{3}(\rho/2)^{3/2}-\sqrt{2\rho}+\log \left(\frac
{\sqrt{\rho}+\sqrt{2}}{\sqrt{\rho}-\sqrt{2}}\right)  \right] +C.\label{coordinate time}\\
\end{equation}
The following figure depicts the trajectory of the particle. 
\begin{figure}[H]
\centering
\includegraphics[scale=0.38]{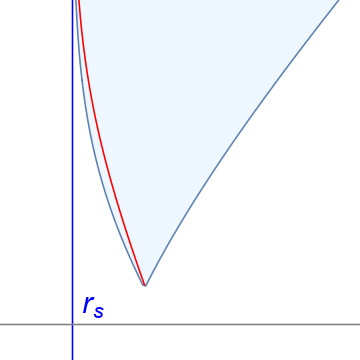}\caption{Worldline of a particle that falls towards the mass.}%
\end{figure}

The red line is the worldline of the particle and the blue region is its 
cronological future, in Schwarzschild coordinates. The path is asymptotic to the line $r=r_s$. 
An observer located very far away from the mass will not see the particle reach the event horizon $r=r_s$.
However, according to equation \eqref{rhotau}, the particle reaches $r=r_s$ for a finite value of the proper time $\tau$.
In fact, it reaches $r=0$ for $\tau=C$. The Schwarzschild coordinates only describe spacetime for $r>r_s$. A more complete 
description of what happens to the particle requires an extension of spacetime that includes the interior of the event horizon. This extension is provided by the Eddington-Finkelstein coordinates, which will be discussed in \S \ref{sedd}.

\subsubsection*{Unbounded orbits}

Radial plunge orbits are examples of unbounded orbits, where the value of $\rho$ is unbounded. These occur when
$E>V$ so that the function $\rho(\tau)$ does not have critical points, and therefore, it is either increasing or decreasing.
In the first case, the particle goes off to infinity. In the second, it falls towards the central mass.
Figure \ref{away} exhibits a potential for this kind of orbit.
\begin{center}
\begin{figure}[h]
\centering
\includegraphics[scale=0.6]{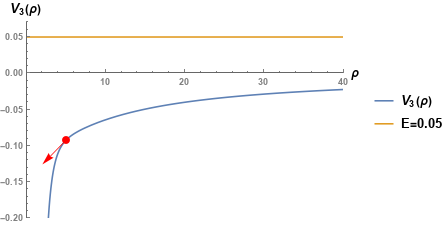}\caption{In this diagram $E>V$ and $\rho'(\tau)<0$, the particle falls towards the mass.}%
\end{figure}
\label{away}
\end{center}

\subsubsection*{Bounded orbits}

\begin{figure}[h]
\centering
\includegraphics[scale=0.6]{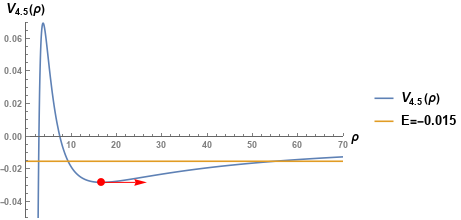}\caption{A bounded orbit. The radius oscillates between a maximum and a minimum.}
\label{periodicalorbit}
\end{figure}

For $h>\sqrt{12}$ there are bounded orbits which are not circles. Figure \ref{periodicalorbit} depicts a potential for this situation.

\section{Precession of Mercury's Perihelion}\label{Mercury}

In Newtonian gravity, the Sun lies on one of the foci of 
the elliptical orbit of each of the planets. The place in the orbit where the distance to the Sun is
minimal is called the Perihelion. The place where the distance is maximal is called the Aphelion. 
In the simplified situation where there are no other planets, the Perihelion
occurs at the same place every year. However, effects such as the presence of other planets cause the Perihelion 
to precess.

\begin{figure}[H]
\centering
\includegraphics[scale=0.6]{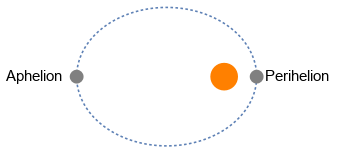}\caption{Perihelion and Aphelion.}
\end{figure}

An anomalous precession of the perihelion
of Mercury had been noticed since 1859. By analyzing observations of transits
of Mercury over the Sun's disk from 1697 to 1848, french astronomer
Urbain Le Verrier showed that the observed rate of precession of Mercury's
perihelion differed from that predicted from Newton's theory by 38" (arc
seconds) per century. This discrepancy was later reestimated at 43". Many ad-hoc
explanations were devised. The existence of another planet, Vulcan, was
postulated. Later, it was suggested that dark dust between the Sun and Mercury
was responsible for this anomaly. None of these hypothesis were consistent
with observations. The phenomenon was explained for the first time by
Einstein, and it was the first empirical evidence of his  theory of
gravitation.  We will next reproduce the relativistic calculation for the precession of Mercury, following \cite{owen}.
For this computation we use units where $G_N=c=1$, so that $r_s=2M$.

\begin{figure}[h]
\centering
\includegraphics[scale=0.5]{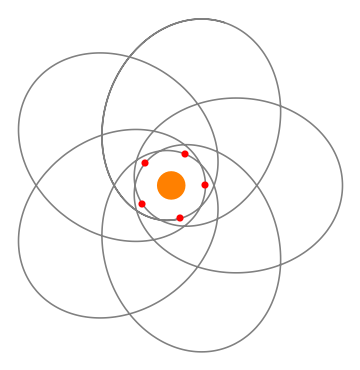}\caption{Precession of an orbit.}
\label{pressorbit}
\end{figure}

Let us define an orbit as the motion between two
successive local minima in the distance to the Sun. The precession of the orbit
is 
\[ \delta\varphi=\Delta\varphi-2\pi,\]
where $\Delta\varphi$ is the angle swept by the orbit. This is depicted in Figure \ref{pressorbit}.

\begin{figure}[H]
\centering
\includegraphics[scale=0.5]{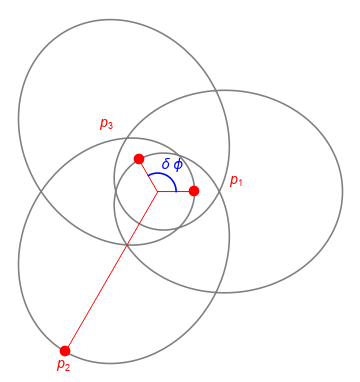}\caption{Precession
of the Perihelion}\label{peri}\end{figure}

\begin{center}
 \begin{figure}[H]
 \centering
\includegraphics[scale=0.6]{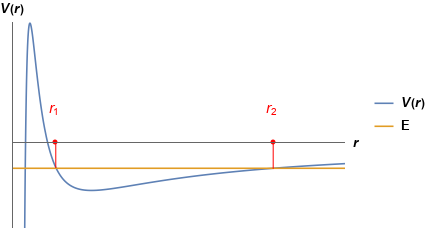}\caption{Shape of the effective potential for Mercury's orbit.
}\label{figurepotM}\end{figure}
\end{center}

We assume that the planet $P$ moves according to an effective potential 
\begin{equation}V(r)=\frac{l^2}{2r^2}- \frac{M}{r}-\frac{ M l^2}{r^3},\end{equation}
which is depicted in Figure \ref{figurepotM}. The points $r_1$ and $r_2$ are called the turning points. They are the places where
$r'(\tau)=0$, the local extrema of the distance to the Sun.

In view of (\ref{lvarphi}), we know that
 \begin{equation}\frac{d\varphi}{d\tau}=\frac{l}{r^{2}},\end{equation} and therefore
 \begin{equation}
\frac{d\varphi}{dr}=\frac{d\varphi}{d\tau} \frac{d\tau}{dr}=\pm\frac{l}{r^2\sqrt{2}\sqrt{E-V(r)}}=\pm \frac{1}{\sqrt{\frac{r^4}{l^2}\big(2E-2V(r)\big)}}.
 \end{equation}
 The expression inside the radical is a polynomial of degree $d=4$:
 \begin{equation}
 P(r)=\frac{r^4}{l^2}\big(2E-2V(r)\big)=\frac{2E}{l^2}r^4+\frac{r_s}{l^2}r^3-r^2+r_s r. \end{equation}
 The turning points $r_{1},r_{2}$ are roots of $E-V(r)$, and therefore of $P(r)$. Also, $r=0$ is a root of $P(r)$. 
We denote by $z$ the remaining root of the polynomial.  We may write
 \[P(r)=\frac{2\mathrm{E}}{l^{2}}r(r-r_{1})(r-r_{2})(r-z).\]
So that
\begin{equation}\label{zeta}z=-r_{1}-r_{2}-\frac{r_{s}}{2E}.\end{equation} 
Notice that $z<r_1$. Otherwise,  $r=0$ and $r=r_1$ would be consecutive roots of $P(r)$, and $r=0$ would be the smallest root. Since
\[ P'(r_1)=\Big(\frac{2r^4}{l^2}(E-V(r))\Big)'(r_1)=-\frac{2r_1^4}{l^2}V'(r_1)>0,\]
then, one would have
\[ P'(0)<0.\]
Since $r=0$ is the smallest root, this would imply
\[ \lim_{r\to -\infty} P(r)=\infty,\]
which is false because $E<0$. We denote by $\varphi
_{1},\varphi_{2},\varphi_{3}$ the angles corresponding to the points $p_{1},p_{2}$
and $p_{3}.$ depicted in Figure \ref{peri}. We want to calculate $\Delta\varphi=\varphi_{3}-\varphi_{1}$. For this, we
write $\Delta\varphi=(\varphi_{2}-\varphi_{1})+(\varphi_{3}-\varphi_{2})$ so that%
\begin{align*}
\Delta\varphi &  =\int_{p_{1}}^{p_{2}}\frac{d\varphi}{dr}dr+\int
\limits_{p_{2}}^{p_{3}}\frac{d\varphi}{dr}dr
 =\int_{r_{1}}^{r_{2}}\frac{+1}{\sqrt{P(r)}}dr+\int_{r_{2}%
}^{r1}\frac{-1}{\sqrt{P(r)}}dr=2\int_{r_{1}}^{r_{2}}\frac{1}%
{\sqrt{P(r)}}dr.
\end{align*}
In order to estimate the later integral we write the integrand as a product of non-negative factors%
\begin{align*}
\frac{1}{\sqrt{P(r)}}  &  =\frac{1}{\sqrt{-\frac{2E}{l^{2}}
r^{2}(r-r_1)(r_{2}-r)(1-\frac{z}{r})}} =\frac{l}{\sqrt{-2E}}\frac{1}{r\sqrt{(r-r_{1})(r_{2}-r)
(1-\frac{z}{r})}}.
\end{align*}
 Since $z<r_1$, we use a linear approximation to obtain
\[
\left( 1-\frac{z}{r}\right)^{-1/2}=1+\frac{z}{2r}+T,
\]
where the error term $T$ can be estimated by using Lagrange's bound as
\begin{equation}
|T|\leq \sup_{r_1<r<r_2}\frac{3}{8}\left(1-\frac{z}{r}\right)^{-5/2}\left(\frac
{z}{r}\right)^{2}\leq\frac{3}{8}\left(1-\frac{z}{r_{2}}\right)^{-5/2}%
\left(\frac{z}{r_{1}}\right)^{2}. \label{error}%
\end{equation}
Using this approximation we write
\[
\frac{1}{\sqrt{P(r)}}=\frac{l}{\sqrt{-2E}}\left(  \frac
{1+T}{r\sqrt{(r-r_{1})(r_{2}-r)}}+\frac{z}{2r^{2}\sqrt
{(r-r_1)(r_{2}-r)}}\right)  .
\]
The integrals of both terms inside the parenthesis can be evaluated in
closed form as
\begin{align*}
\int_{r_{1}}^{r_{2}}\frac{1+T}{r\sqrt{(r-r_{1})(r_{2}-r)}}dr  &
=\frac{\pi(1+T)}{\sqrt{r_{1}r_{2}}},\\
\int_{r_{1}}^{r_{2}}\frac{z}{2r^{2}\sqrt{(r-r_{1})(r_{2}-r
)}}dr  &  =\frac{\pi z}{4\sqrt{r_{1}r_{2}}}\frac{(r_{1}+r_{2})}%
{r_{1}r_{2}}.
\end{align*}
Therefore
\begin{equation}\label{delvar1}
\Delta\varphi=2\int_{r_{1}}^{r_{2}}\frac{1}{\sqrt{P(r)}}dr=\frac{2\pi l 
}{\sqrt{-2E}}\left(  \frac{1+T}{\sqrt{r_{1}r_{2}}}%
+\frac{z}{4\sqrt{r_{1}r_{2}}}\frac{(r_{1}+r_{2})}{r_{1}r_{2}%
}\right)  . 
\end{equation}
We use
the fact that $P(r_{1})=P(r_{2})=0$, to solve for $E$ and $l$ in terms of $r_1$ and $r_2$. This gives
\begin{align}
2E&=\frac{-r_{1}r_{2}r_s+r_s^{2}(r_{1}+r_{2}
)}{r_{1}r_{2}(r_{1}+r_{2}+r_s)-(r_{1}+r_{2})^{2}r_s},\\
l^2&=\frac{r_s r_1^2r_2^2}{r_{1}r_{2}(r_{1}+r_{2}+r_s)-(r_{1}+r_{2})^{2}r_s}.
\end{align}
Hence
\begin{equation}
-\frac{l^2}{2E}=\frac{r_1^2r_2^2}{r_1r_2-r_s(r_1+r_2)}=\frac{r_1r_2}{1-r_s/D},
\end{equation}
where we have put
\begin{equation}
D=\frac{r_1r_2}{r_1+r_2}.
\end{equation}
We can also rewrite (\ref{zeta}) in the form
\begin{equation}z=-r_{1}-r_{2}-\frac{r_{s}}{2E}=\frac{Dr_s}{D-r_s}=\frac{r_s}{1-r_s/D}.\end{equation}
Replacing into (\ref{delvar1}) one obtains
\begin{equation}\label{delvar2}
\Delta\varphi=\frac{2\pi}{\sqrt{1-r_s/D}}\left( 1+T
+\frac{r_s/D}{4(1-r_s/D)}\right) .
\end{equation}
For the planet Mercury, the observed values are $r_{1}=46\times10^{6}\: \mathrm{km}$ and $r_{2}
=69.8\times10^{6}\: \mathrm{km}$ so that $D=27.7\times10^{6}\: \mathrm{km}$. On
the other hand, the Schwarzschild radius of the Sun is approximately $2.95\:\mathrm{km}$. The error bound (\ref{error}) becomes
\[\bigg\vert \frac{2\pi T}{\sqrt{1-r_s/D}}\bigg\vert <10^{-14}.\]
One concludes that \[\Delta
\varphi\approx 2\pi+5.013\times10^{-7},\] so that the precession of its Perihelion is
approximately \[\delta\varphi\approx 5.013\times10^{-7}.\] In one century Mercury
orbits the Sun $415.2$ times that account for a total displacement of its
Perihelion of \[415.2\times\delta\varphi=2081.69\times10^{-7}\text{ radians per century}.\] 
Equivalently,
\[
\frac{360\times3600}{2\pi}\times2081.69\times10^{-7}\approx 43\text{ seconds of
arc per century.}%
\]
According to Blau \cite{Blau}, the observed precession rate in the orbit of Mercury is $5601$ arcseconds per century. 
The plain Newtonian prediction will of course be an elliptical orbit which does not precess. However, a more detailed Newtonian analysis 
that takes into account the gravitational pull of the other planets and the fact that the geocentric coordinate system is not inertial, accounts for a rate of
$5557$ arcseconds per century. That leaves a discrepancy of around $44''$ which is precisely corrected by the relativistic analysis!

\section{Lightlike geodesics in Schwarzschild spacetime}

\begin{figure}[h]
\centering
\includegraphics[scale=0.6]{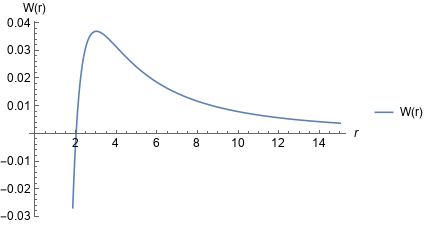}\caption{Effective potential for a massless particle.}%
\label{shapeeffpot}
\end{figure}

In this section we consider the motion of massless particles, such as photons, in Schwarzschild spacetime.
We proceed as in section \S
\ref{nonzero mass in sw}. If $\gamma(\tau)$ is a null geodesic, we set \[e=-\langle \gamma'(\tau),\partial_{t}\rangle,\quad 
l=\langle \gamma'(\tau),\partial_{\varphi}\rangle ,\]
which are conserved quantities
 \begin{align}e&=t'(\tau)\left(1-\frac{r_s}{r(\tau)}\right)c^2\label{ee},\\
 l&=r^{2}(\tau)\sin^{2}\theta(\tau)\varphi'(\tau)\label{ll}.\end{align}
The same argument as in
\S \ref{nonzero mass in sw} shows that the spatial trajectory of a particle $P$ that moves along $\gamma(\tau)$ is contained
in a plane in $\RR^{3}$. We may assume that
this is the equatorial plane $\theta=\pi/2.$ Hence, \begin{equation}l
=r^{2}(\tau)\varphi'(\tau).\end{equation} Since the tangent vector to $\gamma(\tau)$ is null in this case,  equation (\ref{eq material}) becomes
\begin{equation}
-\left(1-\frac{r_s}{r(\tau)}\right)t^{\prime}(\tau)^{2}c^2+\left(1-\frac{r_s}%
{r(\tau)}\right)^{-1}r^{\prime}(\tau)^{2}+r^{2}(\tau)\varphi^{\prime}(\tau)^{2}=0. \label{eq light}%
\end{equation}
Solving for $t'(\tau)$ and $\varphi'(\tau)$ in equations (\ref{ee}) and (\ref{ll}) one obtains
\begin{equation}
-\left(1-\frac{r_s}{r(\tau)}\right)^{-1}\frac{e^2}{c^2}+\left(1-\frac{r_s}%
{r(\tau)}\right)^{-1}r^{\prime}(\tau)^{2}+\frac{l^2}{r^{2}(\tau)}=0, \label{eq light2}
\end{equation}
which is equivalent to
\begin{equation}
\frac{e^2}{c^2}=r^{\prime}(\tau)^{2}+\left(1-\frac{r_s}{r(\tau)}\right)\frac{l^2}{r^{2}(\tau)}. \label{eq light3}%
\end{equation}
 In units where $G_N=c=1$, so that
 \[ W(r)=\left(1-\frac{r_s}{r(\tau)}\right)\frac{l^2}{r^{2}(\tau)}=\left(1-\frac{2M}{r(\tau)}\right)\frac{l^2}{r^{2}(\tau)},\]
equation (\ref{eq light3}) is becomes
\begin{equation}
e^2=r^{\prime}(\tau)^{2}+W(r). \label{eq light4}%
\end{equation}
Therefore, the equation of motion is
\begin{equation}\label{lightmotion}
r'(\tau)=\pm \sqrt{e^2-W(r)}.
\end{equation}
In contrast with the case of massive particles, the effective potential for the massless particle, $W(r)$, has only one critical point, a maximum at $r=3M$.
The only bounded orbits in this case are circular orbits, which are unstable.
Figure \ref{shapeeffpot} shows the shape of the potential.

\subsection*{The circular orbit}

\begin{figure}[H]
\centering
\includegraphics[scale=0.6]{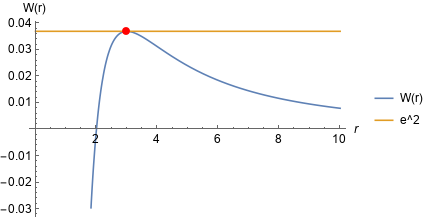}\caption{Circular orbit for a massless particle.}%
\label{lightcir}
\end{figure}

For the radius to be a constant $r(\tau)=r_0$, it is necessary that $e^2=W(r_0)$. Differentiating the equation of motion (\ref{lightmotion}), one obtains
\[ 0=r''(\tau)=\pm \frac{1}{2}\frac{dW}{dr}(r_0),\]
so that $r_0$ is the only critical point of $W(r)$, which is $r_0=3M$.
Notice that there is only one value of $r$, independent of $e$ and $l$, for which it is possible to have a circular orbit.
Figure \ref{lightcir} shows the configuration that corresponds to it.

\subsection*{Absorbing and escaping orbits}

\begin{figure}[H]
\centering
\includegraphics[scale=0.6]{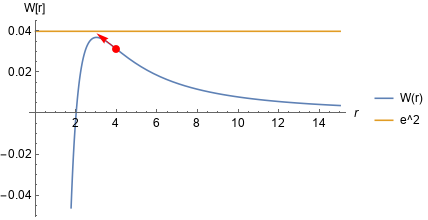}\caption{Absorption of a photon.}%
\label{abslight}
\end{figure}

The effective potential has an absolute maximum at $r_0=3M$ which is
\[ W(3M)=\frac{l^2}{27 M^2}.\]
Therefore, if $e^2>l^2/27 M^2$, then $e^2-W(r)>0$, and therefore $r'(\tau)\neq 0$. Depending on the sign in (\ref{lightmotion}), either the radius is always increasing or always decreasing. In the first case, the particle is absorbed by the mass. In the second, it scapes away from the mass. 
Figure \ref{abslight} shows the potential that corresponds to the absorption of a particle.

\begin{figure}[h]
\centering
\includegraphics[scale=0.5]{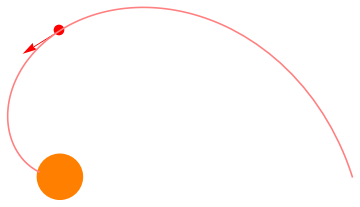}\caption{Absorption
of a photon.}%
\label{fotonab}
\end{figure}

Figure \ref{fotonab} describes the orbit of a photon that is absorbed by a star.

\subsection*{Scattering orbits and reabsorption}

\begin{figure}[H]
\centering
\includegraphics[scale=0.6]{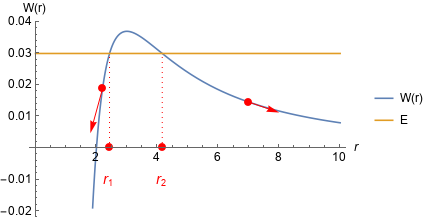}\caption{If $r(0)<r_1$ then the photon can't escape the star and falls back to it. If $r(0)>r_2$ the photon is scattered.}%
\label{twoposs}
\end{figure}

In case $e^2<l^2/27 M^2$, the equation $W(r)=e^2$ has two solutions, the turning points $r_1<r_2$. In this situation, the radius $r(\tau)$ stays
away from the interval $(r_1,r_2)$. There are two possibilities where $r(\tau)$ has a critical point. If $r(0)>r_2$ then the particle approaches the star, turns around and goes away. If $r(0)<r_1$, the particle starts going away from the mass but lacks sufficient speed to escape, and ends up falling back to the star.
The two possibilities are depicted in Figure \ref{twoposs}.

\begin{figure}[H]
\centering
\includegraphics[scale=0.5]{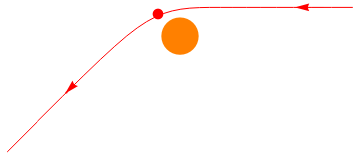}\caption{scattering
orbit}%
\label{fotonscat}
\end{figure}

Figure \ref{fotonscat} describes a scattering orbit.

\section{Gravitational bending of light}\label{bending}

One of the early tests of general relativity was the bending
of light rays caused by the Sun. The first observation of light deflection was performed by Arthur
Eddington and his collaborators during the total solar eclipse of May 29,
1919. Eddington travelled to the island of Pr\'incipe, off the coast of Equatorial Guinea, in West Africa. Another group was sent to Sobral, in Brazil.
Despite unfortunate weather conditions, Eddington was able to take photographs which showed changes in the positions
of the stars that agreed with Einstein's prediction.  The results were reported back to the Royal Society in England, and received with great enthusiasm.
However, some argued that the results had been plagued by systematic errors and confirmation bias.
The validity of Eddington's observations remains a subject of disputes,
although more recent analysis of the data support their accuracy (Ball \cite{wiki2}). Modern experiments have confirmed the relativistic predictions to much higher precision ( Shapiro \emph{et. al} \cite{Shapiro}). In this section we present the calculations that describe the bending of light predicted by general relativity. Again, we use units where $G_N=c=1$.

\begin{figure}[h]
\centering
\includegraphics[scale=0.6]{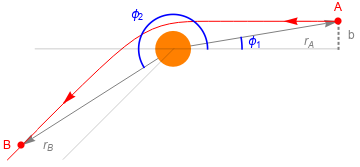}\caption{A
photon approaching a star.}%
\label{impact}
\end{figure}

Consider a photon that is coming from very far away, approaches a star and is deflected by it. The situation is depicted in Figure \ref{impact}.

The quantity $b$ is known as the impact parameter. It can be computed in terms of the other parameters of the orbit as follows.
We suppose that $r=r_A\gg0$ is very large so that $\varphi\approx \sin\varphi$ and $dr/dt\approx-1$. Then
\[ b\approx r\sin\varphi \approx r\varphi ,\]
and therefore
\[ r^2\frac{d\varphi}{dt}=r^2\frac{d}{dt}\left(\frac{b}{r}\right)\approx b.\]
On the other hand
\[ \frac{l}{e}=\frac{r^2\varphi'(\tau)}{t'(\tau)(1-2M/r)}=\frac{r^2 \varphi'(t)}{1-2M/r}=r^2 \varphi'(t).\]
One concludes that
\begin{equation}
b\approx \frac{l}{e}.
\end{equation}
In terms of the impact parameter, equation of motion (\ref{lightmotion}) becomes
\begin{equation}\label{lightmotionb}
r'(\tau)=\pm l \sqrt{\frac{1}{b^2}-\frac{W(r)}{l^2}},
\end{equation}
where the sign is negative before the turning point and positive after it.
On the other hand, since $l=r^{2}\varphi'(\tau)$, one has
\begin{equation}
\frac{d\varphi}{dr}=\frac{d\varphi/d\tau}{dr/d
\tau}=\pm\frac{1}{r^{2}}\left(  \frac
{1}{b^{2}}-\frac{W(r)}{l^2}\right)  ^{-1/2}. \label{rt}%
\end{equation}
Recall that $r_2$ is the minimum of the function $r(\tau)$, and assume that it occurs at $\tau=0$. 
If one sets
\[ \Delta\varphi(\tau)=\varphi(\tau)-\varphi(-\tau),\]
then, the deflection
angle is  \[\delta \varphi=\varphi_2-\varphi_1=\Delta \varphi-\pi,\]
where
$$
\Delta \varphi = \lim_{\tau\to \infty} \Delta\varphi(\tau)
$$
We now compute
\begin{align*}
\Delta\varphi(\tau) &  =\int_{r(-\tau)}^{r(\tau)}\frac{d\varphi}{dr} dr=\int\limits^{r(-\tau)
}_{r_{2}}\frac{1}{r^{2}}\left(  \frac{1}{b^2}-\frac{W(r)}{l^2}\right)
^{-1/2}dr+\int_{r_{2}}^{r(\tau)}\frac{1}{r^{2}}\left(  \frac
{1}{b^2}-\frac{W(r)}{l^2}\right)  ^{-1/2}dr.
\end{align*}
Taking the limit when $\tau \to \infty$ one obtains
\begin{align}
\Delta\varphi=\lim_{\tau \to \infty}\Delta\varphi(\tau) =2\int\limits^{\infty
}_{r_{2}}\frac{1}{r^{2}}\left(  \frac{1}{b^2}-\frac{W(r)}{l^2}\right)
^{-1/2}dr.
\end{align}
We change the variable of integration by letting $r=r_2/u$, so that
$dr=-(r_2/u^{2})du$. Then
\begin{align}\label{deltav}
\Delta\varphi =\frac{2}{r_2}\int_{0}^{1}\Big(  \frac{1}{b^2}-\frac{u^{2}}{r^2_2}\big(1-\frac{2Mu}{r_2
}\big)\Big)  ^{-1/2}du.
\end{align}
Since $r_2$ is a turning point, it satisfies
\[ \frac{1}{b^2}=\left(1-\frac{2M}{r_2}\right)\frac{1}{r_2^2}\]
So that equation (\ref{deltav}) becomes
 \begin{align}\label{deltavv}
\Delta\varphi =2\int_{0}^{1}\left[\left(1-\frac{2M}{r_2}\right)-u^2\left(1-\frac{2Mu}{r_2}\right)\right]^{-1/2}du.
\end{align}
Consider the function
\[ f(x)=  2\int_{0}^{1}\left[(1-2x)-u^2(1-2ux)\right]^{-1/2}du= 2\int_{0}^{1}\left(1-2x -u^2+2u^3 x\right)^{-1/2}du.\]
One is interested in computing $\Delta\varphi=f(M/r_2)$.
Since the minimum radius $r_2$ is greater than the radius of the Sun, which is $R=6.9 \times 10^5 \: \mathrm{km}$ and $M=r_s/2$,
one has
\[\frac{M}{r_2}\leq \frac{r_s}{2R} \leq \frac{2.95}{2 \times 6.9}\times 10^{-5}\approx 2.1 \times 10^{-6}.\]
Therefore, we can use a linear approximation
\[ f\left(\frac{M}{r_2}\right)\approx f(0)+ f'(0)\frac{M}{r_2}.\]
Moreover
\[ f(0)=2\int_{0}^{1}\left(1 -u^2\right)^{-1/2}du=2[\arcsin(1)-\arcsin(0)]=\pi,\]
and
\[ f'(0)=2\int_{0}^{1}\left(1-u^3\right)\left(1 -u^2\right)  ^{-3/2}du=2\left[(-2-u)\frac{\sqrt{1-u}}{\sqrt{1+u}}\right]\bigg \vert^1_0=4.\]
One concludes that
\begin{equation} \delta\varphi=\Delta\varphi-\pi \approx \frac{4M}{r_2}\end{equation}
Since $b\approx R$, it is also true that 
\[\delta\varphi\approx \frac{4M}{b}.\] For the Sun,  $R=6.9 \times 10^5 $ and $M=r_s/2\approx 2.95/2$. So that
 \[ \delta\varphi\approx \frac{4M}{R}\approx \frac{2r_s}{R}\approx \frac{5.9}{6.9} \times 10^{-5}\approx 0.85 \times 10^{-5} \:\mathrm{radians}.\]
 Measured in seconds of arc, this becomes
  \[ \delta\varphi \approx  0.85 \times 10^{-5} \times \frac{360 \times 3600}{2\pi}\approx 1.75 \:\text{seconds of arc}.\]

\section{Conformal maps and Carter-Penrose diagrams}

\begin{figure}[H]
\begin{center}
  \includegraphics[scale=0.38]{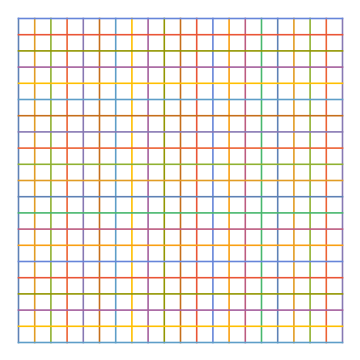}  \quad 
  \includegraphics[scale=0.38]{Figures/flor}
\end{center}
\caption{The holomorphic map $z \mapsto \frac{z}{z-1}$, all lines meet at straight angles.}
\label{flower}
\end{figure}

Conformal geometry is the part of geometry that depends on angles but not on distances.
Let $(M,g)$ and $(N,h)$ be pseudo-Riemannian manifolds. A conformal map is a diffeomorphism
$f: M \rightarrow N$ 
such that
\[ f^*h=e^{\alpha} g,\]
for some smooth function $\alpha$ on $M$. The derivative $Df(p): T_pM \rightarrow T_qN$ of a conformal map preserves angles. Let $v,w \in T_pM$ be non-zero vectors, and denote by $\theta$ the angle between them. If $\theta'$ is the angle between $Df(p)(v)$ and $Df(p)(w)$, then
\[ \cos\theta=\frac{\langle v,w\rangle }{|v||w|}=\frac{\langle Df(p)(v),Df(p)(w)\rangle }{|Df(p)(v)||Df(p)(w)|}=\cos\theta',\]
so the angle between tangent vectors is preserved.
Holomorphic diffeomorphisms are conformal maps. Figure \ref{flower} shows an example where the grid on the left is sent to the lines on the right. Note that in both pictures, all lines intersect orthogonally.
Clearly, the inverse of a conformal map is also conformal.
We say that two manifolds are conformally equivalent if there is a conformal map between them. This is an equivalence relation.
A conformal map between Lorentzian manifolds sends light cones to light cones, therefore, it preserves the causal structure.
In general, it is not the case that a conformal map sends geodesics to geodesics. However, up to reparametrization, lightlike geodesics are preserved by conformal maps, as the following result shows.

\begin{lemma}
Let $f: M \rightarrow N$  be a conformal diffeomorphism between Lorentzian manifolds $(M,g)$ and $(N,h)$. If $\gamma: I \rightarrow M$ is a lightlike geodesic,
then, the curve $f\circ \gamma$ can be reparametrized so that it becomes a lightlike geodesic on $N$.
\end{lemma}
\begin{proof}
Without loss of generality, we may assume that $M=N$, $f=\mathrm{id}_M$ and $g=e^\alpha h$. We denote by $\Gamma^a_{bc}$ and $\tilde{\Gamma}^a_{bc}$ the Christoffel symbols for the metrics $g$ and $h$, respectively. Then, as we know
\begin{equation}
\label{Chris}\tilde{\Gamma}_{bc}^{a}=\frac
{1}{2} \sum_{k} h^{ka}\left(\frac{\partial h_{ck}}{\partial x^{b}}%
+\frac{\partial h_{kb}}{\partial x^{c}}-\frac{\partial h_{bc}}{\partial x^{k}%
}\right).
\end{equation}
Also,
\begin{align*}
\Gamma_{bc}^{a}&=\frac
{1}{2} \sum_{k} g^{ka}\left(\frac{\partial g_{ck}}{\partial x^{b}}%
+\frac{\partial g_{kb}}{\partial x^{c}}-\frac{\partial g_{bc}}{\partial x^{k}%
}\right)\\
&=\frac
{1}{2} \sum_{k}e^{-\alpha} h^{ka}\left[\frac{\partial (e^\alpha h_{ck})}{\partial x^{b}}%
+\frac{\partial (e^\alpha h_{kb})}{\partial x^{c}}-\frac{\partial (e^\alpha h_{bc})}{\partial x^{k}%
}\right]\\
&=\tilde{\Gamma}^a_{bc}+\frac
{1}{2} \sum_{k}h^{ka}\left(\frac{\partial \alpha}{\partial x^b} h_{ck}
+ \frac{\partial \alpha}{\partial x^c}h_{kb}- \frac{\partial \alpha}{\partial x^k}h_{bc}\right).\\
\end{align*}
Since $\gamma(\tau)$ is a geodesic with respect to $g$, it satisfies the equations
\begin{equation}
\frac{d^2\gamma^{a}}{d \tau^2 }+\sum_{b,c}\Gamma_{bc}^{a}\frac{d\gamma^b}{d\tau}\frac{d\gamma^{c}}{d\tau}
=0,
\end{equation}
which are equivalent to
\begin{equation}\label{lightgeo}
\frac{d^2\gamma^{a}}{d \tau^2 }+\sum_{b,c}\tilde{\Gamma}_{bc}^{a}\frac{d\gamma^b}{d\tau}\frac{d\gamma^{c}}{d\tau}
=-\frac{1}{2}\sum_{b,c,k}h^{ka}\left(\frac{\partial \alpha}{\partial x^b} h_{ck}
+ \frac{\partial \alpha}{\partial x^c}h_{kb}- \frac{\partial \alpha}{\partial x^k}h_{bc}\right)\frac{d\gamma^b}{d\tau}\frac{d\gamma^{c}}{d\tau}.
\end{equation}
Since $\gamma(\tau)$ is lightlike, then
\begin{equation}
\sum_{b,c}h_{bc}\frac{d\gamma^b}{d\tau}\frac{d\gamma^{c}}{d\tau}=0,
\end{equation}
so that (\ref{lightgeo}) becomes
\begin{equation}\label{lightgeo2}
\frac{d^2\gamma^{a}}{d \tau^2 }+\sum_{b,c}\tilde{\Gamma}_{bc}^{a}\frac{d\gamma^b}{d\tau}\frac{d\gamma^{c}}{d\tau}
=-\frac{1}{2}\sum_{b,c,k}\left(\frac{\partial \alpha}{\partial x^b}h^{ka} h_{ck}
+ \frac{\partial \alpha}{\partial x^c}h_{kb}\right)\frac{d\gamma^b}{d\tau}\frac{d\gamma^{c}}{d\tau}.
\end{equation}
The right hand side can be computed as follows
\begin{align*}
-\frac{1}{2}\sum_{b,c,k}\frac{d\gamma^b}{d\tau}\frac{d\gamma^{c}}{d\tau}h^{ka}\left(\frac{\partial \alpha}{\partial x^b} h_{ck}
+ \frac{\partial \alpha}{\partial x^c}h_{kb}\right)&=-\sum_{b,c,k}\frac{d\gamma^b}{d\tau}\frac{d\gamma^{c}}{d\tau}\frac{\partial \alpha}{\partial x^b} h^{ka}h_{ck}\\
&=-\sum_{b,k}\frac{d\gamma^b}{d\tau}\frac{d\gamma^{a}}{d\tau}\frac{\partial \alpha}{\partial x^b} 
\\
&=-\frac{d\gamma^{a}}{d\tau}\frac{d\alpha}{d\tau}.
\end{align*}
One concludes that $\gamma(\tau)$ satisfies
\begin{equation}\label{lightgeo3}
\frac{d^2\gamma^{a}}{d \tau^2 }+\sum_{b,c}\tilde{\Gamma}_{bc}^{a}\frac{d\gamma^b}{d\tau}\frac{d\gamma^{c}}{d\tau}
=-\frac{d\gamma^{a}}{d\tau}\frac{d\alpha}{d\tau}.
\end{equation}
Consider a reparametrization $\beta(s)=\gamma(\tau(s))$ of the curve. The condition for $\beta(s)$ to be a geodesic with respect to $h$ is
\begin{equation}\label{lightgeo5}
\frac{d^2\beta^{a}}{d s^2 }+\sum_{b,c}\tilde{\Gamma}_{bc}^{a}\frac{d\beta^b}{ds}\frac{d\beta^{c}}{ds}
=0.
\end{equation}
This is equivalent to
\begin{equation}\label{lightgeo6}
\frac{d^2 \gamma^a}{d\tau^2}\left(\frac{d\tau}{ds}\right)^2+\frac{d\gamma^a}{d\tau}\frac{d^2\tau}{ds^2}+\left(\frac{d\tau}{ds}\right)^2\sum_{b,c}\tilde{\Gamma}_{bc}^{a}\frac{d\gamma^b}{d\tau}\frac{d\gamma^{c}}{d\tau}
=0.
\end{equation}
Using (\ref{lightgeo3}), this becomes
\begin{equation}\label{lightgeo4}
\left(\frac{d\tau}{ds}\right)^2\frac{d\gamma^{a}}{d\tau}\left(\frac{d^2\tau/ds^2}{(d\tau/ds)^2}-\frac{d\alpha}{d\tau}\right)=0.
\end{equation}
Which is satisfied as long as
\begin{equation}
\frac{\tau''(s)}{\tau'(s)^2}=\alpha'(\tau).
\end{equation}
If $(a,b)$ is the domain of $\gamma(\tau)$, we set
\[ s=\int_a^\tau e^{-\alpha (t)}dt,\]
so that $ds/d\tau=e^{-\alpha(\tau)}$
and $d\tau/ds=e^{\alpha(s)}
$. Therefore
\begin{equation}
\frac{\tau''(s)}{\tau'(s)^2}=e^{2\alpha(s)\alpha'(\tau)}{e^{2\alpha(s)}}=\alpha'(\tau).
\end{equation}
One concludes that $\beta(s)=\gamma(\tau(s))$ is a geodesic with respect to $h$.
\end{proof}

\begin{figure}[H]
\centering
\includegraphics[scale=0.38]{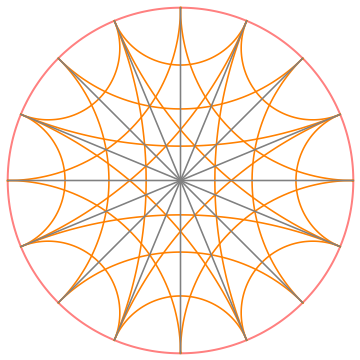}\caption{Geodesics in Poincar\'e's model for the hyperbolic plane.}
\label{geodisk}
\end{figure}

The condition that the geodesics are lightlike is essential in the lemma above. Consider for example Poincar\'e's disk model for the hyperbolic plane. This is the disk in the plane, with metric
\[ h =\frac{4}{(1-x^2-y^2)^2} g,\]
where $g=dx \otimes dx +dy \otimes dy$ is the usual Eucliean metric. Thus, the identity map is a conformal map from the hyperbolic disk to the Euclidean disk. Geodesics for the hyperbolic metric are arcs of circle which are orthogonal to the boundary, and diameters. These arcs cannot be reparametrized to become euclidean geodesics. Figure \ref{geodisk} illustrates the situation.

 Conformal maps provide a tool for representing the causal structure of spacetimes, known as Carter-Penrose diagrams. The idea is quite simple: if one is interested in the causal structure  of a spacetime $X$, it is sometimes convenient to instead describe the structure of another spacetime $X'$, conformal to $X$, where some interesting features are more transparent.
Carter-Penrose diagrams are decorated with symbols that describe the properties of different regions. We will use the following conventions.

\begin{center}
\begin{tabular}{|l|}
\hline \\
{\textbf  \,\,\,\,\,\,\,\,\,\,\,\,\,\,\,\,\,\,\,\,\,\,\,\,\,\,Conventions for Carter-Penrose diagrams} \\  
\\
\hline  \\
$i^+$  denotes a future timelike infinity, where timelike trajectories go. \\
\\
$i^-$  denotes a past timelike infinity, where timelike trajectories come from. \\
\\
$I^+$ denotes a future lightlike infinity, where light goes.\\
\\
$I^-$ denotes a past lightlike infinity, where light comes from. \\
\\
$i^0$ denotes a spacelike infinity. \\
\\
Light travels on straight lines of slope $\pm1$.\\
\\
\hline
\end{tabular}
\end{center}

\subsubsection*{Carter-Penrose diagram for 2d Minkowski spacetime}

\begin{figure}[H]
\centering
\includegraphics[scale=0.45]{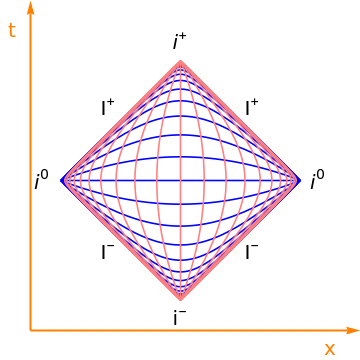}\caption{Carter-Penrose diagram of 2d Minkowski spacetime. The blue lines correspond to constant values of $t$ and the red lines correspond to constant values of $x$.}%
\label{CP2dM}
\end{figure}

Consider the 2-dimensional Minkowski spacetime $\mathbb{M}_2$, with metric
\begin{equation}g= -dt \otimes dt + dx \otimes dx.\end{equation}
Let $\mathbb{M}_2'$ be the region in the $T,R$ plane determined by $|R+T|<2$ and $|R-T|<2$, with metric
\begin{equation} h=-dT \otimes dT + dR \otimes dR.\end{equation}
There is a diffeomorphism $\phi: \mathbb{M}_2 \to \mathbb{M}'_2$
given by
\begin{align}\label{TR}
\begin{split}
T&=\tanh(t-x)+\tanh(t+x) \\ R&=\tanh(t+x)-\tanh(t-x).
\end{split}
\end{align}
One can verify directly that
\[\phi^*h=4\sech(t-x)^2\sech(t+x)^2 g,\]
so $\phi$ is a conformal map. We conclude that $\mathbb{M}_2$ and $\mathbb{M}'_2$ have the same causal structure.
Figure \ref{CP2dM} represents the Carter-Penrose diagram for 2d Minkowski spacetime that arises from this identification.
Let $\gamma(\tau)=(\tau,a \tau )$ be a timelike geodesic on Minkowski spacetime. Since the curve is timelike, $|a|<1$ and therefore
\begin{align*} &\lim_{\tau\to \infty} \phi (\gamma(\tau))\\
&\quad =\lim_{\tau\to \infty}\left(\tanh\big((1-a)\tau\big)+\tanh\big((1+a)\tau\big),\tanh\big((1+a)\tau\big)-\tanh\big((1-a)\tau\big)  \right) \\&\quad =(2,0)\\
&\quad =i^+,\end{align*}
and
\begin{align*} &\lim_{\tau\to -\infty} \phi (\gamma(\tau)) \\
&\quad=\lim_{\tau\to - \infty}\left(\tanh\big((1-a)\tau\big)+\tanh\big((1+a)\tau\big),\tanh\big((1+a)\tau\big)-\tanh\big((1-a)\tau\big)  \right) \\&\quad =(-2,0) \\
&\quad =i^-.\end{align*}
We conclude that timelike geodesics start at $i^-$ and go to $i^+$, as described in the diagram.
For a spacelike geodesic $\gamma(\tau)=(a\tau,\tau),$ with $|a|<1$ one has
\begin{align*} &\lim_{\tau\to \infty} \phi (\gamma(\tau))\\
&\quad =\lim_{\tau\to \infty}\left(\tanh\big((a-1)\tau\big)+\tanh\big((1+a)\tau\big),\tanh\big((1+a)\tau\big)-\tanh\big((a-1)\tau\big)  \right) \\&\quad =(0,2) \\
&\quad =i^0\end{align*}
and
\begin{align*} &\lim_{\tau\to - \infty} \phi (\gamma(\tau)) \\&\quad =\lim_{\tau\to -\infty}\left(\tanh\big((a-1)\tau\big)+\tanh\big((1+a)\tau\big),\tanh\big((1+a)\tau\big)-\tanh\big((a-1)\tau\big)  \right) \\&\quad =(0,-2) \\
&\quad =i^0.\end{align*}
This shows that spacelike geodesics start and end at $i^0$.
Consider also a lightlike geodesic $\gamma(\tau)=(\tau,\tau)$. Then
\begin{align*} \lim_{\tau\to \infty} \phi (\gamma(\tau))&=\lim_{\tau\to \infty}\left(\tanh\big(2\tau\big),\tanh\big(2\tau\big)  \right) =(1,1)\in I^+,\end{align*}
and
\begin{align*} \lim_{\tau\to- \infty} \phi (\gamma(\tau))&=\lim_{\tau\to -\infty}\left(\tanh\big(2\tau\big),\tanh\big(2\tau\big)  \right) =(-1,-1)\in I^-.\end{align*}
Again, as described by the diagram, lightlight geodesics go from $I^-$ to $I^+$.

\subsubsection*{Carter-Penrose diagram for radial Minkowski spacetime}

In polar coordinates, the $4$-dimensional Minkowski metric is
\begin{equation}
g=-dt \otimes dt + dr \otimes dr + r^2 d\Omega.
\end{equation}
On the surface determined by $\theta=\theta_0$ and $\varphi=\varphi_0$, the metric is restricts to
\begin{equation}
g=-dt \otimes dt + dr \otimes dr,
\end{equation}
which is the same as for 2d Minkowski spacetime, except that now $r$ takes only positive values. We call this spacetime the radial Minkowski spacetime. Using again transformation (\ref{TR}) one obtains the following diagram:
\begin{figure}[H]
\centering
\includegraphics[scale=0.45]{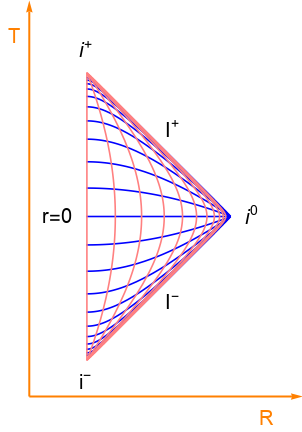}\caption{Carter-Penrose diagram of radial Minkowski spacetime. The blue lines correspond to constant values of $t$ and the red lines correspond to constant values of $r$. Each point in the interior of the triangle represents a sphere.
The segment $r=0$ corresponds to the origin in Minkowski spacetime.}
\end{figure}

\subsection*{Carter-Penrose diagram for Schwarzschild spacetime}
We consider the Schwarzschild spacetime $\mathscr{S}$ in units where $c=G_N=1$, so that $r_s=2M$.
The surface $\mathscr{S}_0$ determined by $\theta=\theta_0$ and $\varphi=\varphi_0$ has the induced metric
\[ g=-\left(1-\frac{r_s}{r}\right)dt \otimes dt +\left(1-\frac{r_s}{r}\right)^{-1} dr \otimes dr.\]
If one defines the tortoise coordinate
\begin{equation}
r^*=r+ r_s\log\left(\frac{r}{r_s}-1\right),
\end{equation}
then
\begin{equation}
dr^*=\left(1-\frac{r_s}{r}\right)^{-1}dr.
\end{equation}
Therefore
\begin{equation}
g=\left(1-\frac{r_s}{r}\right)(-dt \otimes dt + dr^* \otimes dr^*).
\end{equation}
One concludes that the map  $\psi:\mathscr{S}_0 \to \mathbb{M}_2$ given by
$$
\psi(t,r) = (t,r^*),
$$
is a conformal equivalence. Composing with the map $\phi$ defined by (\ref{TR}) one obtains $\xi=\phi \circ \psi:\mathscr{S}_0 \to \mathbb{M}'_2$ which is given by
\begin{align}\label{TRS}
\begin{split}
T&=\tanh(t-r^*)+\tanh(t+r^*),\\
R&=\tanh(t+r^*)-\tanh(t-r^*).\end{split}
\end{align}
This gives the Carter-Penrose diagram for Schwarzschild spacetime.

\begin{figure}[H]
\centering
\includegraphics[scale=0.45]{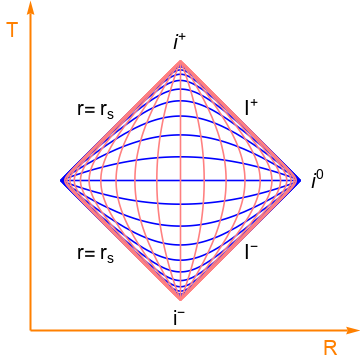}\caption{Carter-Penrose diagram of Schwarzschild spacetime.}
\end{figure}

\section{Incoming Eddington-Finkelstein and black holes}\label{sedd}

The Schwarzschild metric
\begin{equation*}
g=-\left( 1-\frac{r_s}{r}\right)  dt\otimes dt+\left(  1-\frac{r_s}{r}\right)^{-1}dr\otimes dr+r^{2}(d\theta\otimes d\theta
+\sin^{2}\theta\text{ }d\varphi\otimes d\varphi)
\end{equation*}
is not defined at $r=r_s$ because the factor $\left(1-r_s/r\right)^{-1}$ blows up. For this reason, we have so far only considered the region $r>r_s$. It is possible to change the coordinates in Schwarzschild spacetime in such a way that the new coordinates cover a larger region and, in this way, embed the Schwarzschild patch into a larger spacetime.

\begin{figure}[H]
\centering
\includegraphics[scale=0.45
]{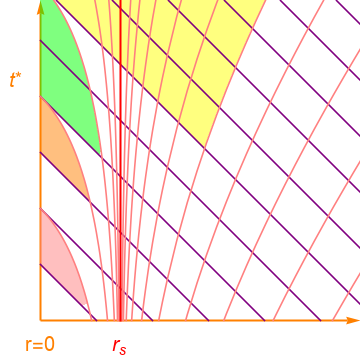}\caption{Incoming Eddington-Finkelstein coordinates. The purple lines are incoming light rays. The pink lines are outgoing light rays. The red line represents the submanifold $r=r_s$. The colored regions represent the chronological futures of particular events.}%
\label{IEF}
\end{figure}

Consider the function $v=t+r^*$ where, as before, $r^*$ is the tortoise coordinate
\begin{equation}
r^*=r+r_s \log \left(\frac{r}{r_s}-1\right).
\end{equation}
In the coordinates $(v,r,\theta,\varphi)$ the Schwarzschild metric takes the form
\begin{equation}\label{EFM} g=-\left( 1-\frac{r_s}{r}\right)  dv\otimes dv+dv\otimes dr +dr \otimes dv+r^{2}(d\theta\otimes d\theta
+\sin^{2}\theta\text{ }d\varphi\otimes d\varphi).\end{equation}
This metric has no singularities. 
Moreover, it has determinant
\[
\det g= \begin{pmatrix}
-\big(1-\frac{r_s}{r}\big)& 1&0&0\\
1&0&0&0\\
0&0&r^2&0\\
0&0&0&r^2 \sin^2(\theta)\\
\end{pmatrix}=-r^4\sin(\theta)<0,
\]
so that it defines a Lorentzian metric everywhere. We will denote by $\mathscr{E}$ the incoming Eddington-Finkelstein spacetime, which is the region $v \in \RR$, $r>0$, $\theta \in (0,\pi)$ and $\varphi \in (0,2\pi)$ with metric given by (\ref{EFM}).
The map $\iota :\mathscr{S} \to \mathscr{E}$ defined by 
$$\iota(t,r,\theta,\varphi)= (t+r^*,r,\theta,\varphi)$$
is an isometric embedding whose image is the region with $r>r_s$. Notice that the coefficients of the metric are analytic functions and therefore, those of the Ricci tensor are too. Since the Ricci tensor vanishes on the Schwarzschild patch, one concludes that $\mathscr{E}$ is Ricci flat. On the radial surface $\mathscr{E}_0$ determined by $\theta=\theta_0$ and $\varphi=\varphi_0$ the metric is
\begin{equation} g=-\left( 1-\frac{r_s}{r}\right)  dv\otimes dv+dv\otimes dr +dr \otimes dv.\end{equation}
A lightlike curve $\gamma(\tau)$ satisfies the equation
\[0= \langle \gamma'(\tau),\gamma'(\tau)\rangle=-\left( 1-\frac{r_s}{r}\right)  v'(\tau)^2+2 v'(\tau) r'(\tau).\]
There are two possibilities
\[ v'(\tau)=0 \quad \text{or} \quad v'(\tau)=\frac{2r'(\tau)r(\tau)}{r(\tau)-r_s}=2r'(\tau)+\frac{2r'(\tau)r_s}{r-r_s}.\]
Integrating on both sides one obtains
\begin{equation}
v(\tau)=\text{const.}\quad \text{or} \quad  v(\tau)=2r(\tau)+2r_s \log \left|\frac{r(\tau)}{r_s}-1\right|+\text{const.}
\end{equation}
We define the new coordinate $t^*$ by $t^*=r^*+t-r$, so that incoming light rays become straight lines of slope $-1$.
Figure \ref{IEF} shows the lightlike trajectories on the incoming Eddington-Finkelstein spacetime.

The diagram above is known as a Finkelstein diagram. It exhibits some of the basic properties of the incoming Eddington-Finkelstein spacetime, which are listed below.


\begin{itemize}

\item The chronological future of an event for which $r<r_s$ is contained in the region $r<r_s$. Moreover,
any timelike curve that starts with $r<r_s$ tends to the singularity $r=r_0$. This means that not even light can scape
the region $r<r_s$, which is called the interior of the black hole. Any object in the interior of the black hole is destined to collapse towards the singularity. The manifold $r=r_s$ is called the event horizon of the black hole. 

\item An observer, Alice, that remains outside of the event horizon has no access to what 
happens inside the black hole. According to equation (\ref{coordinate time}), if she throws a ball towards the mass, Alice will never see it cross the event horizon. The ball does not reach $r=r_s$ in a finite amount of Alice's coordinate time.
%


\end{itemize}
Even though black holes cannot be observed directly, their existence can be inferred from their gravitational effects on visible matter.
There is evidence for thousands of black holes at the center of our galaxy, the Milky Way.

\section{Outgoing Eddington-Finkelstein and white holes}

\begin{figure}[H]
\centering
\includegraphics[scale=0.45
]{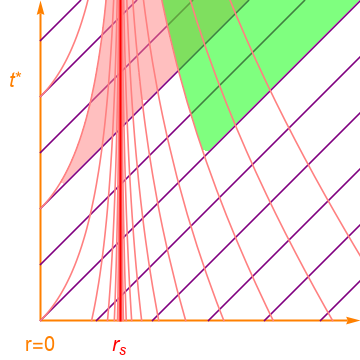}\caption{Outgoing Eddington-Finkelstein coordinates. The purple lines are outgoing light rays. The pink lines are incoming light rays. The red line represents the submanifold $r=r_s$. The colored regions represent the chronological futures of particular events.}
\label{OEF}
\end{figure}

The incoming Eddington-Finkelstein coordinates  were chosen so that incoming light rays become straight lines. It is also possible to choose
coordinates in Schwarzschild spacetime so that outgoing light rays become straight.
Consider the function $u=t-r^*$. In the coordinates $(u,r,\theta,\varphi)$ the Schwarzschild metric takes the form
\begin{equation} g=-\left( 1-\frac{r_s}{r}\right)  dv\otimes dv-dv\otimes dr -dr \otimes dv+r^{2}(d\theta\otimes d\theta
+\sin^{2}\theta\text{ }d\varphi\otimes d\varphi).\end{equation}
This metric has no singularities. 
As in the incoming case, the determinant of the metric is\[
\det g= \begin{pmatrix}
-\big(1-\frac{r_s}{r}\big)& -1&0&0\\
-1&0&0&0\\
0&0&r^2&0\\
0&0&0&r^2 \sin^2\theta\\
\end{pmatrix}=-r^4\sin\theta<0,
\]
so that it defines a Lorentzian metric everywhere. We will denote by $\mathscr{E}^*$ the outgoing Eddington-Finkelstein spacetime, which is the region $u \in \RR$, $r>0$, $\theta \in (0,\pi)$ and $\varphi \in (0,2\pi)$ with metric given by (\ref{EFM}).
The map $\iota :\mathscr{S} \to \mathscr{E}^*$ given by
\[ \iota(t,r,\theta,\varphi)= (t-r^*,r,\theta,\varphi)\]
is an isometric embedding, whose image is the region with $r>r_s$. Again, the coefficients of the metric are analytic functions and the Ricci tensor vanishes on the Schwarzschild patch, so one concludes that $\mathscr{E}^*$ is Ricci flat. On the radial surface $\mathscr{E}^*_0$ determined by $\theta=\theta_0$ and $\varphi=\varphi_0$ the metric is
\begin{equation} g=-\left( 1-\frac{r_s}{r}\right)  du\otimes du-du\otimes dr -dr \otimes du.\end{equation}
A lightlike curve $\gamma(\tau)$ satisfies the equation
\[0= \langle \gamma'(\tau),\gamma'(\tau)\rangle=-\left( 1-\frac{r_s}{r}\right)  u'(\tau)^2-2 u'(\tau) r'(\tau).\]
There are two possibilities
\[ u'(\tau)=0\quad \text{or} \quad u'(\tau)=-\frac{2r'(\tau)r(\tau)}{r(\tau)-r_s}=-2r'(\tau)-\frac{2r'(\tau)r_s}{r-r_s}.\]
Integrating on both sides one obtains
\begin{equation}
u(\tau)=\text{const.}\quad \text{or} \quad  u(\tau)=-2r(\tau)-2r_s \log \left|\frac{r(\tau)}{r_s}-1\right|+\text{const}.
\end{equation}
We define the new coordinate $t^*$ by $t^*=r+t-r^*$, so that outgoing light rays become straight lines of slope $1$.
Figure \ref{OEF} shows the lightlike trajectories on the outgoing Eddington-Finkelstein spacetime.

The outgoing Eddington-Finkelstein spacetime has properties which are opposite to those of the black hole. Naturally, one says that these properties describe a white hole.
\begin{itemize}

\item The chronological future of an event for which $r>r_s$ is contained in the region $r>r_s$. This means that not even light can enter
the region $r<r_s$, which is called the interior of the white hole.

\item The worldline of a particle that starts with $r<r_s$ either leaves the region $r<r_s$ or approaches $r=r_s$. This means that
everything that is in the interior of the white hole tends to leave it.

\end{itemize}
In contrast with black holes, there is no evidence that white holes exist in nature.

\section{Kruskal-Szekeres coordinates}

The Kruskal-Szekeres spacetime contains both the incoming and outgoing Eddington-Finkelstein spacetimes.
In terms of the coordinates $(u,v,\theta,\varphi)$, the Schwarzschild metric takes the form
\begin{equation}
g=-\frac{1}{2}\left(1-\frac{r_s}{r}\right)(dv \otimes du +du \otimes dv)+r^2 d\Omega,
\end{equation}
which is degenerate at $r=r_s$. However, in the coordinates
\begin{align*}
U=-e^{-u/2r_s},\quad V=e^{v/2r_s},
\end{align*}
the metric becomes
\begin{equation}
g=-\frac{2r^3_s}{r e^{r/r_s}}(dV \otimes dU +dU \otimes dV)+r^2 d\Omega,
\end{equation}
where $r$ is regarded as a fuction of $U$ and $V$. Notice that
\[ UV=-e^{(v-u)/2r_s}=-e^{r^*/r_s}=\frac{r_s-r}{r_s}e^{r/r_s} ,\]
and therefore, $r=0$ implies $VU=1$.
The Kruskal-Szekeres spacetime $\mathscr{K}$ is the region of $\RR^4$ with coordinates $(U,V,\theta, \varphi)$ such that $UV<1$, $\theta \in (0, \pi)$ and 
$\varphi \in (0, 2 \pi)$.
The map  $\iota: \mathscr{E} \to \mathscr{K}$ defined by
\[ \iota(v, r,\theta, \varphi)= (U,V,\theta, \varphi)\]
is an isometric embedding of the incoming Eddington-Finkelstein spacetime as the region in $\mathscr{K}$ such that $V>0$.
The map $\iota^*: \mathscr{E}^* \to \mathscr{K}$ defined by
\[ \iota^*(u, r,\theta, \varphi) = (U,V,\theta, \varphi)\]
is an isometric embedding of the outgoing Eddington-Finkelstein spacetime as the region in $\mathscr{K}$ such that $U<0$.
The Schwarzschild spacetime corresponds to the region where $U<0$ and $V>0$.

\begin{figure}[H]
\centering
\includegraphics[scale=0.45]{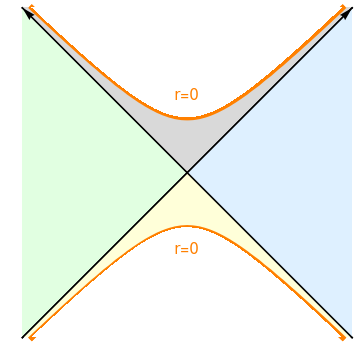}\caption{Kruskal-Szekeres
spacetime. The blue region is the Schwarzschild patch. The gray region is the interior of the black hole. The yellow region is the interior of the white hole. The green region is the mirror image of the Schwarzschild patch. }%
\label{KS}
\end{figure}

 As Figure \ref{KS} illustrates, the Kruskal-Szekeres spacetime has the following properties.

\begin{figure}[H]
\centering
\includegraphics[scale=0.5
]{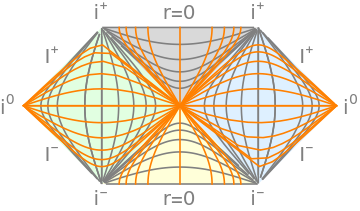}\caption{Carter-Penrose diagram of Kruskal-Szekeres spacetime. The blue region is the Schwarzschild patch. The gray region is the interior of the black hole. The yellow region is the interior of the white hole. The green region is the mirror image of the Schwarzschild patch. Orange line correspond to constant $r$ and gray lines correspond to constant $t$.}\label{penk}%
\end{figure}

\begin{itemize}

\item The blue region is the Schwarzschild patch, the exterior of the black hole.
\item The gray region is the interior of the black hole.
\item The yellow region is the interior of the white hole.
\item  The green region is new. Notice that the map:
\[ (U,V,\theta,\varphi) \mapsto (-U,-V,\theta, \varphi)\]
is an isometry that exchanges the green and blue regions. One concludes that the green region is isometric to the Schwarzschild patch, a second copy of the exterior of the black hole.

\item Recall that the vector field $T=\partial_t$ in the Schwarzschild spacetime is a timelike Killing vector field. In Kruskal-Szekeres coordinates, this vector field is
\[ T=\partial_t=\frac{\partial U}{\partial t} \partial_U+ \frac{\partial V}{\partial t}\partial_V=\frac{1}{2r_s}\big(V \partial_V- U\partial_U\big),\]
and therefore
\[ \langle T,T\rangle=-\frac{(r-r_s)^2}{2r^2_s}e^{r/r_s}=\frac{r_s}{r}-1.\]
This shows that the vector field $T$ is timelike in the Schwarzschild patch, as expected, but it is spacelike in the gray and yellow regions.
\end{itemize}
The causal structure of the Kruskal-Szekeres spacetime can be more transparently described in a Penrose diagram.
Consider the surface $\mathscr{K}_0$ with $\theta=\theta_0$ and $\varphi=\varphi_0$, which has metric
\begin{equation}\label{K0}
g_0=-\frac{2r^3_s}{r e^{r/r_s}}(dV \otimes dU +dU \otimes dV).
\end{equation}
Since a Penrose diagram is conformally invariant, we may consider instead the metric
\begin{equation}\label{K0}
\tilde{g}_0=-(dV \otimes dU +dU \otimes dV),
\end{equation}
in the region $UV<1$. We denote by $\mathscr{P}$ the diamond shaped region in the plane with coordinates $(X,Y)$ determined by the conditions
\[ |Y|<\frac{\pi}{2},\quad X>Y-\pi,\quad X>-Y-\pi,\quad X<Y+\pi,\quad X<-Y+\pi,\]
with metric $h=-dY \otimes dY+dX\otimes dX$. We define a map  $\phi: \mathscr{K}_0 \to \mathscr{P}$ by setting
\[\phi(U,V)=(X,Y) ,\]
where here
\[X=\arctan V-\arctan U,\quad Y=\arctan V+\arctan U.\]
The map $\phi$ is a diffeomorphism. Moreover,
\[ dX=\frac{dV}{1+V^2}-\frac{dU}{1+U^2}, \quad dY=\frac{dV}{1+V^2}+\frac{dU}{1+U^2} ,\]
and therefore
\begin{align*}\phi^*h &=\left(\frac{dV}{1+V^2}-\frac{dU}{1+U^2}\right)\otimes \left(\frac{dV}{1+V^2}-\frac{dU}{1+U^2}\right)\\
&\quad  \, -\left(\frac{dV}{1+V^2}+\frac{dU}{1+U^2}\right)\otimes \left(\frac{dV}{1+V^2}+\frac{dU}{1+U^2}\right)\\
&=-\frac{2}{(1+U^2)(1+V^2)}\left(dU \otimes dV+dV \otimes dU\right)\\
&=\frac{2}{(1+U^2)(1+V^2)} \tilde{g}_0.\end{align*}
One concludes that the map $\phi: \mathscr{K}_0 \to \mathscr{P}$ is conformal with respect to $\tilde{g}_0$ and, therefore, also with respect to $g_0$. This map provides the Carter-Penrose diagram shown in Figure \ref{penk}.
\begin{itemize}
\item An object in the blue region can only go to the gray region, and, once there, it will inevitably go towards the singularity $r=0$.
It can never reach the yellow or green regions.
\item An object in the green region can only go to the gray region, and, once there, it will inevitably go towards the singularity $r=0$.
It can never reach the yellow or blue regions.
\item An object in the gray region will inevitably go towards the singularity $r=0$.
\item An object in the yellow region can go everywhere.
\item The blue and green regions are mirror images of each other, but it is not possible to send information from one region to the other. If Alice is in blue and Beth is in green, they can meet, but only in the interior of the black hole.
\end{itemize}

\begin{figure}[H]
\centering
\includegraphics[scale=0.4
]{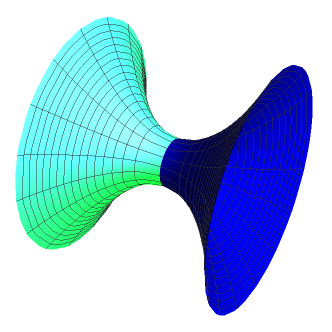}\caption{Einstein-Rosen bridge.}\label{ERB}%
\end{figure}

The green and blue regions in Kruskal-Szekeres spacetime cannot be connected by a timelike curve. However, the spacelike surface $t=0$ and $\theta=\pi/2$, connects these two regions. The geometry of this surface, known as an Einstein-Rosen bridge, is depicted in Figure \ref{ERB}.

\section{\label{shell}Interior of a non rotating star}
In this section we want to analyze the geometry of space-time inside of a
non rotating star, or more generally, inside a spherical shell 
\begin{equation*}
S=\{(r,\theta ,\phi )\mid 0\leq r_{0}<r\leq r_{1},\text{ }0<\theta <\pi
,\text{ }0<\phi <2\pi \}.
\end{equation*}%
We suppose $S$ consists of a perfect fluid of density and
pressure given by smooth functions $\rho (r)$ and $p(r).$ 
Let $M$ denote the $4$-manifold $M=\RR\times S$, with coordinates $%
x=(t,r,\theta ,\phi )$.
The fluid moves in space-time in the direction of the \emph{unitary} vector
field $V= \partial t/\vert \partial_t \vert$. 
 We assume the spacetime is static and spherically
symmetric, so that the metric $g$ can be written as%
\begin{equation*}
g=-e^{2\alpha (r)}dt\otimes dt+e^{2\beta (r)}dr\otimes dr+r^{2}d\Omega.
\end{equation*}%
The components of the vector field $V$ are given by $
v^{0}=1/\left\vert \partial _{t}\right\vert =e^{-\alpha (r)}$, and $v^{i}=0$%
. Associated to $\rho $ and $p$ there is an energy-momentum tensor $T^{\sharp}$
whose components are given by (\ref{perfecto
fluido}), and therefore%
\begin{align*}
T^{00}& =(\rho (r)+p(r))e^{-2\alpha (r)}+(-e^{-2\alpha (r)})p(r)=\rho
(r)e^{-2\alpha (r)} \\
T^{11}& =e^{-2\beta (r)}p(r),\\
T^{22}&=r^{-2}p(r),\\%
T^{33}&=(r^{2}\sin ^{2}\theta )^{-1}p(r),
\end{align*}%
with $T^{ab}= 0$ if $a \neq b$. Lowering indices one gets
\begin{align*}
T_{00}& =g_{00}^{2}T^{00}=e^{4\alpha
(r)}e^{-2\alpha (r)}\rho (r)=e^{2\alpha (r)}\rho (r) \\
T_{11}& =g_{11}^{2}T^{11}=e^{4\beta (r)}e^{-2\beta (r)}p(r)=e^{2\beta
(r)}p(r) \\
T_{22}& =g_{22}^{2}T^{11}=r^{2}r^{-2}p(r)=p(r) \\
T_{33}& =g_{33}^{2}T^{33}=(r^{2}\sin ^{2}\theta )(r^{2}\sin ^{2}\theta
)^{-1}p(r)=p(r).
\end{align*}%
On the other hand, using \eqref{tabla 1}, one obtains that the scalar curvature would be equal to 
\begin{align*}
\Rs& =-e^{-2\alpha (r)}\Ric_{00}+e^{-2\beta (r)}\Ric%
_{11}+r^{-2}\Ric_{22}+(r^{2}\sin ^{2}\theta )^{-1}\Ric_{33}
\\
& =-e^{-2\alpha (r)}\Ric_{00}+e^{-2\beta (r)}\Ric_{11}+\frac{%
2}{r^{2}}\Ric_{22} \\
& =-2e^{-2\beta (r)}\left( \alpha ^{\prime \prime }(r)+\alpha ^{\prime
}(r)^{2}-\alpha ^{\prime }(r)\beta ^{\prime }(r)+\frac{2}{r}\alpha ^{\prime
}(r)-\frac{2}{r}\beta ^{\prime }(r)+\frac{1}{r^{2}}-\frac{1}{r^{2}}e^{2\beta
(r)}\right) .
\end{align*}%
Now, we know the field equations are 
\begin{equation*}
\Ric_{ab}-\frac{1}{2}\Rs g_{ab}=8\pi T_{ab}.
\end{equation*}
For $a=b=0,$ we obtain
\begin{align*}
&\phantom{-}\: e^{2(\alpha (r)-\beta (r))}\left(\alpha ^{\prime \prime }(r)+\alpha ^{\prime
}(r)^{2}-\alpha ^{\prime }(r)\beta ^{\prime }(r)+\frac{2}{r}\alpha ^{\prime
}(r)\right) \\
& -e^{2\alpha (r)-2\beta (r)}\left( \alpha ^{\prime \prime }(r)+\alpha
^{\prime }(r)^{2}-\alpha ^{\prime }(r)\beta ^{\prime }(r)+\frac{2}{r}\alpha
^{\prime }(r)-\frac{2}{r}\beta ^{\prime }(r)+\frac{1}{r^{2}}-\frac{1}{r^{2}}%
e^{2\beta (r)}\right)\\ &\qquad\qquad\qquad\qquad\qquad\qquad\qquad\qquad\qquad\qquad\qquad\qquad\qquad\qquad\qquad=8\pi e^{2\alpha (r)}\rho (r).
\end{align*}%
This simplifies to 
\begin{equation}
1-e^{-2\beta (r)}+2\beta ^{\prime }re^{-2\beta (r)}=8\pi r^{2}\rho (r).
\label{eq00}
\end{equation}
For $a=b=1$, one gets%
\begin{align*}
& -\alpha ^{\prime \prime }(r)-\alpha ^{\prime }(r)^{2}+\alpha ^{\prime
}(r)\beta ^{\prime }(r)+\frac{2}{r}\beta ^{\prime }(r) \\
& +e^{-2\beta (r)}e^{2\beta (r)}\left( \alpha ^{\prime \prime }(r)+\alpha
^{\prime }(r)^{2}-\alpha ^{\prime }(r)\beta ^{\prime }(r)+\frac{2}{r}\alpha
^{\prime }(r)-\frac{2}{r}\beta ^{\prime }(r)+\frac{1}{r^{2}}-\frac{1}{r^{2}}%
e^{2\beta (r)}\right) \\
& \qquad\qquad\qquad\qquad\qquad\qquad\qquad\qquad\qquad\qquad\qquad\qquad\qquad\qquad\qquad=8\pi e^{2\beta (r)}p(r).
\end{align*}%
Simplifying, one obtains
\begin{equation}
\frac{e^{-2\beta (r)}}{r^{2}}\left( 2r\alpha ^{\prime }(r)+1-e^{2\beta
(r)}\right) =8\pi p(r).  \label{eq11}
\end{equation}
In a similar fashion, for $a=b=2$, we get%
\begin{equation}
e^{-2\beta (r)}\left( \alpha ^{\prime \prime }(r)+\alpha ^{\prime
}(r)^{2}-\alpha ^{\prime }(r)\beta ^{\prime }(r)+\frac{\alpha ^{\prime }(r)}{%
r}-\frac{\beta ^{\prime }(r)}{r}\right) =8\pi p(r).  \label{eq22}
\end{equation}%
Let us now use the equation of local conservation of energy $%
\sum_{a}\nabla _{a}T^{ab}=0$. Taking $b=1$ this equation
becomes  
\begin{equation*}
\nabla _{\partial _{a}}T=\sum_{b,c}\frac{\partial T^{bc}}{\partial x^{a}}%
\partial_{x^{b}}\otimes \partial_{x^{c}}+\sum_{b,c}\Gamma
_{ab}^{c}T^{bd}\partial_{x^{c}}\otimes \partial_{x^{d}}+\sum_{b,c}\Gamma
_{ab}^{c}T^{db}\partial_{x^{c}}\otimes \partial_{x^{d}}.
\end{equation*}%
Hence the component $(\nabla _{\partial _{a}}T)^{a1}$ is given by
\begin{equation*}
(\nabla _{\partial _{a}}T)^{a1}=\left\{ 
\begin{array}{c}
\frac{\partial T^{11}}{\partial x^{1}}+2T^{11}\Gamma _{11}^{1},\text{ \ \ \
\ \ \ \ \ \ \ \ \ \ \ \ \ \ \ \ \ \ \ \ \ \ \ \ \ \ \ \ \ \ \ \ \ \ \ \ \ \
\ \ \ if }a=1 \\ 
\\ 
T^{11}(\Gamma _{01}^{0}+\Gamma _{21}^{2}+\Gamma _{31}^{3})+T^{00}\Gamma
_{00}^{1}+T^{22}\Gamma _{22}^{1}+T^{33}\Gamma _{33}^{1}\text{, if }a\neq 1%
\end{array}%
\right. .
\end{equation*}%
Thus, 
\begin{equation*}
(\nabla _{\partial _{a}}T)^{a1}=e^{-2\beta }(p^{\prime }-2\beta ^{\prime
}p+2\beta ^{\prime }p+\alpha ^{\prime }p+\frac{2}{r}+\rho \alpha ^{\prime }-%
\frac{2}{r})=0.
\end{equation*}%
From this we obtain
\begin{equation}
(\rho (r)+p(r))\alpha ^{\prime }(r)+p^{\prime }(r)=0.  \label{eq33}
\end{equation}%
Let us define 
$$
m(r) =\frac{1}{2}(r-re^{-2\beta (r)}),
$$
so that
$$
e^{2\beta (r)} =\left( 1-\frac{2m(r)}{r}\right) ^{-1}.
$$%
Taking the derivative with respect to $r$ we obtain 
\begin{align*}
m'(r)& =\frac{1}{2}(1-e^{-2\beta (r)}+2\beta ^{\prime }re^{-2\beta
(r)}) \\
& =\frac{1}{2}( 8\pi r^{2}\rho (r)) \\
&=4\pi r^{2}\rho (r),
\end{align*}%
where in the second line we have used equation (\ref{eq00}) to substitute
the expression inside the parenthesis for $8\pi r^{2}\rho (r).$ This
immediately gives us 
\begin{equation*}
m(r)=4\pi \int\nolimits_{r_{0}}^{r}r^{2}\rho (r)dr,
\end{equation*}%
with $0\leq
r_{0}<r\leq r_{1}$. Now, in terms of $m(r),$ equation (\ref{eq22}) can be written as:%
\begin{equation}
\alpha ^{\prime }(r)=\frac{m(r)+4\pi r^{3}p(r)}{r(r-2m(r))}.  \label{eq44}
\end{equation}%
Combining (\ref{eq33}) and (\ref{eq44}) one gets the \emph{%
Tolman-Oppenheimer-Volkoff }equation%
\begin{equation}
p^{\prime }(r)=-\frac{(\rho (r)+p(r))(m(r)+4\pi r^{3}p(r))}{r(r-2m(r))}.
\label{eq55}
\end{equation}%
We impose the natural boundary condition $p(r_{1})=0$ since one does not
expect any pressure at the surface of $S$. Assuming $\rho
(r)=\rho$ is \emph{constant}, then, in terms of $p(r),$ the function $%
\alpha (r)$ can be expressed as 
\begin{equation*}
\alpha (r)=-\int_{r_{0}}^{r}\frac{p^{\prime }(s)}{\rho +p(s)}ds=\ln
\left( \frac{\rho +p(r_{0})}{\rho +p(r)}\right) +\alpha (r_{0}),
\end{equation*}%
and therefore 
\begin{equation}
e^{2\alpha (r)}=e^{2\alpha (r_{0})}\left( \frac{\rho +p(r_{0})}{\rho +p(r)}%
\right) ^{2}.  \label{finalmente}
\end{equation}%
For $r=r_{1}$\textrm{\ }the boundary condition\textrm{\ }$p(r_{1})=0$
implies%
\begin{equation*}
e^{2\alpha (r_{1})}=e^{2\alpha (r_{0})}\left( 1+\frac{p(r_{0})}{\rho }%
\right) ^{2}.
\end{equation*}%
In order to match the o Schwarzschild solution on the exterior at the boundary $r=r_{1}$
one needs $e^{2\alpha (r_{1})}$ to be equal to $1-2M/r_{1}$, where $%
M=\frac{4}{3}\pi (r_{1}^{3}-r_{0}^{3})$ is the total Newtonian mass of $S$.
Thus, 
\begin{equation*}
e^{2\alpha (r_{0})}=\frac{e^{2\alpha (r_{1})}}{\left( 1+p(r_{0})/\rho 
\right) ^{2}}=\frac{1-2M/r_{1}}{\left( 1+p(r_{0})/\rho \right) ^{2}}%
.
\end{equation*}%
From (\ref{finalmente}) one obtains 
\begin{equation*}
e^{2\alpha (r)}=\frac{(1-2M/r_{1})\left[ (\rho +p(r_{0}))/(\rho +p(r))%
\right] ^{2}}{\left( 1+p(r_{0})/\rho \right) ^{2}}=\frac{1-2M/r_{1}%
}{(1+p(r)/\rho )^{2}},
\end{equation*}%
and the metric inside $S$ would be given by 
\begin{equation}
g_{S}=\frac{1-2M/r_{1}}{(1+p(r)/\rho )^{2}}dt\otimes dt+\left( 1-%
\frac{2m(r)}{r}\right) ^{-1}dr\otimes dr+r^{2}d\Omega.  \label{elusiva}
\end{equation}%
where $p(r)$ is the unique solution of the Tolman-Oppenheimer-Volkoff equation
with initial condition $p(r_{1})=0.$

\section{Interior of a Uniformly Dense Star}

If our shell \textrm{S} is represented by the interior of a \emph{uniformly
dense} star of radius $R,$ i.e., $\rho (r)=\rho $, $r_{0}=0$, $r_{1}=R$
there is an exact solution to (\ref{eq55}). In this case $m(r)=4/3\pi r^{3}$%
, for $0<r\leq R$, and equation (\ref{eq55}) can be integrated (\cite%
{carroll}) to give%
\begin{equation*}
p(r)=\rho \frac{(1-2M/R)^{1/2}-(1-2Mr^{2}/R^{3})^{1/2}}{%
(1-2Mr^{2}/R^{3})^{1/2}-3(1-2M/R)^{1/2}},
\end{equation*}%
where $M=m(r_{1}).$ We notice that the denominator vanishes if 
\begin{equation*}
r=\frac{R}{\sqrt{M(9M-4R)}},
\end{equation*}%
which is a real number if and only if $9M-4R\geq 0$. Hence, if we assume the
pressure inside the star is finite -a reasonable physical assumption- then
one must have $M<\frac{4R}{9}$ . Even under
the more general assumption that $\rho (r)$ is not constant but
monotonically decreasing, $d\rho /dr\leq 0$, one can show (\cite{wald}, page 130) that for a star to be
physically stable it is required that $M<\frac{4R}{9}$. This is known as  \emph{Buchdahl's
theorem}. If the mass of a star is not too big, once its fuel is exhausted,
and cools down, it will attain a final state of equilibrium and becomes a
white dwarf or a neutron star. However, if the mass of the star is greater
than the \emph{Tolman--Oppenheimer--Volkoff limit, }(three to four times the
mass of the Sun)\emph{\ }then equilibrium will never be achieved. Inner
pressure will not support its own weight and the star will undergo a
complete gravitational collapse, shrinking until its radius becomes smaller
than the corresponding Schwarzschild radius. Once this threshold is
surpassed, it will continue shrinking until it finally disappears, becoming
a black hole.  


\section{Geometry Inside a Spherical Empty Cavity}

We will show that the geometry inside an empty spherical cavity must be
flat. The shell behaves like a
gravitational \emph{Faraday Cage}, where the gravitational forces cancel
out in the interior. One can choose coordinates $(\overline{t},r,\theta
,\phi )$ so the geometry inside the shell is
\begin{equation}
g=-\left( 1-\frac{C}{r}\right) d\overline{t}\otimes d\overline{t}%
+\left( 1-\frac{C}{r}\right) ^{-1}dr\otimes dr+r^{2}d\Omega,  \label{swrevisitada}
\end{equation}%
where $0<r\leq r_{1}$, and $C$ is a suitable constant. For all $0<r<r_{1}$, one
has $p(r)=m(r)=0$ in equation (\ref{eq44}) , and therefore $\alpha
^{\prime }(r)=0$. Consequently, $\alpha (r)$ must be constant$.$ This forces 
$C=0$ in (\ref{swrevisitada}), and therefore the metric is flat in
inside the shell. \emph{However, this does not imply} that
in the Schwarzschild global coordinates the metric is given by 
$$g=-\left(1-\frac{2M}{r}\right)dt\otimes dt+\left(1-\frac{2M}{r}\right)^{-1}dr\otimes dr+r^2d\Omega,
$$
since Birkohff's theorem does not guarantee that the coordinates $(%
\overline{t},r,\theta ,\phi )$ must be the same as those of an observer at
infinity. In fact we will show that $\overline{t}=\kappa t$, with 
\begin{equation*}
\kappa =\frac{(1-2M/r_{1})^{1/2}}{1+p(r_{0})/\rho }.
\end{equation*}%
In \S\ref{shell} we showed that inside the material body
of the shell the metric is given by 
\begin{equation*}
g=-e^{2\alpha (r)}dt\otimes dt+\left( 1-\frac{2m(r)}{r}\right)
^{-1}dr\otimes dr+r^{2}d\Omega.
\end{equation*}%
Since the metric inside the empty cavity determined by $S$ is
constant flat, by continuity it should coincide with the metric at the inner
boundary $r=r_{0}$. Thus, 
\begin{equation*}
g_{\mathrm{int}}=-\kappa\, dt\otimes dt-dr\otimes dr+d\Omega.
\end{equation*}%
We notice that \emph{after the change of coordinates }$\overline{t}=\kappa t$%
\emph{\ the metric becomes the standard Minkowski metric of flat space time.}

\section{Time Machines}

Let us return to our discussion in \S\ref{section 00}. As we
saw there, the frequencies $\omega _{B}$ and $\omega _{A}$ of a pulse of
light as measured by the two observers $B$ and $A$ were given by 
\begin{equation}
\omega _{A}=\omega _{B}\frac{\sqrt{-g_{00}(r_{B})}}{\sqrt{-g_{00}(r_{A})}}%
=\omega _{B}Q(M,r_{B},r_{A}).  \label{eq ya no se}
\end{equation}%
where $Q(M,r_{B},r_{A})>1$  if $r_A<r_B$. In standard units 
\begin{equation}
Q(M,r_{B},r_{A})=\left( \frac{1-2G_{N}M/c^{2}r_{B}}{1-2G_{N}M/c^{2}r_{A}}%
\right) ^{1/2}.  \label{again}
\end{equation}%
We may imagine that each pulse of light emitted by $B$ corresponds to the
ticking of a clock he uses to measure his proper time. The frequency of the
light signal emitted is measured by him to be $2\pi /\Delta s_{B},$ where $%
\Delta s_{B}$ is the corresponding period of the light wave. Now, suppose $A$
receives these signals at intervals $\Delta s_{A}$ (measured in $A$'s proper
time), so that $\omega _{A}=2\pi /\Delta s_{A}$. From (\ref{eq ya no se})
one obtains $\Delta s_{B}=Q(M,r_{B},r_{A})\Delta s_{A}>\Delta s_{A}.$

To see how time dilates, suppose that $A$ hovers very close to the mass $M,$ let's
say just a meter away form the horizon of \emph{V616 Monocerotis}$,$ the
closest known black hole, believe to be located about three thousand light
years away. Its mass is estimated to be eleven times that of our Sun: $%
M=11\times 1.989\times 10^{30}$ \textrm{kg}. Its Schwarzschild radius would
then be $R=2G_{N}M/c^{2}\approx 32.4$ \textrm{km}. On the other hand, we
assume observer $B$ hovers $10$ \textrm{km} away from $A$. One can calculate the factor $Q(M,r_{B},r_{A})$ for these values of the parameters as approximately
equal to $\approx 100$. The entire \textit{movie} of a whole century in $B$%
's world could be watched by $A$ in fast motion in just one year! 

\begin{figure}[tbh]
\centering
\includegraphics[scale=0.6]{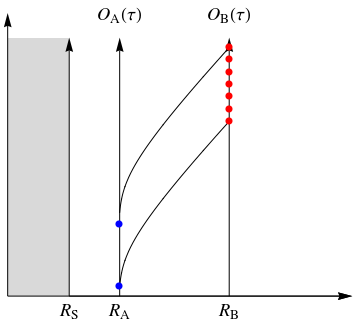}
\caption{Time dilation}
\end{figure}

Let us now find out how much force $A$ would need in order to hover one meter above the horizon of this black hole. $A$'s worldline, parametrized by arc length, would be 
\begin{equation*}
O_{A}(s)=\left( \left(1-\frac{2M}{R_{A}}\right)^{-1/2}s,R_{A},\theta _{A},\phi
_{A}\right)
\end{equation*}%
with  $R_{A},\theta _{A,}\phi _{A}$ are constants. As we discussed in \S\ref{proper acceleration}, at any point $p=O_{A}(s_{0})$ the $4$%
-acceleration $\mathbf{a}_{p}$ of $O_{B}$ at $p$ coincides with $O_{A}$'s $3$%
-acceleration $a_{p}$, as measured in his own frame of reference. An
orthonormal base for $O_{A}$ at $p$ is given by the vectors $e_{a}=\partial
_{a}/\left\vert \partial _{a}\right\vert $. Then the total force he
experiences would then be his rest mass \ times $\left\vert a_{p}\right\vert
.$ Let us compute $\mathbf{a}_{p}$ in the coordinate frame of $O_{A}$ at $p$.
We have 
\begin{align*}
\mathbf{a}_{p}& =\nabla _{\mathbf{u}(s)}\mathbf{u}(s)\\
&=\left(1-\frac{2M}{R_{A}}\right)^{-1}\nabla
_{\partial _{t}}\partial _{t} \\
&=\left(1-\frac{2M}{R_{A}}\right)^{-1}\sum_{c}\Gamma _{00}^{c}\left.
\partial _{r}\right\vert _{p} \\
& =\frac{M}{R_{A}^{2}}\left(1-\frac{2M}{R_{A}}\right)^{-1}\left(1-\frac{2M}{R_{A}}\right)\left.
\partial _{r}\right\vert _{p} \\
& =\frac{M}{R_{A}^{2}}\partial _{r}\vert _{p}  \\
&=\frac{M}{R_{A}^{2}}\left\vert \partial
_{r}\right\vert e_{r} \\
&=\frac{M}{R_{A}^{2}}\left(1-\frac{2M}{R_{A}}\right)^{-1/2}e_{r},
\end{align*}%
since $\left\vert \partial _{r}\right\vert =(1-2M/R_{A})^{-1/2}$ at the
point $p$. In standard units of mass and time 
\begin{equation*}
\left\vert a_{p}\right\vert =\frac{G_{N}M}{R_{A}^{2}}\left(1-\frac{2G_{N}M}{%
c^{2}R_{A}}\right)^{-1/2}\, \frac{\mathrm{m}}{\mathrm{s}}.
\end{equation*}%
This differs from Newtonian acceleration by the factor $%
1-2G_{N}M/c^{2}R_{A}$, which is very small when $R_{A}\gg 2G_{N}M/c^{2}.$
Substituting the values for $G_{N}$ and $M$, 
\begin{equation*}
R_{A}=2G_{N}M+1=32449.98\,
\end{equation*}%
we obtain $\left\vert \mathrm{a}_{p}\right\vert \approx 2.5\times 10^{14}\,
\mathrm{m}/\mathrm{s}^2$. That is, at one meter from the
horizon, $A$ would experience a force exerted by his rocket engines similar
to that he would feel on Earth being under the weight of a mass the size of
mount Everest!

The mental experiment we just discussed tells us that using a black hole as
a time machine does not seem to be feasible. There is, however, one way of
canceling the overwhelming gravitational forces surrounding a big mass: one
could stay inside a homogeneous spherical shell where the total
gravitational force must be zero.

\begin{figure}[H]
\centering
\includegraphics[scale=0.45]{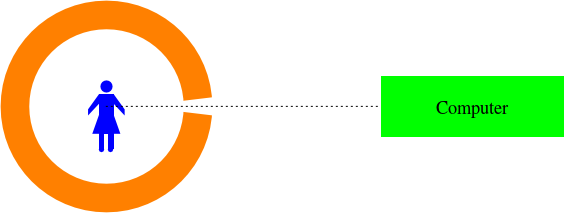}
\caption{Time machine}
\end{figure}
We already know that inside the cavity determined by $\mathrm{S}$ space-time
is flat, where the metric is given in global Schwarzschild coordinates by%
\begin{equation*}
g_{\mathrm{int}}=-\frac{1-2M/r_{1}}{(1+p(r_{0})/\rho )^{2}}%
dt\otimes dt-dr\otimes dr+d\Omega.
\end{equation*}%
As an example, consider a \textquotedblleft thin\textquotedblright\ shell,
let's say with dimensions $r_{0}=100,$ $r_{1}=101$ \textrm{m}. We assume it
has a constant density equal to $\rho =0.000393\, \mathrm{kgg}/\mathrm{m}^{3}$. Then,
its total mass in kilograms would be $6.4\times 10^{28}$ \textrm{kg},
approximately \textrm{60} times the mass of Jupiter. 
If one used the shell as a time machine, it would be possible to observe an entire
century of events in the exterior world in just ten years. The only problem,
of course, would be that to construct such a shell one would need a material
with a density equal to $5.1\times 10^{23}\, \mathrm{kg}/\mathrm{m}^{3}$
approximately a million times more dense than the densest object know in the
universe, a neutron star!


   \clearemptydoublepage 
 \chapter{The FLRW metric and Cosmology\label{expanding metric}}

\vspace{3ex}

\section{The FLRW metric}
The Friedmann-Lemaitre-Robertson-Walker models are solutions to the Einstein equations
that arise in cosmology, the study of the large scale properties of the universe as a whole.
As in the case of Schwarzschild spacetime, symmetry considerations go a long way in determining the FLRW metric.
In cosmology, these symmetries come from an extension of the Copernican principle. Copernicus rejected the idea, predominant at the time, that the Earth plays a special role at the center of the universe. The cosmological principle is a much stronger form of Copernicus' idea. Not only is the Earth not a special place in the universe, there are no special places. Moreover, there are no special directions in space, all directions look the same. Clearly, these assumptions are only reasonable at very large scales. The Earth is a very different place from the Sun. However, at the largest scale, these variations are supposed to average out.

Mathematically, the cosmological principle corresponds to space being homogeneous and isotropic.  A Riemannian manifold $(M,g)$ is homogenous if, given two points $p,q \in M$, there is an isometry that sends $p$ to $q$. It is isotropic if, given two unitary tangent vectors $v,w  \in T_pM$, there is an isometry $\varphi$ such that $
D\varphi(p)(v)=w$.

In three dimensions, a Riemannian manifold that is homogeneous and isotropic has constant curvature.
\begin{lemma}
\label{nice} Let $(M,g)$ be a $3$-dimensional Riemannian manifold which is homogeneous and isotropic. Then,  $M$ has constant curvature $C$.
\end{lemma}
\begin{proof}
Given subspaces $\Pi$ and $\Pi'$  of $T_pM$, consider unitary vectors $v$, $v'$ which are orthogonal to $\Pi$ and $\Pi'$, respectively.
Fix an isometry such that $D\varphi(p)(v)=v'$. This implies that $D\varphi(p)(\Pi)=\Pi'$.
Therefore:
\[ K(p)(\Pi)=K(\varphi(p))(D\varphi (\Pi))=K(p)(\Pi').\]
This implies that the sectional curvature is a scalar function $K:M \to \RR$. Since $M$ is homogenous, this function is independent of $p$, and therefore, it is a constant $C$.
\end{proof}

The Killing-Hopf theorem \ref{KH} states that, if $(M,g)$ is simply connected and geodesically complete, then the metric can be rescaled by a constant factor so that it becomes either Euclidean space, a sphere or hyperbolic space.
 One concludes that a three dimensional manifold which is homogeneous, isotropic, simply connected and complete can be rescaled so that it becomes one of the three model spaces.

\begin{figure}[H]
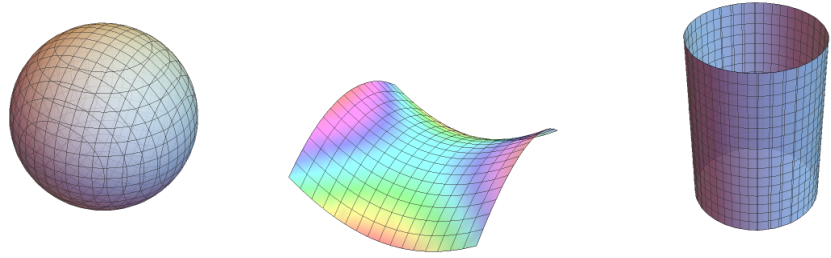

\begin{center}
 \includegraphics[scale=0.3]{Figures/positivec} \quad
 \includegraphics[scale=0.3]{Figures/negativec} \quad
 \includegraphics[scale=0.3]{Figures/zeroc}
\caption{Positive, negative and zero curvature.}
\end{center}
\end{figure}

\begin{center}
\begin{tabular}{ccc}
\hline &&\\
\multicolumn{3}{c}{\textbf  Cosmological Principle} \\  
&&\\
\hline && \\
 {\textbf Physics}&& {\textbf Mathematics} \\  
&&\\
\hline && \\
 There are no preferred places in space. &&  Space is homogeneous.\\
&&\\
There are no prefrerred directions in space. && Space is isotropic.\\
&&\\
\hline
\end{tabular}
\end{center}

It will be convenient to write the metrics for the sphere, Euclidean and hyperbolic spaces in a unified way.
\begin{itemize}
\item \textbf{Euclidean Space:} Let us write the flat metric
\[ g= dx \otimes dx + dy \otimes dy + dz \otimes dz,\]
in spherical coordinates
\begin{align*}
x=r \sin\theta\cos\varphi, \quad
y=r \sin\theta\sin\varphi,\quad
z=r \cos\theta.
\end{align*}
One has
\begin{align*}
dx&= \sin\theta\cos\varphi\, dr - r \sin\theta \sin\varphi \,d \varphi+r \cos\theta\cos\varphi \,d\theta,\\
dy&= \sin\theta\sin\varphi\,dr+ r \cos\theta \sin\varphi\, d\theta+ r \sin\theta \cos\varphi\, d\varphi,\\
dz&= \cos\theta\,dr -r \sin\theta\, d\theta.
\end{align*}
Therefore
\begin{equation}
g=dr \otimes dr + r^2 \left( d\theta \otimes d\theta + \sin^2\theta\, d\varphi \otimes d\varphi \right).
\end{equation}
\item \textbf{Sphere:} Consider four dimensional Euclidean space $\RR^4$ with coordinates $(w,x,y,z)$. We use spherical coordinates 
for $\RR^3$, and keep the coordinate $w$. The sphere is determined by $r^2+ w^2=1$. Therefore:
\[ dw=\pm\frac{r dr}{\sqrt{1-r^2}},\]
so that the Euclidean metric 
\[ g= dx \otimes dx + dy \otimes dy + dz \otimes dz+ dw \otimes dw\]
restricts to the sphere as
\[ g= \frac{1}{1-r^2} dr \otimes dr + r^2 \left( d\theta \otimes d\theta + \sin^2\theta\, d\varphi \otimes d\varphi \right).\]
\item \textbf{Hyperbolic Space:} Recall that hyperbolic space is the subspace of Minkowski spacetime $\MM$ given by
\[ \langle v,v \rangle =x^2+y^2+z^2-w^2=-1,\quad  w>0.\]
Then, $r^2-w^2=-1$, so that
\[ dw=\frac{r dr}{\sqrt{1+r^2}}.\]
Therefore, the Minkowski metric
\[ h= -dw \otimes dw + dx \otimes dx + dy \otimes dy + dz \otimes dz,\]
restricts to hyperbolic space as
\[  g= \frac{1}{1+r^2} dr \otimes dr + r^2 \left( d\theta \otimes d\theta + \sin^2\theta\, d\varphi \otimes d\varphi \right).\]
\end{itemize}

One concludes that, depending on the value of $k$, the metric below describes the geometry of a sphere, Euclidean space or hyperbolic space.
\begin{equation*}
 \frac{1}{1-kr^{2}}dr\otimes dr+r^{2}%
\left(d\theta\otimes d\theta+\sin^{2}\theta\, d\varphi\otimes d\varphi\right)= \begin{cases}
\text{Sphere}& \text{ if } k=1,\\
\text{Euclidean space} & \text{ if } k=0,\\
\text{Hyperbolic space }& \text{ if } k=-1.\\
\end{cases}
\end{equation*}

The cosmological hypothesis assumes that space is homogeneous and isotropic. However, spacetime is not!
For instance, even though space looks the same in all directions, the past may look different from the future. 
Moreover, in the cosmological hypothesis it is implicitly assumed that there are well defined slices of constant time, those which are supposed to be isotropic and homogeneous. There is a global coordinate $t$, the cosmic time. For a fixed value of $t$, the spacelike submanifold $t=t_0$ is, up to scaling factor, one of the spaces of constant curvature.
The Friedmann-Lemaitre-Robertson-Walker metric takes the form
\begin{equation}\label{FLRW}
g=-dt \otimes dt+ A^2(t)\left( \frac{1}{1-kr^{2}}dr\otimes dr+r^{2}
\left(d\theta\otimes d\theta+\sin^{2}\theta\, d\varphi\otimes d\varphi\right)\right),
\end{equation}
where $t>0$, and we take units where $c=G_N=1$. It is always possible to rescale the coordinate $r$ and the curvature parameter
$k$ so that $A(t_0)=1$, where $t_0$ is the present time. Therefore, we will always assume that $A(t_0)=1$.
It remains to determine the function $A(t)$,  so that the Einstein equation is satisfied. 
The main assumption is that matter and energy in the universe move as a perfect
fluid with 4-velocity $U=\partial_{t}$, and with
density and pressure given by two fixed functions $\rho(t)$ and $p(t)$. 
This implies that the energy-momentum tensor is \[T^\sharp
=\big(\rho(t)+p(t)\big) \partial_t \otimes \partial_t+p(t)g^\sharp.\] 
The non-zero components of $T^\sharp$ are
\begin{align*}
T^{00}  & =\rho(t),\quad T^{ii}    =g_{ii}^{-1}p(t).
\end{align*}
The tensor $T$ is obtained by lowering indices. Its non-zero components are
\begin{align*}
T_{00}  &  =\rho(t),\quad
T_{ii}   =g_{ii}p(t).
\end{align*}
Hence, the trace of $T$ is
\[
\mathrm{tr}\, T=g^{00}\rho(t)+
\sum_{i}
g_{ii}^{-1}g_{ii}p(t)=-\rho(t)+3p(t).
\]
We also assume that there is a cosmological constant $\Lambda$. The Einstein equation is 
\begin{equation}\label{EFL}
\Ric-\frac{1}{2}\Rs g=8\pi(T+T^{\Lambda}),
\end{equation}
where 
\[T^{\Lambda}=-\frac{\Lambda }{8\pi}g.\]
The trace of $T^\Lambda$ is
\[ \mathrm{tr}\, T^\Lambda=-\frac{\Lambda }{2\pi }.\]
Taking traces on both sides of (\ref{EFL}) one obtains
\begin{equation}
\Rs=-8 \pi(\mathrm{tr}\, T+ \mathrm{tr}\, T^\Lambda).
\end{equation}
Replacing back into (\ref{EFL}), the Einstein equation becomes
\begin{equation}\label{EFLT}
\Ric=8\pi \left[T+T^{\Lambda}-\left(\frac{\mathrm{tr}\, T+ \mathrm{tr}\, T^{\Lambda}}{2}\right)g\right].
\end{equation}
The non-zero Christoffel symbols of the metric are
\begin{align*}
\Gamma_{11}^{0}  &  =\frac{A(t)A^{\prime}(t)}{1-kr^{2}},\quad 
\Gamma
_{22}^{0}=r^{2}A(t)A^{\prime}(t),\quad \Gamma_{33}^{0}=r^{2}\sin^{2}\theta\,
A(t)A^{\prime}(t), \\  
\Gamma_{11}^{1}    &=\frac{kr}{1-kr^{2}},\quad \Gamma_{22}^{1}
=-r(1-kr^{2}),\quad 
\Gamma_{33}^{1}=-r(1-kr^{2})\sin^{2}\theta,\\
\Gamma_{01}^{1}&=\Gamma_{10}^1=\frac{A^{\prime}(t)}{A(t)},\quad 
\Gamma_{12}^{2}    =\Gamma_{21}^{2}=\frac{1}{r},\quad \Gamma_{33}^{2}=-\sin\theta\cos
\theta,\\
\Gamma_{02}^{2}&=\Gamma_{20}^{2}=\frac{A^{\prime}(t)}{A(t)},\quad \Gamma_{13}^{3}   =\Gamma_{31}^{3} =\frac{1}{r},\quad 
\Gamma_{23}^{3}=\Gamma_{32}^3=\cot\theta,\\
\Gamma_{03}^{3}&=\Gamma_{30}^{3}=\frac{A^{\prime}(t)}{A(t)}.  
\end{align*}
The non-zero components of the Ricci tensor are
\begin{align*}
\Ric_{00}  &  =-3\frac{A^{\prime\prime}(t)}{A(t)},\\
\Ric_{11}  &  =\frac{A(t)A^{\prime\prime}(t)+2(A^{\prime})^{2}%
+2k}{1-kr^{2}},\\
\Ric_{22}  &  =r^{2}\left(  A(t)A^{\prime\prime}(t)+2(A^{\prime}
)^{2}+2k\right), \\
\Ric_{33}  &  =r^{2}\sin^{2}\theta\left(  A(t)A^{\prime\prime
}(t)+2A^{\prime}(t)^{2}+2k\right) .
\end{align*}
Einstein's equation (\ref{EFLT}) for $a=b=0$ is
\begin{equation} \label{RW1}
-3\frac{A^{\prime\prime}(t)}{A(t)}  =4\pi\left(\rho(t)+3p(t)-\frac{\Lambda}{4\pi}\right),%
\end{equation}
or, equivalently,
\begin{equation}
\frac{A^{\prime\prime}(t)}{A(t)}=\frac{-4\pi}{3}\left(\rho(t)+3p(t)-\frac{\Lambda}{4\pi}\right).
\label{RW1.1}%
\end{equation}
For $a=b=i$, Einstein's equation does not depend on $i$. It is
\begin{equation}
\frac{A^{\prime\prime}(t)}{A(t)}+2\left(  \frac{A^{\prime}(t)}{A(t)}\right)
^{2}+\frac{2k}{A(t)^{2}}=4\pi\left(\rho(t)-p(t)+\frac{\Lambda}{4\pi}\right). \label{RW2}%
\end{equation}
Substituting (\ref{RW1.1}) in (\ref{RW2}) one obtains
\begin{equation}
A^{\prime}(t)^{2}=\left(\frac{8\pi\rho(t)}{3}+\frac{\Lambda}{3}\right)A(t)^{2}-k. \label{RW3.1}%
\end{equation}
Equations (\ref{RW1.1}) and (\ref{RW3.1}) are called the Friedmann equations. 
They are the conditions that $A(t)$ must satisfy in order for the metric (\ref{FLRW}) to satisfy the Einstein field equation.
Differentiating both sides of (\ref{RW3.1}) one gets
\begin{equation}\label{RW4}
2A'(t)A''(t)=\frac{8\pi\rho'(t)A(t)^{2}}{3}+2\left(\frac{8\pi\rho(t)}{3}+\frac{\Lambda}{3}\right)A(t)A'(t).
\end{equation}
Solving for $A''(t)$ in (\ref{RW1.1}) and replacing in (\ref{RW4}) one obtains
\begin{equation}
\frac{-8\pi}{3}\left(\rho(t)+3p(t)-\frac{\Lambda}{4\pi}\right)A'(t)A(t)
=\frac{8\pi\rho'(t)A(t)^{2}}{3}+2\left(\frac{8\pi\rho(t)}{3}+\frac{\Lambda}{3}\right)A(t)A'(t).
\end{equation}
The last equation is equivalent to
\begin{equation}
-3\rho(t)A'(t)A(t)-\rho'(t)A(t)^{2}=3p(t)A(t)A'(t).
\end{equation}
Multiplying both sides by $A(t)$ gives
\begin{equation}
-3\rho(t)A'(t)A^2(t)-\rho'(t)A(t)^{3}=3p(t)A^2(t)A'(t),
\end{equation}
which can be rewritten as
\begin{equation}\label{termo}
-\frac{d}{dt}\left(\rho(t) A^3(t)\right)=p(t) \frac{d}{dt}A^3(t).
\end{equation}
Let us now consider some special instances of the Friedmann equations.

\subsection*{Dust} 
This is the case where there is no pressure, $p(t)=0$, and $\Lambda=0$.
Equation (\ref{termo}) then implies
\begin{equation}\rho(t)A^{3}(t)=K.\end{equation}
One can determine the constant $K$ by evaluating at present time $t=t_{0}$, and therefore
\begin{equation}\rho(t)A^{3}(t)=\rho_0,\end{equation}
where $\rho_0=\rho(0)$.
The Friedmann equation (\ref{RW3.1}) reads
\begin{equation}\label{FM}
A^{\prime}(t)^{2}-\frac{8\pi\rho_{0}}{3A(t)}=-k. %
\end{equation}
There are three cases, depending on the value of $k$.
\subsubsection{Flat ($k=0$)}
The Friedmann equations have the explicit solution
\begin{equation}
A(t)= \left(\frac{t}{t_0}\right)^{2/3}.
\end{equation}
 In this case, the function $A(t)$ is always increasing and the universe expands forever.
\subsubsection{Negative curvature ($k<0$)}
The derivative $A'(t)$ cannot vanish, since this would contradict (\ref{FM}). The universe expands forever.
\subsubsection{Positive curvature ($k>0$)}
The function $A(t)$ cannot increase to very large values, since this would contradict (\ref{FM}). There is a big crunch.
\subsection*{Radiation}
The trace of the electromagnetic energy momentum tensor vanishes. For this reason, the case where the trace of $T$ is zero is referred to as radiation. Explicitly, this condition is $3p(t)=\rho(t)$. Equation (\ref{termo}) becomes
\begin{equation}\label{termo1}
-3\frac{d}{dt}\left(p(t) A^3(t)\right)=p(t) \frac{d}{dt}A^3(t).
\end{equation}
which is equivalent to
\begin{equation}\label{termo1}
3p(t)' A^3(t)+12p(t) A^2(t)A'(t)=0.
\end{equation}
Multiplying by $A(t)$ one gets
\begin{equation}\label{termo1}
3p(t)' A^4(t)+12p(t) A^3(t)A'(t)=0,
\end{equation}
which implies
\begin{equation}
\frac{d}{dt}\left(\rho(t) A(t)^4\right)=0.
\end{equation}
Again, after evaluating at $t=t_{0}$, one concludes that 
\begin{equation}
A(t)^4\rho(t)=\rho_{0}.
\end{equation}
The Friedmann equation (\ref{RW3.1}) reads
\begin{equation}\label{FMX}
A^{\prime}(t)^{2}-\frac{8\pi\rho_{0}}{3A^2(t)}=-k. %
\end{equation}
There are three cases, depending on the value of $k$.
\subsubsection{Flat ($k=0$)}
The Friedmann equations have the explicit solution
\begin{equation}
A(t)= \left(\frac{t}{t_0}\right)^{1/2}.
\end{equation}
In this case the function $A(t)$ is always increasing and the universe expands for ever.
\subsubsection{Negative curvature ($k<0$)}
The derivative $A'(t)$ cannot vanish, since this would contradict (\ref{FM}). The universe expands forever.
\subsubsection{Positive curvature ($k>0$)}
The function $A(t)$ cannot increase to arbitrarily large values, since this would contradict (\ref{FM}). There is a big crunch.

\subsection*{Einstein's static universe}

Einstein's original reason for introducing the cosmological constant was the search for a static model of the universe. This was before Hubble's discovery of the expansion of the universe, so  Einstein's goal was a reasonable one. In this model it is assumed that
$A(t)=A_0$, is independent of $t$. Equation (\ref{termo}) then implies that $\rho(t)$ is also constant. Then, (\ref{RW1.1}) implies that $p(t)$ is also independent of time. Moreover
\begin{equation}
\Lambda=4\pi (\rho+ 3p)>0.
\end{equation}
Equation (\ref{RW3.1}) implies that $k=1$ and
\begin{equation}
A_0^2=\frac{1}{4\pi (\rho +p)}.
\end{equation}

\section{Lightlike godesics}

Let us consider the FLRW metric with $k=0$ and scaling factor $A(t)=\varepsilon t^q$, where $q$ is some constant exponent, and $\varepsilon=t_0^{-q}$ is a scaling parameter that makes $A(t_0)=1$.
The metric is
\begin{equation}
g=-dt \otimes dt +  A(t)^2 \left(dx \otimes dx + dy \otimes dy + dz \otimes dz \right).
\end{equation}
The non-zero Christoffel symbols of the metric $g$ are
\begin{align}
\Gamma_{ii}^{0}&=A(t)A'(t)=\varepsilon^{2}qt^{2q-1},\quad \Gamma_{i0}^{i}=\Gamma_{0i}^{i}=\frac{A'(t)}{A(t)}=\frac{q}{t}.
\end{align}
For $a,b,c$ fixed, the curve $\gamma(\tau)=(\tau,a,b,c),$
$s>0,$ defines a timelike geodesic parame-trized by proper time.
Let us now determine the equation of a general null geodesic
$\gamma(\tau)$. We assume that the curve starts at the point
$p=(0,b,0,0)$ and  that $y(\tau)=z(\tau)=0$. The condition of being
null means that
\begin{equation}\label{lgeo0}
-\left(  \frac{dt}{d\tau}\right)  ^{2}+\varepsilon^{2}t(\tau)^{2q}\left(
\frac{dx}{d\tau}\right)  ^{2}=0.
\end{equation}
This is equivalent to 
\begin{equation}\label{lgeo}
 \frac{dt}{d\tau}=\pm \varepsilon t(\tau)^{q}
\frac{dx}{d\tau}.
\end{equation}
On the other hand, the geodesic equations are
\begin{align}
\frac{d^{2}t}{d\tau^2}+\varepsilon^{2}qt(\tau)^{2q-1}\left(\frac{dx}
{d\tau}\right)^{2}  &  =0,\label{ggg}\\
\frac{d^{2}x}{d\tau^{2}}+2\frac{q}{t(\tau)}\frac{dt}{d\tau}\frac{dx}{d\tau}
&  =0.
\end{align}
By replacing (\ref{lgeo0}) into
(\ref{ggg}) one obtains
\begin{align}
\frac{d^{2}t}{d\tau^{2}}+\frac{q}{t(\tau)}\left(\frac{dt}{d\tau}\right)^{2}=0.
\label{E33}%
\end{align}
Integrating (\ref{lgeo}) gives
\begin{equation}
x(\tau)=\pm\frac{t(\tau)^{1-q}}{\epsilon(1-q)}+b. \label{E333}%
\end{equation}
For $q=2/3$, and $\varepsilon=(t_{0})^{-2/3}$, this equation
becomes 
\begin{equation}x(\tau)=\pm3(t_0)^{2/3}t(\tau)^{1/3}+b.\end{equation} The null
geodesic that starts at $p=(0,b,0,0)$ is then given by%
\begin{equation}
t(\tau)=\begin{cases}
\displaystyle\frac{{1}}{27t^2_0}(x(\tau)-b)^{3},&\,\, x>b,\\[2ex]
-\displaystyle\frac{1}{27t^2_0}(x(\tau)-b)^{3},& \,\, x<b.
\end{cases}  \label{geodesicas dos tercios}%
\end{equation}
The lightlike geodesics in FLRW universe are illustrated in the following figure.
\begin{figure}[H]
\centering
\includegraphics[scale=0.45]{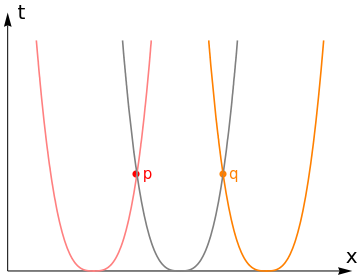}\caption{Lightlike geodesics in the FLRW model. The past cones of events $p$ and $q$ are disjoint.}%
\end{figure}

\begin{figure}[H]
\centering
\includegraphics[scale=0.4]{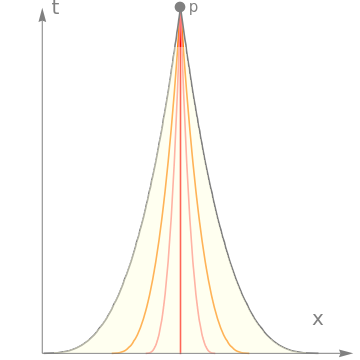}\caption{The red vertical trajectory is the longest timelike curve going to the event $p$. Its length is the cosmic time at $p$. The yellow region is the past cone of $p$. Events that are not in the yellow region can not be seen by an observer at $p$. Not enough time has passed for light to reach $p$. }\label{geotime}
\end{figure}

Let $p$ an event in the FLRW universe.
The cosmic time coordinate $p^{0}=t(p)$  can be defined intrinsically in terms of the geometry of the metric. 
The value $p^0$ is
the maximum of the proper times of all timelike curves that go to $p$.  Clearly, the constant geodesic $\gamma(\tau)=(\tau,p^1,p^2,p^3)$
has proper time $p^0$.
Let us assume $\eta(\tau)$ is a timelike curve that
ends at $p$. This curve can be reparametrized to take the form $\eta(\tau)=(\tau,x^{i}(\tau))$. Then, it has proper time 
\[ L(\eta)=\int_{0}^{p^0}\sqrt{1-
\sum_{i}
\Big(\frac{dx^{i}}{d\tau}\Big)^{2}}d\tau \leq \int_{0}^{p^0}d\tau \leq p^0.\] 
Figure \ref{geotime} illustrates the situation.

\section{Conformal flatness and Penrose diagram}

\begin{figure}[H]
\begin{center}
 \vspace{0pt} \includegraphics[scale=0.35]{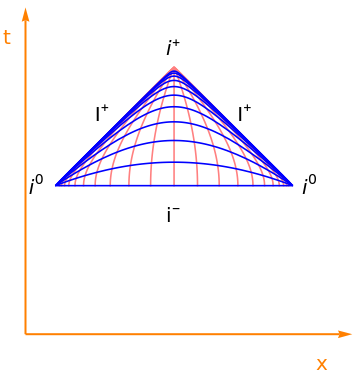} \quad
  \vspace{0pt} \includegraphics[scale=0.35]{Figures/penroseFLRW} \quad
  \vspace{0pt} \includegraphics[scale=0.35]{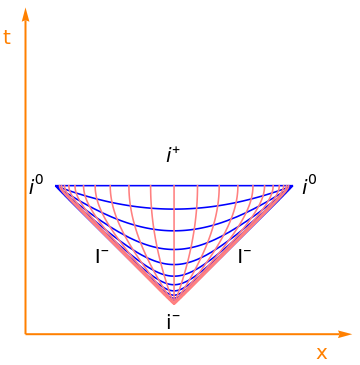}
  \end{center}
\caption{Penrose diagrams for FLRW metric with $k=0$ and $A(t)=\epsilon t^q$. Diagram on the left corresponds to  $0<q<1$, the one in the center to $q=1$, and the one on the right to $q>1$. The blue lines correspond to constant values of $t$ and the red lines to constant values of $x$.}
\label{PenFLRW}
\end{figure}

Consider the FLRW metric
\begin{equation*}
g=-dt \otimes dt+ A^2(t)\left[ \frac{1}{1-kr^{2}}dr\otimes dr+r^{2}
\left(d\theta\otimes d\theta+\sin^{2}\theta\text{ }d\varphi\otimes d\varphi\right)\right],
\end{equation*}
in coordinates $(\eta,r,\theta,\varphi)$ where
\begin{equation} \eta(t)= \int_0^t \frac{ds}{A(s)}.\end{equation}
The function $\eta$, known as conformal time, satisfies $ A(t)d\eta=dt$,
so that the metric takes the form
\begin{equation*}
g=A^2(t)\left[-d\eta \otimes d\eta+  \frac{1}{1-kr^{2}}dr\otimes dr+r^{2}
\left(d\theta\otimes d\theta+\sin^{2}\theta\text{ }d\varphi\otimes d\varphi\right)\right].
\end{equation*}
A metric $h$ is called conformally flat if, locally, it is conformally equivalent to a flat manifold. Spheres and hyperbolic spaces are conformally flat. One concludes that the FLRW is conformally flat.
Let us specialize the discussion to the flat case $k=0$. In order to study the causal structure of $g$ we may disregard the
conformal factor $A(t)^2$, and consider the metric
\begin{equation}
\tilde{g}=-d \eta \otimes d \eta + dx \otimes dx + dy \otimes dy + dz \otimes dz.
\end{equation}
The values taken by the coordinate $\eta$ depend on the scaling factor $A(t)$. Let us assume that $A(t)=\varepsilon t^q,$ for $\varepsilon, q>0$.
Then
\[
\eta(t)=\begin{cases}
\varepsilon \ln(t), & q=1,\\[1ex]
\displaystyle\frac{\varepsilon t^{1-q}}{1-q}, & q \neq 1.
\end{cases}
\]
As $t$ varies in $(0,\infty)$, the values taken by $\eta$ are
\[
\text{Values taken by $\eta$}=\begin{cases}
(0, \infty),& 0<q<1,\\
(-\infty, \infty),& q=1,\\
(-\infty,0),& q>1.
\end{cases}
\]
Therefore, the Penrose diagrams for the FLRW metric is the same as that of the corresponding region in Minkowski spacetime.
These diagrams are depicted Figure \ref{PenFLRW}.

\section{Cosmological red shift and Hubble's law}

\begin{figure}[H]
\centering
\includegraphics[scale=0.52]{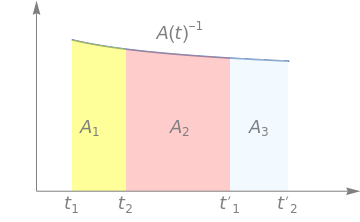}
\caption{}
\label{area1}
\end{figure}

In this section we consider a flat FLRW metric
\begin{equation}
g=-dt \otimes dt +A^2(t)\big(dx \otimes dx + dy \otimes dy + dz\otimes dz \big)
\end{equation}
with a more general scaling function $A(t)$, that
we assume to be positive and increasing. Let $\kappa(\tau)
=(\tau,c,0,0)$ and $\beta(\tau)=(\tau,b,0,0)$, with $b>c$, be the worldlines of two galaxies $C$ and $B$, respectively.  Consider two pulses of light emitted from $B$ at times $t_1$ and $t_{2}=t_{1}+h$. Suppose that  the pulses of light arrive at $C$ at times
 $t_{1}^{\prime}$ and $t_{2}^{\prime}$, respectively. We
want to estimate $h^{\prime}=t_{2}^{\prime}-t_{1}^{\prime}$, as well as the
quotient $h/h^{\prime}$. We know that a null geodesic \[\gamma(\tau)=(t(\tau)
,x(\tau),0,0),\] such that $\gamma(0)=(t_{1},b,0,0)$ must satisfy the equation 
\begin{equation}\Big(\frac{dt}{d\tau}\Big)^{2}=A^{2}(t(\tau))\Big(\frac{dx}{d\tau}
\Big)^{2}.\end{equation}Therefore, 
\begin{equation}\label{geocrs}\frac{dt}{d\tau}=\pm A(t(\tau))\frac{dx}{d\tau}
.\end{equation}
 Suppose that at $\gamma(\tau_{1})=(t_{1}^{\prime},a,0,0).$ Integrating both sides of (\ref{geocrs})
yields
\begin{equation}\label{cmb}
c-b    =x(\tau_{1})-x(0)=\int_{0}^{\tau_{1}}\frac{dx}{d\tau}
d\tau
  =\pm\int_{0}^{\tau_{1}}\frac{1}{A(t(\tau))}\frac{dt}{d\tau}d\tau=-\int_{t_{1}}%
^{t_{1}^{\prime}}\frac{dt}{A(t)},
\end{equation}
where the sign is positive because the left hand side is negative.
Similarly, let $\lambda(\tau)$ be a null geodesic such that
$\lambda(0)=(t_{2},b,0,0),$ and $\lambda(\tau_{2})=(t_{2}^{\prime},c,0,0)$. Then
\begin{equation}
c-b=-\int_{t_{2}}%
^{t_{2}^{\prime}}\frac{dt}{A(t)}.
\end{equation}
If $A_1$, $A_2$ and $A_3$ are the areas depicted in figure \ref{area1}, 
one concludes that $A_{1}+A_{2}=A_{2}+A_{3}$, and consequently $A_{1}=A_{3}$

\begin{figure}[H]
\centering
\includegraphics[scale=0.55]{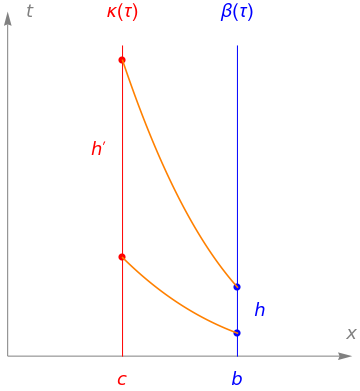}\caption{Cosmological
Redshift}\label{crsg}
\end{figure}

In this case, $h=t_{1}-t_{2}$ and $h^{\prime}=t_{1}^{\prime}-t_{2}^{\prime}$ correspond to the periods of some light signal. Therefore, they are small and one can use the mean value theorem to estimate
$A_{1}$ and $A_{3}$ as:
\[
A_{1}=\int_{t_{1}}^{t_{2}}\frac{dt}{A(t)}\approx \frac{h}{A(t_{1})},
\]
and%
\[
A_{3}=\int_{t_{1}^{\prime}}^{t_{2}^{\prime}}\frac{dt}{A(t)}\approx \frac{h^{\prime}}%
{A(t_{1}^{\prime})}.
\]
One concludes that
\begin{equation}
\frac{h^{\prime}}{h}=\frac{A(t_{1}^{\prime}%
)}{A(t_{1})}>1. \label{red1}%
\end{equation}
This phenomenon is known as
\emph{cosmological redshift}. The wavelength of emitted
radiation is lengthened due to the expansion of the universe, and this shifts
visible light toward the red side of the spectrum. The situation is depicted in figure \ref{crsg}.

The quotient in (\ref{red1}) is usually written as $z+1$, with
\begin{equation}
z=\frac{h^{\prime}-h}{h}=\frac{A(t_{1}^{\prime})}{A(t_{1})}-1=\frac
{A(t_{1}^{\prime})-A(t_{1})}{A(t_{1})}. \label{red shift factor}%
\end{equation}
The quantity $z$ is called the \emph{redshift factor }corresponding to the
celestial object represented by $\beta(\tau).$ The redshift factor can be determined experimentally.
By studying the properties of the received light, it is possible to determine the chemical composition of the emitting object.
From this chemical composition the wavelength in the rest frame of the emitting object can be obtained. The quotient of the wavelengths gives $z$. By measuring $z>0$ for some particular object, one concludes that $A(t'_1)>A(t_1)$, so that the function $A(t)$ is increasing, and the universe is expanding.\\

The expansion of the universe makes the notion of distance between celestial objects rather subtle. We say that an observer $C$ is comoving if its spatial coordinates are constant in comoving coordinates $(t, x,y,z)$. That is, if the worldline of $C$ is $\kappa(\tau)=(\tau,c^1,c^2,c^3)$. We will consider two different notions of distance between comoving observers $C$ and $B$.

\subsubsection{Comoving distance} This is the distance defined by the time it takes a photon to go from $B$ to $C$. Consider a ray of light $\gamma(\tau)$ going from $p=(t_1,b^1,b^2,b^3)$ to $q=(t'_1,c^1,c^2,c^3)$. Since $\gamma(\tau)$ is lightlike it satisfies
\begin{equation}\label{comovingd}
\frac{dt}{d\tau}=A(t)\sqrt{\Big(\frac{dx^1}{d\tau}\Big)^2+\Big(\frac{dx^2}{d\tau}\Big)^2+\Big(\frac{dx^3}{d\tau}\Big)^2}.
\end{equation}
The comoving distance is
\begin{equation}
D_c=\int_{\tau_0}^{\tau_1}\sqrt{\Big(\frac{dx^1}{d\tau}\Big)^2+\Big(\frac{dx^2}{d\tau}\Big)^2+\Big(\frac{dx^3}{d\tau}\Big)^2}d\tau 
\end{equation}

In view of equation (\ref{comovingd}), this is equal to:
\begin{equation}
D_c=\int_{t_1}^{t'_1} \frac{dt}{A(t)}
\end{equation}

\subsubsection{Proper distance} 
\begin{figure}[H]
\centering
\includegraphics[scale=0.44]{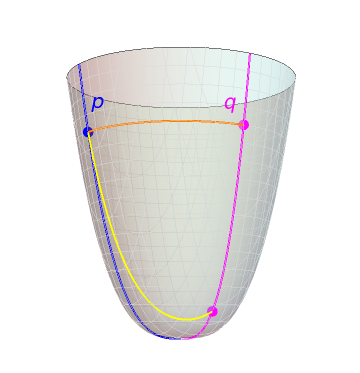}
\caption{The gray surface represents the expanding universe. Cosmic time flows in the vertical direction. The proper distance between $p$ and $q$ is the length of the orange curve. It changes as the universe expands. The yellow line represents the trajectory of a photon.
The time difference between the emision and the reception of the photon is the comoving distance between $p$ and $q$.}
\label{distances}
\end{figure}

Given a fixed value of cosmic time $t=t_1$, the proper distance between $B$ and $C$ is the distance measured in the Riemannian manifold $t=t_1$. It depends on the geometry of space at a specific time. 
Figure \ref{distances} illustrates the two different notions of distance.

The recession velocity of two comoving objects $B$ and $C$ is the rate of change of the proper distance with respect to time
\[ v=\frac{d D_p}{dt}.\]
In terms of the Hubble function \begin{equation}H(t)=\frac{A'(t)}{A(t)},\end{equation}
the recession velocity is
\begin{equation}
v=H(t)D_p(t).
\end{equation}
This relation is known as the Hubble law. Note that the speed of light is not a bound on the recession velocity $v$. This is not in contradiction with special relativity. There is no object whose worldline is not timelike in this situation.

Let us consider the redshift factor of a celestial object that is close to our galaxy.  
In this case, $t'_1=t_0$ and $t_0-t_1\ll t_0$. 
For a scaling factor of the form $A(t)=\epsilon t^q=\Big(\frac{t}{t_0}\Big)^q$ with $0<q<1$, the second derivative
\[ A''(t_0)=\frac{q(q-1)}{t^2_0}<0\]
is small. Therefore, we can use the approximation
\begin{equation}\label{appH}
\frac{A(t_0)-A(t_1)}{t_0-t_1}\approx A'(t_0) .
\end{equation}
Let $b=D_c$ be the comoving distance. We know that
\[ b=\int_{t_1}^{t_0}\frac{dt}{A(t)} =\int_{t_1}^{t_0}\frac{dt}{\epsilon t^q}=\frac{t^{1-q}_0-t^{1-q}_1}{\epsilon(1-q)}.\]
Therefore
\begin{equation}\label{appt0}
t_{0}=\Big((1-q)\epsilon b+(t_{1})^{1-q}\Big)^{1/(1-q)}.
\end{equation}

Using the Taylor expansion for the binomial function
\[
(x+\delta)^{r}=x^{r}+rx^{r-1}\delta+\frac{r(r-1)}{2}\delta^{2}x^{r-2}+\cdots
\]
with $r=1/(1-q)$ one can approximate (\ref{appt0}) as

\begin{equation}\label{apppt0}
t_0 \approx t_1+bA(t_1).
\end{equation}

Relations (\ref{appt0}) and (\ref{appH}) allow one to estimate the redshift factor as follows
\begin{align}\label{approxH}
z=\frac{A(t_0)-A(t_1)}{A(t_1)}\approx \frac{(t_0-t_1)A'(t_0)}{A(t_1)}\approx \frac{b A(t_1) A'(t_0)}{A(t_1)}\approx b H(t_0)A(t_0)\approx H_0 D_p.
\end{align}

The approximation (\ref{approxH}) provides a way to estimate the present value of the Hubble function $H_0=H(t_0)$.
As we mentioned before, the redshift parameter $z$ can be determined experimentally. In some situations it is also possible to determine the proper distance $D_p$. The dimension of the Hubble constant is inverse time, and it is usually measured in units $\frac{\text{km/s}}{\text{ Mpc}}$ where 1 Megaparsec is
\[ 1\:\text{Mpc}\approx 3.08 \times 10^{22} \: \mathrm{m}.\]
The current estimate for the Hubble constant $H_0$ is
\[ H_0 \approx 70\: \frac{\text{kg/s}}{\text{Mpc}}.\]

\section{Age and diameter of the observable universe}

\begin{figure}[h]
\centering
\includegraphics[scale=0.48]{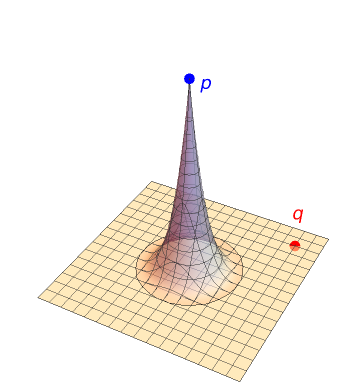}
\caption{ Light emitted from $q$ has not yet had time to reach $p$.}
\label{cosmological horizon}
\end{figure}

The measurement of the Hubble
constant provides an upper bound for the age of the universe. We assume a scaling factor $A(t)$ such that 
$A(0)=0$, so that the big bang occurs at $t=0$. Moreover, we assume that $A''(t)<0$, which is the case
if $A(t)=\epsilon t^q$ for $0<q<1$. Notice that, in the absence of cosmological constant, the Friedmann equation (\ref{RW1.1}) reads
\begin{equation}
\frac{A^{\prime\prime}(t)}{A(t)}=-\frac{4\pi}{3}\big(\rho(t)+3p(t)\big).
\end{equation}
Therefore, the requirement that $A''(t)<0$ is equivalent to $\rho(t)+3p(t)>0$, which is a condition on the energy distribution of the universe.
With this assumption, the function $A'(t)$ is decreasing and therefore, for $t \in [0,t_0]$ one has
\[ A'(t)>A'(t_0).\]
Integrating both sides one gets
\[ A(t) > A'(t_0) t.\]
One concludes that
 \begin{equation}
 t_0 < \frac{A(t_0)}{A'(t_0)}=\frac{1}{H_0}\approx \frac{1}{70}\:\frac{\text{Mpc s}}{\text{km}}\approx \frac{3.08 \times 10^{19}}{70}\:\text{s}\approx 13.95 \text{ Billion years}.
 \end{equation}
 The current estimate for the age of the universe is 
 \begin{equation}
 t_0 \approx 13.8 \text{ Billion years}.
 \end{equation}
Once $t_0$ is known, the size of the observable universe can be determined as follows. Suppose that $\gamma(\tau)$ is the worldline of a photon emitted at the big bang so that $\gamma(0)=(0,b,0,0)$ which reaches our galaxy at present time, $\gamma(\tau_0)=(t_0,0,0,0)$. The size of the observable universe is 
\begin{equation}\label{DH}
D_h= A(t_0) b=b.
\end{equation}
This is a reasonable definition because, for a comoving observer $C$ with comoving distance $c>d$ from our galaxy, not enough time has passed for light from $C$ to reach us. Then, by equation (\ref{cmb}), we know that

\begin{equation}
b    =\int_{0}%
^{t_0}\frac{dt}{A(t)}.
\end{equation}
For instance, if one assumes $A(t)=(t/t_0)^{2/3}$, this can be computed as
\begin{equation}
b =\int_{0}%
^{t_0}\frac{dt}{A(t)}=t^{2/3}_0\int_0^{t_0}t^{-2/3}dt=3t^{2/3}_0 (t^{1/3}_0)=3t_0.
\end{equation}
This gives the following estimate for the size of the observable universe \begin{equation}
D_h=3t_0 \approx 41.8 \text{ Billion light years}.
\end{equation}
The current precise estimate for the size of the observable universe is
\begin{equation}
D_h\approx 93 \text{ Billion light years.}
\end{equation}
Figure \ref{cosmological horizon} illustrates the situation.
   \clearemptydoublepage

  \begin{partwithabstract}{Appendices}
\end{partwithabstract}

 \appendix
 \chapter{Linear algebra and tensors\label{0algebra basicos}}

\begin{center}
\parbox[b]{0.9\textwidth}{\small \sl In this section we review the main notions of multilinear algebra and
introduce the notation used in the text for manipulating tensors. A full treatment of the subject, including complete
proofs for all statements, is available in any book on linear
algebra. We recommend Lang's book \cite{Lang}.
 }
\end{center}

\vspace{3ex}

\section{Linear algebra and matrices}

Let $V$ and $V'$ be real vector spaces of dimensions $n$ and $m$,
respectively. Let us consider a choice of basis $\mathfrak{B}=\{e_{a}\}$ and $\mathfrak{B}'
=\{e_a'\}$ for $V$ and $V'$. Given a linear map $f:V\rightarrow V'$, we denote by $A=[f]_{\mathfrak{B}^{\prime}\mathfrak{B}}$ the matrix associated to $f$ in
the bases $\mathfrak{B}$ and $\mathfrak{B}^{\prime}$. This is defined  by the condition 
\[f(e_{b})=
\sum_{a}
A^{a}_{\phantom{a}b}e'_a.\]
The correspondence between linear transformations and matrices respects the
composition of functions. Consider linear transformations $f:V\rightarrow V^{\prime}$ and
$g:V^{\prime}\rightarrow V^{\prime\prime}$  and let
$\mathfrak{B}$, $\mathfrak{B}^{\prime}$ and $\mathfrak{B}^{\prime\prime}$ be bases for $V,$ $V^{\prime}$ and
$V^{\prime\prime}$, respectively. Then, a direct computation shows that \[[g\circ f]_{\mathfrak{B}^{\prime\prime}\mathfrak{B}}=[g]_{\mathfrak{B}^{\prime\prime
}\mathfrak{B}^{\prime}}[f]_{\mathfrak{B}^{\prime}\mathfrak{B}}.\]
If $V$ and $V^{\prime}$ are vector spaces, $\mathrm{Hom}(V,V^{\prime})$ will denote the vector space of linear maps from $V$ to
$V^{\prime}$. It has dimension $mn$. Moreover, any choice of bases
$\mathfrak{B}$ and $\mathfrak{B}^{\prime}$ determines an isomorphism between $\mathrm{Hom}(V,V^{\prime})$ and $\mathrm{Mat}_{m\times n}(\RR),$ the
space of $m\times n$ matrices with entries in $\RR$, given by the
linear map
\[
\mathrm{Hom}(V,V^{\prime})\rightarrow\mathrm{Mat}_{m\times
n}(\RR),\qquad f\mapsto\lbrack f]_{B^{\prime}B}.
\]

Recall that the \emph{dual} of a vector space $V$, denoted by $V^{\ast},$
is the vector space $\mathrm{Hom}(V,\RR)$ of all linear maps from $V$ to $\RR$.
Any linear map $f:V\rightarrow W$ induces
canonically a linear transformation $f^{\ast}:W^{\ast}\rightarrow
V^{\ast},$ by sending each functional $\alpha\in W^{\ast}$ into $\alpha\circ
f\in V^{\ast}$. For each choice of basis $\mathfrak{B}=\{e_{a}\}$ for $V$ we denote by
$\mathfrak{B}^{\ast}=\{e^{a}\}$ its \textit{dual basis}, where $e^{a}$ is the
functional that takes the value $1$ when evaluated at $e_{a},$ and the value
zero when evaluated at any other vector $e_{b},$ with $b\neq a$. If
$A=[f]_{\mathfrak{B}'\mathfrak{B}}$ represents $f$ with respect to the bases $\mathfrak{B}$ and
$\mathfrak{B}'$ then the linear map $f^{\ast}$ is represented --in the
respective dual bases-- by the \emph{transpose} of $A$, the matrix obtained
from $A$ by interchanging rows and columns. In symbols \[[f^{\ast
}]_{\mathfrak{B}^{\ast}{\mathfrak{B}'}^{\ast}}=A^{\mathrm{T}}.\]
Suppose that $V$ and $W$ are finite dimensional vector spaces.
The map $V^{\ast}\otimes
W\rightarrow\mathrm{Hom}(V,W)$ that sends 
a generator $\alpha\otimes w\in V^{\ast}\otimes W$ to the linear
transformation
$f_{\alpha\otimes w}(v)=\alpha(v)w$, is a linear isomorphism.

A map $g:V\times V\rightarrow\RR$ is called\emph{
bilinear }if it is linear in each of the two arguments separately. Once one fixes
a basis $\mathfrak{B}$ for $V$, a bilinear form is determined by a matrix $G$ whose entries are
$G_{ij}=g(v_i,v_j)$. If $\mathfrak{B}$ and $\mathfrak{B}'$ are two bases for $V$ then the corresponding matrices are related by: 
\begin{equation}
G=P^{\mathrm{T}}G'P \label{E2},%
\end{equation}
where $P$ is the change of basis matrix from $\mathfrak{B}$ to $\mathfrak{B}'$.
The bilinear map $g$ is called \emph{symmetric} if $g(v,u)=g(u,v)$.
A simple exercise shows that $g$ is symmetric if and only if $G=G^{\mathrm{T}}$. The map $g$ is called \emph{non-degenerated} if $g(v,\cdot)$ is the
zero function only if $v=0.$ A symmetric and non-degenerated bilinear map $g$
is called an \emph{inner product}.  When $g$ is an inner product we will often write
$\left\langle v,u\right\rangle$ instead of $g(v,u)$. An inner product is called
\emph{positive-definite} if it also satisfies $g(v,v)\geq0$, for all vectors
$v \in V$. The norm of a vector $v$ is
\[|v|=\sqrt{|\langle v,v\rangle|}.\]

In general, a space $V$ and its dual $V^{\ast}$ are not isomorphic in a
canonical way. However, an inner product $\left\langle \cdot,\cdot\right\rangle$
in $V$ defines an isomorphism between $V$ and $V^{\ast}$ that sends
$v\in V$ to the functional $\alpha_{v}:V\rightarrow\RR$ defined as
$\alpha_{v}(u)=\left\langle v,u\right\rangle$. In any vector space
$V\ $endowed with an inner product $g$ there is always a basis $\mathfrak{B}$ that puts
$G$ in diagonal form (see \cite{Lang}, Chapter XIV). That is, such that,
\begin{equation}
G=\left(\begin{array}{cc}
-\mathrm{id}_k & 0\\
0& \mathrm{id}_{n-k}
\end{array}\right)
\end{equation}
where $k\geq0$ entries in the diagonal equal to $-1,$ and $n-k\geq0$ entries
equal to $1$. The numbers $k$ and $n-k$ depend only on the bilinear form $g$ and are called the signature of $g$. A basis for which $G$ has the form above is called an \emph{orthonormal basis} for $V$ with respect to the bilinear form $g$. The Gram-Schmidt algorithm provides a method for computing an orthonormal basis of the vector space $V$ with respect to $g$ given an arbitrary basis. 

\section{Tensor products}

Let $V$ and $V'$ be two vector spaces. There exists a
vector space $V\otimes V'$ and a bilinear map $\pi
:V\times V' \rightarrow V\otimes V'$ with the following
universal property. Given any bilinear map $f:V\times 
V' \rightarrow W$ into another vector space $W$ there exists a unique linear
transformation $\bar{f}:V\otimes V'\rightarrow W$ such that the
following diagram commutes:%
\[\xymatrix{
V\times V' \ar[r]^\pi \ar[d]_f&  V \otimes V'\ar[dl]^{\bar{f}}\\
W&}
\]
The vector space $V \otimes V'$ is called the tensor product of $V$ and $V'$.
It is an easy exercise to show that if $(V\otimes V,\pi)$ exists,
then it must be unique up to natural isomorphism. This means that if $(U,\pi
^{\prime})$ is another pair satisfying the same universal property, then there
is an isomorphism $h:V\otimes V' \rightarrow U$ making the following
diagram commute:
\[\xymatrix{
V\times V' \ar[r]^\pi \ar[d]_{\pi'}&  V \otimes V'\ar[dl]^{h}\\
U&}
\]
We will now describe the construction of the tensor product of vector spaces.
Let $V$ and $V'$ be real vector spaces and denote by $F$ the
vector space of all real linear combinations of elements of the set
\[B=\{e(v,v')\mid v\in V,v'\in V' \}.\]
Let $H$ be the subspace of $F$ 
generated by vectors of one of the following four forms:
\begin{enumerate}
\item $e(v_{1}+v_{2},v')-e(v_{1},v')-e(_2%
,v')$.

\item $e(c v,v')-c e(v,v')$.

\item $e(v,v'_{1}+v'_{2})-e(v,v'%
_{1})-e(v,v'_{2})$.

\item $e(v,c v')-c e(v,v'),$
\end{enumerate}
The tensor product of $V$ and $V'$, denoted by
$V\otimes V',$ is the quotient vector space $F/H.$ The equivalence
class of $e(v,v')$ in $V \otimes V'$ is denoted by $v\otimes v'$.
It is clear from this definition that each element in the tensor product
can be written (not necessarily in a unique way) as a sum of the form
$\sum_i v_i \otimes v'_i$. One can easily verify that the following relations hold on $V \otimes V'$:
\begin{align*}
v\otimes(v'_{1}+{v'}_{2})&=v\otimes
v'_{1}+v\otimes v'_{2},\\
(v_{1}+v_{2})\otimes v'  &  =v_{1}\otimes
v'+v_{2}\otimes v'\text{,}\\
\alpha(v\otimes v')  &  =(\alpha v)\otimes v'%
=v\otimes(\alpha v').
\end{align*}
The map $\pi:V\times V' \rightarrow V\otimes V'$ is defined by
by \[\pi(v,v')=v\otimes v'.\] This function is
clearly bilinear. It is easy to see that $(V\otimes V'$,
$\pi)$ satisfies the universal property of the tensor
product construction. There exist natural isomorphisms between various tensor products. We leave it as a simple exercise to the reader to prove the following proposition, which can also be found many books on linear algebra, for instance (\cite{Lang},
chapter XVI).

\begin{proposition}
Let $V_1, \cdots, V_n$ be real vector spaces. There exist canonical
isomorphisms of vector spaces defined as follows:

\begin{enumerate}
\item[1.] $V_1 \otimes V_2 \cong V_2 \otimes V_1$, where
$ v_1 \otimes v_2 \mapsto v_2 \otimes v_1$.

\item[2.] $V_1\otimes(V_2 \otimes V_3)\cong(V_1\otimes V_2)\otimes
V_3$, where $v_1 \otimes (v_2 \otimes v_3) \mapsto (v_1 \otimes v_2)\otimes v_3$.

\item[3.] $\RR\otimes V_1\cong V,$ where $c\otimes v \mapsto c v.$

\item[4.] $(V_{1}\oplus V_{2}\oplus\cdots\oplus V_{n-1})\otimes {V_n}%
\cong (V_{1}\otimes V_n)\oplus(V_{2}\otimes V_n)\oplus
\cdots\oplus(V_{n-1}\otimes {V_n}),$ where $(v_{1},\ldots,v_{n-1}
)\otimes v_n\mapsto (v_{1}\otimes v_n,\cdots
,v_{n-1}\otimes v_{n}).$

\item[5.] If $\mathfrak{B}=\{e_{a}\}$ and $\mathfrak{B}'=\{e'_{b}\}$ are bases for
$V$ and $V'$ respectively, then $\mathfrak{B}\otimes \mathfrak{B}'
=\{e_{a}\otimes e'_b\}$ is a basis for $V\otimes V'$. If
$\dim V=m$ and $\dim V'=n$ then $\dim(V\otimes
V')=mn$.
\end{enumerate}
\end{proposition}

Let $f:V\rightarrow V^{\prime}$ and $g:W\rightarrow
W^{\prime}$ be linear transformations. There is an induced  linear map $f\otimes g:V\otimes
W\rightarrow V^{\prime}\otimes W^{\prime}$ defined by
\[(f\otimes g)(v\otimes w)=f(v)\otimes g(w).\]
 Let $\mathfrak{B}$ and $\mathfrak{B}^{\prime}$ be bases for $V$ and $V^{\prime}$, and let $\mathfrak{C}$
and $\mathfrak{C}^{\prime}$ be bases for $W$ and $W^{\prime}$, respectively. If
$A=[f]_{\mathfrak{B}^{\prime}\mathfrak{B}}$ and $B=[g]_{\mathfrak{C}^{\prime}\mathfrak{C}}$ are the matrices that
represent $f$ and $g$ in these bases, the matrix $C=[f\otimes g]_{\mathfrak{B}^{\prime
}\otimes \mathfrak{C}^{\prime}\,\mathfrak{B}\otimes \mathfrak{C}}$ represents $f\otimes g$ in the
bases $\mathfrak{B}\otimes \mathfrak{C}$, and $\mathfrak{B}^{\prime}\otimes \mathfrak{C}^{\prime}$. The matrix $C$ is
known as the Kronecker product of $A$ and $B$, and it is denoted by $A \otimes B$. If $A=(A^{a}_{\phantom{a}b})$ and
$B=(B^{a}_{\phantom{a}b})$ are matrices of sizes $p\times n$ and $q\times m$,
respectively, the Kronecker product is the $pq\times mn$ matrix whose block
form is
\[
A\otimes B=\left(
\begin{array}
[c]{ccc}%
A^{1}_{\phantom{1}1}B & \cdots & A^{1}_{\phantom{1}n}B\\
\vdots &  & \vdots\\
A^{p}_{\phantom{p}1}B & \cdots & A^{p}_{\phantom{p}n}B
\end{array}
\right)
\]

More generally, the tensor product of vector spaces
$V_{1},\ldots,V_{r}$ can be defined as a pair $(V_{1}\otimes
\cdots\otimes V_{r},\pi),$ where $\pi:V_{1}\times\cdots\times V_{r}\rightarrow V_{1}\otimes\cdots\otimes V_{r}$
is a multilinear map that satisfies the
following universal property: Given a multilinear map $T: V_1 \times \cdots \times V_r \rightarrow W$ there is a
unique linear transformation $\bar{T}: V_{1}\otimes\cdots\otimes V_{r} \rightarrow W$ making the following diagram
commutative:
\[
\xymatrix{
V_{1}\times\cdots\times V_{r} \ar[r]^{\pi} \ar[d]_T
&V_{1}\otimes\cdots\otimes V_{r} \ar[dl]^{\bar{T}}\\
Z &
}
\]
The tensor product satisfies the following properties:
\begin{enumerate}
\item The pair $(V_{1}\otimes\cdots\otimes V_{r},\pi)$
exists and is unique up to canonical isomorphisms.

\item The vector space $\mathrm{Mult}(V_1\times\cdots\times V_r,W)$
of all multilinear functions from $V_1\times\cdots\times V_r$ into
$W$ is canonically isomorphic to
$\mathrm{Hom}(V_1\otimes\cdots\otimes V_r,W)$.

\item If $f_{i}:V_{i}\rightarrow Z_{i}$ are linear
transformations, there exists a linear map
\[
f_{1}\otimes \cdots \otimes f_{r}:V_{1}\otimes\cdots\otimes V_{r}\rightarrow
Z_{1}\otimes\cdots\otimes Z_{r},
\] such that
 \[(f_1 \otimes \cdots \otimes f_r)(v_{1}\otimes \cdots \otimes v_{r})
=f_{1}(v_{1})\otimes\cdots\otimes f_{r}(v_{r}).\]

\item Let $\mathfrak{B}_{j}=\{e_{1}^{(j)},\ldots,e_{n_{j}}^{(j)}\}$ be bases for $V_{j}$. Then set $\mathfrak{B}=\mathfrak{B}_{1}\otimes
\cdots\otimes \mathfrak{B}_{r}$ of all elements of the form $e_{j_{1}}^{(1)}\otimes\cdots \otimes
e_{j_{r}}^{(r)}$ with $e_{j_k}^{(k)} \in \mathfrak{B}_{k}$
is a basis for $V_{1}\otimes\cdots\otimes V_{r}.$ 
\end{enumerate}

\section{Tensors \label{engorrosa}}

Let $V$ be a vector space with dual space $V^{\ast}$. A tensor of
type $(p,q)$ is an element of the vector space \[\mathscr{T}^{(p,q)}V=V^{\otimes {p}%
}\otimes V^{\ast \otimes {q}}.\]  A linear map $f:V\rightarrow V$
induces a map $\mathscr{T}^{(p,q)}f:\mathscr{T}^{(p,q)}V\rightarrow \mathscr{T}^{(p,q)}V$  defined as the tensor product \[(f\otimes\cdots\otimes f)\otimes(f^{\ast
}\otimes\cdots\otimes f^{\ast})\] of $p$ copies of $f$ and $q$ copies of
$f^{\ast}$. 

Let $\mathfrak{B}=\{e_{a}\}$ be a basis for $V$ with corresponding dual basis $\mathfrak{B}_{r}^{\ast}=\{e^{a}\}$ for $V^{\ast}$. In the basis for $\mathscr{T}^{\left(  p,q\right)  }V$
given by $\mathfrak{B}^{(p,q)}=\{e_{a_{1}}\otimes
\cdots\otimes e_{a_{p}}\otimes e^{b_{1}}\otimes\cdots\otimes e^{b_{q}}\}$ any $(p,q)$ tensor $T$ can be
written as \[T=
\sum_{a_1,\dots,a_p} \sum_{b_1,\dots,b_q}
T^{a_{1}\ldots a_{p}}_{\phantom{a_{1}\ldots a_{p}}b_{1}\ldots b_{q}}e_{a_{1}}\otimes
\cdots\otimes e_{a_{p}}\otimes e^{b_{1}}\otimes\cdots\otimes e^{b_{q}%
}.
\]
The tensor product of the
$(p,q)$ tensor $T$ and the $(r,s)$ tensor $S$  is the $(p+r, q+s)$ tensor $T \otimes S$ 
whose components are \[(T \otimes S)^{a_{1}\ldots a_{p}c_{1}\ldots c_{r}}_{\phantom{a_{1}\ldots a_{p}c_{1}\ldots c_{r}}b_{1}\ldots
b_{q}d_{1}\ldots d_{s}}=T^{a_{1}\ldots a_{p}}_{\phantom{a_{1}\ldots a_{p}}b_{1}\ldots b_{q}}
S^{c_{1}\ldots c_{r}}_{\phantom{c_{1}\ldots c_{r}}d_{1}\ldots d_{s}}.\] Explicitly,
\begin{align*}
T\otimes S=
\sum_{a_1,\dots,a_p} \sum_{b_1,\dots,b_q}\sum_{c_1,\dots,c_r} \sum_{d_1,\dots,d_s}
& T^{a_{1}\ldots a_{p}}_{\phantom{a_{1}\ldots a_{p}}b_{1}\ldots b_{q}}
S^{c_{1}\ldots c_{r}}_{\phantom{c_{1}\ldots c_{r}}d_{1}\ldots d_{s}}  e_{a_{1}}\otimes
\cdots\otimes e_{a_{p}} \\
&\otimes e_{c_{1}}\otimes
\cdots \otimes e_{c_{r}} \otimes e^{b_{1}}\otimes\cdots\otimes e^{b_{q}} \otimes e^{d_{1}}\otimes\cdots\otimes e^{d_{s}}
\end{align*}
A $(p,q)$ tensor \[T=
\sum_{a_1,\dots,a_p} \sum_{b_1,\dots,b_q}
T^{a_{1}\ldots a_{p}}_{\phantom{a_{1}\ldots a_{p}}b_{1}\ldots b_{q}}e_{a_{1}}\otimes
\cdots\otimes e_{a_{p}}\otimes e^{b_{1}}\otimes\cdots\otimes e^{b_{q}%
}\] can be regarded as a multilinear
map $V^{\ast\otimes p}\otimes V^{\otimes q}\rightarrow\RR$ that sends each element $\alpha_{1}\otimes\cdots\otimes\alpha_{p}\otimes v_{1}%
\otimes\cdots\otimes v_{q}$ to
\[
\sum_{a_1,\dots,a_p} \sum_{b_1,\dots,b_q}
T^{a_{1}\ldots a_{p}}_{\phantom{a_{1}\ldots a_{p}}b_{1}\ldots b_{q}}\alpha_{1}(e_{a_{1}}%
)\cdots\alpha_{p}(e_{a_{p}}) e^{b_{1}}(v_1)\cdots e^{b_{q}}(v_q).
\]

\section{Change of basis\label{change basis}}

Let $\mathfrak{B}=\{e_{a}\}$ and $\mathfrak{B}'=\{e'_{a}\}$ be two bases for
$V$ with corresponding dual bases $\mathfrak{B}^{\ast}=\{e^{b}\}$ and $\mathfrak{B}'^{\ast}=\{
e'^{b}\}$ . We know that
$
\mathfrak{B}^{(p,q)}=\{e_{a_{1}}\otimes
\cdots\otimes e_{a_{p}}\otimes e^{b_{1}}\otimes\cdots\otimes e^{b_{q}}\}$ and $ \mathfrak{B}'^{(p,q)}=\left\{ e'_{a_{1}}\otimes
\cdots\otimes e'_{a_{p}}\otimes e'^{b_{1}}\otimes\cdots\otimes e'^{b_{q}} \right\} $
are bases for $\mathscr{T}^{(p,q)}V$. Let $P=[\mathrm{id}_V]_{\mathfrak{B}'\mathfrak{B}}$ be the
bases change matrix form $\mathfrak{B}$ to $\mathfrak{B}'$. We know that $P^{\mathrm{T}}$  is the bases change matrix from $\mathfrak{B}'^{\ast}$ to $\mathfrak{B}^{\ast}$. Thus,
$Q=(P^{\mathrm{T}})^{-1}=[\mathrm{id}_{V^*}]_{\mathfrak{B}'^{\ast}\mathfrak{B}^{\ast}}$ is the
bases change matrix from $\mathfrak{B}^{\ast}$ to $\mathfrak{B}'^{\ast}.$  Let us  compute the bases change matrix from $\mathfrak{B}^{(p,q)}$ to ${\mathfrak{B}'}%
^{(p,q)}$, that is, the matrix $P^{(p,q)}=[\mathrm{id}_{\mathscr{T}^{(p,q)}V}]_{{\mathfrak{B}'}^{(p,q)}%
\mathfrak{B}^{(p,q)}}$. By definition, the entries $P^{a}_{\phantom{a}c}$ of $P$
satisfy \[e_{c}=\sum_{a}P^{a}_{\phantom{a}c}{e'}_{a}.\] If $Q^{d}_{\phantom{d}b}$ denote
the entries of $Q$, then \[e^{d}=\sum_{b}Q^{d}_{\phantom{d}b}e'^{b}.\]
Hence, $e_{c_{1}}\otimes
\cdots\otimes e_{c_{p}}\otimes e^{d_{1}}\otimes\cdots\otimes e^{d_{q}}$ is equal to
\begin{align}\label{E4}
\begin{split}    
&e_{c_{1}}\otimes
\cdots\otimes e_{c_{p}}\otimes e^{d_{1}}\otimes\cdots\otimes e^{d_{q}} \\
&\quad = \left(\sum_{a}
P^{a}_{\phantom{a}c_1}{e'}_{a} \right)\otimes\cdots\otimes \left(
\sum_{a}
P^{a}_{\phantom{a}c_p}e'_{a}\right)\otimes \left(\sum_{b}
Q^{d}_{\phantom{d}b_1}{e'}^{b}\right)\otimes\cdots \otimes\left (\sum_{b}
Q^{d}_{\phantom{d}b_q}{e'}^{b}\right) \\ 
&\quad =
\sum_{a_1,\dots,a_p} \sum_{b_1,\dots,b_q}
P^{a_1}_{\phantom{a_1}c_1}\cdots P^{a_p}_{\phantom{a_p}c_p}Q^{d_1}_{\phantom{d_1}b_1}\cdots Q^{d_q}_{\phantom{d_q}b_q}
e'_{a_{1}}\otimes
\cdots\otimes e'_{a_{p}}\otimes e'^{b_{1}}\otimes\cdots\otimes e'^{b_{q}}. 
\end{split}
\end{align}
Let $T$ be an arbitrary element of $\mathscr{T}^{(p,q)}V$. One can write
\begin{align*}
T&=
\sum_{c_1,\dots,c_p} \sum_{d_1,\dots,d_q}
T^{c_{1}\ldots c_{p}}_{\phantom{c_{1}\ldots c_{p}}d_{1}\ldots d_{q}}e_{c_{1}}\otimes
\cdots\otimes e_{c_{p}}\otimes e^{d_{1}}\otimes\cdots\otimes e^{d_{q}%
} \\
&=
\sum_{a_1,\dots,a_p} \sum_{b_1,\dots,b_q}
T'^{a_{1}\ldots a_{p}}_{\phantom{a_{1}\ldots a_{p}}b_{1}\ldots b_{q}}e'_{a_{1}}\otimes
\cdots\otimes e'_{a_{p}}\otimes e'^{b_{1}}\otimes\cdots\otimes e'^{b_{q}%
}
\end{align*}
in the bases $\mathfrak{B}^{(p,q)}$and ${\mathfrak{B}'}^{(p,q)}$, respectively. Equation
(\ref{E4}) shows that the coefficients of $T$ in the different basis are related by the following formula:
\begin{equation}\label{E5}
T'^{a_{1}\ldots a_{p}}_{\phantom{a_{1}\ldots a_{p}}b_{1}\ldots b_{q}}=\sum_{c_1,\dots,c_p} \sum_{d_1,\dots,d_q} P^{a_1}_{\phantom{a_1}c_1}\cdots P^{a_p}_{\phantom{a_p}c_p}Q^{d_1}_{\phantom{d_1}b_1}\cdots Q^{d_q}_{\phantom{d_q}b_q}
T^{c_{1}\ldots c_{p}}_{\phantom{c_{1}\ldots c_{p}}d_{1}\ldots d_{q}}.
\end{equation}

\section{Contraction of tensors.}

Let $T$ be a $(p,q)$-tensor and fix a basis $\mathfrak{B}=\{e_a\}$ for $V$. The contraction of $T$ with respect to the indices
$1\leq i\leq p$ and $1\leq j\leq q$ is the $(p-1,q-1)$-tensor with components given by
\[
(C^{i}_{j}T)^{a_{1}\ldots a_{i-1} a_{i+1}\ldots a_{p}}
_{\phantom{a_{1}\ldots a_{i-1} a_{i+1} \ldots a_{p}}b_{1}\ldots b_{j-1} b_{j+1}\ldots b_{q}}=\sum_{c}
T^{a_{1}\cdots a_{i-1} c a_{i+1}\ldots a_{p}}
_{\phantom{a_{1}\ldots a_{i-1} a_{i+1} \ldots a_{p}}b_{1}\ldots b_{j-1} c b_{j+1}\ldots b_{q}}.
\]
The tensor $C^{i}_{j}T$ does not depend on the choice of basis.

Let $V$ be a vector space endowed with an inner product $g=\langle
\cdot,\cdot\rangle$. This bilinear form allows us to identify $V$ and $V^{\ast
}$  by sending a vector $v$ into the functional $\alpha_{v}$
defined as $\alpha_{v}=\langle v,\cdot\rangle$. Let us fix a basis
$\mathfrak{B}=\{e_{a}\}$ for $V$, with dual basis $\mathfrak{B}^{\ast}=\{e^{a}\}$. Denote by $G$ the matrix that represents the bilinear form. It is customary to write the entries of $G^{-1}$ as
$g^{ab}$. If $v=\sum_{a} v^{a}e_{a}$ is a vector, then
$
\alpha_{v}(e_{b})=
\sum_{a}
v^{a}\left\langle e_{a},e_{b}\right\rangle =
\sum_{a}
v^{a}g_{ab}
$.
Thus, $\alpha_{v}$ can be written in the dual basis $\{e^{a}\}$ as
$\alpha_{v}=\sum_{b}v_{b}e^{b}$, with $v_{b}=\sum_{c}g_{bc}v^{c}$. On the other hand, if $\alpha=\sum_{b}\alpha_{b}e^{b}$ is given, then it is
easy to verify that  if $\alpha^a=\sum_{b}g^{ab}\alpha_{b}$ then $v=\sum_{a}
\alpha^{a}e_{a}$ is the unique vector with the property that $\alpha_{v}=\alpha$. One says that $v$ is obtained from $\alpha$ by raising indices or, equivalently, that $\alpha$ is obtained from $v$ by lowering indices.

The isomorphism $V\rightarrow V^{\ast}$ that sends $v$ into
$\alpha_{v}$ can be naturally extended to an identification between $\mathcal{T}^{(p,q)}V$
and $\mathcal{T}^{(p-1,q+1)}V$ as follows. Let  $T$ be a $(p,q)$-tensor. The tensor obtained by lowering the $i$th index,
denoted by $T_{(i)},$ is the $(p-1,q+1)$-tensor

\begin{align*}
T_{(i)}&=\sum_{a_1,\dots,a_p} \sum_{b_1,\dots,b_q} T^{a_{1}\ldots a_{i-1}}{}_{a_{i}}{}^{a_{i+1}\ldots a_{p}}\text{
}_{b_{1}\ldots b_{q}} e_{a_1}\otimes \cdots \otimes e_{a_{i-1}} \otimes e^{a_{i}}\otimes e_{a_{i+1}} \otimes \cdots \otimes e_{a_{p}}  \\
&\qquad \qquad \qquad \qquad\qquad \qquad\qquad \qquad\qquad \qquad \qquad \qquad \quad\: \otimes e^{b_1}\otimes \cdots \otimes e^{b_q}
\end{align*}
where
\[ T^{a_{1}\ldots a_{i-1}}{}_{a_{i}}{}^{a_{i+1}\ldots a_{p}}\text{
}_{b_{1}\ldots b_{q}}=\sum_{c}g_{a_ic} T^{a_{1}\ldots a_{i-1} c a_{i+1} \ldots a_{p}}_{\phantom{a_{1}\ldots a_{i-1} c a_{i+1} \ldots a_{p}}b_{1}\ldots b_{q}}\]
Similarly, the tensor obtained by raising the $j$th index,
denoted by $T^{(j)},$ is the $(p+1,q-1)$-tensor
\begin{align*} T^{(j)}&=\sum_{a_1,\dots,a_p} \sum_{b_1,\dots,b_q} T^{a_{1}\ldots a_{p}}{
}_{b_{1}\ldots b_{j-1}}{}^{b_j} {}_{b_{j+1}\ldots b_{q}} e_{a_1}\otimes \cdots \otimes  e_{a_p}  \\
&\qquad \qquad\qquad\qquad\qquad \qquad\qquad\qquad \:\: \: \: \otimes  e^{b_1}\otimes \cdots\otimes e^{b_{j-1}} \otimes e_{b_j}\otimes e^{b_{j+1}}\cdots  \otimes e^{b_q} \end{align*}
where
\[ T^{a_{1}\ldots a_{p}}{
}_{b_{1}\ldots b_{j-1}}{}^{b_j} {}_{b_{j+1}\ldots b_{q}}=\sum_{c}g^{b_jc}T^{a_{1} \ldots a_{p}}_{\phantom{a_{1} \ldots a_{p}}b_{1}\ldots b_{j-1}c b_{j+1} \ldots b_{q}}.\]
The operations of raising and lowering indices depend only on the inner product and not on the choice of basis.




  \clearemptydoublepage
 \chapter{Topology and analysis}\label{topan}

\vspace{3ex}

\section{The Hodge star operator \label{Hodge}}

Let $V$ be a vector space of dimension $m$ and $g$ an inner product on $V$. There is an induced inner product on
$\Lambda^{k}V$, also denoted by $g$, given by
\[
g(v_{1}\wedge\dots\wedge v_{k},w_{1}\wedge\dots\wedge w_{k})=\det
(g(v_{i},w_{j})).
\]
Let us fix an orientation on $V$ and denote by $\mu_g$ the unique element of
$\Lambda^{m}V$ which is oriented and has unit norm. The Hodge star operator,
denoted by $\ast$ is the linear isomorphism
$\ast:\Lambda^{k}V\rightarrow\Lambda^{m-k}V$, characterized by the property that
\[
\alpha\wedge\ast\beta=g(\alpha,\beta)\mu_g.
\]
More explicitly, if $\mathfrak{B}=\{v_{1},\dots,v_{m}\}$ is an oriented orthonormal basis for $V$
then the star operator is given by
\[
\ast(v_{1}\wedge\dots\wedge v_{k})=g(v_1,v_i) \cdots g(v_k,v_k)v_{k+1}\wedge\dots\wedge v_{m}.
\]
Note that if $V$ is an oriented vector space with an inner product, so is $V^{\ast}$ and therefore the Hodge star operator also induces
isomorphisms $
\ast:\Lambda^{k}V^{\ast}\rightarrow\Lambda^{m-k}V^{\ast}$.

Let $(M,g)$ be an oriented semi-Riemannian manfiold. There is a unique volume
form $\mathrm{vol}_g\in\Omega^{m}(M)$ which in local oriented coordinates can be
written as:
\[
\mathrm{vol}_g=\sqrt{|\det g_{ij}|}dx^{1}\wedge\cdots\wedge dx^{m}.
\]
This form is well defined because the value $\mathrm{vol}_g(p)$ can be characterized
as the unique vector in $\Lambda^{m}T_{p}^{\ast}M$ which is oriented and has
unit norm. The Hodge star operator is the isomorphism $\ast:\Omega^{k}(M)\rightarrow
\Omega^{m-k}(M)$, characterized by the property that
\[
\omega\wedge\ast\eta=g(\omega,\eta)\mathrm{vol}_g,
\]
for any pair of forms $\omega,\eta\in\Omega^{k}(M)$. The formal adjoint $\delta:\Omega^{k}(M)\rightarrow\Omega^{k-1}(M)$ of the
de-Rham operator is defined by \[\delta=(-1)^{k}\ast^{-1}d\ast.\]
The Hodge
Laplacian of $M$ is the differential operator $\Delta:\Omega(M)\rightarrow
\Omega(M)$ defined by \[\Delta=d\delta+\delta d.\]
A differential form $\omega$ is called harmonic if $\Delta\omega=0$.
The fundamental result of Hodge theory states that on a compact oriented Riemannian manifold $M$, each cohomology class admits a unique harmonic representative. Therefore in this case, there is an isomorphism $H^{\bullet}_{\mathrm{DR}}(M) \cong \ker \Delta$.

\section{The Picard-Lindel\"{o}f theorem}\label{PL}

One of the basic existence and uniqueness theorems for ordinary differential
equations is the following.

\begin{theorem}
Let $U\subseteq \RR\times\RR^{n}$ be an open subset and $F:U\rightarrow
\RR^{n}$ a smooth function. Given $(t_{0},p)\in U$ there exists an open
interval $(a,b)$ containing $t_{0}$ and a smooth function $y:(a,b)\rightarrow
\RR^{n}$ such that
\begin{align*}
y^{\prime}(t)&=F(t,y(t)),\\
y(t_{0})&=p.
\end{align*}
Moreover, if $\bar{y}:(c,d)\rightarrow\RR^{n}$ satisfies the same equations,
then $\bar{y}$ and $y$ coincide on the intersection of their domains.
\end{theorem}

The Picard-Lindel\"{o}f theorem guarantees existence and uniqueness of
solutions of first order ordinary differential equations. On the other hand, a
higher order differential equation
\[
y^{(k)}(t)=F(t,y(t),\dots,y^{(k-1)}(t))
\]
with initial conditions
\[
y(t_{0})=p_{1},\: y^{\prime}(t_{0})=p_{2},\dots,\: y^{(k-1)}(t_{0})=p_{k},
\]
can be rewritten as a system of first order equations
\begin{align*}
y_{1}^{\prime}(t)  &  =y_{2}(t),\\
y_{2}^{\prime}(t)  &  =y_{3}(t),\\
&\:\: \vdots\\
y_{k-1}^{\prime}(t)  &  =F(t,y_{1}(t),\dots,y_{k}(t)).
\end{align*}
with initial conditions
\[
y_{1}(t_{0})=p_{1},\: y_{2}(t_{0})=p_{2},\dots,\: y_{k}(t_{0})=p_{k}.
\]
Therefore, the Picard-Lindel\"{o}f theorem implies the following. 
\begin{theorem}
Let $U\subseteq \RR\times\RR^{n}\times\dots\times\RR^{n}$ be an open
subset and $F:U\rightarrow\RR^{n}$ a smooth function. Given
$(t_{0},p_{1},\dots,p_{k})\in U$ there exists an open interval $(a,b)$
containing $t_{0}$ and a smooth function $y:(a,b)\rightarrow\RR^{n}$
such that
\[
y^{(k)}(t)=F(t,y(t),\dots,y^{(k-1)}(t))
\]
and \[y(t_{0})=p_{1},\: y^{\prime}(t_{0})=p_{2},\dots,\: y^{(k-1)}(t_{0})=p_{k}.\]
Moreover, if $\bar{y}:(c,d)\rightarrow
\RR^{n}$ satisfies the same equations, then $\bar{y}$ and $y$ coincide on the
intersection of their domains.
\end{theorem}

\section{Lie groups}
A \emph{Lie group} $G$ is a manifold endowed with a group structure such that the product map $ (g,h)\mapsto gh$ and the inversion map $g \mapsto g^{-1}$ are smooth. Here are some of the examples that are occur in relativity.
 \begin{itemize}
  \item Let $G=\RR$ be the \emph{additive} group of real numbers. This is an abelian Lie group.
  \item Let $G=\RR^+$ be the \emph{multiplicative}  abelian group of positive real numbers. The exponential map is an isomorphism between the additive Lie group of real numbers and the multiplicative Lie group of positive real numbers.
  \item The circle $G=S^1=\{z \in \CC\mid |z|=1\}$ is an abelian Lie group with respect to the multiplication of complex numbers.

  \item If $G_1,\dots,G_n$ are Lie groups, then $G=G_1\times \dots \times G_n$ is a Lie group with the group operations defined component wise. For instance the group \[T^n=\underbrace{S^1\times\dots\times S^1}_n\]  is called the \emph{$n$-torus}.

\item The $3$-sphere $S^3$, seen as the set of all quaternions of norm 1, is a Lie group with respect to quaternionic multiplication.
 
 \item The \emph{general linear group} 
 $$\mathrm{GL}(n,\RR)=\{A \in\mathrm{Mat}_n(\RR)\mid \det A \neq0\}. $$ 
 is a Lie group. Notice that since the determinant is a continuous map,  $\GL(n,\RR)$ is an open subset of the space of all matrices. The product and inverse functions are algebraic and therefore smooth. The subgroups of $\GL(n,\RR)$ that are also smooth submanifolds are called \emph{matrix Lie groups}.
 \item The \emph{special linear group } 
 $$\SL(n,\RR)=\{A \in\mathrm{Mat}_n(\RR)\mid \det A=1\}.$$
 The special linear group is a matrix Lie group. To see this, it is enough to prove that $\SL(n,\RR)$ is a smooth submanifold of $\GL(n,\RR)$. Consider the determinant function $\det\colon\GL(n,\RR)\rightarrow\RR$. 
 Then $\SL(n,\RR)=\det^{-1}(1)$. We assert that for every matrix $A \in\SL(n,\RR)$ the $\RR$-linear map \[D(\det)_{A}\colon T_{A}\GL(n,\RR)\rightarrow T_1\RR\] is surjective. This follows from the fact that every matrix $X$ in $T_{A}\GL(n,\RR)= \mathrm{Mat}_n(\RR)$ satisfies
\begin{align*}
D (\det)_{A} (X) &= \frac{d}{d t}\bigg\vert_{t=0} \det(A + t X) =\frac{d}{d t}\bigg\vert_{t=0}\det[A(I_n + t A^{-1} X)]  \\
&= \frac{d}{d t}\bigg\vert_{t=0} \det A \,\det (I_n + t A^{-1} X)  = \det (A) \frac{d}{d t}\bigg\vert_{t=0} \det(I_n + t A^{-1} X) \\
&=\det A \frac{d}{d t}\bigg\vert_{t=0} [1 + t \mathrm{tr}(A^{-1} X) + O(t^2) ]  =\det A \, \mathrm{tr} (A^{-1} X).
\end{align*}
 Thus, $D(\det)_{A}(A)=n \det(A)$. It follows that $D(\det)_{A}\neq0$. One concludes that $\SL(n,\RR)=\det^{-1}(1)$ is a smooth submanifold of $\GL(n,\RR)$ of dimension $n^2-1$.
 \item The \emph{orthogonal group} is
 $$\O(n)=\{A \in\GL(n,\RR)\mid A A^{\mathrm{T}}=A^{\mathrm{T}}A=I_n\}.$$ 
 One can see that the group $\O(n)$ is also a matrix Lie group. It is enough to prove that $\O(n)$ is a smooth submanifold of $\GL(n,\RR)$. Let us consider the space of symmetric matrices $\Su(n,\RR)=\{M \in\mathrm{Mat}_n(\RR)|M^{\mathrm{T}}=M\}$ which is vector subspace of $\mathrm{Mat}_n(\RR)$ of dimension $n(n+1)/2$. Consider also the smooth map $F\colon\GL(n,\RR)\rightarrow \Su(n,\RR)$ defined by $F(A)=A^{\mathrm{T}}A$. Then $\O(n)=F^{-1}(I_n)$. We assert that for every matrix $A \in\O(n)$ the linear map \[D F_{A}:T_{A}\GL(n,\RR)\rightarrow T_{F(A)}\Su(n,\RR)\] is surjective. Fix $M \in T_{F(A)}\Su(n,\RR)= \Su(n,\RR)$. Then, for every $X \in T_{A}\GL(n,\RR)= \mathrm{Mat}_n(\RR)$ we have that
\begin{align*}
D F_{A}(X) &=  \frac{d}{d t}\bigg\vert_{t=0} F(A + t X) = \frac{d}{d t}\bigg\vert_{t=0} (A + t X)^{\mathrm{T}} (A + t X) \\
&= X^{\mathrm{T}} A + A^{\mathrm{T}} X.
\end{align*}
 This implies
\[
  D F_{A}\left(\tfrac{1}{2}AM\right)=\tfrac{1}{2}M^{\mathrm{T}}A^{\mathrm{T}}A+\tfrac{1}{2}A^{\mathrm{T}}AM=\tfrac{1}{2}M+\tfrac{1}{2}M=M.
 \]
One concludes that $\O(n)=F^{-1}(I_n)$ is a smooth submanifold of $\GL(n,\RR)$ of dimension $n^2-n(n+1)/2=n(n-1)/2$.
Note that $\O(n)$ is a closed and bounded subspace of a vector space and therefore, it is compact.
 \item The \emph{special orthogonal group} is
 \[
\SO(n)=\O(n)\cap\SL(n,\RR).
 \]

  The group $\SO(n)$ is also a matrix Lie group. To see this, consider \[\GL^+(n,\RR)=\{A \in\mathrm{Mat}_n(\RR)\mid\det A>0\}.\] Then $\GL^+(n,\RR)=\det^{-1}[(0,\infty)]$ which means that it is open in $\GL(n,\RR)$. On the other hand, if $A \in\O(n)$, then $A^{\mathrm{T}}A=I_n$ and therefore $
  1=\det(A^{t}A)=(\det A)^2$ which means $\det A=\pm1$. 
 This implies that $\SO(n)=\O(n)\cap\GL^+(n,\RR)$. Then $\SO(n)$ is an open subset of $\O(n)$.  
 
It is easily verified that $\SO(2)$ can be parametrized by $\theta \in[0,2\pi)$ in the following way
  \[
  \SO(2)=\left\{ \left(\begin{array}{cc}
\cos\theta & -\sin\theta\\
\sin\theta & \cos\theta
  \end{array}\right) \mid \theta \in[0,2\pi)\right\}.
    \]
 Using this, it is easy to see that $\SO(2)$ is abelian. However, $\SO(n)$ is not abelian for $n>2$.
 \item The \emph{Lorentz group} is the group of all linear endomorphisms of $\RR^4$ that preserve the Minkowski metric
 $$\O(1,3)=\{\Lambda \in\GL(4,\RR)\mid \langle \Lambda v, \Lambda w \rangle= \langle v , w \rangle \text{ for all } v,w \in \RR^4\}.$$ 
 Here the inner product is the one given by the Minkowski metric.
 It is easy to see that
  $$\O(1,3)=\{\Lambda \in\GL(4,\RR): \Lambda^{\mathrm{T}} g \Lambda=g\},$$ 
  where, as usual,
  \[g=\left(\begin{array}{cccc}
-1&0&0&0\\
0&1&0&0\\
0&0&1&0\\
0&0&0&1
\end{array}\right)
\]
Clearly, $\O(1,3)$ is a subgroup of the general linear group. In order to prove that it is a submanifold one considers the space
$V=\{M \in \mathrm{Mat}_4(\RR)\mid M^{\mathrm{T}}=gMg\}$. 
One easily checks that $V$ is a vector subspace of  $\mathrm{Mat}_4(\RR)$ of dimension $10$. Let us consider the function $F: \GL(4,\RR) \rightarrow V$ defined by $F(M)= g M^{\mathrm{T}} g M$.
Then $ \O(1,3) = F^{-1}(I_4)$. Therefore it suffices to prove that $DF_{\Lambda} \colon T_{\Lambda} GL(4,\RR) \rightarrow T_{F(\Lambda)}V$ is surjective for $\Lambda \in \O(1,3)$.
Let us fix $M \in T_{I_n} V=V$ y $X \in T_{\Lambda}\GL(4,\RR)=\mathrm{Mat}_4(\RR)$. Then,
\begin{align*}
D F_{\Lambda}(X)&= \frac{d}{d t}\bigg\vert_{t=0} F(\Lambda + t X)  \frac{d}{d t}\bigg\vert_{t=0}g(\Lambda+tX)^{\mathrm{T}}g(\Lambda+tX)\\
&=  gX^{\mathrm{T}}g\Lambda+g\Lambda^{\mathrm{T}}gX.
\end{align*}
Therefore,
\begin{align*}
D F_{\Lambda}\left(\tfrac{1}{2}\Lambda M\right)&= \tfrac{1}{2}g(\Lambda M)^{\mathrm{T}}g\Lambda+ \tfrac{1}{2}g\Lambda^{\mathrm{T}}g\Lambda M\\
&=\tfrac{1}{2}gM^{\mathrm{T}}\Lambda^{\mathrm{T}}g\Lambda+ \tfrac{1}{2} gg Y\\
&=\tfrac{1}{2}Mgg+\tfrac{1}{2}ggM=M.
\end{align*}
One concludes that the Lorentz group is a Lie group of dimension $ 16-10=6$. Let us see that if $\Lambda \in \O(1,3)$ then $\Lambda^{0}_{\phantom{0}0}\geq 1$ or $\Lambda^{0}_{\phantom{0}0} \leq -1$. For this we compute
\[ -1=\langle e_0, e_0 \rangle=\langle \Lambda e_0, \Lambda e_0\rangle =-(\Lambda^{0}_{\phantom{0}0})^2+(\Lambda^{1}_{\phantom{1}0})^2+(\Lambda^{2}_{\phantom{2}0})^2+(\Lambda^{3}_{\phantom{3}0})^2\geq -(\Lambda^{0}_{\phantom{0}0})^2.\]
The Lorentz group consists of four connected components, which are
\begin{align*}
 L^{\uparrow}_+&= \{ \Lambda \in \O(1,3)\mid \det \Lambda=1 \text{ and } \Lambda^{0}_{\phantom{0}0}\geq 1\}\\
L^{\uparrow}_-&= \{ \Lambda \in \O(1,3)\mid \det\Lambda=-1 \text{ and } \Lambda^{0}_{\phantom{0}0}\geq 1\}\\
 L^{\downarrow}_+&= \{ \Lambda \in \O(1,3)\mid \det\Lambda=1 \text{ and } \Lambda^{0}_{\phantom{0}0}\leq -1\}\\
 L^{\downarrow}_-&= \{ \Lambda \in \O(1,3)\mid \det\Lambda=-1 \text{ and } \Lambda^{0}_{\phantom{0}0}\leq -1\}
 \end{align*}
The elements of $L^{\uparrow}= L^{\uparrow}_+ \cup  L^{\uparrow}_-$ are called orthochronous Lorentz transformations.
The elements of $L^{\downarrow}= L^{\downarrow}_+ \cup  L^{\downarrow}_-$ are called anorthochronous Lorentz transformations. The group $L^\uparrow_+$ is called the restricted Lorentz group.
  \end{itemize}
  
\section{The linking number}\label{link}

Suppose that $M$ and $N$ are compact connected oriented manifolds of dimension $m$. In this case, the map $H^{m}_{\mathrm{DR}}(M) \rightarrow \RR$ given by
\[\left[ \omega\right]  \mapsto \int_{_{M}}  \omega\]
is an isomorphism.
Let us consider a smooth function $f: M \rightarrow N$. The degree of $f$ is the number \[\deg f= \int_{_{M}} f^{*} \omega, \]
where $[\omega]\in H^m_{\mathrm{DR}}(N)$ is characterized by the property that
\[ \int_N \omega=1.\]
A priori, the degree of $f$ is a real number. Let us see that $\deg f \in \ZZ$. A point $p \in M$ is a critical point of $f$ if $Df(p)$ is singular.
A point $q \in N$ is a regular value if the set $f^{-1}(q)$ contains no critical points. The following result plays an important role in differential topology. A proof can be found in the book by Hirsch \cite{Hirsch}.
\begin{lemma} [Sard's Lemma] Given a function $f: M \rightarrow N$ there exists a regular value $q \in N$.
\end{lemma}

Armed with this, we can now prove what the desired result.

\begin{lemma} The degree of a smooth function $f: M \rightarrow N$ is an integer, that is, $\deg f \in \ZZ$.
\end{lemma}
\begin{proof}
By Sard's lemma, there exists a regular value $q \in N$. For each  $p \in f^{-1}(q)$ we know that  $Df(p)$ is an isomorphism and therefore there exist open subsets  $U_{{p}}$, $V_{{q}}$ such that  $p \in U_{_{p}}$
and $f\vert_{U_p}:U_p \rightarrow V_p$
is a difeomorphism.
Therefore, $f^{-1}({q})$ is a closed discrete subset of $M$. Since $M$ is compact, this implies that the set $f^{-1}(q)$ is finite.
Let us set \[V= \bigcap_{p \in f^{-1}(q)}V_{p}, 
\]
and
\[
  U'_p = \{ x \in U_p\mid f(x)\in W\}.\] 
Then  $K= M \setminus \left(\bigcup_{p \in f^{-1}(q)} {U'_p}\right)$ is compact and $K \cap f^{-1}(q)= \emptyset$.  This implies that $f(K)$ is a closed subset that does not contain $q$. Take $W=N\setminus f(K)$ and $\omega \in \Omega^{m}(N)$ such that \[\int_{{N}} \omega = 1,\] and $\mathrm{supp}\, \omega \subseteq V \cap W$. By definition
\[ \deg f= \int_M f^*\omega.\]
One can also compute
\begin{align*}
\int_{_{M}} f^{*}\omega& = \int_{_{K}} f^{*}\omega + \sum_{{p}} \int_{{U'_{{p}}}} f^{*}\omega\\
 &= \int_K 0+ \sum_{p \in f^{-1} (q)} \int_{U'_p} f^{*}(\omega)\\
 &= \sum_{{p \in f^{-1} (q)}} \mathrm{sign}\left( \det(Df(p))\right)\int_{{V}} \omega\\
 & = \sum_{{p \in f^{-1} (q)}} \mathrm{sign}\left( \det(Df(p))\right) \int_{_{N}} \omega\\
  &= \sum_{{p \in f^{-1} (q)}} \mathrm{sign}\left( \det(Df(p))\right) \in \ZZ.
\end{align*}

\end{proof}

We conclude that if $q$ is a regular value of $f$ then
\[\deg f= \sum_{{p \in f^{-1} (q)}} \mathrm{sign}\left( \det(Df(p))\right).\] Let us illustrate this with an example. 
\begin{example}\label{exs1}
Let $M$ be the circle  $S^{1} \subseteq \CC$  and $f: S^{1} \rightarrow{S}^{1} $ be given by $f(z)=z^n$. The function $f$ is a local diffeomorphism which preserves the orientation and is surjective. Therefore, the degree of $f$ is the number of solutions of the equation $z^n=1$.
We conclude that
\[ \deg f =n.\]
\end{example}

\begin{lemma}
Suppose that $f,g:M \rightarrow N$ are homotopic maps. Then
\[ \deg f=\deg g.\]
\end{lemma}
\begin{proof}
Let $\omega \in \Omega^m(N)$ be a form such that $\int_N \omega=1$.
Let $H: M \times [0,1] \rightarrow N$ be a homotopy.  By Stokes'  theorem we know that:
\[ \int_{M \times [0,1]} d(H^*\omega)= \int_{\partial(M\times [0,1)]}H^*\omega=\int_M f^*\omega-\int_M g^*\omega = \deg f - \deg g.\]
On the other hand, \[d(H^*\omega)=H^* (d\omega) =H^*(0)=0.\] We conclude that $\deg f=\deg g$.
\end{proof}
We will now describe the linking number between two oriented knots.
Let $S^1$ be the unit circle in the plane with the standard counterclockwise orientation.
A parametrised knot in $\RR^3$ is a smooth map $\alpha: S^1 \rightarrow \RR^3$ 
such that for each $z \in S^1$ the derivative $D\alpha(z)$ is injective and
 $\alpha$ is a homeomorphism to its image.
Two parametrised knots $\alpha, \beta$ are equivalent if there exists an orientation preserving diffeomorphism $\varphi: S^1 \rightarrow S^1$ such that $\alpha \circ \varphi=\beta$. An oriented knot is an equivalence class of parametrised knots.

\begin{figure}[H]
\centering
\scalebox{.2}
	\centering
	\includegraphics[scale=0.45]{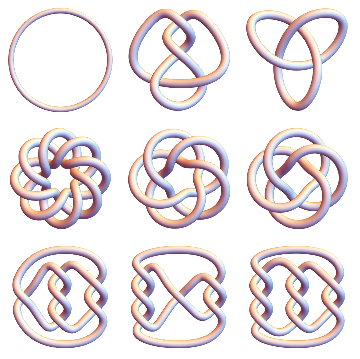}
	\caption{Knots.}
\end{figure}
Let $\alpha$ and $\beta$ be parametrised knots with disjoint images. There is a natural map to the sphere $G: S^1 \times S^1 \rightarrow S^2$ given by
\[ G(t,s)= \frac{\alpha(t)-\beta(s)}{|\alpha(t)-\beta(s)|}.\]
The linking number $L(\alpha, \beta)$ is the degree of the map $G$.
\begin{lemma}\label{link}
The linking number has the following properties:
\begin{enumerate}
\item[(1)] $L(\alpha, \beta)$ is an integer.
\item[(2)] $L(\alpha, \beta)=L(\beta,\alpha)$.
\item[(3)] $L(\alpha, \beta)$ depends only on the equivalence classes of $\alpha$ and $ \beta$.
\item[(4)] $L(\alpha,\beta)$ is deformation invariant. This means that if $H: S^1 \times [0,1] \rightarrow \RR^3$ is such that for each 
$z \in [0,1]$ the map $H_z=H(\cdot,z)$ is a knot and $\beta: S^1 \rightarrow \RR^3$ is another knot which does not intersect the image of $H$ then
\[ L(H_z,\beta)=L(H_{z'}, \beta)\]
for any $z,z'\in [0,1]$.
\end{enumerate}

\begin{proof}
The linking number is an integer because it is the degree of a map. For the second property we note that if the order of the knots is reversed the map $G$ changes to ${a} \circ G \circ \iota$, where $\iota$ interchanges the factors of $S^1\times S^1$ and ${a}$ is the antipodal map in the sphere. Since both ${a}$ and $\iota$ are orientation reversing diffeomorphisms, the integral does not change. For the third statement consider a knot $\gamma= \beta \circ \varphi $ which is equivalent to $\beta$. Then
\[G_{\alpha, \gamma}=G_{\alpha, \beta} \circ (\mathrm{id} \times \varphi).\]
Therefore,
\begin{align*} L(\alpha,\gamma)=\int_{S^1\times S^1}G^*_{\alpha,\gamma}\omega=\int_{S^1\times S^1}(\mathrm{id} \times \varphi)^* G^*_{\alpha,\beta}\omega =\int_{S^1\times S^1} G^*_{\alpha,\beta}\omega=L(\alpha,\beta).\end{align*}
Here we used that $\mathrm{id} \times \varphi$ is an orientation preserving diffeomorphism of $S^1 \times S^1$ and therefore
\[\int_{S^1\times S^1}(\mathrm{id} \times \varphi)^* \eta=\int_{S^1\times S^1}\eta,\]
for all $\eta \in \Omega^2( S^1 \times S^1)$. For the last statement we observe that the corresponding maps $H_z$ and $H_{z'}$ are homotopic and therefore have the same degree.
\end{proof}
\end{lemma}
\begin{figure}[H]
\centering
\scalebox{.2}
	\centering
	\includegraphics[scale=0.38]{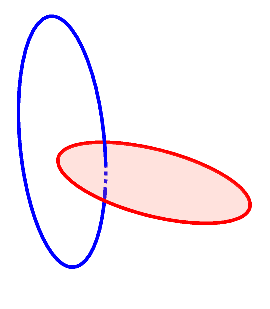}
	\caption{The linking number counts the intersection points of one knot with a surface whose boundary is the other knot.}
\end{figure}
In view of the lemma above, the linking number is defined for oriented knots and one can write $L(C,C')$. This is a topological invariant of the configuration of knots and does not change under continuous deformations. Let us now provide a more explicit formula for the linking number of two knots which uses the language of vector calculus instead of that of differential forms.
\begin{lemma}
Let $\alpha$ and $\beta$ be two disjoint parametrized knots. Then
\[ L( \alpha, \beta)=\frac{1}{4\pi}\int_{S^1 \times S^1}\frac{\det\left(\alpha'(t),\beta'(s), \beta(s)-\alpha(t)\right)}{\left\vert
\alpha(t)-\beta(s)\right\vert ^{3}}dt\, ds.\] 
\end{lemma}
\begin{proof}
Let $\eta$ be the standard volume form on the sphere so that $\eta/4\pi$ integrates to $1$.
It suffices to show that
\[G^*\eta =\frac{\det\left(\alpha'(t),\beta'(s), \beta(s)-\alpha(t)\right)}{\left\vert
\alpha(t)-\beta(s)\right\vert ^{3}}dt\, ds.\]
At a singular point $(t,s) \in S^1 \times S^1$ of $G$, both sides of the equation above vanish.
Let us prove the equation a small open subset $U \subset S^1 \times S^1$ where the map $G$ is an orientation preserving local diffeomorphism. On $U$, the form $G^*\eta$ takes the form
\[  G^*\eta= \sqrt{\det g_{ij}}dt\, ds,\]
where
\[
(g_{ij})=\left( \begin{array}{cc}
\langle \frac{\partial G}{\partial t},\frac{\partial G}{\partial t} \rangle &\langle \frac{\partial G}{\partial t},\frac{\partial G}{\partial s} \rangle\\[1ex]
\langle \frac{\partial G}{\partial s},\frac{\partial G}{\partial t} \rangle&\langle \frac{\partial G}{\partial s},\frac{\partial G}{\partial s} \rangle \end{array}\right).\]
Now, putting $r = \vert \alpha(t)-\beta(s) \vert$, we have
\begin{align*}
 \frac{\partial G}{\partial t}&=\frac{\alpha'(t)}{r}-\frac{(\alpha(t)-\beta(s))}{r^2}\frac{\partial r}{\partial t},\\
 \frac{\partial G}{\partial s}&=-\frac{\beta'(s)}{r}-\frac{(\alpha(t)-\beta(s))}{r^2}\frac{\partial r}{\partial s}
\end{align*}
Thus, by Lagrange's identity,
\begin{align*}
 \sqrt{\det g_{ij}}&=\left\vert \frac{\partial G}{\partial t}\times \frac{\partial G}{\partial s}\right\vert= \left\langle G, \frac{\partial G}{\partial t} \times \frac{\partial G}{\partial s} \right\rangle
 =\det\left(G,\frac{\partial G}{\partial t},\frac{\partial G}{\partial s}\right)\\
 &=\det\left(G,\frac{\alpha'(t)}{r},-\frac{\beta'(s)}{r}\right)=\frac{\det\left(\alpha'(t),\beta'(s), \beta(s)-\alpha(t)\right)}{\left\vert
\alpha(t)-\beta(s)\right\vert ^{3}}.
\end{align*}
The desired conclusion follows at once. 
\end{proof}

  \clearemptydoublepage
\chapter{The future is open}\label{futureopen}

\vspace{3ex}

In this appendix we prove that the chronological future of an event is always open.
This is a technical but important result mentioned in the text. We begin with some preliminary lemmas.
Given a finite dimensional vector space $V$ with an inner product $\langle \cdot,\cdot \rangle$, there is a natural function $\xi: V \to \RR$ given by
\[ \xi(v) = \langle v ,v\rangle. \]
There is also a natural vector field on $V$ sometimes called the Euler vector field, or the position vector field, given by
\[ \tilde{E}= \sum_i x^i \partial_{x^i}.\]
We have the following result. 

\begin{lemma}\label{psiEt}
The derivative of $\xi$ is twice the one form dual with respect to $\langle \cdot,\cdot \rangle$ to the Euler vector field, i.e:
\[d\xi=2\langle \tilde{E}, \cdot\rangle.\]
\end{lemma}

\begin{proof}
Fixing a basis $e_1, \dots ,e_n$ for $V$ so that $v=x^1 e_1 + \dots +x^n e_n$, the function $\xi$ takes the form
\[ \xi=\sum_{ij}x^i x^j g_{ij},\]
where $g_{ij} = \langle e_i ,e_j \rangle$. Thus its derivative is
\[ d\xi=\sum_{i,j}g_{ij} \big(x^i dx^j +x^j dx^i \big),\]
and therefore
\[ d\xi(\partial_{x^k})=\sum_{ij}g_{ij} \big(x^i dx^j +x^j dx^i \big)(\partial_{x^k})=\sum_{i}g_{ik} x^i +\sum_{j}g_{kj} x^j=2\sum_{i}g_{ik} x^i  .\]
On the other hand, bearing in mind that $e_i = \partial_{x^{i}}$, we have
\[2 \langle \tilde{E}, \partial_{x^k}\rangle=2 \Big\langle  \sum_i x^i \partial_{x^i}, \partial_{x^k}\Big\rangle = 2\sum_{i}g_{ik} x^i .\]
On comparing the last two equalities, the result follows. 
\end{proof}

Recall that a normal neighborhood around of $p$ is an open neighborhood $U$ such that there is a convex open neighborhood of zero
$\tilde{U}$ in $T_pM$ such that $\exp(p): \tilde{U} \rightarrow U$ is a diffeomorphism. We denote by $E$ the vector field in $U$ that corresponds to the Euler vector field under this diffeomorphism.
As usual, we denote by $\tilde{C}^+_p$ the set of timelike vectors in $T_pM$ that point to the future. Recall that this
is a convex open set.

\begin{lemma}\label{fut1}
Let $M$ be a time oriented Lorentzian manifold, $p$ a point in $M$, and $U$ a normal neighborhood around $p$. If $\beta: [0,1] \to \tilde{U}$ is a piecewise smooth map that starts at zero such that
$\alpha = \exp(p) \circ \beta$ is timelike, then $\beta(s) \in \tilde{C}^+_p$ for $s >0$.
\end{lemma}
\begin{proof}
Let us first consider the case where $\beta$ is smooth.
Since $\beta'(0)=\alpha'(0)$ is timelike and goes to the future, there exists an $\varepsilon >0$ such that if $0<s<\varepsilon$
then $\beta(s) \in \tilde{C}^+_p$. Therefore, for sufficiently small $s$ one has
\[ \langle \beta'(s),\tilde{E}(\beta(s)\rangle<0.\]
We need to show that $\xi(\beta(s))<0$ for all $s\in (0,1]$.  Since this function is negative for small $s$, it suffices to show that its derivative is negative. Using  lemma \ref{psiEt} and the Gauss lemma one computes
\begin{equation}\label{derivg}\frac{d}{ds}(\xi\circ \beta(s))=2\langle \beta'(s),\tilde{E}(\beta(s)\rangle=2\langle \alpha'(s),E(\alpha(s))\rangle .\end{equation}
As long as $\beta(s)$ is in $\tilde{C}^+_p$, one has
\[ \langle E(\alpha(s)),E(\alpha(s))\rangle=\langle \tilde{E}(\beta(s)),\tilde{E}(\beta(s))\rangle=\langle \beta(s),\beta(s)\rangle<0. \]
This implies that, as long as $\beta(s)$ is in $\tilde{C}^+_p$, the vector $E(\alpha(s))$ is timelike and therefore
\[ \langle \alpha'(s),E(\alpha(s))\rangle\neq 0.\]
We conclude that, as long as $\beta(s)$ stays in $\tilde{C}^+_p$, the  derivative (\ref{derivg})
stays negative and $\psi\circ \beta(s)$ does not vanish for $s>0$.
Let us now consider what happens when $\beta$ is only piecewise smooth. It suffices to show that the sign of the derivative $d(\xi \circ \beta(s))/ds$
does not change at the breaks. Let $s_0$ be a place where the curve is not smooth. Then, the derivative from the left is
\begin{equation}\label{break1} \left.\frac{d }{ds}(\xi\circ \beta(s^-))\right\vert_{s=s_0}=2\langle \alpha'(s^-) \vert_{s=s_0},\mathcal{E}(\alpha(s_0))\rangle, \end{equation}
and the derivative from the right is
\begin{equation}\label{break2} \left.\frac{d }{ds}(\xi\circ \beta(s^+))\right\vert_{s=s_0}=2\langle \alpha'(s^+) \vert_{s=s_0},\mathcal{E}(\alpha(s_0))\rangle.\end{equation}
Since $\alpha$ is timelike and points to the future, the right hand sides of (\ref{break1}) and (\ref{break2}) have the same sign.
\end{proof}

\begin{lemma}\label{fut2}
Let $M$ be a time oriented Lorentzian manifold, $p$ a point in $M$, and $U$ a normal neighborhood around $p$. Then the map
\[ \exp(p): \tilde{C}^+_p \cap \tilde{U} \to C^+_p(U)\]
is a diffeomorphism. In particular, $C^+_p(U)$ is open in $M$.
\end{lemma}
\begin{proof}
Since $\exp(p)\vert_U$ is a diffeomorphism onto its image, so is $\exp(p)\vert_{\tilde{C}^+_p \cap \tilde{U}}$. Therefore, it is enough to show that \[\exp(p)\left(\tilde{C}^+_p \cap \tilde{U}\right)=C^+_p(U).\]
First, assume that $v \in \tilde{C}^+_p \cap \tilde{U}$. The path $\gamma(s)=\exp(p)(tv)$ is a radial geodesic. Since the derivative of the exponential map at zero is the identity, $\gamma'(0)=v$, and therefore
\[ \langle \gamma'(0),\gamma'(0)\rangle=\langle v,v \rangle<0.\]
Using that $\gamma(s)$ is geodesic, one computes
\[ \nabla_{\gamma'(s)} \langle \gamma'(s),\gamma'(s)\rangle= 2 \langle \nabla_{\gamma'(s)}\gamma'(s),\gamma'(s)\rangle=0.\]  
We conclude that $\gamma(s)$ is a timelike curve. Since $v \in \tilde{C}^+_p$, $\gamma(s)$ points to the future. This implies that $\exp(p)(v) \in C^+_p(U)$. Since $v$ is arbitrary, 
 \[\exp(p)\left(\tilde{C}^+_p \cap \tilde{U}\right)\subseteq C^+_p(U).\]
 The fact that 
 \[ C^+_p(U)\subseteq \exp(p)\Big(\tilde{C}^+_p \cap \tilde{U}\Big)\]
 is follows directly from lemma \ref{fut1}.
\end{proof}

\begin{lemma}\label{convex}
Let $M$ be a semi-Riemannian manifold and $p$ a point in $M$. There exists an open neighborhood $U$ of $p$ such that,
for all $q \in U$, there is a normal neighborhood $U_q$ around $q$ that contains $p$.
\end{lemma}
\begin{proof}
Let $W$ be the open set of $TM$ where the map $\exp: W \to M \times M$ given by
\[ \exp(q,X) = \exp(q)(X)\]
is defined. We know that
\[ D(\exp)(p,0)=\begin{pmatrix} \mathrm{id} &0\\
0 & \mathrm{id} \end{pmatrix}.\]
By the implicit function theorem, there exists an open neighborhood $\tilde{W}$ of $(p,0)$ in $TM$ that maps diffeomorphically to a neighborhood $W$ of $(p,p)$ in $M \times M$.
By restricting $\tilde{W}$ if necessary, we may assume that there exists a neighborhood $Z$ of $p$ such that $\tilde{W}$ has the form
\[ \tilde{W}=\{ (q,X) \in TM \mid q \in Z \text{ and } \vert X\vert <\delta\},\]
where, as usual $|X|= \sqrt{\langle X,X \rangle}$. If $U$ is an open neighborhood of $p$ such that $U \times U \subseteq W$, then $U$ has the desired property.
\end{proof}
\begin{proposition}\label{openfuture}
Let $M$ be a time oriented Lorentzian manifold. For any point $p \in M$, the sets $C_p^+(M)$ and $C_p^-(M)$ are open.
\end{proposition}
\begin{proof}
By reversing the time orientation, it is enough to prove the statement for $C_p^+(M)$. Let $q$ be a point in $C_p^+(M)$.
Then, there is a piecewise smooth timelike path to the future $\gamma:[0,1] \to M$ such that $\gamma(0)=p$ and $\gamma(1)=q$. 
By lemma (\ref{convex}), there exists $s_0$ such that $\gamma(s_0)$ has a normal neighborhood $U$ that contains $q$
and $\gamma([s,1]) \subseteq U$.
 Lemma
\ref{fut2} guarantees that $ C^+_{\gamma(s_0)}(U)$ is open. By construction $\gamma(s)\in C^+_p(M)$, and therefore
$ C^+_{\gamma(s)}(U)\subseteq C^+_p(M)$
and $q \in C^+_{\gamma(s)}(U)$.
Since $q$ is arbitrary, one concludes that $C^+_p(M)$ is open.
\end{proof}

  \clearemptydoublepage
\chapter{Other geometric results}

\vspace{3ex}

\section{The theorem of Hopf-Rinow}

Given a piecewise smooth path $\gamma$ from $p$ to $q$ on a Riemannian manifold $(M,g)$, the length of  $\gamma$ is
\[ L(\gamma)=\int_a^b \vert\gamma'(\tau)\vert d\tau .\]
Taking the infimum over all paths from $p$ to $q$ gives a metric $d$ on $M$, which is compatible with the given topology.
The Hopf-Rinow theorem specifies the conditions under which the metric space $(M,d)$ is complete.

To be more precise, let $(M,d)$ be a connected Riemannian manifold and consider the function $d: M \times M \to \RR$
defined by
\[ d(p,q)=\inf\{ L(\gamma):\gamma \in \mathcal{C}(p,q)\} \]
where
\[ \mathcal{C}(p,q)=\{ \gamma:[a,b]\to M \mid \gamma \text{ is piecewise smooth, } \gamma(a)=p \text{ and } \gamma(b)=q \}.\]
We will prove that the function $d$ defines a metric on $M$.
Fix a normal neighborhood $U$ around $p \in M$ so that $\exp(p): \tilde{U}\to U$ is a diffeomorphism.
Consider the function $\tilde{r}: T_pM \to \RR$ defined by
\[ \tilde{r}(v)=\sqrt{\xi(v)}=\sqrt{\langle v,v \rangle},\]
and $r: U \to \RR$ given by $r=\tilde{r}\circ \exp(p)^{-1}$. Using the same notation as in Appendix \ref{futureopen}, we write $E$ for the vector field on $U$ which is the push forward of the Euler vector field under the exponential map. Then, using the Gauss lemma one computes
\begin{align}
\begin{split}
    r(q)&=\tilde{r} \circ \exp(p)^{-1}(q) \\
    &=\sqrt{\langle \tilde{E}\big(\exp(p)^{-1}(q)\big),\tilde{E}\big(\exp(p)^{-1}(q)\big) \rangle} \\
    &=\sqrt{\langle E(q),E(q)\rangle} \\
    &=\vert E(q)\vert.
    \end{split}
    \end{align}
Also, using Lemma \ref{psiEt}, one computes
\begin{equation}
d \tilde{r}= d \sqrt{\xi}=\frac{\langle \tilde{E},\cdot\rangle }{\tilde{r}}.
\end{equation}
In view of the Gauss lemma, this implies
\begin{equation}
d r= \frac{\langle \mathcal{E},\cdot\rangle }{r}.
\end{equation}
If we write $X$ for the unit radial vector field on $U-\{p\}$, that is,
\[ X= \frac{E}{r}=\frac{E}{\vert\mathcal{E}\vert},\]
then
\begin{equation}\label{dr}
dr=\langle X,\cdot \rangle.
\end{equation}

\begin{lemma}\label{local}
Let $U$ be a normal neighborhood of $p$ and $\gamma$ be the radial geodesic from $p$ to $q$.
Then:
\begin{enumerate}
\item[1.] $L(\gamma)=r(q)$.
\item[2.] Given any piecewise smooth path $\alpha$ from $p$ to $q$ in $U$, $L(\alpha) \geq L (\gamma)$.
\item[3.] If $L(\alpha)= L(\gamma)$, then, $\alpha$ is a reparametrization of $\gamma$.
\end{enumerate}
\end{lemma}
\begin{proof}
For the first claim, one computes
\begin{align}
\begin{split}
    L(\gamma)&=\int_0^1 \sqrt{\langle \gamma'(\tau),\gamma'(\tau)\rangle} d\tau \\
&=\int_0^1 \sqrt{\langle \exp(p)( v), \exp(p) ( v)\rangle} d\tau \\
&=\sqrt{\langle v,v\rangle} \\
&=r(q).
\end{split}
\end{align}
Let us now consider the second claim. Away from $p$, one can write
\begin{equation} \alpha'(\tau)= \langle \alpha'(\tau), X \rangle X + Y,\end{equation}
where $Y$ is  orthogonal to the radial direction. Then
\begin{equation} \langle \alpha'(\tau), \alpha'(\tau)\rangle = \langle \alpha'(\tau), X \rangle^2+ \langle Y,Y \rangle^2 \geq  \langle \alpha'(\tau), X \rangle^2. \end{equation}
Therefore
\begin{equation}
L(\alpha)=\int_a^b \sqrt{\langle \alpha'(\tau),\alpha'(\tau)\rangle} d\tau \geq \int_a^b \langle  \alpha'(\tau), X \rangle d\tau.
\end{equation}
Using equation \eqref{dr}  and the fundamental theorem of calculus, one computes
\begin{align}
    \begin{split}
\int_a^b \langle  \alpha'(\tau), X \rangle d\tau &=\int_a^b dr (\alpha(\tau))(\alpha'(\tau))d\tau \\
&=\int_a^b \frac{d}{d\tau}r(\alpha(\tau))d\tau\\
&= r(q)-r(p)\\
&=r(q).
\end{split}
\end{align}
One concludes that $L(\alpha) \geq r(q)=L(\gamma)$.
For the last claim, assume that $L(\alpha)=L(\gamma)$ and that $\vert\gamma'(\tau)\vert=1$ so that $\gamma'(\tau)=X(\gamma(\tau))$ . In this case
\begin{equation*} \alpha'(\tau)= \langle \alpha'(\tau), X \rangle X ,\end{equation*}
and 
\[ (r\circ \alpha)' (\tau)=\langle \alpha'(\tau), X \rangle \geq 0.\]
Consider the curve $\beta(\tau)=\gamma(r(\alpha(\tau)))$,
so that
\[ \beta'(\tau)=\gamma'(r(\alpha(\tau))) \langle \alpha'(\tau), X \rangle =\alpha'(\beta(\tau)). \]
Since $\alpha$ and $\beta$ are solutions to the same differential equation starting at $p$, they are equal. One concludes that
$\alpha$ and $\gamma$ differ by a reparametrization.
\end{proof}

\begin{lemma}
The function $d$ is a metric on $M$ which induces the topology given by the coordinate charts.
\end{lemma}
\begin{proof}
Let us first prove that the function $d$ is a metric.
Since the length of a path does not change when the direction is reversed, the function $d$ is symmetric.
It is also non-negative. By concatenating paths one obtains the triangular inequality. The only nontrivial condition is that the distance between different points is positive. Let $p\neq q$ be two points in $M$ and $U$ a normal neighborhood around $p$  that does not contain $q$, so that $\exp(p): \tilde{U}\to U$ is a diffeomorphism. Choose $\varepsilon>0$ small enough so that $B(0,2\varepsilon) \subseteq \tilde{U}$. We claim that $d(p,q)\geq \varepsilon$. Assume the contrary. Then, there is a path $\alpha$ from $p$ to $q$ in $M$ such that $L(\alpha)<\varepsilon$. Set $B=\exp(p) \left(B(0,\varepsilon)\right)$ and $C=M\setminus\exp(p) \big(\overline{B(0,\varepsilon)}\big)$.  $B$ and $C$ are disjoint open sets in $M$ and $\alpha(a) \in B$ and $\alpha(b) \in C$. Since the interval $[a,b]$ is connected, one concludes that there is 
a $t \in [a,b]$ such that $\alpha(t)\in \exp(p) \left(S(0,\varepsilon)\right)$, where $S(0,\varepsilon)$ is the sphere of radius $\varepsilon$.
Set 
\[ t_0 = \inf \left\{ t \in [a,b] \mid \alpha(t)\in \exp(p)\left( S(0,\varepsilon)\right)\right\}.\]
By continuity, $\alpha(t_0) \in \exp(p) \left(S(0,\varepsilon)\right)$. We claim that $\alpha([a,t_0])\subseteq \exp(p)\big(\overline{B(0,\varepsilon)}\big)$.
Suppose the contrary, then there exists $t_1<t_0$ such that $\alpha(t_1) \in C$. Since the interval $[0,t_1]$ is connected, and the path
$\alpha$ restricted to $[0,t_1]$ starts in $B$ and ends in $C$, there is a $t_2 \in [a,t_1]$ such that
$\alpha(t_2) \in \exp(p)\left( S(0,\varepsilon)\right)$. This contradicts the definition of $t_0$. One concludes that $\alpha ([a,t_0])\subseteq U$.
Lemma \ref{local} implies that
\[ L(\alpha)= \int_a^b \sqrt{\langle \alpha'(\tau),\alpha'(\tau) \rangle} d\tau  \geq \int_a^{t_0} \sqrt{\langle \alpha'(\tau),\alpha'(\tau) \rangle} d\tau \geq 
r(\alpha(t_0))=\varepsilon. \]
It remains to show that the topology induced by the metric coincides with that induced by the charts.
Choose a normal neighborhood $U$ around $p$ so that
$\exp(p):\tilde{U}\to U$ is a diffeomorphism.  Fix $\varepsilon>0$ small enough so that $ B(0,2\varepsilon)\subseteq \tilde{U}$. The argument used in showing that the distance between different points is positive shows that $B(p, \varepsilon)=\exp(p) \left(B(0,\varepsilon)\right)$. This implies that the two topologies are the same.
\end{proof}

\begin{lemma}\label{contg}
Let $(M,g)$ be a Riemannian manifold and $\gamma:[a,b) \to M$ a geodesic. If $\gamma$ can be extended to a continuous function
$\tilde{\gamma}:[a,b] \to M$ then $\tilde{\gamma}$ is a geodesic.
\end{lemma}

\begin{proof}
Since $\gamma$ is a geodesic, $\gamma'(\tau)=C$ is constant.
By Lemma \ref{convex}, there exists  $c<b$ and $\epsilon>0$ such that $\vert c-b\vert<\varepsilon/C$ and $B(\gamma(c),2\varepsilon)$ is a normal neighborhood. Then $\tilde{\gamma}([c,b] )\subseteq B(\gamma(c), 2\varepsilon)$. 
This implies that
$ \exp(\gamma(c))^{-1}\left( \tilde{\gamma}([c,b] )\right)$
is contained in a ray spanned by a vector $w$ in $T_{\gamma(c)}M$. Since $d(\gamma(c), p)<2\varepsilon$, there exists $v$ such that
$p= \exp(\gamma(c))(v)$. Since the rays spanned by $v$ and $w$ intersect in two points, we can rescale so that $v=w$. Then, the curves
$\tilde{\gamma}:[c,b]\to M$ and $\exp(\gamma(c))(\tau v)$ differ by an affine reparametrization. Since the second one is a geodesic, so is the first one.
\end{proof}

\begin{lemma}\label{minimizer}
Suppose that $\gamma:[a,b]\to M$ is a path from $p$ to $q$ with length equal to $d(p,q)$. Then, $\gamma$ can be reparametrized so that it becomes a geodesic. 
\end{lemma}
\begin{proof}
Fix $t_0 \in (a,b)$ and a normal neighborhood $U$ of $\gamma(t_0)$. There is some $t_1>t_0$ such that $\gamma(t_1) \in U$. Since the curve $\gamma$ has minimal length, we conclude that $\gamma:[t_0,t_1]\to M$ can be reparametrized so that it becomes a geodesic. This shows that one can assume that the curve is piecewise geodesic.  It remains to show that the breaks can be removed. By induction, it is enough to consider the case where there is only one break. Therefore we assume that there exists $c\in (a,b)$ so that 
$\gamma:[a,c]\to M$ and $\gamma:[c,b] \to M$ are geodesics. By Lemma \ref{convex} there exists a $u<c $ such that $\gamma(u)$ has a normal neighborhood $U$ that contains $\gamma(c)$. Then, $U$ also contains some $\gamma(s)$ for some $s>c$. Then, by minimality of the path, the curve $\gamma:[u,s]\to M$ can be reparametrized to be a geodesic. This removes the supposed break.
\end{proof}

A Riemannian manifold $(M,g)$ is geodesically complete if for every $p \in M$ the domain of the exponential map $\exp(p)$ is $T_pM$. This is equivalent to the condition that any geodesic $\gamma: [a,b]\to M$ can be extended to a geodesic with domain $\RR$.
\begin{theorem}(Hopf-Rinow)
Let $(M,g)$ be a connected Riemannian manifold. The  following statements are equivalent.
\begin{enumerate}
\item[(a)] $(M,d)$ is a complete metric space.
\item[(b)] $(M,g)$ is geodesically complete.
\item[(c)] There exists a point $p \in M$ such that $\exp(p)$ is defined on $T_pM$.
\item[(d)] Any closed and bounded subspace of $M$ is compact.
\end{enumerate}
Moreover, any of the above conditions implies 
\begin{enumerate}
\item[(e)] Any two points are connected by a geodesic of length equal to the distance between them.
\end{enumerate}
\end{theorem}
\begin{proof}
It is obvious that (b) implies (c). Let us see that (a) implies (b). Suppose that $\gamma:(a,b) \to M$ be a geodesic and set
\[ I= \left\{ x \in \RR\mid \gamma \text{ can be extended to a geodesic on an open interval that contains } x\right\}.\]
We need to prove that $I=\RR$. By construction, $I$ is an open interval and there is a maximal geodesic $\tilde{\gamma}:(c,d)\to M$, that extends $\gamma$. Suppose $(c,d) \neq \RR$, then, either $c$ or $d$ are finite. Without loss of generality, $d<\infty$. Let $d_n$ be an increasing sequence in $(c,d)$ that converges to $d$. Since $\tilde{\gamma}$ is a geodesic, $\vert\tilde{\gamma}'(\tau)\vert=K$, and therefore
\[ d(\tilde{\gamma}(d_n),\tilde{\gamma}(d_m))\leq K \vert  d_n-d_m\vert.\]
Therefore, the sequence $\tilde{\gamma}(d_n)$ is Cauchy. Since $(M,d)$ is complete, it converges to $p \in M$. Then, $\tilde{\gamma}$ can be extended to a continuous function $\overline{\gamma}$ on $(c,d]$ by declaring $\overline{\gamma}(d)=p$. Lemma \ref{contg} implies that $\overline{\gamma}$ is a geodesic, and therefore it can be extended to an open interval that contains $(c,d]$. This contradicts the definition of $(c,d)$. One concludes that $(c,d)=\RR$, and that $(M,g)$ is geodesically complete. Let us now show that (c) implies (e). Fix a point $q \in M$. We need to show that there is a length minimizing geodesic from $p$ to $q$. Consider a normal neighborhood $U_0$ around $p$ so that $\exp(p): \tilde{U}_0 \to U_0$ is a diffeomorphism. Fix $\epsilon>0$ small enough so that $B(0,2\epsilon)\subseteq \tilde{U_0}$ and set $S_0= \exp(p) \left( S(0,\varepsilon)\right)$. Since $S_0$ is compact, there is a point $s_0 \in S_0$ such that $d(s_0,q)=d(S_0,q)$. There exists a unit vector $v \in T_pM$ such that $s_0= \exp(p)(\varepsilon v)$. We claim that if $r=d(p,q)$ then $q=\exp(p)(rv)$.
It suffices to show that the set
\[ B=\left\{ t\in [0,r]\mid d(\exp(p)(tv),q)=r-t\right\}\]
is the interval $[0,r]$. By definition, $0 \in B$. $B$ is closed because it is the place where two continuous functions coincide. Moreover, we claim that if $t \in B$ then $[0,t]\subseteq B$. If $t'<t$ then, by the triangle inequality applied to the points $p, q $ and $\exp(p)(t'v)$, one has
\[ d(\exp(p)(t'v),q)\geq d(p,q)-d(p, \exp(p)(t'v))\geq r-t'.\]
Also, by the triangle inequality applied to the points $ q, \exp(p)(tv)$ and $\exp(p)(t'v)$, one has
\[ d(\exp(p)(t'v),q)\leq d(\exp(p)(t'v), \exp(p)(tv))+ d(\exp(p)(tv),q)\leq t-t' +r-t=r-t'.\]
One concludes that $t' \in B$ and therefore $B=[0,t_1]$. It remains to show that $t_1 =r$. Suppose that $t_1<r$. Fix a normal neighborhood $U_1$ around $p_1= \exp(p)(t_1v)$ so that $\exp(p_1): \tilde{U}_1 \to U_1$ is a diffeomorphism. Fix $\delta>0$ small enough so that $B(0, 2 \delta) \subseteq \tilde{U}_1$ and $ q \notin \exp(p_1)\left(B(0, 2\delta)\right)$. As before, set
$ S_1=\exp(p_1)\left(S(0,\delta)\right)$. Since $S_1$ is compact, there exists $s_1 \in S_1 $ such that $d(s_1,q)=d(S_1,q)$.
Then, $d(s_1, q)=r-t_1-\delta$.
Applying the triangular inequality to the points $p,p_1 $ and $s_1$ one obtains $d(p,s_1)\leq t_1 + \delta$. 
On the other hand, applying the triangular inequality to the points $p, s_1$ and $q$ one gets
$\delta+ t_1 \leq d(p, s_1)$.
One concludes that $d(p,s_1)=\delta+t_1$. Lemma \ref{minimizer} implies that $s_1= \exp(p) (t_1v+\delta v)$. Therefore $t_1 + \delta \in B$. This contradicts the definition of $t_1$. One concludes that $B=[0,r]$.
Let us now prove that (c) implies (d). Let $C$ be a closed and bounded subset of $M$. Since (e) holds, there exists $R\gg0$ such that $C\subseteq \exp(p) \big(\overline{ B(0, R)}\big)$. Then, $C$ is a closed subset of a compact set and therefore, it is compact. It only remains to show that (d) implies (a). Let $\{p_n\}$ be a Cauchy sequence in $M$. Then, there exists some $K \gg 0$ such that $\{p_n\} $ is contained in $ \overline{B(p_1, K)}$, which is closed and bounded, and therefore compact. By compactness, there is a subsequence converging to a point $p \in M$. Since the original sequence was Cauchy, it converges to~$p$.
\end{proof}

An immediate consequence of the Hopf-Rinow theorem is the following.
\begin{corollary}
A compact Riemannian manifold is complete.
\end{corollary}

\section{The theorem of Ambrose}
\begin{theorem}(Ambrose)
Let $(M,g)$ and $(N,h)$ be connected Riemannian manifolds and $f: M \to N$ a local isometry. If $M$ is geodesically complete then $f$ is a covering map and $N$ is geodesically complete.
\end{theorem}
\begin{proof}
The proof will be divided in several claims.\\
\textbf{Claim 1:} Given a geodesic $\gamma:[a,b]\to N$ and a point $\tilde{p}\in M$ such that $f(\tilde{p})=p=\gamma(a)$, there exists a unique geodesic $\tilde{\gamma}(\tau)$ in $M$ such that $\gamma(\tau)=f \circ \tilde{\gamma}(\tau)$ and $\tilde{\gamma}(a)=\tilde{p}$. 
The uniqueness part holds because any two solutions are geodesics with the same initial conditions. Let us show the existence part. Since $f$ is a local diffeomorphism, $\tilde{\gamma}(\tau)$ is defined on an interval $[a, a+ \epsilon)$. Since $(M,g)$ is geodesically complete, this can be extended to the interval $[a,b]$. The geodesics $\gamma(\tau)$ and $f \circ \tilde{\gamma}(\tau)$ have the same initial conditions and therefore, they are equal.\\
 \textbf{Claim 2:} The map $f: M \to N$ is surjective. Since $f$ is a local diffeomorphism, the image of $f$ is open. Since $N$ is connected, it suffices to show that $f(M)$ is closed. Suppose this is not the case, and consider a point $q \in \overline{f(M)}\setminus f(M)$. Let $U$ be a normal neighborhood around $q$ in $N$. Since $ q \in \overline{f(M)}$, there exists $q' \in U \cap f(M)$ and a geodesic $\gamma(\tau)$ in $N$ from $q'$ to $q$. By the previous claim, there exists a lift $\tilde{\gamma}(\tau)$, and therefore $q = \gamma(b)=f (\tilde{\gamma}(b)) \in f(M)$. This contradiction implies that $f(M)$ is closed and, since $N$ is connected, $f$ is surjective. \\
\textbf{Claim 3:} $(N,h)$ is geodesically complete. Let $\gamma(\tau):[a,b]\to N$ be a geodesic. Since $f$ is surjective, there exists a lift
$\tilde{\gamma}(\tau):[a,b]\to M$. Because $(M,g)$ is geodesically complete, $\tilde{\gamma}(\tau)$ can be defined on $\RR$, and therefore, 
$ f \circ \tilde{\gamma}(\tau)$ is an extension of $\gamma(\tau)$ to all of $\RR$.\\
 \textbf{Claim 4:} Fix a point $q \in N$ and choose $\delta>0$ small enough so that $U=B(q,\delta)$ is a normal neighborhood. For each $p \in f^{-1}\{ q\}$ set $U_p=B(p, \delta)$. The map $f: U_p \to U$ is surjective. First, we need to show that $f(U_p) \subseteq U$. Take a point $p' \in U_p$, then, there exists a path $\eta$ of length $l<\delta$ from $p$ to $p'$. Therefore, $f \circ \eta(\tau)$ is a path 
 from $q$ to $f(p')$ which has length $l<\delta$. One concludes that $f(p') \in U$. 
 Let us show that it is surjective. Given $q' \in U$ there is a piecewise geodesic from $q$ to $q'$ which has length $l<\delta$. This can be lifted to a path $\tilde{\gamma}(\tau)$ such that $\gamma(a)=p$. Then $d(p,\tilde{\gamma}(b) )\leq l<\delta$, therefore $\tilde{\gamma}(b) \in U_p$ and $f( \tilde{\gamma}(b))=q'$. One concludes that $f: U_p \to U$ is surjective. \\
 \textbf{Claim 5:} Each $U_p$ is a normal neighborhood of $p$ and $f:U_p \to U$ is a diffeomorphism. By assumption, there exists an open set $W$ in $T_qM$ such that the map
 $\exp(q): W \to U$ is a diffeomorphism. Set $W_p= Df(p)^{-1}(W)$ and consider the commutative diagram  \begin{equation}\label{diagram}
  \xymatrix{
  W_p \ar[r]^{\exp(p)} \ar[d]_{Df(p)}& U_p \ar[d]^f\\
  W \ar[r]^{\exp(q)}&U.
 }\end{equation}
 Since the arrows on the left and the bottom are bijective, then $ \exp(p): W_p \to U_p$ is injective. Let us show that it is also surjective. Fix $p' \in U_p$ and set $q'=f(p')$. By the Hopf-Rinow theorem, there exists a minimal geodesic $\gamma:[0,1]\to M$ from $p$ to $p'$, which has length $l=d(p,p')<\delta$. Then,
 $f \circ \gamma(\tau)$ is a geodesic from $q$ to $q'$ which has legth $l<\delta$. Since $U$ is a normal neighborhood, $f \circ \gamma(\tau)= \exp(q)(\tau w)$, for some $w \in W$. This implies that $\gamma(\tau)= \exp(p)(\tau \tilde{w})$, where $\tilde{w} \in W_p $ is such that $Df(p)(\tilde{w})=w$. In particular, $p' =\gamma(1) =\exp(p)(\tilde{w} )\in \exp(p)(W_p)$.  One concludes that all arrows in diagram (\ref{diagram}) are diffeomorphisms.\\
 \textbf{Claim 6}: $ f^{-1} (U) = \bigcup_{p \in f^{-1}\{ q\}} U_p$. By the previous claim, the right hand side is contained in the left hand side. Let us prove the other contention. Take $x \in f^{-1}(U)$ and set $y =f(x) \in U$. Then, there is a geodesic $\gamma:[a,b]\to U$ from $y$ to $q$ which has length $l< \delta$. Let $\tilde{\gamma}$ be a lift of $\gamma$ such that $\tilde{\gamma}(a)=x$. Set $p=\tilde{\gamma}(b)$. Then $p \in f^{-1}\{q\}$ and the path $\tilde{\gamma}(\tau)$ from $p$ to $x$ has length less that $\delta$. One concludes that $x \in U_p$. \\
 \textbf{Claim 7} If $p \neq p'$ then $U_p \cap U_{p'}=\emptyset$.
 Suppose that $x \in U_p \cap U_{p'}$ and consider radial geodesics $\gamma_p$ from $x$ to $p$ and $\gamma_{p'}$ from 
 $x$ to $p'$. Then the geodesics $f \circ \gamma_p$ and $f\circ \gamma_{p'}$ are radial geodesics from $f(x)$ to $f(p)=f(p')=q$. Since $U$ is a normal neighborhood, they are equal. This implies that $\gamma_p= \gamma_{p'}$ and therefore $p=p'$.
\end{proof}
An immediate consequence of the Ambrose theorem is that complete Riemannian manifolds are maximal.
\begin{corollary}
Let $(M,g)$ and $(N,h)$ be connected Riemannian manifolds and assume that $(M,g)$ is complete. If $\iota: M \to N$ is an open embedding which is a local isometry then $\iota$ is an isometry.
\end{corollary}

\section{Constant curvature and Cartan-Hadamard}
Let $(M,g)$ be a Riemannian manifold and $\Pi \subseteq T_pM$ a two dimensional vector subspace of the tangent space at $p$. The sectional curvature $K$ of $(M,g)$, evaluated at $\Pi$, is the number:
\[ K(p)(\Pi)=\frac{ R(X,Y,Y,X)}{\langle X,X\rangle \langle Y,Y \rangle - \langle X,Y \rangle^2 }=\frac{\langle R(X,Y)(Y),X\rangle}{\langle X,X\rangle \langle Y,Y \rangle - \langle X,Y \rangle^2 },\]
where the vectors $X$ and $Y$ generate $\Pi$. Let us show that the right hand side depends only on the vector subspace $\Pi$. The Bianchi identities imply that the numerator is symmetric on $X$ and $Y$. One concludes that the whole expression also is. It is also clear that the number does not change if $X$ or $Y$ are multiplied by a nonzero scalar. Finally, the antisymmetry of the Riemann tensor implies that the right hand side does not change if $X$ is replaced by $X'=X+ \lambda Y$. The quantity $K$ is known as the sectional curvature of $(M,g)$. A Riemannian manifold is said to have constant curvature if $K(p)(\Pi)$ is independent of $p$ and $\Pi$.
One says that a Riemannian manifold $(M,g)$ is locally isotropic at $p \in M$ if for every pair of unitary tangent vectors $u,v \in T_pM$ there exist open subsets $U,V \subseteq M$ and an isometry $\varphi: U \rightarrow V$ such that $\varphi(p) =p$ and $D\varphi(p)(v)=w$. 
If $(M,g)$ is locally isotropic at every point $p\in M$, one says that it is locally isotropic.

\begin{lemma}\label{ConstantK}
Let $(M,g)$ be a  Riemannian manifold whose sectional curvature is a scalar function, $f(p)=K(p)(\Pi)$, and consider the tensor
\[ S(X,Y,Z) = \langle Y,Z\rangle X - \langle X,Z\rangle Y\]
Then, the Riemann tensor is given by
\[
R=fS.
\]
\end{lemma}
\begin{proof}
By hypothesis, we know that if $X$ and $Y$ are linearly independent, then
\begin{equation}\label{S}
\langle  R(X,Y)Y,X \rangle= f \langle S(X,Y,Y),X \rangle.
\end{equation}
On the other hand, if $X$ and $Y$ are linearly dependent, both sides of (\ref{S}) are zero, and we conclude that it holds for all $X$ and $Y$. Applying this equation to $X=U+V$ one obtains
\begin{align*}
&\langle  R(U+V,Y)Y,U+V \rangle\\
&\quad = f \left( \langle Y,Y\rangle \big( \langle U,U\rangle + \langle V,V \rangle+2\langle V,Z\rangle \big)-\langle Y,U\rangle^2-\langle Y,V\rangle^2-2 \langle Y,U \rangle \langle Y,V \rangle\right).
\end{align*}
On the other hand, using the symmetry of the Riemann tensor one computes
\begin{align*}
&\langle  R(U+V,Y)Y,U+V \rangle\\
&\quad=\langle R(U,Y)Y,U\rangle +\langle R(U,Y)Y,V\rangle+R(V,Y)Y,V\rangle +\langle R(V,Y)Y,U\rangle\\
&\quad=\langle R(U,Y)Y,U\rangle +2\langle R(U,Y)Y,V\rangle+R(V,Y)Y,V\rangle \\
&\quad=f \left( \langle S(U,Y,Y),U \rangle+  \langle S(V,Y,Y),V \rangle \right)+2\langle R(U,Y)Y,V\rangle\\
&\quad=f \left( \langle Y,Y\rangle\langle U,U\rangle-\langle Y,U\rangle^2  + \langle Y,Y\rangle\langle V,V\rangle-\langle Y,V\rangle^2 \right)+2\langle R(U,Y)Y,V\rangle.
\end{align*}
One concludes that
\begin{equation}
\langle R(U,Y)Y,V\rangle=f \left( \langle Y,Y \rangle \langle U,V \rangle-\langle Y,U\rangle\langle Y,V \rangle\right).
\end{equation}
Since $V$ is arbitrary, the above implies
\begin{equation}\label{SS}
 R(U,Y)Y=f S(U,Y,Y).
\end{equation}
Applying \eqref{SS} to $Y=X+V$ one obtains
 \begin{align*}
 &R(U,X+V)(X+V)\\
 &\quad =f \left(\left( \langle X,X \rangle +\langle V,V \rangle +2\langle X, V\rangle \right)U-\left(\langle U,V\rangle +\langle X,U \rangle\right)X -\big(\langle X,U\rangle +\langle U,V \rangle\big)V\right).
 \end{align*}
 On the other hand,
  \begin{align*}
 &R(U,X+V)(X+V) \\
 &\quad=R(U,V)V+R(U,V)X+R(U,X)X+R(U,X)V\\
&\quad=R(U,V)X+R(U,X)V+f\left(\langle V,V\rangle U-\langle U,V\rangle V +\langle X,X\rangle U-\langle U,X\rangle X \right).
\end{align*}
One concludes that
\begin{equation}\label{RZW}
R(U,V)X+R(U,X)V=f\left(2\langle X, V\rangle U- \langle U,V \rangle X-\langle X,U\rangle V\right)
\end{equation}
Substracting the Bianchi identity
\[ R(U,V)X+ R(X,U)V+ R(V,X)U=0,\]
one obtains
\begin{equation}
2R(U,X)V+R(X,V)U=f\left(2\langle X, V\rangle U- \langle U,V \rangle X-\langle X,U\rangle V\right).
\end{equation}
Exchanging $X$ and $U$ this becomes
\begin{equation}\label{XZ}
2R(X,U)V+R(U,V)X=f\left(2\langle U, V\rangle X- \langle V,X \rangle U-\langle X,U\rangle V\right).
\end{equation}
Substracting \eqref{RZW} from \eqref{XZ} one obtains
\begin{equation}
3R(X,U)V=3f\left(\langle U, V\rangle X- \langle X,V\rangle U\right),
\end{equation}
from which the desired result follows.
\end{proof}

\begin{theorem}(Schur's Lemma)\label{SchurLemma}
\label{schur}Let $(M,g)$ be a connected Riemannian manifold of dimension $\geq3$. If there exists a function
$f: M \rightarrow \RR$ such that $f(p)=K(p)(\Pi)$, for all $\Pi \in T_pM$, then $f$ is constant.
\end{theorem}
\begin{proof}
By Lemma \ref{ConstantK} we know that
\begin{equation}\label{rfs}R=f S.\end{equation}
In order to prove that the tensor $S$ is covariantly constant we compute
\begin{align*} &\nabla_{W}S(X,Y,Z) \\
&\quad=\nabla_W(S(X,Y,Z))-S(\nabla_W X,Y,Z)-S(X,\nabla_WY,Z)-S(X,Y,\nabla_WZ)\\
&\quad =\nabla_W(\langle Y,Z \rangle X)-\nabla_W(\langle X,Z \rangle Y)-S(\nabla_WX,Y,Z)-S(X,\nabla_WY,Z)-S(X,Y,\nabla_WZ)\\
&\quad =W\langle Y,Z \rangle X+\langle Y,Z \rangle  \nabla_W X-\nabla_W\langle X,Z \rangle  Y-\langle X,Z \rangle \nabla_WY-\langle Z,Y \rangle \nabla_W X\\
&\quad  \quad \, + \langle Z,\nabla_W X \rangle Y -\langle Z,\nabla_W Y\rangle X
 +\langle Z,X \rangle \nabla_WY- \langle Y, \nabla_W Z\rangle X+ \langle X,\nabla_W Z \rangle Y \\
 &\quad =0.
\end{align*}
Using the second Bianchi identity and (\ref{rfs}) one computes
\begin{align*}
0& =\nabla_XR(Y,Z)W+\nabla_ZR(X,Y)W+\nabla_Y R(Z,X)W\\
&=\nabla_X(fS)(Y,Z,W)+\nabla_Z(fS)(X,Y,W)+\nabla_Y(fS)(Z,X,W)\\
&=(Xf) S(Y,Z,W)+(Zf) S(X,Y,W)+(Yf)  S(Z,X,W).
\end{align*}
Since the dimension of $M$ is at least three, we can choose $X,Y,Z$ to be mutually orthogonal, in particular, linearly independent. Moreover, if we set $W=X$, then 
\begin{align*}
0&=(Xf) S(Y,Z,X)+(Zf)S(X,Y,X)+(Yf)S(Z,X,X).\\
&=-(Zf)\langle X,X\rangle Y+ (Yf) \langle X,X\rangle Z.
\end{align*}
Since $Z$ and $X$ are linearly independent, one concludes that $Yf=0$ and therefore $f$ is locally constant.
\end{proof}

Note that the condition that $d\geq 3$ is necessary. In dimension $d=2$ the statement is false since the Gaussian curvature is typically not constant.

\begin{lemma}\label{Wu}
Let $(M,g)$ be a Riemannian manifold of constant curvature $K$ and $\gamma(\tau):[0,b] \to M$ a geodesic of velocity $l= \sqrt{\langle \gamma'(\tau),\gamma'(\tau)\rangle}$. If
$J(\tau)$ is a Jacobi field along $\gamma(\tau)$ such that $J(0)=0$ and $\langle J(\tau), \gamma'(\tau)\rangle=0$, then, there exists a parallel vector field $E(\tau)$, orthogonal to $\gamma'(\tau)$, such that 
\[ J(\tau)=u(\tau)E(\tau),\]
where
\begin{equation}
u(\tau)=\begin{cases}
l\tau, & \text{ if } K=0,\\
\displaystyle R\sin (l\tau/R),& \text{ if }K=1/R^2>0,\\
R\sinh(l\tau/R),& \text{ if }K=-1/R^2<0.
\end{cases} 
\end{equation}
\end{lemma}
\begin{proof}
Clearly, if $J(\tau)=u(\tau) E(\tau)$ then $J(0)=0$ and $J(\tau)$ is orthogonal to $\gamma'(\tau)$. Let us show that $J(\tau)$ is a Jacobi field. For this we compute
\begin{align*}
\nabla_{\gamma'(\tau)}\big(\nabla_{\gamma'(\tau)} J(\tau)\big)
=\frac{d^2u(\tau)}{d\tau^2 } E(\tau)=-Kl^2u(\tau)E(\tau).
\end{align*}
On the other hand, Lemma \ref{ConstantK} implies that
\begin{align*}
R(\gamma'(\tau),J(\tau))\gamma'(\tau)&=K S(\gamma'(\tau),J(\tau),\gamma'(\tau)) \\
&=-K\langle \gamma'(\tau),\gamma'(\tau)\rangle J(\tau) \\
&=-Kl^2u(\tau)E(\tau).
\end{align*}
One concludes that $J(\tau)=u(\tau)E(\tau)$ is a Jacobi field. The vector space of parallel vector fields that are orthogonal to $\gamma'(\tau)$ has dimension $m-1$. Also, the space of Jacobi fields that vanish at $a$ and are orthogonal to $\gamma'(\tau)$, has dimension $m-1$. The result follows.
\end{proof}

\begin{lemma}\label{nf}
Let $(M,g)$ be a Riemannian manifold of constant curvature $K$, $p$ a point in $M$, and $U=\exp(p)(\tilde{U})$, a normal neighborhood.
For any $q=\exp(p)(v) \in U\setminus\{p\}$, there is a decomposition $T_q M= T^{\perp}_q M \oplus T^{\mathrm{r}}_qM$ into orthogonal and radial directions. The metric satisfies
\begin{equation}
\langle  V,V\rangle=\begin{cases}
\langle \!\langle   V^\perp ,V^\perp \rangle\! \rangle+\langle \!\langle  V^{\mathrm{r}},V^{\mathrm{r}}  \rangle\! \rangle,& \text{ if } K=0,\\
(R^2/l^2) \sin^2(l/R)\langle \!\langle V^\perp,V^\perp \rangle\! \rangle+\langle \!\langle V^{\mathrm{r}},V^{\mathrm{r}} \rangle\! \rangle,& \text{ if }K=1/R^2>0,\\
(R^2/l^2) \sinh^2(l/R)\langle \!\langle V^\perp,V^\perp \rangle\! \rangle+\langle \!\langle V^{\mathrm{r}},V^{\mathrm{r}} \rangle\! \rangle,& \text{ if }K=-1/R^2<0,
\end{cases} 
\end{equation}
where $\langle \!\langle \cdot,\cdot \rangle\! \rangle$ denotes the constant metric induced by the exponential map and $l=|v|$.
In particular, if $(N,h)$ is another manifold of the same dimension with constant curvature $K$ and $q \in N$, then, there is a local isometry 
sending $p$ to $q$.
\end{lemma}

\begin{proof}
We know that $q=\exp(p)(v)= \gamma(1)$, where $\gamma(\tau)=\exp(\tau v)$ is a geodesic of length $l=|v|$.
By the Lemma \ref{Wu} we know that
\[ D \exp(p)(v)(w)=W(1),\]
where $W(\tau)=u(\tau)E(\tau)$ and $E(\tau)$ is a parallel vector field such that $E(0)=w/l$. By the Gauss lemma we know that
\begin{align*}
V^{\perp}&=D\exp(p)(v)(w), \\ V^{\mathrm{r}}&=  D\exp(p)(v)(\lambda v)
\end{align*}
where $\langle v,w\rangle=0$. Then, since the exponential map is a radial isometry, we know that
\[ \langle V^{\mathrm{r}},V^{\mathrm{r}}\rangle =\langle D \exp(v)(\lambda v), D\exp(p) (\lambda v)\rangle=\langle\! \langle V^{\mathrm{r}}, V^{\mathrm{r}}\rangle\! \rangle .\]
On the other hand,
\begin{align*} \langle V^\perp,V^\perp\rangle &=\langle D \exp(v)(w), D\exp(p) (w)\rangle \\
&=\langle W(1),W(1) \rangle  \\
&=u(1)^2\langle E(1) ,E(1) \rangle \\
&=u(1)^2\langle E(0) ,E(0) \rangle \\
&=\frac{u(1)^2}{l^2}\langle w,w\rangle \\
&=\begin{cases}
\langle \!\langle   V^\perp ,V^\perp \rangle\! \rangle,& \text{ if } K=0,\\
(R^2/l^2) \sin^2(l/R)\langle \!\langle V^\perp,V^\perp \rangle\! \rangle,& \text{ if }K=1/R^2>0,\\
(R^2/l^2) \sinh^2(l/R)\langle \!\langle V^\perp,V^\perp \rangle\! \rangle,& \text{ if }K=-1/R^2<0,
\end{cases} \end{align*}
This completes the proof.
\end{proof}

The previous lemma shows that manifolds of constant curvature admit a normal form around each point, and therefore, are locally unique for a given value of $K$. 
The Killing-Hopf theorem provides a global version of this result. In order to have global uniqueness results it is necessary to impose a maximality condition. This is the condition of being geodesically complete.

\begin{lemma}\label{rigidity}
Let $(M,g)$ and $(N,h)$ be connected Rimannian manifolds and let $\varphi,\psi: M \to N$ be isometries.
If $\varphi(p) =\psi(p)$ and $D\varphi(p)=D\psi(p)$ then $\varphi=\psi$.
\end{lemma}
\begin{proof}
Set
\[ X=\{ x \in X \mid \varphi(x)=\psi(x), D\varphi(x)=D\psi(x)\}.\]
Fix a point $q \in M$ and a path $\alpha: [0,1]\to M$ from $p$ to $q$. Define
\[ A=\{ t \in [0,1]\mid \alpha(t) \in X\},\]
and $B=[0,1]\setminus A$. We need to prove that $B=\emptyset$. Suppose the contrary. Then $B$ is nonempy and bounded below,
and therefore, it has an infimum $t_0$. We claim that $t_0>0$. For this, fix a normal neighborhood $U$ of $p$. Given $x$ in $U$ there is a geodesic $\gamma(\tau)$ from $p$ to $x$. Then, the geodesics $\varphi \circ \gamma(\tau)$ and $\psi \circ \gamma(\tau)$ have the same initial conditions, and therefore, they are equal. This shows that $U \subseteq X$ and therefore $V= \alpha^{-1}(U) \subseteq A$ is an open set that contains zero, which implies $t_0>0$. Define $x_0 = \alpha(t_0)$. By Lemma \ref{convex} there exists an open neighborhood $W$ of $x_0$ such that every point in $V$  has a normal neighborhood that contains $x_0$. Then, there exists $0<t_1<t_0$ such that $x_1=\alpha(x_1) \in W$. This implies that there is a normal neighborhood $W_1$ of $x_1$ that contains $x_0$. Given any point $y \in W_1$ there exists a geodesic $\lambda(\tau)$ from $x_1$ to $y$ and therefore $\varphi \circ \lambda(\tau)$ and $\psi \circ \lambda(\tau)$ are equal, in particular
$\varphi(y)=\psi(y)$. This implies that $W_1\subseteq X$, and threfore, $\alpha^{-1}(W_1)$ is an open neighborhood of $t_0$ which is contained in $A$. This contradicts the assumption that $t_0$ is the infimum of $B=[0,1]\setminus A$. One concludes that $A=[0,1]$, and therefore, $X=M$.
\end{proof}

Two points $p$ and $q$ on a Riemannian manifold $M$ are conjugate points if there exists a geodesic $\gamma(\tau)$ form $p$ to $q$ and a nonzero Jacobi field $J(\tau)$ along $\gamma(\tau)$ that vanishes at $p$ and $q$.

\begin{lemma}\label{JV}
Let $J(\tau)$ be a Jacobi field along $\gamma(\tau)$ that vanishes at two different places. Then $J(\tau)$ is orthogonal to $\gamma'(\tau)$.
\end{lemma}

\begin{proof}
Suppose that $J(0)=J(b)=0$. Consider the vector $v=\nabla_{\gamma'(\tau)}J(0)$ which can be written as $v=u+w$ where $u$ is tangent to $\gamma(\tau)$ and $w$ is orthogonal to $\gamma(\tau).$
We denote by $J_u $ and $J_w$ the Jacobi fields with initial conditions
\begin{align*} J_u(0)&=0,\quad \, \nabla_{\gamma'(\tau)}J_u(0)=u,\\
J_w(0)&=0,\quad  \nabla_{\gamma'(\tau)}J_w(0)=w.
\end{align*}
Then $J=J_u + J_w$. Moreover, $J_u =\lambda \tau \gamma'(\tau)$
where $\lambda$ is such that $v=\lambda \gamma'(0)$.
On the other hand, we compute
\[ \langle J_w(\tau),\gamma'(\tau) \rangle'=\langle \nabla_{\gamma'(\tau)} J_w(\tau),\gamma'(\tau) \rangle,\]
so that
\[ \langle J_w(\tau),\gamma'(\tau) \rangle''=\langle\nabla_{\gamma'(\tau)}\big( \nabla_{\gamma'(\tau)} J_w(\tau)\big),\gamma'(\tau) \rangle=\langle R(\gamma'(\tau),J_w(\tau))\gamma'(\tau),\gamma'(\tau) \rangle=0.\]
Therefore
\[ \langle J_w(\tau),\gamma'(\tau) \rangle'=\langle \nabla_{\gamma'(\tau)} J_w(0),\gamma'(0) \rangle=\langle w, \gamma'(0)\rangle =0.\]
This implies that
\begin{equation}\label{jort} \langle J_w(\tau),\gamma'(\tau) \rangle=\langle J_w(0),\gamma'(0) \rangle=0.\end{equation}
Since
\[ 0=J(b)=J_u(b)+J_w(b)=\lambda b \gamma'(b)+ J_w(b),\]
Equation \eqref{jort} implies that $\lambda=0$ and therefore $J(\tau)=J_w(\tau)$ is orthogonal to $\gamma(\tau)$.
\end{proof}

\begin{lemma}\label{critical}
 The point $q$ is conjugate to $p$ if and only if it is a critical value of $\exp(p)$. That is, there exists $v \in T_pM$ such that
 $\exp(p)(v)=q$ and $D \exp(p)(v)$ is singular.
\end{lemma}
\begin{proof}
If $q$ is conjugate to $p$, there exists a geodesic $\gamma:[0,1]\to M$ from $p$ to $q$ and a nonzero Jacobi field $J(\tau)$ that vanishes at $0$ and $1$. If one sets $v=\gamma'(0)$, then
$\exp(p)(v)=q$. Moreover, if we set $w= \nabla_{\gamma'(\tau)}J(0)$ then, $w\neq 0$, and one has
\[ D\exp(p)(v)(w)=J(1)=0.\]
One concludes that $q$ is a critical value of $\exp(p)$. Conversely, let $q=\exp(p)(v)$ be critical value of $\exp(p)$ and $w$ a nonzero
vector in the kernel of $D\exp(p)(v)$. Then, the Jacobi field
along $\exp(p)(\tau v)$ with initial conditions $J(0)=0$ and $\nabla_{\gamma'(\tau)}J(0)=w$ vanishes on $q$.
\end{proof}

\begin{lemma}\label{nconj}
Let $(M,g)$ be a Riemannian manifold of nonpositive curvature. There are no conjugate points in $M$.
\end{lemma}
\begin{proof}
Suppose that $p$ and $q$ are conjugate with geodesic $\gamma(\tau):[a,b]\to M$ and Jacobi field $J(\tau)$. By Lemma \ref{JV} we know that 
$J(\tau)$ is orthogonal to $\gamma(\tau)$.
 Consider the function
\[ f(\tau)=\langle J(\tau),J(\tau)\rangle.\]
Then
\[ f'(\tau)=2 \langle \nabla_{\gamma'(\tau)}J(\tau),J(\tau)\rangle ,\]
and
\begin{align*} f''(\tau)&=2 \langle\nabla_{\gamma'(\tau)}\big( \nabla_{ \gamma'(\tau)}J(\tau)\big),J(\tau)\rangle +2 \langle \nabla_{ \gamma'(\tau)}J(\tau),\nabla_{ \gamma'(\tau)}J(\tau)\rangle\\
&=-2 \langle R(J(\tau),\gamma'(\tau))\gamma'(\tau),J(\tau)\rangle+2 \langle \nabla_{ \gamma'(\tau)}J(\tau),\nabla_{ \gamma'(\tau)}J(\tau)\rangle \geq 0.
\end{align*}
Since $f'(0)=0$, one concludes that $f'(\tau)\geq0$. Since $J(\tau)$ is nonzero, there is a $c \in (a,b)$ where $f(c)>0$. Since $f(b)=0$, the mean value theorem implies that there is some $d$ where $f'(d)<0$. This contradiction implies that there are no conjugate points.
\end{proof}

\begin{theorem}(Cartan-Hadamard)
Let $(M,g)$ be a complete Riemannian manifold of non-positive curvature. For each point $p \in M$ the exponential map
$\exp(p): T_pM \to M$ is a covering map. In particular, if $M$ is simply connected, the exponential map is a diffeomorphism.
\end{theorem}
\begin{proof}
By lemma \ref{nconj}, there are no conjugate points in $M$. Therefore, Lemma \ref{critical} implies that the exponential map $\exp(p):T_pM \to M$ is a locall diffeomorphism. Consider the metric $h$ in $T_pM$ defined by $h= \exp(p)^*(g)$. By construction, the map
$\exp(p): T_pM \to M$ is a local isometry from $(T_pM,h)$ to $(M,g)$. We claim that $(T_pM,h)$ is geodesically complete. By the Hopf-Rinow theorem, it suffices to show that the exponential map is defined on $T_0(T_pM)$. This is true because geodesics starting at zero
are straight lines with respect to the linear structure. One concludes that $(T_pM,h)$ is complete and then, the Ambrose theorem implies that $\exp(p)$ is a covering map.
\end{proof}

\section{Constant curvature and Killing-Hopf}

The Killing-Hopf theorem states that complete simply connected Riemannian manifolds of constant curvature $K$
are determined by the value of $K$. Note that if $(M,g)$ has constant curvature $K$ and $ \lambda>0$, then the manifold
$(M,\lambda g)$ has constant curvature $K/\lambda$. Therefore, it is enough to consider the cases where $K \in \{-1,0,1\}$.
The three possibilities are Euclidean spaces, spheres and hyperbolic spaces. The hyperboloid model for hyperbolic space 
is defined as follows. Consider the vector space $\RR^{n+1}$ with the Minkowski bilinear form $h$ represented by the matrix $\mathrm{diag}(-1,1,\dots,1)$.
Hyperboic space is the submanifold
\[ \HH^n=\{ v =(v^0,\dots, v^n )\in \RR^{n+1}\mid h(v,v)=-1 \text{ and } v^0>0\},\]
with the induced metric. Given a curve $\gamma(\tau)$ in hyperbolic space, one has
\[ 0 =\frac{d}{d\tau}\bigg\vert_{\tau=0} h(\gamma(\tau),\gamma(\tau))=2 h( \gamma'(0),\gamma(0)).\]
One concludes that the tangent space at $v=\gamma(0)$ is $ T_v\HH=v^{\perp}$,  where $v^\perp$ denotes the orthogonal complement of $v$ with respect to $h$. Since $h(v,v)=-1<0$, $v^{\perp}$ is a spacelike subspace.
One concludes that the restriction of $h$ to hyperbolic space is a Riemannian metric. In order to prove that hyperbolic space has constant curvature $C=-1$, it will be convenient to discuss totally geodesic submanifolds. A submanifold $\iota: N \hookrightarrow M$ of a Riemannian manifold $(M,g)$ is called totally geodesic if any geodesic in $N$ is also a geodesic in $M$. The condition of being totally geodesic can be expressed in terms of the relationship between the Levi-Civita connection of $N$ and that of $M$.

\begin{lemma}
Let $\iota: N \hookrightarrow M$ be a submanifold of a Riemannian manifold $(M,g)$. The vector bundle $\iota^*TM$ decomposes as a direct sum
\[ \iota^*TM= TN \oplus TN^{\perp},\]
and we denote by $\pi_1$ and $\pi_2$ the two projections. We denote by $\nabla^*$ the pullback of the Levi-Civita connection
 $\nabla $ on $M$ to $\iota^*TM$ and by $\nabla^N$ the Levi-Civita connection of $N$. Then
\[ \nabla^N=\pi_1 \circ \nabla^*.\]
\end{lemma}
\begin{proof}
Let us first show that $\pi_1 \circ \nabla^*$ defines a connection on $TN$. The expression
$ \pi_1 ( \nabla^*_X Y)$ is linear with respect to functions on the variable $Y$. Also,
\[ \pi_1 ( \nabla^*_X(fY))= \pi_1 (f \nabla^*_X Y)+ \pi_1((Xf) Y)=f \pi_1 ( \nabla^*_X Y)+ (Xf) Y.\]
One concludes that $\pi_1 \circ \nabla^*$ is a connection on $TN$. In order to show that this is the Levi-Civita connection, it is
enough to show that it is torsion free and metric preserving. For the torsion free part, we compute
\[ \pi_1 ( \nabla^*_X Y)-\pi_1 ( \nabla^*_Y X)-[X,Y]=\pi_1  \left( \nabla^*_X Y-  \nabla^*_Y X-[X,Y]\right) =0.\]
And for the metric preserving part,
\[ X \langle Y,Z \rangle -\langle \pi_1 (\nabla^*_X Y),Z \rangle -\langle Y, \pi_1(\nabla^*_X Z)\rangle= X \langle Y,Z \rangle -\langle \nabla^*_X Y,Z \rangle -\langle Y, \nabla^*_X Z\rangle =0. \]
This completes the proof.
\end{proof}

The second fundamental form of a submanifold $\iota: N \hookrightarrow M$ is the tensor $\mathrm{II} \in \Gamma(T^*N\otimes T^*N \otimes TN^{\perp})$ defined by
\begin{equation}
\mathrm{II}(X,Y)= \pi_2( \nabla^*_X Y)= \nabla^*_X Y-\nabla^N_X Y.
\end{equation}
It is easy to check that $\mathrm{II}$ is symmetric on $X$ and $Y$.

\begin{lemma}\label{tgg}
Let $\iota: N \hookrightarrow M$ be a submanifold of $(M,g)$. Then:
\begin{enumerate}
\item[1.] $N$ is totally geodesic if and only if $\mathrm{II}=0$.

\item[2.] If $N$ is totally geodesic, then the curvature tensor of $N$ is the restriction of the curvature tensor of $M$.
\end{enumerate}
\end{lemma}
\begin{proof}
Suppose that $\mathrm{II}=0$ and $\gamma(\tau)$ is a geodesic on $N$. Then
\[ 0 =\nabla^{N}_{\gamma'(\tau)}\gamma'(\tau)=\nabla^*_{\gamma'(\tau)}\gamma'(\tau)=\nabla_{\gamma'(\tau)}\gamma'(\tau).\]
One concludes that $\gamma(\tau)$ is a geodesic in $M$. Conversely, suppose that $N$ is totally geodesic. Fix a point $p \in N$ and a tangent vector $X \in T_pN$. Consider a geodesic $\gamma(\tau)$ in $N$ such that $\gamma'(0)=X$. Then
\[ 0 = \nabla^N_{\gamma'(\tau)}\gamma'(\tau)=\nabla^*_{\gamma'(\tau)}\gamma'(\tau)-\mathrm{II}(\gamma'(\tau),\gamma'(\tau))=-\mathrm{II}(\gamma'(\tau),\gamma'(\tau)).\]
Evaluating at $\tau=0$ one concludes that $\mathrm{II}(X,X)=0$. Since $X$ is arbitrary and $\mathrm{II}$ is symmetric, one concludes that $\mathrm{II}=0$. This proves (1). For (2) we compute
\begin{align*}
\langle R^M(X,Y)Z,W\rangle&= \langle \nabla^*_X \nabla^*_Y Z- \nabla^*_Y \nabla^*_X Z- \nabla^*_{[X,Y]}Z ,W\rangle \\
&= \langle \nabla^N_X \nabla^N_Y Z- \nabla^N_Y \nabla^N_X Z- \nabla^N_{[X,Y]}Z ,W\rangle,
\end{align*}
which shows that the result holds.
\end{proof}

An involution on a Riemannian manifold $(M,g)$ is an isometry $\varphi: M \to M$ such that $\varphi^2=\mathrm{id}_M$.
\begin{lemma}\label{tg}
If a submanifold $\iota: N \hookrightarrow M$ is the set of fixed points of an involution $\varphi \colon M \rightarrow M$, then $N$ is totally geodesic.
\end{lemma}
\begin{proof}
Fix a point $p \in N$. We claim that the derivative of $\varphi \colon M \rightarrow M$ at $p$ takes the form
\begin{equation}\label{invo} D\varphi(p)=\begin{pmatrix}
\mathrm{id}_{T_pN} &0\\
0 & -\mathrm{id}_{T_pN^{\perp}}
\end{pmatrix}\end{equation}
with respect to the decomposition $T_pM=T_pN \oplus T_pN^{\perp}$. Clearly, $D\varphi(p)\big \vert _{T_pN}=\mathrm{id}_{T_pN}$. Let us show that
$D\varphi(p)(T_pN^{\perp})\subseteq T_pN^{\perp}$. Take $X \in T_pN$ and $w \in T_pN^\perp$. Then, we have
\[\langle D\varphi(p)(w), v \rangle=\langle D\varphi (p)D\varphi(p)(w), D\varphi(p)v \rangle=\langle w, v \rangle=0.\]
One concludes that 
\[ D\varphi(p)=\begin{pmatrix}
\mathrm{id}_{T_pN} &0\\
0 & A
\end{pmatrix}\]
The matrix $A$ satisfies $A^2-1=0$
and therefore it diagonalizes with eigenvalues $\pm1$. It remains to show that there is no $w \in T_pN^\perp$ such that $D\varphi(p)(w)=w$. Suppose this were the case. Consider the geodesic, $\gamma(\tau): [a,b]\to M$, such that $\gamma'(0)=w$.  Then
\[ \varphi (\gamma(\tau))=\varphi (\exp(p)(\tau w))=\exp(p)(d\varphi(p)(\tau w))=\exp(p)(\tau w)=\gamma(\tau).\]
This implies that $\gamma(\tau)$ is a curve in $N$ and therefore $w \in T_pN$, which is a contradiction. One concludes that \eqref{invo} holds. Let us now prove that $\mathrm{II}=0$. Using that $\varphi$ is an isometry we compute
\[ D\varphi (p) (\mathrm{II}(X,Y))= D\varphi (p)( \pi_2 (\nabla^*_X Y))=\pi_2(\nabla^*_{D\varphi X}D\varphi (Y))= \pi_2 (\nabla^*_X Y)=\mathrm{II}(X,Y).\]
On the other hand, in view of \eqref{invo},
\[ D\varphi(p)(\mathrm{II}(X,Y))=-\mathrm{II}(X,Y).\]
This implies that $\mathrm{II}=0$.
\end{proof}

\begin{proposition}
The following holds for all $n\geq 2$:
\begin{enumerate}
\item[1.] Euclidean space $\RR^n$ is a complete simply connected Riemannian manifold of constant curvature $K=0$.
\item[2.] The sphere $S^n$ is a complete simply connected Riemannian manifold of constant curvature $K=1$.
\item[3.] Hyperbolic space $\HH^n$ is a complete simply connected Riemannian manifold of constant curvature $K=-1$,
\end{enumerate}
\end{proposition}
\begin{proof}
The first statement is clear. Let us consider (2). Since spheres are compact, by the Hopf-Rinow theorem, they are geodesically complete. 
They are also simply connected. It remains to show the statement about curvature. The group $\mathrm{SO}(n+1)$ acts transitively by isometries on $S^n$. Moreover, given $p \in S^n$ and  two dimensional subspaces $\Pi , \Pi' \subseteq T_p S^n$ there exists an element $A \in \mathrm{SO}(n+1)$ such that $A(p)=p$ and $DA(p)(\Pi)=\Pi'$. This symmetry implies that the sphere $S^n$ has constant curvature $C_n$. We will argue by induction that $C_n=1$. For $n=2$, this is an explicit computation that we will omit. Consider the involution $\varphi: S^n \to S^n$ given by
$ \varphi(x^0,\dots , x^n)=(x^0,\dots ,-x^n)$. The sphere $S^{n-1}$ is the set of fixed points of $\varphi$, and therefore, by Lemma \ref{tg} it is totally geodesic. By induction hypothesis $C_{n-1}=1$. Lemma \ref{tgg} implies that $C_n=C_{n-1}=1$. Let us prove (3). The projection $\pi: \HH^n \to \RR^n$ given by $\pi(v^0,\dots, v^n)=(v^1,\dots, v^n)$ is a diffeomorphism. In particular, $\HH^n$ is simply connected. The group $\mathrm{O}(1,n)^\uparrow $, of symmetries of $(\RR^{n+1},h)$ that preserve the positive cone, acts transitively by isometries on $\HH^n$. Moreover, given $p \in \HH^n$ and two dimensional subspaces $\Pi , \Pi' \subseteq T_p \HH^n$ there exists an element $A \in \mathrm{O}(1,n)^\uparrow$ such that $A(p)=p$ and $DA(p)(\Pi)=\Pi'$. This symmetry implies that hyperbolic space $\HH^n$ has constant curvature $C_n$. Again, we argue by induction to show that $C_n=-1$. A direct computation proves the case $n=2$. Since $\HH^{n-1}$ is the set of fixed points of the involution $\varphi(v^0,\dots, v^n)=(v_0,\dots,-v^n)$, it is totally geodesic. Therefore $C_n=C_{n-1}=-1$. By the symmetry of the situation and the fact that $\HH^2$ is totally geodesic in $\HH^n$, it is enough to exhibit a geodesic in $\HH^2$ defined on $\RR$. Consider the involution $\varphi(v^0,v^1,v^2)=(v^0,v^1,-v^2)$. The fixed points of this involution is an embedded line $L$ in $\HH^2$, which can be parametrized by $\gamma(\tau)= (\sinh(\tau),\cosh(\tau),0)$. Since $L$ is totally geodesic, it is the image of a geodesic. It only remains to show that $\gamma'(\tau)$ has constant velocity. For this we compute
\[ \langle \gamma'(\tau),\gamma'(\tau)\rangle= -\sinh^2(\tau)^2+\cosh^2(\tau)=1.\]
One concludes that $\gamma(\tau)$ is a geodesic defined on $\RR$, and that $\HH^n$ is geodesically complete.
\end{proof}

The following figure illustrates geodesics in $\HH^2$.
\begin{figure}[H]
\centering
\includegraphics[scale=0.38
]{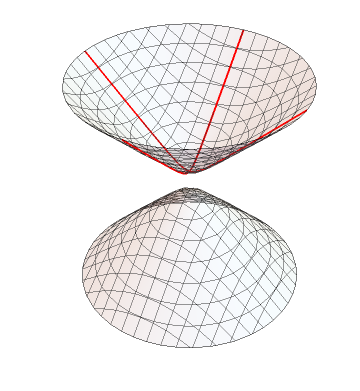}\caption{Geodesics in two dimensional hyperbolic space.}
\end{figure}

We will use the following lemma in the proof of the Killing-Hopf theorem.
\begin{lemma}\label{isom}
Let $(M,g)$ and $(N,h)$ be manifolds of constant curvature $C$, $U$ a normal neighborhood of $p \in M$  and $f: T_pM \to T_qN$ an
isometry. If $\exp(q)$ is a local diffeomorphism on $f (\exp(p)^{-1}(U))$, then the map
\[ \varphi=\exp(q)\circ f \circ \exp(p)^{-1}:U \to N\]
is a local isometry.
\end{lemma}
\begin{proof}
This is a consequence of Lemma \ref{nf}.
\end{proof}

\begin{theorem}(Killing-Hopf)\label{KH}
Let $(M,g)$ be a geodesically complete simply connected Riemannian manifold of constant curvature $K$.
\begin{enumerate}
\item[1.] If $K=0$ then $(M,g)$ is isometric to the Euclidean space $\RR^n$.
\item[2.] If $K=1$ then $(M,g)$ is isometric to the sphere $S^n$.
\item[3.]  If $K=-1$ then $(M,g)$ is isometric to the hyperbolic space $\HH^n$.
\end{enumerate}
\end{theorem}
\begin{proof}
Let us first consider the case where $K< 0$. Fix $p \in \HH^n$, $q \in M$ and an isometry $f: T_p \HH^n \to T_q M$.
By the Cartan-Hadamard theorem, the exponential maps at $p$ and $q$ are global diffeomorphisms. Therefore, the map
$\varphi=\exp(q) \circ f \circ \exp^{-1}(p): \HH^n \to M$ 
is a diffeomorphism. Moreover, by Lemma \ref{isom}, $\varphi$ is an isometry.
Exactly the same proof works in the case $K=0$. We are left with the case $K=1$. Let $p$ be the north pole $p=(0, \dots 0,1)$ and $U= S^n\setminus\{-p\}$ be the complement of the south pole. Since geodesics are maximal circles, $U$ is a normal neighborhood around $p$.
Fix a point $q \in M$ and linear isometry $f: T_pS^n \to T_q M$. By Lemma \ref{isom} the map:
$\varphi : \exp(q) \circ f \circ \exp(p)^{-1}:U \to M$ is a local isometry around $p$. Fix another point $x \in S^m$ which is neither of the poles and  $v=D\varphi(x): T_x S^n \to T_{y}M$, where $y=\varphi(x)$. By the same argument, the map
$ \psi =\exp(y) \circ v \circ \exp(x)^{-1}:W \to M$,
where $W=S^{n}\setminus\{-x\}$, is a local isometry. Moreover,
$ \psi(x)=y=\varphi(x)$ and
 \begin{align*}
D\psi(x)&=D(\exp(y))(0)\circ Dv(0)\circ D(\exp(x)^{-1})(x)\\
&=D(\exp(y))(0)\circ D\varphi(x)\circ D(\exp(x)^{-1})(x) \\
&=D\varphi(x).
\end{align*}
By Lemma \ref{rigidity}, the functions $\psi$ and $\varphi$ coincide in the intersection of their domains. Therefore, together they define a function $\xi: S^n \to M$.
Moreover, since $x$ is arbitrary, $\xi$ is a local isometry and therefore an open map. Since $S^n$ is compact and $M$ is Hausdorff, the image of $\xi$ is closed.
Since $M$ is connected we conclude that $\xi$ is surjective and $M$ is compact. Since $M$ is simply connected, the Ambrose theorem implies that $\xi$ is an isometry.
\end{proof}
  \clearemptydoublepage
   
\titleformat{\chapter}[display]
     {\sc  \huge}
     {} 
     {0pt} {\titlerule[1pt] \vspace{1ex}%
  }[ \vspace{1ex}{\titlerule[1pt]}] 
\begin{small}

\addcontentsline{toc}{chapter}{Bibliography}
\printindex
\end{small}

\end{document}